\documentclass[12pt]{amsbook}

\usepackage{hyperref}

\usepackage[polutonikogreek, english]{babel}
\usepackage[babel=true]{csquotes}

\usepackage[a4paper]{geometry}
\usepackage{setspace}

\usepackage{lscape, enumerate, fancybox, fullpage,color}
\usepackage{cases}
\usepackage{tikz}
\usepackage{frcursive}

\usetikzlibrary{shapes}
\usetikzlibrary{positioning,arrows}

\DeclareMathAlphabet{\mathpzc}{OT1}{pzc}{m}{it}

\usepackage{bm, dsfont, amssymb, graphicx} 
\usepackage{color}
\def\CoulA{0. 0. 0.}
\def\CoulB{1. 0. 0.}
\def\CoulC{0. 0. 1.}

\definecolor{mg}{rgb}   {0.85,  0.7,    0.85}
\definecolor{bl}{rgb}   {0.5,  0.3,    1}
\newcommand{\Bk}{\color{black}}

\newcommand{\Bl}{\color{blue}}

\definecolor{webred}{rgb}{0.75,0,0}
\definecolor{webgreen}{rgb}{0,0.75,0}

\newtheorem{theo}{Theorem}[chapter]
\newtheorem{prop}[theo]{Proposition}
\newtheorem{lem}[theo]{Lemma}
\newtheorem{cor}[theo]{Corollary}
\newtheorem{definition}[theo]{Definition}
\newtheorem{notation}[theo]{Notation}
\newtheorem{conj}[theo]{Conjecture}
\newtheorem{hyp}[theo]{Assumption}
\theoremstyle{remark}
\newtheorem{rem}[theo]{Remark}
\newtheorem{exe}[theo]{Exercise}
\newtheorem{assumption}[theo]{Assumption}

\makeatletter

\@addtoreset{equation}{section}
\makeatother

\ifx\figforTeXisloaded\relax \else\global\let\figforTeXisloaded=\relax\fi
\catcode`\@=11
\ifx\ctr@ln@m\undefined\else%
    \immediate\write16{*** Fig4TeX WARNING : \string\ctr@ln@m\space already defined.}\fi
\def\ctr@ln@m#1{\ifx#1\undefined\else%
    \immediate\write16{*** Fig4TeX WARNING : \string#1 already defined.}\fi}
\ctr@ln@m\ctr@ld@f
\def\ctr@ld@f#1#2{\ctr@ln@m#2#1#2}
\ctr@ld@f\def\ctr@ln@w#1#2{\ctr@ln@m#2\csname#1\endcsname#2}
{\catcode`\/=0 \catcode`/\=12 /ctr@ld@f/gdef/BS@{\}}
\ctr@ld@f\def\ctr@lcsn@m#1{\expandafter\ifx\csname#1\endcsname\relax\else%
    \immediate\write16{*** Fig4TeX WARNING : \BS@\expandafter\string#1\space already defined.}\fi}
\ctr@ld@f\edef\colonc@tcode{\the\catcode`\:}
\ctr@ld@f\edef\semicolonc@tcode{\the\catcode`\;}
\ctr@ld@f\def\t@stc@tcodech@nge{{\let\c@tcodech@nged=\z@%
    \ifnum\colonc@tcode=\the\catcode`\:\else\let\c@tcodech@nged=\@ne\fi%
    \ifnum\semicolonc@tcode=\the\catcode`\;\else\let\c@tcodech@nged=\@ne\fi%
    \ifx\c@tcodech@nged\@ne%
    \immediate\write16{}
    \immediate\write16{!!!=============================================================!!!}
    \immediate\write16{ Fig4TeX WARNING :}
    \immediate\write16{ The category code of some characters has been changed, which will}
    \immediate\write16{ result in an error (message "Runaway argument?").}
    \immediate\write16{ This probably comes from another package that changed the category}
    \immediate\write16{ code after Fig4TeX was loaded. If that proves to be exact, the}
    \immediate\write16{ solution is to exchange the loading commands on top of your file}
    \immediate\write16{ so that Fig4TeX is loaded last. For example, in LaTeX, we should}
    \immediate\write16{ say :}
    \immediate\write16{\BS@ usepackage[french]{babel}}
    \immediate\write16{\BS@ usepackage{fig4tex}}
    \immediate\write16{!!!=============================================================!!!}
    \immediate\write16{}
    \fi}}
\ctr@ld@f\def\FigforTeX{F\kern-.05em i\kern-.05em g\kern-.1em\raise-.14em\hbox{4}\kern-.19em\TeX}
\ctr@ln@w{newdimen}\epsil@n\epsil@n=0.00005pt
\ctr@ln@w{newdimen}\Cepsil@n\Cepsil@n=0.005pt
\ctr@ln@w{newdimen}\dcq@\dcq@=254pt
\ctr@ln@w{newdimen}\PI@\PI@=3.141592pt
\ctr@ln@w{newdimen}\DemiPI@deg\DemiPI@deg=90pt
\ctr@ln@w{newdimen}\PI@deg\PI@deg=180pt
\ctr@ln@w{newdimen}\DePI@deg\DePI@deg=360pt
\ctr@ld@f\chardef\t@n=10
\ctr@ld@f\chardef\c@nt=100
\ctr@ld@f\chardef\@lxxiv=74
\ctr@ld@f\chardef\@xci=91
\ctr@ld@f\mathchardef\@nMnCQn=9949
\ctr@ld@f\chardef\@vi=6
\ctr@ld@f\chardef\@xxx=30
\ctr@ld@f\chardef\@lvi=56
\ctr@ld@f\chardef\@@lxxi=71
\ctr@ld@f\chardef\@lxxxv=85
\ctr@ld@f\mathchardef\@@mmmmlxviii=4068
\ctr@ld@f\mathchardef\@ccclx=360
\ctr@ld@f\mathchardef\@dccxx=720
\ctr@ln@w{newcount}\p@rtent \ctr@ln@w{newcount}\f@ctech \ctr@ln@w{newcount}\result@tent
\ctr@ln@w{newdimen}\v@lmin \ctr@ln@w{newdimen}\v@lmax \ctr@ln@w{newdimen}\v@leur
\ctr@ln@w{newdimen}\result@t\ctr@ln@w{newdimen}\result@@t
\ctr@ln@w{newdimen}\mili@u \ctr@ln@w{newdimen}\c@rre \ctr@ln@w{newdimen}\delt@
\ctr@ld@f\def\degT@rd{0.017453 }  % pi/180
\ctr@ld@f\def\rdT@deg{57.295779 } % 180/pi
\ctr@ln@m\v@leurseule
{\catcode`p=12 \catcode`t=12 \gdef\v@leurseule#1pt{#1}}
\ctr@ld@f\def\repdecn@mb#1{\expandafter\v@leurseule\the#1\space}
\ctr@ld@f\def\arct@n#1(#2,#3){{\v@lmin=#2\v@lmax=#3%
    \maxim@m{\mili@u}{-\v@lmin}{\v@lmin}\maxim@m{\c@rre}{-\v@lmax}{\v@lmax}%
    \delt@=\mili@u\m@ech\mili@u%
    \ifdim\c@rre>\@nMnCQn\mili@u\divide\v@lmax\tw@\c@lATAN\v@leur(\z@,\v@lmax)% DY > 9949 DX
    \else%
    \maxim@m{\mili@u}{-\v@lmin}{\v@lmin}\maxim@m{\c@rre}{-\v@lmax}{\v@lmax}%
    \m@ech\c@rre%
    \ifdim\mili@u>\@nMnCQn\c@rre\divide\v@lmin\tw@% DX > 9949 DY
    \maxim@m{\mili@u}{-\v@lmin}{\v@lmin}\c@lATAN\v@leur(\mili@u,\z@)%
    \else\c@lATAN\v@leur(\delt@,\v@lmax)\fi\fi%
    \ifdim\v@lmin<\z@\v@leur=-\v@leur\ifdim\v@lmax<\z@\advance\v@leur-\PI@%
    \else\advance\v@leur\PI@\fi\fi%
    \global\result@t=\v@leur}#1=\result@t}
\ctr@ld@f\def\m@ech#1{\ifdim#1>1.646pt\divide\mili@u\t@n\divide\c@rre\t@n\m@ech#1\fi}
\ctr@ld@f\def\c@lATAN#1(#2,#3){{\v@lmin=#2\v@lmax=#3\v@leur=\z@\delt@=\tw@ pt%
    \un@iter{0.785398}{\v@lmax<}%
    \un@iter{0.463648}{\v@lmax<}%
    \un@iter{0.244979}{\v@lmax<}%
    \un@iter{0.124355}{\v@lmax<}%
    \un@iter{0.062419}{\v@lmax<}%
    \un@iter{0.031240}{\v@lmax<}%
    \un@iter{0.015624}{\v@lmax<}%
    \un@iter{0.007812}{\v@lmax<}%
    \un@iter{0.003906}{\v@lmax<}%
    \un@iter{0.001953}{\v@lmax<}%
    \un@iter{0.000976}{\v@lmax<}%
    \un@iter{0.000488}{\v@lmax<}%
    \un@iter{0.000244}{\v@lmax<}%
    \un@iter{0.000122}{\v@lmax<}%
    \un@iter{0.000061}{\v@lmax<}%
    \un@iter{0.000030}{\v@lmax<}%
    \un@iter{0.000015}{\v@lmax<}%
    \global\result@t=\v@leur}#1=\result@t}
\ctr@ld@f\def\un@iter#1#2{%
    \divide\delt@\tw@\edef\dpmn@{\repdecn@mb{\delt@}}%
    \mili@u=\v@lmin%
    \ifdim#2\z@%
      \advance\v@lmin-\dpmn@\v@lmax\advance\v@lmax\dpmn@\mili@u%
      \advance\v@leur-#1pt%
    \else%
      \advance\v@lmin\dpmn@\v@lmax\advance\v@lmax-\dpmn@\mili@u%
      \advance\v@leur#1pt%
    \fi}
\ctr@ld@f\def\c@ssin#1#2#3{\expandafter\ifx\csname COS@\number#3\endcsname\relax\c@lCS{#3pt}%
    \expandafter\xdef\csname COS@\number#3\endcsname{\repdecn@mb\result@t}%
    \expandafter\xdef\csname SIN@\number#3\endcsname{\repdecn@mb\result@@t}\fi%
    \edef#1{\csname COS@\number#3\endcsname}\edef#2{\csname SIN@\number#3\endcsname}}
\ctr@ld@f\def\c@lCS#1{{\mili@u=#1\p@rtent=\@ne%
    \relax\ifdim\mili@u<\z@\red@ng<-\else\red@ng>+\fi\f@ctech=\p@rtent%
    \relax\ifdim\mili@u<\z@\mili@u=-\mili@u\f@ctech=-\f@ctech\fi\c@@lCS}}
\ctr@ld@f\def\c@@lCS{\v@lmin=\mili@u\c@rre=-\mili@u\advance\c@rre\DemiPI@deg\v@lmax=\c@rre%
    \mili@u\@@lxxi\mili@u\divide\mili@u\@@mmmmlxviii%
    \edef\v@larg{\repdecn@mb{\mili@u}}\mili@u=-\v@larg\mili@u%
    \edef\v@lmxde{\repdecn@mb{\mili@u}}%
    \c@rre\@@lxxi\c@rre\divide\c@rre\@@mmmmlxviii%
    \edef\v@largC{\repdecn@mb{\c@rre}}\c@rre=-\v@largC\c@rre%
    \edef\v@lmxdeC{\repdecn@mb{\c@rre}}%
    \fctc@s\mili@u\v@lmin\global\result@t\p@rtent\v@leur%
    \let\t@mp=\v@larg\let\v@larg=\v@largC\let\v@largC=\t@mp%
    \let\t@mp=\v@lmxde\let\v@lmxde=\v@lmxdeC\let\v@lmxdeC=\t@mp%
    \fctc@s\c@rre\v@lmax\global\result@@t\f@ctech\v@leur}
\ctr@ld@f\def\fctc@s#1#2{\v@leur=#1\relax\ifdim#2<\@lxxxv\p@\cosser@h\else\sinser@t\fi}
\ctr@ld@f\def\cosser@h{\advance\v@leur\@lvi\p@\divide\v@leur\@lvi%
    \v@leur=\v@lmxde\v@leur\advance\v@leur\@xxx\p@%
    \v@leur=\v@lmxde\v@leur\advance\v@leur\@ccclx\p@%
    \v@leur=\v@lmxde\v@leur\advance\v@leur\@dccxx\p@\divide\v@leur\@dccxx}
\ctr@ld@f\def\sinser@t{\v@leur=\v@lmxdeC\p@\advance\v@leur\@vi\p@%
    \v@leur=\v@largC\v@leur\divide\v@leur\@vi}
\ctr@ld@f\def\red@ng#1#2{\relax\ifdim\mili@u#1#2\DemiPI@deg\advance\mili@u#2-\PI@deg%
    \p@rtent=-\p@rtent\red@ng#1#2\fi}
\ctr@ld@f\def\pr@c@lCS#1#2#3{\ctr@lcsn@m{COS@\number#3 }%
    \expandafter\xdef\csname COS@\number#3\endcsname{#1}%
    \expandafter\xdef\csname SIN@\number#3\endcsname{#2}}
\pr@c@lCS{1}{0}{0}
\pr@c@lCS{0.7071}{0.7071}{45}\pr@c@lCS{0.7071}{-0.7071}{-45}
\pr@c@lCS{0}{1}{90}          \pr@c@lCS{0}{-1}{-90}
\pr@c@lCS{-1}{0}{180}        \pr@c@lCS{-1}{0}{-180}
\pr@c@lCS{0}{-1}{270}        \pr@c@lCS{0}{1}{-270}
\ctr@ld@f\def\invers@#1#2{{\v@leur=#2\maxim@m{\v@lmax}{-\v@leur}{\v@leur}%
    \f@ctech=\@ne\m@inv@rs%
    \multiply\v@leur\f@ctech\edef\v@lv@leur{\repdecn@mb{\v@leur}}%
    \p@rtentiere{\p@rtent}{\v@leur}\v@lmin=\p@\divide\v@lmin\p@rtent%
    \inv@rs@\multiply\v@lmax\f@ctech\global\result@t=\v@lmax}#1=\result@t}
\ctr@ld@f\def\m@inv@rs{\ifdim\v@lmax<\p@\multiply\v@lmax\t@n\multiply\f@ctech\t@n\m@inv@rs\fi}
\ctr@ld@f\def\inv@rs@{\v@lmax=-\v@lmin\v@lmax=\v@lv@leur\v@lmax%
    \advance\v@lmax\tw@ pt\v@lmax=\repdecn@mb{\v@lmin}\v@lmax%
    \delt@=\v@lmax\advance\delt@-\v@lmin\ifdim\delt@<\z@\delt@=-\delt@\fi%
    \ifdim\delt@>\epsil@n\v@lmin=\v@lmax\inv@rs@\fi}
\ctr@ld@f\def\minim@m#1#2#3{\relax\ifdim#2<#3#1=#2\else#1=#3\fi}
\ctr@ld@f\def\maxim@m#1#2#3{\relax\ifdim#2>#3#1=#2\else#1=#3\fi}
\ctr@ld@f\def\p@rtentiere#1#2{#1=#2\divide#1by65536 }
\ctr@ld@f\def\r@undint#1#2{{\v@leur=#2\divide\v@leur\t@n\p@rtentiere{\p@rtent}{\v@leur}%
    \v@leur=\p@rtent pt\global\result@t=\t@n\v@leur}#1=\result@t}
\ctr@ld@f\def\sqrt@#1#2{{\v@leur=#2%
    \minim@m{\v@lmin}{\p@}{\v@leur}\maxim@m{\v@lmax}{\p@}{\v@leur}%
    \f@ctech=\@ne\m@sqrt@\sqrt@@%
    \mili@u=\v@lmin\advance\mili@u\v@lmax\divide\mili@u\tw@\multiply\mili@u\f@ctech%
    \global\result@t=\mili@u}#1=\result@t}
\ctr@ld@f\def\m@sqrt@{\ifdim\v@leur>\dcq@\divide\v@leur\c@nt\v@lmax=\v@leur%
    \multiply\f@ctech\t@n\m@sqrt@\fi}
\ctr@ld@f\def\sqrt@@{\mili@u=\v@lmin\advance\mili@u\v@lmax\divide\mili@u\tw@%
    \c@rre=\repdecn@mb{\mili@u}\mili@u%
    \ifdim\c@rre<\v@leur\v@lmin=\mili@u\else\v@lmax=\mili@u\fi%
    \delt@=\v@lmax\advance\delt@-\v@lmin\ifdim\delt@>\epsil@n\sqrt@@\fi}
\ctr@ld@f\def\extrairelepremi@r#1\de#2{\expandafter\lepremi@r#2@#1#2}
\ctr@ld@f\def\lepremi@r#1,#2@#3#4{\def#3{#1}\def#4{#2}\ignorespaces}
\ctr@ld@f\def\@cfor#1:=#2\do#3{%
  \edef\@fortemp{#2}%
  \ifx\@fortemp\empty\else\@cforloop#2,\@nil,\@nil\@@#1{#3}\fi}
\ctr@ln@m\@nextwhile
\ctr@ld@f\def\@cforloop#1,#2\@@#3#4{%
  \def#3{#1}%
  \ifx#3\Fig@nnil\let\@nextwhile=\Fig@fornoop\else#4\relax\let\@nextwhile=\@cforloop\fi%
  \@nextwhile#2\@@#3{#4}}

\ctr@ld@f\def\@ecfor#1:=#2\do#3{%
  \def\@@cfor{\@cfor#1:=}%
  \edef\@@@cfor{#2}%
  \expandafter\@@cfor\@@@cfor\do{#3}}
\ctr@ld@f\def\Fig@nnil{\@nil}
\ctr@ld@f\def\Fig@fornoop#1\@@#2#3{}
\ctr@ln@m\list@@rg
\ctr@ld@f\def\trtlis@rg#1#2{\def\list@@rg{#1}%
    \@ecfor\p@rv@l:=\list@@rg\do{\expandafter#2\p@rv@l|}}
\ctr@ln@w{newbox}\b@xvisu
\ctr@ln@w{newtoks}\let@xte
\ctr@ln@w{newif}\ifitis@K
\ctr@ln@w{newcount}\s@mme
\ctr@ln@w{newcount}\l@mbd@un \ctr@ln@w{newcount}\l@mbd@de
\ctr@ln@w{newcount}\superc@ntr@l\superc@ntr@l=\@ne        % Controle impose
\ctr@ln@w{newcount}\typec@ntr@l\typec@ntr@l=\superc@ntr@l % Controle souhaite
\ctr@ln@w{newdimen}\v@lX  \ctr@ln@w{newdimen}\v@lY  \ctr@ln@w{newdimen}\v@lZ
\ctr@ln@w{newdimen}\v@lXa \ctr@ln@w{newdimen}\v@lYa \ctr@ln@w{newdimen}\v@lZa
\ctr@ln@w{newdimen}\unit@\unit@=\p@ % Initialisation a la valeur par defaut.
\ctr@ld@f\def\unit@util{pt}
\ctr@ld@f\def\ptT@ptps{0.996264 }
\ctr@ld@f\def\ptpsT@pt{1.00375 }
\ctr@ld@f\def\ptT@unit@{1} % Initialisation correspondant a la valeur par defaut de \unit@
\ctr@ld@f\def\setunit@#1{\def\unit@util{#1}\setunit@@#1:\invers@{\result@t}{\unit@}%
    \edef\ptT@unit@{\repdecn@mb\result@t}}
\ctr@ld@f\def\setunit@@#1#2:{\ifcat#1a\unit@=\@ne#1#2\else\unit@=#1#2\fi}
\ctr@ld@f\def\d@fm@cdim#1#2{{\v@leur=#2\v@leur=\ptT@unit@\v@leur\xdef#1{\repdecn@mb\v@leur}}}
\ctr@ln@w{newif}\ifBdingB@x\BdingB@xtrue
\ctr@ln@w{newdimen}\c@@rdXmin \ctr@ln@w{newdimen}\c@@rdYmin  % Dimensions de la BoundingBox
\ctr@ln@w{newdimen}\c@@rdXmax \ctr@ln@w{newdimen}\c@@rdYmax
\ctr@ld@f\def\b@undb@x#1#2{\ifBdingB@x%
    \relax\ifdim#1<\c@@rdXmin\global\c@@rdXmin=#1\fi%
    \relax\ifdim#2<\c@@rdYmin\global\c@@rdYmin=#2\fi%
    \relax\ifdim#1>\c@@rdXmax\global\c@@rdXmax=#1\fi%
    \relax\ifdim#2>\c@@rdYmax\global\c@@rdYmax=#2\fi\fi}
\ctr@ld@f\def\b@undb@xP#1{{\Figg@tXY{#1}\b@undb@x{\v@lX}{\v@lY}}}
\ctr@ld@f\def\ellBB@x#1;#2,#3(#4,#5,#6){{\s@uvc@ntr@l\et@tellBB@x%
    \setc@ntr@l{2}\figptell-2::#1;#2,#3(#4,#6)\b@undb@xP{-2}%
    \figptell-2::#1;#2,#3(#5,#6)\b@undb@xP{-2}%
    \c@ssin{\C@}{\S@}{#6}\v@lmin=\C@ pt\v@lmax=\S@ pt%
    \mili@u=#3\v@lmin\delt@=#2\v@lmax\arct@n\v@leur(\delt@,\mili@u)%
    \mili@u=-#3\v@lmax\delt@=#2\v@lmin\arct@n\c@rre(\delt@,\mili@u)%
    \v@leur=\rdT@deg\v@leur\advance\v@leur-\DePI@deg%
    \c@rre=\rdT@deg\c@rre\advance\c@rre-\DePI@deg%
    \v@lmin=#4pt\v@lmax=#5pt%
    \loop\ifdim\v@leur<\v@lmax\ifdim\v@leur>\v@lmin%
    \edef\@ngle{\repdecn@mb\v@leur}\figptell-2::#1;#2,#3(\@ngle,#6)%
    \b@undb@xP{-2}\fi\advance\v@leur\PI@deg\repeat%
    \loop\ifdim\c@rre<\v@lmax\ifdim\c@rre>\v@lmin%
    \edef\@ngle{\repdecn@mb\c@rre}\figptell-2::#1;#2,#3(\@ngle,#6)%
    \b@undb@xP{-2}\fi\advance\c@rre\PI@deg\repeat%
    \resetc@ntr@l\et@tellBB@x}\ignorespaces}
\ctr@ld@f\def\initb@undb@x{\c@@rdXmin=\maxdimen\c@@rdYmin=\maxdimen%
    \c@@rdXmax=-\maxdimen\c@@rdYmax=-\maxdimen}
\ctr@ld@f\def\c@ntr@lnum#1{%
    \relax\ifnum\typec@ntr@l=\@ne%
    \ifnum#1<\z@%
    \immediate\write16{*** Forbidden point number (#1). Abort.}\end\fi\fi%
    \set@bjc@de{#1}}
\ctr@ln@m\objc@de
\ctr@ld@f\def\set@bjc@de#1{\edef\objc@de{@BJ\ifnum#1<\z@ M\romannumeral-#1\else\romannumeral#1\fi}}
\s@mme=\m@ne\loop\ifnum\s@mme>-19
  \set@bjc@de{\s@mme}\ctr@lcsn@m\objc@de\ctr@lcsn@m{\objc@de T}
\advance\s@mme\m@ne\repeat
\s@mme=\@ne\loop\ifnum\s@mme<6
  \set@bjc@de{\s@mme}\ctr@lcsn@m\objc@de\ctr@lcsn@m{\objc@de T}
\advance\s@mme\@ne\repeat
\ctr@ld@f\def\setc@ntr@l#1{\ifnum\superc@ntr@l>#1\typec@ntr@l=\superc@ntr@l%
    \else\typec@ntr@l=#1\fi}
\ctr@ld@f\def\resetc@ntr@l#1{\global\superc@ntr@l=#1\setc@ntr@l{#1}}
\ctr@ld@f\def\s@uvc@ntr@l#1{\edef#1{\the\superc@ntr@l}}
\ctr@ln@m\c@lproscal
\ctr@ld@f\def\c@lproscalDD#1[#2,#3]{{\Figg@tXY{#2}%
    \edef\Xu@{\repdecn@mb{\v@lX}}\edef\Yu@{\repdecn@mb{\v@lY}}\Figg@tXY{#3}%
    \global\result@t=\Xu@\v@lX\global\advance\result@t\Yu@\v@lY}#1=\result@t}
\ctr@ld@f\def\c@lproscalTD#1[#2,#3]{{\Figg@tXY{#2}\edef\Xu@{\repdecn@mb{\v@lX}}%
    \edef\Yu@{\repdecn@mb{\v@lY}}\edef\Zu@{\repdecn@mb{\v@lZ}}%
    \Figg@tXY{#3}\global\result@t=\Xu@\v@lX\global\advance\result@t\Yu@\v@lY%
    \global\advance\result@t\Zu@\v@lZ}#1=\result@t}
\ctr@ld@f\def\c@lprovec#1{%
    \det@rmC\v@lZa(\v@lX,\v@lY,\v@lmin,\v@lmax)%
    \det@rmC\v@lXa(\v@lY,\v@lZ,\v@lmax,\v@leur)%
    \det@rmC\v@lYa(\v@lZ,\v@lX,\v@leur,\v@lmin)%
    \Figv@ctCreg#1(\v@lXa,\v@lYa,\v@lZa)}
\ctr@ld@f\def\det@rm#1[#2,#3]{{\Figg@tXY{#2}\Figg@tXYa{#3}%
    \delt@=\repdecn@mb{\v@lX}\v@lYa\advance\delt@-\repdecn@mb{\v@lY}\v@lXa%
    \global\result@t=\delt@}#1=\result@t}
\ctr@ld@f\def\det@rmC#1(#2,#3,#4,#5){{\global\result@t=\repdecn@mb{#2}#5%
    \global\advance\result@t-\repdecn@mb{#3}#4}#1=\result@t}
\ctr@ld@f\def\getredf@ctDD#1(#2,#3){{\maxim@m{\v@lXa}{-#2}{#2}\maxim@m{\v@lYa}{-#3}{#3}%
    \maxim@m{\v@lXa}{\v@lXa}{\v@lYa}% \v@lXa = ||X||inf
    \ifdim\v@lXa>\@xci pt\divide\v@lXa\@xci%
    \p@rtentiere{\p@rtent}{\v@lXa}\advance\p@rtent\@ne\else\p@rtent=\@ne\fi%
    \global\result@tent=\p@rtent}#1=\result@tent\ignorespaces}
\ctr@ld@f\def\getredf@ctTD#1(#2,#3,#4){{\maxim@m{\v@lXa}{-#2}{#2}\maxim@m{\v@lYa}{-#3}{#3}%
    \maxim@m{\v@lZa}{-#4}{#4}\maxim@m{\v@lXa}{\v@lXa}{\v@lYa}%
    \maxim@m{\v@lXa}{\v@lXa}{\v@lZa}% \v@lXa = ||X||inf
    \ifdim\v@lXa>\@lxxiv pt\divide\v@lXa\@lxxiv%
    \p@rtentiere{\p@rtent}{\v@lXa}\advance\p@rtent\@ne\else\p@rtent=\@ne\fi%
    \global\result@tent=\p@rtent}#1=\result@tent\ignorespaces}
\ctr@ld@f\def\FigptintercircB@zDD#1:#2:#3,#4[#5,#6,#7,#8]{{\s@uvc@ntr@l\et@tfigptintercircB@zDD%
    \setc@ntr@l{2}\figvectPDD-1[#5,#8]\Figg@tXY{-1}\getredf@ctDD\f@ctech(\v@lX,\v@lY)%
    \mili@u=#4\unit@\divide\mili@u\f@ctech\c@rre=\repdecn@mb{\mili@u}\mili@u%
    \figptBezierDD-5::#3[#5,#6,#7,#8]%
    \v@lmin=#3\p@\v@lmax=\v@lmin\advance\v@lmax0.1\p@%
    \loop\edef\T@{\repdecn@mb{\v@lmax}}\figptBezierDD-2::\T@[#5,#6,#7,#8]%
    \figvectPDD-1[-5,-2]\n@rmeucCDD{\delt@}{-1}\ifdim\delt@<\c@rre\v@lmin=\v@lmax%
    \advance\v@lmax0.1\p@\repeat%
    \loop\mili@u=\v@lmin\advance\mili@u\v@lmax%
    \divide\mili@u\tw@\edef\T@{\repdecn@mb{\mili@u}}\figptBezierDD-2::\T@[#5,#6,#7,#8]%
    \figvectPDD-1[-5,-2]\n@rmeucCDD{\delt@}{-1}\ifdim\delt@>\c@rre\v@lmax=\mili@u%
    \else\v@lmin=\mili@u\fi\v@leur=\v@lmax\advance\v@leur-\v@lmin%
    \ifdim\v@leur>\epsil@n\repeat\figptcopyDD#1:#2/-2/%
    \resetc@ntr@l\et@tfigptintercircB@zDD}\ignorespaces}
\ctr@ln@m\figptinterlines
\ctr@ld@f\def\inters@cDD#1:#2[#3,#4;#5,#6]{{\s@uvc@ntr@l\et@tinters@cDD%
    \setc@ntr@l{2}\vecunit@{-1}{#4}\vecunit@{-2}{#6}%
    \Figg@tXY{-1}\setc@ntr@l{1}\Figg@tXYa{#3}%
    \edef\A@{\repdecn@mb{\v@lX}}\edef\B@{\repdecn@mb{\v@lY}}%
    \v@lmin=\B@\v@lXa\advance\v@lmin-\A@\v@lYa%
    \Figg@tXYa{#5}\setc@ntr@l{2}\Figg@tXY{-2}%
    \edef\C@{\repdecn@mb{\v@lX}}\edef\D@{\repdecn@mb{\v@lY}}%
    \v@lmax=\D@\v@lXa\advance\v@lmax-\C@\v@lYa%
    \delt@=\A@\v@lY\advance\delt@-\B@\v@lX%
    \invers@{\v@leur}{\delt@}\edef\v@ldelta{\repdecn@mb{\v@leur}}%
    \v@lXa=\A@\v@lmax\advance\v@lXa-\C@\v@lmin%
    \v@lYa=\B@\v@lmax\advance\v@lYa-\D@\v@lmin%
    \v@lXa=\v@ldelta\v@lXa\v@lYa=\v@ldelta\v@lYa%
    \setc@ntr@l{1}\Figp@intregDD#1:{#2}(\v@lXa,\v@lYa)%
    \resetc@ntr@l\et@tinters@cDD}\ignorespaces}
\ctr@ld@f\def\inters@cTD#1:#2[#3,#4;#5,#6]{{\s@uvc@ntr@l\et@tinters@cTD%
    \setc@ntr@l{2}\figvectNVTD-1[#4,#6]\figvectNVTD-2[#6,-1]\figvectPTD-1[#3,#5]%
    \r@pPSTD\v@leur[-2,-1,#4]\edef\v@lcoef{\repdecn@mb{\v@leur}}%
    \figpttraTD#1:{#2}=#3/\v@lcoef,#4/\resetc@ntr@l\et@tinters@cTD}\ignorespaces}
\ctr@ld@f\def\r@pPSTD#1[#2,#3,#4]{{\Figg@tXY{#2}\edef\Xu@{\repdecn@mb{\v@lX}}%
    \edef\Yu@{\repdecn@mb{\v@lY}}\edef\Zu@{\repdecn@mb{\v@lZ}}%
    \Figg@tXY{#3}\v@lmin=\Xu@\v@lX\advance\v@lmin\Yu@\v@lY\advance\v@lmin\Zu@\v@lZ%
    \Figg@tXY{#4}\v@lmax=\Xu@\v@lX\advance\v@lmax\Yu@\v@lY\advance\v@lmax\Zu@\v@lZ%
    \invers@{\v@leur}{\v@lmax}\global\result@t=\repdecn@mb{\v@leur}\v@lmin}%
    #1=\result@t}
\ctr@ln@m\n@rminf
\ctr@ld@f\def\n@rminfDD#1#2{{\Figg@tXY{#2}\maxim@m{\v@lX}{\v@lX}{-\v@lX}%
    \maxim@m{\v@lY}{\v@lY}{-\v@lY}\maxim@m{\global\result@t}{\v@lX}{\v@lY}}%
    #1=\result@t}
\ctr@ld@f\def\n@rminfTD#1#2{{\Figg@tXY{#2}\maxim@m{\v@lX}{\v@lX}{-\v@lX}%
    \maxim@m{\v@lY}{\v@lY}{-\v@lY}\maxim@m{\v@lZ}{\v@lZ}{-\v@lZ}%
    \maxim@m{\v@lX}{\v@lX}{\v@lY}\maxim@m{\global\result@t}{\v@lX}{\v@lZ}}%
    #1=\result@t}
\ctr@ld@f\def\n@rmeucCDD#1#2{\Figg@tXY{#2}\divide\v@lX\f@ctech\divide\v@lY\f@ctech%
    #1=\repdecn@mb{\v@lX}\v@lX\v@lX=\repdecn@mb{\v@lY}\v@lY\advance#1\v@lX}
\ctr@ld@f\def\n@rmeucCTD#1#2{\Figg@tXY{#2}%
    \divide\v@lX\f@ctech\divide\v@lY\f@ctech\divide\v@lZ\f@ctech%
    #1=\repdecn@mb{\v@lX}\v@lX\v@lX=\repdecn@mb{\v@lY}\v@lY\advance#1\v@lX%
    \v@lX=\repdecn@mb{\v@lZ}\v@lZ\advance#1\v@lX}
\ctr@ln@m\n@rmeucSV
\ctr@ld@f\def\n@rmeucSVDD#1#2{{\Figg@tXY{#2}%
    \v@lXa=\repdecn@mb{\v@lX}\v@lX\v@lYa=\repdecn@mb{\v@lY}\v@lY%
    \advance\v@lXa\v@lYa\sqrt@{\global\result@t}{\v@lXa}}#1=\result@t}
\ctr@ld@f\def\n@rmeucSVTD#1#2{{\Figg@tXY{#2}\v@lXa=\repdecn@mb{\v@lX}\v@lX%
    \v@lYa=\repdecn@mb{\v@lY}\v@lY\v@lZa=\repdecn@mb{\v@lZ}\v@lZ%
    \advance\v@lXa\v@lYa\advance\v@lXa\v@lZa\sqrt@{\global\result@t}{\v@lXa}}#1=\result@t}
\ctr@ln@m\n@rmeuc
\ctr@ld@f\def\n@rmeucDD#1#2{{\Figg@tXY{#2}\getredf@ctDD\f@ctech(\v@lX,\v@lY)%
    \divide\v@lX\f@ctech\divide\v@lY\f@ctech%
    \v@lXa=\repdecn@mb{\v@lX}\v@lX\v@lYa=\repdecn@mb{\v@lY}\v@lY%
    \advance\v@lXa\v@lYa\sqrt@{\global\result@t}{\v@lXa}%
    \global\multiply\result@t\f@ctech}#1=\result@t}
\ctr@ld@f\def\n@rmeucTD#1#2{{\Figg@tXY{#2}\getredf@ctTD\f@ctech(\v@lX,\v@lY,\v@lZ)%
    \divide\v@lX\f@ctech\divide\v@lY\f@ctech\divide\v@lZ\f@ctech%
    \v@lXa=\repdecn@mb{\v@lX}\v@lX%
    \v@lYa=\repdecn@mb{\v@lY}\v@lY\v@lZa=\repdecn@mb{\v@lZ}\v@lZ%
    \advance\v@lXa\v@lYa\advance\v@lXa\v@lZa\sqrt@{\global\result@t}{\v@lXa}%
    \global\multiply\result@t\f@ctech}#1=\result@t}
\ctr@ln@m\vecunit@
\ctr@ld@f\def\vecunit@DD#1#2{{\Figg@tXY{#2}\getredf@ctDD\f@ctech(\v@lX,\v@lY)%
    \divide\v@lX\f@ctech\divide\v@lY\f@ctech%
    \Figv@ctCreg#1(\v@lX,\v@lY)\n@rmeucSV{\v@lYa}{#1}%
    \invers@{\v@lXa}{\v@lYa}\edef\v@lv@lXa{\repdecn@mb{\v@lXa}}%
    \v@lX=\v@lv@lXa\v@lX\v@lY=\v@lv@lXa\v@lY%
    \Figv@ctCreg#1(\v@lX,\v@lY)\multiply\v@lYa\f@ctech\global\result@t=\v@lYa}}
\ctr@ld@f\def\vecunit@TD#1#2{{\Figg@tXY{#2}\getredf@ctTD\f@ctech(\v@lX,\v@lY,\v@lZ)%
    \divide\v@lX\f@ctech\divide\v@lY\f@ctech\divide\v@lZ\f@ctech%
    \Figv@ctCreg#1(\v@lX,\v@lY,\v@lZ)\n@rmeucSV{\v@lYa}{#1}%
    \invers@{\v@lXa}{\v@lYa}\edef\v@lv@lXa{\repdecn@mb{\v@lXa}}%
    \v@lX=\v@lv@lXa\v@lX\v@lY=\v@lv@lXa\v@lY\v@lZ=\v@lv@lXa\v@lZ%
    \Figv@ctCreg#1(\v@lX,\v@lY,\v@lZ)\multiply\v@lYa\f@ctech\global\result@t=\v@lYa}}
\ctr@ld@f\def\vecunitC@TD[#1,#2]{\Figg@tXYa{#1}\Figg@tXY{#2}%
    \advance\v@lX-\v@lXa\advance\v@lY-\v@lYa\advance\v@lZ-\v@lZa\c@lvecunitTD}
\ctr@ld@f\def\vecunitCV@TD#1{\Figg@tXY{#1}\c@lvecunitTD}
\ctr@ld@f\def\c@lvecunitTD{\getredf@ctTD\f@ctech(\v@lX,\v@lY,\v@lZ)%
    \divide\v@lX\f@ctech\divide\v@lY\f@ctech\divide\v@lZ\f@ctech%
    \v@lXa=\repdecn@mb{\v@lX}\v@lX%
    \v@lYa=\repdecn@mb{\v@lY}\v@lY\v@lZa=\repdecn@mb{\v@lZ}\v@lZ%
    \advance\v@lXa\v@lYa\advance\v@lXa\v@lZa\sqrt@{\v@lYa}{\v@lXa}%
    \invers@{\v@lXa}{\v@lYa}\edef\v@lv@lXa{\repdecn@mb{\v@lXa}}%
    \v@lX=\v@lv@lXa\v@lX\v@lY=\v@lv@lXa\v@lY\v@lZ=\v@lv@lXa\v@lZ}
\ctr@ln@m\figgetangle
\ctr@ld@f\def\figgetangleDD#1[#2,#3,#4]{\ifps@cri{\s@uvc@ntr@l\et@tfiggetangleDD\setc@ntr@l{2}%
    \figvectPDD-1[#2,#3]\figvectPDD-2[#2,#4]\vecunit@{-1}{-1}%
    \c@lproscalDD\delt@[-2,-1]\figvectNVDD-1[-1]\c@lproscalDD\v@leur[-2,-1]%
    \arct@n\v@lmax(\delt@,\v@leur)\v@lmax=\rdT@deg\v@lmax%
    \ifdim\v@lmax<\z@\advance\v@lmax\DePI@deg\fi\xdef#1{\repdecn@mb{\v@lmax}}%
    \resetc@ntr@l\et@tfiggetangleDD}\ignorespaces\fi}
\ctr@ld@f\def\figgetangleTD#1[#2,#3,#4,#5]{\ifps@cri{\s@uvc@ntr@l\et@tfiggetangleTD\setc@ntr@l{2}%
    \figvectPTD-1[#2,#3]\figvectPTD-2[#2,#5]\figvectNVTD-3[-1,-2]%
    \figvectPTD-2[#2,#4]\figvectNVTD-4[-3,-1]%
    \vecunit@{-1}{-1}\c@lproscalTD\delt@[-2,-1]\c@lproscalTD\v@leur[-2,-4]%
    \arct@n\v@lmax(\delt@,\v@leur)\v@lmax=\rdT@deg\v@lmax%
    \ifdim\v@lmax<\z@\advance\v@lmax\DePI@deg\fi\xdef#1{\repdecn@mb{\v@lmax}}%
    \resetc@ntr@l\et@tfiggetangleTD}\ignorespaces\fi}    
\ctr@ld@f\def\figgetdist#1[#2,#3]{\ifps@cri{\s@uvc@ntr@l\et@tfiggetdist\setc@ntr@l{2}%
    \figvectP-1[#2,#3]\n@rmeuc{\v@lX}{-1}\v@lX=\ptT@unit@\v@lX\xdef#1{\repdecn@mb{\v@lX}}%
    \resetc@ntr@l\et@tfiggetdist}\ignorespaces\fi}
\ctr@ld@f\def\Figg@tT#1{\c@ntr@lnum{#1}%
    {\expandafter\expandafter\expandafter\extr@ctT\csname\objc@de\endcsname:%
     \ifnum\B@@ltxt=\z@\ptn@me{#1}\else\csname\objc@de T\endcsname\fi}}
\ctr@ld@f\def\extr@ctT#1,#2,#3/#4:{\def\B@@ltxt{#3}}
\ctr@ld@f\def\Figg@tXY#1{\c@ntr@lnum{#1}%
    \expandafter\expandafter\expandafter\extr@ctC\csname\objc@de\endcsname:}
\ctr@ln@m\extr@ctC
\ctr@ld@f\def\extr@ctCDD#1/#2,#3,#4:{\v@lX=#2\v@lY=#3}
\ctr@ld@f\def\extr@ctCTD#1/#2,#3,#4:{\v@lX=#2\v@lY=#3\v@lZ=#4}
\ctr@ld@f\def\Figg@tXYa#1{\c@ntr@lnum{#1}%
    \expandafter\expandafter\expandafter\extr@ctCa\csname\objc@de\endcsname:}
\ctr@ln@m\extr@ctCa
\ctr@ld@f\def\extr@ctCaDD#1/#2,#3,#4:{\v@lXa=#2\v@lYa=#3}
\ctr@ld@f\def\extr@ctCaTD#1/#2,#3,#4:{\v@lXa=#2\v@lYa=#3\v@lZa=#4}
\ctr@ln@m\t@xt@
\ctr@ld@f\def\figinit#1{\t@stc@tcodech@nge\initpr@lim\Figinit@#1,:\initpss@ttings\ignorespaces}
\ctr@ld@f\def\Figinit@#1,#2:{\setunit@{#1}\def\t@xt@{#2}\ifx\t@xt@\empty\else\Figinit@@#2:\fi}
\ctr@ld@f\def\Figinit@@#1#2:{\if#12 \else\Figs@tproj{#1}\initTD@\fi}
\ctr@ln@w{newif}\ifTr@isDim
\ctr@ld@f\def\UnD@fined{UNDEFINED}
\ctr@ld@f\def\ifundefined#1{\expandafter\ifx\csname#1\endcsname\relax}
\ctr@ln@m\@utoFN
\ctr@ln@m\@utoFInDone
\ctr@ln@m\disob@unit
\ctr@ld@f\def\initpr@lim{\initb@undb@x\figsetmark{}\figsetptname{$A_{##1}$}\def\Sc@leFact{1}%
    \initDD@\figsetroundcoord{yes}\ps@critrue\expandafter\setupd@te\defaultupdate:%
    \edef\disob@unit{\UnD@fined}\edef\t@rgetpt{\UnD@fined}\gdef\@utoFInDone{1}\gdef\@utoFN{0}}
\ctr@ld@f\def\initDD@{\Tr@isDimfalse%
    \ifPDFm@ke%
     \let\Ps@rcerc=\Ps@rcercBz%
     \let\Ps@rell=\Ps@rellBz%
    \fi
    \let\c@lDCUn=\c@lDCUnDD%
    \let\c@lDCDeux=\c@lDCDeuxDD%
    \let\c@ldefproj=\relax%
    \let\c@lproscal=\c@lproscalDD%
    \let\c@lprojSP=\relax%
    \let\extr@ctC=\extr@ctCDD%
    \let\extr@ctCa=\extr@ctCaDD%
    \let\extr@ctCF=\extr@ctCFDD%
    \let\Figp@intreg=\Figp@intregDD%
    \let\Figpts@xes=\Figpts@xesDD%
    \let\n@rmeucSV=\n@rmeucSVDD\let\n@rmeuc=\n@rmeucDD\let\n@rminf=\n@rminfDD%
    \let\pr@dMatV=\pr@dMatVDD%
    \let\ps@xes=\ps@xesDD%
    \let\vecunit@=\vecunit@DD%
    \let\figcoord=\figcoordDD%
    \let\figgetangle=\figgetangleDD%
    \let\figpt=\figptDD%
    \let\figptBezier=\figptBezierDD%
    \let\figptbary=\figptbaryDD%
    \let\figptcirc=\figptcircDD%
    \let\figptcircumcenter=\figptcircumcenterDD%
    \let\figptcopy=\figptcopyDD%
    \let\figptcurvcenter=\figptcurvcenterDD%
    \let\figptell=\figptellDD%
    \let\figptendnormal=\figptendnormalDD%
    \let\figptinterlineplane=\figptinterlineplaneDD%
    \let\figptinterlines=\inters@cDD%
    \let\figptorthocenter=\figptorthocenterDD%
    \let\figptorthoprojline=\figptorthoprojlineDD%
    \let\figptorthoprojplane=\figptorthoprojplaneDD%
    \let\figptrot=\figptrotDD%
    \let\figptscontrol=\figptscontrolDD%
    \let\figptsintercirc=\figptsintercircDD%
    \let\figptsinterlinell=\figptsinterlinellDD%
    \let\figptsorthoprojline=\figptsorthoprojlineDD%
    \let\figptorthoprojplane=\figptorthoprojplaneDD%
    \let\figptsrot=\figptsrotDD%
    \let\figptssym=\figptssymDD%
    \let\figptstra=\figptstraDD%
    \let\figptsym=\figptsymDD%
    \let\figpttraC=\figpttraCDD%
    \let\figpttra=\figpttraDD%
    \let\figptvisilimSL=\figptvisilimSLDD%
    \let\figsetobdist=\figsetobdistDD%
    \let\figsettarget=\figsettargetDD%
    \let\figsetview=\figsetviewDD%
    \let\figvectDBezier=\figvectDBezierDD%
    \let\figvectN=\figvectNDD%
    \let\figvectNV=\figvectNVDD%
    \let\figvectP=\figvectPDD%
    \let\figvectU=\figvectUDD%
    \let\psarccircP=\psarccircPDD%
    \let\psarccirc=\psarccircDD%
    \let\psarcell=\psarcellDD%
    \let\psarcellPA=\psarcellPADD%
    \let\psarrowBezier=\psarrowBezierDD%
    \let\psarrowcircP=\psarrowcircPDD%
    \let\psarrowcirc=\psarrowcircDD%
    \let\psarrowhead=\psarrowheadDD%
    \let\psarrow=\psarrowDD%
    \let\psBezier=\psBezierDD%
    \let\pscirc=\pscircDD%
    \let\pscurve=\pscurveDD%
    \let\psnormal=\psnormalDD%
    }
\ctr@ld@f\def\initTD@{\Tr@isDimtrue\initb@undb@xTD\newt@rgetptfalse\newdis@bfalse%
    \let\c@lDCUn=\c@lDCUnTD%
    \let\c@lDCDeux=\c@lDCDeuxTD%
    \let\c@ldefproj=\c@ldefprojTD%
    \let\c@lproscal=\c@lproscalTD%
    \let\extr@ctC=\extr@ctCTD%
    \let\extr@ctCa=\extr@ctCaTD%
    \let\extr@ctCF=\extr@ctCFTD%
    \let\Figp@intreg=\Figp@intregTD%
    \let\Figpts@xes=\Figpts@xesTD%
    \let\n@rmeucSV=\n@rmeucSVTD\let\n@rmeuc=\n@rmeucTD\let\n@rminf=\n@rminfTD%
    \let\pr@dMatV=\pr@dMatVTD%
    \let\ps@xes=\ps@xesTD%
    \let\vecunit@=\vecunit@TD%
    \let\figcoord=\figcoordTD%
    \let\figgetangle=\figgetangleTD%
    \let\figpt=\figptTD%
    \let\figptBezier=\figptBezierTD%
    \let\figptbary=\figptbaryTD%
    \let\figptcirc=\figptcircTD%
    \let\figptcircumcenter=\figptcircumcenterTD%
    \let\figptcopy=\figptcopyTD%
    \let\figptcurvcenter=\figptcurvcenterTD%
    \let\figptinterlineplane=\figptinterlineplaneTD%
    \let\figptinterlines=\inters@cTD%
    \let\figptorthocenter=\figptorthocenterTD%
    \let\figptorthoprojline=\figptorthoprojlineTD%
    \let\figptorthoprojplane=\figptorthoprojplaneTD%
    \let\figptrot=\figptrotTD%
    \let\figptscontrol=\figptscontrolTD%
    \let\figptsintercirc=\figptsintercircTD%
    \let\figptsorthoprojline=\figptsorthoprojlineTD%
    \let\figptsorthoprojplane=\figptsorthoprojplaneTD%
    \let\figptsrot=\figptsrotTD%
    \let\figptssym=\figptssymTD%
    \let\figptstra=\figptstraTD%
    \let\figptsym=\figptsymTD%
    \let\figpttraC=\figpttraCTD%
    \let\figpttra=\figpttraTD%
    \let\figptvisilimSL=\figptvisilimSLTD%
    \let\figsetobdist=\figsetobdistTD%
    \let\figsettarget=\figsettargetTD%
    \let\figsetview=\figsetviewTD%
    \let\figvectDBezier=\figvectDBezierTD%
    \let\figvectN=\figvectNTD%
    \let\figvectNV=\figvectNVTD%
    \let\figvectP=\figvectPTD%
    \let\figvectU=\figvectUTD%
    \let\psarccircP=\psarccircPTD%
    \let\psarccirc=\psarccircTD%
    \let\psarcell=\psarcellTD%
    \let\psarcellPA=\psarcellPATD%
    \let\psarrowBezier=\psarrowBezierTD%
    \let\psarrowcircP=\psarrowcircPTD%
    \let\psarrowcirc=\psarrowcircTD%
    \let\psarrowhead=\psarrowheadTD%
    \let\psarrow=\psarrowTD%
    \let\psBezier=\psBezierTD%
    \let\pscirc=\pscircTD%
    \let\pscurve=\pscurveTD%
    }
\ctr@ld@f\def\un@v@ilable#1{\immediate\write16{*** The macro #1 is not available in the current context.}}
\ctr@ld@f\def\figinsert#1{{\def\t@xt@{#1}\relax%
    \ifx\t@xt@\empty\ifnum\@utoFInDone>\z@\Figinsert@\DefGIfilen@me,:\fi%
    \else\expandafter\FiginsertNu@#1 :\fi}\ignorespaces}
\ctr@ld@f\def\FiginsertNu@#1 #2:{\def\t@xt@{#1}\relax\ifx\t@xt@\empty\def\t@xt@{#2}%
    \ifx\t@xt@\empty\ifnum\@utoFInDone>\z@\Figinsert@\DefGIfilen@me,:\fi%
    \else\FiginsertNu@#2:\fi\else\expandafter\FiginsertNd@#1 #2:\fi}
\ctr@ld@f\def\FiginsertNd@#1#2:{\ifcat#1a\Figinsert@#1#2,:\else%
    \ifnum\@utoFInDone>\z@\Figinsert@\DefGIfilen@me,#1#2,:\fi\fi}
\ctr@ln@m\Sc@leFact
\ctr@ld@f\def\Figinsert@#1,#2:{\def\t@xt@{#2}\ifx\t@xt@\empty\xdef\Sc@leFact{1}\else%
    \X@rgdeux@#2\xdef\Sc@leFact{\@rgdeux}\fi%
    \Figdisc@rdLTS{#1}{\t@xt@}\@psfgetbb{\t@xt@}%
    \v@lX=\@psfllx\p@\v@lX=\ptpsT@pt\v@lX\v@lX=\Sc@leFact\v@lX%
    \v@lY=\@psflly\p@\v@lY=\ptpsT@pt\v@lY\v@lY=\Sc@leFact\v@lY%
    \b@undb@x{\v@lX}{\v@lY}%
    \v@lX=\@psfurx\p@\v@lX=\ptpsT@pt\v@lX\v@lX=\Sc@leFact\v@lX%
    \v@lY=\@psfury\p@\v@lY=\ptpsT@pt\v@lY\v@lY=\Sc@leFact\v@lY%
    \b@undb@x{\v@lX}{\v@lY}%
    \ifPDFm@ke\Figinclud@PDF{\t@xt@}{\Sc@leFact}\else%
    \v@lX=\c@nt pt\v@lX=\Sc@leFact\v@lX\edef\F@ct{\repdecn@mb{\v@lX}}%
    \ifx\TeXturesonMacOSltX\special{postscriptfile #1 vscale=\F@ct\space hscale=\F@ct}%
    \else\includegraphics{#1}\fi\fi%
    \message{[\t@xt@]}\ignorespaces}
\ctr@ld@f\def\Figdisc@rdLTS#1#2{\expandafter\Figdisc@rdLTS@#1 :#2}
\ctr@ld@f\def\Figdisc@rdLTS@#1 #2:#3{\def#3{#1}\relax\ifx#3\empty\expandafter\Figdisc@rdLTS@#2:#3\fi}
\ctr@ld@f\def\figinsertE#1{\FiginsertE@#1,:\ignorespaces}
\ctr@ld@f\def\FiginsertE@#1,#2:{{\def\t@xt@{#2}\ifx\t@xt@\empty\xdef\Sc@leFact{1}\else%
    \X@rgdeux@#2\xdef\Sc@leFact{\@rgdeux}\fi%
    \Figdisc@rdLTS{#1}{\t@xt@}\pdfximage{\t@xt@}%
    \setbox\Gb@x=\hbox{\pdfrefximage\pdflastximage}%
    \v@lX=\z@\v@lY=-\Sc@leFact\dp\Gb@x\b@undb@x{\v@lX}{\v@lY}%
    \advance\v@lX\Sc@leFact\wd\Gb@x\advance\v@lY\Sc@leFact\dp\Gb@x%
    \advance\v@lY\Sc@leFact\ht\Gb@x\b@undb@x{\v@lX}{\v@lY}%
    \v@lX=\Sc@leFact\wd\Gb@x\pdfximage width \v@lX {\t@xt@}%
    \rlap{\pdfrefximage\pdflastximage}\message{[\t@xt@]}}\ignorespaces}
\ctr@ld@f\def\X@rgdeux@#1,{\edef\@rgdeux{#1}}
\ctr@ln@m\figpt
\ctr@ld@f\def\figptDD#1:#2(#3,#4){\ifps@cri\c@ntr@lnum{#1}%
    {\v@lX=#3\unit@\v@lY=#4\unit@\Fig@dmpt{#2}{\z@}}\ignorespaces\fi}
\ctr@ld@f\def\Fig@dmpt#1#2{\def\t@xt@{#1}\ifx\t@xt@\empty\def\B@@ltxt{\z@}%
    \else\expandafter\gdef\csname\objc@de T\endcsname{#1}\def\B@@ltxt{\@ne}\fi%
    \expandafter\xdef\csname\objc@de\endcsname{\ifitis@vect@r\C@dCl@svect%
    \else\C@dCl@spt\fi,\z@,\B@@ltxt/\the\v@lX,\the\v@lY,#2}}
\ctr@ld@f\def\C@dCl@spt{P}
\ctr@ld@f\def\C@dCl@svect{V}
\ctr@ln@m\c@@rdYZ
\ctr@ln@m\c@@rdY
\ctr@ld@f\def\figptTD#1:#2(#3,#4){\ifps@cri\c@ntr@lnum{#1}%
    \def\c@@rdYZ{#4,0,0}\extrairelepremi@r\c@@rdY\de\c@@rdYZ%
    \extrairelepremi@r\c@@rdZ\de\c@@rdYZ%
    {\v@lX=#3\unit@\v@lY=\c@@rdY\unit@\v@lZ=\c@@rdZ\unit@\Fig@dmpt{#2}{\the\v@lZ}%
    \b@undb@xTD{\v@lX}{\v@lY}{\v@lZ}}\ignorespaces\fi}
\ctr@ln@m\Figp@intreg
\ctr@ld@f\def\Figp@intregDD#1:#2(#3,#4){\c@ntr@lnum{#1}%
    {\result@t=#4\v@lX=#3\v@lY=\result@t\Fig@dmpt{#2}{\z@}}\ignorespaces}
\ctr@ld@f\def\Figp@intregTD#1:#2(#3,#4){\c@ntr@lnum{#1}%
    \def\c@@rdYZ{#4,\z@,\z@}\extrairelepremi@r\c@@rdY\de\c@@rdYZ%
    \extrairelepremi@r\c@@rdZ\de\c@@rdYZ%
    {\v@lX=#3\v@lY=\c@@rdY\v@lZ=\c@@rdZ\Fig@dmpt{#2}{\the\v@lZ}%
    \b@undb@xTD{\v@lX}{\v@lY}{\v@lZ}}\ignorespaces}
\ctr@ln@m\figptBezier
\ctr@ld@f\def\figptBezierDD#1:#2:#3[#4,#5,#6,#7]{\ifps@cri{\s@uvc@ntr@l\et@tfigptBezierDD%
    \FigptBezier@#3[#4,#5,#6,#7]\Figp@intregDD#1:{#2}(\v@lX,\v@lY)%
    \resetc@ntr@l\et@tfigptBezierDD}\ignorespaces\fi}
\ctr@ld@f\def\figptBezierTD#1:#2:#3[#4,#5,#6,#7]{\ifps@cri{\s@uvc@ntr@l\et@tfigptBezierTD%
    \FigptBezier@#3[#4,#5,#6,#7]\Figp@intregTD#1:{#2}(\v@lX,\v@lY,\v@lZ)%
    \resetc@ntr@l\et@tfigptBezierTD}\ignorespaces\fi}
\ctr@ld@f\def\FigptBezier@#1[#2,#3,#4,#5]{\setc@ntr@l{2}%
    \edef\T@{#1}\v@leur=\p@\advance\v@leur-#1pt\edef\UNmT@{\repdecn@mb{\v@leur}}%
    \figptcopy-4:/#2/\figptcopy-3:/#3/\figptcopy-2:/#4/\figptcopy-1:/#5/%
    \l@mbd@un=-4 \l@mbd@de=-\thr@@\p@rtent=\m@ne\c@lDecast%
    \l@mbd@un=-4 \l@mbd@de=-\thr@@\p@rtent=-\tw@\c@lDecast%
    \l@mbd@un=-4 \l@mbd@de=-\thr@@\p@rtent=-\thr@@\c@lDecast\Figg@tXY{-4}}
\ctr@ln@m\c@lDCUn
\ctr@ld@f\def\c@lDCUnDD#1#2{\Figg@tXY{#1}\v@lX=\UNmT@\v@lX\v@lY=\UNmT@\v@lY%
    \Figg@tXYa{#2}\advance\v@lX\T@\v@lXa\advance\v@lY\T@\v@lYa%
    \Figp@intregDD#1:(\v@lX,\v@lY)}
\ctr@ld@f\def\c@lDCUnTD#1#2{\Figg@tXY{#1}\v@lX=\UNmT@\v@lX\v@lY=\UNmT@\v@lY\v@lZ=\UNmT@\v@lZ%
    \Figg@tXYa{#2}\advance\v@lX\T@\v@lXa\advance\v@lY\T@\v@lYa\advance\v@lZ\T@\v@lZa%
    \Figp@intregTD#1:(\v@lX,\v@lY,\v@lZ)}
\ctr@ld@f\def\c@lDecast{\relax\ifnum\l@mbd@un<\p@rtent\c@lDCUn{\l@mbd@un}{\l@mbd@de}%
    \advance\l@mbd@un\@ne\advance\l@mbd@de\@ne\c@lDecast\fi}
\ctr@ld@f\def\figptmap#1:#2=#3/#4/#5/{\ifps@cri{\s@uvc@ntr@l\et@tfigptmap%
    \setc@ntr@l{2}\figvectP-1[#4,#3]\Figg@tXY{-1}%
    \pr@dMatV/#5/\figpttra#1:{#2}=#4/1,-1/%
    \resetc@ntr@l\et@tfigptmap}\ignorespaces\fi}
\ctr@ln@m\pr@dMatV
\ctr@ld@f\def\pr@dMatVDD/#1,#2;#3,#4/{\v@lXa=#1\v@lX\advance\v@lXa#2\v@lY%
    \v@lYa=#3\v@lX\advance\v@lYa#4\v@lY\Figv@ctCreg-1(\v@lXa,\v@lYa)}
\ctr@ld@f\def\pr@dMatVTD/#1,#2,#3;#4,#5,#6;#7,#8,#9/{%
    \v@lXa=#1\v@lX\advance\v@lXa#2\v@lY\advance\v@lXa#3\v@lZ%
    \v@lYa=#4\v@lX\advance\v@lYa#5\v@lY\advance\v@lYa#6\v@lZ%
    \v@lZa=#7\v@lX\advance\v@lZa#8\v@lY\advance\v@lZa#9\v@lZ%
    \Figv@ctCreg-1(\v@lXa,\v@lYa,\v@lZa)}
\ctr@ln@m\figptbary
\ctr@ld@f\def\figptbaryDD#1:#2[#3;#4]{\ifps@cri{\edef\list@num{#3}\extrairelepremi@r\p@int\de\list@num%
    \s@mme=\z@\@ecfor\c@ef:=#4\do{\advance\s@mme\c@ef}%
    \edef\listec@ef{#4,0}\extrairelepremi@r\c@ef\de\listec@ef%
    \Figg@tXY{\p@int}\divide\v@lX\s@mme\divide\v@lY\s@mme%
    \multiply\v@lX\c@ef\multiply\v@lY\c@ef%
    \@ecfor\p@int:=\list@num\do{\extrairelepremi@r\c@ef\de\listec@ef%
           \Figg@tXYa{\p@int}\divide\v@lXa\s@mme\divide\v@lYa\s@mme%
           \multiply\v@lXa\c@ef\multiply\v@lYa\c@ef%
           \advance\v@lX\v@lXa\advance\v@lY\v@lYa}%
    \Figp@intregDD#1:{#2}(\v@lX,\v@lY)}\ignorespaces\fi}
\ctr@ld@f\def\figptbaryTD#1:#2[#3;#4]{\ifps@cri{\edef\list@num{#3}\extrairelepremi@r\p@int\de\list@num%
    \s@mme=\z@\@ecfor\c@ef:=#4\do{\advance\s@mme\c@ef}%
    \edef\listec@ef{#4,0}\extrairelepremi@r\c@ef\de\listec@ef%
    \Figg@tXY{\p@int}\divide\v@lX\s@mme\divide\v@lY\s@mme\divide\v@lZ\s@mme%
    \multiply\v@lX\c@ef\multiply\v@lY\c@ef\multiply\v@lZ\c@ef%
    \@ecfor\p@int:=\list@num\do{\extrairelepremi@r\c@ef\de\listec@ef%
           \Figg@tXYa{\p@int}\divide\v@lXa\s@mme\divide\v@lYa\s@mme\divide\v@lZa\s@mme%
           \multiply\v@lXa\c@ef\multiply\v@lYa\c@ef\multiply\v@lZa\c@ef%
           \advance\v@lX\v@lXa\advance\v@lY\v@lYa\advance\v@lZ\v@lZa}%
    \Figp@intregTD#1:{#2}(\v@lX,\v@lY,\v@lZ)}\ignorespaces\fi}
\ctr@ld@f\def\figptbaryR#1:#2[#3;#4]{\ifps@cri{%
    \v@leur=\z@\@ecfor\c@ef:=#4\do{\maxim@m{\v@lmax}{\c@ef pt}{-\c@ef pt}%
    \ifdim\v@lmax>\v@leur\v@leur=\v@lmax\fi}%
    \ifdim\v@leur<\p@\f@ctech=\@M\else\ifdim\v@leur<\t@n\p@\f@ctech=\@m\else%
    \ifdim\v@leur<\c@nt\p@\f@ctech=\c@nt\else\ifdim\v@leur<\@m\p@\f@ctech=\t@n\else%
    \f@ctech=\@ne\fi\fi\fi\fi%
    \def\listec@ef{0}%
    \@ecfor\c@ef:=#4\do{\sc@lec@nvRI{\c@ef pt}\edef\listec@ef{\listec@ef,\the\s@mme}}%
    \extrairelepremi@r\c@ef\de\listec@ef\figptbary#1:#2[#3;\listec@ef]}\ignorespaces\fi}
\ctr@ld@f\def\sc@lec@nvRI#1{\v@leur=#1\p@rtentiere{\s@mme}{\v@leur}\advance\v@leur-\s@mme\p@%
    \multiply\v@leur\f@ctech\p@rtentiere{\p@rtent}{\v@leur}%
    \multiply\s@mme\f@ctech\advance\s@mme\p@rtent}
\ctr@ln@m\figptcirc
\ctr@ld@f\def\figptcircDD#1:#2:#3;#4(#5){\ifps@cri{\s@uvc@ntr@l\et@tfigptcircDD%
    \c@lptellDD#1:{#2}:#3;#4,#4(#5)\resetc@ntr@l\et@tfigptcircDD}\ignorespaces\fi}
\ctr@ld@f\def\figptcircTD#1:#2:#3,#4,#5;#6(#7){\ifps@cri{\s@uvc@ntr@l\et@tfigptcircTD%
    \setc@ntr@l{2}\c@lExtAxes#3,#4,#5(#6)\figptellP#1:{#2}:#3,-4,-5(#7)%
    \resetc@ntr@l\et@tfigptcircTD}\ignorespaces\fi}
\ctr@ln@m\figptcircumcenter
\ctr@ld@f\def\figptcircumcenterDD#1:#2[#3,#4,#5]{\ifps@cri{\s@uvc@ntr@l\et@tfigptcircumcenterDD%
    \setc@ntr@l{2}\figvectNDD-5[#3,#4]\figptbaryDD-3:[#3,#4;1,1]%
                  \figvectNDD-6[#4,#5]\figptbaryDD-4:[#4,#5;1,1]%
    \resetc@ntr@l{2}\inters@cDD#1:{#2}[-3,-5;-4,-6]%
    \resetc@ntr@l\et@tfigptcircumcenterDD}\ignorespaces\fi}
\ctr@ld@f\def\figptcircumcenterTD#1:#2[#3,#4,#5]{\ifps@cri{\s@uvc@ntr@l\et@tfigptcircumcenterTD%
    \setc@ntr@l{2}\figvectNTD-1[#3,#4,#5]%
    \figvectPTD-3[#3,#4]\figvectNVTD-5[-1,-3]\figptbaryTD-3:[#3,#4;1,1]%
    \figvectPTD-4[#4,#5]\figvectNVTD-6[-1,-4]\figptbaryTD-4:[#4,#5;1,1]%
    \resetc@ntr@l{2}\inters@cTD#1:{#2}[-3,-5;-4,-6]%
    \resetc@ntr@l\et@tfigptcircumcenterTD}\ignorespaces\fi}
\ctr@ln@m\figptcopy
\ctr@ld@f\def\figptcopyDD#1:#2/#3/{\ifps@cri{\Figg@tXY{#3}%
    \Figp@intregDD#1:{#2}(\v@lX,\v@lY)}\ignorespaces\fi}
\ctr@ld@f\def\figptcopyTD#1:#2/#3/{\ifps@cri{\Figg@tXY{#3}%
    \Figp@intregTD#1:{#2}(\v@lX,\v@lY,\v@lZ)}\ignorespaces\fi}
\ctr@ln@m\figptcurvcenter
\ctr@ld@f\def\figptcurvcenterDD#1:#2:#3[#4,#5,#6,#7]{\ifps@cri{\s@uvc@ntr@l\et@tfigptcurvcenterDD%
    \setc@ntr@l{2}\c@lcurvradDD#3[#4,#5,#6,#7]\edef\Sprim@{\repdecn@mb{\result@t}}%
    \figptBezierDD-1::#3[#4,#5,#6,#7]\figpttraDD#1:{#2}=-1/\Sprim@,-5/%
    \resetc@ntr@l\et@tfigptcurvcenterDD}\ignorespaces\fi}
\ctr@ld@f\def\figptcurvcenterTD#1:#2:#3[#4,#5,#6,#7]{\ifps@cri{\s@uvc@ntr@l\et@tfigptcurvcenterTD%
    \setc@ntr@l{2}\figvectDBezierTD -5:1,#3[#4,#5,#6,#7]%
    \figvectDBezierTD -6:2,#3[#4,#5,#6,#7]\vecunit@TD{-5}{-5}%
    \edef\Sprim@{\repdecn@mb{\result@t}}\figvectNVTD-1[-6,-5]%
    \figvectNVTD-5[-5,-1]\c@lproscalTD\v@leur[-6,-5]%
    \invers@{\v@leur}{\v@leur}\v@leur=\Sprim@\v@leur\v@leur=\Sprim@\v@leur%
    \figptBezierTD-1::#3[#4,#5,#6,#7]\edef\Sprim@{\repdecn@mb{\v@leur}}%
    \figpttraTD#1:{#2}=-1/\Sprim@,-5/\resetc@ntr@l\et@tfigptcurvcenterTD}\ignorespaces\fi}
\ctr@ld@f\def\c@lcurvradDD#1[#2,#3,#4,#5]{{\figvectDBezierDD -5:1,#1[#2,#3,#4,#5]%
    \figvectDBezierDD -6:2,#1[#2,#3,#4,#5]\vecunit@DD{-5}{-5}%
    \edef\Sprim@{\repdecn@mb{\result@t}}\figvectNVDD-5[-5]\c@lproscalDD\v@leur[-6,-5]%
    \invers@{\v@leur}{\v@leur}\v@leur=\Sprim@\v@leur\v@leur=\Sprim@\v@leur%
    \global\result@t=\v@leur}}
\ctr@ln@m\figptell
\ctr@ld@f\def\figptellDD#1:#2:#3;#4,#5(#6,#7){\ifps@cri{\s@uvc@ntr@l\et@tfigptell%
    \c@lptellDD#1::#3;#4,#5(#6)\figptrotDD#1:{#2}=#1/#3,#7/%
    \resetc@ntr@l\et@tfigptell}\ignorespaces\fi}
\ctr@ld@f\def\c@lptellDD#1:#2:#3;#4,#5(#6){\c@ssin{\C@}{\S@}{#6}\v@lmin=\C@ pt\v@lmax=\S@ pt%
    \v@lmin=#4\v@lmin\v@lmax=#5\v@lmax%
    \edef\Xc@mp{\repdecn@mb{\v@lmin}}\edef\Yc@mp{\repdecn@mb{\v@lmax}}%
    \setc@ntr@l{2}\figvectC-1(\Xc@mp,\Yc@mp)\figpttraDD#1:{#2}=#3/1,-1/}
\ctr@ld@f\def\figptellP#1:#2:#3,#4,#5(#6){\ifps@cri{\s@uvc@ntr@l\et@tfigptellP%
    \setc@ntr@l{2}\figvectP-1[#3,#4]\figvectP-2[#3,#5]%
    \v@leur=#6pt\c@lptellP{#3}{-1}{-2}\figptcopy#1:{#2}/-3/%
    \resetc@ntr@l\et@tfigptellP}\ignorespaces\fi}
\ctr@ln@m\@ngle
\ctr@ld@f\def\c@lptellP#1#2#3{\edef\@ngle{\repdecn@mb\v@leur}\c@ssin{\C@}{\S@}{\@ngle}%
    \figpttra-3:=#1/\C@,#2/\figpttra-3:=-3/\S@,#3/}
\ctr@ln@m\figptendnormal
\ctr@ld@f\def\figptendnormalDD#1:#2:#3,#4[#5,#6]{\ifps@cri{\s@uvc@ntr@l\et@tfigptendnormal%
    \Figg@tXYa{#5}\Figg@tXY{#6}%
    \advance\v@lX-\v@lXa\advance\v@lY-\v@lYa%
    \setc@ntr@l{2}\Figv@ctCreg-1(\v@lX,\v@lY)\vecunit@{-1}{-1}\Figg@tXY{-1}%
    \delt@=#3\unit@\maxim@m{\delt@}{\delt@}{-\delt@}\edef\l@ngueur{\repdecn@mb{\delt@}}%
    \v@lX=\l@ngueur\v@lX\v@lY=\l@ngueur\v@lY%
    \delt@=\p@\advance\delt@-#4pt\edef\l@ngueur{\repdecn@mb{\delt@}}%
    \figptbaryR-1:[#5,#6;#4,\l@ngueur]\Figg@tXYa{-1}%
    \advance\v@lXa\v@lY\advance\v@lYa-\v@lX%
    \setc@ntr@l{1}\Figp@intregDD#1:{#2}(\v@lXa,\v@lYa)\resetc@ntr@l\et@tfigptendnormal}%
    \ignorespaces\fi}
\ctr@ld@f\def\figptexcenter#1:#2[#3,#4,#5]{\ifps@cri{\let@xte={-}%
    \Figptexinsc@nter#1:#2[#3,#4,#5]}\ignorespaces\fi}
\ctr@ld@f\def\figptincenter#1:#2[#3,#4,#5]{\ifps@cri{\let@xte={}%
    \Figptexinsc@nter#1:#2[#3,#4,#5]}\ignorespaces\fi}
\ctr@ld@f% pour compatibilite avec anciennes versions
\ctr@ld@f\def\Figptexinsc@nter#1:#2[#3,#4,#5]{%
    \figgetdist\LA@[#4,#5]\figgetdist\LB@[#3,#5]\figgetdist\LC@[#3,#4]%
    \figptbaryR#1:{#2}[#3,#4,#5;\the\let@xte\LA@,\LB@,\LC@]}
\ctr@ln@m\figptinterlineplane
\ctr@ld@f\def\figptinterlineplaneDD{\un@v@ilable{figptinterlineplane}}
\ctr@ld@f\def\figptinterlineplaneTD#1:#2[#3,#4;#5,#6]{\ifps@cri{\s@uvc@ntr@l\et@tfigptinterlineplane%
    \setc@ntr@l{2}\figvectPTD-1[#3,#5]\vecunit@TD{-2}{#6}%
    \r@pPSTD\v@leur[-2,-1,#4]\edef\v@lcoef{\repdecn@mb{\v@leur}}%
    \figpttraTD#1:{#2}=#3/\v@lcoef,#4/\resetc@ntr@l\et@tfigptinterlineplane}\ignorespaces\fi}
\ctr@ln@m\figptorthocenter
\ctr@ld@f\def\figptorthocenterDD#1:#2[#3,#4,#5]{\ifps@cri{\s@uvc@ntr@l\et@tfigptorthocenterDD%
    \setc@ntr@l{2}\figvectNDD-3[#3,#4]\figvectNDD-4[#4,#5]%
    \resetc@ntr@l{2}\inters@cDD#1:{#2}[#5,-3;#3,-4]%
    \resetc@ntr@l\et@tfigptorthocenterDD}\ignorespaces\fi}
\ctr@ld@f\def\figptorthocenterTD#1:#2[#3,#4,#5]{\ifps@cri{\s@uvc@ntr@l\et@tfigptorthocenterTD%
    \setc@ntr@l{2}\figvectNTD-1[#3,#4,#5]%
    \figvectPTD-2[#3,#4]\figvectNVTD-3[-1,-2]%
    \figvectPTD-2[#4,#5]\figvectNVTD-4[-1,-2]%
    \resetc@ntr@l{2}\inters@cTD#1:{#2}[#5,-3;#3,-4]%
    \resetc@ntr@l\et@tfigptorthocenterTD}\ignorespaces\fi}
\ctr@ln@m\figptorthoprojline
\ctr@ld@f\def\figptorthoprojlineDD#1:#2=#3/#4,#5/{\ifps@cri{\s@uvc@ntr@l\et@tfigptorthoprojlineDD%
    \setc@ntr@l{2}\figvectPDD-3[#4,#5]\figvectNVDD-4[-3]\resetc@ntr@l{2}%
    \inters@cDD#1:{#2}[#3,-4;#4,-3]\resetc@ntr@l\et@tfigptorthoprojlineDD}\ignorespaces\fi}
\ctr@ld@f\def\figptorthoprojlineTD#1:#2=#3/#4,#5/{\ifps@cri{\s@uvc@ntr@l\et@tfigptorthoprojlineTD%
    \setc@ntr@l{2}\figvectPTD-1[#4,#3]\figvectPTD-2[#4,#5]\vecunit@TD{-2}{-2}%
    \c@lproscalTD\v@leur[-1,-2]\edef\v@lcoef{\repdecn@mb{\v@leur}}%
    \figpttraTD#1:{#2}=#4/\v@lcoef,-2/\resetc@ntr@l\et@tfigptorthoprojlineTD}\ignorespaces\fi}
\ctr@ln@m\figptorthoprojplane
\ctr@ld@f\def\figptorthoprojplaneDD{\un@v@ilable{figptorthoprojplane}}
\ctr@ld@f\def\figptorthoprojplaneTD#1:#2=#3/#4,#5/{\ifps@cri{\s@uvc@ntr@l\et@tfigptorthoprojplane%
    \setc@ntr@l{2}\figvectPTD-1[#3,#4]\vecunit@TD{-2}{#5}%
    \c@lproscalTD\v@leur[-1,-2]\edef\v@lcoef{\repdecn@mb{\v@leur}}%
    \figpttraTD#1:{#2}=#3/\v@lcoef,-2/\resetc@ntr@l\et@tfigptorthoprojplane}\ignorespaces\fi}
\ctr@ld@f\def\figpthom#1:#2=#3/#4,#5/{\ifps@cri{\s@uvc@ntr@l\et@tfigpthom%
    \setc@ntr@l{2}\figvectP-1[#4,#3]\figpttra#1:{#2}=#4/#5,-1/%
    \resetc@ntr@l\et@tfigpthom}\ignorespaces\fi}
\ctr@ln@m\figptrot
\ctr@ld@f\def\figptrotDD#1:#2=#3/#4,#5/{\ifps@cri{\s@uvc@ntr@l\et@tfigptrotDD%
    \c@ssin{\C@}{\S@}{#5}\setc@ntr@l{2}\figvectPDD-1[#4,#3]\Figg@tXY{-1}%
    \v@lXa=\C@\v@lX\advance\v@lXa-\S@\v@lY%
    \v@lYa=\S@\v@lX\advance\v@lYa\C@\v@lY%
    \Figv@ctCreg-1(\v@lXa,\v@lYa)\figpttraDD#1:{#2}=#4/1,-1/%
    \resetc@ntr@l\et@tfigptrotDD}\ignorespaces\fi}
\ctr@ld@f\def\figptrotTD#1:#2=#3/#4,#5,#6/{\ifps@cri{\s@uvc@ntr@l\et@tfigptrotTD%
    \c@ssin{\C@}{\S@}{#5}%
    \setc@ntr@l{2}\figptorthoprojplaneTD-3:=#4/#3,#6/\figvectPTD-2[-3,#3]%
    \n@rmeucTD\v@leur{-2}\ifdim\v@leur<\Cepsil@n\Figg@tXYa{#3}\else%
    \edef\v@lcoef{\repdecn@mb{\v@leur}}\figvectNVTD-1[#6,-2]%
    \Figg@tXYa{-1}\v@lXa=\v@lcoef\v@lXa\v@lYa=\v@lcoef\v@lYa\v@lZa=\v@lcoef\v@lZa%
    \v@lXa=\S@\v@lXa\v@lYa=\S@\v@lYa\v@lZa=\S@\v@lZa\Figg@tXY{-2}%
    \advance\v@lXa\C@\v@lX\advance\v@lYa\C@\v@lY\advance\v@lZa\C@\v@lZ%
    \Figg@tXY{-3}\advance\v@lXa\v@lX\advance\v@lYa\v@lY\advance\v@lZa\v@lZ\fi%
    \Figp@intregTD#1:{#2}(\v@lXa,\v@lYa,\v@lZa)\resetc@ntr@l\et@tfigptrotTD}\ignorespaces\fi}
\ctr@ln@m\figptsym
\ctr@ld@f\def\figptsymDD#1:#2=#3/#4,#5/{\ifps@cri{\s@uvc@ntr@l\et@tfigptsymDD%
    \resetc@ntr@l{2}\figptorthoprojlineDD-5:=#3/#4,#5/\figvectPDD-2[#3,-5]%
    \figpttraDD#1:{#2}=#3/2,-2/\resetc@ntr@l\et@tfigptsymDD}\ignorespaces\fi}
\ctr@ld@f\def\figptsymTD#1:#2=#3/#4,#5/{\ifps@cri{\s@uvc@ntr@l\et@tfigptsymTD%
    \resetc@ntr@l{2}\figptorthoprojplaneTD-3:=#3/#4,#5/\figvectPTD-2[#3,-3]%
    \figpttraTD#1:{#2}=#3/2,-2/\resetc@ntr@l\et@tfigptsymTD}\ignorespaces\fi}
\ctr@ln@m\figpttra
\ctr@ld@f\def\figpttraDD#1:#2=#3/#4,#5/{\ifps@cri{\Figg@tXYa{#5}\v@lXa=#4\v@lXa\v@lYa=#4\v@lYa%
    \Figg@tXY{#3}\advance\v@lX\v@lXa\advance\v@lY\v@lYa%
    \Figp@intregDD#1:{#2}(\v@lX,\v@lY)}\ignorespaces\fi}
\ctr@ld@f\def\figpttraTD#1:#2=#3/#4,#5/{\ifps@cri{\Figg@tXYa{#5}\v@lXa=#4\v@lXa\v@lYa=#4\v@lYa%
    \v@lZa=#4\v@lZa\Figg@tXY{#3}\advance\v@lX\v@lXa\advance\v@lY\v@lYa%
    \advance\v@lZ\v@lZa\Figp@intregTD#1:{#2}(\v@lX,\v@lY,\v@lZ)}\ignorespaces\fi}
\ctr@ln@m\figpttraC
\ctr@ld@f\def\figpttraCDD#1:#2=#3/#4,#5/{\ifps@cri{\v@lXa=#4\unit@\v@lYa=#5\unit@%
    \Figg@tXY{#3}\advance\v@lX\v@lXa\advance\v@lY\v@lYa%
    \Figp@intregDD#1:{#2}(\v@lX,\v@lY)}\ignorespaces\fi}
\ctr@ld@f\def\figpttraCTD#1:#2=#3/#4,#5,#6/{\ifps@cri{\v@lXa=#4\unit@\v@lYa=#5\unit@\v@lZa=#6\unit@%
    \Figg@tXY{#3}\advance\v@lX\v@lXa\advance\v@lY\v@lYa\advance\v@lZ\v@lZa%
    \Figp@intregTD#1:{#2}(\v@lX,\v@lY,\v@lZ)}\ignorespaces\fi}
\ctr@ld@f\def\figptsaxes#1:#2(#3){\ifps@cri{\an@lys@xes#3,:\ifx\t@xt@\empty%
    \ifTr@isDim\Figpts@xes#1:#2(0,#3,0,#3,0,#3)\else\Figpts@xes#1:#2(0,#3,0,#3)\fi%
    \else\Figpts@xes#1:#2(#3)\fi}\ignorespaces\fi}
\ctr@ln@m\Figpts@xes
\ctr@ld@f\def\Figpts@xesDD#1:#2(#3,#4,#5,#6){%
    \s@mme=#1\figpttraC\the\s@mme:$x$=#2/#4,0/%
    \advance\s@mme\@ne\figpttraC\the\s@mme:$y$=#2/0,#6/}
\ctr@ld@f\def\Figpts@xesTD#1:#2(#3,#4,#5,#6,#7,#8){%
    \s@mme=#1\figpttraC\the\s@mme:$x$=#2/#4,0,0/%
    \advance\s@mme\@ne\figpttraC\the\s@mme:$y$=#2/0,#6,0/%
    \advance\s@mme\@ne\figpttraC\the\s@mme:$z$=#2/0,0,#8/}
\ctr@ld@f\def\figptsmap#1=#2/#3/#4/{\ifps@cri{\s@uvc@ntr@l\et@tfigptsmap%
    \setc@ntr@l{2}\def\list@num{#2}\s@mme=#1%
    \@ecfor\p@int:=\list@num\do{\figvectP-1[#3,\p@int]\Figg@tXY{-1}%
    \pr@dMatV/#4/\figpttra\the\s@mme:=#3/1,-1/\advance\s@mme\@ne}%
    \resetc@ntr@l\et@tfigptsmap}\ignorespaces\fi}
\ctr@ln@m\figptscontrol
\ctr@ld@f\def\figptscontrolDD#1[#2,#3,#4,#5]{\ifps@cri{\s@uvc@ntr@l\et@tfigptscontrolDD\setc@ntr@l{2}%
    \v@lX=\z@\v@lY=\z@\Figtr@nptDD{-5}{#2}\Figtr@nptDD{2}{#5}%
    \divide\v@lX\@vi\divide\v@lY\@vi%
    \Figtr@nptDD{3}{#3}\Figtr@nptDD{-1.5}{#4}\Figp@intregDD-1:(\v@lX,\v@lY)%
    \v@lX=\z@\v@lY=\z@\Figtr@nptDD{2}{#2}\Figtr@nptDD{-5}{#5}%
    \divide\v@lX\@vi\divide\v@lY\@vi\Figtr@nptDD{-1.5}{#3}\Figtr@nptDD{3}{#4}%
    \s@mme=#1\advance\s@mme\@ne\Figp@intregDD\the\s@mme:(\v@lX,\v@lY)%
    \figptcopyDD#1:/-1/\resetc@ntr@l\et@tfigptscontrolDD}\ignorespaces\fi}
\ctr@ld@f\def\figptscontrolTD#1[#2,#3,#4,#5]{\ifps@cri{\s@uvc@ntr@l\et@tfigptscontrolTD\setc@ntr@l{2}%
    \v@lX=\z@\v@lY=\z@\v@lZ=\z@\Figtr@nptTD{-5}{#2}\Figtr@nptTD{2}{#5}%
    \divide\v@lX\@vi\divide\v@lY\@vi\divide\v@lZ\@vi%
    \Figtr@nptTD{3}{#3}\Figtr@nptTD{-1.5}{#4}\Figp@intregTD-1:(\v@lX,\v@lY,\v@lZ)%
    \v@lX=\z@\v@lY=\z@\v@lZ=\z@\Figtr@nptTD{2}{#2}\Figtr@nptTD{-5}{#5}%
    \divide\v@lX\@vi\divide\v@lY\@vi\divide\v@lZ\@vi\Figtr@nptTD{-1.5}{#3}\Figtr@nptTD{3}{#4}%
    \s@mme=#1\advance\s@mme\@ne\Figp@intregTD\the\s@mme:(\v@lX,\v@lY,\v@lZ)%
    \figptcopyTD#1:/-1/\resetc@ntr@l\et@tfigptscontrolTD}\ignorespaces\fi}
\ctr@ld@f\def\Figtr@nptDD#1#2{\Figg@tXYa{#2}\v@lXa=#1\v@lXa\v@lYa=#1\v@lYa%
    \advance\v@lX\v@lXa\advance\v@lY\v@lYa}
\ctr@ld@f\def\Figtr@nptTD#1#2{\Figg@tXYa{#2}\v@lXa=#1\v@lXa\v@lYa=#1\v@lYa\v@lZa=#1\v@lZa%
    \advance\v@lX\v@lXa\advance\v@lY\v@lYa\advance\v@lZ\v@lZa}
\ctr@ld@f\def\figptscontrolcurve#1,#2[#3]{\ifps@cri{\s@uvc@ntr@l\et@tfigptscontrolcurve%
    \def\list@num{#3}\extrairelepremi@r\Ak@\de\list@num%
    \extrairelepremi@r\Ai@\de\list@num\extrairelepremi@r\Aj@\de\list@num%
    \s@mme=#1\figptcopy\the\s@mme:/\Ai@/%
    \setc@ntr@l{2}\figvectP -1[\Ak@,\Aj@]%
    \@ecfor\Ak@:=\list@num\do{\advance\s@mme\@ne\figpttra\the\s@mme:=\Ai@/\curv@roundness,-1/%
       \figvectP -1[\Ai@,\Ak@]\advance\s@mme\@ne\figpttra\the\s@mme:=\Aj@/-\curv@roundness,-1/%
       \advance\s@mme\@ne\figptcopy\the\s@mme:/\Aj@/%
       \edef\Ai@{\Aj@}\edef\Aj@{\Ak@}}\advance\s@mme-#1\divide\s@mme\thr@@%
       \xdef#2{\the\s@mme}%
    \resetc@ntr@l\et@tfigptscontrolcurve}\ignorespaces\fi}
\ctr@ln@m\figptsintercirc
\ctr@ld@f\def\figptsintercircDD#1[#2,#3;#4,#5]{\ifps@cri{\s@uvc@ntr@l\et@tfigptsintercircDD%
    \setc@ntr@l{2}\let\c@lNVintc=\c@lNVintcDD\Figptsintercirc@#1[#2,#3;#4,#5]%    
    \resetc@ntr@l\et@tfigptsintercircDD}\ignorespaces\fi}
\ctr@ld@f\def\figptsintercircTD#1[#2,#3;#4,#5;#6]{\ifps@cri{\s@uvc@ntr@l\et@tfigptsintercircTD%
    \setc@ntr@l{2}\let\c@lNVintc=\c@lNVintcTD\vecunitC@TD[#2,#6]%
    \Figv@ctCreg-3(\v@lX,\v@lY,\v@lZ)\Figptsintercirc@#1[#2,#3;#4,#5]%
    \resetc@ntr@l\et@tfigptsintercircTD}\ignorespaces\fi}
\ctr@ld@f\def\Figptsintercirc@#1[#2,#3;#4,#5]{\figvectP-1[#2,#4]%
    \vecunit@{-1}{-1}\delt@=\result@t\f@ctech=\result@tent%
    \s@mme=#1\advance\s@mme\@ne\figptcopy#1:/#2/\figptcopy\the\s@mme:/#4/%
    \ifdim\delt@=\z@\else%
    \v@lmin=#3\unit@\v@lmax=#5\unit@\v@leur=\v@lmin\advance\v@leur\v@lmax%
    \ifdim\v@leur>\delt@%
    \v@leur=\v@lmin\advance\v@leur-\v@lmax\maxim@m{\v@leur}{\v@leur}{-\v@leur}%
    \ifdim\v@leur<\delt@%
    \divide\v@lmin\f@ctech\divide\v@lmax\f@ctech\divide\delt@\f@ctech%
    \v@lmin=\repdecn@mb{\v@lmin}\v@lmin\v@lmax=\repdecn@mb{\v@lmax}\v@lmax%
    \invers@{\v@leur}{\delt@}\advance\v@lmax-\v@lmin%
    \v@lmax=-\repdecn@mb{\v@leur}\v@lmax\advance\delt@\v@lmax\delt@=.5\delt@%
    \v@lmax=\delt@\multiply\v@lmax\f@ctech%
    \edef\t@ille{\repdecn@mb{\v@lmax}}\figpttra-2:=#2/\t@ille,-1/%
    \delt@=\repdecn@mb{\delt@}\delt@\advance\v@lmin-\delt@%
    \sqrt@{\v@leur}{\v@lmin}\multiply\v@leur\f@ctech\edef\t@ille{\repdecn@mb{\v@leur}}%
    \c@lNVintc\figpttra#1:=-2/-\t@ille,-1/\figpttra\the\s@mme:=-2/\t@ille,-1/\fi\fi\fi}
\ctr@ld@f\def\c@lNVintcDD{\Figg@tXY{-1}\Figv@ctCreg-1(-\v@lY,\v@lX)} % <=> \figvectNVDD-1[-1]
\ctr@ld@f\def\c@lNVintcTD{{\Figg@tXY{-3}\v@lmin=\v@lX\v@lmax=\v@lY\v@leur=\v@lZ%
    \Figg@tXY{-1}\c@lprovec{-3}\vecunit@{-3}{-3}% <=> \figvectNVTD-3[-1,-3]\vecunit@{-3}{-3}
    \Figg@tXY{-1}\v@lmin=\v@lX\v@lmax=\v@lY%
    \v@leur=\v@lZ\Figg@tXY{-3}\c@lprovec{-1}}} % <=> \figvectNVTD-1[-3,-1]
\ctr@ln@m\figptsinterlinell
\ctr@ld@f\def\figptsinterlinellDD#1[#2,#3,#4,#5;#6,#7]{\ifps@cri{\s@uvc@ntr@l\et@tfigptsinterlinellDD%
    \figptcopy#1:/#6/\s@mme=#1\advance\s@mme\@ne\figptcopy\the\s@mme:/#7/%
    \v@lmin=#3\unit@\v@lmax=#4\unit@% a, b
    \setc@ntr@l{2}\figptbaryDD-4:[#6,#7;1,1]\figptsrotDD-3=-4,#7/#2,-#5/% D et rotation
    \Figg@tXY{-3}\Figg@tXYa{#2}\advance\v@lX-\v@lXa\advance\v@lY-\v@lYa% alpha, beta
    \figvectP-1[-3,-2]\Figg@tXYa{-1}\figvectP-3[-4,#7]\Figptsint@rLE{#1}% u1, u2
    \resetc@ntr@l\et@tfigptsinterlinellDD}\ignorespaces\fi}
\ctr@ld@f\def\figptsinterlinellP#1[#2,#3,#4;#5,#6]{\ifps@cri{\s@uvc@ntr@l\et@tfigptsinterlinellP%
    \figptcopy#1:/#5/\s@mme=#1\advance\s@mme\@ne\figptcopy\the\s@mme:/#6/\setc@ntr@l{2}%
    \figvectP-1[#2,#3]\vecunit@{-1}{-1}\v@lmin=\result@t% a
    \figvectP-2[#2,#4]\vecunit@{-2}{-2}\v@lmax=\result@t% b
    \figptbary-4:[#5,#6;1,1]% D
    \figvectP-3[#2,-4]\c@lproscal\v@lX[-3,-1]\c@lproscal\v@lY[-3,-2]% alpha, beta
    \figvectP-3[-4,#6]\c@lproscal\v@lXa[-3,-1]\c@lproscal\v@lYa[-3,-2]% u1, u2
    \Figptsint@rLE{#1}\resetc@ntr@l\et@tfigptsinterlinellP}\ignorespaces\fi}
\ctr@ld@f\def\Figptsint@rLE#1{%
    \getredf@ctDD\f@ctech(\v@lmin,\v@lmax)%
    \getredf@ctDD\p@rtent(\v@lX,\v@lY)\ifnum\p@rtent>\f@ctech\f@ctech=\p@rtent\fi%
    \getredf@ctDD\p@rtent(\v@lXa,\v@lYa)\ifnum\p@rtent>\f@ctech\f@ctech=\p@rtent\fi%
    \divide\v@lmin\f@ctech\divide\v@lmax\f@ctech\divide\v@lX\f@ctech\divide\v@lY\f@ctech%
    \divide\v@lXa\f@ctech\divide\v@lYa\f@ctech%
    \c@rre=\repdecn@mb\v@lXa\v@lmax\mili@u=\repdecn@mb\v@lYa\v@lmin%
    \getredf@ctDD\f@ctech(\c@rre,\mili@u)%
    \c@rre=\repdecn@mb\v@lX\v@lmax\mili@u=\repdecn@mb\v@lY\v@lmin%
    \getredf@ctDD\p@rtent(\c@rre,\mili@u)\ifnum\p@rtent>\f@ctech\f@ctech=\p@rtent\fi%
    \divide\v@lmin\f@ctech\divide\v@lmax\f@ctech\divide\v@lX\f@ctech\divide\v@lY\f@ctech%
    \divide\v@lXa\f@ctech\divide\v@lYa\f@ctech%
    \v@lmin=\repdecn@mb{\v@lmin}\v@lmin\v@lmax=\repdecn@mb{\v@lmax}\v@lmax%
    \edef\G@xde{\repdecn@mb\v@lmin}\edef\P@xde{\repdecn@mb\v@lmax}%
    \c@rre=-\v@lmax\v@leur=\repdecn@mb\v@lY\v@lY\advance\c@rre\v@leur\c@rre=\G@xde\c@rre%
    \v@leur=\repdecn@mb\v@lX\v@lX\v@leur=\P@xde\v@leur\advance\c@rre\v@leur% C
    \v@lmin=\repdecn@mb\v@lYa\v@lmin\v@lmax=\repdecn@mb\v@lXa\v@lmax%
    \mili@u=\repdecn@mb\v@lX\v@lmax\advance\mili@u\repdecn@mb\v@lY\v@lmin% B
    \v@lmax=\repdecn@mb\v@lXa\v@lmax\advance\v@lmax\repdecn@mb\v@lYa\v@lmin% A
    \ifdim\v@lmax>\epsil@n%
    \maxim@m{\v@leur}{\c@rre}{-\c@rre}\maxim@m{\v@lmin}{\mili@u}{-\mili@u}%
    \maxim@m{\v@leur}{\v@leur}{\v@lmin}\maxim@m{\v@lmin}{\v@lmax}{-\v@lmax}%
    \maxim@m{\v@leur}{\v@leur}{\v@lmin}\p@rtentiere{\p@rtent}{\v@leur}\advance\p@rtent\@ne%
    \divide\c@rre\p@rtent\divide\mili@u\p@rtent\divide\v@lmax\p@rtent%
    \delt@=\repdecn@mb{\mili@u}\mili@u\v@leur=\repdecn@mb{\v@lmax}\c@rre%
    \advance\delt@-\v@leur\ifdim\delt@<\z@\else\sqrt@\delt@\delt@%
    \invers@\v@lmax\v@lmax\edef\Uns@rAp{\repdecn@mb\v@lmax}%
    \v@leur=-\mili@u\advance\v@leur-\delt@\v@leur=\Uns@rAp\v@leur%
    \edef\t@ille{\repdecn@mb\v@leur}\figpttra#1:=-4/\t@ille,-3/\s@mme=#1\advance\s@mme\@ne%
    \v@leur=-\mili@u\advance\v@leur\delt@\v@leur=\Uns@rAp\v@leur%
    \edef\t@ille{\repdecn@mb\v@leur}\figpttra\the\s@mme:=-4/\t@ille,-3/\fi\fi}
\ctr@ln@m\figptsorthoprojline
\ctr@ld@f\def\figptsorthoprojlineDD#1=#2/#3,#4/{\ifps@cri{\s@uvc@ntr@l\et@tfigptsorthoprojlineDD%
    \setc@ntr@l{2}\figvectPDD-3[#3,#4]\figvectNVDD-4[-3]\resetc@ntr@l{2}%
    \def\list@num{#2}\s@mme=#1\@ecfor\p@int:=\list@num\do{%
    \inters@cDD\the\s@mme:[\p@int,-4;#3,-3]\advance\s@mme\@ne}%
    \resetc@ntr@l\et@tfigptsorthoprojlineDD}\ignorespaces\fi}
\ctr@ld@f\def\figptsorthoprojlineTD#1=#2/#3,#4/{\ifps@cri{\s@uvc@ntr@l\et@tfigptsorthoprojlineTD%
    \setc@ntr@l{2}\figvectPTD-2[#3,#4]\vecunit@TD{-2}{-2}%
    \def\list@num{#2}\s@mme=#1\@ecfor\p@int:=\list@num\do{%
    \figvectPTD-1[#3,\p@int]\c@lproscalTD\v@leur[-1,-2]%
    \edef\v@lcoef{\repdecn@mb{\v@leur}}\figpttraTD\the\s@mme:=#3/\v@lcoef,-2/%
    \advance\s@mme\@ne}\resetc@ntr@l\et@tfigptsorthoprojlineTD}\ignorespaces\fi}
\ctr@ln@m\figptsorthoprojplane
\ctr@ld@f\def\figptsorthoprojplaneDD{\un@v@ilable{figptsorthoprojplane}}
\ctr@ld@f\def\figptsorthoprojplaneTD#1=#2/#3,#4/{\ifps@cri{\s@uvc@ntr@l\et@tfigptsorthoprojplane%
    \setc@ntr@l{2}\vecunit@TD{-2}{#4}%
    \def\list@num{#2}\s@mme=#1\@ecfor\p@int:=\list@num\do{\figvectPTD-1[\p@int,#3]%
    \c@lproscalTD\v@leur[-1,-2]\edef\v@lcoef{\repdecn@mb{\v@leur}}%
    \figpttraTD\the\s@mme:=\p@int/\v@lcoef,-2/\advance\s@mme\@ne}%
    \resetc@ntr@l\et@tfigptsorthoprojplane}\ignorespaces\fi}
\ctr@ld@f\def\figptshom#1=#2/#3,#4/{\ifps@cri{\s@uvc@ntr@l\et@tfigptshom%
    \setc@ntr@l{2}\def\list@num{#2}\s@mme=#1%
    \@ecfor\p@int:=\list@num\do{\figvectP-1[#3,\p@int]%
    \figpttra\the\s@mme:=#3/#4,-1/\advance\s@mme\@ne}%
    \resetc@ntr@l\et@tfigptshom}\ignorespaces\fi}
\ctr@ln@m\figptsrot
\ctr@ld@f\def\figptsrotDD#1=#2/#3,#4/{\ifps@cri{\s@uvc@ntr@l\et@tfigptsrotDD%
    \c@ssin{\C@}{\S@}{#4}\setc@ntr@l{2}\def\list@num{#2}\s@mme=#1%
    \@ecfor\p@int:=\list@num\do{\figvectPDD-1[#3,\p@int]\Figg@tXY{-1}%
    \v@lXa=\C@\v@lX\advance\v@lXa-\S@\v@lY%
    \v@lYa=\S@\v@lX\advance\v@lYa\C@\v@lY%
    \Figv@ctCreg-1(\v@lXa,\v@lYa)\figpttraDD\the\s@mme:=#3/1,-1/\advance\s@mme\@ne}%
    \resetc@ntr@l\et@tfigptsrotDD}\ignorespaces\fi}
\ctr@ld@f\def\figptsrotTD#1=#2/#3,#4,#5/{\ifps@cri{\s@uvc@ntr@l\et@tfigptsrotTD%
    \c@ssin{\C@}{\S@}{#4}%
    \setc@ntr@l{2}\def\list@num{#2}\s@mme=#1%
    \@ecfor\p@int:=\list@num\do{\figptorthoprojplaneTD-3:=#3/\p@int,#5/%
    \figvectPTD-2[-3,\p@int]%
    \figvectNVTD-1[#5,-2]\n@rmeucTD\v@leur{-2}\edef\v@lcoef{\repdecn@mb{\v@leur}}%
    \Figg@tXYa{-1}\v@lXa=\v@lcoef\v@lXa\v@lYa=\v@lcoef\v@lYa\v@lZa=\v@lcoef\v@lZa%
    \v@lXa=\S@\v@lXa\v@lYa=\S@\v@lYa\v@lZa=\S@\v@lZa\Figg@tXY{-2}%
    \advance\v@lXa\C@\v@lX\advance\v@lYa\C@\v@lY\advance\v@lZa\C@\v@lZ%
    \Figg@tXY{-3}\advance\v@lXa\v@lX\advance\v@lYa\v@lY\advance\v@lZa\v@lZ%
    \Figp@intregTD\the\s@mme:(\v@lXa,\v@lYa,\v@lZa)\advance\s@mme\@ne}%
    \resetc@ntr@l\et@tfigptsrotTD}\ignorespaces\fi}
\ctr@ln@m\figptssym
\ctr@ld@f\def\figptssymDD#1=#2/#3,#4/{\ifps@cri{\s@uvc@ntr@l\et@tfigptssymDD%
    \setc@ntr@l{2}\figvectPDD-3[#3,#4]\Figg@tXY{-3}\Figv@ctCreg-4(-\v@lY,\v@lX)%
    \resetc@ntr@l{2}\def\list@num{#2}\s@mme=#1%
    \@ecfor\p@int:=\list@num\do{\inters@cDD-5:[#3,-3;\p@int,-4]\figvectPDD-2[\p@int,-5]%
    \figpttraDD\the\s@mme:=\p@int/2,-2/\advance\s@mme\@ne}%
    \resetc@ntr@l\et@tfigptssymDD}\ignorespaces\fi}
\ctr@ld@f\def\figptssymTD#1=#2/#3,#4/{\ifps@cri{\s@uvc@ntr@l\et@tfigptssymTD%
    \setc@ntr@l{2}\vecunit@TD{-2}{#4}\def\list@num{#2}\s@mme=#1%
    \@ecfor\p@int:=\list@num\do{\figvectPTD-1[\p@int,#3]%
    \c@lproscalTD\v@leur[-1,-2]\v@leur=2\v@leur\edef\v@lcoef{\repdecn@mb{\v@leur}}%
    \figpttraTD\the\s@mme:=\p@int/\v@lcoef,-2/\advance\s@mme\@ne}%
    \resetc@ntr@l\et@tfigptssymTD}\ignorespaces\fi}
\ctr@ln@m\figptstra
\ctr@ld@f\def\figptstraDD#1=#2/#3,#4/{\ifps@cri{\Figg@tXYa{#4}\v@lXa=#3\v@lXa\v@lYa=#3\v@lYa%
    \def\list@num{#2}\s@mme=#1\@ecfor\p@int:=\list@num\do{\Figg@tXY{\p@int}%
    \advance\v@lX\v@lXa\advance\v@lY\v@lYa%
    \Figp@intregDD\the\s@mme:(\v@lX,\v@lY)\advance\s@mme\@ne}}\ignorespaces\fi}
\ctr@ld@f\def\figptstraTD#1=#2/#3,#4/{\ifps@cri{\Figg@tXYa{#4}\v@lXa=#3\v@lXa\v@lYa=#3\v@lYa%
    \v@lZa=#3\v@lZa\def\list@num{#2}\s@mme=#1\@ecfor\p@int:=\list@num\do{\Figg@tXY{\p@int}%
    \advance\v@lX\v@lXa\advance\v@lY\v@lYa\advance\v@lZ\v@lZa%
    \Figp@intregTD\the\s@mme:(\v@lX,\v@lY,\v@lZ)\advance\s@mme\@ne}}\ignorespaces\fi}
\ctr@ln@m\figptvisilimSL
\ctr@ld@f\def\figptvisilimSLDD{\un@v@ilable{figptvisilimSL}}
\ctr@ld@f\def\figptvisilimSLTD#1:#2[#3,#4;#5,#6]{\ifps@cri{\s@uvc@ntr@l\et@tfigptvisilimSLTD%
    \setc@ntr@l{2}\figvectP-1[#3,#4]\n@rminf{\delt@}{-1}%
    \ifcase\curr@ntproj\v@lX=\cxa@\p@\v@lY=-\p@\v@lZ=\cxb@\p@% Proj cav
    \Figv@ctCreg-2(\v@lX,\v@lY,\v@lZ)\figvectP-3[#5,#6]\figvectNV-1[-2,-3]%
    \or\figvectP-1[#5,#6]\vecunitCV@TD{-1}\v@lmin=\v@lX\v@lmax=\v@lY% Proj ortho
    \v@leur=\v@lZ\v@lX=\cza@\p@\v@lY=\czb@\p@\v@lZ=\czc@\p@\c@lprovec{-1}%
    \or\c@ley@pt{-2}\figvectN-1[#5,#6,-2]\fi% Proj rea
    \edef\Ai@{#3}\edef\Aj@{#4}\figvectP-2[#5,\Ai@]\c@lproscal\v@leur[-1,-2]%
    \ifdim\v@leur>\z@\p@rtent=\@ne\else\p@rtent=\m@ne\fi%
    \figvectP-2[#5,\Aj@]\c@lproscal\v@leur[-1,-2]%
    \ifdim\p@rtent\v@leur>\z@\figptcopy#1:#2/#3/%
    \message{*** \BS@ figptvisilimSL: points are on the same side.}\else%
    \figptcopy-3:/#3/\figptcopy-4:/#4/%
    \loop\figptbary-5:[-3,-4;1,1]\figvectP-2[#5,-5]\c@lproscal\v@leur[-1,-2]%
    \ifdim\p@rtent\v@leur>\z@\figptcopy-3:/-5/\else\figptcopy-4:/-5/\fi%
    \divide\delt@\tw@\ifdim\delt@>\epsil@n\repeat%
    \figptbary#1:#2[-3,-4;1,1]\fi\resetc@ntr@l\et@tfigptvisilimSLTD}\ignorespaces\fi}
\ctr@ld@f\def\c@ley@pt#1{\t@stp@r\ifitis@K\v@lX=\cza@\p@\v@lY=\czb@\p@\v@lZ=\czc@\p@%
    \Figv@ctCreg-1(\v@lX,\v@lY,\v@lZ)\Figp@intreg-2:(\wd\Bt@rget,\ht\Bt@rget,\dp\Bt@rget)%
    \figpttra#1:=-2/-\disob@intern,-1/\else\end\fi}
\ctr@ld@f\def\t@stp@r{\itis@Ktrue\ifnewt@rgetpt\else\itis@Kfalse%
    \message{*** \BS@ figptvisilimXX: target point undefined.}\fi\ifnewdis@b\else%
    \itis@Kfalse\message{*** \BS@ figptvisilimXX: observation distance undefined.}\fi%
    \ifitis@K\else\message{*** This macro must be called after \BS@ psbeginfig or after
    having set the missing parameter(s) with \BS@ figset proj()}\fi}
\ctr@ld@f\def\figscan#1(#2,#3){{\s@uvc@ntr@l\et@tfigscan\@psfgetbb{#1}\if@psfbbfound\else%
    \def\@psfllx{0}\def\@psflly{20}\def\@psfurx{540}\def\@psfury{640}\fi\figscan@{#2}{#3}%
    \resetc@ntr@l\et@tfigscan}\ignorespaces}
\ctr@ld@f\def\figscan@#1#2{%
    \unit@=\@ne bp\setc@ntr@l{2}\figsetmark{}%
    \def\minst@p{20pt}%
    \v@lX=\@psfllx\p@\v@lX=\Sc@leFact\v@lX\r@undint\v@lX\v@lX%
    \v@lY=\@psflly\p@\v@lY=\Sc@leFact\v@lY\ifdim\v@lY>\z@\r@undint\v@lY\v@lY\fi%
    \delt@=\@psfury\p@\delt@=\Sc@leFact\delt@%
    \advance\delt@-\v@lY\v@lXa=\@psfurx\p@\v@lXa=\Sc@leFact\v@lXa\v@leur=\minst@p%
    \edef\valv@lY{\repdecn@mb{\v@lY}}\edef\LgTr@it{\the\delt@}%
    \loop\ifdim\v@lX<\v@lXa\edef\valv@lX{\repdecn@mb{\v@lX}}%
    \figptDD -1:(\valv@lX,\valv@lY)\figwriten -1:\hbox{\vrule height\LgTr@it}(0)%
    \ifdim\v@leur<\minst@p\else\figsetmark{\raise-8bp\hbox{$\scriptscriptstyle\triangle$}}%
    \figwrites -1:\@ffichnb{0}{\valv@lX}(6)\v@leur=\z@\figsetmark{}\fi%
    \advance\v@leur#1pt\advance\v@lX#1pt\repeat%
    \def\minst@p{10pt}%
    \v@lX=\@psfllx\p@\v@lX=\Sc@leFact\v@lX\ifdim\v@lX>\z@\r@undint\v@lX\v@lX\fi%
    \v@lY=\@psflly\p@\v@lY=\Sc@leFact\v@lY\r@undint\v@lY\v@lY%
    \delt@=\@psfurx\p@\delt@=\Sc@leFact\delt@%
    \advance\delt@-\v@lX\v@lYa=\@psfury\p@\v@lYa=\Sc@leFact\v@lYa\v@leur=\minst@p%
    \edef\valv@lX{\repdecn@mb{\v@lX}}\edef\LgTr@it{\the\delt@}%
    \loop\ifdim\v@lY<\v@lYa\edef\valv@lY{\repdecn@mb{\v@lY}}%
    \figptDD -1:(\valv@lX,\valv@lY)\figwritee -1:\vbox{\hrule width\LgTr@it}(0)%
    \ifdim\v@leur<\minst@p\else\figsetmark{$\triangleright$\kern4bp}%
    \figwritew -1:\@ffichnb{0}{\valv@lY}(6)\v@leur=\z@\figsetmark{}\fi%
    \advance\v@leur#2pt\advance\v@lY#2pt\repeat}
\ctr@ld@f
\ctr@ld@f\def\figscan@E#1(#2,#3){{\s@uvc@ntr@l\et@tfigscan@E%
    \Figdisc@rdLTS{#1}{\t@xt@}\pdfximage{\t@xt@}%
    \setbox\Gb@x=\hbox{\pdfrefximage\pdflastximage}%
    \edef\@psfllx{0}\v@lY=-\dp\Gb@x\edef\@psflly{\repdecn@mb{\v@lY}}%
    \edef\@psfurx{\repdecn@mb{\wd\Gb@x}}%
    \v@lY=\dp\Gb@x\advance\v@lY\ht\Gb@x\edef\@psfury{\repdecn@mb{\v@lY}}%
    \figscan@{#2}{#3}\resetc@ntr@l\et@tfigscan@E}\ignorespaces}
\ctr@ld@f\def\figshowpts[#1,#2]{{\figsetmark{$\bullet$}\figsetptname{\bf ##1}%
    \p@rtent=#2\relax\ifnum\p@rtent<\z@\p@rtent=\z@\fi%
    \s@mme=#1\relax\ifnum\s@mme<\z@\s@mme=\z@\fi%
    \loop\ifnum\s@mme<\p@rtent\pt@rvect{\s@mme}%
    \ifitis@K\figwriten{\the\s@mme}:(4pt)\fi\advance\s@mme\@ne\repeat%
    \pt@rvect{\s@mme}\ifitis@K\figwriten{\the\s@mme}:(4pt)\fi}\ignorespaces}
\ctr@ld@f\def\pt@rvect#1{\set@bjc@de{#1}%
    \expandafter\expandafter\expandafter\inqpt@rvec\csname\objc@de\endcsname:}
\ctr@ld@f\def\inqpt@rvec#1#2:{\if#1\C@dCl@spt\itis@Ktrue\else\itis@Kfalse\fi}
\ctr@ld@f\def\figshowsettings{{%
    \immediate\write16{====================================================================}%
    \immediate\write16{ Current settings about:}%
    \immediate\write16{ --- GENERAL ---}%
    \immediate\write16{Scale factor and Unit = \unit@util\space (\the\unit@)
     \space -> \BS@ figinit{ScaleFactorUnit}}%
    \immediate\write16{Update mode = \ifpsupdatem@de yes\else no\fi
     \space-> \BS@ psset(update=yes/no) or \BS@ pssetdefault(update=yes/no)}%
    \immediate\write16{ --- PRINTING ---}%
    \immediate\write16{Implicit point name = \ptn@me{i} \space-> \BS@ figsetptname{Name}}%
    \immediate\write16{Point marker = \the\c@nsymb \space -> \BS@ figsetmark{Mark}}%
    \immediate\write16{Print rounded coordinates = \ifr@undcoord yes\else no\fi
     \space-> \BS@ figsetroundcoord{yes/no}}%
    \immediate\write16{ --- GRAPHICAL (general) ---}%
    \immediate\write16{First-level (or primary) settings:}%
    \immediate\write16{ Color = \curr@ntcolor \space-> \BS@ psset(color=ColorDefinition)}%
    \immediate\write16{ Filling mode = \iffillm@de yes\else no\fi
     \space-> \BS@ psset(fillmode=yes/no)}%
    \immediate\write16{ Line join = \curr@ntjoin \space-> \BS@ psset(join=miter/round/bevel)}%
    \immediate\write16{ Line style = \curr@ntdash \space-> \BS@ psset(dash=Index/Pattern)}%
    \immediate\write16{ Line width = \curr@ntwidth
     \space-> \BS@ psset(width=real in PostScript units)}%
    \immediate\write16{Second-level (or secondary) settings:}%
    \immediate\write16{ Color = \sec@ndcolor \space-> \BS@ psset second(color=ColorDefinition)}%
    \immediate\write16{ Line style = \curr@ntseconddash
     \space-> \BS@ psset second(dash=Index/Pattern)}%
    \immediate\write16{ Line width = \curr@ntsecondwidth
     \space-> \BS@ psset second(width=real in PostScript units)}%
    \immediate\write16{Third-level (or ternary) settings:}%
    \immediate\write16{ Color = \th@rdcolor \space-> \BS@ psset third(color=ColorDefinition)}%
    \immediate\write16{ --- GRAPHICAL (specific) ---}%
    \immediate\write16{Arrow-head:}%
    \immediate\write16{ (half-)Angle = \@rrowheadangle
     \space-> \BS@ psset arrowhead(angle=real in degrees)}%
    \immediate\write16{ Filling mode = \if@rrowhfill yes\else no\fi
     \space-> \BS@ psset arrowhead(fillmode=yes/no)}%
    \immediate\write16{ "Outside" = \if@rrowhout yes\else no\fi
     \space-> \BS@ psset arrowhead(out=yes/no)}%
    \immediate\write16{ Length = \@rrowheadlength
     \if@rrowratio\space(not active)\else\space(active)\fi
     \space-> \BS@ psset arrowhead(length=real in user coord.)}%
    \immediate\write16{ Ratio = \@rrowheadratio
     \if@rrowratio\space(active)\else\space(not active)\fi
     \space-> \BS@ psset arrowhead(ratio=real in [0,1])}%
    \immediate\write16{Curve: Roundness = \curv@roundness
     \space-> \BS@ psset curve(roundness=real in [0,0.5])}%
    \immediate\write16{Mesh: Diagonal = \c@ntrolmesh
     \space-> \BS@ psset mesh(diag=integer in {-1,0,1})}%
    \immediate\write16{Flow chart:}%
    \immediate\write16{ Arrow position = \@rrowp@s
     \space-> \BS@ psset flowchart(arrowposition=real in [0,1])}%
    \immediate\write16{ Arrow reference point = \ifcase\@rrowr@fpt start\else end\fi
     \space-> \BS@ psset flowchart(arrowrefpt = start/end)}%     
    \immediate\write16{ Line type = \ifcase\fclin@typ@ curve\else polygon\fi
     \space-> \BS@ psset flowchart(line=polygon/curve)}%
    \immediate\write16{ Padding = (\Xp@dd, \Yp@dd)
     \space-> \BS@ psset flowchart(padding = real in user coord.)}%
    \immediate\write16{\space\space\space\space(or
     \BS@ psset flowchart(xpadding=real, ypadding=real) )}%
    \immediate\write16{ Radius = \fclin@r@d
     \space-> \BS@ psset flowchart(radius=positive real in user coord.)}%
    \immediate\write16{ Shape = \fcsh@pe
     \space-> \BS@ psset flowchart(shape = rectangle, ellipse or lozenge)}%
    \immediate\write16{ Thickness = \thickn@ss
     \space-> \BS@ psset flowchart(thickness = real in user coord.)}%
    \ifTr@isDim%
    \immediate\write16{ --- 3D to 2D PROJECTION ---}%
    \immediate\write16{Projection : \typ@proj \space-> \BS@ figinit{ScaleFactorUnit, ProjType}}%
    \immediate\write16{Longitude (psi) = \v@lPsi \space-> \BS@ figset proj(psi=real in degrees)}%
    \ifcase\curr@ntproj\immediate\write16{Depth coeff. (Lambda)
     \space = \v@lTheta \space-> \BS@ figset proj(lambda=real in [0,1])}%
    \else\immediate\write16{Latitude (theta)
     \space = \v@lTheta \space-> \BS@ figset proj(theta=real in degrees)}%
    \fi%
    \ifnum\curr@ntproj=\tw@%
    \immediate\write16{Observation distance = \disob@unit
     \space-> \BS@ figset proj(dist=real in user coord.)}%
    \immediate\write16{Target point = \t@rgetpt \space-> \BS@ figset proj(targetpt=pt number)}%
     \v@lX=\ptT@unit@\wd\Bt@rget\v@lY=\ptT@unit@\ht\Bt@rget\v@lZ=\ptT@unit@\dp\Bt@rget%
    \immediate\write16{ Its coordinates are
     (\repdecn@mb{\v@lX}, \repdecn@mb{\v@lY}, \repdecn@mb{\v@lZ})}%
    \fi%
    \fi%
    \immediate\write16{====================================================================}%
    \ignorespaces}}
\ctr@ln@w{newif}\ifitis@vect@r
\ctr@ld@f\def\figvectC#1(#2,#3){{\itis@vect@rtrue\figpt#1:(#2,#3)}\ignorespaces}
\ctr@ld@f\def\Figv@ctCreg#1(#2,#3){{\itis@vect@rtrue\Figp@intreg#1:(#2,#3)}\ignorespaces}
\ctr@ln@m\figvectDBezier
\ctr@ld@f\def\figvectDBezierDD#1:#2,#3[#4,#5,#6,#7]{\ifps@cri{\s@uvc@ntr@l\et@tfigvectDBezierDD%
    \FigvectDBezier@#2,#3[#4,#5,#6,#7]\v@lX=\c@ef\v@lX\v@lY=\c@ef\v@lY%
    \Figv@ctCreg#1(\v@lX,\v@lY)\resetc@ntr@l\et@tfigvectDBezierDD}\ignorespaces\fi}
\ctr@ld@f\def\figvectDBezierTD#1:#2,#3[#4,#5,#6,#7]{\ifps@cri{\s@uvc@ntr@l\et@tfigvectDBezierTD%
    \FigvectDBezier@#2,#3[#4,#5,#6,#7]\v@lX=\c@ef\v@lX\v@lY=\c@ef\v@lY\v@lZ=\c@ef\v@lZ%
    \Figv@ctCreg#1(\v@lX,\v@lY,\v@lZ)\resetc@ntr@l\et@tfigvectDBezierTD}\ignorespaces\fi}
\ctr@ld@f\def\FigvectDBezier@#1,#2[#3,#4,#5,#6]{\setc@ntr@l{2}%
    \edef\T@{#2}\v@leur=\p@\advance\v@leur-#2pt\edef\UNmT@{\repdecn@mb{\v@leur}}%
    \ifnum#1=\tw@\def\c@ef{6}\else\def\c@ef{3}\fi%
    \figptcopy-4:/#3/\figptcopy-3:/#4/\figptcopy-2:/#5/\figptcopy-1:/#6/%
    \l@mbd@un=-4 \l@mbd@de=-\thr@@\p@rtent=\m@ne\c@lDecast%
    \ifnum#1=\tw@\c@lDCDeux{-4}{-3}\c@lDCDeux{-3}{-2}\c@lDCDeux{-4}{-3}\else%
    \l@mbd@un=-4 \l@mbd@de=-\thr@@\p@rtent=-\tw@\c@lDecast%
    \c@lDCDeux{-4}{-3}\fi\Figg@tXY{-4}}
\ctr@ln@m\c@lDCDeux
\ctr@ld@f\def\c@lDCDeuxDD#1#2{\Figg@tXY{#2}\Figg@tXYa{#1}%
    \advance\v@lX-\v@lXa\advance\v@lY-\v@lYa\Figp@intregDD#1:(\v@lX,\v@lY)}
\ctr@ld@f\def\c@lDCDeuxTD#1#2{\Figg@tXY{#2}\Figg@tXYa{#1}\advance\v@lX-\v@lXa%
    \advance\v@lY-\v@lYa\advance\v@lZ-\v@lZa\Figp@intregTD#1:(\v@lX,\v@lY,\v@lZ)}
\ctr@ln@m\figvectN
\ctr@ld@f\def\figvectNDD#1[#2,#3]{\ifps@cri{\Figg@tXYa{#2}\Figg@tXY{#3}%
    \advance\v@lX-\v@lXa\advance\v@lY-\v@lYa%
    \Figv@ctCreg#1(-\v@lY,\v@lX)}\ignorespaces\fi}
\ctr@ld@f\def\figvectNTD#1[#2,#3,#4]{\ifps@cri{\vecunitC@TD[#2,#4]\v@lmin=\v@lX\v@lmax=\v@lY%
    \v@leur=\v@lZ\vecunitC@TD[#2,#3]\c@lprovec{#1}}\ignorespaces\fi}
\ctr@ln@m\figvectNV
\ctr@ld@f\def\figvectNVDD#1[#2]{\ifps@cri{\Figg@tXY{#2}\Figv@ctCreg#1(-\v@lY,\v@lX)}\ignorespaces\fi}
\ctr@ld@f\def\figvectNVTD#1[#2,#3]{\ifps@cri{\vecunitCV@TD{#3}\v@lmin=\v@lX\v@lmax=\v@lY%
    \v@leur=\v@lZ\vecunitCV@TD{#2}\c@lprovec{#1}}\ignorespaces\fi}
\ctr@ln@m\figvectP
\ctr@ld@f\def\figvectPDD#1[#2,#3]{\ifps@cri{\Figg@tXYa{#2}\Figg@tXY{#3}%
    \advance\v@lX-\v@lXa\advance\v@lY-\v@lYa%
    \Figv@ctCreg#1(\v@lX,\v@lY)}\ignorespaces\fi}
\ctr@ld@f\def\figvectPTD#1[#2,#3]{\ifps@cri{\Figg@tXYa{#2}\Figg@tXY{#3}%
    \advance\v@lX-\v@lXa\advance\v@lY-\v@lYa\advance\v@lZ-\v@lZa%
    \Figv@ctCreg#1(\v@lX,\v@lY,\v@lZ)}\ignorespaces\fi}
\ctr@ln@m\figvectU
\ctr@ld@f\def\figvectUDD#1[#2]{\ifps@cri{\n@rmeuc\v@leur{#2}\invers@\v@leur\v@leur%
    \delt@=\repdecn@mb{\v@leur}\unit@\edef\v@ldelt@{\repdecn@mb{\delt@}}%
    \Figg@tXY{#2}\v@lX=\v@ldelt@\v@lX\v@lY=\v@ldelt@\v@lY%
    \Figv@ctCreg#1(\v@lX,\v@lY)}\ignorespaces\fi}
\ctr@ld@f\def\figvectUTD#1[#2]{\ifps@cri{\n@rmeuc\v@leur{#2}\invers@\v@leur\v@leur%
    \delt@=\repdecn@mb{\v@leur}\unit@\edef\v@ldelt@{\repdecn@mb{\delt@}}%
    \Figg@tXY{#2}\v@lX=\v@ldelt@\v@lX\v@lY=\v@ldelt@\v@lY\v@lZ=\v@ldelt@\v@lZ%
    \Figv@ctCreg#1(\v@lX,\v@lY,\v@lZ)}\ignorespaces\fi}
\ctr@ld@f\def\figvisu#1#2#3{\c@ldefproj\initb@undb@x\xdef\figforTeXFigno{\figforTeXnextFigno}%
    \s@mme=\figforTeXnextFigno\advance\s@mme\@ne\xdef\figforTeXnextFigno{\number\s@mme}%
    \setbox\b@xvisu=\hbox{\ifnum\@utoFN>\z@\figinsert{}\gdef\@utoFInDone{0}\fi\ignorespaces#3}%
    \gdef\@utoFInDone{1}\gdef\@utoFN{0}%
    \v@lXa=-\c@@rdYmin\v@lYa=\c@@rdYmax\advance\v@lYa-\c@@rdYmin%
    \v@lX=\c@@rdXmax\advance\v@lX-\c@@rdXmin%
    \setbox#1=\hbox{#2}\v@lY=-\v@lX\maxim@m{\v@lX}{\v@lX}{\wd#1}%
    \advance\v@lY\v@lX\divide\v@lY\tw@\advance\v@lY-\c@@rdXmin%
    \setbox#1=\vbox{\parindent0mm\hsize=\v@lX\vskip\v@lYa%
    \rlap{\hskip\v@lY\smash{\raise\v@lXa\box\b@xvisu}}%
    \def\t@xt@{#2}\ifx\t@xt@\empty\else\medskip\centerline{#2}\fi}\wd#1=\v@lX}
\ctr@ld@f\def\figDecrementFigno{{\xdef\figforTeXnextFigno{\figforTeXFigno}%
    \s@mme=\figforTeXFigno\advance\s@mme\m@ne\xdef\figforTeXFigno{\number\s@mme}}}
\ctr@ln@w{newbox}\Bt@rget\setbox\Bt@rget=\null
\ctr@ln@w{newbox}\BminTD@\setbox\BminTD@=\null
\ctr@ln@w{newbox}\BmaxTD@\setbox\BmaxTD@=\null
\ctr@ln@w{newif}\ifnewt@rgetpt\ctr@ln@w{newif}\ifnewdis@b
\ctr@ld@f\def\b@undb@xTD#1#2#3{%
    \relax\ifdim#1<\wd\BminTD@\global\wd\BminTD@=#1\fi%
    \relax\ifdim#2<\ht\BminTD@\global\ht\BminTD@=#2\fi%
    \relax\ifdim#3<\dp\BminTD@\global\dp\BminTD@=#3\fi%
    \relax\ifdim#1>\wd\BmaxTD@\global\wd\BmaxTD@=#1\fi%
    \relax\ifdim#2>\ht\BmaxTD@\global\ht\BmaxTD@=#2\fi%
    \relax\ifdim#3>\dp\BmaxTD@\global\dp\BmaxTD@=#3\fi}
\ctr@ld@f\def\c@ldefdisob{{\ifdim\wd\BminTD@<\maxdimen\v@leur=\wd\BmaxTD@\advance\v@leur-\wd\BminTD@%
    \delt@=\ht\BmaxTD@\advance\delt@-\ht\BminTD@\maxim@m{\v@leur}{\v@leur}{\delt@}%
    \delt@=\dp\BmaxTD@\advance\delt@-\dp\BminTD@\maxim@m{\v@leur}{\v@leur}{\delt@}%
    \v@leur=5\v@leur\else\v@leur=800pt\fi\c@ldefdisob@{\v@leur}}}
\ctr@ln@m\disob@intern
\ctr@ln@m\disob@
\ctr@ln@m\divf@ctproj
\ctr@ld@f\def\c@ldefdisob@#1{{\v@leur=#1\ifdim\v@leur<\p@\v@leur=800pt\fi%
    \xdef\disob@intern{\repdecn@mb{\v@leur}}%
    \delt@=\ptT@unit@\v@leur\xdef\disob@unit{\repdecn@mb{\delt@}}%
    \f@ctech=\@ne\loop\ifdim\v@leur>\t@n pt\divide\v@leur\t@n\multiply\f@ctech\t@n\repeat%
    \xdef\disob@{\repdecn@mb{\v@leur}}\xdef\divf@ctproj{\the\f@ctech}}%
    \global\newdis@btrue}
\ctr@ln@m\t@rgetpt
\ctr@ld@f\def\c@ldeft@rgetpt{\newt@rgetpttrue\def\t@rgetpt{CenterBoundBox}{%
    \delt@=\wd\BmaxTD@\advance\delt@-\wd\BminTD@\divide\delt@\tw@%
    \v@leur=\wd\BminTD@\advance\v@leur\delt@\global\wd\Bt@rget=\v@leur%
    \delt@=\ht\BmaxTD@\advance\delt@-\ht\BminTD@\divide\delt@\tw@%
    \v@leur=\ht\BminTD@\advance\v@leur\delt@\global\ht\Bt@rget=\v@leur%
    \delt@=\dp\BmaxTD@\advance\delt@-\dp\BminTD@\divide\delt@\tw@%
    \v@leur=\dp\BminTD@\advance\v@leur\delt@\global\dp\Bt@rget=\v@leur}}
\ctr@ln@m\c@ldefproj
\ctr@ld@f\def\c@ldefprojTD{\ifnewt@rgetpt\else\c@ldeft@rgetpt\fi\ifnewdis@b\else\c@ldefdisob\fi}
\ctr@ld@f\def\c@lprojcav{% Projection cavaliere : X = x + y L cos t, Y = z + y L sin t
    \v@lZa=\cxa@\v@lY\advance\v@lX\v@lZa%
    \v@lZa=\cxb@\v@lY\v@lY=\v@lZ\advance\v@lY\v@lZa\ignorespaces}
\ctr@ln@m\v@lcoef
\ctr@ld@f\def\c@lprojrea{% Projection realiste
    \advance\v@lX-\wd\Bt@rget\advance\v@lY-\ht\Bt@rget\advance\v@lZ-\dp\Bt@rget%
    \v@lZa=\cza@\v@lX\advance\v@lZa\czb@\v@lY\advance\v@lZa\czc@\v@lZ%
    \divide\v@lZa\divf@ctproj\advance\v@lZa\disob@ pt\invers@{\v@lZa}{\v@lZa}%
    \v@lZa=\disob@\v@lZa\edef\v@lcoef{\repdecn@mb{\v@lZa}}%
    \v@lXa=\cxa@\v@lX\advance\v@lXa\cxb@\v@lY\v@lXa=\v@lcoef\v@lXa%
    \v@lY=\cyb@\v@lY\advance\v@lY\cya@\v@lX\advance\v@lY\cyc@\v@lZ%
    \v@lY=\v@lcoef\v@lY\v@lX=\v@lXa\ignorespaces}
\ctr@ld@f\def\c@lprojort{% Projection orthogonale
    \v@lXa=\cxa@\v@lX\advance\v@lXa\cxb@\v@lY%
    \v@lY=\cyb@\v@lY\advance\v@lY\cya@\v@lX\advance\v@lY\cyc@\v@lZ%
    \v@lX=\v@lXa\ignorespaces}
\ctr@ld@f\def\Figptpr@j#1:#2/#3/{{\Figg@tXY{#3}\superc@lprojSP%
    \Figp@intregDD#1:{#2}(\v@lX,\v@lY)}\ignorespaces}
\ctr@ln@m\figsetobdist
\ctr@ld@f\def\figsetobdistDD{\un@v@ilable{figsetobdist}}
\ctr@ld@f\def\figsetobdistTD(#1){{\ifcurr@ntPS%
    \immediate\write16{*** \BS@ figsetobdist is ignored inside a
     \BS@ psbeginfig-\BS@ psendfig block.}%
    \else\v@leur=#1\unit@\c@ldefdisob@{\v@leur}\fi}\ignorespaces}
\ctr@ln@m\c@lprojSP
\ctr@ln@m\curr@ntproj
\ctr@ln@m\typ@proj
\ctr@ln@m\superc@lprojSP
\ctr@ld@f\def\Figs@tproj#1{%
    \if#13 \d@faultproj\else\if#1c\d@faultproj%
    \else\if#1o\xdef\curr@ntproj{1}\xdef\typ@proj{orthogonal}%
         \figsetviewTD(\def@ultpsi,\def@ulttheta)%
         \global\let\c@lprojSP=\c@lprojort\global\let\superc@lprojSP=\c@lprojort%
    \else\if#1r\xdef\curr@ntproj{2}\xdef\typ@proj{realistic}%
         \figsetviewTD(\def@ultpsi,\def@ulttheta)%
         \global\let\c@lprojSP=\c@lprojrea\global\let\superc@lprojSP=\c@lprojrea%
    \else\d@faultproj\message{*** Unknown projection. Cavalier projection assumed.}%
    \fi\fi\fi\fi}
\ctr@ld@f\def\d@faultproj{\xdef\curr@ntproj{0}\xdef\typ@proj{cavalier}\figsetviewTD(\def@ultpsi,0.5)%
         \global\let\c@lprojSP=\c@lprojcav\global\let\superc@lprojSP=\c@lprojcav}
\ctr@ln@m\figsettarget
\ctr@ld@f\def\figsettargetDD{\un@v@ilable{figsettarget}}
\ctr@ld@f\def\figsettargetTD[#1]{{\ifcurr@ntPS%
    \immediate\write16{*** \BS@ figsettarget is ignored inside a
     \BS@ psbeginfig-\BS@ psendfig block.}%
    \else\global\newt@rgetpttrue\xdef\t@rgetpt{#1}\Figg@tXY{#1}\global\wd\Bt@rget=\v@lX%
    \global\ht\Bt@rget=\v@lY\global\dp\Bt@rget=\v@lZ\fi}\ignorespaces}
\ctr@ln@m\figsetview
\ctr@ld@f\def\figsetviewDD{\un@v@ilable{figsetview}}
\ctr@ld@f\def\figsetviewTD(#1){\ifcurr@ntPS%
     \immediate\write16{*** \BS@ figsetview is ignored inside a
     \BS@ psbeginfig-\BS@ psendfig block.}\else\Figsetview@#1,:\fi\ignorespaces}
\ctr@ld@f\def\Figsetview@#1,#2:{{\xdef\v@lPsi{#1}\def\t@xt@{#2}%
    \ifx\t@xt@\empty\def\@rgdeux{\v@lTheta}\else\X@rgdeux@#2\fi%
    \c@ssin{\costhet@}{\sinthet@}{#1}\v@lmin=\costhet@ pt\v@lmax=\sinthet@ pt%
    \ifcase\curr@ntproj%
    \v@leur=\@rgdeux\v@lmin\xdef\cxa@{\repdecn@mb{\v@leur}}%
    \v@leur=\@rgdeux\v@lmax\xdef\cxb@{\repdecn@mb{\v@leur}}\v@leur=\@rgdeux pt%
    \relax\ifdim\v@leur>\p@\message{*** Lambda too large ! See \BS@ figset proj() !}\fi%
    \else%
    \v@lmax=-\v@lmax\xdef\cxa@{\repdecn@mb{\v@lmax}}\xdef\cxb@{\costhet@}%
    \ifx\t@xt@\empty\edef\@rgdeux{\def@ulttheta}\fi\c@ssin{\C@}{\S@}{\@rgdeux}%
    \v@lmax=-\S@ pt%
    \v@leur=\v@lmax\v@leur=\costhet@\v@leur\xdef\cya@{\repdecn@mb{\v@leur}}%
    \v@leur=\v@lmax\v@leur=\sinthet@\v@leur\xdef\cyb@{\repdecn@mb{\v@leur}}%
    \xdef\cyc@{\C@}\v@lmin=-\C@ pt%
    \v@leur=\v@lmin\v@leur=\costhet@\v@leur\xdef\cza@{\repdecn@mb{\v@leur}}%
    \v@leur=\v@lmin\v@leur=\sinthet@\v@leur\xdef\czb@{\repdecn@mb{\v@leur}}%
    \xdef\czc@{\repdecn@mb{\v@lmax}}\fi%
    \xdef\v@lTheta{\@rgdeux}}}
\ctr@ld@f\def\def@ultpsi{40}
\ctr@ld@f\def\def@ulttheta{25}
\ctr@ln@m\l@debut
\ctr@ln@m\n@mref
\ctr@ld@f\def\figset#1(#2){\def\t@xt@{#1}\ifx\t@xt@\empty\trtlis@rg{#2}{\Figsetwr@te}% write
    \else\keln@mde#1|%
    \def\n@mref{pr}\ifx\l@debut\n@mref\ifcurr@ntPS% projection
     \immediate\write16{*** \BS@ figset proj(...) is ignored inside a
     \BS@ psbeginfig-\BS@ psendfig block.}\else\trtlis@rg{#2}{\Figsetpr@j}\fi\else%
    \def\n@mref{wr}\ifx\l@debut\n@mref\trtlis@rg{#2}{\Figsetwr@te}\else% write
    \immediate\write16{*** Unknown keyword: \BS@ figset #1(...)}%
    \fi\fi\fi\ignorespaces}
\ctr@ld@f\def\Figsetpr@j#1=#2|{\keln@mtr#1|%
    \def\n@mref{dep}\ifx\l@debut\n@mref\Figsetd@p{#2}\else% depth (lambda)
    \def\n@mref{dis}\ifx\l@debut\n@mref%
     \ifnum\curr@ntproj=\tw@\figsetobdist(#2)\else\Figset@rr\fi\else% dist
    \def\n@mref{lam}\ifx\l@debut\n@mref\Figsetd@p{#2}\else% depth (lambda)
    \def\n@mref{lat}\ifx\l@debut\n@mref\Figsetth@{#2}\else% latitude (theta)
    \def\n@mref{lon}\ifx\l@debut\n@mref\figsetview(#2)\else% longitude (psi)
    \def\n@mref{psi}\ifx\l@debut\n@mref\figsetview(#2)\else% longitude (psi)
    \def\n@mref{tar}\ifx\l@debut\n@mref%
     \ifnum\curr@ntproj=\tw@\figsettarget[#2]\else\Figset@rr\fi\else% target point
    \def\n@mref{the}\ifx\l@debut\n@mref\Figsetth@{#2}\else% latitude (theta)
    \immediate\write16{*** Unknown attribute: \BS@ figset proj(..., #1=...).}%
    \fi\fi\fi\fi\fi\fi\fi\fi}
\ctr@ld@f\def\Figsetd@p#1{\ifnum\curr@ntproj=\z@\figsetview(\v@lPsi,#1)\else\Figset@rr\fi}
\ctr@ld@f\def\Figsetth@#1{\ifnum\curr@ntproj=\z@\Figset@rr\else\figsetview(\v@lPsi,#1)\fi}
\ctr@ld@f\def\Figset@rr{\message{*** \BS@ figset proj(): Attribute "\n@mref" ignored, incompatible
    with current projection}}
\ctr@ld@f\def\initb@undb@xTD{\wd\BminTD@=\maxdimen\ht\BminTD@=\maxdimen\dp\BminTD@=\maxdimen%
    \wd\BmaxTD@=-\maxdimen\ht\BmaxTD@=-\maxdimen\dp\BmaxTD@=-\maxdimen}
\ctr@ln@w{newbox}\Gb@x      % boite a tout faire
\ctr@ln@w{newbox}\Gb@xSC    % boite qui contient le point marker
\ctr@ln@w{newtoks}\c@nsymb  % the point marker
\ctr@ln@w{newif}\ifr@undcoord\ctr@ln@w{newif}\ifunitpr@sent
\ctr@ld@f\def\unssqrttw@{0.707106 }
\ctr@ld@f\def\figAst{\raise-1.15ex\hbox{$\ast$}}
\ctr@ld@f\def\figBullet{\raise-1.15ex\hbox{$\bullet$}}
\ctr@ld@f\def\figCirc{\raise-1.15ex\hbox{$\circ$}}
\ctr@ld@f\def\figDiamond{\raise-1.15ex\hbox{$\diamond$}}%
\ctr@ld@f\def\boxit#1#2{\leavevmode\hbox{\vrule\vbox{\hrule\vglue#1%
    \vtop{\hbox{\kern#1{#2}\kern#1}\vglue#1\hrule}}\vrule}}
\ctr@ld@f
\ctr@ld@f
\ctr@ld@f\def\c@nterpt{\ignorespaces%
    \kern-.5\wd\Gb@xSC%
    \raise-.5\ht\Gb@xSC\rlap{\hbox{\raise.5\dp\Gb@xSC\hbox{\copy\Gb@xSC}}}%
    \kern .5\wd\Gb@xSC\ignorespaces}
\ctr@ld@f\def\b@undb@xSC#1#2{{\v@lXa=#1\v@lYa=#2%
    \v@leur=\ht\Gb@xSC\advance\v@leur\dp\Gb@xSC%
    \advance\v@lXa-.5\wd\Gb@xSC\advance\v@lYa-.5\v@leur\b@undb@x{\v@lXa}{\v@lYa}%
    \advance\v@lXa\wd\Gb@xSC\advance\v@lYa\v@leur\b@undb@x{\v@lXa}{\v@lYa}}}
\ctr@ln@m\Dist@n
\ctr@ln@m\l@suite
\ctr@ld@f\def\@keldist#1#2{\edef\Dist@n{#2}\y@tiunit{\Dist@n}%
    \ifunitpr@sent#1=\Dist@n\else#1=\Dist@n\unit@\fi}
\ctr@ld@f\def\y@tiunit#1{\unitpr@sentfalse\expandafter\y@tiunit@#1:}
\ctr@ld@f\def\y@tiunit@#1#2:{\ifcat#1a\unitpr@senttrue\else\def\l@suite{#2}%
    \ifx\l@suite\empty\else\y@tiunit@#2:\fi\fi}
\ctr@ln@m\figcoord
\ctr@ld@f\def\figcoordDD#1{{\v@lX=\ptT@unit@\v@lX\v@lY=\ptT@unit@\v@lY%
    \ifr@undcoord\ifcase#1\v@leur=0.5pt\or\v@leur=0.05pt\or\v@leur=0.005pt%
    \or\v@leur=0.0005pt\else\v@leur=\z@\fi%
    \ifdim\v@lX<\z@\advance\v@lX-\v@leur\else\advance\v@lX\v@leur\fi%
    \ifdim\v@lY<\z@\advance\v@lY-\v@leur\else\advance\v@lY\v@leur\fi\fi%
    (\@ffichnb{#1}{\repdecn@mb{\v@lX}},\ifmmode\else\thinspace\fi%
    \@ffichnb{#1}{\repdecn@mb{\v@lY}})}}
\ctr@ld@f\def\@ffichnb#1#2{{\def\@@ffich{\@ffich#1(}\edef\n@mbre{#2}%
    \expandafter\@@ffich\n@mbre)}}
\ctr@ld@f\def\@ffich#1(#2.#3){{#2\ifnum#1>\z@.\fi\def\dig@ts{#3}\s@mme=\z@%
    \loop\ifnum\s@mme<#1\expandafter\@ffichdec\dig@ts:\advance\s@mme\@ne\repeat}}
\ctr@ld@f\def\@ffichdec#1#2:{\relax#1\def\dig@ts{#20}}
\ctr@ld@f\def\figcoordTD#1{{\v@lX=\ptT@unit@\v@lX\v@lY=\ptT@unit@\v@lY\v@lZ=\ptT@unit@\v@lZ%
    \ifr@undcoord\ifcase#1\v@leur=0.5pt\or\v@leur=0.05pt\or\v@leur=0.005pt%
    \or\v@leur=0.0005pt\else\v@leur=\z@\fi%
    \ifdim\v@lX<\z@\advance\v@lX-\v@leur\else\advance\v@lX\v@leur\fi%
    \ifdim\v@lY<\z@\advance\v@lY-\v@leur\else\advance\v@lY\v@leur\fi%
    \ifdim\v@lZ<\z@\advance\v@lZ-\v@leur\else\advance\v@lZ\v@leur\fi\fi%
    (\@ffichnb{#1}{\repdecn@mb{\v@lX}},\ifmmode\else\thinspace\fi%
     \@ffichnb{#1}{\repdecn@mb{\v@lY}},\ifmmode\else\thinspace\fi%
     \@ffichnb{#1}{\repdecn@mb{\v@lZ}})}}
\ctr@ld@f\def\figsetroundcoord#1{\expandafter\Figsetr@undcoord#1:\ignorespaces}
\ctr@ld@f\def\Figsetr@undcoord#1#2:{\if#1n\r@undcoordfalse\else\r@undcoordtrue\fi}
\ctr@ld@f\def\Figsetwr@te#1=#2|{\keln@mun#1|%
    \def\n@mref{m}\ifx\l@debut\n@mref\figsetmark{#2}\else% mark
    \immediate\write16{*** Unknown attribute: \BS@ figset (..., #1=...)}%
    \fi}
\ctr@ld@f\def\figsetmark#1{\c@nsymb={#1}\setbox\Gb@xSC=\hbox{\the\c@nsymb}\ignorespaces}
\ctr@ln@m\ptn@me
\ctr@ld@f\def\figsetptname#1{\def\ptn@me##1{#1}\ignorespaces}
\ctr@ld@f\def\FigWrit@L#1:#2(#3,#4){\ignorespaces\@keldist\v@leur{#3}\@keldist\delt@{#4}%
    \C@rp@r@m\def\list@num{#1}\@ecfor\p@int:=\list@num\do{\FigWrit@pt{\p@int}{#2}}}
\ctr@ld@f\def\FigWrit@pt#1#2{\FigWp@r@m{#1}{#2}\Vc@rrect\figWp@si%
    \ifdim\wd\Gb@xSC>\z@\b@undb@xSC{\v@lX}{\v@lY}\fi\figWBB@x}
\ctr@ld@f\def\FigWp@r@m#1#2{\Figg@tXY{#1}%
    \setbox\Gb@x=\hbox{\def\t@xt@{#2}\ifx\t@xt@\empty\Figg@tT{#1}\else#2\fi}\c@lprojSP}
\ctr@ld@f\let\Vc@rrect=\relax
\ctr@ld@f\let\C@rp@r@m=\relax
\ctr@ld@f\def\figwrite[#1]#2{{\ignorespaces\def\list@num{#1}\@ecfor\p@int:=\list@num\do{%
    \setbox\Gb@x=\hbox{\def\t@xt@{#2}\ifx\t@xt@\empty\Figg@tT{\p@int}\else#2\fi}%
    \Figwrit@{\p@int}}}\ignorespaces}
\ctr@ld@f\def\Figwrit@#1{\Figg@tXY{#1}\c@lprojSP%
    \rlap{\kern\v@lX\raise\v@lY\hbox{\unhcopy\Gb@x}}\v@leur=\v@lY%
    \advance\v@lY\ht\Gb@x\b@undb@x{\v@lX}{\v@lY}\advance\v@lX\wd\Gb@x%
    \v@lY=\v@leur\advance\v@lY-\dp\Gb@x\b@undb@x{\v@lX}{\v@lY}}
\ctr@ld@f\def\figwritec[#1]#2{{\ignorespaces\def\list@num{#1}%
    \@ecfor\p@int:=\list@num\do{\Figwrit@c{\p@int}{#2}}}\ignorespaces}
\ctr@ld@f\def\Figwrit@c#1#2{\FigWp@r@m{#1}{#2}%
    \rlap{\kern\v@lX\raise\v@lY\hbox{\rlap{\kern-.5\wd\Gb@x%
    \raise-.5\ht\Gb@x\hbox{\raise.5\dp\Gb@x\hbox{\unhcopy\Gb@x}}}}}%
    \v@leur=\ht\Gb@x\advance\v@leur\dp\Gb@x%
    \advance\v@lX-.5\wd\Gb@x\advance\v@lY-.5\v@leur\b@undb@x{\v@lX}{\v@lY}%
    \advance\v@lX\wd\Gb@x\advance\v@lY\v@leur\b@undb@x{\v@lX}{\v@lY}}
\ctr@ld@f\def\figwritep[#1]{{\ignorespaces\def\list@num{#1}\setbox\Gb@x=\hbox{\c@nterpt}%
    \@ecfor\p@int:=\list@num\do{\Figwrit@{\p@int}}}\ignorespaces}
\ctr@ld@f\def\figwritew#1:#2(#3){\figwritegcw#1:{#2}(#3,0pt)}
\ctr@ld@f\def\figwritee#1:#2(#3){\figwritegce#1:{#2}(#3,0pt)}
\ctr@ld@f\def\figwriten#1:#2(#3){{\def\Vc@rrect{\v@lZ=\v@leur\advance\v@lZ\dp\Gb@x}%
    \Figwrit@NS#1:{#2}(#3)}\ignorespaces}
\ctr@ld@f\def\figwrites#1:#2(#3){{\def\Vc@rrect{\v@lZ=-\v@leur\advance\v@lZ-\ht\Gb@x}%
    \Figwrit@NS#1:{#2}(#3)}\ignorespaces}
\ctr@ld@f\def\Figwrit@NS#1:#2(#3){\let\figWp@si=\FigWp@siNS\let\figWBB@x=\FigWBB@xNS%
    \FigWrit@L#1:{#2}(#3,0pt)}
\ctr@ld@f\def\FigWp@siNS{\rlap{\kern\v@lX\raise\v@lY\hbox{\rlap{\kern-.5\wd\Gb@x%
    \raise\v@lZ\hbox{\unhcopy\Gb@x}}\c@nterpt}}}
\ctr@ld@f\def\FigWBB@xNS{\advance\v@lY\v@lZ%
    \advance\v@lY-\dp\Gb@x\advance\v@lX-.5\wd\Gb@x\b@undb@x{\v@lX}{\v@lY}%
    \advance\v@lY\ht\Gb@x\advance\v@lY\dp\Gb@x%
    \advance\v@lX\wd\Gb@x\b@undb@x{\v@lX}{\v@lY}}
\ctr@ld@f\def\figwritenw#1:#2(#3){{\let\figWp@si=\FigWp@sigW\let\figWBB@x=\FigWBB@xgWE%
    \def\C@rp@r@m{\v@leur=\unssqrttw@\v@leur\delt@=\v@leur%
    \ifdim\delt@=\z@\delt@=\epsil@n\fi}\let@xte={-}\FigWrit@L#1:{#2}(#3,0pt)}\ignorespaces}
\ctr@ld@f\def\figwritesw#1:#2(#3){{\let\figWp@si=\FigWp@sigW\let\figWBB@x=\FigWBB@xgWE%
    \def\C@rp@r@m{\v@leur=\unssqrttw@\v@leur\delt@=-\v@leur%
    \ifdim\delt@=\z@\delt@=-\epsil@n\fi}\let@xte={-}\FigWrit@L#1:{#2}(#3,0pt)}\ignorespaces}
\ctr@ld@f\def\figwritene#1:#2(#3){{\let\figWp@si=\FigWp@sigE\let\figWBB@x=\FigWBB@xgWE%
    \def\C@rp@r@m{\v@leur=\unssqrttw@\v@leur\delt@=\v@leur%
    \ifdim\delt@=\z@\delt@=\epsil@n\fi}\let@xte={}\FigWrit@L#1:{#2}(#3,0pt)}\ignorespaces}
\ctr@ld@f\def\figwritese#1:#2(#3){{\let\figWp@si=\FigWp@sigE\let\figWBB@x=\FigWBB@xgWE%
    \def\C@rp@r@m{\v@leur=\unssqrttw@\v@leur\delt@=-\v@leur%
    \ifdim\delt@=\z@\delt@=-\epsil@n\fi}\let@xte={}\FigWrit@L#1:{#2}(#3,0pt)}\ignorespaces}
\ctr@ld@f\def\figwritegw#1:#2(#3,#4){{\let\figWp@si=\FigWp@sigW\let\figWBB@x=\FigWBB@xgWE%
    \let@xte={-}\FigWrit@L#1:{#2}(#3,#4)}\ignorespaces}
\ctr@ld@f\def\figwritege#1:#2(#3,#4){{\let\figWp@si=\FigWp@sigE\let\figWBB@x=\FigWBB@xgWE%
    \let@xte={}\FigWrit@L#1:{#2}(#3,#4)}\ignorespaces}
\ctr@ld@f\def\FigWp@sigW{\v@lXa=\z@\v@lYa=\ht\Gb@x\advance\v@lYa\dp\Gb@x%
    \ifdim\delt@>\z@\relax%
    \rlap{\kern\v@lX\raise\v@lY\hbox{\rlap{\kern-\wd\Gb@x\kern-\v@leur%
          \raise\delt@\hbox{\raise\dp\Gb@x\hbox{\unhcopy\Gb@x}}}\c@nterpt}}%
    \else\ifdim\delt@<\z@\relax\v@lYa=-\v@lYa%
    \rlap{\kern\v@lX\raise\v@lY\hbox{\rlap{\kern-\wd\Gb@x\kern-\v@leur%
          \raise\delt@\hbox{\raise-\ht\Gb@x\hbox{\unhcopy\Gb@x}}}\c@nterpt}}%
    \else\v@lXa=-.5\v@lYa%
    \rlap{\kern\v@lX\raise\v@lY\hbox{\rlap{\kern-\wd\Gb@x\kern-\v@leur%
          \raise-.5\ht\Gb@x\hbox{\raise.5\dp\Gb@x\hbox{\unhcopy\Gb@x}}}\c@nterpt}}%
    \fi\fi}
\ctr@ld@f\def\FigWp@sigE{\v@lXa=\z@\v@lYa=\ht\Gb@x\advance\v@lYa\dp\Gb@x%
    \ifdim\delt@>\z@\relax%
    \rlap{\kern\v@lX\raise\v@lY\hbox{\c@nterpt\kern\v@leur%
          \raise\delt@\hbox{\raise\dp\Gb@x\hbox{\unhcopy\Gb@x}}}}%
    \else\ifdim\delt@<\z@\relax\v@lYa=-\v@lYa%
    \rlap{\kern\v@lX\raise\v@lY\hbox{\c@nterpt\kern\v@leur%
          \raise\delt@\hbox{\raise-\ht\Gb@x\hbox{\unhcopy\Gb@x}}}}%
    \else\v@lXa=-.5\v@lYa%
    \rlap{\kern\v@lX\raise\v@lY\hbox{\c@nterpt\kern\v@leur%
          \raise-.5\ht\Gb@x\hbox{\raise.5\dp\Gb@x\hbox{\unhcopy\Gb@x}}}}%
    \fi\fi}
\ctr@ld@f\def\FigWBB@xgWE{\advance\v@lY\delt@%
    \advance\v@lX\the\let@xte\v@leur\advance\v@lY\v@lXa\b@undb@x{\v@lX}{\v@lY}%
    \advance\v@lX\the\let@xte\wd\Gb@x\advance\v@lY\v@lYa\b@undb@x{\v@lX}{\v@lY}}
\ctr@ld@f\def\figwritegcw#1:#2(#3,#4){{\let\figWp@si=\FigWp@sigcW\let\figWBB@x=\FigWBB@xgcWE%
    \let@xte={-}\FigWrit@L#1:{#2}(#3,#4)}\ignorespaces}
\ctr@ld@f\def\figwritegce#1:#2(#3,#4){{\let\figWp@si=\FigWp@sigcE\let\figWBB@x=\FigWBB@xgcWE%
    \let@xte={}\FigWrit@L#1:{#2}(#3,#4)}\ignorespaces}
\ctr@ld@f\def\FigWp@sigcW{\rlap{\kern\v@lX\raise\v@lY\hbox{\rlap{\kern-\wd\Gb@x\kern-\v@leur%
     \raise-.5\ht\Gb@x\hbox{\raise\delt@\hbox{\raise.5\dp\Gb@x\hbox{\unhcopy\Gb@x}}}}%
     \c@nterpt}}}
\ctr@ld@f\def\FigWp@sigcE{\rlap{\kern\v@lX\raise\v@lY\hbox{\c@nterpt\kern\v@leur%
    \raise-.5\ht\Gb@x\hbox{\raise\delt@\hbox{\raise.5\dp\Gb@x\hbox{\unhcopy\Gb@x}}}}}}
\ctr@ld@f\def\FigWBB@xgcWE{\v@lZ=\ht\Gb@x\advance\v@lZ\dp\Gb@x%
    \advance\v@lX\the\let@xte\v@leur\advance\v@lY\delt@\advance\v@lY.5\v@lZ%
    \b@undb@x{\v@lX}{\v@lY}%
    \advance\v@lX\the\let@xte\wd\Gb@x\advance\v@lY-\v@lZ\b@undb@x{\v@lX}{\v@lY}}
\ctr@ld@f\def\figwritebn#1:#2(#3){{\def\Vc@rrect{\v@lZ=\v@leur}\Figwrit@NS#1:{#2}(#3)}\ignorespaces}
\ctr@ld@f\def\figwritebs#1:#2(#3){{\def\Vc@rrect{\v@lZ=-\v@leur}\Figwrit@NS#1:{#2}(#3)}\ignorespaces}
\ctr@ld@f\def\figwritebw#1:#2(#3){{\let\figWp@si=\FigWp@sibW\let\figWBB@x=\FigWBB@xbWE%
    \let@xte={-}\FigWrit@L#1:{#2}(#3,0pt)}\ignorespaces}
\ctr@ld@f\def\figwritebe#1:#2(#3){{\let\figWp@si=\FigWp@sibE\let\figWBB@x=\FigWBB@xbWE%
    \let@xte={}\FigWrit@L#1:{#2}(#3,0pt)}\ignorespaces}
\ctr@ld@f\def\FigWp@sibW{\rlap{\kern\v@lX\raise\v@lY\hbox{\rlap{\kern-\wd\Gb@x\kern-\v@leur%
          \hbox{\unhcopy\Gb@x}}\c@nterpt}}}
\ctr@ld@f\def\FigWp@sibE{\rlap{\kern\v@lX\raise\v@lY\hbox{\c@nterpt\kern\v@leur%
          \hbox{\unhcopy\Gb@x}}}}
\ctr@ld@f\def\FigWBB@xbWE{\v@lZ=\ht\Gb@x\advance\v@lZ\dp\Gb@x%
    \advance\v@lX\the\let@xte\v@leur\advance\v@lY\ht\Gb@x\b@undb@x{\v@lX}{\v@lY}%
    \advance\v@lX\the\let@xte\wd\Gb@x\advance\v@lY-\v@lZ\b@undb@x{\v@lX}{\v@lY}}
\ctr@ln@w{newread}\frf@g  \ctr@ln@w{newwrite}\fwf@g
\ctr@ln@w{newif}\ifcurr@ntPS
\ctr@ln@w{newif}\ifps@cri
\ctr@ln@w{newif}\ifUse@llipse
\ctr@ln@w{newif}\ifpsdebugmode \psdebugmodefalse 
\ctr@ln@w{newif}\ifPDFm@ke
\ifx\pdfliteral\undefined\else\ifnum\pdfoutput>\z@\PDFm@ketrue\fi\fi
\ctr@ld@f\def\initPDF@rDVI{%
\ifPDFm@ke
 \let\figscan=\figscan@E
 \let\newGr@FN=\newGr@FNPDF
 \ctr@ld@f\def\c@mcurveto{c}
 \ctr@ld@f\def\c@mfill{f}
 \ctr@ld@f\def\c@mgsave{q}
 \ctr@ld@f\def\c@mgrestore{Q}
 \ctr@ld@f\def\c@mlineto{l}
 \ctr@ld@f\def\c@mmoveto{m}
 \ctr@ld@f\def\c@msetgray{g}     \ctr@ld@f\def\c@msetgrayStroke{G}
 \ctr@ld@f\def\c@msetcmykcolor{k}\ctr@ld@f\def\c@msetcmykcolorStroke{K}
 \ctr@ld@f\def\c@msetrgbcolor{rg}\ctr@ld@f\def\c@msetrgbcolorStroke{RG}
 \ctr@ld@f\def\d@fprimarC@lor{\curr@ntcolor\space\curr@ntcolorc@md%
               \space\curr@ntcolor\space\curr@ntcolorc@mdStroke}
 \ctr@ld@f\def\d@fsecondC@lor{\sec@ndcolor\space\sec@ndcolorc@md%
               \space\sec@ndcolor\space\sec@ndcolorc@mdStroke}
 \ctr@ld@f\def\d@fthirdC@lor{\th@rdcolor\space\th@rdcolorc@md%
              \space\th@rdcolor\space\th@rdcolorc@mdStroke}
 \ctr@ld@f\def\c@msetdash{d}
 \ctr@ld@f\def\c@msetlinejoin{j}
 \ctr@ld@f\def\c@msetlinewidth{w}
 \ctr@ld@f\def\f@gclosestroke{\immediate\write\fwf@g{s}}
 \ctr@ld@f\def\f@gfill{\immediate\write\fwf@g{\fillc@md}}% Voir la def de \fillc@md ****
 \ctr@ld@f\def\f@gnewpath{}
 \ctr@ld@f\def\f@gstroke{\immediate\write\fwf@g{S}}
\else
 \let\figinsertE=\figinsert
 \let\newGr@FN=\newGr@FNDVI
 \ctr@ld@f\def\c@mcurveto{curveto}
 \ctr@ld@f\def\c@mfill{fill}
 \ctr@ld@f\def\c@mgsave{gsave}
 \ctr@ld@f\def\c@mgrestore{grestore}
 \ctr@ld@f\def\c@mlineto{lineto}
 \ctr@ld@f\def\c@mmoveto{moveto}
 \ctr@ld@f\def\c@msetgray{setgray}          \ctr@ld@f\def\c@msetgrayStroke{}
 \ctr@ld@f\def\c@msetcmykcolor{setcmykcolor}\ctr@ld@f\def\c@msetcmykcolorStroke{}
 \ctr@ld@f\def\c@msetrgbcolor{setrgbcolor}  \ctr@ld@f\def\c@msetrgbcolorStroke{}
 \ctr@ld@f\def\d@fprimarC@lor{\curr@ntcolor\space\curr@ntcolorc@md}
 \ctr@ld@f\def\d@fsecondC@lor{\sec@ndcolor\space\sec@ndcolorc@md}
 \ctr@ld@f\def\d@fthirdC@lor{\th@rdcolor\space\th@rdcolorc@md}
 \ctr@ld@f\def\c@msetdash{setdash}
 \ctr@ld@f\def\c@msetlinejoin{setlinejoin}
 \ctr@ld@f\def\c@msetlinewidth{setlinewidth}
 \ctr@ld@f\def\f@gclosestroke{\immediate\write\fwf@g{closepath\space stroke}}
 \ctr@ld@f\def\f@gfill{\immediate\write\fwf@g{\fillc@md}}
 \ctr@ld@f\def\f@gnewpath{\immediate\write\fwf@g{newpath}}
 \ctr@ld@f\def\f@gstroke{\immediate\write\fwf@g{stroke}}
\fi}
\ctr@ld@f\def\c@pypsfile#1#2{\c@pyfil@{\immediate\write#1}{#2}}
\ctr@ld@f\def\Figinclud@PDF#1#2{\openin\frf@g=#1\pdfliteral{q #2 0 0 #2 0 0 cm}%
    \c@pyfil@{\pdfliteral}{\frf@g}\pdfliteral{Q}\closein\frf@g}
\ctr@ln@w{newif}\ifmored@ta
\ctr@ln@m\bl@nkline
\ctr@ld@f\def\c@pyfil@#1#2{\def\bl@nkline{\par}{\catcode`\%=12
    \loop\ifeof#2\mored@tafalse\else\mored@tatrue\immediate\read#2 to\tr@c
    \ifx\tr@c\bl@nkline\else#1{\tr@c}\fi\fi\ifmored@ta\repeat}}
\ctr@ld@f\def\keln@mun#1#2|{\def\l@debut{#1}\def\l@suite{#2}}
\ctr@ld@f\def\keln@mde#1#2#3|{\def\l@debut{#1#2}\def\l@suite{#3}}
\ctr@ld@f\def\keln@mtr#1#2#3#4|{\def\l@debut{#1#2#3}\def\l@suite{#4}}
\ctr@ld@f\def\keln@mqu#1#2#3#4#5|{\def\l@debut{#1#2#3#4}\def\l@suite{#5}}
\ctr@ld@f\let\@psffilein=\frf@g % file to \read
\ctr@ln@w{newif}\if@psffileok    % continue looking for the bounding box?
\ctr@ln@w{newif}\if@psfbbfound   % success?
\ctr@ln@w{newif}\if@psfverbose   % report what you're making?
\@psfverbosetrue
\ctr@ln@m\@psfllx \ctr@ln@m\@psflly
\ctr@ln@m\@psfurx \ctr@ln@m\@psfury
\ctr@ln@m\resetcolonc@tcode
\ctr@ld@f\def\@psfgetbb#1{\global\@psfbbfoundfalse%
\global\def\@psfllx{0}\global\def\@psflly{0}%
\global\def\@psfurx{30}\global\def\@psfury{30}%
\openin\@psffilein=#1\relax
\ifeof\@psffilein\errmessage{I couldn't open #1, will ignore it}\else
   \edef\resetcolonc@tcode{\catcode`\noexpand\:\the\catcode`\:\relax}%
   {\@psffileoktrue \chardef\other=12
    \def\do##1{\catcode`##1=\other}\dospecials \catcode`\ =10 \resetcolonc@tcode
    \loop
       \read\@psffilein to \@psffileline
       \ifeof\@psffilein\@psffileokfalse\else
          \expandafter\@psfaux\@psffileline:. \\%
       \fi
   \if@psffileok\repeat
   \if@psfbbfound\else
    \if@psfverbose\message{No bounding box comment in #1; using defaults}\fi\fi
   }\closein\@psffilein\fi}%
\ctr@ln@m\@psfbblit
\ctr@ln@m\@psfpercent
{\catcode`\%=12 \global\let\@psfpercent=%\global\def\@psfbblit{%BoundingBox}}%
\ctr@ln@m\@psfaux
\long\def\@psfaux#1#2:#3\\{\ifx#1\@psfpercent
   \def\testit{#2}\ifx\testit\@psfbblit
      \@psfgrab #3 . . . \\%
      \@psffileokfalse
      \global\@psfbbfoundtrue
   \fi\else\ifx#1\par\else\@psffileokfalse\fi\fi}%
\ctr@ld@f\def\@psfempty{}%
\ctr@ld@f\def\@psfgrab #1 #2 #3 #4 #5\\{%
\global\def\@psfllx{#1}\ifx\@psfllx\@psfempty
      \@psfgrab #2 #3 #4 #5 .\\\else
   \global\def\@psflly{#2}%
   \global\def\@psfurx{#3}\global\def\@psfury{#4}\fi}%
\ctr@ld@f\def\PSwrit@cmd#1#2#3{{\Figg@tXY{#1}\c@lprojSP\b@undb@x{\v@lX}{\v@lY}%
    \v@lX=\ptT@ptps\v@lX\v@lY=\ptT@ptps\v@lY%
    \immediate\write#3{\repdecn@mb{\v@lX}\space\repdecn@mb{\v@lY}\space#2}}}
\ctr@ld@f\def\PSwrit@cmdS#1#2#3#4#5{{\Figg@tXY{#1}\c@lprojSP\b@undb@x{\v@lX}{\v@lY}%
    \global\result@t=\v@lX\global\result@@t=\v@lY%
    \v@lX=\ptT@ptps\v@lX\v@lY=\ptT@ptps\v@lY%
    \immediate\write#3{\repdecn@mb{\v@lX}\space\repdecn@mb{\v@lY}\space#2}}%
    \edef#4{\the\result@t}\edef#5{\the\result@@t}}
\ctr@ld@f\def\psaltitude#1[#2,#3,#4]{{\ifcurr@ntPS\ifps@cri%
    \PSc@mment{psaltitude Square Dim=#1, Triangle=[#2 / #3,#4]}%
    \s@uvc@ntr@l\et@tpsaltitude\resetc@ntr@l{2}\figptorthoprojline-5:=#2/#3,#4/%
    \figvectP -1[#3,#4]\n@rminf{\v@leur}{-1}\vecunit@{-3}{-1}%
    \figvectP -1[-5,#3]\n@rminf{\v@lmin}{-1}\figvectP -2[-5,#4]\n@rminf{\v@lmax}{-2}%
    \ifdim\v@lmin<\v@lmax\s@mme=#3\else\v@lmax=\v@lmin\s@mme=#4\fi%
    \figvectP -4[-5,#2]\vecunit@{-4}{-4}\delt@=#1\unit@%
    \edef\t@ille{\repdecn@mb{\delt@}}\figpttra-1:=-5/\t@ille,-3/%
    \figptstra-3=-5,-1/\t@ille,-4/\psline[#2,-5]\s@uvdash{\typ@dash}%
    \pssetdash{\defaultdash}\psline[-1,-2,-3]\pssetdash{\typ@dash}%
    \ifdim\v@leur<\v@lmax\Pss@tsecondSt\psline[-5,\the\s@mme]\Psrest@reSt\fi%
    \PSc@mment{End psaltitude}\resetc@ntr@l\et@tpsaltitude\fi\fi}}
\ctr@ld@f\def\Ps@rcerc#1;#2(#3,#4){\ellBB@x#1;#2,#2(#3,#4,0)%
    \f@gnewpath{\delt@=#2\unit@\delt@=\ptT@ptps\delt@%
    \BdingB@xfalse%
    \PSwrit@cmd{#1}{\repdecn@mb{\delt@}\space #3\space #4\space arc}{\fwf@g}}}
\ctr@ln@m\psarccirc
\ctr@ld@f\def\psarccircDD#1;#2(#3,#4){\ifcurr@ntPS\ifps@cri%
    \PSc@mment{psarccircDD Center=#1 ; Radius=#2 (Ang1=#3, Ang2=#4)}%
    \iffillm@de\Ps@rcerc#1;#2(#3,#4)%
    \f@gfill%
    \else\Ps@rcerc#1;#2(#3,#4)\f@gstroke\fi%
    \PSc@mment{End psarccircDD}\fi\fi}
\ctr@ld@f\def\psarccircTD#1,#2,#3;#4(#5,#6){{\ifcurr@ntPS\ifps@cri\s@uvc@ntr@l\et@tpsarccircTD%
    \PSc@mment{psarccircTD Center=#1,P1=#2,P2=#3 ; Radius=#4 (Ang1=#5, Ang2=#6)}%
    \setc@ntr@l{2}\c@lExtAxes#1,#2,#3(#4)\psarcellPATD#1,-4,-5(#5,#6)%
    \PSc@mment{End psarccircTD}\resetc@ntr@l\et@tpsarccircTD\fi\fi}}
\ctr@ld@f\def\c@lExtAxes#1,#2,#3(#4){%
    \figvectPTD-5[#1,#2]\vecunit@{-5}{-5}\figvectNTD-4[#1,#2,#3]\vecunit@{-4}{-4}%
    \figvectNVTD-3[-4,-5]\delt@=#4\unit@\edef\r@yon{\repdecn@mb{\delt@}}%
    \figpttra-4:=#1/\r@yon,-5/\figpttra-5:=#1/\r@yon,-3/}
\ctr@ln@m\psarccircP
\ctr@ld@f\def\psarccircPDD#1;#2[#3,#4]{{\ifcurr@ntPS\ifps@cri\s@uvc@ntr@l\et@tpsarccircPDD%
    \PSc@mment{psarccircPDD Center=#1; Radius=#2, [P1=#3, P2=#4]}%
    \Ps@ngleparam#1;#2[#3,#4]\ifdim\v@lmin>\v@lmax\advance\v@lmax\DePI@deg\fi%
    \edef\@ngdeb{\repdecn@mb{\v@lmin}}\edef\@ngfin{\repdecn@mb{\v@lmax}}%
    \psarccirc#1;\r@dius(\@ngdeb,\@ngfin)%
    \PSc@mment{End psarccircPDD}\resetc@ntr@l\et@tpsarccircPDD\fi\fi}}
\ctr@ld@f\def\psarccircPTD#1;#2[#3,#4,#5]{{\ifcurr@ntPS\ifps@cri\s@uvc@ntr@l\et@tpsarccircPTD%
    \PSc@mment{psarccircPTD Center=#1; Radius=#2, [P1=#3, P2=#4, P3=#5]}%
    \setc@ntr@l{2}\c@lExtAxes#1,#3,#5(#2)\psarcellPP#1,-4,-5[#3,#4]%
    \PSc@mment{End psarccircPTD}\resetc@ntr@l\et@tpsarccircPTD\fi\fi}}
\ctr@ld@f\def\Ps@ngleparam#1;#2[#3,#4]{\setc@ntr@l{2}%
    \figvectPDD-1[#1,#3]\vecunit@{-1}{-1}\Figg@tXY{-1}\arct@n\v@lmin(\v@lX,\v@lY)%
    \figvectPDD-2[#1,#4]\vecunit@{-2}{-2}\Figg@tXY{-2}\arct@n\v@lmax(\v@lX,\v@lY)%
    \v@lmin=\rdT@deg\v@lmin\v@lmax=\rdT@deg\v@lmax%
    \v@leur=#2pt\maxim@m{\mili@u}{-\v@leur}{\v@leur}%
    \edef\r@dius{\repdecn@mb{\mili@u}}}
\ctr@ld@f\def\Ps@rcercBz#1;#2(#3,#4){\Ps@rellBz#1;#2,#2(#3,#4,0)}
\ctr@ld@f\def\Ps@rellBz#1;#2,#3(#4,#5,#6){%
    \ellBB@x#1;#2,#3(#4,#5,#6)\BdingB@xfalse%
    \c@lNbarcs{#4}{#5}\v@leur=#4pt\setc@ntr@l{2}\figptell-13::#1;#2,#3(#4,#6)%
    \f@gnewpath\PSwrit@cmd{-13}{\c@mmoveto}{\fwf@g}%
    \s@mme=\z@\bcl@rellBz#1;#2,#3(#6)\BdingB@xtrue}
\ctr@ld@f\def\bcl@rellBz#1;#2,#3(#4){\relax%
    \ifnum\s@mme<\p@rtent\advance\s@mme\@ne%
    \advance\v@leur\delt@\edef\@ngle{\repdecn@mb\v@leur}\figptell-14::#1;#2,#3(\@ngle,#4)%
    \advance\v@leur\delt@\edef\@ngle{\repdecn@mb\v@leur}\figptell-15::#1;#2,#3(\@ngle,#4)%
    \advance\v@leur\delt@\edef\@ngle{\repdecn@mb\v@leur}\figptell-16::#1;#2,#3(\@ngle,#4)%
    \figptscontrolDD-18[-13,-14,-15,-16]%
    \PSwrit@cmd{-18}{}{\fwf@g}\PSwrit@cmd{-17}{}{\fwf@g}%
    \PSwrit@cmd{-16}{\c@mcurveto}{\fwf@g}%
    \figptcopyDD-13:/-16/\bcl@rellBz#1;#2,#3(#4)\fi}
\ctr@ld@f\def\Ps@rell#1;#2,#3(#4,#5,#6){\ellBB@x#1;#2,#3(#4,#5,#6)%
    \f@gnewpath{\v@lmin=#2\unit@\v@lmin=\ptT@ptps\v@lmin%
    \v@lmax=#3\unit@\v@lmax=\ptT@ptps\v@lmax\BdingB@xfalse%
    \PSwrit@cmd{#1}%
    {#6\space\repdecn@mb{\v@lmin}\space\repdecn@mb{\v@lmax}\space #4\space #5\space ellipse}{\fwf@g}}%
    \global\Use@llipsetrue}
\ctr@ln@m\psarcell
\ctr@ld@f\def\psarcellDD#1;#2,#3(#4,#5,#6){{\ifcurr@ntPS\ifps@cri%
    \PSc@mment{psarcellDD Center=#1 ; XRad=#2, YRad=#3 (Ang1=#4, Ang2=#5, Inclination=#6)}%
    \iffillm@de\Ps@rell#1;#2,#3(#4,#5,#6)%
    \f@gfill%
    \else\Ps@rell#1;#2,#3(#4,#5,#6)\f@gstroke\fi%
    \PSc@mment{End psarcellDD}\fi\fi}}
\ctr@ld@f\def\psarcellTD#1;#2,#3(#4,#5,#6){{\ifcurr@ntPS\ifps@cri\s@uvc@ntr@l\et@tpsarcellTD%
    \PSc@mment{psarcellTD Center=#1 ; XRad=#2, YRad=#3 (Ang1=#4, Ang2=#5, Inclination=#6)}%
    \setc@ntr@l{2}\figpttraC -8:=#1/#2,0,0/\figpttraC -7:=#1/0,#3,0/%
    \figvectC -4(0,0,1)\figptsrot -8=-8,-7/#1,#6,-4/\psarcellPATD#1,-8,-7(#4,#5)%
    \PSc@mment{End psarcellTD}\resetc@ntr@l\et@tpsarcellTD\fi\fi}}
\ctr@ln@m\psarcellPA
\ctr@ld@f\def\psarcellPADD#1,#2,#3(#4,#5){{\ifcurr@ntPS\ifps@cri\s@uvc@ntr@l\et@tpsarcellPADD%
    \PSc@mment{psarcellPADD Center=#1,PtAxis1=#2,PtAxis2=#3 (Ang1=#4, Ang2=#5)}%
    \setc@ntr@l{2}\figvectPDD-1[#1,#2]\vecunit@DD{-1}{-1}\v@lX=\ptT@unit@\result@t%
    \edef\XR@d{\repdecn@mb{\v@lX}}\Figg@tXY{-1}\arct@n\v@lmin(\v@lX,\v@lY)%
    \v@lmin=\rdT@deg\v@lmin\edef\Inclin@{\repdecn@mb{\v@lmin}}%
    \figgetdist\YR@d[#1,#3]\psarcellDD#1;\XR@d,\YR@d(#4,#5,\Inclin@)%
    \PSc@mment{End psarcellPADD}\resetc@ntr@l\et@tpsarcellPADD\fi\fi}}
\ctr@ld@f\def\psarcellPATD#1,#2,#3(#4,#5){{\ifcurr@ntPS\ifps@cri\s@uvc@ntr@l\et@tpsarcellPATD%
    \PSc@mment{psarcellPATD Center=#1,PtAxis1=#2,PtAxis2=#3 (Ang1=#4, Ang2=#5)}%
    \iffillm@de\Ps@rellPATD#1,#2,#3(#4,#5)%
    \f@gfill%
    \else\Ps@rellPATD#1,#2,#3(#4,#5)\f@gstroke\fi%
    \PSc@mment{End psarcellPATD}\resetc@ntr@l\et@tpsarcellPATD\fi\fi}}
\ctr@ld@f\def\Ps@rellPATD#1,#2,#3(#4,#5){\let\c@lprojSP=\relax%
    \setc@ntr@l{2}\figvectPTD-1[#1,#2]\figvectPTD-2[#1,#3]\c@lNbarcs{#4}{#5}%
    \v@leur=#4pt\c@lptellP{#1}{-1}{-2}\Figptpr@j-5:/-3/%
    \f@gnewpath\PSwrit@cmdS{-5}{\c@mmoveto}{\fwf@g}{\X@un}{\Y@un}%
    \edef\C@nt@r{#1}\s@mme=\z@\bcl@rellPATD}
\ctr@ld@f\def\bcl@rellPATD{\relax%
    \ifnum\s@mme<\p@rtent\advance\s@mme\@ne%
    \advance\v@leur\delt@\c@lptellP{\C@nt@r}{-1}{-2}\Figptpr@j-4:/-3/%
    \advance\v@leur\delt@\c@lptellP{\C@nt@r}{-1}{-2}\Figptpr@j-6:/-3/%
    \advance\v@leur\delt@\c@lptellP{\C@nt@r}{-1}{-2}\Figptpr@j-3:/-3/%
    \v@lX=\z@\v@lY=\z@\Figtr@nptDD{-5}{-5}\Figtr@nptDD{2}{-3}%
    \divide\v@lX\@vi\divide\v@lY\@vi%
    \Figtr@nptDD{3}{-4}\Figtr@nptDD{-1.5}{-6}\v@lmin=\v@lX\v@lmax=\v@lY%
    \v@lX=\z@\v@lY=\z@\Figtr@nptDD{2}{-5}\Figtr@nptDD{-5}{-3}%
    \divide\v@lX\@vi\divide\v@lY\@vi\Figtr@nptDD{-1.5}{-4}\Figtr@nptDD{3}{-6}%
    \BdingB@xfalse%
    \Figp@intregDD-4:(\v@lmin,\v@lmax)\PSwrit@cmdS{-4}{}{\fwf@g}{\X@de}{\Y@de}%
    \Figp@intregDD-4:(\v@lX,\v@lY)\PSwrit@cmdS{-4}{}{\fwf@g}{\X@tr}{\Y@tr}%
    \BdingB@xtrue\PSwrit@cmdS{-3}{\c@mcurveto}{\fwf@g}{\X@qu}{\Y@qu}%
    \B@zierBB@x{1}{\Y@un}(\X@un,\X@de,\X@tr,\X@qu)%
    \B@zierBB@x{2}{\X@un}(\Y@un,\Y@de,\Y@tr,\Y@qu)%
    \edef\X@un{\X@qu}\edef\Y@un{\Y@qu}\figptcopyDD-5:/-3/\bcl@rellPATD\fi}
\ctr@ld@f\def\c@lNbarcs#1#2{%
    \delt@=#2pt\advance\delt@-#1pt\maxim@m{\v@lmax}{\delt@}{-\delt@}%
    \v@leur=\v@lmax\divide\v@leur45 \p@rtentiere{\p@rtent}{\v@leur}\advance\p@rtent\@ne%
    \s@mme=\p@rtent\multiply\s@mme\thr@@\divide\delt@\s@mme}
\ctr@ld@f\def\psarcellPP#1,#2,#3[#4,#5]{{\ifcurr@ntPS\ifps@cri\s@uvc@ntr@l\et@tpsarcellPP%
    \PSc@mment{psarcellPP Center=#1,PtAxis1=#2,PtAxis2=#3 [Point1=#4, Point2=#5]}%
    \setc@ntr@l{2}\figvectP-2[#1,#3]\vecunit@{-2}{-2}\v@lmin=\result@t%
    \invers@{\v@lmax}{\v@lmin}%
    \figvectP-1[#1,#2]\vecunit@{-1}{-1}\v@leur=\result@t%
    \v@leur=\repdecn@mb{\v@lmax}\v@leur\edef\AsB@{\repdecn@mb{\v@leur}}% a/b
    \c@lAngle{#1}{#4}{\v@lmin}\edef\@ngdeb{\repdecn@mb{\v@lmin}}%
    \c@lAngle{#1}{#5}{\v@lmax}\ifdim\v@lmin>\v@lmax\advance\v@lmax\DePI@deg\fi%
    \edef\@ngfin{\repdecn@mb{\v@lmax}}\psarcellPA#1,#2,#3(\@ngdeb,\@ngfin)%
    \PSc@mment{End psarcellPP}\resetc@ntr@l\et@tpsarcellPP\fi\fi}}
\ctr@ld@f\def\c@lAngle#1#2#3{\figvectP-3[#1,#2]%
    \c@lproscal\delt@[-3,-1]\c@lproscal\v@leur[-3,-2]%
    \v@leur=\AsB@\v@leur\arct@n#3(\delt@,\v@leur)#3=\rdT@deg#3}
\ctr@ln@w{newif}\if@rrowratio\@rrowratiotrue
\ctr@ln@w{newif}\if@rrowhfill
\ctr@ln@w{newif}\if@rrowhout
\ctr@ld@f\def\Psset@rrowhe@d#1=#2|{\keln@mun#1|%
    \def\n@mref{a}\ifx\l@debut\n@mref\pssetarrowheadangle{#2}\else% angle
    \def\n@mref{f}\ifx\l@debut\n@mref\pssetarrowheadfill{#2}\else% fillmode
    \def\n@mref{l}\ifx\l@debut\n@mref\pssetarrowheadlength{#2}\else% length
    \def\n@mref{o}\ifx\l@debut\n@mref\pssetarrowheadout{#2}\else% out
    \def\n@mref{r}\ifx\l@debut\n@mref\pssetarrowheadratio{#2}\else% ratio
    \immediate\write16{*** Unknown attribute: \BS@ psset arrowhead(..., #1=...)}%
    \fi\fi\fi\fi\fi}
\ctr@ln@m\@rrowheadangle
\ctr@ln@m\C@AHANG \ctr@ln@m\S@AHANG \ctr@ln@m\UNSS@N
\ctr@ld@f\def\pssetarrowheadangle#1{\edef\@rrowheadangle{#1}{\c@ssin{\C@}{\S@}{#1}%
    \xdef\C@AHANG{\C@}\xdef\S@AHANG{\S@}\v@lmax=\S@ pt%
    \invers@{\v@leur}{\v@lmax}\maxim@m{\v@leur}{\v@leur}{-\v@leur}%
    \xdef\UNSS@N{\the\v@leur}}}
\ctr@ld@f\def\pssetarrowheadfill#1{\expandafter\set@rrowhfill#1:}
\ctr@ld@f\def\set@rrowhfill#1#2:{\if#1n\@rrowhfillfalse\else\@rrowhfilltrue\fi}
\ctr@ld@f\def\pssetarrowheadout#1{\expandafter\set@rrowhout#1:}
\ctr@ld@f\def\set@rrowhout#1#2:{\if#1n\@rrowhoutfalse\else\@rrowhouttrue\fi}
\ctr@ln@m\@rrowheadlength
\ctr@ld@f\def\pssetarrowheadlength#1{\edef\@rrowheadlength{#1}\@rrowratiofalse}
\ctr@ln@m\@rrowheadratio
\ctr@ld@f\def\pssetarrowheadratio#1{\edef\@rrowheadratio{#1}\@rrowratiotrue}
\ctr@ln@m\defaultarrowheadlength
\ctr@ld@f\def\psresetarrowhead{%
    \pssetarrowheadangle{\defaultarrowheadangle}%
    \pssetarrowheadfill{\defaultarrowheadfill}%
    \pssetarrowheadout{\defaultarrowheadout}%
    \pssetarrowheadratio{\defaultarrowheadratio}%
    \d@fm@cdim\defaultarrowheadlength{\defaulth@rdahlength}% Valeur par defaut...
    \pssetarrowheadlength{\defaultarrowheadlength}}
\ctr@ld@f\def\defaultarrowheadratio{0.1}
\ctr@ld@f\def\defaultarrowheadangle{20}
\ctr@ld@f\def\defaultarrowheadfill{no}
\ctr@ld@f\def\defaultarrowheadout{no}
\ctr@ld@f\def\defaulth@rdahlength{8pt}
\ctr@ln@m\psarrow
\ctr@ld@f\def\psarrowDD[#1,#2]{{\ifcurr@ntPS\ifps@cri\s@uvc@ntr@l\et@tpsarrow%
    \PSc@mment{psarrowDD [Pt1,Pt2]=[#1,#2]}\pssetfillmode{no}%
    \psarrowheadDD[#1,#2]\setc@ntr@l{2}\psline[#1,-3]%
    \PSc@mment{End psarrowDD}\resetc@ntr@l\et@tpsarrow\fi\fi}}
\ctr@ld@f\def\psarrowTD[#1,#2]{{\ifcurr@ntPS\ifps@cri\s@uvc@ntr@l\et@tpsarrowTD%
    \PSc@mment{psarrowTD [Pt1,Pt2]=[#1,#2]}\resetc@ntr@l{2}%
    \Figptpr@j-5:/#1/\Figptpr@j-6:/#2/\let\c@lprojSP=\relax\psarrowDD[-5,-6]%
    \PSc@mment{End psarrowTD}\resetc@ntr@l\et@tpsarrowTD\fi\fi}}
\ctr@ln@m\psarrowhead
\ctr@ld@f\def\psarrowheadDD[#1,#2]{{\ifcurr@ntPS\ifps@cri\s@uvc@ntr@l\et@tpsarrowheadDD%
    \if@rrowhfill\def\@hangle{-\@rrowheadangle}\else\def\@hangle{\@rrowheadangle}\fi%
    \if@rrowratio%
    \if@rrowhout\def\@hratio{-\@rrowheadratio}\else\def\@hratio{\@rrowheadratio}\fi%
    \PSc@mment{psarrowheadDD Ratio=\@hratio, Angle=\@hangle, [Pt1,Pt2]=[#1,#2]}%
    \Ps@rrowhead\@hratio,\@hangle[#1,#2]%
    \else%
    \if@rrowhout\def\@hlength{-\@rrowheadlength}\else\def\@hlength{\@rrowheadlength}\fi%
    \PSc@mment{psarrowheadDD Length=\@hlength, Angle=\@hangle, [Pt1,Pt2]=[#1,#2]}%
    \Ps@rrowheadfd\@hlength,\@hangle[#1,#2]%
    \fi%
    \PSc@mment{End psarrowheadDD}\resetc@ntr@l\et@tpsarrowheadDD\fi\fi}}
\ctr@ld@f\def\psarrowheadTD[#1,#2]{{\ifcurr@ntPS\ifps@cri\s@uvc@ntr@l\et@tpsarrowheadTD%
    \PSc@mment{psarrowheadTD [Pt1,Pt2]=[#1,#2]}\resetc@ntr@l{2}%
    \Figptpr@j-5:/#1/\Figptpr@j-6:/#2/\let\c@lprojSP=\relax\psarrowheadDD[-5,-6]%
    \PSc@mment{End psarrowheadTD}\resetc@ntr@l\et@tpsarrowheadTD\fi\fi}}
\ctr@ld@f\def\Ps@rrowhead#1,#2[#3,#4]{\v@leur=#1\p@\maxim@m{\v@leur}{\v@leur}{-\v@leur}%
    \ifdim\v@leur>\Cepsil@n{% Arrow is not degenerated
    \PSc@mment{ps@rrowhead Ratio=#1, Angle=#2, [Pt1,Pt2]=[#3,#4]}\v@leur=\UNSS@N%
    \v@leur=\curr@ntwidth\v@leur\v@leur=\ptpsT@pt\v@leur\delt@=.5\v@leur% = width / (2 sin(Angle))
    \setc@ntr@l{2}\figvectPDD-3[#4,#3]%
    \Figg@tXY{-3}\v@lX=#1\v@lX\v@lY=#1\v@lY\Figv@ctCreg-3(\v@lX,\v@lY)%
    \vecunit@{-4}{-3}\mili@u=\result@t%
    \ifdim#2pt>\z@\v@lXa=-\C@AHANG\delt@%
     \edef\c@ef{\repdecn@mb{\v@lXa}}\figpttraDD-3:=-3/\c@ef,-4/\fi%
    \edef\c@ef{\repdecn@mb{\delt@}}%
    \v@lXa=\mili@u\v@lXa=\C@AHANG\v@lXa%
    \v@lYa=\ptpsT@pt\p@\v@lYa=\curr@ntwidth\v@lYa\v@lYa=\sDcc@ngle\v@lYa%
    \advance\v@lXa-\v@lYa\gdef\sDcc@ngle{0}%
    \ifdim\v@lXa>\v@leur\edef\c@efendpt{\repdecn@mb{\v@leur}}%
    \else\edef\c@efendpt{\repdecn@mb{\v@lXa}}\fi%
    \Figg@tXY{-3}\v@lmin=\v@lX\v@lmax=\v@lY%
    \v@lXa=\C@AHANG\v@lmin\v@lYa=\S@AHANG\v@lmax\advance\v@lXa\v@lYa%
    \v@lYa=-\S@AHANG\v@lmin\v@lX=\C@AHANG\v@lmax\advance\v@lYa\v@lX%
    \setc@ntr@l{1}\Figg@tXY{#4}\advance\v@lX\v@lXa\advance\v@lY\v@lYa%
    \setc@ntr@l{2}\Figp@intregDD-2:(\v@lX,\v@lY)%
    \v@lXa=\C@AHANG\v@lmin\v@lYa=-\S@AHANG\v@lmax\advance\v@lXa\v@lYa%
    \v@lYa=\S@AHANG\v@lmin\v@lX=\C@AHANG\v@lmax\advance\v@lYa\v@lX%
    \setc@ntr@l{1}\Figg@tXY{#4}\advance\v@lX\v@lXa\advance\v@lY\v@lYa%
    \setc@ntr@l{2}\Figp@intregDD-1:(\v@lX,\v@lY)%
    \ifdim#2pt<\z@\fillm@detrue\psline[-2,#4,-1]% fill
    \else\figptstraDD-3=#4,-2,-1/\c@ef,-4/\psline[-2,-3,-1]\fi% no fill
    \ifdim#1pt>\z@\figpttraDD-3:=#4/\c@efendpt,-4/\else\figptcopyDD-3:/#4/\fi%
    \PSc@mment{End ps@rrowhead}}\fi}
\ctr@ld@f\def\sDcc@ngle{0}% Initialisation
\ctr@ld@f\def\Ps@rrowheadfd#1,#2[#3,#4]{{%
    \PSc@mment{ps@rrowheadfd Length=#1, Angle=#2, [Pt1,Pt2]=[#3,#4]}%
    \setc@ntr@l{2}\figvectPDD-1[#3,#4]\n@rmeucDD{\v@leur}{-1}\v@leur=\ptT@unit@\v@leur%
    \invers@{\v@leur}{\v@leur}\v@leur=#1\v@leur\edef\R@tio{\repdecn@mb{\v@leur}}%
    \Ps@rrowhead\R@tio,#2[#3,#4]\PSc@mment{End ps@rrowheadfd}}}
\ctr@ln@m\psarrowBezier
\ctr@ld@f\def\psarrowBezierDD[#1,#2,#3,#4]{{\ifcurr@ntPS\ifps@cri\s@uvc@ntr@l\et@tpsarrowBezierDD%
    \PSc@mment{psarrowBezierDD Control points=#1,#2,#3,#4}\setc@ntr@l{2}%
    \if@rrowratio\c@larclengthDD\v@leur,10[#1,#2,#3,#4]\else\v@leur=\z@\fi%
    \Ps@rrowB@zDD\v@leur[#1,#2,#3,#4]%
    \PSc@mment{End psarrowBezierDD}\resetc@ntr@l\et@tpsarrowBezierDD\fi\fi}}
\ctr@ld@f\def\psarrowBezierTD[#1,#2,#3,#4]{{\ifcurr@ntPS\ifps@cri\s@uvc@ntr@l\et@tpsarrowBezierTD%
    \PSc@mment{psarrowBezierTD Control points=#1,#2,#3,#4}\resetc@ntr@l{2}%
    \Figptpr@j-7:/#1/\Figptpr@j-8:/#2/\Figptpr@j-9:/#3/\Figptpr@j-10:/#4/%
    \let\c@lprojSP=\relax\ifnum\curr@ntproj<\tw@\psarrowBezierDD[-7,-8,-9,-10]%
    \else\f@gnewpath\PSwrit@cmd{-7}{\c@mmoveto}{\fwf@g}%
    \if@rrowratio\c@larclengthDD\mili@u,10[-7,-8,-9,-10]\else\mili@u=\z@\fi%
    \p@rtent=\NBz@rcs\advance\p@rtent\m@ne\subB@zierTD\p@rtent[#1,#2,#3,#4]%
    \f@gstroke%
    \advance\v@lmin\p@rtent\delt@% Initialized in \subB@zierTD
    \v@leur=\v@lmin\advance\v@leur0.33333 \delt@\edef\unti@rs{\repdecn@mb{\v@leur}}%
    \v@leur=\v@lmin\advance\v@leur0.66666 \delt@\edef\deti@rs{\repdecn@mb{\v@leur}}%
    \figptcopyDD-8:/-10/\c@lsubBzarc\unti@rs,\deti@rs[#1,#2,#3,#4]%
    \figptcopyDD-8:/-4/\figptcopyDD-9:/-3/\Ps@rrowB@zDD\mili@u[-7,-8,-9,-10]\fi%
    \PSc@mment{End psarrowBezierTD}\resetc@ntr@l\et@tpsarrowBezierTD\fi\fi}}
\ctr@ld@f\def\c@larclengthDD#1,#2[#3,#4,#5,#6]{{\p@rtent=#2\figptcopyDD-5:/#3/%
    \delt@=\p@\divide\delt@\p@rtent\c@rre=\z@\v@leur=\z@\s@mme=\z@%
    \loop\ifnum\s@mme<\p@rtent\advance\s@mme\@ne\advance\v@leur\delt@%
    \edef\T@{\repdecn@mb{\v@leur}}\figptBezierDD-6::\T@[#3,#4,#5,#6]%
    \figvectPDD-1[-5,-6]\n@rmeucDD{\mili@u}{-1}\advance\c@rre\mili@u%
    \figptcopyDD-5:/-6/\repeat\global\result@t=\ptT@unit@\c@rre}#1=\result@t}
\ctr@ld@f\def\Ps@rrowB@zDD#1[#2,#3,#4,#5]{{\pssetfillmode{no}%
    \if@rrowratio\delt@=\@rrowheadratio#1\else\delt@=\@rrowheadlength pt\fi%
    \v@leur=\C@AHANG\delt@\edef\R@dius{\repdecn@mb{\v@leur}}%
    \FigptintercircB@zDD-5::0,\R@dius[#5,#4,#3,#2]%
    \pssetarrowheadlength{\repdecn@mb{\delt@}}\psarrowheadDD[-5,#5]%
    \let\n@rmeuc=\n@rmeucDD\figgetdist\R@dius[#5,-3]%
    \FigptintercircB@zDD-6::0,\R@dius[#5,#4,#3,#2]%
    \figptBezierDD-5::0.33333[#5,#4,#3,#2]\figptBezierDD-3::0.66666[#5,#4,#3,#2]%
    \figptscontrolDD-5[-6,-5,-3,#2]\psBezierDD1[-6,-5,-4,#2]}}
\ctr@ln@m\psarrowcirc
\ctr@ld@f\def\psarrowcircDD#1;#2(#3,#4){{\ifcurr@ntPS\ifps@cri\s@uvc@ntr@l\et@tpsarrowcircDD%
    \PSc@mment{psarrowcircDD Center=#1 ; Radius=#2 (Ang1=#3,Ang2=#4)}%
    \pssetfillmode{no}\Pscirc@rrowhead#1;#2(#3,#4)%
    \setc@ntr@l{2}\figvectPDD -4[#1,-3]\vecunit@{-4}{-4}%
    \Figg@tXY{-4}\arct@n\v@lmin(\v@lX,\v@lY)%
    \v@lmin=\rdT@deg\v@lmin\v@leur=#4pt\advance\v@leur-\v@lmin%
    \maxim@m{\v@leur}{\v@leur}{-\v@leur}%
    \ifdim\v@leur>\DemiPI@deg\relax\ifdim\v@lmin<#4pt\advance\v@lmin\DePI@deg%
    \else\advance\v@lmin-\DePI@deg\fi\fi\edef\ar@ngle{\repdecn@mb{\v@lmin}}%
    \ifdim#3pt<#4pt\psarccirc#1;#2(#3,\ar@ngle)\else\psarccirc#1;#2(\ar@ngle,#3)\fi%
    \PSc@mment{End psarrowcircDD}\resetc@ntr@l\et@tpsarrowcircDD\fi\fi}}
\ctr@ld@f\def\psarrowcircTD#1,#2,#3;#4(#5,#6){{\ifcurr@ntPS\ifps@cri\s@uvc@ntr@l\et@tpsarrowcircTD%
    \PSc@mment{psarrowcircTD Center=#1,P1=#2,P2=#3 ; Radius=#4 (Ang1=#5, Ang2=#6)}%
    \resetc@ntr@l{2}\c@lExtAxes#1,#2,#3(#4)\let\c@lprojSP=\relax%
    \figvectPTD-11[#1,-4]\figvectPTD-12[#1,-5]\c@lNbarcs{#5}{#6}%
    \if@rrowratio\v@lmax=\degT@rd\v@lmax\edef\D@lpha{\repdecn@mb{\v@lmax}}\fi%
    \advance\p@rtent\m@ne\mili@u=\z@%
    \v@leur=#5pt\c@lptellP{#1}{-11}{-12}\Figptpr@j-9:/-3/%
    \f@gnewpath\PSwrit@cmdS{-9}{\c@mmoveto}{\fwf@g}{\X@un}{\Y@un}%
    \edef\C@nt@r{#1}\s@mme=\z@\bcl@rcircTD\f@gstroke%
    \advance\v@leur\delt@\c@lptellP{#1}{-11}{-12}\Figptpr@j-5:/-3/%
    \advance\v@leur\delt@\c@lptellP{#1}{-11}{-12}\Figptpr@j-6:/-3/%
    \advance\v@leur\delt@\c@lptellP{#1}{-11}{-12}\Figptpr@j-10:/-3/%
    \figptscontrolDD-8[-9,-5,-6,-10]%
    \if@rrowratio\c@lcurvradDD0.5[-9,-8,-7,-10]\advance\mili@u\result@t%
    \maxim@m{\mili@u}{\mili@u}{-\mili@u}\mili@u=\ptT@unit@\mili@u%
    \mili@u=\D@lpha\mili@u\advance\p@rtent\@ne\divide\mili@u\p@rtent\fi%
    \Ps@rrowB@zDD\mili@u[-9,-8,-7,-10]%
    \PSc@mment{End psarrowcircTD}\resetc@ntr@l\et@tpsarrowcircTD\fi\fi}}
\ctr@ld@f\def\bcl@rcircTD{\relax%
    \ifnum\s@mme<\p@rtent\advance\s@mme\@ne%
    \advance\v@leur\delt@\c@lptellP{\C@nt@r}{-11}{-12}\Figptpr@j-5:/-3/%
    \advance\v@leur\delt@\c@lptellP{\C@nt@r}{-11}{-12}\Figptpr@j-6:/-3/%
    \advance\v@leur\delt@\c@lptellP{\C@nt@r}{-11}{-12}\Figptpr@j-10:/-3/%
    \figptscontrolDD-8[-9,-5,-6,-10]\BdingB@xfalse%
    \PSwrit@cmdS{-8}{}{\fwf@g}{\X@de}{\Y@de}\PSwrit@cmdS{-7}{}{\fwf@g}{\X@tr}{\Y@tr}%
    \BdingB@xtrue\PSwrit@cmdS{-10}{\c@mcurveto}{\fwf@g}{\X@qu}{\Y@qu}%
    \if@rrowratio\c@lcurvradDD0.5[-9,-8,-7,-10]\advance\mili@u\result@t\fi%
    \B@zierBB@x{1}{\Y@un}(\X@un,\X@de,\X@tr,\X@qu)%
    \B@zierBB@x{2}{\X@un}(\Y@un,\Y@de,\Y@tr,\Y@qu)%
    \edef\X@un{\X@qu}\edef\Y@un{\Y@qu}\figptcopyDD-9:/-10/\bcl@rcircTD\fi}
\ctr@ld@f\def\Pscirc@rrowhead#1;#2(#3,#4){{%
    \PSc@mment{pscirc@rrowhead Center=#1 ; Radius=#2 (Ang1=#3,Ang2=#4)}%
    \v@leur=#2\unit@\edef\s@glen{\repdecn@mb{\v@leur}}\v@lY=\z@\v@lX=\v@leur%
    \resetc@ntr@l{2}\Figv@ctCreg-3(\v@lX,\v@lY)\figpttraDD-5:=#1/1,-3/%
    \figptrotDD-5:=-5/#1,#4/%
    \figvectPDD-3[#1,-5]\Figg@tXY{-3}\v@leur=\v@lX%
    \ifdim#3pt<#4pt\v@lX=\v@lY\v@lY=-\v@leur\else\v@lX=-\v@lY\v@lY=\v@leur\fi%
    \Figv@ctCreg-3(\v@lX,\v@lY)\vecunit@{-3}{-3}%
    \if@rrowratio\v@leur=#4pt\advance\v@leur-#3pt\maxim@m{\mili@u}{-\v@leur}{\v@leur}%
    \mili@u=\degT@rd\mili@u\v@leur=\s@glen\mili@u\edef\s@glen{\repdecn@mb{\v@leur}}%
    \mili@u=#2\mili@u\mili@u=\@rrowheadratio\mili@u\else\mili@u=\@rrowheadlength pt\fi%
    \figpttraDD-6:=-5/\s@glen,-3/\v@leur=#2pt\v@leur=2\v@leur%
    \invers@{\v@leur}{\v@leur}\c@rre=\repdecn@mb{\v@leur}\mili@u% = sin = L/(2R)
    \mili@u=\c@rre\mili@u=\repdecn@mb{\c@rre}\mili@u%
    \v@leur=\p@\advance\v@leur-\mili@u% \v@leur = cos*cos
    \invers@{\mili@u}{2\v@leur}\delt@=\c@rre\delt@=\repdecn@mb{\mili@u}\delt@%
    \xdef\sDcc@ngle{\repdecn@mb{\delt@}}% sin/(2*cos*cos) used in \Ps@rrowhead
    \sqrt@{\mili@u}{\v@leur}\arct@n\v@leur(\mili@u,\c@rre)%
    \v@leur=\rdT@deg\v@leur% \cor@ngle = atan(L/sqrt(4R*R-L*L))
    \ifdim#3pt<#4pt\v@leur=-\v@leur\fi%
    \if@rrowhout\v@leur=-\v@leur\fi\edef\cor@ngle{\repdecn@mb{\v@leur}}%
    \figptrotDD-6:=-6/-5,\cor@ngle/\psarrowheadDD[-6,-5]%
    \PSc@mment{End pscirc@rrowhead}}}
\ctr@ln@m\psarrowcircP
\ctr@ld@f\def\psarrowcircPDD#1;#2[#3,#4]{{\ifcurr@ntPS\ifps@cri%
    \PSc@mment{psarrowcircPDD Center=#1; Radius=#2, [P1=#3,P2=#4]}%
    \s@uvc@ntr@l\et@tpsarrowcircPDD\Ps@ngleparam#1;#2[#3,#4]%
    \ifdim\v@leur>\z@\ifdim\v@lmin>\v@lmax\advance\v@lmax\DePI@deg\fi%
    \else\ifdim\v@lmin<\v@lmax\advance\v@lmin\DePI@deg\fi\fi%
    \edef\@ngdeb{\repdecn@mb{\v@lmin}}\edef\@ngfin{\repdecn@mb{\v@lmax}}%
    \psarrowcirc#1;\r@dius(\@ngdeb,\@ngfin)%
    \PSc@mment{End psarrowcircPDD}\resetc@ntr@l\et@tpsarrowcircPDD\fi\fi}}
\ctr@ld@f\def\psarrowcircPTD#1;#2[#3,#4,#5]{{\ifcurr@ntPS\ifps@cri\s@uvc@ntr@l\et@tpsarrowcircPTD%
    \PSc@mment{psarrowcircPTD Center=#1; Radius=#2, [P1=#3,P2=#4,P3=#5]}%
    \figgetangleTD\@ngfin[#1,#3,#4,#5]\v@leur=#2pt%
    \maxim@m{\mili@u}{-\v@leur}{\v@leur}\edef\r@dius{\repdecn@mb{\mili@u}}%
    \ifdim\v@leur<\z@\v@lmax=\@ngfin pt\advance\v@lmax-\DePI@deg%
    \edef\@ngfin{\repdecn@mb{\v@lmax}}\fi\psarrowcircTD#1,#3,#5;\r@dius(0,\@ngfin)%
    \PSc@mment{End psarrowcircPTD}\resetc@ntr@l\et@tpsarrowcircPTD\fi\fi}}
\ctr@ld@f\def\psaxes#1(#2){{\ifcurr@ntPS\ifps@cri\s@uvc@ntr@l\et@tpsaxes%
    \PSc@mment{psaxes Origin=#1 Range=(#2)}\an@lys@xes#2,:\resetc@ntr@l{2}%
    \ifx\t@xt@\empty\ifTr@isDim\ps@xes#1(0,#2,0,#2,0,#2)\else\ps@xes#1(0,#2,0,#2)\fi%
    \else\ps@xes#1(#2)\fi\PSc@mment{End psaxes}\resetc@ntr@l\et@tpsaxes\fi\fi}}
\ctr@ld@f\def\an@lys@xes#1,#2:{\def\t@xt@{#2}}
\ctr@ln@m\ps@xes
\ctr@ld@f\def\ps@xesDD#1(#2,#3,#4,#5){%
    \figpttraC-5:=#1/#2,0/\figpttraC-6:=#1/#3,0/\psarrowDD[-5,-6]%
    \figpttraC-5:=#1/0,#4/\figpttraC-6:=#1/0,#5/\psarrowDD[-5,-6]}
\ctr@ld@f\def\ps@xesTD#1(#2,#3,#4,#5,#6,#7){%
    \figpttraC-7:=#1/#2,0,0/\figpttraC-8:=#1/#3,0,0/\psarrowTD[-7,-8]%
    \figpttraC-7:=#1/0,#4,0/\figpttraC-8:=#1/0,#5,0/\psarrowTD[-7,-8]%
    \figpttraC-7:=#1/0,0,#6/\figpttraC-8:=#1/0,0,#7/\psarrowTD[-7,-8]}
\ctr@ln@m\newGr@FN
\ctr@ld@f\def\newGr@FNPDF#1{\s@mme=\Gr@FNb\advance\s@mme\@ne\xdef\Gr@FNb{\number\s@mme}}
\ctr@ld@f\def\newGr@FNDVI#1{\newGr@FNPDF{}\xdef#1{\jobname GI\Gr@FNb.anx}}
\ctr@ld@f\def\psbeginfig#1{\newGr@FN\DefGIfilen@me\gdef\@utoFN{0}%
    \def\t@xt@{#1}\relax\ifx\t@xt@\empty\psupdatem@detrue%
    \gdef\@utoFN{1}\Psb@ginfig\DefGIfilen@me\else\expandafter\Psb@ginfigNu@#1 :\fi}
\ctr@ld@f\def\Psb@ginfigNu@#1 #2:{\def\t@xt@{#1}\relax\ifx\t@xt@\empty\def\t@xt@{#2}%
    \ifx\t@xt@\empty\psupdatem@detrue\gdef\@utoFN{1}\Psb@ginfig\DefGIfilen@me%
    \else\Psb@ginfigNu@#2:\fi\else\Psb@ginfig{#1}\fi}
\ctr@ln@m\PSfilen@me \ctr@ln@m\auxfilen@me
\ctr@ld@f\def\Psb@ginfig#1{\ifcurr@ntPS\else%
    \edef\PSfilen@me{#1}\edef\auxfilen@me{\jobname.anx}%
    \ifpsupdatem@de\ps@critrue\else\openin\frf@g=\PSfilen@me\relax%
    \ifeof\frf@g\ps@critrue\else\ps@crifalse\fi\closein\frf@g\fi%
    \curr@ntPStrue\c@ldefproj\expandafter\setupd@te\defaultupdate:%
    \ifps@cri\initb@undb@x%
    \immediate\openout\fwf@g=\auxfilen@me\initpss@ttings\fi%
    \fi}
\ctr@ld@f\def\Gr@FNb{0}
\ctr@ld@f\def\figforTeXFileno{\Gr@FNb}
\ctr@ld@f\def\figforTeXFigno{0 }
\ctr@ld@f\def\figforTeXnextFigno{1 }
\ctr@ld@f\edef\DefGIfilen@me{\jobname GI.anx}
\ctr@ld@f\def\initpss@ttings{\psreset{arrowhead,curve,first,flowchart,mesh,second,third}%
    \Use@llipsefalse}
\ctr@ld@f\def\B@zierBB@x#1#2(#3,#4,#5,#6){{\c@rre=\t@n\epsil@n% Do not reduce this value
    \v@lmax=#4\advance\v@lmax-#5\v@lmax=\thr@@\v@lmax\advance\v@lmax#6\advance\v@lmax-#3%
    \mili@u=#4\mili@u=-\tw@\mili@u\advance\mili@u#3\advance\mili@u#5%
    \v@lmin=#4\advance\v@lmin-#3\maxim@m{\v@leur}{-\v@lmax}{\v@lmax}%
    \maxim@m{\delt@}{-\mili@u}{\mili@u}\maxim@m{\v@leur}{\v@leur}{\delt@}%
    \maxim@m{\delt@}{-\v@lmin}{\v@lmin}\maxim@m{\v@leur}{\v@leur}{\delt@}%
    \ifdim\v@leur>\c@rre\invers@{\v@leur}{\v@leur}\edef\Uns@rM@x{\repdecn@mb{\v@leur}}%
    \v@lmax=\Uns@rM@x\v@lmax\mili@u=\Uns@rM@x\mili@u\v@lmin=\Uns@rM@x\v@lmin%
    \maxim@m{\v@leur}{-\v@lmax}{\v@lmax}\ifdim\v@leur<\c@rre%
    \maxim@m{\v@leur}{-\mili@u}{\mili@u}\ifdim\v@leur<\c@rre\else%
    \invers@{\mili@u}{\mili@u}\v@leur=-0.5\v@lmin%
    \v@leur=\repdecn@mb{\mili@u}\v@leur\m@jBBB@x{\v@leur}{#1}{#2}(#3,#4,#5,#6)\fi%
    \else\delt@=\repdecn@mb{\mili@u}\mili@u\v@leur=\repdecn@mb{\v@lmax}\v@lmin%
    \advance\delt@-\v@leur\ifdim\delt@<\z@\else\invers@{\v@lmax}{\v@lmax}%
    \edef\Uns@rAp{\repdecn@mb{\v@lmax}}\sqrt@{\delt@}{\delt@}%
    \v@leur=-\mili@u\advance\v@leur\delt@\v@leur=\Uns@rAp\v@leur%
    \m@jBBB@x{\v@leur}{#1}{#2}(#3,#4,#5,#6)%
    \v@leur=-\mili@u\advance\v@leur-\delt@\v@leur=\Uns@rAp\v@leur%
    \m@jBBB@x{\v@leur}{#1}{#2}(#3,#4,#5,#6)\fi\fi\fi}}
\ctr@ld@f\def\m@jBBB@x#1#2#3(#4,#5,#6,#7){{\relax\ifdim#1>\z@\ifdim#1<\p@%
    \edef\T@{\repdecn@mb{#1}}\v@lX=\p@\advance\v@lX-#1\edef\UNmT@{\repdecn@mb{\v@lX}}%
    \v@lX=#4\v@lY=#5\v@lZ=#6\v@lXa=#7\v@lX=\UNmT@\v@lX\advance\v@lX\T@\v@lY%
    \v@lY=\UNmT@\v@lY\advance\v@lY\T@\v@lZ\v@lZ=\UNmT@\v@lZ\advance\v@lZ\T@\v@lXa%
    \v@lX=\UNmT@\v@lX\advance\v@lX\T@\v@lY\v@lY=\UNmT@\v@lY\advance\v@lY\T@\v@lZ%
    \v@lX=\UNmT@\v@lX\advance\v@lX\T@\v@lY%
    \ifcase#2\or\v@lY=#3\or\v@lY=\v@lX\v@lX=#3\fi\b@undb@x{\v@lX}{\v@lY}\fi\fi}}
\ctr@ld@f\def\PsB@zier#1[#2]{{\f@gnewpath%
    \s@mme=\z@\def\list@num{#2,0}\extrairelepremi@r\p@int\de\list@num%
    \PSwrit@cmdS{\p@int}{\c@mmoveto}{\fwf@g}{\X@un}{\Y@un}\p@rtent=#1\bclB@zier}}
\ctr@ld@f\def\bclB@zier{\relax%
    \ifnum\s@mme<\p@rtent\advance\s@mme\@ne\BdingB@xfalse%
    \extrairelepremi@r\p@int\de\list@num\PSwrit@cmdS{\p@int}{}{\fwf@g}{\X@de}{\Y@de}%
    \extrairelepremi@r\p@int\de\list@num\PSwrit@cmdS{\p@int}{}{\fwf@g}{\X@tr}{\Y@tr}%
    \BdingB@xtrue%
    \extrairelepremi@r\p@int\de\list@num\PSwrit@cmdS{\p@int}{\c@mcurveto}{\fwf@g}{\X@qu}{\Y@qu}%
    \B@zierBB@x{1}{\Y@un}(\X@un,\X@de,\X@tr,\X@qu)%
    \B@zierBB@x{2}{\X@un}(\Y@un,\Y@de,\Y@tr,\Y@qu)%
    \edef\X@un{\X@qu}\edef\Y@un{\Y@qu}\bclB@zier\fi}
\ctr@ln@m\psBezier
\ctr@ld@f\def\psBezierDD#1[#2]{\ifcurr@ntPS\ifps@cri%
    \PSc@mment{psBezierDD N arcs=#1, Control points=#2}%
    \iffillm@de\PsB@zier#1[#2]%
    \f@gfill%
    \else\PsB@zier#1[#2]\f@gstroke\fi%
    \PSc@mment{End psBezierDD}\fi\fi}
\ctr@ln@m\et@tpsBezierTD% ou doubler les {}
\ctr@ld@f\def\psBezierTD#1[#2]{\ifcurr@ntPS\ifps@cri\s@uvc@ntr@l\et@tpsBezierTD%
    \PSc@mment{psBezierTD N arcs=#1, Control points=#2}%
    \iffillm@de\PsB@zierTD#1[#2]%
    \f@gfill%
    \else\PsB@zierTD#1[#2]\f@gstroke\fi%
    \PSc@mment{End psBezierTD}\resetc@ntr@l\et@tpsBezierTD\fi\fi}
\ctr@ld@f\def\PsB@zierTD#1[#2]{\ifnum\curr@ntproj<\tw@\PsB@zier#1[#2]\else\PsB@zier@TD#1[#2]\fi}
\ctr@ld@f\def\PsB@zier@TD#1[#2]{{\f@gnewpath%
    \s@mme=\z@\def\list@num{#2,0}\extrairelepremi@r\p@int\de\list@num%
    \let\c@lprojSP=\relax\setc@ntr@l{2}\Figptpr@j-7:/\p@int/%
    \PSwrit@cmd{-7}{\c@mmoveto}{\fwf@g}%
    \loop\ifnum\s@mme<#1\advance\s@mme\@ne\extrairelepremi@r\p@intun\de\list@num%
    \extrairelepremi@r\p@intde\de\list@num\extrairelepremi@r\p@inttr\de\list@num%
    \subB@zierTD\NBz@rcs[\p@int,\p@intun,\p@intde,\p@inttr]\edef\p@int{\p@inttr}\repeat}}
\ctr@ld@f\def\subB@zierTD#1[#2,#3,#4,#5]{\delt@=\p@\divide\delt@\NBz@rcs\v@lmin=\z@%
    {\Figg@tXY{-7}\edef\X@un{\the\v@lX}\edef\Y@un{\the\v@lY}%
    \s@mme=\z@\loop\ifnum\s@mme<#1\advance\s@mme\@ne%
    \v@leur=\v@lmin\advance\v@leur0.33333 \delt@\edef\unti@rs{\repdecn@mb{\v@leur}}%
    \v@leur=\v@lmin\advance\v@leur0.66666 \delt@\edef\deti@rs{\repdecn@mb{\v@leur}}%
    \advance\v@lmin\delt@\edef\trti@rs{\repdecn@mb{\v@lmin}}%
    \figptBezierTD-8::\trti@rs[#2,#3,#4,#5]\Figptpr@j-8:/-8/%
    \c@lsubBzarc\unti@rs,\deti@rs[#2,#3,#4,#5]\BdingB@xfalse%
    \PSwrit@cmdS{-4}{}{\fwf@g}{\X@de}{\Y@de}\PSwrit@cmdS{-3}{}{\fwf@g}{\X@tr}{\Y@tr}%
    \BdingB@xtrue\PSwrit@cmdS{-8}{\c@mcurveto}{\fwf@g}{\X@qu}{\Y@qu}%
    \B@zierBB@x{1}{\Y@un}(\X@un,\X@de,\X@tr,\X@qu)%
    \B@zierBB@x{2}{\X@un}(\Y@un,\Y@de,\Y@tr,\Y@qu)%
    \edef\X@un{\X@qu}\edef\Y@un{\Y@qu}\figptcopyDD-7:/-8/\repeat}}
\ctr@ld@f\def\NBz@rcs{2}
\ctr@ld@f\def\c@lsubBzarc#1,#2[#3,#4,#5,#6]{\figptBezierTD-5::#1[#3,#4,#5,#6]%
    \figptBezierTD-6::#2[#3,#4,#5,#6]\Figptpr@j-4:/-5/\Figptpr@j-5:/-6/%
    \figptscontrolDD-4[-7,-4,-5,-8]}
\ctr@ln@m\pscirc
\ctr@ld@f\def\pscircDD#1(#2){\ifcurr@ntPS\ifps@cri\PSc@mment{pscircDD Center=#1 (Radius=#2)}%
    \psarccircDD#1;#2(0,360)\PSc@mment{End pscircDD}\fi\fi}
\ctr@ld@f\def\pscircTD#1,#2,#3(#4){\ifcurr@ntPS\ifps@cri%
    \PSc@mment{pscircTD Center=#1,P1=#2,P2=#3 (Radius=#4)}%
    \psarccircTD#1,#2,#3;#4(0,360)\PSc@mment{End pscircTD}\fi\fi}
\ctr@ln@m\p@urcent
{\catcode`\%=12\gdef\p@urcent{%}}
\ctr@ld@f\def\PSc@mment#1{\ifpsdebugmode\immediate\write\fwf@g{\p@urcent\space#1}\fi}
\ctr@ln@m\acc@louv \ctr@ln@m\acc@lfer
{\catcode`\[=1\catcode`\{=12\gdef\acc@louv[{}}
{\catcode`\]=2\catcode`\}=12\gdef\acc@lfer{}]]
\ctr@ld@f\def\PSdict@{\ifUse@llipse%
    \immediate\write\fwf@g{/ellipsedict 9 dict def ellipsedict /mtrx matrix put}%
    \immediate\write\fwf@g{/ellipse \acc@louv ellipsedict begin}%
    \immediate\write\fwf@g{ /endangle exch def /startangle exch def}%
    \immediate\write\fwf@g{ /yrad exch def /xrad exch def}%
    \immediate\write\fwf@g{ /rotangle exch def /y exch def /x exch def}%
    \immediate\write\fwf@g{ /savematrix mtrx currentmatrix def}%
    \immediate\write\fwf@g{ x y translate rotangle rotate xrad yrad scale}%
    \immediate\write\fwf@g{ 0 0 1 startangle endangle arc}%
    \immediate\write\fwf@g{ savematrix setmatrix end\acc@lfer def}%
    \fi\PShe@der{EndProlog}}
\ctr@ld@f\def\Pssetc@rve#1=#2|{\keln@mun#1|%
    \def\n@mref{r}\ifx\l@debut\n@mref\pssetroundness{#2}\else% roundness
    \immediate\write16{*** Unknown attribute: \BS@ psset curve(..., #1=...)}%
    \fi}
\ctr@ln@m\curv@roundness
\ctr@ld@f\def\pssetroundness#1{\edef\curv@roundness{#1}}
\ctr@ld@f\def\defaultroundness{0.2} % Valeur par defaut
\ctr@ln@m\pscurve
\ctr@ld@f\def\pscurveDD[#1]{{\ifcurr@ntPS\ifps@cri\PSc@mment{pscurveDD Points=#1}%
    \s@uvc@ntr@l\et@tpscurveDD%
    \iffillm@de\Psc@rveDD\curv@roundness[#1]%
    \f@gfill%
    \else\Psc@rveDD\curv@roundness[#1]\f@gstroke\fi%
    \PSc@mment{End pscurveDD}\resetc@ntr@l\et@tpscurveDD\fi\fi}}
\ctr@ld@f\def\pscurveTD[#1]{{\ifcurr@ntPS\ifps@cri%
    \PSc@mment{pscurveTD Points=#1}\s@uvc@ntr@l\et@tpscurveTD\let\c@lprojSP=\relax%
    \iffillm@de\Psc@rveTD\curv@roundness[#1]%
    \f@gfill%
    \else\Psc@rveTD\curv@roundness[#1]\f@gstroke\fi%
    \PSc@mment{End pscurveTD}\resetc@ntr@l\et@tpscurveTD\fi\fi}}
\ctr@ld@f\def\Psc@rveDD#1[#2]{%
    \def\list@num{#2}\extrairelepremi@r\Ak@\de\list@num%
    \extrairelepremi@r\Ai@\de\list@num\extrairelepremi@r\Aj@\de\list@num%
    \f@gnewpath\PSwrit@cmdS{\Ai@}{\c@mmoveto}{\fwf@g}{\X@un}{\Y@un}%
    \setc@ntr@l{2}\figvectPDD -1[\Ak@,\Aj@]%
    \@ecfor\Ak@:=\list@num\do{\figpttraDD-2:=\Ai@/#1,-1/\BdingB@xfalse%
       \PSwrit@cmdS{-2}{}{\fwf@g}{\X@de}{\Y@de}%
       \figvectPDD -1[\Ai@,\Ak@]\figpttraDD-2:=\Aj@/-#1,-1/%
       \PSwrit@cmdS{-2}{}{\fwf@g}{\X@tr}{\Y@tr}\BdingB@xtrue%
       \PSwrit@cmdS{\Aj@}{\c@mcurveto}{\fwf@g}{\X@qu}{\Y@qu}%
       \B@zierBB@x{1}{\Y@un}(\X@un,\X@de,\X@tr,\X@qu)%
       \B@zierBB@x{2}{\X@un}(\Y@un,\Y@de,\Y@tr,\Y@qu)%
       \edef\X@un{\X@qu}\edef\Y@un{\Y@qu}\edef\Ai@{\Aj@}\edef\Aj@{\Ak@}}}
\ctr@ld@f\def\Psc@rveTD#1[#2]{\ifnum\curr@ntproj<\tw@\Psc@rvePPTD#1[#2]\else\Psc@rveCPTD#1[#2]\fi}
\ctr@ld@f\def\Psc@rvePPTD#1[#2]{\setc@ntr@l{2}%
    \def\list@num{#2}\extrairelepremi@r\Ak@\de\list@num\Figptpr@j-5:/\Ak@/%
    \extrairelepremi@r\Ai@\de\list@num\Figptpr@j-3:/\Ai@/%
    \extrairelepremi@r\Aj@\de\list@num\Figptpr@j-4:/\Aj@/%
    \f@gnewpath\PSwrit@cmdS{-3}{\c@mmoveto}{\fwf@g}{\X@un}{\Y@un}%
    \figvectPDD -1[-5,-4]%
    \@ecfor\Ak@:=\list@num\do{\Figptpr@j-5:/\Ak@/\figpttraDD-2:=-3/#1,-1/%
       \BdingB@xfalse\PSwrit@cmdS{-2}{}{\fwf@g}{\X@de}{\Y@de}%
       \figvectPDD -1[-3,-5]\figpttraDD-2:=-4/-#1,-1/%
       \PSwrit@cmdS{-2}{}{\fwf@g}{\X@tr}{\Y@tr}\BdingB@xtrue%
       \PSwrit@cmdS{-4}{\c@mcurveto}{\fwf@g}{\X@qu}{\Y@qu}%
       \B@zierBB@x{1}{\Y@un}(\X@un,\X@de,\X@tr,\X@qu)%
       \B@zierBB@x{2}{\X@un}(\Y@un,\Y@de,\Y@tr,\Y@qu)%
       \edef\X@un{\X@qu}\edef\Y@un{\Y@qu}\figptcopyDD-3:/-4/\figptcopyDD-4:/-5/}}
\ctr@ld@f\def\Psc@rveCPTD#1[#2]{\setc@ntr@l{2}%
    \def\list@num{#2}\extrairelepremi@r\Ak@\de\list@num%
    \extrairelepremi@r\Ai@\de\list@num\extrairelepremi@r\Aj@\de\list@num%
    \Figptpr@j-7:/\Ai@/%
    \f@gnewpath\PSwrit@cmd{-7}{\c@mmoveto}{\fwf@g}%
    \figvectPTD -9[\Ak@,\Aj@]%
    \@ecfor\Ak@:=\list@num\do{\figpttraTD-10:=\Ai@/#1,-9/%
       \figvectPTD -9[\Ai@,\Ak@]\figpttraTD-11:=\Aj@/-#1,-9/%
       \subB@zierTD\NBz@rcs[\Ai@,-10,-11,\Aj@]\edef\Ai@{\Aj@}\edef\Aj@{\Ak@}}}
\ctr@ld@f\def\psendfig{\ifcurr@ntPS\ifps@cri\immediate\closeout\fwf@g%
    \immediate\openout\fwf@g=\PSfilen@me\relax%
    \ifPDFm@ke\PSBdingB@x\else%
    \immediate\write\fwf@g{\p@urcent\string!PS-Adobe-2.0 EPSF-2.0}%
    \PShe@der{Creator\string: TeX (fig4tex.tex)}%
    \PShe@der{Title\string: \PSfilen@me}%
    \PShe@der{CreationDate\string: \the\day/\the\month/\the\year}%
    \PSBdingB@x%
    \PShe@der{EndComments}\PSdict@\fi%
    \immediate\write\fwf@g{\c@mgsave}%
    \openin\frf@g=\auxfilen@me\c@pypsfile\fwf@g\frf@g\closein\frf@g%
    \immediate\write\fwf@g{\c@mgrestore}%
    \PSc@mment{End of file.}\immediate\closeout\fwf@g%
    \immediate\openout\fwf@g=\auxfilen@me\immediate\closeout\fwf@g%
    \immediate\write16{File \PSfilen@me\space created.}\fi\fi\curr@ntPSfalse\ps@critrue}
\ctr@ld@f\def\PShe@der#1{\immediate\write\fwf@g{\p@urcent\p@urcent#1}}
\ctr@ld@f\def\PSBdingB@x{{\v@lX=\ptT@ptps\c@@rdXmin\v@lY=\ptT@ptps\c@@rdYmin%
     \v@lXa=\ptT@ptps\c@@rdXmax\v@lYa=\ptT@ptps\c@@rdYmax%
     \PShe@der{BoundingBox\string: \repdecn@mb{\v@lX}\space\repdecn@mb{\v@lY}%
     \space\repdecn@mb{\v@lXa}\space\repdecn@mb{\v@lYa}}}}
\ctr@ld@f\def\psfcconnect[#1]{{\ifcurr@ntPS\ifps@cri\PSc@mment{psfcconnect Points=#1}%
    \pssetfillmode{no}\s@uvc@ntr@l\et@tpsfcconnect\resetc@ntr@l{2}%
    \fcc@nnect@[#1]\resetc@ntr@l\et@tpsfcconnect\PSc@mment{End psfcconnect}\fi\fi}}
\ctr@ld@f\def\fcc@nnect@[#1]{\let\N@rm=\n@rmeucDD\def\list@num{#1}%
    \extrairelepremi@r\Ai@\de\list@num\edef\pr@m{\Ai@}\v@leur=\z@\p@rtent=\@ne\c@llgtot%
    \ifcase\fclin@typ@\edef\list@num{[\pr@m,#1,\Ai@}\expandafter\pscurve\list@num]%
    \else\ifdim\fclin@r@d\p@>\z@\Pslin@conge[#1]\else\psline[#1]\fi\fi%
    \v@leur=\@rrowp@s\v@leur\edef\list@num{#1,\Ai@,0}%
    \extrairelepremi@r\Ai@\de\list@num\mili@u=\epsil@n\c@llgpart%
    \advance\mili@u-\epsil@n\advance\mili@u-\delt@\advance\v@leur-\mili@u%
    \ifcase\fclin@typ@\invers@\mili@u\delt@%
    \ifnum\@rrowr@fpt>\z@\advance\delt@-\v@leur\v@leur=\delt@\fi%
    \v@leur=\repdecn@mb\v@leur\mili@u\edef\v@lt{\repdecn@mb\v@leur}%
    \extrairelepremi@r\Ak@\de\list@num%
    \figvectPDD-1[\pr@m,\Aj@]\figpttraDD-6:=\Ai@/\curv@roundness,-1/%
    \figvectPDD-1[\Ak@,\Ai@]\figpttraDD-7:=\Aj@/\curv@roundness,-1/%
    \delt@=\@rrowheadlength\p@\delt@=\C@AHANG\delt@\edef\R@dius{\repdecn@mb{\delt@}}%
    \ifcase\@rrowr@fpt%
    \FigptintercircB@zDD-8::\v@lt,\R@dius[\Ai@,-6,-7,\Aj@]\psarrowheadDD[-5,-8]\else%
    \FigptintercircB@zDD-8::\v@lt,\R@dius[\Aj@,-7,-6,\Ai@]\psarrowheadDD[-8,-5]\fi%
    \else\advance\delt@-\v@leur%
    \p@rtentiere{\p@rtent}{\delt@}\edef\C@efun{\the\p@rtent}%
    \p@rtentiere{\p@rtent}{\v@leur}\edef\C@efde{\the\p@rtent}%
    \figptbaryDD-5:[\Ai@,\Aj@;\C@efun,\C@efde]\ifcase\@rrowr@fpt%
    \delt@=\@rrowheadlength\unit@\delt@=\C@AHANG\delt@\edef\t@ille{\repdecn@mb{\delt@}}%
    \figvectPDD-2[\Ai@,\Aj@]\vecunit@{-2}{-2}\figpttraDD-5:=-5/\t@ille,-2/\fi%
    \psarrowheadDD[\Ai@,-5]\fi}
\ctr@ld@f\def\c@llgtot{\@ecfor\Aj@:=\list@num\do{\figvectP-1[\Ai@,\Aj@]\N@rm\delt@{-1}%
    \advance\v@leur\delt@\advance\p@rtent\@ne\edef\Ai@{\Aj@}}}
\ctr@ld@f\def\c@llgpart{\extrairelepremi@r\Aj@\de\list@num\figvectP-1[\Ai@,\Aj@]\N@rm\delt@{-1}%
    \advance\mili@u\delt@\ifdim\mili@u<\v@leur\edef\pr@m{\Ai@}\edef\Ai@{\Aj@}\c@llgpart\fi}
\ctr@ld@f\def\Pslin@conge[#1]{\ifnum\p@rtent>\tw@{\def\list@num{#1}%
    \extrairelepremi@r\Ai@\de\list@num\extrairelepremi@r\Aj@\de\list@num%
    \figptcopy-6:/\Ai@/\figvectP-3[\Ai@,\Aj@]\vecunit@{-3}{-3}\v@lmax=\result@t%
    \@ecfor\Ak@:=\list@num\do{\figvectP-4[\Aj@,\Ak@]\vecunit@{-4}{-4}%
    \minim@m\v@lmin\v@lmax\result@t\v@lmax=\result@t%
    \det@rm\delt@[-3,-4]\maxim@m\mili@u{\delt@}{-\delt@}\ifdim\mili@u>\Cepsil@n%
    \ifdim\delt@>\z@\figgetangleDD\Angl@[\Aj@,\Ak@,\Ai@]\else%
    \figgetangleDD\Angl@[\Aj@,\Ai@,\Ak@]\fi%
    \v@leur=\PI@deg\advance\v@leur-\Angl@\p@\divide\v@leur\tw@%
    \edef\Angl@{\repdecn@mb\v@leur}\c@ssin{\C@}{\S@}{\Angl@}\v@leur=\fclin@r@d\unit@%
    \v@leur=\S@\v@leur\mili@u=\C@\p@\invers@\mili@u\mili@u%
    \v@leur=\repdecn@mb{\mili@u}\v@leur%
    \minim@m\v@leur\v@leur\v@lmin\edef\t@ille{\repdecn@mb{\v@leur}}%
    \figpttra-5:=\Aj@/-\t@ille,-3/\psline[-6,-5]\figpttra-6:=\Aj@/\t@ille,-4/%
    \figvectNVDD-3[-3]\figvectNVDD-8[-4]\inters@cDD-7:[-5,-3;-6,-8]%
    \ifdim\delt@>\z@\psarccircP-7;\fclin@r@d[-5,-6]\else\psarccircP-7;\fclin@r@d[-6,-5]\fi%
    \else\psline[-6,\Aj@]\figptcopy-6:/\Aj@/\fi% Points alignes
    \edef\Ai@{\Aj@}\edef\Aj@{\Ak@}\figptcopy-3:/-4/}\psline[-6,\Aj@]}\else\psline[#1]\fi}
\ctr@ld@f\def\psfcnode[#1]#2{{\ifcurr@ntPS\ifps@cri\PSc@mment{psfcnode Points=#1}%
    \s@uvc@ntr@l\et@tpsfcnode\resetc@ntr@l{2}%
    \def\t@xt@{#2}\ifx\t@xt@\empty\def\g@tt@xt{\setbox\Gb@x=\hbox{\Figg@tT{\p@int}}}%
    \else\def\g@tt@xt{\setbox\Gb@x=\hbox{#2}}\fi%
    \v@lmin=\h@rdfcXp@dd\advance\v@lmin\Xp@dd\unit@\multiply\v@lmin\tw@%
    \v@lmax=\h@rdfcYp@dd\advance\v@lmax\Yp@dd\unit@\multiply\v@lmax\tw@%
    \Figv@ctCreg-8(\unit@,-\unit@)\def\list@num{#1}%
    \delt@=\curr@ntwidth bp\divide\delt@\tw@%
    \fcn@de\PSc@mment{End psfcnode}\resetc@ntr@l\et@tpsfcnode\fi\fi}}
\ctr@ld@f\def\d@butn@de{\g@tt@xt\v@lX=\wd\Gb@x%
    \v@lY=\ht\Gb@x\advance\v@lY\dp\Gb@x\advance\v@lX\v@lmin\advance\v@lY\v@lmax}
\ctr@ld@f\def\fcn@deE{%
    \@ecfor\p@int:=\list@num\do{\d@butn@de\v@lX=\unssqrttw@\v@lX\v@lY=\unssqrttw@\v@lY%
    \ifdim\thickn@ss\p@>\z@% Shadow
    \v@lXa=\v@lX\advance\v@lXa\delt@\v@lXa=\ptT@unit@\v@lXa\edef\XR@d{\repdecn@mb\v@lXa}%
    \v@lYa=\v@lY\advance\v@lYa\delt@\v@lYa=\ptT@unit@\v@lYa\edef\YR@d{\repdecn@mb\v@lYa}%
    \arct@n\v@leur(\v@lXa,\v@lYa)\v@leur=\rdT@deg\v@leur\edef\@nglde{\repdecn@mb\v@leur}%
    {\c@lptellDD-2::\p@int;\XR@d,\YR@d(\@nglde)}% \v@lmin & \v@lmax modified in \c@lptellDD
    \advance\v@leur-\PI@deg\edef\@nglun{\repdecn@mb\v@leur}%
    {\c@lptellDD-3::\p@int;\XR@d,\YR@d(\@nglun)}%
    \figptstra-6=-3,-2,\p@int/\thickn@ss,-8/\pssetfillmode{yes}\us@secondC@lor%
    \psline[-2,-3,-6,-5]\psarcell-4;\XR@d,\YR@d(\@nglun,\@nglde,0)\fi% End shadow
    \v@lX=\ptT@unit@\v@lX\v@lY=\ptT@unit@\v@lY%
    \edef\XR@d{\repdecn@mb\v@lX}\edef\YR@d{\repdecn@mb\v@lY}%
    \pssetfillmode{yes}\us@thirdC@lor\psarcell\p@int;\XR@d,\YR@d(0,360,0)%
    \pssetfillmode{no}\us@primarC@lor\psarcell\p@int;\XR@d,\YR@d(0,360,0)}}
\ctr@ld@f\def\fcn@deL{\delt@=\ptT@unit@\delt@\edef\t@ille{\repdecn@mb\delt@}%
    \@ecfor\p@int:=\list@num\do{\Figg@tXYa{\p@int}\d@butn@de%
    \ifdim\v@lX>\v@lY\itis@Ktrue\else\itis@Kfalse\fi%
    \advance\v@lXa-\v@lX\Figp@intreg-1:(\v@lXa,\v@lYa)%
    \advance\v@lXa\v@lX\advance\v@lYa-\v@lY\Figp@intreg-2:(\v@lXa,\v@lYa)%
    \advance\v@lXa\v@lX\advance\v@lYa\v@lY\Figp@intreg-3:(\v@lXa,\v@lYa)%
    \advance\v@lXa-\v@lX\advance\v@lYa\v@lY\Figp@intreg-4:(\v@lXa,\v@lYa)%
    \ifdim\thickn@ss\p@>\z@\Figg@tXYa{\p@int}\pssetfillmode{yes}\us@secondC@lor% Shadow
    \c@lpt@xt{-1}{-4}\c@lpt@xt@\v@lXa\v@lYa\v@lX\v@lY\c@rre\delt@%
    \Figp@intregDD-9:(\v@lZ,\v@lYa)\Figp@intregDD-11:(\v@lZa,\v@lYa)%
    \c@lpt@xt{-4}{-3}\c@lpt@xt@\v@lYa\v@lXa\v@lY\v@lX\delt@\c@rre%
    \Figp@intregDD-12:(\v@lXa,\v@lZ)\Figp@intregDD-10:(\v@lXa,\v@lZa)%
    \ifitis@K\figptstra-7=-9,-10,-11/\thickn@ss,-8/\psline[-9,-11,-5,-6,-7]\else%
    \figptstra-7=-10,-11,-12/\thickn@ss,-8/\psline[-10,-12,-5,-6,-7]\fi\fi% End shadow
    \pssetfillmode{yes}\us@thirdC@lor\psline[-1,-2,-3,-4]%
    \pssetfillmode{no}\us@primarC@lor\psline[-1,-2,-3,-4,-1]}}
\ctr@ld@f\def\c@lpt@xt#1#2{\figvectN-7[#1,#2]\vecunit@{-7}{-7}\figpttra-5:=#1/\t@ille,-7/%
    \figvectP-7[#1,#2]\Figg@tXY{-7}\c@rre=\v@lX\delt@=\v@lY\Figg@tXY{-5}}
\ctr@ld@f\def\c@lpt@xt@#1#2#3#4#5#6{\v@lZ=#6\invers@{\v@lZ}{\v@lZ}\v@leur=\repdecn@mb{#5}\v@lZ%
    \v@lZ=#2\advance\v@lZ-#4\mili@u=\repdecn@mb{\v@leur}\v@lZ%
    \v@lZ=#3\advance\v@lZ\mili@u\v@lZa=-\v@lZ\advance\v@lZa\tw@#1}
\ctr@ld@f\def\fcn@deR{\@ecfor\p@int:=\list@num\do{\Figg@tXYa{\p@int}\d@butn@de%
    \advance\v@lXa-0.5\v@lX\advance\v@lYa-0.5\v@lY\Figp@intreg-1:(\v@lXa,\v@lYa)%
    \advance\v@lXa\v@lX\Figp@intreg-2:(\v@lXa,\v@lYa)%
    \advance\v@lYa\v@lY\Figp@intreg-3:(\v@lXa,\v@lYa)%
    \advance\v@lXa-\v@lX\Figp@intreg-4:(\v@lXa,\v@lYa)%
    \ifdim\thickn@ss\p@>\z@\pssetfillmode{yes}\us@secondC@lor% Shadow
    \Figv@ctCreg-5(-\delt@,-\delt@)\figpttra-9:=-1/1,-5/%
    \Figv@ctCreg-5(\delt@,-\delt@)\figpttra-10:=-2/1,-5/%
    \Figv@ctCreg-5(\delt@,\delt@)\figpttra-11:=-3/1,-5/%
    \figptstra-7=-9,-10,-11/\thickn@ss,-8/\psline[-9,-11,-5,-6,-7]\fi% End shadow
    \pssetfillmode{yes}\us@thirdC@lor\psline[-1,-2,-3,-4]%
    \pssetfillmode{no}\us@primarC@lor\psline[-1,-2,-3,-4,-1]}}
\ctr@ln@m\@rrowp@s
\ctr@ln@m\Xp@dd     \ctr@ln@m\Yp@dd
\ctr@ln@m\fclin@r@d \ctr@ln@m\thickn@ss
\ctr@ld@f\def\Pssetfl@wchart#1=#2|{\keln@mtr#1|%
    \def\n@mref{arr}\ifx\l@debut\n@mref\expandafter\keln@mtr\l@suite|%
     \def\n@mref{owp}\ifx\l@debut\n@mref\edef\@rrowp@s{#2}\else% arrowposition
     \def\n@mref{owr}\ifx\l@debut\n@mref\setfcr@fpt#2|\else% arrowrefpt
     \immediate\write16{*** Unknown attribute: \BS@ psset flowchart(..., #1=...)}%
     \fi\fi\else%
    \def\n@mref{lin}\ifx\l@debut\n@mref\setfccurv@#2|\else% line
    \def\n@mref{pad}\ifx\l@debut\n@mref\edef\Xp@dd{#2}\edef\Yp@dd{#2}\else% padding
    \def\n@mref{rad}\ifx\l@debut\n@mref\edef\fclin@r@d{#2}\else% connection radius
    \def\n@mref{sha}\ifx\l@debut\n@mref\setfcshap@#2|\else% shape
    \def\n@mref{thi}\ifx\l@debut\n@mref\edef\thickn@ss{#2}\else% thickness
    \def\n@mref{xpa}\ifx\l@debut\n@mref\edef\Xp@dd{#2}\else% xpadding
    \def\n@mref{ypa}\ifx\l@debut\n@mref\edef\Yp@dd{#2}\else% ypadding
    \immediate\write16{*** Unknown attribute: \BS@ psset flowchart(..., #1=...)}%
    \fi\fi\fi\fi\fi\fi\fi\fi}
\ctr@ln@m\@rrowr@fpt \ctr@ln@m\fclin@typ@
\ctr@ld@f\def\setfcr@fpt#1#2|{\if#1e\def\@rrowr@fpt{1}\else\def\@rrowr@fpt{0}\fi}
\ctr@ld@f\def\setfccurv@#1#2|{\if#1c\def\fclin@typ@{0}\else\def\fclin@typ@{1}\fi}
\ctr@ln@m\h@rdfcXp@dd \ctr@ln@m\h@rdfcYp@dd
\ctr@ln@m\fcn@de \ctr@ln@m\fcsh@pe
\ctr@ld@f\def\setfcshap@#1#2|{%
    \if#1e\let\fcn@de=\fcn@deE\def\h@rdfcXp@dd{4pt}\def\h@rdfcYp@dd{4pt}%
     \edef\fcsh@pe{ellipse}\else%
    \if#1l\let\fcn@de=\fcn@deL\def\h@rdfcXp@dd{4pt}\def\h@rdfcYp@dd{4pt}%
     \edef\fcsh@pe{lozenge}\else%
          \let\fcn@de=\fcn@deR\def\h@rdfcXp@dd{6pt}\def\h@rdfcYp@dd{6pt}%
     \edef\fcsh@pe{rectangle}\fi\fi}
\ctr@ld@f\def\psline[#1]{{\ifcurr@ntPS\ifps@cri\PSc@mment{psline Points=#1}%
    \let\pslign@=\Pslign@P\Pslin@{#1}\PSc@mment{End psline}\fi\fi}}
\ctr@ld@f\def\pslineF#1{{\ifcurr@ntPS\ifps@cri\PSc@mment{pslineF Filename=#1}%
    \let\pslign@=\Pslign@F\Pslin@{#1}\PSc@mment{End pslineF}\fi\fi}}
\ctr@ld@f\def\pslineC(#1){{\ifcurr@ntPS\ifps@cri\PSc@mment{pslineC}%
    \let\pslign@=\Pslign@C\Pslin@{#1}\PSc@mment{End pslineC}\fi\fi}}
\ctr@ld@f\def\Pslin@#1{\iffillm@de\pslign@{#1}%
    \f@gfill%
    \else\pslign@{#1}\ifx\derp@int\premp@int%
    \f@gclosestroke%
    \else\f@gstroke\fi\fi}
\ctr@ld@f\def\Pslign@P#1{\def\list@num{#1}\extrairelepremi@r\p@int\de\list@num%
    \edef\premp@int{\p@int}\f@gnewpath%
    \PSwrit@cmd{\p@int}{\c@mmoveto}{\fwf@g}%
    \@ecfor\p@int:=\list@num\do{\PSwrit@cmd{\p@int}{\c@mlineto}{\fwf@g}%
    \edef\derp@int{\p@int}}}
\ctr@ld@f\def\Pslign@F#1{\s@uvc@ntr@l\et@tPslign@F\setc@ntr@l{2}\openin\frf@g=#1\relax%
    \ifeof\frf@g\message{*** File #1 not found !}\end\else%
    \read\frf@g to\tr@c\edef\premp@int{\tr@c}\expandafter\extr@ctCF\tr@c:%
    \f@gnewpath\PSwrit@cmd{-1}{\c@mmoveto}{\fwf@g}%
    \loop\read\frf@g to\tr@c\ifeof\frf@g\mored@tafalse\else\mored@tatrue\fi%
    \ifmored@ta\expandafter\extr@ctCF\tr@c:\PSwrit@cmd{-1}{\c@mlineto}{\fwf@g}%
    \edef\derp@int{\tr@c}\repeat\fi\closein\frf@g\resetc@ntr@l\et@tPslign@F}
\ctr@ln@m\extr@ctCF
\ctr@ld@f\def\extr@ctCFDD#1 #2:{\v@lX=#1\unit@\v@lY=#2\unit@\Figp@intregDD-1:(\v@lX,\v@lY)}
\ctr@ld@f\def\extr@ctCFTD#1 #2 #3:{\v@lX=#1\unit@\v@lY=#2\unit@\v@lZ=#3\unit@%
    \Figp@intregTD-1:(\v@lX,\v@lY,\v@lZ)}
\ctr@ld@f\def\Pslign@C#1{\s@uvc@ntr@l\et@tPslign@C\setc@ntr@l{2}%
    \def\list@num{#1}\extrairelepremi@r\p@int\de\list@num%
    \edef\premp@int{\p@int}\f@gnewpath%
    \expandafter\Pslign@C@\p@int:\PSwrit@cmd{-1}{\c@mmoveto}{\fwf@g}%
    \@ecfor\p@int:=\list@num\do{\expandafter\Pslign@C@\p@int:%
    \PSwrit@cmd{-1}{\c@mlineto}{\fwf@g}\edef\derp@int{\p@int}}%
    \resetc@ntr@l\et@tPslign@C}
\ctr@ld@f\def\Pslign@C@#1 #2:{{\def\t@xt@{#1}\ifx\t@xt@\empty\Pslign@C@#2:% Discard leading spaces
    \else\extr@ctCF#1 #2:\fi}}
\ctr@ln@m\c@ntrolmesh
\ctr@ld@f\def\Pssetm@sh#1=#2|{\keln@mun#1|%
    \def\n@mref{d}\ifx\l@debut\n@mref\pssetmeshdiag{#2}\else% diag
    \immediate\write16{*** Unknown attribute: \BS@ psset mesh(..., #1=...)}%
    \fi}
\ctr@ld@f\def\pssetmeshdiag#1{\edef\c@ntrolmesh{#1}}
\ctr@ld@f\def\defaultmeshdiag{0}    % Valeur par defaut
\ctr@ld@f\def\psmesh#1,#2[#3,#4,#5,#6]{{\ifcurr@ntPS\ifps@cri%
    \PSc@mment{psmesh N1=#1, N2=#2, Quadrangle=[#3,#4,#5,#6]}%
    \s@uvc@ntr@l\et@tpsmesh\Pss@tsecondSt\setc@ntr@l{2}%
    \ifnum#1>\@ne\Psmeshp@rt#1[#3,#4,#5,#6]\fi%
    \ifnum#2>\@ne\Psmeshp@rt#2[#4,#5,#6,#3]\fi%
    \ifnum\c@ntrolmesh>\z@\Psmeshdi@g#1,#2[#3,#4,#5,#6]\fi%
    \ifnum\c@ntrolmesh<\z@\Psmeshdi@g#2,#1[#4,#5,#6,#3]\fi\Psrest@reSt%
    \psline[#3,#4,#5,#6,#3]\PSc@mment{End psmesh}\resetc@ntr@l\et@tpsmesh\fi\fi}}
\ctr@ld@f\def\Psmeshp@rt#1[#2,#3,#4,#5]{{\l@mbd@un=\@ne\l@mbd@de=#1\loop%
    \ifnum\l@mbd@un<#1\advance\l@mbd@de\m@ne\figptbary-1:[#2,#3;\l@mbd@de,\l@mbd@un]%
    \figptbary-2:[#5,#4;\l@mbd@de,\l@mbd@un]\psline[-1,-2]\advance\l@mbd@un\@ne\repeat}}
\ctr@ld@f\def\Psmeshdi@g#1,#2[#3,#4,#5,#6]{\figptcopy-2:/#3/\figptcopy-3:/#6/%
    \l@mbd@un=\z@\l@mbd@de=#1\loop\ifnum\l@mbd@un<#1%
    \advance\l@mbd@un\@ne\advance\l@mbd@de\m@ne\figptcopy-1:/-2/\figptcopy-4:/-3/%
    \figptbary-2:[#3,#4;\l@mbd@de,\l@mbd@un]%
    \figptbary-3:[#6,#5;\l@mbd@de,\l@mbd@un]\Psmeshdi@gp@rt#2[-1,-2,-3,-4]\repeat}
\ctr@ld@f\def\Psmeshdi@gp@rt#1[#2,#3,#4,#5]{{\l@mbd@un=\z@\l@mbd@de=#1\loop%
    \ifnum\l@mbd@un<#1\figptbary-5:[#2,#5;\l@mbd@de,\l@mbd@un]%
    \advance\l@mbd@de\m@ne\advance\l@mbd@un\@ne%
    \figptbary-6:[#3,#4;\l@mbd@de,\l@mbd@un]\psline[-5,-6]\repeat}}
\ctr@ln@m\psnormal
\ctr@ld@f\def\psnormalDD#1,#2[#3,#4]{{\ifcurr@ntPS\ifps@cri%
    \PSc@mment{psnormal Length=#1, Lambda=#2 [Pt1,Pt2]=[#3,#4]}%
    \s@uvc@ntr@l\et@tpsnormal\resetc@ntr@l{2}\figptendnormal-6::#1,#2[#3,#4]%
    \figptcopyDD-5:/-1/\psarrow[-5,-6]%
    \PSc@mment{End psnormal}\resetc@ntr@l\et@tpsnormal\fi\fi}}
\ctr@ld@f\def\psreset#1{\trtlis@rg{#1}{\Psreset@}}
\ctr@ld@f\def\Psreset@#1|{\keln@mde#1|%
    \def\n@mref{ar}\ifx\l@debut\n@mref\psresetarrowhead\else% arrowhead
    \def\n@mref{cu}\ifx\l@debut\n@mref\psset curve(roundness=\defaultroundness)\else% curve
    \def\n@mref{fi}\ifx\l@debut\n@mref\psset (color=\defaultcolor,dash=\defaultdash,%
         fill=\defaultfill,join=\defaultjoin,width=\defaultwidth)\else% primary settings
    \def\n@mref{fl}\ifx\l@debut\n@mref\psset flowchart(arrowp=\defaultfcarrowposition,%
	arrowr=\defaultfcarrowrefpt,line=\defaultfcline,xpadd=\defaultfcxpadding,%
	ypadd=\defaultfcypadding,radius=\defaultfcradius,shape=\defaultfcshape,%
	thick=\defaultfcthickness)\else% flow chart
    \def\n@mref{me}\ifx\l@debut\n@mref\psset mesh(diag=\defaultmeshdiag)\else% mesh
    \def\n@mref{se}\ifx\l@debut\n@mref\psresetsecondsettings\else% secondary
    \def\n@mref{th}\ifx\l@debut\n@mref\psset third(color=\defaultthirdcolor)\else% ternary
    \immediate\write16{*** Unknown keyword #1 (\BS@ psreset).}%
    \fi\fi\fi\fi\fi\fi\fi}
\ctr@ld@f\def\psset#1(#2){\def\t@xt@{#1}\ifx\t@xt@\empty\trtlis@rg{#2}{\Pssetf@rst}% primary settings
    \else\keln@mde#1|%
    \def\n@mref{ar}\ifx\l@debut\n@mref\trtlis@rg{#2}{\Psset@rrowhe@d}\else% arrow-head
    \def\n@mref{cu}\ifx\l@debut\n@mref\trtlis@rg{#2}{\Pssetc@rve}\else% curve
    \def\n@mref{fi}\ifx\l@debut\n@mref\trtlis@rg{#2}{\Pssetf@rst}\else% primary settings
    \def\n@mref{fl}\ifx\l@debut\n@mref\trtlis@rg{#2}{\Pssetfl@wchart}\else% flow chart
    \def\n@mref{me}\ifx\l@debut\n@mref\trtlis@rg{#2}{\Pssetm@sh}\else% mesh
    \def\n@mref{se}\ifx\l@debut\n@mref\trtlis@rg{#2}{\Pssets@cond}\else% secondary settings
    \def\n@mref{th}\ifx\l@debut\n@mref\trtlis@rg{#2}{\Pssetth@rd}\else% ternary settings
    \immediate\write16{*** Unknown keyword: \BS@ psset #1(...)}%
    \fi\fi\fi\fi\fi\fi\fi\fi}
\ctr@ld@f\def\pssetdefault#1(#2){\ifcurr@ntPS\immediate\write16{*** \BS@ pssetdefault is ignored
    inside a \BS@ psbeginfig-\BS@ psendfig block.}%
    \immediate\write16{*** It must be called before \BS@ psbeginfig.}\else%
    \def\t@xt@{#1}\ifx\t@xt@\empty\trtlis@rg{#2}{\Pssd@f@rst}\else\keln@mde#1|%
    \def\n@mref{ar}\ifx\l@debut\n@mref\trtlis@rg{#2}{\Pssd@@rrowhe@d}\else% arrow-head
    \def\n@mref{cu}\ifx\l@debut\n@mref\trtlis@rg{#2}{\Pssd@c@rve}\else% curve
    \def\n@mref{fi}\ifx\l@debut\n@mref\trtlis@rg{#2}{\Pssd@f@rst}\else% primary settings
    \def\n@mref{fl}\ifx\l@debut\n@mref\trtlis@rg{#2}{\Pssd@fl@wchart}\else% flow chart
    \def\n@mref{me}\ifx\l@debut\n@mref\trtlis@rg{#2}{\Pssd@m@sh}\else% mesh
    \def\n@mref{se}\ifx\l@debut\n@mref\trtlis@rg{#2}{\Pssd@s@cond}\else% secondary settings
    \def\n@mref{th}\ifx\l@debut\n@mref\trtlis@rg{#2}{\Pssd@th@rd}\else% ternary settings
    \immediate\write16{*** Unknown keyword: \BS@ pssetdefault #1(...)}%
    \fi\fi\fi\fi\fi\fi\fi\fi\initpss@ttings\fi}
\ctr@ld@f\def\Pssd@f@rst#1=#2|{\keln@mun#1|%
    \def\n@mref{c}\ifx\l@debut\n@mref\edef\defaultcolor{#2}\else% color
    \def\n@mref{d}\ifx\l@debut\n@mref\edef\defaultdash{#2}\else% dash
    \def\n@mref{f}\ifx\l@debut\n@mref\edef\defaultfill{#2}\else% fillmode
    \def\n@mref{j}\ifx\l@debut\n@mref\edef\defaultjoin{#2}\else% line join
    \def\n@mref{u}\ifx\l@debut\n@mref\edef\defaultupdate{#2}\pssetupdate{#2}\else% update
    \def\n@mref{w}\ifx\l@debut\n@mref\edef\defaultwidth{#2}\else% line width
    \immediate\write16{*** Unknown attribute: \BS@ pssetdefault (..., #1=...)}%
    \fi\fi\fi\fi\fi\fi}
\ctr@ld@f\def\Pssd@@rrowhe@d#1=#2|{\keln@mun#1|%
    \def\n@mref{a}\ifx\l@debut\n@mref\edef\defaultarrowheadangle{#2}\else% angle
    \def\n@mref{f}\ifx\l@debut\n@mref\edef\defaultarrowheadangle{#2}\else% fillmode
    \def\n@mref{l}\ifx\l@debut\n@mref\y@tiunit{#2}\ifunitpr@sent%
     \edef\defaulth@rdahlength{#2}\else\edef\defaulth@rdahlength{#2pt}%
     \message{*** \BS@ pssetdefault (..., #1=#2, ...) : unit is missing, pt is assumed.}%
     \fi\else% length
    \def\n@mref{o}\ifx\l@debut\n@mref\edef\defaultarrowheadout{#2}\else% out
    \def\n@mref{r}\ifx\l@debut\n@mref\edef\defaultarrowheadratio{#2}\else% ratio
    \immediate\write16{*** Unknown attribute: \BS@ pssetdefault arrowhead(..., #1=...)}%
    \fi\fi\fi\fi\fi}
\ctr@ld@f\def\Pssd@c@rve#1=#2|{\keln@mun#1|%
    \def\n@mref{r}\ifx\l@debut\n@mref\edef\defaultroundness{#2}\else%
    \immediate\write16{*** Unknown attribute: \BS@ pssetdefault curve(..., #1=...)}%
    \fi}
\ctr@ld@f\def\Pssd@fl@wchart#1=#2|{\keln@mtr#1|%
    \def\n@mref{arr}\ifx\l@debut\n@mref\expandafter\keln@mtr\l@suite|%
     \def\n@mref{owp}\ifx\l@debut\n@mref\edef\defaultfcarrowposition{#2}\else% arrowposition
     \def\n@mref{owr}\ifx\l@debut\n@mref\edef\defaultfcarrowrefpt{#2}\else% arrowrefpt
     \immediate\write16{*** Unknown attribute: \BS@ pssetdefault flowchart(..., #1=...)}%
     \fi\fi\else%
    \def\n@mref{lin}\ifx\l@debut\n@mref\edef\defaultfcline{#2}\else% line
    \def\n@mref{pad}\ifx\l@debut\n@mref\edef\defaultfcxpadding{#2}%
                    \edef\defaultfcypadding{#2}\else% padding
    \def\n@mref{rad}\ifx\l@debut\n@mref\edef\defaultfcradius{#2}\else% connection radius
    \def\n@mref{sha}\ifx\l@debut\n@mref\edef\defaultfcshape{#2}\else% shape
    \def\n@mref{thi}\ifx\l@debut\n@mref\edef\defaultfcthickness{#2}\else% thickness
    \def\n@mref{xpa}\ifx\l@debut\n@mref\edef\defaultfcxpadding{#2}\else% xpadding
    \def\n@mref{ypa}\ifx\l@debut\n@mref\edef\defaultfcypadding{#2}\else% ypadding
    \immediate\write16{*** Unknown attribute: \BS@ pssetdefault flowchart(..., #1=...)}%
    \fi\fi\fi\fi\fi\fi\fi\fi}
\ctr@ld@f\def\defaultfcarrowposition{0.5}%\ctr@ld@f\let\defaultfcarrowpos=\defaultfcarrowposition
\ctr@ld@f\def\defaultfcarrowrefpt{start}
\ctr@ld@f\def\defaultfcline{polygon}
\ctr@ld@f\def\defaultfcradius{0}
\ctr@ld@f\def\defaultfcshape{rectangle}
\ctr@ld@f\def\defaultfcthickness{0}%\ctr@ld@f\let\defaultfcthick=\defaultfcthickness
\ctr@ld@f\def\defaultfcxpadding{0}%\ctr@ld@f\let\defaultfcxpad=\defaultfcxpadding
\ctr@ld@f\def\defaultfcypadding{0}%\ctr@ld@f\let\defaultfcypad=\defaultfcypadding
\ctr@ld@f\def\Pssd@m@sh#1=#2|{\keln@mun#1|%
    \def\n@mref{d}\ifx\l@debut\n@mref\edef\defaultmeshdiag{#2}\else%
    \immediate\write16{*** Unknown attribute: \BS@ pssetdefault mesh(..., #1=...)}%
    \fi}
\ctr@ld@f\def\Pssd@s@cond#1=#2|{\keln@mun#1|%
    \def\n@mref{c}\ifx\l@debut\n@mref\edef\defaultsecondcolor{#2}\else%
    \def\n@mref{d}\ifx\l@debut\n@mref\edef\defaultseconddash{#2}\else%
    \def\n@mref{w}\ifx\l@debut\n@mref\edef\defaultsecondwidth{#2}\else%
    \immediate\write16{*** Unknown attribute: \BS@ pssetdefault second(..., #1=...)}%
    \fi\fi\fi}
\ctr@ld@f\def\Pssd@th@rd#1=#2|{\keln@mun#1|%
    \def\n@mref{c}\ifx\l@debut\n@mref\edef\defaultthirdcolor{#2}\else%
    \immediate\write16{*** Unknown attribute: \BS@ pssetdefault third(..., #1=...)}%
    \fi}
\ctr@ln@w{newif}\iffillm@de
\ctr@ld@f\def\pssetfillmode#1{\expandafter\setfillm@de#1:}
\ctr@ld@f\def\setfillm@de#1#2:{\if#1n\fillm@defalse\else\fillm@detrue\fi}
\ctr@ld@f\def\defaultfill{no}     % Valeur par defaut
\ctr@ln@w{newif}\ifpsupdatem@de
\ctr@ld@f\def\pssetupdate#1{\ifcurr@ntPS\immediate\write16{*** \BS@ pssetupdate is ignored inside a
     \BS@ psbeginfig-\BS@ psendfig block.}%
    \immediate\write16{*** It must be called before \BS@ psbeginfig.}%
    \else\expandafter\setupd@te#1:\fi}
\ctr@ld@f\def\setupd@te#1#2:{\if#1n\psupdatem@defalse\else\psupdatem@detrue\fi}
\ctr@ld@f\def\defaultupdate{no}     % Valeur par defaut
\ctr@ln@m\curr@ntcolor \ctr@ln@m\curr@ntcolorc@md
\ctr@ld@f\def\Pssetc@lor#1{\ifps@cri\result@tent=\@ne\expandafter\c@lnbV@l#1 :%
    \def\curr@ntcolor{}\def\curr@ntcolorc@md{}%
    \ifcase\result@tent\or\pssetgray{#1}\or\or\pssetrgb{#1}\or\pssetcmyk{#1}\fi\fi}
\ctr@ln@m\curr@ntcolorc@mdStroke
\ctr@ld@f\def\pssetcmyk#1{\ifps@cri\def\curr@ntcolor{#1}\def\curr@ntcolorc@md{\c@msetcmykcolor}%
    \def\curr@ntcolorc@mdStroke{\c@msetcmykcolorStroke}%
    \ifcurr@ntPS\PSc@mment{pssetcmyk Color=#1}\us@primarC@lor\fi\fi}
\ctr@ld@f\def\pssetrgb#1{\ifps@cri\def\curr@ntcolor{#1}\def\curr@ntcolorc@md{\c@msetrgbcolor}%
    \def\curr@ntcolorc@mdStroke{\c@msetrgbcolorStroke}%
    \ifcurr@ntPS\PSc@mment{pssetrgb Color=#1}\us@primarC@lor\fi\fi}
\ctr@ld@f\def\pssetgray#1{\ifps@cri\def\curr@ntcolor{#1}\def\curr@ntcolorc@md{\c@msetgray}%
    \def\curr@ntcolorc@mdStroke{\c@msetgrayStroke}%
    \ifcurr@ntPS\PSc@mment{pssetgray Gray level=#1}\us@primarC@lor\fi\fi}
\ctr@ln@m\fillc@md
\ctr@ld@f\def\us@primarC@lor{\immediate\write\fwf@g{\d@fprimarC@lor}%
    \let\fillc@md=\prfillc@md}
\ctr@ld@f\def\prfillc@md{\d@fprimarC@lor\space\c@mfill}
\ctr@ld@f\def\defaultcolor{0}       % Valeur par defaut
\ctr@ld@f\def\c@lnbV@l#1 #2:{\def\t@xt@{#1}\relax\ifx\t@xt@\empty\c@lnbV@l#2:% Discard leading spaces
    \else\c@lnbV@l@#1 #2:\fi}
\ctr@ld@f\def\c@lnbV@l@#1 #2:{\def\t@xt@{#2}\ifx\t@xt@\empty%
    \def\t@xt@{#1}\ifx\t@xt@\empty\advance\result@tent\m@ne\fi% Discard trailing spaces
    \else\advance\result@tent\@ne\c@lnbV@l@#2:\fi}
\ctr@ld@f\def\Blackcmyk{0 0 0 1}
\ctr@ld@f\def\Whitecmyk{0 0 0 0}
\ctr@ld@f\def\Cyancmyk{1 0 0 0}
\ctr@ld@f\def\Magentacmyk{0 1 0 0}
\ctr@ld@f\def\Yellowcmyk{0 0 1 0}
\ctr@ld@f\def\Redcmyk{0 1 1 0}
\ctr@ld@f\def\Greencmyk{1 0 1 0}
\ctr@ld@f\def\Bluecmyk{1 1 0 0}
\ctr@ld@f\def\Graycmyk{0 0 0 0.50}
\ctr@ld@f\def\BrickRedcmyk{0 0.89 0.94 0.28} % PANTONE 1805
\ctr@ld@f\def\Browncmyk{0 0.81 1 0.60} % PANTONE 1615
\ctr@ld@f\def\ForestGreencmyk{0.91 0 0.88 0.12} % PANTONE 349
\ctr@ld@f\def\Goldenrodcmyk{ 0 0.10 0.84 0} % PANTONE 109
\ctr@ld@f\def\Marooncmyk{0 0.87 0.68 0.32} % PANTONE 201
\ctr@ld@f\def\Orangecmyk{0 0.61 0.87 0} % PANTONE ORANGE-021
\ctr@ld@f\def\Purplecmyk{0.45 0.86 0 0} % PANTONE PURPLE
\ctr@ld@f\def\RoyalBluecmyk{1. 0.50 0 0} % No PANTONE match
\ctr@ld@f\def\Violetcmyk{0.79 0.88 0 0} % PANTONE VIOLET
\ctr@ld@f\def\Blackrgb{0 0 0}
\ctr@ld@f\def\Whitergb{1 1 1}
\ctr@ld@f\def\Redrgb{1 0 0}
\ctr@ld@f\def\Greenrgb{0 1 0}
\ctr@ld@f\def\Bluergb{0 0 1}
\ctr@ld@f\def\Cyanrgb{0 1 1}
\ctr@ld@f\def\Magentargb{1 0 1}
\ctr@ld@f\def\Yellowrgb{1 1 0}
\ctr@ld@f\def\Grayrgb{0.5 0.5 0.5}
\ctr@ld@f\def\Chocolatergb{0.824 0.412 0.118}
\ctr@ld@f\def\DarkGoldenrodrgb{0.722 0.525 0.043}
\ctr@ld@f\def\DarkOrangergb{1 0.549 0}
\ctr@ld@f\def\Firebrickrgb{0.698 0.133 0.133}
\ctr@ld@f\def\ForestGreenrgb{0.133 0.545 0.133}
\ctr@ld@f\def\Goldrgb{1 0.843 0}
\ctr@ld@f\def\HotPinkrgb{1 0.412 0.706}
\ctr@ld@f\def\Maroonrgb{0.690 0.188 0.376}
\ctr@ld@f\def\Pinkrgb{1 0.753 0.796}
\ctr@ld@f\def\RoyalBluergb{0.255 0.412 0.882}
\ctr@ld@f\def\Pssetf@rst#1=#2|{\keln@mun#1|%
    \def\n@mref{c}\ifx\l@debut\n@mref\Pssetc@lor{#2}\else% color
    \def\n@mref{d}\ifx\l@debut\n@mref\pssetdash{#2}\else% dash
    \def\n@mref{f}\ifx\l@debut\n@mref\pssetfillmode{#2}\else% fillmode
    \def\n@mref{j}\ifx\l@debut\n@mref\pssetjoin{#2}\else% line join
    \def\n@mref{u}\ifx\l@debut\n@mref\pssetupdate{#2}\else% update
    \def\n@mref{w}\ifx\l@debut\n@mref\pssetwidth{#2}\else% line width
    \immediate\write16{*** Unknown attribute: \BS@ psset (..., #1=...)}%
    \fi\fi\fi\fi\fi\fi}
\ctr@ln@m\curr@ntdash
\ctr@ld@f\def\s@uvdash#1{\edef#1{\curr@ntdash}}
\ctr@ld@f\def\defaultdash{1}        % Valeur par defaut (numero sans espace)
\ctr@ld@f\def\pssetdash#1{\ifps@cri\edef\curr@ntdash{#1}\ifcurr@ntPS\expandafter\Pssetd@sh#1 :\fi\fi}
\ctr@ld@f\def\Pssetd@shI#1{\PSc@mment{pssetdash Index=#1}\ifcase#1%
    \or\immediate\write\fwf@g{[] 0 \c@msetdash}%         Index=1
    \or\immediate\write\fwf@g{[6 2] 0 \c@msetdash}%      Index=2
    \or\immediate\write\fwf@g{[4 2] 0 \c@msetdash}%      Index=3
    \or\immediate\write\fwf@g{[2 2] 0 \c@msetdash}%      Index=4
    \or\immediate\write\fwf@g{[1 2] 0 \c@msetdash}%      Index=5
    \or\immediate\write\fwf@g{[2 4] 0 \c@msetdash}%      Index=6
    \or\immediate\write\fwf@g{[3 5] 0 \c@msetdash}%      Index=7
    \or\immediate\write\fwf@g{[3 3] 0 \c@msetdash}%      Index=8
    \or\immediate\write\fwf@g{[3 5 1 5] 0 \c@msetdash}%  Index=9
    \or\immediate\write\fwf@g{[6 4 2 4] 0 \c@msetdash}%  Index=10
    \fi}
\ctr@ld@f\def\Pssetd@sh#1 #2:{{\def\t@xt@{#1}\ifx\t@xt@\empty\Pssetd@sh#2:% Discard leading spaces
    \else\def\t@xt@{#2}\ifx\t@xt@\empty\Pssetd@shI{#1}\else\s@mme=\@ne\def\debutp@t{#1}%
    \an@lysd@sh#2:\ifodd\s@mme\edef\debutp@t{\debutp@t\space\finp@t}\def\finp@t{0}\fi%
    \PSc@mment{pssetdash Pattern=#1 #2}%
    \immediate\write\fwf@g{[\debutp@t] \finp@t\space\c@msetdash}\fi\fi}}
\ctr@ld@f\def\an@lysd@sh#1 #2:{\def\t@xt@{#2}\ifx\t@xt@\empty\def\finp@t{#1}\else%
    \edef\debutp@t{\debutp@t\space#1}\advance\s@mme\@ne\an@lysd@sh#2:\fi}
\ctr@ln@m\curr@ntwidth
\ctr@ld@f\def\s@uvwidth#1{\edef#1{\curr@ntwidth}}
\ctr@ld@f\def\defaultwidth{0.4}     % Valeur par defaut
\ctr@ld@f\def\pssetwidth#1{\ifps@cri\edef\curr@ntwidth{#1}\ifcurr@ntPS%
    \PSc@mment{pssetwidth Width=#1}\immediate\write\fwf@g{#1 \c@msetlinewidth}\fi\fi}
\ctr@ln@m\curr@ntjoin
\ctr@ld@f\def\pssetjoin#1{\ifps@cri\edef\curr@ntjoin{#1}\ifcurr@ntPS\expandafter\Pssetj@in#1:\fi\fi}
\ctr@ld@f\def\Pssetj@in#1#2:{\PSc@mment{pssetjoin join=#1}%
    \if#1r\def\t@xt@{1}\else\if#1b\def\t@xt@{2}\else\def\t@xt@{0}\fi\fi%
    \immediate\write\fwf@g{\t@xt@\space\c@msetlinejoin}}
\ctr@ld@f\def\defaultjoin{miter}   % Valeur par defaut
\ctr@ld@f\def\Pssets@cond#1=#2|{\keln@mun#1|%
    \def\n@mref{c}\ifx\l@debut\n@mref\Pssets@condcolor{#2}\else%
    \def\n@mref{d}\ifx\l@debut\n@mref\pssetseconddash{#2}\else%
    \def\n@mref{w}\ifx\l@debut\n@mref\pssetsecondwidth{#2}\else%
    \immediate\write16{*** Unknown attribute: \BS@ psset second(..., #1=...)}%
    \fi\fi\fi}
\ctr@ln@m\curr@ntseconddash
\ctr@ld@f\def\pssetseconddash#1{\edef\curr@ntseconddash{#1}}
\ctr@ld@f\def\defaultseconddash{4}  % Valeur par defaut (numero sans espace)
\ctr@ln@m\curr@ntsecondwidth
\ctr@ld@f\def\pssetsecondwidth#1{\edef\curr@ntsecondwidth{#1}}
\ctr@ld@f\edef\defaultsecondwidth{\defaultwidth} % Valeur par defaut
\ctr@ld@f\def\psresetsecondsettings{%
    \pssetseconddash{\defaultseconddash}\pssetsecondwidth{\defaultsecondwidth}%
    \Pssets@condcolor{\defaultsecondcolor}}
\ctr@ln@m\sec@ndcolor \ctr@ln@m\sec@ndcolorc@md
\ctr@ld@f\def\Pssets@condcolor#1{\ifps@cri\result@tent=\@ne\expandafter\c@lnbV@l#1 :%
    \def\sec@ndcolor{}\def\sec@ndcolorc@md{}%
    \ifcase\result@tent\or\pssetsecondgray{#1}\or\or\pssetsecondrgb{#1}%
    \or\pssetsecondcmyk{#1}\fi\fi}
\ctr@ln@m\sec@ndcolorc@mdStroke
\ctr@ld@f\def\pssetsecondcmyk#1{\def\sec@ndcolor{#1}\def\sec@ndcolorc@md{\c@msetcmykcolor}%
    \def\sec@ndcolorc@mdStroke{\c@msetcmykcolorStroke}}
\ctr@ld@f\def\pssetsecondrgb#1{\def\sec@ndcolor{#1}\def\sec@ndcolorc@md{\c@msetrgbcolor}%
    \def\sec@ndcolorc@mdStroke{\c@msetrgbcolorStroke}}
\ctr@ld@f\def\pssetsecondgray#1{\def\sec@ndcolor{#1}\def\sec@ndcolorc@md{\c@msetgray}%
    \def\sec@ndcolorc@mdStroke{\c@msetgrayStroke}}
\ctr@ld@f\def\us@secondC@lor{\immediate\write\fwf@g{\d@fsecondC@lor}%
    \let\fillc@md=\sdfillc@md}
\ctr@ld@f\def\sdfillc@md{\d@fsecondC@lor\space\c@mfill}
\ctr@ld@f\edef\defaultsecondcolor{\defaultcolor} % Valeur par defaut
\ctr@ld@f\def\Pss@tsecondSt{%
    \s@uvdash{\typ@dash}\pssetdash{\curr@ntseconddash}%
    \s@uvwidth{\typ@width}\pssetwidth{\curr@ntsecondwidth}\us@secondC@lor}
\ctr@ld@f\def\Psrest@reSt{\pssetwidth{\typ@width}\pssetdash{\typ@dash}\us@primarC@lor}
\ctr@ld@f\def\Pssetth@rd#1=#2|{\keln@mun#1|%
    \def\n@mref{c}\ifx\l@debut\n@mref\Pssetth@rdcolor{#2}\else%
    \immediate\write16{*** Unknown attribute: \BS@ psset third(..., #1=...)}%
    \fi}
\ctr@ln@m\th@rdcolor \ctr@ln@m\th@rdcolorc@md
\ctr@ld@f\def\Pssetth@rdcolor#1{\ifps@cri\result@tent=\@ne\expandafter\c@lnbV@l#1 :%
    \def\th@rdcolor{}\def\th@rdcolorc@md{}%
    \ifcase\result@tent\or\Pssetth@rdgray{#1}\or\or\Pssetth@rdrgb{#1}%
    \or\Pssetth@rdcmyk{#1}\fi\fi}
\ctr@ln@m\th@rdcolorc@mdStroke
\ctr@ld@f\def\Pssetth@rdcmyk#1{\def\th@rdcolor{#1}\def\th@rdcolorc@md{\c@msetcmykcolor}%
    \def\th@rdcolorc@mdStroke{\c@msetcmykcolorStroke}}
\ctr@ld@f\def\Pssetth@rdrgb#1{\def\th@rdcolor{#1}\def\th@rdcolorc@md{\c@msetrgbcolor}%
    \def\th@rdcolorc@mdStroke{\c@msetrgbcolorStroke}}
\ctr@ld@f\def\Pssetth@rdgray#1{\def\th@rdcolor{#1}\def\th@rdcolorc@md{\c@msetgray}%
    \def\th@rdcolorc@mdStroke{\c@msetgrayStroke}}
\ctr@ld@f\def\us@thirdC@lor{\immediate\write\fwf@g{\d@fthirdC@lor}%
    \let\fillc@md=\thfillc@md}
\ctr@ld@f\def\thfillc@md{\d@fthirdC@lor\space\c@mfill}
\ctr@ld@f\def\defaultthirdcolor{1}  % Valeur par defaut
\ctr@ld@f\def\pstrimesh#1[#2,#3,#4]{{\ifcurr@ntPS\ifps@cri%
    \PSc@mment{pstrimesh Type=#1, Triangle=[#2,#3,#4]}%
    \s@uvc@ntr@l\et@tpstrimesh\ifnum#1>\@ne\Pss@tsecondSt\setc@ntr@l{2}%
    \Pstrimeshp@rt#1[#2,#3,#4]\Pstrimeshp@rt#1[#3,#4,#2]%
    \Pstrimeshp@rt#1[#4,#2,#3]\Psrest@reSt\fi\psline[#2,#3,#4,#2]%
    \PSc@mment{End pstrimesh}\resetc@ntr@l\et@tpstrimesh\fi\fi}}
\ctr@ld@f\def\Pstrimeshp@rt#1[#2,#3,#4]{{\l@mbd@un=\@ne\l@mbd@de=#1\loop\ifnum\l@mbd@de>\@ne%
    \advance\l@mbd@de\m@ne\figptbary-1:[#2,#3;\l@mbd@de,\l@mbd@un]%
    \figptbary-2:[#2,#4;\l@mbd@de,\l@mbd@un]\psline[-1,-2]%
    \advance\l@mbd@un\@ne\repeat}}
\initpr@lim\initpss@ttings\initPDF@rDVI% Initialisation preliminaire
\ctr@ln@w{newbox}\figBoxA
\ctr@ln@w{newbox}\figBoxB
\ctr@ln@w{newbox}\figBoxC
\catcode`\@=12

\pssetdefault(update=yes)
\newbox\figbox
\newbox\figbox
\def\courb#1#2#3#4#5{
    \figptssym 102 = #2 /#1, #3 /
    \figgetdist\rayon[#1,#2]
    \figvectN 111 [#2,#1]
    \figvectU 112 [111]
    \figptstra 104 = #2 /#4, 112/
    \figptssym 105 = 104 /#1, #3 /
    \psset(color=#5)
    \psset(width=1.2)
    \psarccircP #1 ; \rayon [#2,102]   
    \psline[#2,104]
    \psline[102,105]
}

\renewcommand{\Re}{\mathsf{Re}\,}
\renewcommand{\Im}{\mathsf{Im}\,}
\newcommand{\dG}{\mathsf{dG}}
\newcommand{\Po}{\mathsf{e}}
\newcommand{\LP}{\mathsf{LP}}

\newcommand{\sH}{\mathsf{H}}
\newcommand{\sV}{\mathsf{V}}
\newcommand{\sW}{\mathsf{W}}
\newcommand{\sL}{\mathsf{L}}
\newcommand{\sB}{\mathsf{B}}
\newcommand{\Dom}{\mathsf{Dom}\,}
\newcommand{\dist}{\mathrm{dist}}
\newcommand{\inff}{\mathop{\operatorname{\vphantom{p}inf}}}
\newcommand{\ess}{\mathsf{ess}}
\newcommand{\dis}{\mathsf{dis}}

\newcommand{\N}{\mathbb{N}}
\newcommand{\Q}{\mathbb{Q}}
\newcommand{\M}{\mathbb{M}}
\newcommand{\Z}{\mathbb{Z}}
\newcommand{\R}{\mathbb{R}}
\newcommand{\C}{\mathbb{C}}

\newcommand{\A}{\textbf{A}}
\newcommand{\B}{\textbf{B}}

\newcommand{\eps}{\varepsilon}
\newcommand{\dr}{\partial}
\newcommand{\phy}{\varphi}

\def\dx{\, {\rm d}}

\newcommand{\x}{\bm{\mathsf{x}}}
\newcommand{\q}{\bm{\mathsf{q}}}
\newcommand{\ex}{\mathsf{ex}}
\newcommand{\lin}{\mathsf{lin}}
\newcommand{\cc}{c}
\newcommand{\van}{\mathsf{vf}}
\newcommand{\gh}{\mathfrak{h}}
\newcommand{\gq}{\mathfrak{q}}
\newcommand{\gS}{\mathfrak{S}}
\newcommand{\one}{\mathds{1}}
\newcommand{\bA}{{\bf{A}}}
\newcommand{\bB}{{\bf{B}}}
\newcommand{\gb}{\mathfrak{b}}
\newcommand{\loc}{\mathsf{loc}}

\newcommand{\kb}{\mathbf{k}}
\newcommand{\y}{\mathsf{y}}
\newcommand{\e}{\mathsf{e}}

\def\a{\tfrac\alpha2}
\def\Ca{\mathcal{C}_{\alpha}} 
\newcommand{\cut}{\mathsf{cut}}
\def\spann{{\sf{span}}}
\def\model{\mathsf{model}}

\def\Mont{{\mathsf{Mo}}}
\def\spe{{\mathsf{sp}_{\mathsf{ess}}}}
\def\spd{{\mathsf{sp}_{\mathsf{dis}}}}
\def\sp{{\mathsf{sp}}}
\def\sgn{{\mathsf{sgn}}}
\def\supp{{\mathsf{supp}}}

\newcommand{\BO}{\mathsf{BO}}
\newcommand{\Id}{\mathsf{Id}}
\newcommand{\mode}{\mathsf{mod}}
\newcommand{\tens}{\mathsf{tens}}

\newcommand{\Hess}{\mathsf{Hess}}
\newcommand{\W}{\mathsf{W}}
\newcommand{\neww}{\mathsf{new}}
\newcommand{\wgt}{\mathsf{wgt}}
\newcommand{\Bhe}{{\mathcal B}(h^{-1/2}\varepsilon_{0})} 
\newcommand{\CBhe}{\complement{\mathcal B}(h^{-1/2}\varepsilon_{0})}
\newcommand{\ga}{\mathfrak{a}}
\newcommand{\gA}{\mathfrak{A}}

\newcommand{\FH}{\mathsf{c}}
\newcommand{\PR}{\mathsf{e}}
\newcommand{\bfn}{\bm{n}}
\newcommand{\an}{a}
\newcommand{\bn}{b}
\newcommand{\Vc}{\mathcal{V}}

\newcommand{\resc}{\mathsf{resc}}
\newcommand{\rad}{\mathsf{rad}}
\newcommand{\nor}{\mathsf{No}}
\newcommand{\harm}{\mathsf{harm}}

\newcommand{\cz}{{\scriptscriptstyle{\mathcal{Z}}}}
\newcommand{\n}{\bm{n}}
\newcommand{\Neu}{\mathsf{Neu}}
\newcommand{\Dir}{\mathsf{Dir}}
\newcommand{\Wedge}{\mathcal{W}}
\newcommand{\fla}{\mathsf{flat}}
\newcommand{\qm}{\mathsf{qm}}
\newcommand{\lens}{\mathsf{edge}}

\newcommand{\pscal}[2]{\langle #1,#2\rangle}
\newcommand{\norm}[1]{\left\|#1\right\|}
\newcommand{\deriv}[2]{\frac{\partial #1}{\partial #2}}
\newcommand{\abs}[1]{\left|#1\right|}
\newcommand{\ham}[1]{\mathcal{X}_{#1}}
\newcommand{\Op}{\mathsf{Op}}
\newcommand{\trsp}{\raisebox{.6ex}{${\scriptstyle t}$}}
\newcommand{\ad}{\mathsf{ad}}
\newcommand{\T}{\mathsf{T}}

\newcommand{\defo}{\mathsf{def}}
\newcommand{\para}{\parallel}
\newcommand{\app}{\mathsf{app}}
\newcommand{\appp}{\mathsf{app2}}
\newcommand{\eff}{\mathsf{eff}}
\newcommand{\curl}{\mathsf{curl}\,}
\newcommand{\F}{\mathsf{F}}
\newcommand{\shi}{\mathsf{sh}}

\newcommand{\toy}{\mathsf{toy}}
\newcommand{\Tri}{\mathsf{Tri}}
\newcommand{\Rec}{\mathsf{Rec}}
\newcommand{\Gui}{\mathsf{Gui}}
\newcommand{\Hst}{\mathsf{Hst}}
\newcommand{\Stlef}{\mathsf{Hlef}}
\newcommand{\Stri}{\mathsf{Hrig}}
\newcommand{\sing}{\mathsf{sing}}
\newcommand{\lef}{\mathsf{lef}}
\newcommand{\ri}{\mathsf{rig}}
\newcommand{\Ai}{\mathsf{Ai}^{\mathsf{rev}}}

\newcommand{\Mix}{\mathsf{Mix}}
\newcommand{\comp}{\mathsf{comp}}

\renewcommand\textgreek[1]{\foreignlanguage{greek}{#1}}

\begin{document}
\frontmatter
\title{Little Magnetic Book \\
{\small{Geometry and Bound States of the Magnetic Schr\"odinger Operator}}}

\author{Nicolas Raymond\footnote{IRMAR, Universit\'e de Rennes 1, Campus de Beaulieu, F-35042 Rennes cedex, France;
e-mail: \texttt{nicolas.raymond@univ-rennes1.fr}\begin{flushright}\today\end{flushright}}}

%\author{\textgreek{O{~>u}t'is}}

\maketitle
\newpage
\thispagestyle{empty}
\vspace*{6cm}
\begin{flushright}
\begin{minipage}{0.35\textwidth}
\textgreek{t`o a>ut`o noe\~in >estin te ka`i e{\~{>i}}nai\\
\begin{flushright}
Parmen'idhs
\end{flushright}
}
\end{minipage}
\end{flushright}
\newpage

\section*[]{Prol\'egom\`enes francophones}

Toute oeuvre qui se destine aux hommes ne devrait jamais \^etre \'ecrite que sous le nom de \textgreek{O{~>u}t'is}. C'est le nom par lequel \textgreek{\'{O}dusse'us} (Ulysse) s'est pr\'esent\'e au cyclope Polyph\`eme dont il venait de crever l'oeil. Rares sont les moments de l'Odyss\'ee o\`u \textgreek{\'{O}dusse'us} communique son v\'eritable nom ; il est le voyageur anonyme par excellence et ne sera reconnu qu'\`a la fin de son p\'eriple par ceux qui ont fid\`element pr\'eserv\'e sa m\'emoire. Mais que vient faire un tel commentaire au d\'ebut d'un livre de math\'ematiques~? Toutes les activit\'es de pens\'ee nous am\`enent, un jour ou l'autre, \`a nous demander si nous sommes bien les propri\'etaires de nos pens\'ees. Peut-on seulement les enfermer dans un livre et y associer notre nom~? N'en va-t-il pas pour elles comme il en va de l'amour~? Aussit\^ot poss\'ed\'ees, elles perdent leur attrait, aussit\^ot enferm\'ees elles perdent vie. Plus on touche \`a l'universel, moins la possession n'a de sens. Les Id\'ees n'appartiennent \`a personne et la v\'erit\'e est ingrate : elle n'a que faire de ceux qui la disent. \^O lecteur ! Fuis la renomm\'ee~! Car, aussit\^ot une reconnaissance obtenue, tu craindras de la perdre et, tel Don Quichotte, tu t'agiteras \`a nouveau pour te placer dans une vaine lumi\`ere. C'est un plaisir tellement plus d\'elicat de laisser aller et venir les Id\'ees, de constater que les plus belles d'entre elles trouvent leur profondeur dans l'\'eph\'em\`ere et que, \`a peine saisies, elles ne sont d\'ej\`a plus tout \`a fait ce qu'on croit. Le doute est essentiel \`a toute activit\'e de recherche. Il s'agit non seulement de v\'erifier nos affirmations, mais aussi de s'\'etonner devant ce qui se pr\'esente. Sans le doute, nous nous contenterions d'arguments d'autorit\'e et nous passerions devant les probl\`emes les plus profonds avec indiff\'erence. On \'ecrit rarement toutes les interrogations qui ont jalonn\'e la preuve d'un th\'eor\`eme. Une fois une preuve correcte \'etablie, pourquoi se souviendrait-on de nos errements ? Il est si reposant de passer d'une cause \`a une cons\'equence, de voir dans le pr\'esent l'expression m\'ecanique du pass\'e et de se lib\'erer ainsi du fardeau de la m\'emoire. Dans la vie morale, personne n'oserait pourtant penser ainsi et cette paresse d\'emonstrative passerait pour une terrible insouciance. Ce Petit Livre Magn\'etique pr\'esente une oeuvre continue et tiss\'ee par la m\'emoire de son auteur au cours de trois ann\'ees de m\'editation. L'id\'ee qui l'a constamment irrigu\'e est sans doute qu'une intuition a plus de valeur qu'un discours abstrait et parfaitement rigoureux. \`A l'instar de Bergson, on peut en effet penser que les abstractions \'enoncent du monde ce qu'il a de plus insignifiant. Avec lui, on peut aussi croire qu'un discours trop bien rod\'e et trop syst\'ematique peut \^etre le signe d'un manque d'id\'ees et d'intuitions. Une fois d\'eshabill\'e, ce type de discours, aussi paresseux que soporifique, exprime, dans sa perfection m\^eme, une recherche d'approbation. Et quoi de plus absurde que de rechercher des suffrages quand on s'int\'eresse authentiquement \`a la v\'erit\'e~? Ce livre fait ainsi le pari que la singularit\'e des exemples peut \^etre une source d'intuitions fertiles et que, depuis cette singularit\'e, on peut graduellement progresser vers quelques \'enonc\'es pr\'ecis dont la g\'en\'eralit\'e est \`a la mesure des exemples. Ici, d\'emarches scientifique et existentielle co\"{\i}ncident. Quelle diff\'erence en effet entre une psychologie enrichie par des \'epreuves et des th\'eor\`emes fa\c{c}onn\'es par des exemples ? Quelle diff\'erence entre une existence pass\'ee \`a l'imitation des conventions et des th\'eor\`emes sans \^ames ? Il est de bon ton, en notre temps, de faire montre de nos capacit\'es \`a changer sans cesse de th\`eme de r\'eflexion et \`a butiner ici et l\`a ce qui se d\'ecolle sans effort de la surface des choses ; mais pourquoi courir apr\`es les modes, si nous voulons durer~? Pourquoi vouloir changer, puisque la r\'ealit\'e elle-m\^eme est changement~?  \^O lecteur, prends le temps de juger des articulations et du d\'eveloppement des concepts pour t'en forger une id\'ee vivante~! Si ce livre fait na\^{\i}tre le doute et l'\'etonnement, c'est qu'il aura rempli son oeuvre.\newline
\begin{flushright}
\`A Aarhus, le 10 juin 2015
\end{flushright}
\newpage
\section*[]{Preface}
This little book was born in September 2012 during a summer school in Tunisia organized by H. Najar. I would like to thank him very much for this exciting invitation! This book also (strictly) contains my lecture notes for a master's degree. It is aimed to be a synthesis of recent advances in the spectral theory of the magnetic Schr\"odinger operator. It is also the opportunity for the author to rethink, simplify and sometimes correct the ideas of his papers and to present them in a more unified way. Therefore this book can be considered as a catalog of concrete examples of magnetic spectral asymptotics. Part~\ref{Part1} is devoted to an overview of some known results and to the statement of the main theorems proved in this book. Many point of views are used to describe the discrete spectrum, as well as the eigenfunctions, of the magnetic Laplacian in function of the (non necessarily) semiclassical parameter: naive powers series expansions, Feshbash-Grushin reductions, WKB constructions, coherent states decompositions, normal forms, etc. It turns out that, despite of the simplicity in the expression of the magnetic Laplacian, the influence of the geometry (smooth or not) and of the space variation of the magnetic field often give rise to completely different semiclassical structures that are governed by effective Hamiltonians reflecting the \textit{magnetic geometry}. In this spirit, two generic examples are presented in Part \ref{Part.MW} for the two dimensional case and three canonical examples involving a boundary in three dimensions are given in Part \ref{Part.BMW}. A feature underlined in this book is that many asymptotic problems related to the magnetic Laplacian lead to a dimensional reduction in the spirit of the famous Born-Oppenheimer approximation and therefore Part \ref{Part.Spec.Red} is devoted to a simplified theory to get access to the essential ideas. Actually, in the attempt to understand the normal forms of the magnetic Schr\"odinger operator, one may be tempted to make an analogy with spectral problems coming from the waveguides framework: this is the aim of Part \ref{Part.Wave}. Since this book is involved with many notions from Spectral Theory, Part \ref{Part.Models} provides a concise presentation of the main concepts and strategies used in this book as well as many examples. The students are especially invited to read this part first, even before the introduction. The reader is warned that this book gravitates towards ideas so that, at some point, part of the arguments might stay in the shadow to avoid too heavy technical details.

Last but not least, I would like to thank my collaborators, colleagues or students for all our magnetic discusssions: Z. Ammari, V. Bonnaillie-No\"el, B. Boutin, C.~Cheverry, M.~Dauge, N.~Dombrowski, V. Duch\^ene, F. Faure, S. Fournais, B. Helffer, F.~H\'erau, P.~Hislop, Y. Kordyukov, D. Krej{\v{c}}i{\v{r}}{\'{\i}}k, Y. Lafranche, L. Le Treust, F.~M\'ehats, J-P.~Miqueu, T.~Ourmi\`eres-Bonafos, M. P. Sundqvist, N. Popoff, M. Tu\v{s}ek, J. Van Schaftingen and S. V\~u Ng\d{o}c. This book is the story of our discussions.

\newpage

\tableofcontents

\mainmatter

\part{Main topics}\label{Part1}

\chapter{A magnetic story}\label{intro}

\begin{flushright}

\begin{minipage}{0.25\textwidth}
\textgreek{Gn\~wji seaut'on.} 
\vspace*{0.5cm}
\end{minipage}

\end{flushright}

\section{A magnetic realm }

\subsection{Once upon a time...}

Let us present two reasons which lead to the analysis of the magnetic Laplacian.

The first motivation arises in the mathematical theory of superconductivity. A model for this theory (see \cite{SJST}) is given by the Ginzburg-Landau functional:
$$\mathcal{G}(\psi,\A)=\int_{\Omega}|(-i\nabla+\kappa\sigma\A)\psi|^2-\kappa^2|\psi|^2+\frac{\kappa^2}{2}|\psi|^4\dx x+\kappa^2\int_{\Omega}|\sigma\nabla\times\A-\sigma\B|^2\dx x\,,$$
where $\Omega\subset\R^d$ is the place occupied by the superconductor, $\psi$ is the so-called order parameter ($|\psi|^2$ is the density of Cooper pairs), $\A$ is a magnetic potential and $\B$ the applied magnetic field. The parameter $\kappa$ is characteristic of the sample (the superconductors of type II  are such that $\kappa>> 1$) and $\sigma$ corresponds to the intensity of the applied magnetic field. Roughly speaking, the question is to determine the nature of the minimizers. Are they normal, that is $(\psi,\A)=(0, {\bf{F}})$ with $\nabla\times {\bf{F}}=\B$ (and $\nabla\cdot  {\bf{F}}=0$), or not?
We can mention the important result of Giorgi-Phillips \cite{GP99} which states that, if the applied magnetic field does not vanish, then, for $\sigma$ large enough, the normal state is the unique minimizer of $\mathcal{G}$ (with the divergence free condition). When analyzing the local minimality of $(0, {\bf{F}})$, we are led to compute the Hessian of $\mathcal{G}$ at $(0, {\bf{F}})$ and to analyze the positivity of:
$$(-i\nabla+\kappa\sigma\A)^2-\kappa^2\,.$$
For further details, we refer to the book by Fournais and Helffer \cite{FouHel10} and to the papers by Lu and Pan \cite{LuPan99a, LuPan00a}.
Therefore the theory of superconductivity leads to investigate the lowest eigenvalue $\lambda_{1}(h)$ of the Neumann realization of the \emph{magnetic Laplacian}, that is $(-ih\nabla+\A)^2$, where $h>0$ is small ($\kappa$ is assumed to be large).

The second motivation is to understand at which point there is an analogy between the electric Laplacian $-h^2\Delta+V(x)$ and the magnetic Laplacian $(-ih\nabla+\A)^2$. For instance, in the electric case (and in dimension one), when $V$ admits a unique and non-degenerate minimum at $0$ and satisfies $\displaystyle{\liminf_{|x|\to+\infty}V(x)>V(0)}$, we know that the $n$-th eigenvalue $\lambda_{n}(h)$ exists and satisfies:
\begin{equation}\label{eaa}
\lambda_{n}(h)=V(0)+(2n-1)\sqrt{\frac{V''(0)}{2}} h+\mathcal{O}(h^2)\,.
\end{equation}
Therefore a natural question arises:
\begin{center}
\enquote{Are there similar results to \eqref{eaa} in pure magnetic cases?}
\end{center}
In order to answer this question this book develops a theory of the \emph{Magnetic Harmonic Approximation}. Concerning the Schr\"odinger equation in presence of magnetic field the reader may consult \cite{AHS} (see also \cite{CSS78}) and the surveys \cite{MR94}, \cite{E07} and \cite{HelKo14}.

Jointly with \eqref{eaa} it is also well-known that we can perform WKB constructions for the electric Laplacian (see the book of Dimassi and Sj\"ostrand \cite[Chapter 3]{DiSj99}). Unfortunately such constructions do not seem to be possible \textit{in full generality} for the pure magnetic case (see the course of Helffer \cite[Section 6]{Hel05} and the paper by Martinez and Sordoni \cite{MS99}) and the naive localization estimates of Agmon are no more optimal (see \cite{HelSj87}, the paper by Erd\H{o}s \cite{Er96b} or the papers by Nakamura \cite{N96, N99}). For the magnetic situation, such accurate expansions of the eigenvalues (and eigenfunctions) are difficult to obtain. In fact, the more we know about the expansion of the eigenpairs, the more we can estimate the tunnel effect in the spirit of the electric tunnel effect of Helffer and Sj\"ostrand (see for instance \cite{HelSj84, HelSj85} and the papers by Simon \cite{S83, S84}) when there are symmetries. Estimating the magnetic tunnel effect is still a widely open question directly related to the approximation of the eigenfunctions (see \cite{HelSj87} and \cite{BDN13} for electric tunneling in presence of magnetic field and \cite{BDMV07} in the case with corners). Hopefully the main philosophy living throughout this book will prepare the future investigations on this fascinating subject. In particular we will provide the first examples of magnetic WKB constructions inspired by the recent work \cite{BHR14}. 
Anyway this book proposes a change of perspective in the study of the magnetic Laplacian. In fact, during the past decades, the philosophy behind the spectral analysis was essentially variational. Many papers dealt with the construction of \textit{quasimodes} used as test functions for the quadratic form associated with the magnetic Laplacian. In any case the attention was focused on the functions of the domain more than on the operator itself. In this book we systematically try to inverse the point of view: the main problem is no more to find \textit{appropriate quasimodes} but an \textit{appropriate (and sometimes microlocal) representation of the operator}. By doing this we will partially leave the min-max principle and the variational theory for the spectral theorem and the microlocal and hypoelliptic spirit.

\subsection{What is the magnetic Laplacian?}
Let $\Omega$ be a Lipschitzian domain in $\R^d$. Let us denote $\A=(A_{1},\cdots,A_{d})$ a smooth vector potential on $\overline{\Omega}$. We consider the $1$-form (see \cite[Chapter 7]{Arnold} for a brief introduction to differential forms):
\[\omega_{\A}=\sum_{k=1}^dA_{k} \dx x_{k}\,.\]
We introduce the exterior derivative of $\omega_{\A}$:
\[\sigma_{\B}=\dx \omega_{\A}=\sum_{1\leq k<\ell\leq d}B_{k\ell}\dx x_{k}\wedge\dx x_{\ell}\,,\]
with \[B_{k\ell}=\partial_{k}A_{\ell}-\partial_{\ell}A_{k}\,.\]
For further use, let us also introduce the magnetic matrix $M_{\B}=(B_{k\ell})_{1\leq k, \ell\leq d}$. In dimension two, the only coefficient is $B_{12}=\partial_{x_{1}}A_{2}-\partial_{x_{2}}A_{1}$. In dimension three, the magnetic field is defined as
\[\B=(B_{1}, B_{2}, B_{3})=(B_{23},-B_{13}, B_{12})=\nabla\times\A\,.\]
We will discuss in this book the spectral properties of some self-adjoint realizations of the magnetic operator:
\[\mathfrak{L}_{h,\A,\Omega}=\sum_{k=1}^d (-ih\dr_{k}+A_{k})^2\,,\]
where $h>0$ is a parameter (related to the Planck constant). We notice the fundamental property, called gauge invariance:
\[e^{-i\phi/h} (-ih\nabla+\A)e^{i\phi/h}=-ih\nabla+\A+\nabla\phi\]
so that
\begin{equation}\label{gauge-invariance}
e^{-i\phi/h} (-ih\nabla+\A)^2e^{i\phi/h}=(-ih\nabla+\A+\nabla\phi)^2\,,
\end{equation}
where $\phi\in \sH^1(\Omega,\R)$. 

Before describing important spectral results obtained in the last twenty years, let us discuss some basic properties of the magnetic Laplacian when $\Omega=\R^d$. 

First, we can observe that the presence of a magnetic field increases the energy of the system in the following sense.
\begin{theo}\label{dia}
Let $\A : \R^d\to\R^d$ be in $\sL^2_{\loc}(\R^d)$ and suppose that $f\in\sL^2_{\loc}(\R^d)$ is such that $(-i\nabla+\A)f\in\sL^2_{\loc}(\R^d)$. Then $|f|\in\sH^1_{\loc}(\R^d)$ and
$$|\nabla|f||\leq |(-i\nabla+\A)f|\,,\qquad\mbox{almost everywhere.}$$
\end{theo}
The inequality of Theorem \ref{dia} is called \textit{diamagnetic inequality} and a proof may be found for instance in \cite[Chapter 2]{FouHel10}.

The following proposition also gives an idea of the effect of the magnetic on the magnetic energy.
\begin{prop}\label{minoration-B}
Let $\A\in\mathcal{C}^\infty(\R^d,\R^d)$. Then, for all $\varphi\in\mathcal{C}^\infty_{0}(\R^d)$, we have, for all $k, \ell\in\{1,\cdots, d\}$,
\[\mathfrak{Q}_{\A}(\varphi):=\int_{\R^d}|(-i\nabla+\A)\varphi|^2\dx\x\geq \left|\int_{\R^d}B_{k\ell}|\varphi|^2\dx\x\right|\,.\]
\end{prop}

\begin{proof}
We have 
\[[D_{x_{k}}+A_{k},D_{x_{\ell}}+A_{\ell}]=-iB_{k\ell}\,,\]
and thus, for all $\varphi\in\mathcal{C}^\infty_{0}(\R^d)$,
\[\langle [D_{x_{k}}+A_{k},D_{x_{\ell}}+A_{\ell}]\varphi,\varphi \rangle_{\sL^2(\R^d)}=-i\int_{\R^d}B_{k\ell}|\varphi|^2\dx\x\,.\]
By integration by parts, it follows that
\[\left|\langle [D_{x_{k}}+A_{k},D_{x_{\ell}}+A_{\ell}]\varphi,\varphi \rangle_{\sL^2(\R^d)}\right|\leq \|(D_{x_{k}}+A_{k})\varphi\|_{\sL^2(\R^d)} \|(D_{x_{\ell}}+A_{\ell})\varphi\|_{\sL^2(\R^d)}\]
and thus
\[\left|\langle [D_{x_{k}}+A_{k},D_{x_{\ell}}+A_{\ell}]\varphi,\varphi \rangle_{\sL^2(\R^d)}\right|\leq  \|(D_{x_{k}}+A_{k})\varphi\|^2_{\sL^2(\R^d)}+\|(D_{x_{\ell}}+A_{\ell})\varphi\|^2_{\sL^2(\R^d)}\,.\]
The conclusion easily follows.
\end{proof}

It is classical that $\mathfrak{L}_{h,\A,\R^d}=(-ih\nabla+\A)^2$, acting on $\mathcal{C}^\infty_{0}(\R^d)$, is essentially self-adjoint (see \cite[Theorem 1.2.2]{FouHel10}). Let us describe its spectrum when $d=2, 3$ and when the magnetic field is constant. The reader may find some reminders about spectral theory in Chapter \ref{chapter-spectral-theory}.

\subsubsection{Where is the magnetic field?}
We started with a given $1$-form and then we defined the magnetic field as its exterior derivative. The reason for this comes from the expression of the magnetic Laplacian, involving only the vector potential. In fact, one could start from a closed $2$-form $\sigma$ and define a $1$-form $\omega$ such that $\dx \omega=\sigma$. Let us recall how we can do this with the help of classical concepts from differential geometry. We summarize this with the following lemma.
\begin{lem}[Poincar\'e's lemma]\label{lem.Poincare}
Let $p\geq 1$ and $\sigma$ be a closed $p$-form defined (and smooth) in a neighborhood of $0$ and define
\[\omega_{\x}(\cdot)=\int_{0}^1t^{p-1} \sigma_{t\x}(\x,\cdot)\dx t\,.\]
Then, we have $\dx \omega=\sigma$.
\end{lem} 
\begin{proof}
The reader may skip this proof and read instead the forthcoming examples. Nevertheless, we recall these classical details for further use (especially, see Chapter \ref{chapter-appendix} where we recall some basic concepts). Note that the proof may done thanks to a direct computation.

We introduce the family $\varphi_{t}(\x)=t\x$, for $t\in[0,1]$. For $t\in(0,1]$, this is a family of smooth diffeomorphisms. Introducing $\mathcal{X}_{t}(\x)=t^{-1}\x$, we have
\[\frac{d }{d t}\varphi_{t}=\mathcal{X}_{t}(\varphi_{t})\,.\]
We notice that
\[\sigma_{\x}=\varphi_{1}^*\sigma-\varphi_{0}^*\sigma=\int_{0}^1 \frac{d}{dt}\varphi^*_{t}\sigma \dx t\,,\]
where $*$ denotes the pull back of the form. Then, by definition of the Lie derivative, we find
\[\sigma_{\x}=\int_{0}^1 \varphi^*_{t}\mathcal{L}_{\mathcal{X}_{t}}\sigma \dx t\,,\]
We apply the general Cartan formula
\[\mathcal{L}_{X}\sigma=\dx (\iota_{X}\sigma)+\iota_{X}\dx\sigma\,,\]
where $\iota_{X}$ means that we replace the first entry of the form by $X$.
Since $\sigma$ is closed ($\dx \sigma=0$), we get
\[\sigma_{\x}=\int_{0}^1 \varphi^*_{t}\dx(\iota_{\mathcal{X}_{t}}\sigma) \dx t\,,\]
and we deduce (by commuting $\dx$ with the pull back and the integration):
\[\sigma_{\x}=\dx\int_{0}^1 \varphi^*_{t}\iota_{\mathcal{X}_{t}}\sigma \dx t\,.\]
Then, by homogeneity, we find
\[\int_{0}^1 \varphi^*_{t}\iota_{\mathcal{X}_{t}}\sigma \dx t=\int_{0}^1t^{p-1} \sigma_{t\x}(\x,\cdot)\dx t\,.\]
\end{proof}
When the magnetic $2$-form is constant, a possible vector potential is given by
\[\langle\bA(\x),\cdot\rangle_{\R^d}=\int_{0}^1 \sigma_{\bB}(t\x,\cdot) \dx t=\frac{1}{2}\sigma_{\bB}(\x,\cdot)\,.\]
This choice of vector potential is called \textit{Lorentz gauge}. Explicitly we have
\[\bA(\x)=\frac{1}{2}M_{\bB}\x\,,\]
where $M_{\bB}$ is the $d\times d$ anti-symmetric matrix $(B_{k\ell})$. 

\subsubsection{From the magnetic matrix to the magnetic field}\label{sec.MBtoB}
Note that, in dimension three, we have, with the usual vector product:
\[M_{\bB}\x=\bB\times\x=\bB\x\,.\]
Let us discuss here the effect of changes of coordinates on the magnetic form. If $\Phi$ is a local diffeomorphism, we let $\x=\Phi(\y)$ and
\[\Phi^*\omega_{\A}=\sum_{j=1}^d \mathcal{A}_{j}\dx y_{j}\,,\quad\mbox{ where } \mathcal{A}=(d\Phi)^{\mathsf{T}}\A(\Phi)\,.\]
Since the exterior derivative commutes with the pull-back, we get
\[\dx(\Phi^*\omega_{\A})=\Phi^{*}\sigma_{\B}\,.\]
In the new coordinates $\y$, the new magnetic matrix is given by
\[M_{\mathcal{B}}=(d\Phi)^{\mathsf{T}}M_{\B}d\Phi\,.\]
In the case of dimension three, we may write the relation between the field $\B$ and the field $\mathcal{B}$. We have
\[\langle M_{\mathcal{B}}\y,\mathsf{z}\rangle_{\R^3}=\langle\mathcal{B}\times \y,\mathsf{z}\rangle_{\R^3}=\langle\y\times\mathsf{z},\mathcal{B}\rangle_{\R^3}\,,\]
and also
\[\langle M_{\mathcal{B}}\y,\mathsf{z}\rangle_{\R^3}=\langle d\Phi(\y)\times d\Phi(\mathsf{z}),\B\rangle_{\R^3}\,.\]
It is a classical exercise to see that
\[\langle d\Phi(\y)\times d\Phi(\mathsf{z}),\B\rangle_{\R^3}=\det(d\Phi)\langle \y\times\mathsf{z},(d\Phi)^{-1}\B\rangle_{\R^3}\,.\]
Thus we get the formula
\[\nabla_{\mathsf{y}}\times\mathcal{A}=\mathcal{B}=\det(d\Phi)(d\Phi)^{-1}\B\,,\mbox{ or }\quad \mathcal{B}=\widetilde{d\Phi}\B\,,\]
where $\widetilde{d\Phi}$ is the adjugate matrix of $d\Phi$.
\subsubsection{Constant magnetic field in dimension two}\label{ssec.2}
In dimension two, thanks to the gauge invariance \eqref{gauge-invariance}, when $B=1$, we may assume that the vector potential is given by
\[\A(x_{1},x_{2})=(0,x_{1})\,,\]
so that
\[\mathfrak{L}_{h, \A, \R^2}=h^2D_{x_{1}}^2+(hD_{x_{2}}+x_{1})^2\,,\qquad \mbox{ with the notation }\quad D=-i\partial\,.\]
By using the partial Fourier transform $\mathcal{F}_{x_{2}\mapsto\xi_{2}}$ (normalized to be unitary), we get
\[\mathcal{F}_{x_{2}\mapsto\xi_{2}}\mathfrak{L}_{h, \A, \R^2}\mathcal{F}^{-1}_{x_{2}\mapsto\xi_{2}}=h^2D_{x_{1}}^2+(h\xi_{2}+x_{1})^2\,.\]
Then, we introduce the unitary transform
\[Tf(\tilde x_{1},\tilde x_{2})=f(\tilde x_{1}-h\tilde\xi_{2},\tilde\xi_{2})\,,\]
and we get the operator, acting on $\sL^2(\R^2_{\tilde x_{1},\tilde\xi_{2}})$,
\[T\mathcal{F}_{x_{2}\mapsto\xi_{2}}\mathfrak{L}_{h, \A, \R^2}\mathcal{F}^{-1}_{x_{2}\mapsto\xi_{2}}T^{-1}=h^2D_{\tilde x_{1}}^2+\tilde x^2_{1}\,.\]
We recognize a rescaled version of the harmonic oscillator (see for instance Chapter \ref{chapter-examples}, Section \ref{Harmonic}) and we deduce that the spectrum of $\mathfrak{L}_{h, \A, \R^2}$ is essential and given by the set of the \textit{Landau levels}
\[\{(2n-1)h,\quad n\in\N^*\}\,.\]
Let us underline that each element of the spectrum is an eigenvalue of \textit{infinite} multiplicity.

\subsubsection{Constant magnetic field in dimension three}\label{ssec.3}
In dimension three, we are easily reduced to the investigation of
\begin{equation}\label{formula.3D}
\mathfrak{L}_{h,\A,\R^3}=h^2D_{x_{1}}^2+(hD_{x_{2}}+x_{1})^2+h^2D_{x_{3}}^2\,,
\end{equation}
and, thanks to partial Fourier transforms with respect to $x_{2}$ and $x_{3}$ and then to a transvection with respect to $x_{1}$, we again get that the magnetic Laplacian is unitary equivalent to the operator, acting on $\sL^2(\R^3_{\tilde x_{1},\tilde\xi_{2},\xi_{3}})$,
\[h^2D_{\tilde x_{1}}^2+\tilde x^2_{1}+h^2\xi_{3}^2\,.\]
In this case, the spectrum of the magnetic Laplacian is essential and given by the interval
\[[h,+\infty)\,.\]
This can be seen by using appropriate Weyl's sequences.

\subsubsection{Higher dimensions}
Let us briefly discuss the case of higher dimension. We would like to generalize the simplified form given in \eqref{formula.3D}.

For $\mathsf{Q}\in O(d)$, we let $\x=\mathsf{Q}\y$ and, modulo a unitary transform, the magnetic Laplacian becomes
\[\left(-ih\nabla_{\y}+\frac{1}{2}\mathsf{Q}^\mathsf{T}\bB \mathsf{Q}\y\right)^2\,.\]
By the classical diagonalization result for anti-symmetric matrices, there exists an element $\mathsf{Q}\in O(d)$ such that $\mathsf{Q}^\mathsf{T}\bB \mathsf{Q}$ is bloc diagonal, with $2$ by $2$ blocs in the form $\begin{pmatrix}0&\beta_{j}\\ -\beta_{j}&0\end{pmatrix}$, with $j=1,\ldots,\left\lfloor\frac{d}{2}\right\rfloor$ and $\beta_{j}>0$. By applying the analysis of dimension two, we get, by separation of variables that the bottom of the spectrum is given by $h\mathsf{Tr}^+\B$ where
\[\mathsf{Tr}^+\B=\sum_{j=1}^{\left\lfloor\frac{d}{2}\right\rfloor}|\beta_{j}|\,.\]
When $d=3$, since the Hilbert-Schmidt norm is preserved by rotation, we have $\mathsf{Tr}^+\B=\|\bB\|$.

\subsection{Magnetic wells}\label{sec.magw}
When the magnetic field is variable (say in dimension two or three), it is possible to approximate the spectrum thanks to a local approximation of the magnetic field by the constant field. From the classical point of view, this means that, locally, the motion of the particle is well described (on a small time scale) by the \textit{cyclotron motion} (see the discussion in Chapter \ref{intro-van-birk}, Section \ref{sec.classical-dyn}). In particular, if the magnetic field is large enough at infinity and if its norm admits a positive minimum, we have the estimate
\begin{equation}\label{eq.helmo}
\lambda_{1}(h)=b_{0}h+o(h)\,,
\end{equation}
where $b_{0}>0$ is either the minimum of $|B|$ in dimension two, or the minimum of $\|\B\|$ in dimension three. This result was proved by Helffer and Morame in \cite[Theorem 1.1]{HelMo01}. One calls the point where the minimum is obtained a \enquote{magnetic well}.

As suggested a few lines ago, the semi\textit{classical} limit should have something to do with the \textit{classical} mechanics. At some point, one should be able to interpret the semiclassical approximations of the magnetic eigenvalues from a classical point of view. In many cases, the classical interpretation turns out to be difficult in the magnetic case (in presence of a boundary for instance). The main term in the asymptotic expansion of $\lambda_{1}(h)$ is related to the cyclotron motion or equivalently to the approximation by the constant magnetic field. But, in the classical world (see for instance \cite{benettin-sempio} or \cite{cheverry14} in a nonlinear context), it is known that the cyclotron motion is not enough to describe the fancy dynamics in variable magnetic fields that give rise to \textit{magnetic bottles}, \textit{magnetic bananas} or \textit{magnetic mirror points}. The moral of these rough classical considerations is that, to get the \textit{classical-quantum correspondence}, one should go further in the semiclassical expansion of $\lambda_{1}(h)$ and also consider the next eigenvalues. Roughly speaking, the magnetic motion, in dimension three, can be decomposed into three elementary motions: the cyclotron motion, the oscillation along the field lines and the oscillation within the space of field lines.
The concept of \textit{magnetic harmonic approximation} developed through out this book is an attempt to reveal, at the quantum level, these three motions in various geometric settings without a deep understanding of the classical dynamics (one could call this a \textit{semiquantum} approximation). To stimulate the reader, let us give two examples of semiclassical expansions tackling these issues.
In dimension two, if the magnetic field admits a unique minimum at $q_{0}$ that is non degenerate and that the magnetic field is large enough at infinity, we have
\[
\lambda_{n}(h) =b_{0}{h} +\left[\theta^{\mathsf{2D}}(q_{0})\left(n-\frac{1}{2}\right)+\zeta^{\mathsf{2D}}(q_{0})\right] {h}^2 + \mathcal{O}({h}^3)
\]
where
\begin{equation}\label{eq.magin2}
b_{0}=\min_{\R^2} B\,,\qquad\theta^{\mathsf{2D}}(q_{0})=\sqrt{\frac{\det\mathsf{Hess}_{q_{0}} B}{b_{0}^2}}\,,
\end{equation}
and where $\zeta^{\mathsf{2D}}(q_{0})$ is another explicit constant. Here the term $b_{0}h$ is related to the cyclotron motion and $\theta^{\mathsf{2D}}(q_{0})\left(n-\frac{1}{2}\right)h^2$ is related to the magnetic drift motion (the oscillation in the space of field lines). This expansion has been obtained by different means in \cite{HelKo11, HelKo13b, RVN13}. We will present one of them in this book.

In dimension three, by denoting $b=\|\B\|$ and assuming again the uniqueness and non-degeneracy of the minimum at $q_{0}$, we have the following striking asymptotic expansion
\[\lambda_{n}(h) = b_0h +\sigma^\mathsf{3D}(q_{0})h^{\frac{3}{2}}+
  \left[\theta^{\mathsf{3D}}(q_{0})\left(n-\tfrac{1}{2}\right)+\zeta^{\mathsf{3D}}(q_{0})\right]{h}^2+\mathcal{O}({h}^{\frac{5}{2}})
\]
where
\begin{equation}\label{eq.magin3}
b_{0}=\min_{\R^3} b\,,\quad  \sigma^\mathsf{3D}(q_{0})=\sqrt{\frac{\mathsf{Hess}_{q_0} b \,(\mathbf{B},  \mathbf{B})}{2b_0^2}},\quad \theta^{\mathsf{3D}}(q_{0})=\sqrt{\frac{\det {\mathsf{Hess}}_{q_0} b}{\mathsf{Hess}_{q_0} b \,(\mathbf{B},  \mathbf{B})}}\,,
\end{equation}
and where $\zeta^{\mathsf{3D}}(q_{0})$ is again an explicit constant. In this case, $b_{0}h$ is related with the cyclotron motion, the term $\sigma^\mathsf{3D}(q_{0})h^{\frac{3}{2}}$ with the oscillation along field lines and $\theta^{\mathsf{3D}}(q_{0})h^2$ within the oscillation in the space of field lines. This asymptotic expansion in dimension three has been obtained in \cite{HKRVN14}. We will not provide the proof of this one (that is largely beyond the scope of this book).

\subsection{The magnetic curvature}\label{fascination}

Let us now discuss the influence of geometry (and especially of a boundary) on the spectrum of the magnetic Laplacian, in the semiclassical limit. Before introducing the definition of the concrete model operators, let us first present the nature of some known results.

\subsubsection{Constant magnetic field}
In dimension two the constant magnetic field (with intensity $1$) case is treated when $\Omega$ is the unit disk (with Neumann condition) by Bauman, Phillips and Tang in \cite{BPT98} (see \cite{BH} and \cite{Er96} for the Dirichlet case). In particular, they prove a two terms expansion in the form
\begin{equation}\label{eq.BPT-result}
\lambda_{1}(h)=\Theta_{0}h-\frac{C_{1}}{R} h^{3/2}+o(h^{3/2})\,,
\end{equation}
where $\Theta_{0}\in(0,1)$ and $C_{1}>0$ are universal constants.
This result, which was conjectured in \cite{BeSt, PFS00}, is generalized to smooth and bounded domains by Helffer and Morame in \cite{HelMo01} where it is proved that:
\begin{equation}\label{HM-result}
\lambda_{1}(h)=\Theta_{0} h-C_{1}\kappa_{max} h^{3/2}+o(h^{3/2})\,,
\end{equation}
where $\kappa_{max}$ is the maximal curvature of the boundary. Let us emphasize that, in these papers, the authors are only concerned by the first terms of the asymptotic expansion of $\lambda_{1}(h)$. In the case of smooth domains the complete asymptotic expansion of all the eigenvalues is done by Fournais and Helffer in \cite{FouHel06a}. When the boundary is not smooth, we may mention the papers of Jadallah and Pan \cite{Jad01, Pan02}. In the semiclassical regime, we refer to the papers of Bonnaillie-No\"el, Dauge and Fournais \cite{Bon05, BD06, BonFou07} where perturbation theory is used in relation with the estimates of Agmon. For numerical investigations the reader may consider the paper \cite{BDMV07}.

In dimension three the constant magnetic field case (with intensity $1$) is treated by Helffer and Morame in \cite{HelMo04} under generic assumptions on the (smooth) boundary of $\Omega$:
$$\lambda_{1}(h)=\Theta_{0}h+\hat\gamma_{0}h^{4/3}+o(h^{4/3})\,,$$
where the constant $\hat\gamma_{0}$ is related to the magnetic curvature of a curve in the boundary along which the magnetic field is tangent to the boundary. The case of the ball is analyzed in details by Fournais and Persson in \cite{FouPer11}.

\subsubsection{Variable magnetic field}
The case when the magnetic field is not constant can be motivated by anisotropic superconductors (see for instance \cite{CDG95, ABG10}) or the liquid crystal theory (see \cite{HelPan08, HelPan09, Ray10a, Ray10b}). For the case with a non vanishing variable magnetic field, we refer to \cite{LuPan99a, Ray09} for the first terms of the lowest eigenvalue. In particular the paper \cite{Ray09} provides (under a generic condition) an asymptotic expansion with two terms in the form:
$$\lambda_{1}(h)=\Theta_{0} b' h+C^{\mathsf{2D}}_{1}(\x_{0},\B,\partial\Omega )h^{3/2}+o(h^{3/2})\,,$$
where $C^{\mathsf{2D}}_{1}(\x_{0},\B,\partial\Omega )$ depends on the geometry of the boundary and on the magnetic field at $\x_{0}$ and where $\displaystyle{b'=\min_{\partial\Omega}B=B(\x_{0})}$. When the magnetic field vanishes, the first analysis of the lowest eigenvalue is due to Montgomery in \cite{Montgomery95} followed by Helffer and Morame in \cite{HelMo96} (see also \cite{PK02, HelKo09, HelKo12a}). 

In dimension three (with Neumann condition on a smooth boundary), the first term of $\lambda_{1}(h)$ is given by Lu and Pan in \cite{LuPan00a}. The next terms in the expansion are investigated in \cite{Ray10c} where we can find in particular an upper bound in the form
$$\lambda_{1}(h)\leq \|\B(\x_{0})\|\mathfrak{s}(\theta(\x_{0}))h+C^{\mathsf{3D}}_{1}(\x_{0},\B,\partial\Omega)h^{3/2}+C^{\mathsf{3D}}_{2}(\x_{0},\B,\partial\Omega)h^2+Ch^{5/2}\,,$$
where $\mathfrak{s}$ is a spectral invariant defined in the next section, $\theta(\x_{0})$ is the angle of $\B(\x_{0})$ with the boundary at $\x_{0}$ and the constants $C^{\mathsf{3D}}_{j}(\x_{0},\B,\partial\Omega)$ are related to the geometry and the magnetic field at $\x_{0}\in\partial\Omega$. Let us finally mention the recent paper by Bonnaillie-No\"el-Dauge-Popoff \cite{BDP13} which establishes a one term asymptotics in the case of Neumann boundaries with corners.

\subsection{Some model operators}
It turns out that the results recalled in Section \ref{fascination} are related to many model operators. Let us introduce some of them.

\subsubsection{De Gennes operator}
The analysis of the magnetic Laplacian with Neumann condition on $\R^2_{+}$ makes the so-called de Gennes operator to appear. We refer to \cite{DauHel} where this model is studied in details (see also \cite{FouHel10}). For $\zeta\in\R$, we consider the Neumann realization on $\sL^2(\R_+)$ of
\begin{equation}
\mathfrak{L}^{[0]}_{\zeta}=D_{t}^2+(\zeta-t)^2\,.
\end{equation}
We denote by $\nu_{1}^{[0]}(\zeta)$ the lowest eigenvalue of $\mathfrak{L}^{[0]}(\zeta)$. It is possible to prove that the function $\zeta\mapsto \nu_{1}^{[0]}(\zeta)$ admits a unique and non-degenerate minimum at a point $\zeta^{[0]}_{0}>0$, shortly denoted by $\zeta_{0}$, and that we have
\begin{equation}\label{Theta0}
\Theta_{0}:=\min_{\xi\in\R}\nu_{1}^{[0]}(\zeta)\in(0,1)\,.
\end{equation}
The proof is recalled in Chapter \ref{chapter-examples}, Section \ref{Sec:deGennes}.

\subsubsection{Montgomery operator}
Let us now introduce another important model. This one was introduced by Montgomery in \cite{Montgomery95} to study the case of vanishing magnetic fields in dimension two (see also \cite{PK02} and \cite[Section 2.4]{HelMo04}). This model was revisited by Helffer in \cite{Hel10}, generalized by Helffer and Persson in \cite{HelPer10} and Fournais and Persson in \cite{FouPer13}. The Montgomery operator with parameter $\zeta\in\R$ is the self-adjoint realization on $\R$ of:
\begin{equation}
 \mathfrak{L}^{[1]}_{\zeta}=D_{t}^2+\left(\zeta-\frac{t^2}{2}\right)^2\,.
\end{equation}

\subsubsection{Popoff operator} The investigation of the magnetic Laplacian on dihedral domains (see \cite{Popoff}) leads to the analysis of the Neumann realization on $\sL^2(\mathcal{S}_{\alpha}, \dx t \dx z)$ of:
\begin{equation}
\mathfrak{L}^\Po_{\alpha,\zeta}=D_{ t}^2+D_{ z}^2+(t-\zeta)^2\,,
\end{equation}
where $\mathcal{S}_{\alpha}$ is the sector with angle $\alpha$,
\[\mathcal{S}_{\alpha}=\left\{( t, z)\in \R^2 : |z|< t\tan\left(\frac{\alpha}{2}\right)\right\}\,.\]
\subsubsection{Lu-Pan operator}\label{intro:Lu-Pan}
Let us present a last model operator appearing in dimension three in the case of smooth Neumann boundary (see \cite{LuPan00a, HelMo02, BDPR11} and \eqref{formula.3D}).
We denote by $(s,t)$ the coordinates in $\R^2$ and by $ \R^2_+$ the half-plane:
\[\R^2_+=\{(s,t)\in\R^2, \ t>0\}\,.\]
We introduce the self-adjoint Neumann realization on the half-plane $ \R^2_+$ of the Schr\"odinger operator $\mathfrak{L}^\LP_\theta$ with potential $V_\theta$:
\begin{equation}\label{Ltheta}
\mathfrak{L}^\LP_\theta=-\Delta+V_{\theta}=D_{s}^2+D_{t}^2+V^2_{\theta}\,,
\end{equation}
 where $V_{\theta}$ is defined for any $\theta\in (0,\tfrac\pi2)$ by
\[V_\theta : (s,t)\in \R^2_+ \longmapsto  t\cos\theta-s\sin\theta\,.\]
We can notice that $V^2_\theta$ reaches its minimum $0$ all along the line $t\cos\theta=s\sin\theta$, which makes the angle $\theta$ with  $\partial \R^2_+$. We denote by $\mathfrak{s}_{1}(\theta)$ or simply $\mathfrak{s}(\theta)$ the infimum of the spectrum of $\mathfrak{L}^\LP_{\theta}$. In \cite{FouHel10} (and \cite{HelMo02, LuPan00a}), it is proved that $\mathfrak{s}$ is analytic and strictly increasing on $\left(0,\frac{\pi}{2}\right)$.

\section{A connection with waveguides}

\subsection{Existence of a bound state of $\mathfrak{L}^\LP_{\theta}$}
Among other things one can prove (cf. \cite{HelMo02, LuPan00a}):
\begin{lem}
For all $\theta\in\left(0,\frac \pi 2\right)$ there exists an eigenvalue of $\mathfrak{L}^\LP_{\theta}$ below the essential spectrum which equals $[1,+\infty)$.
\end{lem}
A classical result combining an estimate of Agmon (cf. \cite{Agmon82}) and a theorem due to Persson (cf. \cite{Persson60}) implies that the corresponding eigenfunctions are localized near $(0,0)$. This result is slightly surprising since the existence of the discrete spectrum is related to the association between the Neumann condition and the partial confinement of $V_{\theta}$.
After translation and rescaling, we are led to a new operator:
\[h D_{s}^2+D_{t}^2+(t-\zeta_0-s h^{1/2})^2-\Theta_0\,,\]
where $h=\tan\theta$. Then one can reduce the (semiclassical) analysis to the so-called \emph{Born-Oppenheimer}  approximation:
\[h D_{s}^2+\nu_{1}^{[0]}(\zeta_0+s h^{1/2})-\Theta_0\,.\]
This last operator is very easy to analyze with the classical theory of the harmonic approximation and we get (see \cite{BDPR11}):
\begin{theo}
The lowest eigenvalues of  $\mathfrak{L}^\LP_{\theta}$ admit the following expansions:
\begin{equation}
\label{asymptoticsigman}
\mathfrak{s}_{n}(\theta)\underset{\theta\to 0}{\sim}\sum_{j\geq 0} \gamma_{j,n} \theta^{j}
\end{equation}
with $\gamma_{0,n}=\Theta_{0}$ et $\gamma_{1,n}=(2n-1)\, \sqrt{\frac{(\nu_{1}^{[0]})''(\zeta_0)}{2}}\,.$
\end{theo}
\begin{figure}[ht]
\begin{center}

\scalebox{0.6}{
\begin{tabular}{ccccc}
&
\includegraphics[keepaspectratio=true,width=3cm]{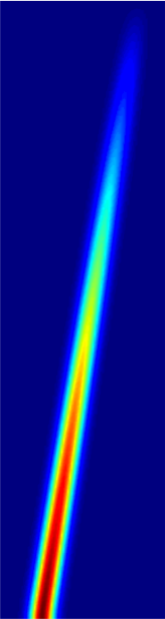}
&
\includegraphics[keepaspectratio=true,width=3cm]{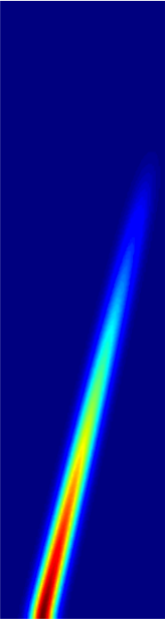}  
&
\includegraphics[keepaspectratio=true,width=3cm]{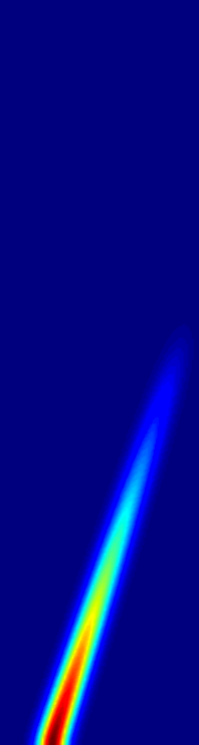}  
&
\includegraphics[keepaspectratio=true,width=3cm]{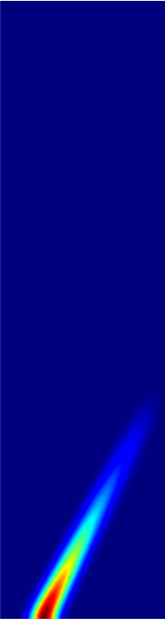}  \\
$\mathfrak{s}_1(\theta)$ & $1.0001656284$ & $0.99987798948$ & $0.99910390126$ &
$0.99445407220$
\end{tabular}
}
\caption{First eigenfunction of $\mathfrak{L}^\LP_{\theta}$ for $\theta=\vartheta\pi/2$ with $\vartheta=0.9$, $0.85$, $0.8$ et $0.7$.}
\label{F1-intro}
\end{center}
\end{figure}
\subsection{A result of Duclos and Exner}
\begin{figure}[h]
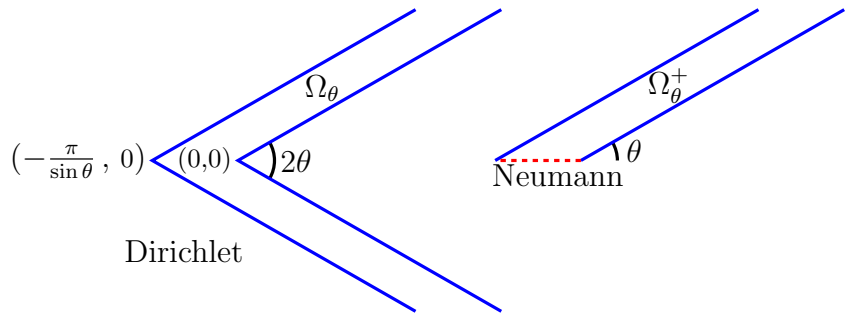

\begin{center}
% 1. Definition of characteristic points
    \figinit{0.8mm}
    \figpt 1:(  0,  0)
    \figpt 2:( 50,  0)
    \figptrot 3: = 2 /1, 30/
    \figpt 4:(-14.1,  0)
    \figvectP 101 [1,4]
    \figptstra 14 = 3/1, 101/
    \figptssym 23 = 3, 14 /1, 2/
    \figptstra 50 = 1, 3, 4, 14, 2 /-4, 101/

% 2. Creation of the postscript file
    \def\MyPSfile{}
    \psbeginfig{\MyPSfile}
    \psset (width=1.2)
    \psarccircP 1 ; 6 [23,3]
    \psarccircP 50 ; 6 [54,51]
    \psset (color=\CoulC)
    \psline[23,1,3]
    \psline[24,4,14]
    \psline[50,51]
    \psline[52,53]
    \psset (color=\CoulB)
    \psset (dash=4)
    \psline[50,52]
  \psendfig

% 3. Writing text on the figure
    \figvisu{\figbox}{}{%
    \figinsert{\MyPSfile}
    \figwritew 1 :{\footnotesize(0,0)}(1.0)
    \figwritee 1 :{$2\theta$}(7.0)
    \figwritegce 50 :{$\theta$}(7.5,1.5)
    \figwritegcw 4 :{($-\frac{\pi}{\sin\theta}$\,, {\footnotesize 0})}(1,0)
    \figwritegcw 4 :{ Dirichlet}(-15,-15)
    \figwritegce 1 :{$\Omega_{\theta}$}(11,12)
    \figwritegce 50 :{$\Omega_{\theta}^+$}(11,12.4)
    \figwritegce 52 :{Neumann}(-0.5,-2.5)
    }
\centerline{\box\figbox}
\caption{ Waveguide with corner $\Omega_{\theta}$ and half-waceguide $\Omega^+_\theta$.}
\label{Coin}
\end{center}
\end{figure}

Figure \ref{F1-intro} can make us think to a \emph{broken waveguide} (see \cite{Ray12a}). Indeed, if one uses the Neumann condition to symmetrize $\mathfrak{L}^\LP_{\theta}$ and if one replaces the confinement property of $V_{\theta}$ by a Dirichlet condition, we are led to the situation described in Figure \ref{Coin}. 
This heuristic comparison reminds the famous paper \cite{Duclos95} where Duclos and Exner introduce a definition of standard (and smooth) waveguides and perform a spectral analysis. For example, in dimension two (see Figure \ref{tube}), a waveguide of width $\eps$ is determined by a smooth curve $s\mapsto c(s)\in\R^2$ as the subset of $\R^2$ given by:
\[\left\{c(s)+t\n(s),\quad (s,t)\in\R\times(-\eps,\eps)\right\}\,,\]
where $\n(s)$ is the normal to the curve $c(\R)$ at the point $c(s)$.

\begin{figure}[ht]
\begin{minipage}{0.4\textwidth}
% 1. Definition of characteristic points
    \figinit{1mm}
    \figpt 1:( 0,  0)
    \figpt 2:( -5, 10)
    \figpt 3:( 50, 0)
    \figptshom 12 = 2 /1, 0.75/
    \figptshom 22 = 2 /1, 1.25/
% 2. Creation of the postscript file
    \def\MyPSfile{}
    \psbeginfig{\MyPSfile}
    \courb{1}{2}{3}{20}{\CoulB}
    \courb{1}{12}{3}{20}{\CoulA}
    \courb{1}{22}{3}{20}{\CoulA}
    \psendfig

% 3. Writing text on the figure
    \figvisu{\figbox}{}{%
    \figinsert{\MyPSfile}
    }
    \centerline{\box\figbox}
    \caption{Waveguide}\label{tube}
\end{minipage}
\begin{minipage}{0.4\textwidth}
% 1. Definition of characteristic points
    \figinit{1mm}
    \figpt 1:( 0,  0)
    \figpt 2:( -0.05, 0.1)
    \figpt 3:( 50, 0)
    \figpt 4:( 25, 0)
    \figvectC 30 (1,0)
    \figptstra 11 = 1,2,3 /6, 30/
    \figptstra 21 = 1,2,3 /12, 30/
% 2. Creation of the postscript file
    \def\MyPSfile{}
    \psbeginfig{\MyPSfile}
    \courb{1}{2}{3}{48}{\CoulA}
    \courb{11}{12}{13}{48}{\CoulB}
    \courb{21}{22}{23}{48}{\CoulA}
    \psendfig

% 3. Writing text on the figure
    \figvisu{\figbox}{}{%
    \figinsert{\MyPSfile}
    \figsetmark{$\bullet$}
    %\figwritee 4 :{${\bf 0}$}(2)
    }
    \centerline{\box\figbox}
    \caption{Broken guide}
\label{F:0b}
\end{minipage}

\end{figure}
Assuming that the waveguide is straight at infinity but not everywhere, Duclos and Exner prove that there is always an eigenvalue below the essential spectrum (in the case of a circular cross section in dimensions two and three). Let us notice that the essential spectrum is $[\lambda,+\infty)$ where $\lambda$ is the lowest eigenvalue of the Dirichlet Laplacian on the cross section. The proof of the existence of discrete spectrum is elementary and relies on the min-max principle. Letting, for $\psi\in \sH^{1}_{0}(\Omega)$,
$$q(\psi)=\int_{\Omega}|\nabla\psi|^2\dx \x\,,$$
it is enough to find $\psi_{0}$ such that $q(\psi_{0})<\lambda\|\psi_{0}\|_{\sL^2(\Omega)}.$
Such a function can be constructed by considering a perturbed Weyl sequence associated with $\lambda$.

\subsection{Waveguides and magnetic fields}
Bending a waveguide induces discrete spectrum below the essential spectrum, but what about twisting a waveguide? This question arises for instance in the papers \cite{K08, KZ1,EKK08} where it is proved that twisting a waveguide plays against the existence of the discrete spectrum. In the case without curvature, the quadratic form is defined for  $\psi\in \sH^1_{0}(\R\times\omega)$ by:
$$q(\psi)=\|\dr_{1}\psi-\rho(s)(t_{3}\dr_{2}-t_{2}\dr_{3})\psi\|^2+\|\dr_{2}\psi\|^2+\|\dr_{3}\psi\|^2\,,$$
where $s\mapsto\rho(s)$ represents the effect of twisting the cross section $\omega$ and $(t_{2}, t_{3})$ are coordinates in $\omega$. From a heuristic point of view, the twisting perturbation seems to act \enquote{as} a magnetic field. This leads to the natural question:
\begin{center}
\enquote{Is the spectral effect of a torsion the same as the effect of a magnetic field?}
\end{center}
If the geometry of a waveguide can formally generate a magnetic field, we can conversely wonder if a magnetic field can generate a waveguide. This remark partially appears in \cite{DGR11} where the discontinuity of a magnetic field along a line plays the role of a waveguide. More generally it appears that, when the magnetic field cancels along a curve, this curve becomes an effective waveguide. 

\section{General organization of the book}\label{orga}

\subsection{Spectral analysis of model operators and spectral reductions}

Chapter~\ref{intro-models} deals with model operators. This notion of model operators is fundamental in the theory of the magnetic Laplacian. We have already introduced some important and historical examples. There are essentially two natural ways to meet reductions to model operators. The first approach can be done thanks to a (space) partition of unity which reduces the spectral analysis to the one of localized and simplified models (we straighten the geometry and freeze the magnetic field). The second approach, which involves an analysis in the phase space, is to identify the possible different scales of the problem, that is the fast and slow variables. This often involves an investigation in the microlocal spirit: we shall analyze the properties of symbols and deduce a microlocal reduction to a spectral problem in lower dimension. In Chapter \ref{intro-models} we provide explicit examples of models and provide their spectral analysis. In Chapter \ref{intro-models}, Section \ref{Sec.vmf} we present a model related to vanishing magnetic fields in dimension two. Due to an inhomogeneity of the magnetic operator, this other model leads to a microlocal reduction and therefore to the investigation of an effective symbol. In fact, the pedagogical example of Section \ref{Sec.vmf} can lead (and actually has led, in the last years) to a more general framework. In Chapter \ref{intro-models}, Section \ref{Sec.BOM} we provide a general and elementary theory of the \enquote{magnetic Born-Oppenheimer approximation} which is a systematic semiclassical reduction to model operators (under generic assumptions on some effective symbols). We also provide the first known examples of pure magnetic WKB constructions.

We provide basic arguments and examples in Chapters \ref{chapter-spectral-theory} and \ref{chapter-models} in relation with model operators. The methods are cast into a non linear framework in Chapter \ref{chapter-non-linear} where the $p$-eigenvalues of the magnetic Laplacian are analyzed in two dimensions. In Chapter \ref{chapter-appendix}, we explain the elementary ideas behind the semiclassical Birkhoff normal form, in an electric case, as a preparation for the magnetic Birkhoff normal form of Chapter \ref{chapter-birk}. This will lead us to use standard tools from microlocal analysis. Our presentation of these technics will be minimalist, the aim being to give the flavor of these tools and to see how they can be used in practice.

The Born-Oppenheimer approximation is discussed in Chapters \ref{chapter-BOE} and  \ref{chapter-BOM}, whereas elementary WKB constructions are analyzed in Chapter \ref{chapter-WKB} in the spirit of dimensional reduction.

\subsection{Normal forms philosophy and the magnetic semi-excited states}
As we have seen there is a non trivial connection between the discrete spectrum, the possible magnetic field and the possible boundary. In fact \emph{normal form} procedures are often deeply rooted in the different proofs, not only in the semiclassical framework. We present in Chapters \ref{intro-van-birk} and \ref{intro-semi} \ the results of five studies \cite{DomRay12}, \cite{RVN13}, \cite{Ray12}, \cite{PoRay12}, \cite{BR13b} which are respectively detailed in Chapters \ref{chapter-vanishing}, \ref{chapter-birk}, \ref{chapter-variable3D}, \ref{chapter-edge}, \ref{chapter-cones}. These studies are concerned by the (often semiclassical) asymptotics of the magnetic eigenvalues and eigenfunctions. 

\subsubsection{Towards the magnetic semi-excited states}\label{intro:semi-ex}
We now describe the philosophy of the proofs of asymptotic expansions for the magnetic Laplacian with respect to a parameter $h$. Let us distinguish between the different conceptual levels of the analysis. Our analysis uses the standard construction of quasimodes, localization techniques and \textit{a priori} estimates of Agmon type satisfied by the eigenfunctions. These \enquote{standard} tools, which are used in most of the papers dealing with $\lambda_{1}(h)$, are not enough to investigate $\lambda_{n}(h)$ due to the spectral splitting arising sometimes in the subprincipal terms. In fact such a fine behavior is the sign of a microlocal effect. In order to investigate this effect, we use normal form procedures \emph{in the spirit of the Egorov theorem}. It turns out that this normal form strategy also strongly simplifies the construction of quasimodes. Once the behavior of the eigenfunctions in the phase space is established, we use the Feshbach-Grushin approach to reduce our operator to an electric Laplacian. Let us comment more in details the whole strategy.

The first step to analyze such problems is to perform an accurate construction of quasimodes and to apply the spectral theorem. In other words we look for pairs $(\lambda,\psi)$ such that we have $\|(\mathcal{L}_{h}-\lambda)\psi\|\leq\eps\|\psi\|$. Such pairs are constructed through an homogenization procedure involving different scales with respect to the different variables. In particular the construction uses a formal power series expansion of the operator and an Ansatz in the same form for $(\lambda,\psi)$. The main difficulty in order to succeed is to choose the appropriate scalings. 

The second step aims at giving \emph{a priori} estimates satisfied by the eigenfunctions.These are localization estimates \emph{\`a la Agmon} (see \cite{Agmon82}). To prove them one generally needs to have \emph{a priori} estimates for the eigenvalues which can be obtained with a partition of unity and local comparisons with model operators. Then such \emph{a priori} estimates, which are in general not optimal, involve an improvement in the asymptotic expansion of the eigenvalues. If we are just interested in the first terms of $\lambda_{1}(h)$, these classical tools are enough.

In fact, the major difference with the electric Laplacian arises precisely in the analysis of the spectral splitting between the lowest eigenvalues. Let us describe what is done in \cite{FouHel06a} (dimension two, constant magnetic field) and in \cite{Ray11b} (non constant magnetic field). In \cite{FouHel06a, Ray11b} quasimodes are constructed and the usual localization estimates are proved. Then the behavior with respect to a phase variable needs to be determined to allow a dimensional reduction. Let us underline here that this phenomenon of phase localization is characteristic of the magnetic Laplacian and is intimately related to the structure of the low lying spectrum. In \cite{FouHel06a} Fournais and Helffer are led to use the pseudo-differential calculus and the Grushin formalism. In \cite{Ray11b} the approach is structurally not the same. In \cite{Ray11b}, in the spirit of the Egorov theorem (see \cite{Egorov71, Ro87, Martinez02}), we use successive canonical transforms of the symbol of the operator corresponding to unitary transforms (change of gauge, change of variable, Fourier transform) and we reduce the operator, modulo remainders which are controlled thanks to the \emph{a priori} estimates, to an electric Laplacian being in the Born-Oppenheimer form (see \cite{CDS81, Martinez89} and more recently \cite{BDPR11}). This reduction enlightens the crucial idea that the inhomogeneity of the magnetic operator is responsible for its spectral structure.

\subsubsection{... which leads to the Birkhoff procedure}
As we suggested above, our magnetic normal forms are close to the Birkhoff procedure and it is rather surprising that it has never been
implemented to enlighten the effect of magnetic fields on the low
lying eigenvalues of the magnetic Laplacian. A reason might be that,
compared to the case of a Schr\"odinger operator with an electric potential, the pure magnetic case presents a specific feature: the symbol \enquote{itself} is not enough to generate a localization of the eigenfunctions. This difficulty can be seen in the recent papers by Helffer and Kordyukov \cite{HelKo11} (dimension
two) and \cite{HelKo13} (dimension three) which treat cases without
boundary. In dimension two, they prove that if the magnetic field has a unique and non-degenerate
minimum, the $n$-th eigenvalue admits an expansion in powers of
$h^{\frac{1}{2}}$ of the form:
\[\lambda_n(h) \sim h \min_{\R^2} B(q) + h^2(c_1(2n-1)+c_0) + \mathcal{O}(h^{\frac{5}{2}})\,,\]
where $c_{0}$ and $c_{1}$ are constants depending on the magnetic field (see the discussion in Section \ref{sec.magw}). 
In Chapter \ref{chapter-birk} (whose main ideas are presented in Chapter \ref{chapter-appendix}), we extend their result by obtaining a complete
asymptotic expansion which actually applies to more general
magnetic wells and allows to describe larger eigenvalues.

\subsection{The spectrum of waveguides}
In Chapter \ref{intro-wg} we present some results occurring in the theory of waveguides. In particular we consider the question:
\begin{center}
\enquote{What is the spectral influence of a magnetic field on a waveguide ?}
\end{center}
We answer this question in Chapter \ref{chapter-mwg}. Then, when there is no magnetic field, we would also like to analyze the effect of a corner on the spectrum and present a non smooth version of the result of Duclos and Exner (see Chapter \ref{chapter-triangles}). For that purpose we also present some results concerning the \emph{semiclassical triangles} in Chapter \ref{chapter-triangles}.

Finally, in Chapter \ref{NLWG}, we cast the linear technics into a non linear framework to investigate the existence of global solutions to the cubic non linear Schr\"odinger equation in a bidimensional waveguide.

\begin{landscape}
\thispagestyle{empty}

\hspace*{-3.4cm}
\scalebox{0.9}{

\begin{tikzpicture}[every node/.append style={fill=mg, rounded corners, draw=bl, inner sep=2.5ex, minimum size=12pt, line width=1mm}]
\begin{scope}[xscale=2.7,yscale=2]

\node (MC) at (8,0) [rectangle,draw] {
\begin{tabular}{c}
\textbf{Magnetic cone}\\
\\
\begin{Bitemize}
\item Conical singularity (Ch. \ref{intro-models}, Sec. \ref{intro-conical}),
\item Small aperture limit (Sec. \ref{intro-cone-asymp}).
\end{Bitemize}
\end{tabular} 
};

\node (VMF) at (0,3)  
{
\begin{tabular}{c}\textbf{"Vanishing" magnetic fields in two dimensions}\\
\\
\begin{Bitemize}
\item Montgomery operators (Ch. \ref{intro-models}, Sec. \ref{Sec.vmf}),
\item Model for vanishing fields with boundary (Section \ref{intro-bvmf}),
\item Normal form (Ch. \ref{intro-semi}, Sec. \ref{intro-vanishing}).
\end{Bitemize}

\end{tabular}
};
\node (WKB) at (4,3)
{
\begin{tabular}{c}
\textbf{Normal Forms}\\
\\ 
\textbf{Adiabatic Reductions}\\
\\
\textbf{WKB Analysis} \\
\end{tabular}
};

\node (SG) at (0,6)
{
\begin{tabular}{c}
\textbf{Non vanishing magnetic field}\\
\textbf{in two dimensions}\\ 
\\
\begin{Bitemize}
\item Ch. \ref{intro-semi}, Sec. \ref{intro-birk},
\item Symplectic geometry,
\item Birkhoff normal form,
\item Pseudo-differential calculus.
\end{Bitemize}
\end{tabular}
};

\node (ML) at (8,3)
{
\begin{tabular}{c}
\textbf{Magnetic wedge} \\
\\
\begin{Bitemize}
\item Edge singularity (Ch. \ref{intro-semi}, Sec. \ref{intro-edge}),
\item Normal form (Ch. \ref{intro-semi}, Sec. \ref{intro-edge}).
\end{Bitemize}
\end{tabular}
};

\node (MTO) at (0,0) 
{
\begin{tabular}{c}
\textbf{Magnetic half-space}\\
\\
\begin{Bitemize}
\item Ch. \ref{intro-semi}, Sec. \ref{intro-variable3D},
\item Polynomial estimates in the phase space.
\end{Bitemize}
\end{tabular}
};

\node (ExWKB) at (4,0) 
{
\begin{tabular}{c}
\textbf{Born-Oppenheimer approximation}\\
\\
\begin{Bitemize}
\item Ch. \ref{intro-models}, Sec. \ref{Sec.BOM},
\item WKB constructions (Sec. \ref{ssec.WKB}),
\item Coherent states (Sec. \ref{intro-coherent}).
\end{Bitemize}
\end{tabular}
};

\node (BW) at (8,6)
{
\begin{tabular}{c}
\textbf{Broken waveguides} \\
\\
\begin{Bitemize}
\item Ch. \ref{intro-wg},
\item Semiclassical triangles (Sec. \ref{intro-wg-triangles}),
\item Boundary layer.
\end{Bitemize}
\end{tabular}
};

\node (MW) at (4,6)
{
\begin{tabular}{c}
\textbf{Magnetic waveguides}\\
\\
\begin{Bitemize}
\item Ch. \ref{intro-wg},
\item Effective Hamiltonians \textit{\`a la} Duclos-Exner,
\item Norm resolvent convergence (Sec. \ref{intro-NRC}).
\end{Bitemize}

\end{tabular}
};
\draw[-, bl,line width=1mm] (SG.south east)-- (WKB.north west);
\draw[-, bl,line width=1mm] (MW)-- (WKB);
\draw[-, bl,line width=1mm] (BW.south west)-- (WKB.north east);
\draw[-, bl,line width=1mm] (ML)-- (WKB);
\draw[-, bl,line width=1mm] (MC.north west)-- (WKB.south east);
\draw[-, bl,line width=1mm] (ExWKB)-- (WKB);
\draw[-, bl,line width=1mm] (MTO.north east)-- (WKB.south west);
\draw[-, bl,line width=1mm] (VMF)-- (WKB);
\end{scope}

\end{tikzpicture}

}

\end{landscape}

\thispagestyle{empty}

\chapter{Models and spectral reductions}\label{intro-models}

\begin{flushright}
\begin{minipage}{0.6\textwidth}
The soul unfolds itself, like a lotus of countless petals.
\begin{flushright}
\textit{The Prophet}, Self-Knowledge, Khalil Gibran
\end{flushright}
\end{minipage}
\vspace*{0.5cm}
\end{flushright}

In this chapter we introduce a model operator (depending on parameters). It appears in dimension two when studying vanishing magnetic fields in the case when the cancellation line of the field intersects the boundary. Though this model seems very specific, we will see how it can lead to a quite general strategy: the (magnetic) Born-Oppenheimer approximation and its relations to WKB constructions.

\section{Vanishing magnetic fields and boundary}\label{Sec.vmf}

\subsection{Why considering vanishing magnetic fields?}\label{intro-vmf}
A motivation is related to the papers of R. Montgomery \cite{Montgomery95}, X-B. Pan and K-H. Kwek \cite{PK02} and B. Helffer and Y. Kordyukov \cite{HelKo09} (see also \cite{HelMo96}, \cite{Hel05} and the thesis of Miqueu \cite{Miqueu}) where the authors analyze the spectral influence of the cancellation of the magnetic field in the semiclassical limit. Another motivation appears in the paper \cite{DGR11} where the authors are concerned with the \enquote{magnetic waveguides} and inspired by the physical considerations  \cite{RMCP02,HEKI04} (see also \cite{HW11}). In any case the case of vanishing magnetic fields can inspire the analysis of non trivial examples of magnetic normal forms, as we will see later.

\subsection{Montgomery operator}\label{intro-mont}
Without going into the details let us describe the model operator introduced in \cite{Montgomery95}. Montgomery was concerned by the magnetic Laplacian $(-ih\nabla+\A)^2$ on $\sL^2(\R^2)$ in the case when the magnetic field $\B=\nabla\times\A$ vanishes along a smooth curve $\Gamma$. Assuming that the magnetic field non degenerately vanishes, he was led to consider the self-adjoint realization on $\sL^2(\R^2)$ of
$$\mathfrak{L}=D_{t}^2+(D_{s}-st)^2\,.$$
In this case the magnetic field is given by $\beta(s,t)=s$ so that the zero locus of $\beta$ is the line $s=0$. Let us write the following change of gauge:
$$\mathfrak{L}^\Mont=e^{-i\frac{s^2t}{2}}\,\mathfrak{L}\,e^{i\frac{s^2t}{2}}=D_{s}^2+\left(D_{t}+\frac{s^2}{2}\right)^2\,.$$
The Fourier transform (after changing $\zeta$ in $-\zeta$) with respect to $t$ gives the direct integral:
$$\mathfrak{L}^\Mont=\int^\oplus \mathfrak{L}^{[1]}_{\zeta}\,\dx\zeta,\quad \mbox{ where }\quad \mathfrak{L}^{[1]}_{\zeta}=D_{s}^2+\left(\zeta-\frac{s^2}{2}\right)^2\,.$$
Note that this family of model operators will be seen as special case of a more general family in Section \ref{BOM-mag}.
Let us recall a few important properties of the lowest eigenvalue $\nu_{1}^{[1]}({\zeta})$ of $\mathfrak{L}^{[1]}_{\zeta}$ (for the proofs, see \cite{PK02, Hel10, HelPer10}).
\begin{prop}\label{Montgomery}
The following properties hold:
\begin{enumerate}
\item For all $\zeta\in\R$,  $\nu_{1}^{[1]}({\zeta})$ is simple.
\item  The function $\zeta\mapsto\nu_{1}^{[1]}({\zeta})$ is analytic.
\item  We have: $\displaystyle{\lim_{|\zeta|\to+\infty}\nu_{1}^{[1]}({\zeta})=+\infty}$.
\item  The function $\zeta\mapsto\nu_{1}^{[1]}({\zeta})$ admits a unique minimum at a point $\zeta^{[1]}_{0}$ and it is non degenerate.
\end{enumerate}
\end{prop}
We have:
\begin{equation}\label{Montgomery-R2}
\sp(\mathfrak{L})=\spe(\mathfrak{L})=\left[\nu_{\Mont},+\infty\right)\,,
\end{equation}
with $\nu_{\Mont}=\nu_{1}^{[1]}(\zeta^{[1]}_{0})$. With a finite element method and Dirichlet condition on the artificial boundary, a upper-bound of the minimum is obtained in \cite[Table 1]{HelPer10} and the 
numerical simulations provide $\nu_{\Mont}\simeq 0.5698$ reached for $\zeta^{[1]}_{0}\simeq 0.3467$ with a discretization step at $10^{-4}$ for the parameter $\zeta$. 
This numerical estimate is already mentioned in \cite{Montgomery95}. In fact we can prove the following lower bound (see \cite{BR13} for a proof using the Temple inequality).
\begin{prop}\label{conjecture-mont}
We have: $\nu_{\Mont}\geq 0.5.$
\end{prop}
If we consider the Neumann realization $\mathfrak{L}^{[1],+}_{\zeta}$ of $D_{s}^2+\left(\zeta-\frac{s^2}2\right)^2$ on $\R^+$, then, by symmetry, the bottom of the spectrum of this operator is linked to the Montgomery operator:
\begin{prop}
If we denote by $\nu_{1}^{[1],+}({\zeta})$ the bottom of the spectrum of $\mathfrak{L}^{[1],+}_{\zeta}$, then
$$\nu_{1}^{[1],+}({\zeta})=\nu_{1}^{[1]}({\zeta})\,.$$
\end{prop}

\subsection{Generalized Montgomery operators}
It turns out that we can generalize the Montgomery operator by allowing an higher order of degeneracy of the magnetic field.
Let $k$ be a positive integer. The generalized Montgomery operator of order $k$ is the self-adjoint realization on $\R$ defined by:
$$\mathfrak{L}^{[k]}_{\zeta}=D_{t}^2+\left(\zeta-\frac{t^{k+1}}{k+1}\right)^2\,.$$
The following theorem (which generalizes Proposition \ref{Montgomery}) is proved in \cite[Theorem 1.3]{FouPer13}.
\begin{theo}\label{theo:FouPer}
$\zeta\mapsto\nu_{1}^{[k]}(\zeta)$ admits a unique and non-degenerate minimum at $\zeta=\zeta_{0}^{[k]}$.
\end{theo}
\begin{notation}
For real $\zeta$, the lowest eigenvalue of $\mathfrak{L}^{[k]}_{\zeta}$ is denoted by $\nu_{1}^{[k]}(\zeta)$ and we denote by $u_{\zeta}^{[k]}$ the positive and $L^2$-normalized eigenfunction associated with $\nu_{1}^{[k]}(\zeta)$. We denote in the same way its holomorphic extension near $\zeta^{[k]}_{0}$.
\end{notation}

\subsection{A broken Montgomery operator}\label{intro-bvmf}
\subsubsection{Heuristics and motivation}
As mentioned above, the bottom of the spectrum of $\mathfrak{L}$ is essential. This fact is due to the translation invariance along the zero locus of $\B$. This situation reminds what happens in the waveguides framework. Guided by the ideas developed for the waveguides, we aim at analyzing the effect of breaking the zero locus of $\B$. Introducing the \enquote{breaking parameter} $\theta\in(-\pi,\pi]$, we will break the invariance of the zero locus in two different ways:
\begin{enumerate}
\item{Case with Dirichlet boundary: $\mathfrak{L}_{\theta}^\Dir$. }
We let $\R^2_{+}=\{(s,t)\in\R^2, t>0\}$ and consider $\mathfrak{L}_{\theta}^\Dir$ the Dirichlet realization, defined as a Friedrichs extension, on $\sL^2(\R^2_{+})$ of:
$$D_{t}^2+\left(D_{s}+\frac{t^2}{2}\cos\theta-st\sin\theta\right)^2\,.$$

\item{Case with Neumann boundary: $\mathfrak{L}_{\theta}^\Neu$. }
We consider $\mathfrak{L}_{\theta}^\Neu$ the Neumann realization, defined as a Friedrichs extension, on $\sL^2(\R^2_{+})$ of:
$$D_{t}^2+\left(D_{s}+\frac{t^2}{2}\cos\theta-st\sin\theta\right)^2\,.$$
The corresponding magnetic field is $\B(s,t)=t\cos\theta-s\sin\theta$. It cancels along the half-line $t=s\tan\theta$.

\end{enumerate}

\begin{notation}
We use the notation $\mathfrak{L}_{\theta}^{\bullet}$ where $\bullet$ can be $\Dir$, $\Neu$. 
\end{notation}

\subsubsection{Properties of the spectra}
Let us analyze the dependence of the spectra of $\mathfrak{L}_{\theta}^\bullet$  on the parameter $\theta$. Denoting by $S$ the axial symmetry $(s,t)\mapsto (-s,t)$, we get:
$$\mathfrak{L}_{-\theta}^\bullet =S\overline{\mathfrak{L}^\bullet_{\theta}}S\,,$$
where the line denotes the complex conjugation. Then, we notice that $\mathfrak{L}^\bullet_{\theta}$ and $\overline{\mathfrak{L}^\bullet_{\theta}}$ are isospectral. Therefore, the analysis is reduced to $\theta\in[0,\pi)$. Moreover, we get:
$$S \mathfrak{L}^\bullet_{\theta}S= \mathfrak{L}^\bullet_{\pi-\theta}\,.$$
The study is reduced to $\theta\in\left[0,\frac{\pi}{2}\right]$. 

We observe that at $\theta=0$ and $\theta=\frac{\pi}{2}$ the domain of $\mathfrak{L}^\bullet_{\theta}$ is not continuous.
\begin{lem}
The family $\left(\mathfrak{L}^\bullet_{\theta}\right)_{\theta\in\left(0,\frac{\pi}{2}\right)}$ is analytic of type (A).
\end{lem}
The following proposition states that the infimum of the essential spectrum is the same for $\mathfrak{L}_{\theta}^{\Dir}$, $\mathfrak{L}_{\theta}^{\Neu}$ and $\mathfrak{L}_{\theta}$.
\begin{prop}\label{essentials}
For $\theta\in\left(0,\frac{\pi}{2}\right)$, we have
$\inf\spe(\mathfrak{L}_{\theta}^{\bullet})=\nu_{\Mont}.$
\end{prop}
In the Dirichlet case, the spectrum is essential.
\begin{prop}\label{Dirichlet}
For all $\theta\in\left(0,\frac{\pi}{2}\right)$, we have $\sp(\mathfrak{L}_{\theta}^\Dir)=[\nu_\Mont,+\infty)$.
\end{prop}
\begin{notation}
Let us denote by $\lambda_{n}^\bullet(\theta)$ the $n$-th number in the sense of the Rayleigh variational formula for $\mathfrak{L}^\bullet_{\theta}$.
\end{notation}
The following proposition (the proof of which can be found in \cite[Lemma 5.2]{PK02}) states that $\mathfrak{L}^\Neu_{\theta}$ admits an eigenvalue below its essential spectrum when $\theta\in\left(0,\frac{\pi}{2}\right]$. 
\begin{prop}\label{bound-state}
For all $\theta\in\left(0,\frac{\pi}{2}\right]$, $\lambda_{1}^\Neu(\theta)<\nu_{\Mont}$.
\end{prop}

\subsection{Singular limit $\theta\to 0$}
\subsubsection{Renormalization}
Thanks to Proposition \ref{bound-state}, one knows that breaking the invariance of the zero locus of the magnetic field with a Neumann boundary creates a bound state. We also would like to tackle this question for $\mathfrak{L}_{\theta}$ and in any case to estimate more quantitatively this effect. A way to do this is to consider the limit $\theta\to 0$ which reveals new model operators. 
\begin{notation}
We let $h=\tan\theta$.
\end{notation}
First, we perform a scaling:
\begin{equation}\label{scaling-tan-theta}
s=h^{-1}(\cos\theta)^{-1/3}\sigma,\quad t=(\cos\theta)^{-1/3}\tau\,.
\end{equation}
The operator $\mathfrak{L}_{\theta}^\Neu$ is thus unitarily equivalent to $(\cos\theta)^{2/3}\hat{\mathfrak{L}}^\Neu_{\tan\theta}$, where the expression of $\hat{\mathfrak{L}}^\Neu_{\tan\theta}$ is given by:
\begin{equation}\label{VBeps}
D_{\tau}^2+\left(h D_{\sigma}+\frac{\tau^2}{2}-\sigma\tau\right)^2\,.
\end{equation}

\subsubsection{New model operators}
We are led to two families of one dimensional operators on $\sL^2(\R^2_{\Neu})$ with two parameters $(x,\xi)\in\R^2$:
\[\mathcal{M}_{x,\xi}^\Neu=D_{\tau}^2+\left(-\xi-x \tau+\frac{\tau^2}{2}\right)^2\,.\]
These operators have compact resolvents and are analytic families with respect to the variables $(x,\xi)\in\R^2$. 
\begin{notation}
We denote by $\mu_{n}^\Neu(x,\xi)$ the $n$-th eigenvalue of $\mathcal{M}_{x,\xi}^\Neu$.
\end{notation}
Roughly speaking $\mathcal{M}_{x,-\xi}^\Neu$ is the operator valued symbol of \eqref{VBeps}, so that we expect that the behavior of the so-called \enquote{band function} $(x,\xi)\mapsto\mu_{1}^\Neu(x,\xi)$ determines the structure of the low lying spectrum of $\mathfrak{M}_{h,x,\xi}^\Neu$ in the limit $h\to 0$.

The following theorem (proved in Chapter \ref{chapter-models}, Section \ref{sec.theo-Rp}) states that the band function admits a minimum and was initially proved in \cite{BR13}.
\begin{theo}\label{theo-Rp}
The function $\R\times\R\ni (x,\xi)\mapsto \mu_{1}^\Neu(x,\xi)$ admits a minimum denoted by $\underline\mu_{1}^\Neu$. 
Moreover we have
\[\liminf_{|x|+|\xi|\to+\infty}\mu^\Neu_{1}(x,\xi)\geq \nu_{\Mont}>\min_{(x,\xi)\in\R^2}\mu^\Neu_{1}(x,\xi)=\underline\mu_{1}^\Neu\,.\]
\end{theo}

\begin{rem}
We have:
\begin{equation}\label{comparison-minima}
\underline\mu_{1}^\Neu\leq \underline\mu_{1}\,.
\end{equation}

\end{rem}
Numerical experiments lead to the following conjecture.
\begin{conj}\label{minimum}
\begin{itemize}
\item The inequality \eqref{comparison-minima} is strict.
\item The minimum $\underline\mu_{1}^\Neu$
is unique and non-degenerate. 
\end{itemize}
\end{conj}

\begin{rem}
Under Conjecture \ref{minimum}, it is possible to prove complete asymptotic expansions of the first eigenvalues of $\mathfrak{L}^\Neu_{\theta}$. In fact, this can be done by using the magnetic Born-Oppenheimer approximation (see Section \ref{Sec.BOM}).
\end{rem}

\section{Magnetic Born-Oppenheimer approximation}\label{Sec.BOM}
This section is devoted to the analysis of the operator on $\sL^2(\R_{s}^m\times\R_{t}^n,\dx s \dx t)$:
\begin{equation}\label{Lfh}
\mathfrak{L}_{h}=(-ih\nabla_{s}+A_{1}(s,t))^2+(-i\nabla_{t}+A_{2}(s,t))^2\,,
\end{equation}
We will assume that $A_{1}$ and $A_{2}$ are real analytic. We would like to describe the lowest eigenvalues of this operator in the limit $h\to 0$ under elementary confining assumptions.
The problem of considering partial semiclassical problems appears for instance in the context of \cite{Martinez89, KMSW92} where the main issue is to approximate the eigenvalues and eigenfunctions of operators in the form:
\begin{equation}\label{original}
-h^2\Delta_{s}-\Delta_{t}+V(s,t)\,.
\end{equation}
The main idea, due to Born and Oppenheimer in \cite{BO27}, is to replace, for fixed $s$, the operator $-\Delta_{t}+V(s,t)$ by its eigenvalues $\mu_{k}(s)$. Then we are led  to consider for instance the reduced operator (called Born-Oppenheimer approximation):
$$-h^2\Delta_{s}+\mu_{1}(s)$$
and to apply the semiclassical techniques \textit{\`a la} Helffer-Sj\"ostrand \cite{HelSj84, HelSj85} to analyze in particular the tunnel effect when the potential $\mu_{1}$ admits symmetries. The main point it to make the reduction of dimension rigorous. Note that we have always the following lower bound:
\begin{equation}\label{lbBOE}
-h^2\Delta_{s}-\Delta_{t}+V(s,t)\geq -h^2\Delta_{s}+\mu_{1}(s)\,,
\end{equation}
which involves accurate estimates of Agmon with respect to $s$.

\subsection{Electric Born-Oppenheimer approximation and low lying spectrum}
Before dealing with the so-called Born-Oppenheimer approximation in presence of magnetic fields, we shall recall the philosophy in a simplified electric case.
\subsubsection{Electric result}
Let us explain the question in which we are interested. We shall study operators in $\sL^2(\R\times\Omega)$ (with $\Omega\subset\R^{n}$) in the form
$$\mathfrak{H}_{h}=h^2 D_{s}^2-\Delta_{t}+V(s,t)\,,$$
where $V\in\mathcal{C}^\infty(\R\times\overline{\Omega})$ is a non negative potential (with $V$ as a polynomial for simplicity). The operator is defined as the self-adjoint extension associated with the quadratic form
$$\mathfrak{Q}_{h}(\psi)=\int_{\R\times\Omega} h^2|\partial_{s}\psi|^2+|\nabla_{t}\psi|^2+V(s,t)|\psi|^2\dx s \dx t\,.$$
We will also need the partial operator $\mathcal{V}(s)=-\Delta_{t}+V(t,s)$ defined in the same way by its quadratic form
$$q_{s}(\varphi)=\int_{\Omega} |\nabla_{t}\varphi|^2+V(s,t)|\varphi|^2 \dx t\,.$$
We will assume that $V(s,t)\underset{|t|\to+\infty}{\to}+\infty$. Moreover we will assume that $\left(\mathcal{V}(s)\right)_{s\in\R}$ is an analytic family of type $(A)$ in the sense of Kato.

It can be shown that the lowest eigenvalue of $\mathcal{V}(s)$ denoted by $\nu(s)$ is simple (and thus it is analytic). 

\begin{assumption}\label{liminfnu}
The function $\nu(s)$ admits a unique and non degenerate minimum $\nu_{0}$ at $s_{0}$. Moreover, we have
$$\liminf_{|s|\to+\infty} \nu(s)>\nu_{0}\,.$$
\end{assumption}

We want to analyze the low lying eigenvalues of $\mathfrak{H}_{h}$ and we now try to understand the heuristics. We hope that $\mathfrak{H}_{h}$ can be described by its \enquote{Born-Oppenheimer} approximation:
$$\mathfrak{H}^{\BO}_{h}=h^2 D_{s}^2+\nu_{1}(s)\,,$$
which is an electric Laplacian in dimension one. Then, we guess that $\mathfrak{H}^{\BO}_{h}$ is well approximated by its Taylor expansion:
$$h^2 D_{s}^2+\nu(s_{0})+\frac{\nu_{1}''(s_{0})}{2}(s-s_{0})^2\,.$$
In fact this heuristics can be made rigorous.

\begin{assumption}\label{essential}
For $R\geq 0$, we let $\Omega_{R}=\R^{1+n}\setminus \overline{\mathcal{B}(0,R)}$. We denote by $\mathfrak{H}_{h}^{\Dir,\Omega_{R}}$ the Dirichlet realization on $\Omega_{R}$ of $h^2D_{s}^2+D^2_{t}+V(s,t)$. We assume that there exist $R_{0}\geq 0$, $h_{0}>0$ and $\nu_{0}^*>\nu_{0}$ such that, for all $h\in(0,h_{0})$,
\[\lambda_{1}^{\Dir,\Omega_{R_{0}}}(h)\geq \nu_{0}^*\,.\]
\end{assumption}
\begin{rem}\label{confining2-el}
In particular, due to the monotonicity of the Dirichlet realization with respect to the domain, Assumption \ref{confining} implies that there exist $R_{0}>0$ and $h_{0}>0$ such that for all $R\geq R_{0}$ and $h\in(0,h_{0})$:
\[\lambda_{1}^{\Dir,\Omega_{R}}(h)\geq \lambda_{1}^{\Dir,\Omega_{R_{0}}}(h)\geq \nu_{0}^*\,.\]
\end{rem}
By using the Persson's theorem (see Chapter \ref{chapter-spectral-theory}, Proposition \ref{Persson}), we have the following proposition.
\begin{prop}\label{essential-BOE}
Let us assume Assumption \ref{confining}. There exists $h_{0}>0$ such that for all $h\in(0,h_{0})$:
\[\inf \spe(\mathfrak{H}_{h})\geq \nu_{0}^*\,.\]
\end{prop}
The following theorem is proved in Chapter \ref{chapter-BOE}.
\begin{theo}\label{BOelec}
Under Assumptions \ref{liminfnu} and \ref{essential}, the $n$-th eigenvalue of $\mathfrak{H}_{h}$ has the expansion
\[\lambda_{n}(h)=\nu(s_{0})+h(2n-1)\left(\frac{\nu_{1}''(s_{0})}{2}\right)^{1/2}+\mathcal{O}(h^{\frac{3}{2}})\,.\]
\end{theo}
\subsubsection{Counting function}
In the last theorem we are only interested in the low lying spectrum. It turns out that the so-called Born-Oppenheimer reduction is a slightly more general procedure (see \cite{Martinez89, KMSW92}) which provides in general an effective Hamiltonian which describes the spectrum below some fixed energy level (and allows for instance to estimate the counting function). 

\begin{notation}\label{N.number}
Given $\mathfrak H$ a semi-bounded self-adjoint operator and $a<\inf\spe(\mathfrak H)$, we denote
\[\mathsf{N}(\mathfrak H,a)\ = \ \#\{ \lambda\in \sp(\mathfrak H) \ : \ \lambda \leq a \}<+\infty\,.\]
The eigenvalues are counted with multiplicity.
\end{notation}
The following theorem (see the proof in Chapter \ref{chapter-BOE}, Section \ref{sec.chapter-BOE-number}) provides the asymptotics of the number of bound states (see the related works \cite{B85, MoTr05, DR14}).
\begin{theo}\label{number}
Let us assume that $\nu_{1}$ is bounded, that $\displaystyle{\liminf_{|s|\to+\infty}\nu_{1}(s)>\nu_{0}}$. In addition, if $u_{s}$ denotes the positive and $\sL^2$-normalized eigenfunction of $\mathcal{V}(s)$ associated with $\nu_{1}(s)$, we assume that $R(s):=\|\partial_{s}u_{s}\|_{\sL^2(\R_{t})}$ is bounded. Then, for $E\in\left(\nu_{0},\displaystyle{\liminf_{|s|\to+\infty}\nu_{1}(s)}\right)$ and if $\nu_{2}\geq E'>E$, we have
\[\mathsf{N}\left(\mathfrak{H}_{h}, E\right)\underset{h\to 0}{\sim}\frac{1}{\pi h}\int_{\R}\sqrt{\left(E-\nu_{1}(s)\right)_{+}}\dx s\,.\]
\end{theo}

\subsection{Magnetic case}\label{BOM-mag}
We would like to understand the analogy between \eqref{Lfh} and \eqref{original}. In particular even the formal dimensional reduction does not seem to be as clear as in the electric case. Let us write the operator valued symbol of $\mathfrak{L}_{h}$. For $(x,\xi)\in\R^n \times\R^n$, we introduce the electro-magnetic Laplacian acting on $\sL^2(\R^n, \dx t)$:
\[\mathcal{M}_{x,\xi}=(-i\nabla_{t}+A_{2}(x,t))^2+(\xi+A_{1}(x,t))^2\,.\]
Denoting by $\mu_{1}(x,\xi)=\mu(x,\xi)$ its lowest eigenvalue we would like to replace $\mathfrak{L}_{h}$ by the $m$-dimensional pseudo-differential operator:
\[\mu(s,-ih\nabla_{s})\,.\]
This can be done modulo $\mathcal{O}(h)$ (see \cite{Martinez07}). Nevertheless we do not have an obvious comparison as in \eqref{lbBOE} so that the microlocal behavior of the eigenfunctions with respect to $s$ is not directly reachable (we can not directly apply the exponential estimates of \cite{Martinez94} due to the possible essential spectrum, see Assumption \ref{confining}). In particular we shall prove that the remainder $\mathcal{O}(h)$ is indeed small when acting on the eigenfunctions and then estimate it precisely. In addition, the point of view presented below is rather self-contained and do not assume more that the elements of pseudo-differential calculus.

\subsubsection{Eigenvalue asymptotics in the magnetic Born-Oppenheimer approximation}
We will work under the following assumptions. The first assumption states that the lowest eigenvalue of the operator symbol of $\mathfrak{L}_{h}$ admits a unique and non-degenerate minimum.
\begin{assumption}\label{hyp-gen}
\begin{itemize}
\item[-] The family $(\mathcal{M}_{x,\xi})_{(x,\xi)\in\R^m\times\R^m}$ is analytic of type (B) in the sense of Kato \cite[Chapter VII]{Kato66}. 
\item[-] For all $(x,\xi)\in\R^m\times\R^m$, the bottom of the spectrum of $\mathcal{M}_{x,\xi}$ is a simple eigenvalue denoted by $\mu(x,\xi)$ (in particular it is an analytic function thanks to Kato's theory) and associated with a $\sL^2$-normalized eigenfunction $u_{x,\xi}\in\mathcal{S}(\R^n)$ which also analytically depends on $(x,\xi)$. 
\item[-] The function $\mu$ admits a unique and non degenerate minimum $\mu_{0}$ at point denoted by $(x_{0},\xi_{0})$ and such that $\liminf_{|x|+|\xi|\to+\infty}\mu(x,\xi)>\mu_{0}$. 
\item[-] The family $(\mathcal{M}_{x,\xi})_{(x,\xi)\in\R^m\times\R^m}$ can be analytically extended in a complex neighborhood of $(x_{0},\xi_{0})$.
\end{itemize}
\end{assumption}
\begin{assumption}\label{hyp-gen'}
Under Assumption \ref{hyp-gen}, let us denote by $\Hess\,\mu_{1}(x_{0},\xi_{0})$ the Hessian matrix of $\mu_{1}$ at $(x_{0},\xi_{0})$. We assume that the spectrum of $\Hess\,\mu_{1}(x_{0},\xi_{0})(\sigma, D_{\sigma})$ is simple.
\end{assumption}
The next assumption is a spectral confinement.
\begin{assumption}\label{confining}
For $R\geq 0$, we let $\Omega_{R}=\R^{m+n}\setminus \overline{\mathcal{B}(0,R)}$. We denote by $\mathfrak{L}_{h}^{\Dir,\Omega_{R}}$ the Dirichlet realization on $\Omega_{R}$ of $(-i\nabla_{t}+A_{2}(s,t))^2+(-ih\nabla_{s}+A_{1}(s,t))^2$. We assume that there exist $R_{0}\geq 0$, $h_{0}>0$ and $\mu_{0}^*>\mu_{0}$ such that for all $h\in(0,h_{0})$:
\[\lambda_{1}^{\Dir,\Omega_{R_{0}}}(h)\geq \mu_{0}^*\,.\]
\end{assumption}

We have the following proposition.
\begin{prop}\label{essential-BOM}
Let us assume Assumption \ref{confining}. There exists $h_{0}>0$ such that, for all $h\in(0,h_{0})$,
\[\inf \spe(\mathfrak{L}_{h})\geq \mu_{0}^*\,.\]
\end{prop}
We can now state the theorem concerning the spectral asymptotics (see Chapter \ref{chapter-BOM} and \cite{BHR14}). 
\begin{theo}\label{main-theorem-BOM}
We assume that $A_{1}$ and $A_{2}$ are polynomials and Assumptions \ref{hyp-gen}, \ref{hyp-gen'} and \ref{confining}. For all $n\geq 1$, there exist a sequence $(\gamma_{j,n})_{j\geq 0}$ and $h_{0}>0$ such that for all $h\in(0,h_{0})$ the $n$-th eigenvalue of $\mathfrak{L}_{h}$ exists and satifies:
\[\lambda_{n}(h)\underset{h\to 0}{\sim} \sum_{j\geq 0} \gamma_{j,n}h^{j/2}\,,\]
where $\gamma_{0,n}=\mu_{0},\quad \gamma_{1,n}=0$ and $\mu_{2,n}$ is the $n$-th eigenvalue of $\frac{1}{2}\Hess_{x_{0},\xi_{0}}\,\mu_{1}(\sigma,D_{\sigma})$.
\end{theo}

\subsubsection{Coherent states}\label{intro-coherent}
Let us recall the formalism of coherent states which play a central role in the proof of Theorem \ref{main-theorem-BOM}. We refer to the books \cite{Fol89} and \cite{CR12} for details (see also \cite{Ray13}). We let:
\[g_{0}(\sigma)=\pi^{-1/4}e^{-|\sigma|^2/2}\]
and the usual creation and annihilation operators:
\[\ga_{j}=\frac{1}{\sqrt{2}}(\sigma_{j}+\dr_{\sigma_{j}}),\qquad \ga_{j}^*=\frac{1}{\sqrt{2}}(\sigma_{j}-\dr_{\sigma_{j}})\]
which satisfy the commutator identities:
\[[\ga_{j},\ga_{j}^*]=1,\qquad [\ga_{j}, \ga_{k}^*]=0 \mbox{ if } k\neq j\,.\]
We notice that
\[\sigma_{j}=\frac{\ga_{j}+\ga_{j}^*}{\sqrt{2}},\quad \dr_{\sigma_{j}}=\frac{\ga_{j}-\ga_{j}^*}{\sqrt{2}},\quad \ga_{j}\ga_{j}^*=\frac{1}{2}(D_{\sigma_{j}}^2+\sigma_{j}^2+1)\,.\]
For $(u,p)\in\R^m\times \R^m$, we introduce the coherent state
\[f_{u,p}(\sigma)=e^{ip\cdot \sigma} g_{0}(\sigma-u)\,,\]
and the associated projection
\[\Pi_{u,p}\psi=\langle\psi, f_{u,p}\rangle_{\sL^2(\R^m)} f_{u,p}=\psi_{u,p}f_{u,p}\,,\]
which satisfies
\[\psi=\int_{\R^{2m}}\Pi_{u,p}\psi\dx u\dx p\,,\]
and the Parseval formula
\[\|\psi\|^2=\int_{\R^n}\int_{\R^{2m}}|\psi_{u,p}|^2\dx u\dx p\dx\tau\,.\]
We recall that
\[\ga_{j}f_{u,p}=\frac{u_{j}+ip_{j}}{\sqrt{2}}f_{u,p}\]
and 
\[(\ga_{j})^\ell (\ga_{k}^*)^q\psi=\int_{\R^{2m}}\left(\frac{u_{j}+ip_{j}}{\sqrt{2}}\right)^\ell\left(\frac{u_{k}-ip_{k}}{\sqrt{2}}\right)^q \Pi_{u,p}\psi\dx u\dx p\,.\]
We recall that (see \eqref{Lch}):
\[\mathcal{L}_{h}=(-i\nabla_{\tau}+A_{2}(x_{0}+h^{1/2}\sigma,\tau))^2+(\xi_{0}-ih^{1/2}\nabla_{\sigma}+A_{1}(x_{0}+h^{1/2}\sigma,\tau))^2\]
and, assuming that $A_{1}$ and $A_{2}$ are polynomial:
\[\mathcal{L}_{h}=\mathcal{L}_{0}+h^{1/2}\mathcal{L}_{1}+h\mathcal{L}_{2}+\ldots +(h^{1/2})^{M}\mathcal{L}_{M}\,.\]
If we write the Wick ordered operator, we get:
\begin{equation}\label{ordering}
\mathcal{L}_{h}=\underbrace{\mathcal{L}_{0}+h^{1/2}\mathcal{L}_{1}+h\mathcal{L}^\W_{2}+\ldots +(h^{1/2})^{M}\mathcal{L}^\W_{M}}_{\mathcal{L}_{h}^\W}+\underbrace{hR_{2}+\ldots+(h^{1/2})^{M}R_{M}}_{\mathcal{R}_{h}}\,,
\end{equation}
where the $R_{j}$ satisfy, for $j\geq 2$:
\begin{equation}\label{remainders}
h^{j/2}R_{j}=h^{j/2}\mathcal{O}_{j-2}(\sigma, D_{\sigma})
\end{equation}
and are the remainders in the so-called Wick ordering. In the last formula the notation $\mathcal{O}_{k}(\sigma,D_{\sigma})$ stands for a polynomial operator with total degree in $(\sigma,D_{\sigma})$ less than $k$. We recall that
\[\mathcal{L}_{h}^\W=\int_{\R^{2m}}\mathcal{M}_{x_{0}+h^{1/2}u, \xi_{0}+h^{1/2}p}\dx u \dx p\,.\]

\subsubsection{A family of examples}
In order to make our Assumptions \ref{hyp-gen} and \ref{confining} more concrete, let us provide a family of examples in dimension two which is related to \cite{HelPer10} and the more recent result by Fournais and Persson \cite{FouPer13}. Our examples are strongly connected with \cite[Conjecture 1.1 and below]{HelKo09}.

For $k\in\N\setminus\{0\}$, we consider the operator the following magnetic Laplacian on $\sL^2(\R^2,\dx x\dx s)$:
\[\mathfrak{L}_{h,\A^{[k]}}=h^2 D_{\mathsf{t}}^2+\left(h D_{\mathsf{s}}-\gamma(\mathsf{s})\frac{\mathsf{t}^{k+1}}{k+1}\right)^2\,.\]
Let us perform the rescaling:
\[\mathsf{s}=s,\quad \mathsf{t}=h^{\frac{1}{1+k}}t\,.\]
The operator becomes
\[h^{\frac{2k+2}{k+2}}\left(D_{t}^2+\left(h^{\frac{1}{k+2}}D_{s}-\gamma(s)\frac{t^{k+1}}{k+1}\right)^2\right)\,.\]
and the investigation is reduced to the one of
\[\mathfrak{L}_{h}^{\van, [k]}=D_{t}^2+\left(h^{\frac{1}{k+2}}D_{s}-\gamma(s)\frac{t^{k+1}}{k+1}\right)^2\,.\]

\begin{prop}\label{verify-assumptions}
Let us assume that either $\gamma$ is polynomial and admits a unique minimum $\gamma_{0}>0$ at $s_{0}=0$ which is non degenerate, or $\gamma$ is analytic and such that  $\liminf_{x\to\pm\infty}\gamma=\gamma_{\infty}\in(\gamma_{0},+\infty)$.
For $k\in\N\setminus\{0\}$, the operator $\mathfrak{L}_{h}^{[k]}$ satisfies Assumptions \ref{hyp-gen}, \ref{hyp-gen'} and \ref{confining}. Moreover we can choose
$\mu_{0}^*> \mu_{0}$.
\end{prop}
\begin{proof}
Let us verify Assumption \ref{hyp-gen}. The $h^{\frac{1}{k+2}}$-symbol of $\mathfrak{L}_{h}^{[k]}$ with respect to $s$ is:
\[\mathcal{M}^{[k]}_{x,\xi}=D_{t}^2+\left(\xi-\gamma(x)\frac{t^{k+1}}{k+1}\right)^2\,.\]
The lowest eigenvalue of $\mathcal{M}^{[k]}_{x,\xi}$, denoted by $\mu_{1}^{[k]}(x,\xi)$, satisfies:
\[\mu_{1}^{[k]}(x,\xi)=(\gamma(x))^{\frac{2}{k+2}}\nu_{1}^{[k]}\left((\gamma(x))^{-\frac{1}{k+2}}\xi\right)\,,\]
where $\nu_{1}^{[k]}(\zeta)$ denotes the first eigenvalue of
\[\mathfrak{L}^{[k]}_{\zeta}=D_{t}^2+\left(\zeta-\frac{t^{k+1}}{k+1}\right)^2\,.\]
We recall that $\zeta\mapsto\nu_{1}^{[k]}(\zeta)$ admits a unique and non-degenerate minimum at $\zeta=\zeta_{0}^{[k]}$ (see Theorem \ref{theo:FouPer}). Therefore Assumption \ref{hyp-gen} is satisfied. This is much more delicate (and beyond the scope of this book) to verify Assumption \ref{confining} and this relies on a basic normal form procedure that we will use for our magnetic WKB constructions.
\end{proof}
\section{Magnetic WKB expansions: examples}\label{ssec.WKB}

\subsection{WKB analysis and estimates of Agmon}
As we explained in Chapter \ref{intro}, Section \ref{intro:semi-ex}, in many papers about asymptotic expansions of the magnetic eigenfunctions, one of the methods consists in using a formal power series expansion. It turns out that these constructions are never in the famous WKB form, but in a weaker and somehow more flexible one. When there is an additional electric potential, the WKB expansions are possible as we can see in \cite{HelSj87} and \cite{MS99}. The reason for which we would like to have a WKB description of the eigenfunctions is to get a precise estimate of the magnetic tunnel effect in the case of symmetries. Until now, such estimates are only investigated in two dimensional corner domains in \cite{BD06} and \cite{BDMV07} for the numerical counterpart. It turns out that the crucial point to get an accurate estimate of the exponentially small splitting of the eigenvalues is to establish exponential decay estimates of Agmon type. These localization estimates are rather easy to obtain (at least to get the good scale in the exponential decay) in the corner cases due to the fact that the operator is \enquote{more elliptic} than in the regular case in the following sense: the spectral asymptotics is completely drifted by the principal symbol. Nevertheless, let us notice here that, on the one hand, the numerics suggests that the eigenvalues do not seem to be simple and, on the other hand, establishing the optimal estimates of Agmon is still an open problem. In smooth cases, due to a lack of ellipticity and to the multiple scales, the localization estimates obtained in the literature are in  general not optimal or rely on the presence of an electric potential (see \cite{N96, N99}): the principal symbol provides only a partial confinement whereas the precise localization of the eigenfunctions seems to be determined by the subprincipal terms. Our WKB analysis (inspired by our paper \cite{BHR14}), in the explicit cases discussed in this book, will give some hints for the optimal candidate to be the effective Agmon distance.

\subsection{WKB expansions for $\mathfrak{L}_{h}^{\van, [k]}$}
The following theorem states that the first eigenfunctions of $\mathfrak{L}_{h}^{\van, [k]}$ are in the WKB form. It turns out that this property is very general and verified for the general $\mathfrak{L}_{h}$ under our generic assumptions. Nevertheless this general and fundamental result is beyond the scope of this book. We will only give the flavor of such constructions for our explicit model. As far as we know such a result was not even known on an example. Let us state one of the main results of this book concerning the WKB expansions (see Chapter \ref{chapter-WKB} and \cite{BHR14} for a more general statement about $\mathfrak{L}_{h}$).
\begin{theo}\label{WKB-van}
Let us assume that either $\gamma$ is polynomial and admits a unique minimum $\gamma_{0}>0$ at $s_{0}=0$ which is non degenerate, either $\gamma$ is analytic and such that  $\liminf_{x\to\pm\infty}\gamma=\gamma_{\infty}\in(\gamma_{0},+\infty)$. There exist a function $\Phi=\Phi(s)$ defined in a neighborhood $\Vc$ of $0$ with $\Re \Phi''(0) >0$ and a sequence of real numbers $(\lambda^{\van}_{n,j})_{j\geq 0}$ such that the $n$-th eigenvalue of $\mathfrak{L}_{h}^{\van, [k]}$ satisfies
\[\lambda^\van_n(h) \underset{h\to 0}{\sim}\sum_{j\geq 0}\lambda^\van_{n,j} h^{\frac{j}{k+2}}\]
in the sense of formal series, with $\lambda^\van_{n,0}=\mu_{0}=\nu_{1}^{[k]}(\zeta_{0}^{[k]})$. Besides there exists a formal
series of smooth functions on $ \Vc \times \R^n_t$
\[\an^\van_{n}(.,h)\underset{h\to 0}{\sim}\sum_{j\geq 0}\an^\van_{n,j} h^{\frac{j}{k+2}}\]
with $\an^\van_{n,0} \neq 0$ such that
\[\left(\mathfrak{L}^{\van,[k]}_{h}-\lambda_{n}(h)\right)\left( \an^\van_{n}( \cdot,h) e^{-\Phi/h^{\frac{1}{k+2}}}\right)=\mathcal{O}\left(h^{\infty}\right)  e^{-\Phi/h^{\frac{1}{k+2}}}\,,\]
In addition,  there exists $c_0>0$ such that for all $h\in(0,h_{0})$
\[\mathcal{B}\Big(\lambda^\van_{n,0} + \lambda^\van_{n,1} h^{\frac{1}{k+2}},c_0h^{\frac{2}{k+2}}\Big)\cap \sp\left(\mathfrak{L}^{\van,[k]}_{h}\right)=\{\lambda^\van_{n}(h)\}\,,\]
and $\lambda^\van_{n}(h)$ is a simple eigenvalue.
\end{theo}
\begin{rem}
In fact, if $\gamma(s)^{-1}\gamma(0)-1$ is small enough (weak magnetic barrier), our construction of $\Phi$ can be made global, that is $\Vc=\R$. In this book, we will provide a proof of this theorem when $\gamma$ is a polynomial.
\end{rem}
We will prove Theorem \ref{WKB-van} in Chapter \ref{chapter-WKB}, Section \ref{WKB-vanishing}.
\subsection{Curvature induced magnetic bound states}
As we have seen, in many situations the spectral splitting appears in the second term of the asymptotic expansion of the eigenvalues. It turns out that we can also deal with more degenerate situations. The next lines are motivated by the initial paper \cite{HelMo01} whose main result is recalled in \eqref{HM-result}. Their fundamental result establishes that a smooth Neumann boundary can trap the lowest eigenfunctions near the points of maximal curvature. These considerations are generalized in \cite[Theorem 1.1]{FouHel06a} where the complete asymptotic expansion of the $n$-th eigenvalue of $\mathfrak{L}^\FH_{h,\A}=(-ih\nabla+\A)^2$ is provided and satisfies in particular:
\begin{equation}\label{splitting-FH}
\Theta_{0}h-C_{1}\kappa_{\max}h^{3/2}+(2n-1)C_{1}\Theta_{0}^{1/4}\sqrt{\frac{3k_{2}}{2}} h^{7/4}+o(h^{7/4})\,,
\end{equation}
where $k_{2}=-\kappa''(0)$. In this book, as in \cite{FouHel06a}, we will consider the magnetic Neumann Laplacian on a smooth domain $\Omega$ such that the algebraic curvature $\kappa$ satisfies the following assumption.
\begin{assumption}\label{kappa-max}
The function $\kappa$ is smooth and admits a unique and non-degenerate maximum.
\end{assumption}
In Chapter \ref{chapter-WKB}, Section \ref{WKB-FouHel} we prove that the lowest eigenfunctions are approximated by local WKB expansions which can be made global when for instance $\partial\Omega$ is the graph of a smooth function. In particular we recover the term $C_{1}\Theta_{0}^{1/4}\sqrt{\frac{3k_{2}}{2}}$ by a method different from the one of Fournais and Helffer and we explicitly provide a candidate to be the optimal distance of Agmon in the boundary. Since it is quite unusual to exhibit a pure magnetic Agmon distance, let us provide a precise statement. For that purpose, let us consider the following Neumann realization on $\sL^2(\R^2_{+}, m(s,t)\dx s\dx t)$, which is nothing but the expression of the magnetic Laplacian in curvilinear coordinates,
\begin{multline}
\mathcal{L}^{\FH}_{h}=m(s,t)^{-1}h D_{t}m(s,t)h D_{t}\\
+m(s,t)^{-1}\left(h D_{s}+\zeta_{0}h^{\frac 1 2}-t+\kappa(s)\frac{t^2}{2}\right)m(s,t)^{-1}\left(h D_{s}+\zeta_{0}h^{\frac 1 2}-t+\kappa(s)\frac{t^2}{2}\right),
\end{multline}
 where $m(s,t)=1-t\kappa(s)$. Thanks to the rescaling
\[t=h^{1/2}\tau,\qquad s=\sigma\,,\]
 and after division by $h$ the operator $\mathcal{L}^{\FH}_{h}$ becomes
\[\mathfrak{L}^{\FH}_{h}=m(\sigma,h^{1/2}\tau)^{-1}D_{\tau}m(\sigma,h^{1/2}\tau)D_{\tau}+m(\sigma,h^{1/2}\tau)^{-1}P_{h}m(\sigma,h^{1/2}\tau)^{-1}P_{h}\,,\]
on the space $\sL^2(m(\sigma,h^{1/2}\tau)\dx \sigma\dx\tau)$ and where
\[P_{h}=h^{1/2}D_{\sigma}+\zeta_{0}-\tau+h^{1/2}\kappa(\sigma)\frac{\tau^2}{2}\,.\]
\begin{theo}\label{WKB-FH}
Under Assumption \ref{alpha-max}, there exist a function 
\[\Phi=\Phi(\sigma)=\left(\frac{2C_{1}}{\nu_{1}''(\zeta_{0})}\right)^{1/2}\left|\int_{0}^\sigma (\kappa(0)-\kappa(s))^{1/2}\dx s\right|\]
defined in a neighborhood $\Vc$ of $(0,0)$ such that $\Re\Phi''(0)>0$, and a sequence of real numbers $(\lambda^{\FH}_{n,j})$ such that 
\[\lambda_{n}^{\FH}(h)\underset{h\to 0}{\sim} \sum_{j\geq 0} \lambda^{\FH}_{n,j} h^{\frac j 4}\,.\]
Besides there exists a formal series of smooth functions on $ \Vc$,
\[\an_{n}^{\FH} \underset{h\to 0}{\sim} \sum_{j\geq 0} \an^{\FH}_{n,j} h^{\frac j 4}\]
such that 
\[\left(\mathfrak{L}^{\FH}_{h}-\lambda_{n}^{\FH}(h)\right)\left(\an_{n}^{\FH} e^{-\Phi/h^{\frac14}}\right)=\mathcal{O}\left(h^{\infty} \right) e^{-\Phi/h^{\frac14}}.
\]
We also have that  $\lambda^{\FH}_{n,0}=\Theta_{0}$, $\lambda^{\FH}_{n,1}=0$, $\lambda^{\FH}_{n,2}=-C_{1}\kappa_{\max}$ and  $\lambda^{\FH}_{n,3}=(2n-1)C_{1}\Theta_{0}^{1/4}\sqrt{\frac{3k_{2}}{2}}$. The main term in the Ansatz is in the form
\[\an^{\FH}_{n,0}(\sigma,\tau)=f^\FH_{n,0}(\sigma)u_{\zeta_{0}}(\tau)\,.\]
Moreover, for all $n\geq 1$, there exist $h_{0}>0$, $c>0$ such that for all $h\in(0,h_{0})$,  we have
\[\mathcal{B}\Big(\lambda^{\FH}_{n,0}+ \lambda^{\FH}_{n,2}h^{1/2} + \lambda^{\FH}_{n,3}h^{\frac{ 3}{ 4}}, ch^{\frac 3 4}\Big)\cap \sp\left(\mathfrak{L}^\FH_{h}\right)=\{\lambda^{\FH}_{n}(h)\}\,,\]
and $\lambda^{\FH}_{n}(h)$ is a simple eigenvalue. 
\end{theo}
\begin{rem}
In particular, Theorem \ref{WKB-FH} proves that there are no odd powers of $h^{\frac{1}{8}}$ in the expansion of the eigenvalues (see \cite[Theorem 1.1]{FouHel06a}).
\end{rem}

\chapter{Magnetic wells in dimension two}\label{intro-van-birk}
\begin{flushright}
\begin{minipage}{0.5\textwidth}
Ce n'est pas assez d'avoir l'esprit bon, mais le principal est de l'appliquer bien.
\begin{flushright}
\textit{Discours de la m\'ethode}, Descartes
\end{flushright}
\end{minipage}
\vspace*{0.5cm}
\end{flushright}
This chapter is devoted to the semiclassical analysis with magnetic fields in dimension two in the following situations:
\begin{enumerate}[(i)]
\item the case when the magnetic field vanishes along a smooth curve,
\item the case when it does not vanish.
\end{enumerate}
Each situation leads to different semiclassical behaviors and technics:
\begin{enumerate}[(i)]
\item a dimensional reduction in the spirit of the Born-Oppenheimer approximation,
\item a semiclassical Birkhoff normal form.
\end{enumerate}

\section{Vanishing magnetic fields}\label{intro-vanishing}
In this section we study the influence of the cancellation of the magnetic field along a smooth curve in dimension two.
\subsection{Framework}
We consider a vector potential $\A\in\mathcal{C}^{\infty}(\R^2,\R^2)$ and we consider the self-adjoint operator on $\sL^2(\R^2)$ defined by:
\[\mathfrak{L}_{h,\A}=(-ih\nabla+\A)^2\,.\]
\begin{notation}
We will denote by $\lambda_{n}(h)$  the $n$-th eigenvalue of $\mathfrak{L}_{h,\A}$.
\end{notation}

\subsubsection{How does $\B$ vanish?}
In order $\mathfrak{L}_{h,\A}$ to have compact resolvent, we will assume that:
\begin{equation}\label{limit-beta}
\B(x)\underset{|x|\to+\infty}\to+\infty.
\end{equation}
As in \cite{PK02, HelKo09}, we will investigate the case when $\B$ cancels along a closed and smooth curve $\mathcal{C}$ in $\R^2$. We have already discussed the motivation in Chapter \ref{intro-models}, Section \ref{Sec.vmf}.
Let us notice that the assumption \eqref{limit-beta} could clearly be relaxed so that one could also consider a smooth, bounded and simply connected domain of $\R^2$ with Dirichlet or Neumann condition on the boundary as far as the magnetic field does not vanish near the boundary (in this case one should meet a model presented in Chapter \ref{intro-models}, Section \ref{Sec.vmf}). We let:
\[\mathcal{C}=\{\cc(s), s\in\R\}\,.\]
We assume that $\B$ is positive inside $\mathcal{C}$ and negative outside. We introduce the standard tubular coordinates $(s,t)$ near $\mathcal{C}$ defined by the map
\[(s,t)\mapsto\cc(s)+t\bfn(s)\,,\]
where $\bfn(s)$ denotes the inward pointing normal to $\mathcal{C}$ at $\cc(s)$.
The function $\tilde\B$ will denote $\B$ in the coordinates $(s,t)$, so that $\tilde\B(s,0)=0$.

\subsubsection{Heuristics and leading operator} Let us adopt first a heuristic point of view to introduce the leading operator of the analysis presented in this section. We want to describe the operator   $\mathfrak{L}_{h,\A}$ near the cancellation line of $\B$, that is near $\mathcal{C}.$ In a rough approximation, near $(s_{0},0)$, we can imagine that the line is straight ($t=0$) and that the magnetic field cancels linearly so that we can consider $\tilde\B(s,t)=\gamma(s_{0})t$ where $\gamma(s_{0})$ is the derivative of $\tilde\B$ with respect to $t$. Therefore the operator to which we are reduced at the leading order near $s_{0}$ is: 
\[h^2D_{t}^2+\left(hD_{s}-\gamma(s_{0})\frac{t^2}{2}\right)^2\,.\]
This operator is a special case of the larger class introduced in Chapter \ref{intro-models}, see also Chapter \ref{chapter-BOM}, Section \ref{BOM-mag}.

\subsection{Montgomery operator and rescaling}\label{subsec.Mont}
We will be led to use the Montgomery operator with parameters $\eta\in\R$ and $\gamma>0$:
\begin{equation}\label{H-eta-delta-vanishing}
\mathfrak{L}^{[1]}_{\gamma,\zeta}=D_{t}^2+\left(\zeta-\frac \gamma 2 t^2\right)^2\,.
\end{equation}
The Montgomery operator has clearly compact resolvent and we can consider its lowest eigenvalue denoted by $\nu^{[1]}_{1}(\gamma, \zeta)$. In fact one can take $\gamma=1$ up to the rescaling $t=\gamma^{-1/3}\tau$ and $\mathfrak{L}^{[1]}_{\gamma, \zeta}$ is unitarily equivalent to:
\[\gamma^{2/3}\left(D_{\tau}^2+(-\eta \gamma^{-1/3}+\frac{1}{2}\tau^2)^2\right)=\gamma^{2/3}\mathfrak{L}^{[1]}_{1,\zeta\gamma^{-1/3}}\,.\]
Let us emphasize that this rescaling is related with the normal form analysis that we will use in the semiclassical spectral asymptotics.
For all $\gamma>0$, we have (see Chapter \ref{intro-models}, Proposition \ref{Montgomery}):
\begin{equation}\label{nu1-min-vanishing}
\zeta\mapsto \nu^{[1]}_{1}(\gamma, \zeta) \mbox{ admits a unique and non-degenerate minimum at a point } \zeta^{[1]}_{0}(\gamma)\,.
\end{equation}
If $\gamma=1$, we have $\zeta^{[1]}_{0}(1)=\zeta^{[1]}_{0}$. We may write:
\begin{equation}\label{nu-delta-vanishing}
\inf_{\zeta\in\R}\nu^{[1]}_{1}(\gamma, \zeta)=\gamma^{2/3}\nu^{[1]}_{1}(\zeta^{[1]}_{0})\,.
\end{equation}
Let us recall some notation.
\begin{notation}
We notice that $\mathfrak{L}^{[1]}_{\zeta}=\mathfrak{L}^{[1]}_{1,\zeta}$ and we denote by $u^{[1]}_{\zeta}$ the $L^2$-normalized and positive eigenfunction associated with $\nu^{[1]}_{1}(\zeta)$.
\end{notation}
For fixed $\gamma>0$, the family $(\mathfrak{L}^{[1]}_{\gamma, \zeta})_{\eta\in\R}$ is an analytic family of type $(A)$ so that $(\nu^{[1]}_{1}(\zeta), u^{[1]}_{\zeta})$ has an analytic dependence on $\zeta$ (see  Chapter \ref{chapter-examples}, Section \ref{Sec.analytic} and also \cite{Kato66}).

\subsection{Semiclassical asymptotics with vanishing magnetic fields}
We consider the normal derivative of $\B$ on $\mathcal{C}$, i.e. the function $\gamma : s\mapsto\dr_{t}\tilde\B(s,0)$. We will assume the following.
\begin{hyp}\label{Assumption-vanishing}
$\gamma$ admits a unique, non-degenerate and positive minimum at $x_{0}$.
\end{hyp}
We let $\gamma_{0}=\gamma(0)$ and assume without loss of generality that $x_{0}=(0,0)$.
Let us state the main result of this section and proved in Chapter \ref{chapter-vanishing}.
\begin{theo}\label{main-result-vanishing}
We assume Assumption \ref{Assumption-vanishing}. For all $n\geq 1$, there exists a sequence $(\theta_{j}^n)_{j\geq 0}$ such that we have:
\[\lambda_{n}(h)\underset{h\to 0}{\sim} h^{4/3}\sum_{j\geq 0} \theta^n_{j} h^{j/6}\]
where:
\[\theta_{0}^{n}=\gamma_{0}^{2/3}\nu^{[1]}_{1}(\zeta^{[1]}_{0}), \quad \theta_{1}^n=0, \quad \theta_{2}^n=\gamma_{0}^{2/3}C_{0}+\gamma_{0}^{2/3}(2n-1)\left(\frac{\alpha\nu^{[1]}_{1}(\eta_{0})(\nu^{[1]}_{1})''(\zeta^{[1]}_{0})}{3}\right)^{1/2}\,,\]
where we have let
\begin{equation}\label{alpha-vanishing}
\alpha=\frac 1 2 \gamma_{0}^{-1}\gamma''(0)>0
\end{equation}
and
\begin{align}\label{C0-vanishing}
C_{0}=\langle L u^{[1]}_{\zeta_{0}^{[1]}},u^{[1]}_{\zeta_{0}^{[1]}}\rangle_{\sL^2(\R_{\hat\tau})},
\end{align}
where
\[L=2k(0)\gamma_{0}^{-4/3}\left(\frac{\hat\tau^2}{2}-\zeta^{[1]}_{0}\right)\hat\tau^3+2\hat\tau\gamma_{0}^{-1/3}\kappa(0)\left(-\zeta^{[1]}_{0}+\frac{\hat\tau^2}{2}\right)^2\,,\]
and
\[k(0)=\frac{1}{6}\dr^2_{t}\tilde{\B}(0,0)-\frac{\kappa(0)}{3}\gamma_{0}\,.\]
\end{theo}
\begin{rem}
This theorem is mainly motivated by the paper of Helffer and Kordyukov \cite{HelKo09} (see also \cite[Section 5.2]{Hel05} where the above result  is presented as a conjecture and the paper \cite{HelMo96} where the case of discrete wells is analyzed) where the authors prove a one term asymptotics for all the eigenvalues (see \cite[Corollary 1.1]{HelKo09}). Moreover, they also prove an accurate upper bound in \cite[Theorem 1.4]{HelKo09} thanks to a Grushin type method (see \cite{Gru72}). This result could be generalized to the case when the magnetic vanishes on hypersurfaces at a given order.
\end{rem}

\section{Non vanishing magnetic fields}\label{intro-birk}
As we will see, the result of Section \ref{intro-vanishing} is essentially a consequence of a normal form investigation. Other examples, in three dimensions, will be given in Chapter \ref{intro-semi}. For each example, we will introduce an appropriate change of variable or equivalently a \enquote{Fourier integral operator} and we will \emph{normalize} the magnetic Laplacian by transferring the magnetic geometry into the coefficients of the operator. We can interpret this normalization as a very explicit application of the Egorov theorem. Then, in the investigation, we are led to use the Feshbach projection to simplified again the situation. This projection method can also be heuristically interpreted as a normal form in the spirit of Egorov: taking the average of the operator in a certain quantum state is nothing but the quantum analog of averaging a full Hamiltonian with respect to a reduced Hamiltonian. In problems with boundaries or with vanishing magnetic fields it appears that the dynamics of the reduced Hamiltonian is less understood (due to the boundary conditions for instance) than the spectral theory of its quantization. Keeping this remark in mind it now naturally appears that we should implement a general normal form for instance in the simplest situation of dimension two, without boundary and with a non vanishing magnetic field.
\subsection{Classical dynamics}\label{sec.classical-dyn}

Let us recall a basic example from classical mechanics. After a normalization, Newton's equation of a mass on a spring is given by the Hook law (the classical harmonic oscillator)
\[\frac{d^2 q}{dt^2}=-q\,.\]
Of course, it can be easily solved, but it can also be put into the Hamiltonian form:
\[
\left\{\begin{array}{ccc}
\frac{d q}{dt}&=&\partial_{p}H\,,\\
\frac{d p}{dt}&=&-\partial_{q}H\,,
\end{array}\right.
\]
where $H(q,\xi)=\frac{1}{2}(q^2+p^2)$. Note that it is also the flow of the Hamiltonian vector field $\mathcal{X}_{H}$ defined by $dH=\omega_{0}(\mathcal{X}_{H},\cdot)$ where $\omega_{0}$ is the canonical symplectic form on $\R^2$, that is 
\[\forall (u,v)\in\R^2\times\R^2\,,\qquad\omega_{0}(u,v)=v_{1}u_{2}-u_{1}v_{2}\,.\]
If we let $z=q+ip$, we get $\frac{d z}{dt}=-iz$ and thus $z(t)=z_{0}e^{-it}$.

Let us now investigate the case of constant magnetic field in dimension two. Newton's equation is now
\begin{equation}\label{equ:newton}
\frac{d^2 \q}{dt^2}=\frac{d\q}{dt}\times\B\,,
\end{equation}
where $\B=B(0,0,1)=Be_{3}$ and where the right hand side is the Lorentz force. Here we have $\q=(q_{1}, q_{2}, 0)$. The equation becomes
\[
\left\{\begin{array}{ccc}
\frac{d q_{1}}{dt}&=&p_{1}\,,\\
\frac{d q_{2}}{dt}&=&p_{2}\,,\\
\frac{d p_{1}}{dt}&=&Bp_{2}\,,\\
\frac{d p_{2}}{dt}&=&-Bp_{1}\,.\\
\end{array}\right.
\]
The last two equations are in a Hamiltonian form, as for the harmonic oscillator and we let $v=p_{1}+ip_{2}$ so that the evolution of the velocity is given by $v(t)=v(0)e^{-iBt}$. Letting $q=q_{1}+iq_{2}$, it follows that $\frac{dq}{dt}=v(0)e^{-iBt}$ and thus
\[q(t)=q(0)-\frac{i}{B}v(0)+\frac{i}{B}v(0)e^{-iBt}\,.\]
The particle rotates at a distance (the cyclotron radius) $\frac{|v(0)|}{|B|}$ of the center $q(0)-\frac{i}{B}v(0)$. The frequency of the rotation is $B$ so that the large field limit is also a high frequency regime (the semiclassical regime).

In fact, the general equation \eqref{equ:newton} may be put in a Hamiltonian form. To see this, we introduce $\A\in \mathcal{C}^\infty(\R^d,\R^d)$ (the source of the magnetic field) such that
\[\B = \dx \A\,,\]
where we used the identification 
\[\A=\sum_{j=1}^d A_j\dx q_j\,.\]
We recall that Equation \eqref{equ:newton} may also be put in the more general form
\begin{equation}\label{equ:newton'}
\frac{d^2 q}{dt^2}=-M_{\B}\left(\frac{dq}{dt}\right)\,,
\end{equation}
where $M_{\B}$ is the (antisymmetric) magnetic matrix $(B_{k\ell})$. The matrix $M_{\B}$ is also the antisymmetric part of the differential (not to confuse with the exterior derivative $\dx\A$) $d\A$:
\[M_{\B}=d\A-(d\A)^{\mathsf{T}}\,.\]
Thus Equation \eqref{equ:newton'} becomes 
\[\frac{d^2 q}{dt^2}+d\A\left(\frac{dq}{dt}\right)=(d\A)^{\mathsf{T}}\left(\frac{dq}{dt}\right)\,.\]
If we let $\xi=\frac{dq}{dt}$, this becomes
\[\frac{d}{dt}\left(\xi+\A\left(q\right)\right)=(d\A)^{\mathsf{T}}\left(\xi\right)\,,\]
and we get the new system
\[
\left\{\begin{array}{ccc}
\frac{dq}{dt}&=&p-\A\,,\\
\frac{dp}{dt}&=&(d\A)^{\mathsf{T}}\left(p-\A\right)\,.
\end{array}\right.
\]
It is easy to see that the
Hamiltonian of our system is
\begin{equation}
  \frac{1}{2}\norm{p-\A(q)}^2\,.
  \label{equ:hamiltonian}
\end{equation}

\subsection{Classical magnetic normal forms}
From now on we use the Euclidean norm on $\R^2$, which allows the
identification of $\R^2$ with $(\R^2)^*$ by
\begin{equation}
  \forall (v,p)\in\R^2\times (\R^2)^*, \qquad p(v) = \pscal{p}{v}\,.
  \label{equ:identification}
\end{equation}
Thus, the canonical symplectic structure $\omega$ on $T^*\R^2$ is
given by
\begin{equation}
  \omega_{0}((Q_1,P_1), (Q_2,P_2)) = \pscal{P_1}{Q_2} - \pscal{P_2}{Q_1}\,.
  \label{equ:omega}
\end{equation}
Before considering the semiclassical magnetic Laplacian we shall briefly discuss some results concerning the classical dynamics for large time. We will not discuss the proofs in this book, but these considerations will give some insights to answer the semiclassical questions. As we have already mentioned in the introduction, the large time dynamics problem has to face the issue that the conservation of the energy $H$ is not enough to confine the trajectories in a compact set .

\begin{figure}[h]
  \centering
  \includegraphics[width=0.5\textwidth]{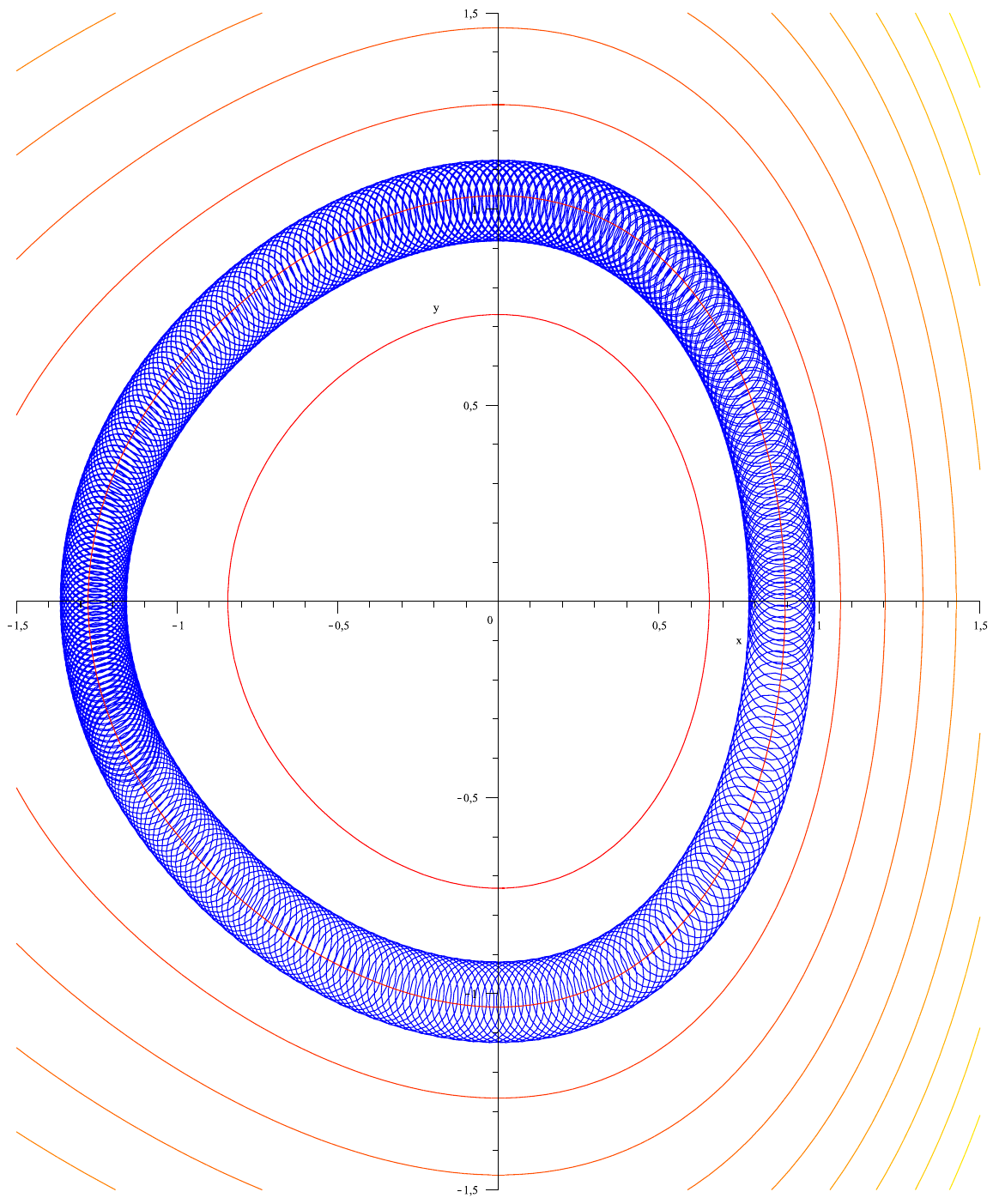}
  \caption{Numerical simulation of the flow of $H$ when the magnetic
    field is given by
    $B(x,y)=2+x^2+y^2+\frac{x^3}3+\frac{x^4}{20}$, and
    $\epsilon=0.05$, $t\in[0,500]$. The picture also displays in red some
    level sets of $B$. Graph courtesy of S. V\~u Ng\d{o}c}
  \label{fig:numerics}
\end{figure}

The first result (see Chapter \ref{chapter-birk} for a proof) shows the existence of a smooth symplectic
diffeomorphism that transforms the initial Hamiltonian into a normal
form, up to any order in the distance to the zero energy surface.
\begin{theo}
  \label{theo:classical}
  Let
  \[
  H(q,p)=\norm{p-\A(q)}^2, \quad (q,p)\in T^* \R^2 =
  \R^2\times \R^2,
  \]
  where the magnetic potential $\A:\R^2\to\R^2$ is smooth. Let
  $B:=\deriv{A_2}{q_1} - \deriv{A_1}{q_2}$ be the corresponding
  magnetic field.  Let $\Omega\subset\R^2$ be a bounded open set where $B$
  does not vanish. Then there exists a symplectic diffeomorphism
  $\Phi$, defined in an open set
  $\tilde{\Omega}\subset\C_{z_1}\times\R^2_{z_2}$, with values in $T^*\R^2$,
  which sends the plane $\{z_1=0\}$ to the surface $\{H=0\}$, and such
  that
  \begin{equation}
    H\circ \Phi = \abs{z_1}^2 f(z_2,\abs{z_1}^2) +
    \mathcal{O}(\abs{z_1}^\infty)\,,
    \label{equ:forme-normale}  
  \end{equation}
  where $f:\R^2\times\R \to \R$ is smooth.  Moreover, the map
  \begin{equation}
    \varphi: \Omega \ni q \mapsto  \Phi^{-1}(q,\A(q)) \in
    (\{0\} \times \R^2_{z_2} ) \cap \tilde\Omega
    \label{equ:phy}
  \end{equation}
  is a local diffeomorphism and
  \[
  f\circ(\varphi(q),0) = \abs{B(q)}\,.
  \]
\end{theo}
In the following theorem we denote by $K=\abs{z_1}^2
f(z_2,\abs{z_1}^2) \circ\Phi^{-1}$ the (completely integrable) normal
form of $H$ given be Theorem~\ref{theo:classical} above. Let
$\varphi_H^t$ be the Hamiltonian flow of $H$, and let $\varphi_K^t$ be the
Hamiltonian flow of $K$. Let us state, without proofs, the important dynamical consequences of Theorem \ref{theo:classical} (see Figure~\ref{fig:numerics}).
\begin{theo}
  \label{theo:confining}
  Assume that the magnetic field $B>0$ is confining: there exists
  $C>0$ and $M>0$ such that $B(q)\geq C$ if $\norm{q}\geq M$. Let $C_0
  < C$. Then 
  \begin{enumerate}[(i)]
  \item The flow $\varphi_H^t$ is uniformly bounded for all starting
    points $(q,p)$ such that $B(q)\leq C_0$ and
    $H(q,p)=\mathcal{O}(\epsilon)$ and for times of order
    $\mathcal{O}(1/\epsilon^N)$, where $N$ is arbitrary.
  \item Up to a time of order
    $T_\epsilon=\mathcal{O}(\abs{\ln\epsilon})$, we have
    \begin{equation}
      \norm{\varphi_H^t(q,p) - \varphi_K^t(q,p)} = \mathcal{O}(\epsilon^\infty)
      \label{equ:two-flows}  
    \end{equation}
    for all starting points $(q,p)$ such that $B(q)\leq C_0$ and
    $H(q,p)=\mathcal{O}(\epsilon)$.
  \end{enumerate}
\end{theo}
It is interesting to notice that, if one restricts to regular values
of $B$, one obtains the same control for a much longer time, as stated
below.
\begin{theo}\label{theo:confining2}
  Under the same confinement hypothesis as
  Theorem~\ref{theo:confining}, let $J\subset(0,C_0)$ be a closed
  interval such that $dB$ does not vanish on $B^{-1}(J)$. Then up to a
  time of order $T=\mathcal{O}(1/\epsilon^N)$, for an arbitrary $N>0$,
  we have
  \[
  \norm{\varphi_H^t(q,p) - \varphi_K^t(q,p)} = \mathcal{O}(\epsilon^\infty)
  \]
  for all starting points $(q,p)$ such that $B(q)\in J$ and
  $H(q,p)=\mathcal{O}(\epsilon)$.
\end{theo}

\subsection{Semiclassical magnetic normal forms}
We turn now to the quantum counterpart of these results.  Let
$\mathfrak{L}_{h,\A}=(-ih\nabla-\A)^2$ be the magnetic Laplacian on
$\R^2$, where the potential $\A:\R^2\to\R^2$ is smooth, and
such that $\mathfrak{L}_{h,\A}\in S(m)$ for some order function $m$ on
$\R^4$ (see Chapter \ref{chapter-appendix} for a brief reminder and \cite[Chapter 7]{DiSj99}).  We will work with the Weyl
quantization; for a classical symbol $a=a(x,\xi)\in S(m)$ , it is
defined as:
$$\Op_{h}^w a\, \psi(x)=\frac{1}{(2\pi h)^2}\int\int e^{i(x-y)\cdot\xi/h} a\left(\frac{x+y}{2},\xi\right)\psi(y)\dx y\dx \xi\,,\quad \forall \psi\in\mathcal{S}(\R^2)\,.$$

The first result (see Chapter \ref{chapter-birk}, Sections \ref{BNF}, \ref{spec}) shows that the spectral theory of
$\mathfrak{L}_{h,\A}$ is governed at first order by the magnetic field
itself, viewed as a symbol.
\begin{theo}\label{spectrum}
  Assume that the magnetic field $B$ is non vanishing on $\R^2$ and confining:
  there exist constants $\tilde{C}_1>0$, $M_0>0$ such that
  \begin{equation}\label{conf}
  B(q)\geq \tilde{C}_{1} \quad \text{ for } \quad |q|\geq M_{0}\,.
  \end{equation} 
  Let $\mathcal{H}^0_h=\Op_{h}^w (H^0)$, where
  $H^0=B(\varphi^{-1}(z_{2}))|z_{1}|^2$ where $\varphi:\R^2\to\R^2$ is a diffeomorphism.  Then there exists a bounded classical
  pseudo-differential operator $Q_h$ on $\R^2$, such that
  \begin{enumerate}[(i)]
  \item  $Q_h$ commutes with $\Op_{h}^w(\abs{z_1}^2)$;
  \item $Q_h$ is relatively bounded with respect to
    $\mathcal{H}^0_h$ with an arbitrarily small relative bound;
  \item its Weyl symbol is $\mathcal O_{z_2}(h^2+h\abs{z_1}^2+\abs{z_1}^4)$,
  \end{enumerate}
so that
  the following holds. 
  Let $0<C_1<\tilde{C}_1$. Then the spectra of $\mathfrak{L}_{h,\A}$
  and $\mathcal{L}^{\nor}_{h}:=\mathcal{H}^0_h+ Q_h$ in $(-\infty,C_1h]$
  are discrete. We denote by $0<\lambda_1(h)\leq \lambda_2(h)\leq
  \cdots$ the eigenvalues of $\mathfrak{L}_{h,\A}$ and by
  $0<\mu_1(h)\leq \mu_2(h)\leq \cdots$ the eigenvalues of
  $\mathcal{L}^{\nor}_{h}$. Then for all $j\in\N^*$ such that
  $\lambda_j(h)\leq C_1h$ and $\mu_j(h)\leq C_1h$, we have
\[
\abs{\lambda_j(h) - \mu_j(h)} = \mathcal{O}(h^\infty)\,.
\]
\end{theo}
As we see in the proof, Theorem \ref{spectrum} is a consequence of the following theorem (see \cite{I98} where a close form of this theorem appears),
which provides in particular an accurate description of $Q_h$. In the
statement, we use the notation of Theorem~\ref{theo:classical}. We
recall that $\Sigma$ is the zero set of the classical Hamiltonian $H$.
\begin{theo}\label{main-theo-Bir}
   For $h$ small enough there exists a unitary operator
  $U_h$ such that
%\[
%U_h^* U_h = \Id + Z_h, \qquad U_h U_h^* = \Id + Z'_h\,,
%\]
%where $Z_h, Z'_h$ are pseudo-differential operators that
%microlocally vanish in a neighborhood of $\tilde\Omega\cap\Sigma$, and
\begin{equation}\label{formal}
  U_h^* \mathfrak{L}_{h,\A} U_h =  \mathcal{L}^\nor_h +
  R_h+S_{h}\,,
 \end{equation}
  where
  \begin{enumerate}[(i)]
  \item $\mathcal{L}^{\nor}_h$ is a classical pseudo-differential operator in $S(m)$
    that commutes with 
    \[ 
    \mathcal{I}_{h}:= -h^2\frac{\partial^2}{\partial x_1^2} +x_1^2\,
    \]
  \item \label{item:hermite} For any Hermite function $ e_{n,h}(x_1)$ such that
    $\mathcal{I}_{h}  e_{n,h} = h(2n-1) e_{n,h}$, the operator $\mathcal{L}^{\nor, (n)}_h$
    acting on $\sL^2(\R_{x_2})$ by
    \[
    e_{n,h}\otimes \mathcal{L}^{\nor, (n)}_h (u) = \mathcal{L}^{\nor}_h(e_{n, h}\otimes u)
    \]
    is a classical pseudo-differential operator in $S_{\R^2}(m)$ of $h$-order 1 with
    principal symbol
    \[
    F^{(n)}(x_2,\xi_2) = h(2n-1)B(q)\,,
    \]
    where $(0,x_2+ i \xi_2)=\varphi(q)$ as in~\eqref{equ:phy};
  \item \label{item:R} the pseudo-differential operators $R_{h}$ and $S_{h}$ have a symbols in $S(m)$. The Taylor series of the symbol of $R_{h}$ with respect to $(x_{1}, \xi_{1}, h)$ vanishes in  a neighborhood of $\Sigma$ and the symbol of $S_{h}$ vanishes in  a neighborhood of $\tilde\Omega\cap\Sigma$.
  \item \label{item:Q} $\mathcal{L}^{\nor}_h= \mathcal{H}_{h}^0+
    Q_h$, where $\mathcal{H}^0_h=\Op_{h}^w (H^0)$, $H^0=B(\varphi^{-1}(
    z_{2}))|z_{1}|^2$, and the operator $Q_h$ is relatively bounded
    with respect to $\mathcal{H}^0_h$ with an arbitrarily small
    relative bound.

  \end{enumerate}
\end{theo} 
We recover the result of \cite{HelKo11}, adding the fact that no odd
power of $h^{\frac 1 2}$ can show up in the asymptotic expansion (see the recent work \cite{HelKo13b} where a Grushin type method is used to obtain a close result).
\begin{cor}[Low lying eigenvalues]
\label{cor:low}
  Assume that $B$ has a unique non-degenerate minimum at $q_{0}$. Then there
  exists a constant $c_0$ such that for any $j$, the eigenvalue
  $\lambda_j(h)$ has a full asymptotic expansion in integral powers
  of $h$ whose first terms have the following form:
\[
\lambda_j(h) \sim h\min B + h^2(c_1(2j-1)+c_0) + \mathcal{O}(h^3)\,,
\]
with
$c_1=\frac{\sqrt{\det(\Hess_{q_{0}}B)}}{2b_{0}}$, where $b_{0}=B(q_{0})$.
\end{cor}
\begin{proof}
  The first eigenvalues of $\mathfrak{L}_{h,A}$ are equal to the eigenvalues of $\mathcal{L}_h^{\nor, (1)}$ (in point~\eqref{item:hermite}
  of Theorem~\ref{main-theo-Bir}). Since $B$ has a non-degenerate minimum,
  the symbol of  $\mathcal{L}_h^{\nor, (1)}$ has a non-degenerate minimum, and the
  spectral asymptotics of the low-lying eigenvalues for such a 1D
  pseudo-differential operator are well known. We get
  \[
  \lambda_j(h) \sim h\min B + h^2(c_1(2j-1)+c_0) + \mathcal{O}(h^3)\,,
  \]
  with $c_1=\frac{1}{2}\sqrt{\det(\Hess_{0}(B\circ \varphi^{-1})}$. One can easily
  compute
  \[
  c_1 =
  \frac{\sqrt{\det\Hess_{q_{0}} B}}{2\abs{\det(d\varphi^{-1}(0))}}
  = \frac{\sqrt{\det\Hess_{q_{0}} B}}{2B\circ\varphi^{-1}(0)}\,,
  \]
  where we used the definition of $\varphi$ in~\eqref{equ:phy} (it is a diffeomorphism that transforms the 2-form $B\dx q_{1}\wedge \dx q_{2}$ into $\dx q_{1}\wedge \dx q_{2}$).
\end{proof}

\chapter{Boundary magnetic wells in dimension three}\label{intro-semi}

\begin{flushright}
\begin{minipage}{0.50\textwidth}
Now do you imagine he would have attempted to inquire or learn what he thought he knew, when he did not know it, until he had been reduced to the perplexity of realizing that he did not know, and had felt a craving to know?
\begin{flushright}
\textit{Meno}, Plato
\end{flushright}
\vspace*{0.5cm}
\end{minipage}
\end{flushright}

In this chapter we enlighten the normal form philosophy explained in Chapter \ref{intro}, Section \ref{orga} by presenting three results of \emph{magnetic harmonic approximation} induced by the presence of a boundary in dimension three:
\begin{enumerate}[(i)]
\item when the boundary is a half-space,
\item when it is a wedge,
\item when it is a cone.
\end{enumerate}
We will see that the semiclassical structures are different from each other.

\section[Magnetic half-space]{Magnetic half-space}\label{intro-variable3D}
This section is devoted to the investigation of the relation between a smooth (Neumann) boundary and the magnetic field in dimension three. 

\subsection{A toy model}
Let us introduce the geometric domain
\[\Omega_0=\{(x,y,z)\in\R^3 : |x|\leq x_0,\quad |y|\leq y_0\quad \mbox{ and }\quad 0<z\leq z_0\}\,,\]
where $x_0,y_0,z_0>0$. The part of the boundary which carries the Dirichlet condition is given by
\[\dr_{\Dir}\Omega_{0}=\{(x,y,z)\in\Omega_{0} : |x|=x_{0}\mbox{ or } |y|=y_{0} \mbox{ or } z=z_{0}\}\,.\]
\subsubsection{Definition of the operator}
For $h>0$, $\alpha\geq 0$ and $\theta\in\left(0,\frac{\pi}{2}\right)$, we consider the self-adjoint operator:
\begin{equation}\label{op}
\mathfrak{L}_{h,\alpha,\theta}=h^2 D_y^2+h^2 D_z^2+(hD_x+z\cos\theta-y\sin\theta+\alpha z(x^2+y^2))^2\,,
\end{equation}
with domain
\begin{align*}
\Dom(\mathfrak{L}_{h,\alpha,\theta}) =
\{\psi\in \sL^2(\Omega_{0}) : \quad&\mathfrak{L}_{h, \alpha,\theta}\psi\in\sL^2(\Omega_{0}),\\
&\psi=0 \ \mbox{ on } \ \partial_\Dir\Omega_{0}
\quad\mbox{and} \quad \partial_{z}\psi=0 \mbox{ on } z=0 \}.
\end{align*}
Since $\alpha$ and $\theta$ are fixed, we let $\mathfrak{L}_{h}=\mathfrak{L}_{h, \alpha,\theta}$. The vector potential is expressed as
\[\A(x,y,z)=(V_{\theta}(y,z)+\alpha z(x^2+y^2),0,0)\]
where 
\begin{equation}\label{Vtheta}
V_{\theta}(y,z)=z\cos\theta-y\sin\theta\,.
\end{equation}
The associated magnetic field is given by
\begin{equation}\label{magnetic}
\nabla\times\A=\B=(0,\cos\theta+\alpha(x^2+y^2),\sin\theta-2\alpha yz)\,.
\end{equation}
In particular $\theta$ is the angle between $\B(0,0,0)$ and the Neumann boundary $z=0$.
\subsubsection{Constant magnetic field ($\alpha=0$)}
Let us examine the case of constant magnetic field. In this case, we have
\[\mathfrak{L}_{h,0,\theta}=h^2D_y^2+h^2 D_z^2+(hD_x+V_\theta(y,z))^2\,,\]
viewed as an operator on $\sL^2(\R^3_{+})$.
We perform the rescaling:
\begin{equation}\label{scaling-variable3D}
x=h^{\frac 1 2}r,\quad y=h^{\frac 1 2}s,\quad z=h^{\frac 1 2}t
\end{equation}
and the operator becomes (after division by $h$):
\[\mathfrak{L}_{1,0,\theta}=D_s^2+D_t^2+(D_r+V_\theta(s,t))^2\,.\]
Making a Fourier transform in the variable $r$ denoted by $\mathcal{F}_{r\to\eta}$, we get
\begin{equation}\label{Fourier}
\mathcal{F}_{r\to\eta}\mathfrak{L}_{1,0,\theta}\mathcal{F}_{r\to\eta}^{-1}=D_s^2+D_t^2+(\eta+V_\theta(s,t))^2\,.
\end{equation}
Then, we use a change of coordinates:
\begin{equation}\label{Trans}
U_{\theta}(\rho,s,t)=\left(\rho,\sigma,\tau\right)=\left(\eta,s-\frac{\eta}{\sin\theta},t\right)
\end{equation}
and we obtain
\[\mathfrak{H}^{\Neu}_{\theta}=U_{\theta}\mathcal{F}_{r\to\eta}\mathfrak{L}_{1,0,\theta}\mathcal{F}_{r\to\eta}^{-1}U_{\theta}^{-1}=D_{\sigma}^2+D_{\tau}^2+V_\theta(\sigma,\tau)^2\,.\]
\begin{notation}
We denote by $\mathfrak{Q}^\Neu_{\theta}$ the quadratic form associated with $\mathfrak{H}^{\Neu}_{\theta}$.
\end{notation}
The operator $\mathfrak{H}^{\Neu}_{\theta}$ viewed as an operator acting on $\sL^2(\R^2_{+})$ is nothing but $\mathfrak{L}^{\LP}_{\theta}$ (see Chapter \ref{intro}, Section \ref{intro:Lu-Pan}). Let us also recall that the lower bound of the essential spectrum is related, through the Persson's theorem (see Chapter \ref{chapter-spectral-theory}), to the following estimate:
\[\mathfrak{q}^{\LP}_\theta(\chi_R u)\geq (1-\eps(R))\|\chi_R u\|^2,\quad\forall u\in\Dom(\mathfrak{q}^{\LP}_{\theta})\,,\]
where $\mathfrak{q}^{\LP}_{\theta}$ is the quadratic form associated with $\mathfrak{L}^{\LP}_{\theta}$, where $\chi_R$ is a cutoff function away from the ball $B(0,R)$ and $\eps(R)$ is tending to zero when $R$ tends to infinity. Moreover, if we consider the Dirichlet realization $\mathfrak{L}^{\LP, \Dir}_{\theta}$, we have
\begin{equation}\label{ineqDir}
\mathfrak{q}^{\LP,\Dir}_{\theta}(u)\geq \|u\|^2,\quad\forall u\in\Dom(\mathfrak{q}^{\LP, \Dir}_{\theta})\,.
\end{equation}

\subsection{A \enquote{generic} model}
Let us explain why we are led to consider this model. Let us introduce the variable angle $\theta(x,y)$ that is the angle of $\B(x,y,0)$ with the boundary $z=0$ and defined by the relation
\[\|\B(x,y,0)\|\sin\theta(x,y)=\B(x,y,0)\cdot\bfn(x,y)\,.\]
If we make the approximation of the magnetic field by the constant magnetic field near the boundary, the Lu-Pan operator $\mathfrak{L}^{\LP}_{\theta}$ appears and this leads to introduce
\[\B_{\mathfrak{s}}(x,y)=\mathfrak{s}(\theta(x,y))\|\B(x,y,0)\|\,,\]
where $\bfn(x,y)$ is the inward normal at $(x,y,0)$. It is proved in \cite{LuPan00a} that the semiclassical asymptotics of the lowest eigenvalue is given by
\[\lambda_{1}(h)=\min\left\{\inf \B_{\mathfrak{s}},\inf_{\Omega_{0}}\|\B\|\right\} h+o(h)\,.\]
We are interested in the case when the following generic assumption is satisfied.
\begin{assumption}
We assume that we are in the case of \enquote{boundary attraction}:
\begin{equation}\label{A1}
\inf \B_{\mathfrak{s}}<\inf_{\Omega_{0}}\|\B\|\,.
\end{equation}
and in the case of \enquote{boundary magnetic well}:
\begin{equation}\label{A2}
\B_{\mathfrak{s}} \mbox{ admits a unique and non degenerate minimum.}
\end{equation}
\end{assumption}
Under these assumptions, a three terms upper bound is proved for $\lambda_{1}(h)$ in \cite{Ray10c} and the corresponding lower bound, for a general domain, is still an open problem. 

For $\alpha>0$, the toy operator (\ref{op}) is the simplest example of Schr\"odinger operator with variable magnetic field satisfying Assumptions (\ref{A1}) and (\ref{A2}). 
We have the Taylor expansion:
\begin{equation}\label{nondeg}
\B_{\mathfrak{s}}(x,y)=\mathfrak{s}(\theta)+\alpha C(\theta)(x^2+y^2)+\mathcal{O}(|x|^3+|y|^3)\,.
\end{equation}
with
\[C(\theta)=\cos\theta\,\mathfrak{s}(\theta)-\sin\theta\, \mathfrak{s}'(\theta)\,.\]
Moreover, it is proved in Chapter \ref{chapter-models}, Proposition \ref{Ctheta>0} that $C(\theta)>0$, for $\theta\in\left(0,\frac{\pi}{2}\right)$. Thus, Assumption (\ref{A2}) is verified if $x_{0}$, $y_{0}$ and $z_{0}$ are fixed small enough. Using $\mathfrak{s}(\theta)<1$ when $\theta\in\left(0,\frac{\pi}{2}\right)$ and $\|\B(0,0,0)\|=1$, we get Assumption (\ref{A1}).

\subsubsection{Remark on the function $\B_{\mathfrak{s}}$}
Using the explicit expression of the magnetic field, we have
\[\B_{\mathfrak{s}}(x,y)=\B_{\mathfrak{s}, \rad}(R), \quad R=\alpha(x^2+y^2)\]
and an easy computation gives
\[\B_{\mathfrak{s}, \rad}(R)=\|\B_{\rad}(R)\|\mathfrak{s}\left(\arctan\left(\frac{\sin\theta}{\cos\theta+R}\right)\right)\,,\]
with 
\[\|\B_{\rad}(R)\|=\sqrt{(\cos\theta+R)^2+\sin^2\theta}\,.\]
The results of Chapter \ref{chapter-models} imply that $\B_{\mathfrak{s}, \rad}$ is strictly increasing and 
\[{\dr_{R}\B_{\mathfrak{s}, \rad}(R=0)=C(\theta)>0}\,.\]
Consequently, $\B_{\mathfrak{s}}$ admits a unique and non degenerate minimum on $\R^3_{+}$ and tends to infinity far from $0$.
This is easy to see that
\[\inf_{\R^3_{+}}\|\B\|=\cos\theta\,.\]
We deduce that, as long as $\mathfrak{s}(\theta)<\cos\theta$, the generic assumptions are satisfied with $\Omega_{0}=\R_{+}^3$.

\subsubsection{Three dimensional magnetic wells induced by the magnetic field and the (smooth) boundary}
Let us introduce the fundamental operator 
\[\mathfrak{S}_{\theta}(D_{\rho},\rho)=\left(2\int_{\R^2_{+}} \tau V_{\theta} (u^\LP_{\theta})^2\dx \sigma \dx \tau\right)\mathcal{H}_{\harm}+\left(\frac{2}{\sin\theta}\int_{\R^2_{+}} \tau V_{\theta}(u^\LP_{\theta})^2\dx\sigma \dx\tau\right)\rho+d(\theta)\,,\]
where 
\[\mathcal{H}_{\harm}=D_{\rho}^2+\frac{\rho^2}{\sin^2\theta}\]
and
\[d(\theta)=\sin^{-2}\theta\langle \tau(D_{\sigma}^2V_{\theta}+V_{\theta}D_{\sigma}^2)u^\LP_{\theta},u^\LP_{\theta}\rangle+2\int_{\R^2_{+}} \tau\sigma^2V_{\theta}(u^\LP_{\theta})^2\dx \sigma \dx \tau\,.\]
By using the perturbation theory, we can establish the following formula (see \cite[Formula (2.31)]{Ray10c}):
\[2\int_{\R^2_{+}} \tau V_{\theta} (u_{\theta}^\LP)^2\dx \sigma\dx \tau=C(\theta)>0\,,\]
so that $\mathfrak{S}_{\theta}(D_{\rho},\rho)$ can be viewed as the harmonic oscillator up to a dilation and translations.

We can now state the main result of this section: a complete semiclassical expansion of the $n$-th eigenvalue. The proof is given in Chapter \ref{chapter-variable3D}.
\begin{theo}\label{maintheo-variable3D}
For all $\alpha>0$, $\theta\in\left(0,\frac{\pi}{2}\right)$, there exist a sequence $(\mu_{j,n})_{j\geq 0}$ and $\eps_{0}>0$ such that, for $|x_{0}|+|y_{0}|+|z_{0}|\leq \eps_{0}$, 
\[\lambda_{n}(h)\sim h\sum_{j\geq 0} \mu_{j,n}h^j\]
and we have $\mu_{0,n}=\mathfrak{s}(\theta)$ and $\mu_{1,n}$ is the $n$-th eigenvalue of $\alpha\mathfrak{S}_{\theta}(D_{\rho},\rho)$.
\end{theo}

\section{Magnetic wedge}\label{intro-edge}
We analyze here the effect of an edge in the boundary and how its combines with the magnetic field to produce a spectral asymptotics. 
\subsection{ Geometry and local models}
 \label{SS:geometry}
We consider the magnetic Laplacian on a wedge of aperture $\alpha$, denoted by $\mathcal{W}_{\alpha}$. In our situation the magnetic field $ {\B}=(0,0,1)$ is normal to the plane where the edge lies. 

Here we are concerned with the case when the domain is a wedge with varying aperture, that is with the Neumann magnetic Laplacian $\mathfrak{L}^\PR_{h,\A}=(-ih\nabla+\A)^2$ on the space $\sL^2(\mathcal{W}_{s\mapsto \alpha(s)},\dx s \dx t \dx z)$.

\subsubsection{Properties of the magnetic wedge}\label{SS:descriptionlens}
Let us recall the definition of the magnetic wedge with constant aperture $\alpha$. Many properties of this operator can be found in the thesis of Popoff \cite{Popoff}.
We let
\[\mathcal{W}_{\alpha}=\R\times \mathcal{S}_{\alpha}\,,\]
where the bidemensional corner with fixed angle $\alpha\in (0,\pi)$ is defined by
\[\mathcal{S}_{\alpha}=\left\{( t, z)\in \R^2 : |z|< t\tan\left(\frac{\alpha}{2}\right)\right\}\,.\]
\begin{definition}
\label{D:opmodel}
Let $\mathfrak{L}^\Po_{\alpha}$ be the Neumann realization on $\sL^2(\mathcal{W}_{\alpha}, \dx s \dx t \dx z)$ of
\begin{equation}
\label{D:Lgot}
D_{t}^2+D_{z}^2+(D_{s}- t)^2\,.
\end{equation}
We denote by $\nu^\Po_{1}(\alpha)$ the bottom of the spectrum of $\mathfrak{L}^\Po_{\alpha}$.
\end{definition}
Using the Fourier transform with respect to $\hat s$, we have the decomposition:
\begin{equation}
\label{E:rel3d2d}
\mathfrak{L}^\Po_{\alpha}=\int^{\oplus} \mathfrak{L}^\Po_{\alpha,\zeta}\dx\zeta\,,
\end{equation}
where $\mathfrak{L}^\Po_{\alpha,\zeta}$ is the \Bk following Neumann realization on $\sL^2(\mathcal{S}_{\alpha}, \dx t \dx z)$:
\begin{equation}
\label{D:Lmodel}
\mathfrak{L}^\Po_{\alpha,\zeta}=D_{t}^2+D_{z}^2+(\zeta- t)^2\,,
\end{equation}
where $\zeta\in\R$ is the Fourier parameter. As 
$$\lim_{\substack{|(t,z)|\to +\infty \\ (t,z)\in \mathcal{S}_{\alpha}}}(\zeta-t)^2=+\infty , $$
the Schr¬\"odinger operator $\mathfrak{L}^\Po_{\alpha,\zeta}$ has compact resolvent for all $(\alpha,\zeta)\in (0,\pi)\times \R$.
\begin{notation}
For each $\alpha\in(0,\pi)$, we denote by $\nu^\Po_{1}(\alpha,\eta)$ the lowest eigenvalue of $\mathfrak{L}^\Po_{\alpha,\zeta}$ and we denote by $u^\Po_{\alpha,\zeta}$ a normalized corresponding eigenfunction.
\end{notation} 
Using \eqref{E:rel3d2d} we have: 
\begin{equation}
\label{E:rel3d2dinf}
\nu_{1}^\Po(\alpha)=\inf_{\zeta\in\R}\nu_{1}^\Po(\alpha,\zeta)\,.
\end{equation}
Let us gather a few elementary properties.
\begin{lem}
We have:
\begin{enumerate}
\item For all $(\alpha,\zeta)\in(0,\pi)\times \R$, $\nu_{1}^\Po(\alpha,\zeta)$ is a simple eigenvalue of $\mathfrak{L}^\Po_{\alpha,\zeta}$.
\item The function $(0,\pi)\times\R\ni(\alpha,\zeta)\mapsto\nu_{1}^\Po(\alpha,\zeta)$ is analytic.
\item
\label{It:decroit1} For all $\zeta\in\R$, the function $(0,\pi)\ni\alpha\mapsto\nu_{1}^\Po(\alpha,\zeta)$ is decreasing.
\item
\label{It:decroit2} 
 The function $(0,\pi)\ni\alpha\mapsto\nu_{1}^\Po(\alpha)$ is non increasing.
\item For all $\alpha\in (0,\pi)$, we have
\begin{equation}
\label{E:calcullimite}
\lim_{\eta\to-\infty} \nu_{1}^\Po(\alpha,\zeta)=+\infty \quad \text{and} \quad \lim_{\zeta\to+\infty}\nu_{1}^\Po(\alpha,\zeta)=\mathfrak{s}(\tfrac{\pi-\alpha}{2})\,.
\end{equation}
\end{enumerate}
\end{lem}
\begin{proof}
We refer to \cite[Section 3]{Popoff} for the first two statements. The monoticity comes from \cite[Proposition 8.14]{Popoff} and the limits as $\zeta$ goes to $\pm\infty$ are computed in \cite[Theorem 5.2]{Popoff}.
\end{proof}
\begin{rem}
\label{R:compnutheta}
As $\nu_{1}^\Po(\pi)=\Theta_{0}$, we have: 
\begin{equation}
\label{E:compnuettheta0}
\forall\alpha\in (0,\pi), \quad \nu_{1}^\Po(\alpha)\geq \Theta_{0}\,.
\end{equation}
Let us note that it is proved in \cite[Proposition 8.13]{Popoff} that $\nu_{1}^\Po(\alpha)>\Theta_{0}$ for all $\alpha\in(0,\pi)$. 
\end{rem}
\begin{prop}\label{prop:comparaisonnusigma}
There exists $\tilde{\alpha}\in (0,\pi)$ such that for $\alpha\in(0,\tilde{\alpha})$, the function $\zeta \mapsto \nu^\Po_{1}(\alpha,\zeta)$ reaches its infimum and  
\begin{equation}
\label{E:comparaisonnusigma}
\nu_{1}^\Po(\alpha) < \mathfrak{s}\left(\frac{\pi-\alpha}{2}\right)\,,
\end{equation}
where the spectral function $\mathfrak{s}$ is defined in Chapter \ref{intro}, Section \ref{intro:Lu-Pan}.
\end{prop}
\begin{rem}
Numerical computations show that in fact \eqref{E:comparaisonnusigma} seems to hold for all $\alpha\in(0,\pi)$.
\end{rem}
We will work under the following conjecture: 
\begin{conj}\label{min-nu-edge}
For all $\alpha\in(0,\pi)$, $\zeta\mapsto \nu_{1}^\Po(\alpha,\zeta)$ has a unique critical point denoted by $\zeta^\Po_{0}(\alpha)$ and it is a non degenerate minimum.
\end{conj}
\begin{rem}
A numerical analysis seems to indicate that Conjecture \ref{min-nu-edge} is true (see \cite[Subsection 6.4.1]{Popoff}).
\end{rem}
Under this conjecture and using the analytic implicit functions theorem, we deduce the following lemma.
\begin{lem}
Under Conjecture \ref{min-nu-edge}, the function $(0,\pi)\ni\alpha\mapsto\zeta^\Po_{0}(\alpha)$ is analytic and so is $(0,\pi)\ni\alpha\mapsto\nu_{1}^\Po(\alpha)$. Moreover the function $(0,\pi)\ni\alpha\mapsto\nu_{1}^\Po(\alpha)$ is decreasing.
\end{lem}
We will assume that there is a unique point of maximal aperture (which is non-degenerate).
\begin{assumption}\label{alpha-max}
The function $s\mapsto \alpha(s)$ is analytic and admits a unique and non-degenerate maximum $\alpha_{0}$ at $s=0$.
\end{assumption}
\begin{notation}
We let $\mathcal{T}(s)=\tan\left(\frac{\alpha(s)}{2}\right)$.
\end{notation}

\subsubsection{Assumptions}\label{leading-operator}
For $\x\in \partial\Omega\setminus E$ we introduce the angle $\theta(\x)$ defined by:
\begin{equation}
\label{D:theta(x)}
\B\cdot \n(\x)=\sin\theta(\x)\,.
\end{equation}
We have 
 \begin{equation} 
 \label{E:casmodele}
 \forall \x\in\partial\Omega\setminus E, \quad \frac{\pi-\alpha}{2} < \theta(\x)  
 \end{equation}
 where $\alpha\in(0,\pi)$ is the opening angle of the lens and we notice that the magnetic field is nowhere tangent to the boundary. We will assume that the opening angle of the lens is variable. For a given point $\x$ of the boundary, we analyze the localized (in a neighborhood of $\x$) magnetic Laplacian and we distinguish between $\x$ belonging to the edge and $\x$ belonging to the smooth part of the boundary.

\begin{figure}[ht]
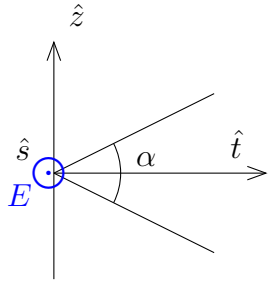

\begin{center}

   \figinit{pt}
\figpt 1:(0,0)
\figpt 11:(0,6)
\figpt 12:(28,0)
\figpt 2:(30,0)
\figpt 20:(80,0)
\figpt 3:(0,50)
\figpt 30:(0,-40)
\figpt 4:(60,30)
\figpt 5:(60,-30)

\psbeginfig{}
\psline[1,4]
\psline[1,5]
\psarrow [30,3]
\psarrow[1,20]
\psarccircP 1;25 [5,4]
\psendfig

\figvisu{\figBoxB}{}
{
\figwrites11:$\Bl \bigodot \Bk$ (0.001)
\figwritenw1 :$\hat{s}$ (7)
\figwritesw1 :$\Bl E$ (6)
\figwritene3 :$\hat{z}$ (8)
\figwritenw20 :$\hat{t}$ (8)
\figwritene12:$\alpha$ (4)
}

\centerline{\box\figBoxA\hfil\box\figBoxB}
\caption{Coordinates $(\hat{s},\hat{t},\hat{z})$.}
\label{F:lentilledemihauteur}
\end{center}
\end{figure}

The model situations (magnetic wedge and smooth boundary) lead to compare the following quantities
\[\inf_{\x\in  E} \nu^\Po_{1}(\alpha(\x)),\quad \inf_{\x\in\partial\Omega\setminus E}\mathfrak{s}_{1}(\theta(\x))\,.\]
Let us state the different assumptions under which we work. The first assumption could be called the \enquote{edge concentration} assumption.
\begin{hyp}\label{comparison}
\begin{equation}
\label{H:comparaisonspectre}
\inf_{\x\in  E} \nu_{1}^\Po(\alpha(\x))<\quad \inf_{\x\in\partial\Omega\setminus E}\mathfrak{s}_{1}(\theta(\x))\,.
\end{equation}
\end{hyp}
From the properties of the leading operator we will be led to work near the point of the edge of maximal opening. Therefore we will assume the following generic assumption.
\begin{hyp}\label{generic-assumption}
We denote by $\alpha:E \mapsto (0,\pi)$ the opening angle of the lens. We assume that $\alpha$ admits a unique and non degenerate maximum  at the point $\x_{0}$ and we let
\[\alpha_{0}=\max_{E}\alpha \,.\]
We denote $\mathcal{T}=\tan\frac{\alpha}{2}$ and $\mathcal{T}_{0}=\tan\frac{\alpha_{0}}{2}$.
\end{hyp}
In particular, under this assumption and Conjecture \ref{min-nu-edge}, the function $s\mapsto \nu_{1}^\Po(\alpha(s))$ admits a unique and non-degenerate minimum.

\subsection{Normal form}
This is \enquote{classical} that Assumption \ref{comparison} leads to localization properties of the eigenfunctions near the edge $E$ and more precisely near the points of the edge where $E\ni\x\mapsto \nu(\alpha(\x))$ is minimal. Therefore, since $\nu$ is decreasing and thanks to Assumption \ref{generic-assumption}, we expect that the first eigenfunctions concentrate near the point $\x_{0}$ where the opening is maximal. 

Let us write below the expression of the magnetic Laplacian in the new local coordinates $(\check s, \check t, \check z)$ where $\check{s}$ is a curvilinear abscissa of the edge. The normal form of the magnetic Laplacian $\mathfrak{L}^\lens_{h}$ is given by $\check{\mathfrak{L}}^\lens_{h}:=\check\nabla_{h}^2$ where:
\begin{equation}
\label{D:gradientcheck}
\check\nabla_{h}=\begin{pmatrix}    
hD_{\check s}\\
hD_{\check t}\\
h\mathcal{T}(\check s)^{-1}\mathcal{T}(0)D_{\check z}
\end{pmatrix}+
\begin{pmatrix}    
-\check t+\zeta^\Po_{0}h^{1/2}-h\frac{\mathcal{T}'}{2\mathcal{T}}(\check z D_{\check z}+D_{\check z}\check z)\\
0\\
0
\end{pmatrix}\,.
\end{equation}

\begin{rem}
Such a normal form allows us to describe the leading structure of this magnetic Laplace-Beltrami operator. Indeed, modulo some remainders, our operator takes the simpler form:
\[(hD_{\check s}-\check t+\zeta^\Po_{0}h^{1/2})^2+h^2D_{\check t}^2+h^2\mathcal{T}(0)^{2}\mathcal{T}(\check s)^{-2}D_{\check z}^2\,.\]
Performing another formal Taylor expansion near $\check s=0$, we are led to the following operator:
\[(hD_{\check s}-\check t+\zeta^\Po_{0}h^{1/2})^2+h^2D_{\check t}^2+h^2D_{\check z}^2+ch^2\check s^2D_{\check z}^2\,,\]
where $c>0$.
Using a scaling, we get a rescaled operator $\mathcal{L}_{h}$ whose first term is the leading operator $\mathfrak{L}^\Po_{\alpha_{0}}$ and which allows to construct quasimodes. Moreover this form is suitable to establish microlocalization properties of the eigenfunctions with respect to $D_{\check{s}}$.
\end{rem}
\subsection{Magnetic wells induced by the variations of a singular geometry}
The main result of this section is a complete asymptotic expansion of all the first eigenvalues of $\mathfrak{L}^\lens_{h}$ (see the proof in Chapter \ref{chapter-edge}).
\begin{theo}\label{main-result-edge}
We assume that Conjecture \ref{min-nu-edge} is true. We also assume Assumptions \ref{comparison} and \ref{generic-assumption}. For all $n\geq 1$ there exists $(\mu_{j,n})_{j\geq0}$ such that we have
\[\lambda_{n}(h)\underset{h\to 0}{\sim}h\sum_{j\geq 0}\mu_{j,n}h^{j/4}\,.\]
Moreover, we have
\[\mu_{0,n}=\nu^\Po_{1}(\alpha_{0}), \quad \mu_{1,n}=0,\quad \mu_{2,n}=(2n-1)\sqrt{\kappa \tau_{0}^{-1}\|D_{\hat z}u^\Po_{\zeta^\Po_{0}}\|^2 \dr_{\zeta}^2\nu^\Po_{1}(\alpha_{0},\zeta^\Po_{0})}\,.\]
where
\begin{equation}\label{kappa}
\kappa=-\frac{\mathcal{T}''(0)}{2}>0\,.
\end{equation}
\end{theo}

\begin{rem}
We observe that, for all $n\geq 1$, $\lambda_{n}(h)$ is simple for $h$ small enough. This simplicity, jointly with a quasimodes construction, also provides an approximation of the corresponding normalized eigenfunction.
\end{rem}

\section{Magnetic cone}\label{Sec.pop}
\begin{figure}[h!tb]
\begin{center}
\includegraphics[height=6cm]{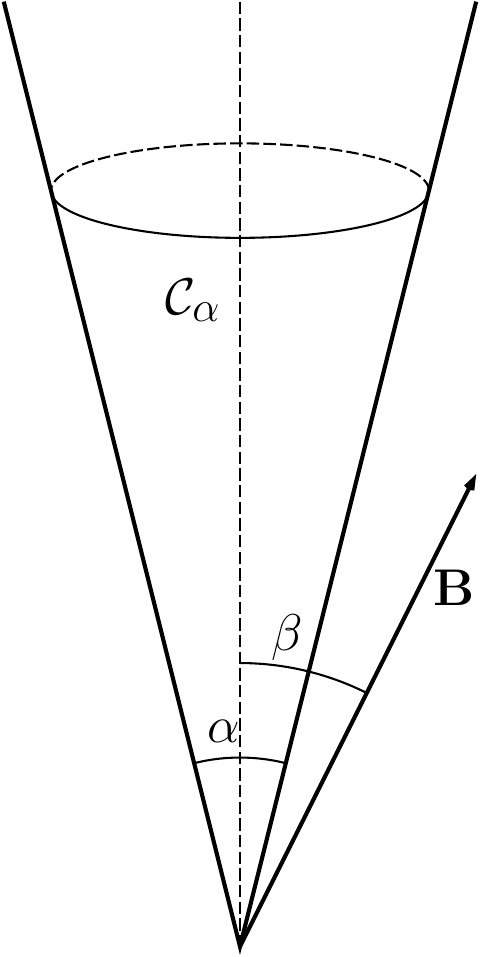}
\caption{Geometric setting.\label{fig.cone}}
\end{center}
\end{figure}

We are now interested in the low-lying eigenvalues of the magnetic Neumann Laplacian with a constant magnetic field applied to a \enquote{ peak }, i.e. a right circular cone $\Ca$. The right circular cone $\Ca$ of angular opening $\alpha\in\left(0,\pi\right)$ (see Figure~\ref{fig.cone}) is defined  in the Cartesian coordinates $(x,y,z)$ by
\[\Ca=\{(x,y,z)\in\R^3,\ z>0,\ x^2+y^2< z^2\tan^2\a\}\,.\]
\noindent Let ${\bf B}$ be the constant magnetic field
\[{\bf B}(x,y,z)=(0,\sin\beta,\cos\beta)^{\sf T}\,,\]
where $\beta\in\left[0,\frac{\pi}{2}\right]$.
We choose the following magnetic potential $\A$:
\[\A(x,y,z) =\frac{1}{2}{\bf{B}}\times {\bf{x}}=\frac 12(z\sin\beta-y\cos\beta,x\cos\beta,-x\sin\beta)^{\sf T}\,.\]
We consider $\mathfrak{L}_{\alpha,\beta}$ the Friedrichs extension associated with the quadratic form 
\[\mathcal{Q}_{\A}(\psi)=\|(-i\nabla+\A)\psi\|^2_{\sL^2(\Ca)}\,,\]
defined for $\psi\in \sH^1_{\A}(\Ca)$ with
\[\sH^1_{\A}(\Ca)=\{u\in \sL^2(\Ca), (-i\nabla+\A)u\in \sL^2(\Ca)\}\,.\]
The operator $\mathfrak{L}_{\alpha}$ is $(-i\nabla+\A)^2$ with domain:
\[\sH^2_{\A}(\Ca)=\{u\in \sH^1_{\A}(\Ca), (-i\nabla+\A)^2 u\in \sL^2(\Ca), (-i\nabla+\A)u\cdot\nu=0\, \mbox{ on } \dr\Ca\}\,.\]
Note that, here, we have $h=1$: we are easily reduced to this case by homogeneity. Thus there is no semiclassical effect and the only parameter with which we can play is $\alpha$.
We define the $n$-th eigenvalue $\lambda_{n}(\alpha,\beta)$ of $\mathfrak{L}_{\alpha,\beta}$ as the $n$-th Rayleigh quotient (see Chapter \ref{chapter-spectral-theory}).
Let $\psi_{n}(\alpha,\beta)$ be a normalized associated eigenvector (if it exists).

\subsection{Why studying magnetic cones?}
One of the most interesting results of the last fifteen years is provided by Helffer and Morame in \cite{HelMo01} where they prove that the magnetic eigenfunctions, in 2D, concentrates near the points of the boundary where the (algebraic) curvature is maximal, see \eqref{HM-result}. This property aroused interest in domains with corners, which somehow correspond to points of the boundary where the curvature becomes infinite (see \cite{Jad01,Pan02} for the quarter plane and \cite{Bon05, BD06} for more general domains). Denoting by $\mathcal{S}_{\alpha}$ the sector in $\R^2$ with angle $\alpha$ and considering the magnetic Neumann Laplacian with constant magnetic field of intensity $1$, it is proved in \cite{Bon05} that, as soon as $\alpha$ is small enough, a bound state exists. Its energy is denoted by $\mu(\alpha)$. An asymptotic expansion at any order is even provided (see \cite[Theorem 1.1]{Bon05}):
\begin{equation}\label{secteur}
\mu(\alpha)\sim\alpha\sum_{j\geq 0} m_{j}\alpha^{2j},\qquad \mbox{Êwith }\quad m_{0}=\frac{1}{\sqrt{3}}\,.
\end{equation}
In particular, this proves that $\mu(\alpha)$ becomes smaller than the lowest eigenvalue of the magnetic Neumann Laplacian in the half-plane with constant magnetic field (of intensity $1$), that is:
\[\mu(\alpha)<\Theta_{0},\qquad \alpha\in(0,\alpha_{0})\,,\]
where $\Theta_{0}$ is defined in \eqref{Theta0}.This motivates the study of dihedral domains (see \cite{Popoff, Po13}). Another possibility of investigation, in dimension three, is the case of a conical singularity of the boundary. We would especially like to answer the following questions: Can we go below $\mu(\alpha)$  and can we describe the structure of the spectrum when the aperture of the cone goes to zero?

\subsection{The magnetic Laplacian in spherical coordinates}\label{intro-conical}
Since the spherical coordinates are naturally adapted to the geometry, we consider the change of variable:
\[\Phi(\tau,\theta,\varphi):=(x,y,z)=\alpha^{-1/2}(\tau\cos\theta\sin\alpha\varphi,\ \tau\sin\theta\sin\alpha\varphi,\ \tau\cos\alpha\varphi)\,.\]
This change of coordinates is nothing but a first normal form. We denote by $\mathcal{P}$ the semi-infinite rectangular parallelepiped
\[\mathcal{P}:=\{(\tau,\theta,\varphi)\in\R^3,\ \tau>0,\ \theta\in [0,2\pi),\ \varphi\in(0,\tfrac12)\}\,.\]
Let $\psi\in\sH^1_{\A}(\Ca)$. We write $\psi(\Phi(\tau,\theta,\varphi))= \alpha^{1/4}\tilde\psi(\tau,\theta,\varphi)$ for any $(\tau,\theta,\varphi)\in\mathcal{P}$ in these new coordinates and we have
\[\|\psi\|^2_{\sL^2(\Ca)}=
\int_{\mathcal{P}}|\tilde\psi(\tau,\theta,\varphi)|^2\, \tau^2 \sin\alpha\varphi \dx \tau \dx\theta \dx\varphi\,,\]
and:
\[ \mathfrak{Q}_{\A}(\psi)  =\alpha \mathcal{Q}_{\alpha,\beta}(\tilde\psi)\,,\]
where the quadratic form $\mathcal{Q}_{\alpha,\beta}$ is defined on the transformed form domain $\sH^1_{\tilde\A}(\mathcal{P})$ by
\begin{equation}\label{eq.Qalpha3D}
\mathcal{Q}_{\alpha,\beta}(\psi):=\int_{\mathcal{P}} \left(|P_{1}\psi|^2+|P_{2}\psi|^2+|P_{3}\psi|^2\right)\dx\tilde\mu\,,
\end{equation}
with the measure 
\[\dx\tilde\mu=  \tau^2\sin\alpha\varphi \dx t \dx\theta \dx\varphi\,,\]
and:
\[\sH^1_{\tilde\A}(\mathcal{P}) =\{\psi\in \sL^2(\mathcal{P},\dx\tilde\mu), (-i\nabla+\tilde\A)\psi\in\sL^2(\mathcal{P},\dx\tilde\mu)\}\,.\]
We also have:
\begin{align*}
&P_{1}=D_{\tau}-\tau\varphi\cos\theta\sin\beta)\tau^2(D_{\tau}-\tau\varphi\cos\theta\sin\beta\,,\\
&P_{2}=(\tau\sin(\alpha\varphi))^{-1}\left(D_{\theta}+\frac{\tau^2}{2\alpha}\sin^2(\alpha\varphi)\cos\beta
+\frac{\tau^2\varphi}{2}\left(1-\frac{\sin(2\alpha\varphi)}{2\alpha\varphi}\right)\sin\beta\sin\theta\right)\,,\\
&P_{3}=(\tau\sin(\alpha\varphi))^{-1}D_{\varphi}\,.
\end{align*}
We consider $\mathcal{L}_{\alpha,\beta}$ the Friedrichs extension associated with the quadratic form $\mathcal{Q}_{\alpha,\beta}$:
\begin{eqnarray*}
\mathcal{L}_{\alpha,\beta}
&=& \tau^{-2}(D_{\tau}-\tau\varphi\cos\theta\sin\beta)\tau^2(D_{\tau}-\tau\varphi\cos\theta\sin\beta)\\
&&+\frac{1}{\tau^2\sin^2(\alpha\varphi)}\left(D_{\theta}+\frac{\tau^2}{2\alpha}\sin^2(\alpha\varphi)\cos\beta
+\frac{\tau^2\varphi}{2}\left(1-\frac{\sin(2\alpha\varphi)}{2\alpha\varphi}\right)\sin\beta\sin\theta\right)^2\\
&&+\frac{1}{\alpha^2 \tau^2\sin(\alpha\varphi)}D_{\varphi}\sin(\alpha\varphi) D_{\varphi}\,.
\end{eqnarray*}

We define $\tilde\lambda_{n}(\alpha,\beta)$ the $n$-th Rayleigh quotient of $\mathcal{L}_{\alpha,\beta}$.

\subsection{Spectrum of the magnetic cone in the small angle limit}\label{intro-cone-asymp}
\subsubsection{Eigenvalues in the small angle limit}
We aim at estimating the discrete spectrum, if it exists, of $\mathfrak{L}_{\alpha,\beta}$. For that purpose, we shall first determine the bottom of its essential spectrum. From Persson's characterization of the infimum of the essential spectrum, it is enough to consider the behavior at infinity and it is possible to establish the following proposition.
\begin{prop}\label{inf-ess-sp}
Let us denote by $\spe(\mathfrak{L}_{\alpha,\beta})$ the essential spectrum of $\mathfrak{L}_{\alpha,\beta}$. We have:
\[\inf\spe(\mathfrak{L}_{\alpha,\beta})\in[\Theta_{0},1]\,,\]
where $\Theta_{0}>0$ is defined in \eqref{Theta0}.
\end{prop}
At this stage we still do not know that discrete spectrum exists. As it is the case in dimension two (see \cite{Bon05}) or in the case on the infinite wedge (see \cite{Popoff}), there is hope to prove such an existence in the limit $\alpha\to 0$. Here is the main theorem of this section (see Chapter \ref{chapter-cones} for elements of the proof and the papers \cite{BR12, BR13b} for all the details).

\begin{theo}\label{main-theo-cone}
For all $n\geq 1$, there exist $\alpha_{0}(n)>0$ and a sequence $(\gamma_{j,n})_{j\geq 0}$ such that, for all $\alpha\in(0,\alpha_{0}(n))$, the $n$-th eigenvalue exists and satisfies:
\[\lambda_{n}(\alpha,\beta)\underset{\alpha\to 0}{\sim}\alpha\sum_{j\geq 0}  \gamma_{j,n} \alpha^{j},\qquad
\mbox{ with }\quad \gamma_{0,n}=\frac{\sqrt{1+\sin^2\beta}}{2^{5/2}} (4n-1)\,.\]
\end{theo}

\begin{rem}
In particular the main term is minimum when $\beta=0$ and in this case Theorem \ref{main-theo-cone} states that $\lambda_{1}(\alpha)\sim \frac{3}{2^{5/2}}\alpha$. We have $ \frac{3}{2^{5/2}}<\frac{1}{\sqrt{3}}$ so that the lowest eigenvalue of the magnetic cone goes below the lowest eigenvalue of the two dimensional magnetic sector (see \eqref{secteur}). 
\end{rem}

\begin{rem}
As a consequence of Theorem \ref{main-theo-cone}, we deduce that the lowest eigenvalues are simple as soon as $\alpha$ is small enough. Therefore, the spectral theorem implies that the quasimodes constructed in the proof are approximations of the eigenfunctions of $\mathcal{L}_{\alpha,\beta}$. In particular, using the rescaled spherical coordinates, for all $n\geq 1$, there exist $\alpha_{n}>0$ and $C_{n}$ such that, for $\alpha\in(0,\alpha_{n})$:
\[\|\tilde\psi_{n}(\alpha,\beta)-\mathfrak{f}_{n}\|_{\sL^2(\mathcal{P},\dx\tilde\mu)}\leq C_{n}\alpha^2\,,\]
where $\mathfrak{f}_{n}$ (which is $\beta$ dependent) is related to the $n$-th Laguerre's function and $\tilde\psi_{n}(\alpha,\beta)$ is the $n$-th normalized eigenfunction.
\end{rem}
Let us now sketch the proof of Theorem \ref{main-theo-cone}.
\subsubsection{Axissymmetric case: $\beta=0$}
We apply the strategy presented in Chapter \ref{intro}, Section \ref{orga}. In this situation, the phase variable that we should understand is the dual variable of $\theta$ given by a Fourier series decomposition and denoted by $m\in\Z$. In other words, we make a Fourier decomposition of $\mathcal{L}_{\alpha,0}$ with respect to $\theta$ and we introduce the family of 2D-operators $(\mathcal{L}_{\alpha,0,m})_{m\in\Z}$ acting on $\sL^2(\mathcal{R}, \dx\mu)$:
\[
\mathcal{L}_{\alpha,0,m}
=-\frac{1}{\tau^{2}}\dr_{\tau}\tau^2\dr_{\tau}
+\frac{1}{\tau^2\sin^2(\alpha \varphi)}\left(m+\frac{\sin^2(\alpha\varphi)}{2\alpha}\tau^2\right)^2
-\frac{1}{\alpha^2\ \tau^2 \sin(\alpha\varphi)}\dr_{\varphi}\sin(\alpha\varphi)\dr_{\varphi}
\,,\]
with
\[\mathcal{R} = \{(\tau,\varphi)\in\R^2,\ \tau>0,\ \varphi\in(0,\tfrac12)\}\,,\]
and
\[\dx\mu=\tau^2 \sin(\alpha\varphi)\dx \tau \dx \varphi\,.\]
We denote by $\mathcal{Q}_{\alpha,0,m}$ the quadratic form associated with $\mathcal{L}_{\alpha,0,m}$. This  normal form is also the suitable form  to construct quasimodes.
Then an integrability argument proves that the eigenfunctions are microlocalized in $m=0$, i.e. they are axisymmetric. Thus this allows a first reduction of dimension. It remains to notice that  the last term in $\mathcal{L}_{\alpha,0,0}$ is penalized by $\alpha^{-2}$ so that the Feshbach-Grushin projection on the groundstate of $-{\alpha^{-2} (\sin(\alpha\varphi))^{-1}}\dr_{\varphi}\sin(\alpha\varphi)\dr_{\varphi}$ (the constant function) acts as an approximation of the identity on the eigenfunctions. Therefore the spectrum of $\mathcal{L}_{\alpha,0,0}$ is described modulo lower order terms by the spectrum of the average of $\mathcal{L}_{\alpha,0}$ with respect to $\varphi$ which involves the so-called Laguerre operator (radial harmonic oscillator).

\subsubsection{Case $\beta\in\left[0,\frac{\pi}{2}\right]$}
In this case we cannot use the axisymmetry, but we are still able to construct formal series and prove localization estimates of Agmon type. Moreover we notice that the magnetic momentum with respect to $\theta$ is strongly penalized by  $(\tau^2\sin^2(\alpha\varphi))^{-1}$ so that, jointly with the localization estimates it is possible to prove that the eigenfunctions are asymptotically independent from $\theta$ and we are reduced to the situation $\beta=0$.

\chapter{Waveguides}\label{intro-wg}

\begin{flushright}
\begin{minipage}{0.50\textwidth}
Si on me presse de dire pourquoi je l'aimais, je sens que cela ne se peut exprimer qu'en r\'epondant : Parce que c'\'etait lui : parce que c'\'etait moi.
\begin{flushright}
\textit{Les Essais}, Livre I, Chapitre XXVIII, Montaigne
\end{flushright}
\vspace*{0.5cm}
\end{minipage}
\end{flushright}

This chapter presents recent progress in the spectral theory of waveguides. In Section~\ref{intro-intro-wg} we describe magnetic waveguides in dimensions two and three and we analyze the spectral influence of the width $\eps$ of the waveguide and the intensity $b$ if the magnetic field. In particular we investigate the limit $\eps\to 0$. In Section \ref{intro-wg-layers} we describe the same problem in the case of layers. In Sections \ref{intro-wg-triangles} and \ref{intro-wg-broken} the effect of a corner in dimension two is tackled.

\section{Magnetic waveguides}\label{intro-intro-wg}

This section is concerned with spectral properties of 
a curved quantum waveguide when a magnetic field is applied. We will give a precise definition of what a waveguide is in Sections \ref{intro-wg2} and \ref{intro-wg3}. Without going into the details we can already mention that we will use the definition given in the famous (non magnetic) paper of Duclos and Exner \cite{Duclos95} and its generalizations \cite{Duclos05, KK05, FK08}. The waveguide is nothing but a tube $\Omega_{\eps}$ about an unbounded curve $\gamma$ in the Euclidean space $\R^d$, with $d \geq 2$, where $\eps$ is a positive shrinking parameter and the cross section is defined as $\eps\omega=\{\eps\tau: \tau \in \omega\}$.

More precisely this section is devoted to the spectral analysis of the magnetic operator with Dirichlet boundary conditions $\mathfrak{L}_{\eps,b\A}^{[d]}$ defined as
\begin{equation}\label{Laplace}
  (-i\nabla_{x}+b\;\!\A(x))^2
  \qquad \mbox{on} \qquad
  \sL^2(\Omega_{\eps},\dx x)\,.
\end{equation}
where $b>0$ is a positive parameter and $\A$ a smooth vector potential
associated with a given magnetic field~$\B$.

\subsection{The result of Duclos and Exner}\label{DE}
One of the deep facts which is proved by Duclos and Exner is that the Dirichlet Laplacian on $\Omega_{\eps}$ always has discrete spectrum 
below its essential spectrum when the waveguide is not straight and asymptotically straight. Let us sketch the proof of this result in the case of two dimensional waveguides.

Let us consider a smooth and injective curve $\gamma$: $\R\ni s\mapsto \gamma(s)$ which is parameterized by its arc length $s$. The normal 
to the curve at $\gamma(s)$ is defined as the unique unit vector $\n(s)$ such that $\gamma'(s)\cdot\nu(s)=0$ and $\det(\gamma',\nu)=1$. We have the relation $\gamma''(s)=\kappa(s)\n(s)$ where $\kappa(s)$ denotes the algebraic curvature at the point $\gamma(s)$. We can now define standard tubular coordinates. We consider:
$$\R\times(-\eps,\eps)\ni(s,t)\mapsto\Phi(s,t)=\gamma(s)+t\n(s)\,.$$
We always assume 
\begin{equation}\label{not-overlap}
\mbox{$\Phi$ is injective}
\qquad\mbox{and}\qquad 
\eps\sup_{s\in\R} |\kappa(s)|<1\,.
\end{equation}
Then it is well known (see \cite{KK05}) that $\Phi$ defines a smooth diffeomorphism from $\R\times(-\eps,\eps)$ 
onto the image $\Omega_{\eps}=\Phi(\R\times(-\eps,\eps))$,
which we identify with our waveguide. In these new coordinates, the operator becomes (exercise)
$$\mathfrak{L}_{\eps,0}^{[2]}=-m^{-1}\dr_{s}m^{-1}\dr_{s}-m^{-1}\dr_{t}m\dr_{t},\quad m(s,t)=1-t\kappa(s)\,,$$
which is acting in the weighted space $\sL^2(\R\times(-\eps,\eps),m(s,t)\dx s \dx t)$. 
We introduce the shifted quadratic form:
$$\mathcal{Q}_{\eps,0}^{[2],\shi}(\phi)=\int_{\R\times(-\eps,\eps)}\left(m^{-2}|\dr_{s}(\phi)|^2+|\dr_{t}\phi|^2-\frac{\pi^2}{4\eps^2}|\phi|^2\right)m\dx s \dx t$$
and we let:
$$\phi_{n}(s,t)=\chi_{0}(n^{-1}s)\cos\left(\frac{\pi}{2\eps}t\right),$$
where $\chi_{0}$ is a smooth cutoff function which is $1$ near $0$. We can check that $\mathcal{Q}_{\eps,0}^{[2],\shi}(\phi_{n})\underset{n\to+\infty}{\to} 0$. Let us now consider a smooth cutoff function $\chi_{1}$ which is $1$ near a point where $\kappa$ is not zero and define $\tilde\phi(s,t)=-\chi^2_{1}(s,t)\mathcal{L}_{\eps,0}^{[2],\shi}\phi_{n}(s,t)$ which does not depend on $n$ as soon as $n$ is large enough. Then we have:
$$\mathcal{Q}_{\eps,0}^{[2],\shi}(\phi_{n}+\eta\tilde\phi)=\mathcal{Q}_{\eps,0}^{[2],\shi}(\phi_{n})-2\eta \mathcal{B}_{\eps,0}^{[2],\shi}(\phi_{n},\chi_{1}(s)\mathcal{L}_{\eps,0}^{[2],\shi}\phi_{n})+\eta^2 \mathcal{Q}_{\eps,0}^{[2],\shi}(\tilde\phi)\,.$$
For $n$ large enough, the quantity $\mathcal{B}_{\eps,0}^{[2],\shi}(\phi_{n},\chi_{1}(s)\mathcal{L}_{\eps,0}^{[2],\shi}\phi_{n})$ does not depend on $n$ and is positive. For such an $n$, we take $\eta$ small enough and we find:
$$\mathcal{Q}_{\eps,0}^{[2],\shi}(\phi_{n}+\eta\tilde\phi)<0\,.$$
Therefore the bottom of the spectrum is an eigenvalue due to the min-max principle.

Duclos and Exner also investigate the limit $\eps\to0$ to show that the Dirichlet Laplacian on the tube $\Omega_{\eps}$ converges in a suitable sense to the effective one dimensional operator
\begin{equation*}
  \mathcal{L}^{{\eff}}=-\dr_{s}^2-\frac{\kappa(s)^2}{4}
  \qquad \mbox{on} \qquad
  \sL^2(\gamma,\dx s)\,.
\end{equation*}
In addition it is proved in \cite{Duclos95} that each eigenvalue of this effective operator generates an eigenvalue of the Dirichlet Laplacian on the tube. 

As Duclos and Exner we are interested in approximations of $\mathfrak{L}_{\eps,b\A}^{[d]}$ in the small cross section limit $\eps\to 0$. Such an approximation might non trivially depends on the intensity of the magnetic field $b$ especially if it is allowed to depend on $\eps$. 
\subsection{Waveguides with more geometry}
In dimension three it is also possible to twist the waveguide 
by allowing the cross section of the waveguide to non-trivially 
rotate by an angle function $\theta$
with respect to a relatively parallel frame of $\gamma$
(then the velocity $\theta'$ can be interpreted 
as a \enquote{torsion}). 
It is proved in \cite{EKK08} that, 
whereas the curvature is favourable to discrete spectrum, 
the torsion plays against it. 
In particular, the spectrum of a straight 
twisted waveguide is stable under small perturbations (such as local electric field or bending). 
This repulsive effect of twisting is quantified in \cite{EKK08} (see also \cite{K08,KZ1}) by means of a Hardy type inequality. The limit $\eps\to 0$ permits to compare the effects bending and twisting (\cite{BMLT07,deOliveira_2006,KS12}) and the effective operator is given by 
\begin{equation*}
  \mathcal{L}^{\eff}=-\dr_{s}^2-\frac{\kappa(s)^2}{4}
  +C(\omega)\;\!\theta'(s)^2
  \qquad \mbox{on} \qquad
  \sL^2(\gamma,\dx s)\,,
\end{equation*}
where $C(\omega)$ is a positive constant 
whenever $\omega$ is not a disk or annulus. 
\begin{figure}[h]
\begin{center}
\scalebox{0.7}{\includegraphics{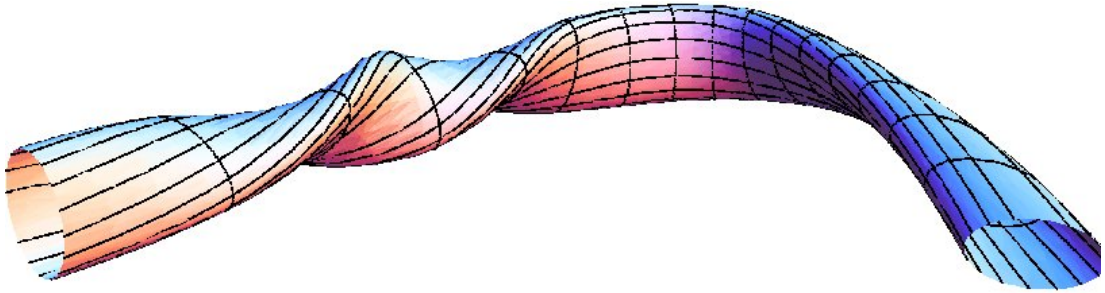}}
\end{center}
\caption{Torsion on the left and curvature on the right}
\end{figure}
Writing \eqref{Laplace} in suitable curvilinear coordinates
(see \eqref{L3} below),
one may notice similarities in the appearance of 
the torsion and the magnetic field in the coefficients 
of the operator and it therefore seems natural to
ask the following question: 
\begin{center}
\enquote{Does the magnetic field act as the torsion ?} 
\end{center}
In order to define our effective operators in the limit $\eps\to 0$ we shall describe more accurately the geometry of our waveguides. This is the aim of the next two sections in which we will always assume that the geometry (curvature and twist) and the magnetic field are compactly supported.

\subsection{Two-dimensional waveguides}\label{intro-wg2} 
Up to changing the gauge, 
the Laplace-Beltrami expression of $\mathfrak{L}_{\eps,b\A}^{[2]}$ in these coordinates is given by
$$\mathfrak{L}^{[2]}_{\eps,b\mathcal{A}}=(1-t\kappa(s))^{-1}(i\dr_{s}+b\mathcal{A}_{1})(1-t\kappa(s))^{-1}(i\dr_{s}+b\mathcal{A}_{1})-(1-t\kappa(s))^{-1}\dr_{t}(1-t\kappa(s))\dr_{t}\,,$$
with the gauge:
\[\mathcal{A}(s,t)=(\mathcal{A}_{1}(s,t),0)  ,
\quad \mathcal{A}_{1}(s,t)=\int_{0}^t (1-t'\kappa(s))\B(\Phi(s,t'))\dx t'\,.\]
We let
\[m(s,t)=1-t\kappa(s)\,.\]
The self-adjoint operator $\mathfrak{L}^{[2]}_{\eps,b\mathcal{A}}$ on $\sL^2(\R\times(-\eps,\eps),m \dx  s\dx  t)$ is unitarily equivalent to the self-adjoint operator on $\sL^2(\R\times(-\eps,\eps),\dx s\dx t)$:
\[\mathcal{L}^{[2]}_{\eps,b\mathcal{A}}=m^{1/2}\mathfrak{L}^{[2]}_{\eps,b\A}m^{-1/2}\,.\]
Introducing the rescaling
\begin{equation}\label{rescaling}
t=\eps \tau,
\end{equation}
we let
\[\mathcal{A}_{\eps}(s,\tau)=(\mathcal{A}_{1,\eps}(s,\tau),0)=(\mathcal{A}_{1}(s,\eps\tau),0)\]
and denote by $\mathcal{L}^{[2]}_{\eps,b \mathcal{A}_{\eps}}$ the homogenized operator on $\sL^2(\R\times(-1,1),\dx s\dx \tau)$:
\begin{equation}\label{L2}
\mathcal{L}^{[2]}_{\eps, b\mathcal{A}_{\eps}}=m^{-1/2}_{\eps}(i\dr_{s}+b\mathcal{A}_{1,\eps})m_{\eps}^{-1}(i\dr_{s}+b\mathcal{A}_{1,\eps})m^{-1/2}_{\eps}-\eps^{-2}\dr_{\tau}^2+V_{\eps}(s,\tau)\,,
\end{equation}
with
\[m_{\eps}(s,\tau)=m(s,\eps\tau),\quad V_{\eps}(s,\tau)=-\frac{\kappa(s)^2}{4}(1-\eps\kappa(s)\tau)^{-2}\,.\]
It is easy to verify that 
$\mathcal{L}^{[2]}_{\eps, b\mathcal{A}}$,
defined as Friedrich extension of the operator
initially defined on $\mathcal{C}^\infty_{0}(\R\times(-\eps,\eps))$,
has form domain $\sH_0^1(\R\times(-\eps,\eps))$.
Similarly, the form domain of
$\mathcal{L}^{[2]}_{\eps, b\mathcal{A}_{\eps}}$
is $\sH_0^1(\R\times(-1,1))$.

\subsection{Three-dimensional waveguides}\label{intro-wg3}
The situation is geometrically more complicated in dimension~3. 
We consider a smooth curve~$\gamma$ 
which is parameterized by its arc length~$s$
and does not overlap itself. 
We use the so-called Tang frame 
(or the relatively parallel frame, see for instance \cite{KS12}) 
to describe the geometry of the tubular neighborhood of~$\gamma$. 
Denoting the (unit) tangent vector by $T(s)=\gamma'(s)$, 
the Tang frame $(T(s), M_{2}(s), M_{3}(s))$
satisfies the relations:
\begin{eqnarray*}
T'&=&\kappa_{2} M_{2}+\kappa_{3} M_{3}\,,\\
M_{2}'&=&-\kappa_{2} T\,,\\
M_{3}'&=&-\kappa_{3}T\,.
\end{eqnarray*}
The functions $\kappa_2$ and $\kappa_3$ are the curvatures related to the choice of the normal fields $M_2$ and $M_3$.  
We can notice that $\kappa^2=\kappa_{2}^2+\kappa_{3}^2=|\gamma''|^2$ is the square of the usual curvature of $\gamma$. 

Let $\theta:\R\to\R$ a smooth function (twisting). 
We introduce the map $\Phi : \R\times(\eps\omega)\to\Omega_{\eps}$ defined by:
\begin{equation}\label{Phi}
x=\Phi(s,t_{2},t_{3})=\gamma(s)+t_{2}(\cos\theta M_{2}(s)+\sin\theta M_{3}(s))+t_{3}(-\sin\theta M_{2}(s)+\cos\theta M_{3}(s))\,.
\end{equation}
Let us notice that $s$ will often be denoted by $t_{1}$. 
As in dimension two, we always assume:
\begin{equation}\label{not-overlap3}
\mbox{$\Phi$ is injective}
\qquad\mbox{and}\qquad 
\eps\sup_{(\tau_{2},\tau_{3})\in\omega}(|\tau_{2}|+|\tau_{3}|)\,\sup_{s\in\R} |\kappa(s)|<1\,.
\end{equation}
Sufficient conditions ensuring the injectivity hypothesis can be found in~\cite[App.~A]{EKK08}.
We define 
$\mathcal{A}= (d\Phi){^\mathsf{T}} \A(\Phi)=(\mathcal{A}_{1}, \mathcal{A}_{2}, \mathcal{A}_{3})$, 
\begin{eqnarray*}
h&=&1-t_{2}(\kappa_{2}\cos\theta+\kappa_{3}\sin\theta)-t_{3}(-\kappa_{2}\sin\theta+\kappa_{3}\cos\theta)\,,\\
h_{2}&=&-t_{2}\theta'\,,\\
h_{3}&=&t_{3}\theta'\,,
\end{eqnarray*}
and $\mathcal{R}=h_{3}b\mathcal{A}_{2}+h_{2}b\mathcal{A}_{3}$. We also introduce the angular derivative $\dr_{\alpha}=t_{3}\dr_{t_{2}}-t_{2}\dr_{t_{3}}$. We will see in Chapter \ref{chapter-mwg}, Section \ref{3D} that the magnetic operator $\mathfrak{L}^{[3]}_{\eps,b\A}$ is unitarily equivalent to the operator on $\sL^2(\Omega_{\eps},h\dx t)$ given by
\begin{multline}\label{L-frak3}
\mathfrak{L}^{[3]}_{\eps,b\mathcal{A}}=\sum_{j=2,3}h^{-1}(-i\dr_{t_{j}}+b\mathcal{A}_{j})h(-i\dr_{t_{j}}+b\mathcal{A}_{j})\\
+h^{-1}(-i\dr_{s}+b\mathcal{A}_{1}-i\theta'\dr_{\alpha}+\mathcal{R})h^{-1}(-i\dr_{s}+b\mathcal{A}_{1}-i\theta'\dr_{\alpha}+\mathcal{R})\,.
\end{multline}
By considering the conjugate operator $h^{1/2}\mathfrak{L}^{[3]}_{\eps,b\mathcal{A}} h^{-1/2}$, we find that $\mathfrak{L}^{[3]}_{\eps,b\mathcal{A}}$ is unitarily equivalent to the operator defined on 
$\sL^2(\R\times(\eps\omega),\dx s \dx t_2 \dx t_3)$ 
given by:
\begin{multline}
\mathcal{L}^{[3]}_{\eps,b\mathcal{A}}=\sum_{j=2,3} (-i\dr_{t_{j}}+b\mathcal{A}_{j})^2-\frac{\kappa^2}{4h^2}\\
+h^{-1/2}(-i\dr_{s}+b\mathcal{A}_{1}-i\theta'\dr_{\alpha}+\mathcal{R})h^{-1}(-i\dr_{s}+b\mathcal{A}_{1}-i\theta'\dr_{\alpha}+\mathcal{R})h^{-1/2}\,.
\end{multline}
Finally, introducing the rescaling
$$(t_{2}, t_{3})=\eps(\tau_{2}, \tau_{3})=\eps\tau\,,$$
we define the homogenized operator on $\sL^2(\R\times\omega,\dx s \dx \tau)$:
\begin{multline}\label{L3}
\mathcal{L}^{[3]}_{\eps,b\mathcal{A}_{\eps}}=\sum_{j=2,3} (-i\eps^{-1}\dr_{\tau_{j}}+b\mathcal{A}_{j,\eps})^2-\frac{\kappa^2}{4h_{\eps}^2}\\ 
+h_{\eps}^{-1/2}(-i\dr_{s}+b\mathcal{A}_{1,\eps}-i\theta'\dr_{\alpha}+\mathcal{R}_{\eps})h_{\eps}^{-1}(-i\dr_{s}+b\mathcal{A}_{1,\eps}-i\theta'\dr_{\alpha}+\mathcal{R}_{\eps})h_{\eps}^{-1/2}\,,
\end{multline}
where $\mathcal{A}_{\eps}(s,\tau)=\mathcal{A}(s,\eps\tau)$, $h_{\eps}(s,\tau)=h(s,\eps\tau)$ and $\mathcal{R}_{\eps}=\mathcal{R}(s,\eps\tau)$.

We leave as an exercise the verification that the form domains of 
$\mathcal{L}^{[3]}_{\eps, b\mathcal{A}}$
and $\mathcal{L}^{[3]}_{\eps, b\mathcal{A}_{\eps}}$
are $\sH_0^1(\R\times(-\eps,\eps))$ and $\sH_0^1(\R\times(-1,1))$,
respectively.

\subsection{Limiting models and asymptotic expansions}
We can now state our main results concerning the effective models in the limit $\eps\to 0$. We will denote by $\lambda_{n}^\Dir(\omega)$ the $n$-th eigenvalue of the Dirichlet Laplacian $-\Delta^\Dir_{\omega}$ on $\sL^2(\omega)$. The first positive and $\sL^2$-normalized eigenfunction will be denoted by $J_{1}$. 

\begin{definition}[Case $d=2$] 
For $\delta\in(-\infty,1)$, we define:
$$\mathcal{L}^{\eff, [2]}_{\eps,\delta}=-\eps^{-2}\Delta^\Dir_{\omega}-\dr_{s}^2-\frac{\kappa(s)^2}{4}$$
and for $\delta=1$, we let:
$$\mathcal{L}^{\eff,[2]}_{\eps,1}=-\eps^{-2}\Delta^\Dir_{\omega}+\mathcal{T}^{[2]}\,,$$
where
$$\mathcal{T}^{[2]}=-\dr_{s}^2+\left(\frac{1}{3}+\frac{2}{\pi^2}\right) \B(\gamma(s))^2-\frac{\kappa(s)^2}{4}\,.$$
\end{definition}
The following theorem is proved in Chapter \ref{chapter-mwg}, Section \ref{2D}.
\begin{theo}[Case $d=2$]\label{Thm-2D}
There exists $K$ such that, for all $\delta\in(-\infty,1]$, there exist $\eps_{0}>0, C>0$ such that, for all $\eps\in(0,\eps_{0})$,
\begin{multline*}
\left\|\left(\mathcal{L}^{[2]}_{\eps,\eps^{-\delta}\mathcal{A}_{\eps}}-\eps^{-2}\lambda_{1}^\Dir(\omega)+K\right)^{-1}-\left(\mathcal{L}^{\eff,[2]}_{\eps,\delta}-\eps^{-2}\lambda_{1}^\Dir(\omega)+K\right)^{-1}\right\|\leq C\max(\eps^{1-\delta},\eps)\,,\\ 
\mbox{ for } \delta<1
\end{multline*}
and
\[\left\|\left(\mathcal{L}^{[2]}_{\eps,\eps^{-1}\mathcal{A}_{\eps}}-\eps^{-2}\lambda_{1}^\Dir(\omega)+K\right)^{-1}-\left(\mathcal{L}^{\eff,[2]}_{\eps,1}-\eps^{-2}\lambda_{1}^\Dir(\omega)+K\right)^{-1}\right\|\leq C\eps\,.\]
\end{theo}

In the critical regime $\delta=1$, we deduce the following corollary providing the asymptotic expansions of the lowest eigenvalues $\lambda_{n}^{[2]}(\eps)$ of $\mathcal{L}^{[2]}_{\eps,\eps^{-1}\mathcal{A}_{\eps}}$.

\begin{cor}[Case $d=2$ and $\delta=1$]\label{expansion-eigenvalues-2D}
Let us assume that $\mathcal{T}^{[2]}$ admits $N$ (simple) eigenvalues $\mu_{0},\cdots, \mu_{N}$ below the threshold of the essential spectrum. Then, for all $n\in\{1,\cdots N\}$, there exist $(\gamma_{j,n})_{j\geq 0}$ and $\eps_{0}>0$ such that for all $\eps\in(0,\eps_{0})$:
$$\lambda_{n}^{[2]}(\eps)\underset{\eps\to 0}{\sim}\sum_{j\geq 0} \gamma_{j,n}\eps^{-2+j}\,,$$
with 
$$\gamma_{0,n}=\frac{\pi^2}{4},\quad \gamma_{1,n}=0,\quad \gamma_{2,n}=\mu_{n}\,.$$
\end{cor}
Thanks to the spectral theorem, we also get the approximation of the corresponding eigenfunctions at any order (see our quasimodes in \eqref{quasi}). 

In order to present analogous results in dimension three,
we introduce supplementary notation.
The norm and the inner product in $\sL^2(\omega)$
will be denoted by $\|\cdot\|_{\omega}$
and $\langle\cdot,\cdot\rangle_{\omega}$, respectively.

\begin{definition}[Case $d=3$] For $\delta\in(-\infty,1)$, we define:
$$\mathcal{L}^{\eff, [3]}_{\eps,\delta}=-\eps^{-2}\Delta^\Dir_{\omega}-\dr_{s}^2-\frac{\kappa(s)^2}{4}+\|\dr_{\alpha}J_{1}\|_{\omega}^2\theta'^2$$
and for $\delta=1$, we let:
$$\mathcal{L}^{\eff,[3]}_{\eps,1}=-\eps^{-2}\Delta^\Dir_{\omega}+\mathcal{T}^{[3]}\,,$$
where $\mathcal{T}^{[3]}$ is defined by:
\begin{multline*}
\mathcal{T}^{[3]}=\langle(-i\dr_{s}-i\theta'\dr_{\alpha}-\mathcal{B}_{12}(s,0,0)\tau_{2}-\mathcal{B}_{13}(s,0,0)\tau_{3})^2 \Id(s)\otimes J_{1}, \Id(s)\otimes J_{1}\rangle_{\omega}\\
+\mathcal{B}_{23}^2(s,0,0)\left(\frac{\|\tau J_{1}\|_{\omega}^2}{4}-\langle D_{\alpha}R_{\omega},J_{1}\rangle_{\omega}\right)-\frac{\kappa^2(s)}{4}\,,
\end{multline*}
with $R_{\omega}$ being given in \eqref{J1} and 
\begin{eqnarray*}
\mathcal{B}_{23}(s,0,0)&=&\B(\gamma (s))\cdot T(s)\,,\\
\mathcal{B}_{13}(s,0,0)&=&\B(\gamma (s))\cdot(\cos\theta\, M_{2}(s)-\sin\theta\, M_{3}(s))\,,\\
\mathcal{B}_{12}(s,0,0)&=&\B(\gamma (s))\cdot(-\sin\theta\, M_{2}(s)+\cos\theta\, M_{3}(s))\,.
\end{eqnarray*}
\end{definition}
The following theorem is proved in Chapter \ref{chapter-mwg}, Section \ref{3D}.
\begin{theo}[Case $d=3$]\label{Thm-3D}
There exists $K$ such that for all $\delta\in(-\infty,1]$, there exist $\eps_{0}>0, C>0$ such that, for all $\eps\in(0,\eps_{0})$,
\begin{multline*}
\left\|\left(\mathcal{L}^{[3]}_{\eps,\eps^{-\delta}\mathcal{A}_{\eps}}-\eps^{-2}\lambda_{1}^\Dir(\omega)+K\right)^{-1}-\left(\mathcal{L}^{\eff,[3]}_{\eps,\delta}-\eps^{-2}\lambda_{1}^\Dir(\omega)+K\right)^{-1}\right\|\leq C\max(\eps^{1-\delta},\eps)\,,\\ 
\mbox{ for } \delta<1
\end{multline*}
and
$$\left\|\left(\mathcal{L}^{[3]}_{\eps,\eps^{-1}\mathcal{A}_{\eps}}-\eps^{-2}\lambda_{1}^\Dir(\omega)+K\right)^{-1}-\left(\mathcal{L}^{\eff,[3]}_{\eps,1}-\eps^{-2}\lambda_{1}^\Dir(\omega)+K\right)^{-1}\right\|\leq C\eps\,.$$
\end{theo}
In the same way, this theorem implies asymptotic expansions of eigenvalues $\lambda_{n}^{[3]}(\eps)$ of $\mathcal{L}^{[3]}_{\eps,\eps^{-1}\mathcal{A}_{\eps}}$.
\begin{cor}[Case $d=3$ and $\delta=1$]\label{expansion-eigenvalues-3D}
Let us assume that $\mathcal{T}^{[3]}$ admits $N$ (simple) eigenvalues $\nu_{0},\cdots, \nu_{N}$ below the threshold of the essential spectrum. Then, for all $n\in\{1,\cdots N\}$, there exist $(\gamma_{j,n})_{j\geq 0}$ and $\eps_{0}>0$ such that for all $\eps\in(0,\eps_{0})$:
$$\lambda_{n}^{[3]}(\eps)\underset{\eps\to 0}{\sim}\sum_{j\geq 0} \gamma_{j,n}\eps^{-2+j}\,,$$
with 
$$\gamma_{0,n}=\lambda_{1}^\Dir(\omega),\quad \gamma_{1,n}=0,\quad \gamma_{2,n}=\nu_{n}\,.$$
\end{cor}
As in two dimensions, we also get the corresponding expansion for the eigenfunctions. Complete asymptotic expansions for eigenvalues in finite 
three-dimensional waveguides
without magnetic field are also previously established in 
\cite{Grushin_2009, Borisov-Cardone_2011}. Such expansions were also obtained in \cite{Grushin_2008} in the case $\delta=0$ in a periodic framework.

\begin{rem}
As expected, when $\delta=0$ that is when $b$ is kept fixed, the magnetic field does not persists in the limit $\eps\to 0$ as well in dimension two as in dimension three. Indeed, in this limit $\Omega_{\eps}$ converges to the one dimensional curve $\gamma$ and there is no magnetic field in dimension $1$. 
\end{rem}

\subsection{Norm resolvent convergence}\label{intro-NRC}
Let us state an auxiliary result, 
inspired by the approach of \cite{FriSolo09},
which tells us that, in order to estimate the difference between two resolvents, 
it is sufficient to analyse the difference between the corresponding sesquilinear forms as soon as their domains are the same.
\begin{lem}\label{NRC}
Let $\mathfrak{L}_{1}$ and $\mathfrak{L}_{2}$ be two positive self-adjoint operators on a Hilbert space $\mathsf{H}$. Let $\mathfrak{B}_{1}$ and $\mathfrak{B}_{2}$ be their associated sesquilinear forms. We assume that $\Dom(\mathfrak{B}_{1})=\Dom(\mathfrak{B}_{2})$. Assume that there exists $\eta>0$ such that for all $\phi,\psi\in\Dom(\mathfrak{B}_{1})$:
$$\left|\mathfrak{B}_{1}(\phi,\psi)-\mathfrak{B}_{2}(\phi,\psi)\right|\leq \eta \sqrt{\mathfrak{Q}_{1}(\psi)}\sqrt{\mathfrak{Q}_{2}(\phi)}\,,$$
where $\mathfrak{Q}_{j}(\varphi)=\mathfrak{B}_{j}(\varphi,\varphi)$ for $j=1,2$ and $\varphi\in\Dom(\mathfrak{B}_{1})$.
Then, we have:
$$\|\mathfrak{L}_{1}^{-1}-\mathfrak{L}_{2}^{-1}\|\leq \eta\|\mathfrak{L}_{1}^{-1}\|^{1/2}\|\mathfrak{L}_{2}^{-1}\|^{1/2}\,.$$
\end{lem}
\begin{proof}
The original proof can be found in \cite[Prop. 5.3]{KS12}. Let us consider $\tilde\phi,\tilde\psi\in \mathsf{H}$. We let $\phi=\mathfrak{L}_{2}^{-1}\tilde\phi$ and $\psi=\mathfrak{L}_{1}^{-1}\tilde\psi$. We have $\phi,\psi\in\Dom(\mathfrak{B}_{1})=\Dom(\mathfrak{B}_{2})$. We notice that:
$$\mathfrak{B}_{1}(\phi,\psi)=\langle \mathfrak{L}_{2}^{-1}\tilde\phi,\tilde\psi\rangle, \quad \mathfrak{B}_{2}(\phi,\psi)=\langle \mathfrak{L}_{1}^{-1}\tilde\phi,\tilde\psi\rangle$$
and:
$$\mathfrak{Q}_{1}(\psi)=\langle\tilde\psi,\mathfrak{L}_{1}^{-1}\tilde\psi\rangle,\quad \mathfrak{Q}_{2}(\phi)=\langle\tilde\phi, \mathfrak{L}_{2}^{-1}\tilde\phi\rangle\,.$$
We infer that:
$$\left|\langle (\mathfrak{L}_{1}^{-1}- \mathfrak{L}_{2}^{-1})\tilde\phi,\tilde\psi\rangle\right|\leq\eta \|\mathfrak{L}_{1}^{-1}\|^{1/2}\|\mathfrak{L}_{2}^{-1}\|^{1/2} \|\tilde\phi\|\|\tilde\psi\|$$
and the result elementarily follows.
\end{proof}

\subsection{A magnetic Hardy inequality}
In dimension $2$, the limiting model (with $\delta=1$) enlightens the fact that the magnetic field plays against the curvature, whereas in dimension $3$ this repulsive effect is not obvious (it can be seen that $\langle D_{\alpha}R_{\omega},J_{1}\rangle_{\omega}\geq 0$). Nevertheless, if $\omega$ is a disk, we have $\langle D_{\alpha}R_{\omega},J_{1}\rangle_{\omega}=0$ and thus the component of the magnetic field parallel to~$\gamma$ plays against the curvature (in comparison, a pure torsion has no effect when the cross section is a disk). In the flat case ($\kappa=0$), we can quantify this repulsive effect by means of a magnetic Hardy inequality (see \cite{EK05} where this inequality is discussed in dimension two). We will not discuss the proof of this inequality in this book.
\begin{theo}\label{stability}
Let $d\geq 2$. Let us consider $\Omega=\R\times \omega$. 
For $R>0$, we let:
$$\Omega(R)=\{t\in\Omega : |t_{1}|< R\}\,.$$
Let $\A$ be a smooth vector potential such that $\sigma_{\B}$ is not zero on $\Omega(R_{0})$ for some $R_{0}>0$. Then, there exists $C>0$ such that, for all $R\geq R_{0}$, there exists $c_{R}(\B)>0$ such that, we have:
\begin{equation}\label{Hardy}
\int_{\Omega} |(-i\nabla+\A)\psi|^2-\lambda_{1}^\Dir(\omega)|\psi|^2\,\dx t\geq \int _{\Omega}\frac{c_{R}(\B)}{1+s^2}|\psi|^2\dx t,\quad \forall\psi\in \mathcal{C}^\infty_{0}(\Omega)\,.
\end{equation}
Moreover we can take:
$$c_{R}(\B)=\left(1+CR^{-2}\right)^{-1}\min\left(\frac{1}{4},\lambda_{1}^{\Dir,\Neu}(\B,\Omega(R))-\lambda_{1}^\Dir(\omega)\right),$$
where $\lambda_{1}^{\Dir,\Neu}(\B,\Omega(R))$ denotes the first eigenvalue of the magnetic Laplacian on $\Omega(R)$, with Dirichlet condition on $\R\times\dr\omega$ and Neumann condition on $\{|s|=R\}\times\omega$.
\end{theo}
The inequality of Theorem \ref{stability} can be applied to prove 
certain stability of the spectrum of the magnetic Laplacian on $\Omega$ under local and small deformations of $\Omega$. Let us fix $\eps>0$ and describe a generic deformation of the straight tube $\Omega$. 
We consider the local diffeomorphism:
$$\Phi_{\eps}(t)=\Phi_{\eps}(s,t_{2},t_{3})=(s,0,\cdots,0)+\sum_{j=2}^d (t_{j}+\eps_{j}(s))M_{j}+\mathcal{E}_{1}(s)\,,$$
where $(M_{j})_{j=2}^d$ is the canonical basis of $\{0\}\times\R^{d-1}$. The functions $\eps_{j}$ and $\mathcal{E}_{1}$ are smooth and compactly supported in a compact set $K$. As previously we assume that $\Phi_{\eps}$ is a global diffeomorphism and we consider the deformed tube $\Omega^{\defo,\eps}=\Phi_{\eps}(\R\times\omega)$.
\begin{prop}\label{perturbed}
Let $d\geq 2$. There exists $\eps_{0}>0$ such that for $\eps\in(0,\eps_{0})$, the spectrum of the Dirichlet realization of $(-i\nabla+\A)^2$ on $\Omega^{\defo,\eps}$ coincides with the spectrum of the Dirichlet realization of $(-i\nabla+\A)^2$ on $\Omega$. The spectrum is given by $[\lambda_{1}^\Dir(\omega),+\infty)$.
\end{prop}
By using a semiclassical argument, it is possible to prove a stability result which does not use the Hardy inequality.
\begin{prop}\label{strong-b}
Let $R_{0}>0$ and $\Omega(R_{0})=\{t\in\R\times\omega : |t_{1}|\leq R_{0}\}$. Let us assume that $\sigma_\B=\dx\xi_\A$ does not vanish on $\Phi(\Omega(R_{0}))$ and that on $\Omega_{1}\setminus \Phi(\Omega(R_{0}))$ the curvature is zero. Then, there exists $b_{0}>0$ such that for $b\geq b_{0}$, the discrete spectrum of $\mathfrak{L}^{[d]}_{1,b\A}$ is empty.
\end{prop}

\section{Magnetic layers}\label{intro-wg-layers}
As we will sketch below, the philosophy of Duclos and Exner may also apply to thin quantum layers as we can see in the contributions \cite{DEK01, Carron04, LL06, LL07, LL07b, Ro12, KRT13} and the related  papers \cite{JK71,daC1,daC2,Tol88,M01, FrH01, FK08,Wittich_2008, TW13, LTW11, KS12}.

Let us consider $\Sigma$ an hypersurface embedded in $\R^d$ with $d \geq 2$,
and define a tubular neighbourhood about $\Sigma$,
\begin{equation}\label{layer.intro}
  \Omega_\eps := \big\{x+t\n \in \R^d \ \big| \ 
  (x,t) \in \Sigma \times (-\eps,\eps) \big\}
  \,,
\end{equation}
where $\n$ denotes a unit normal vector field of $\Sigma$. We investigate:
\begin{equation}\label{Laplacian.intro}
\mathcal{L}_{\A,\Omega_{\eps}}=(-i\nabla+\A)^2 
  \qquad \mbox{on} \qquad 
 \sL^2(\Omega_\eps)
  \,,
\end{equation}
with Dirichlet boundary conditions on $\partial\Omega_\eps$.

\subsection{Normal form}
As usual the game is to find an appropriate normal form for the magnetic Laplacian.
Given $I:=(-1,1)$ and $\eps>0$,
we define a layer $\Omega_\eps$ of width $2\eps$
along $\Sigma$ as the image of the mapping
\begin{equation}\label{layer}
\Phi:\Sigma\times I\to\R^d: \ 
  \big\{(x,u)\mapsto x+\eps u \n\big\}\,.
\end{equation}
Let us denote by $\tilde{\A}$ the components of the vector potential
expressed in the curvilinear coordinates induced by the embedding \eqref{layer}.
Moreover, assume 
\begin{equation}\label{gauge}
  \tilde{A}_{d} = 0\,.
\end{equation}
Thanks to the diffeomorphism $\Phi:\Sigma \times I \to \Omega_\eps$, 
we may identify $\mathcal{L}_{\A,\Omega_{\eps}}$ 
with an operator $\hat{H}$ on $\sL^{2}(\Sigma\times I,\dx\Omega_\eps)$ 
that acts, in the form sense, as 
\begin{equation*}
  \hat{H}
  =|G|^{-1/2}(-i\partial_{x^\mu}+\tilde{A}_{\mu})
  |G|^{1/2}G^{\mu\nu}(-i\partial_{x^\nu}
  +\tilde{A}_{\nu})-\varepsilon^{-2}|G|^{-1/2}
  \partial_{u}|G|^{1/2}\partial_{u}
  \,.
\end{equation*}
Let us define 
$$
  J := \frac{1}{4} \ln\frac{|G|}{|g|}
  = \frac{1}{2}\sum_{\mu=1}^{d-1}\ln(1-\varepsilon u\kappa_{\mu})
  =\frac{1}{2}\ln\left[1+\sum_{\mu=1}^{d-1}(-\varepsilon u)^{\mu}
  \binom{d-1}{\mu}K_{\mu}\right]\,.
$$
Using the unitary transform 
$$
  U:\sL^{2}(\Sigma\times I,\dx\Omega_\eps)
  \to \sL^{2}(\Sigma\times I,\dx\Sigma \wedge \dx u) : \
  \left\{
  \psi\mapsto e^{J} \psi
  \right\}
  \,,
$$ 
we arrive at the unitarily equivalent operator 
\begin{equation*}
  H: = U\hat{H}U^{-1}
  = |g|^{-1/2}(-i\partial_{x^\mu}+\tilde{A}_{\mu})
  |g|^{1/2}G^{\mu\nu}(-i\partial_{x^\nu}+\tilde{A}_{\nu})
  -\varepsilon^{-2}\partial_{u}^{2}+V
  \,,
\end{equation*}
where
$$
  V := |g|^{-1/2} \, 
  \partial_{x^i}\big( |g|^{1/2} G^{ij} (\partial_{x^j} J) \big)
  + (\partial_{x^i} J) G^{ij} (\partial_{x^j} J)
  \,.
$$
We get
$$
  H = U\hat{U}(-\Delta_{D,A}^{\Omega_\eps})\hat{U}^{-1}U^{-1}
  \,.
$$
\subsection{The effective operator}
$H$ is approximated in the norm resolvent sense (see \cite{KRT13} for the details) by
\begin{equation}\label{H0-Matej}
  H_{0}=h_{\mathrm{eff}}-\varepsilon^{-2}\partial_{u}^{2}
  \ \simeq \ h_{\mathrm{eff}} \otimes 1 
  + 1 \otimes (-\varepsilon^{-2}\partial_{u}^{2})
\end{equation}
on 
$
  \sL^{2}(\Sigma \times I,\dx\Sigma\wedge \dx u)
  \simeq
  \sL^{2}(\Sigma,\dx\Sigma)\otimes \sL^{2}(I,\dx u)
$
with the effective Hamiltonian
\begin{equation}\label{H.eff}
  h_{\mathrm{eff}}
  := |g|^{-1/2}\big(-i\partial_{x^\mu}+\tilde{A}_{\mu}(.,0)\big)
  |g|^{1/2}g^{\mu\nu}\big(-i\partial_{x^\nu}+\tilde{A}_{\nu}(.,0)\big)
  +V_{\mathrm{eff}}\,,
\end{equation}
where
\begin{equation}\label{V.eff}
  V_{\mathrm{eff}}
  :=-\frac{1}{2}\sum_{\mu=1}^{d-1}\kappa_{\mu}^{2}
  +\frac{1}{4}\left(\sum_{\mu=1}^{d-1}\kappa_{\mu}\right)^2
  \,.
\end{equation}
~\\
\section{Broken waveguides}

\subsection{Semiclassical triangles}\label{intro-wg-triangles}
As we would like to analyze the spectrum of broken waveguides (that is waveguides with an angle), this is natural to prepare the investigation by studying the Dirichlet eigenvalues of the Laplacian on some special shrinking triangles. This subject is already dealt with in \cite[Theorem 1]{Fre07} where four-term asymptotics is proved for the lowest eigenvalue, whereas a three-term asymptotics for the second eigenvalue is provided in \cite[Section 2]{Fre07}. We can mention the papers \cite{FriSolo08, FriSolo09} whose results provide two-term asymptotics for the thin rhombi and also \cite{BoFre09} which deals with a regular case (thin ellipse for instance), see also \cite{BoFre10}. We also invite the reader to take a look at \cite{HJ11}. For a complete description of the low lying spectrum of general shrinking triangles, one may consult the paper by Ourmi\`eres \cite{Ourm14} (especially the existence of a boundary layer living near the shrinking height is proved, see also \cite{DauRay12, LR13}) where tunnel effect estimates are also established. In dimension three the generalization to cones with small aperture is done in \cite{Ourm12} and which is motivated by \cite{ExTa10}.  

Let us define the isosceles triangle in which we are interested:
\begin{equation}
\label{Tri}
   \Tri_{\theta}=\left\{(x_{1},x_{2})\in\mathbb{R}_{-}\times\R : x_{1}\tan\theta<|x_{2}|<\left(x_{1}+\frac{\pi}{\sin\theta}\right)\tan\theta\right\}\,.
\end{equation}
We will use the coordinates
\begin{equation}
\label{E:xy-0}
   x=x_{1}\sqrt{2}\sin\theta, \quad y=x_{2}\sqrt{2}\cos\theta\,,
\end{equation}
which transform $\Tri_{\theta}$ into $\Tri_{\pi/4}.$
The operator becomes:
$$\mathcal{D}_{\Tri}(h)=2\sin^2\!\theta\,\dr_{x}^2-2\cos^2\!\theta\,\dr_{y}^2\,,$$
with Dirichlet condition on the boundary of $\Tri$. 
We let $h=\tan\theta$ ; after a division by $2\cos^2\theta$, we get the new operator:
\begin{equation}
\label{E:LTri}
   \mathcal{L}_{\Tri}(h) = -h^2\dr_{x}^2-\dr_{y}^2\,.
\end{equation}
This operator is thus in the \enquote{Born-Oppenheimer form} and we shall introduce its Born-Oppenheimer approximation which is the Dirichlet realization on $\sL^2((-\pi\sqrt{2},0))$ of:
\begin{equation}\label{HT}
\mathcal{H}_{\BO, \Tri}(h)=-h^2\dr_{x}^2+\frac{\pi^2}{4(x+\pi\sqrt{2})^2}\,.
\end{equation}
The following theorem is a consequence of the Born-Oppenheimer strategy (see Chapter \ref{intro-models}, Section \ref{Sec.BOM}).
\begin{theo}\label{spectrumBOT}
The eigenvalues of $\mathcal{H}_{\BO, \Tri}(h)$, denoted by $\lambda_{\BO, \Tri,n}(h)$, admit the expansions:
$$\lambda_{\BO, \Tri,n}(h)\underset{h\to 0}{\sim}\sum_{j\geq 0}\hat{\beta}_{j,n}h^{2j/3}\,,
\quad\mbox{with}\quad
\hat{\beta}_{0,n}=\frac{1}{8} \ \ \mbox{and}\ \ \hat{\beta}_{1,n}=(4\pi\sqrt{2})^{-2/3}z_{\Ai}(n)\,,
$$
where $z_{\Ai}(n)$ is the $n$-th zero of the reversed Airy function $\Ai(x)=\mathsf{Ai}(-x)$.
\end{theo}

We state the main result of this section for the scaled operator $\mathcal{L}_{\Tri}(h)$. A proof may be found in Chapter \ref{chapter-triangles}.

\begin{theo}\label{spectrumtriangle}
The eigenvalues of $\mathcal{L}_{\Tri}(h)$, denoted by $\lambda_{\Tri,n}(h)$, admit the expansions:
$$
   \lambda_{\Tri,n}(h)\underset{h\to 0}{\sim}\sum_{j\ge0}\beta_{j,n}h^{j/3}
   \quad
   \mbox{with} \ \ \beta_{0,n}=\frac{1}{8}, \ \ \beta_{1,n}=0,
   \ \ \mbox{and}\ \ \beta_{2,n}=(4\pi\sqrt{2})^{-2/3}z_{\Ai}(n)\,,
$$
the terms of odd rank being zero for $j\leq 8$.
The corresponding eigenvectors have expansions in powers of $h^{1/3}$ with both scales $x/h^{2/3}$ and $x/h$.
\end{theo}

\subsection{Broken waveguides}\label{intro-wg-broken}

\subsubsection{Physical motivation}
As we have already recalled at the beginning of this chapter, it has been proved in \cite{Duclos95} that a curved, smooth and asymptotically straight waveguide has discrete spectrum below its essential spectrum. Now we would like to explain the influence of a corner which is somehow an infinite curvature and extend the philosophy of the smooth case. This question is investigated with the $L$-shape waveguide in \cite{Exner89} where the existence of discrete spectrum is proved. For an arbitrary angle too, this existence is proved in \cite{ABGM91} and an asymptotic study of the ground energy is done when $\theta$ goes to $\frac{\pi}{2}$ (where $\theta$ is the semi-opening of the waveguide). Another question which arises is the estimate of the lowest eigenvalues in the regime $\theta\to 0$. This problem is analyzed in \cite{CLMM93} where a waveguide with corner is the model chosen to describe some electromagnetic experiments (see the experimental results in \cite{CLMM93}). We also refer to our work \cite{DaLafRa11, DauRay12}.
\subsubsection{Geometric description}
\begin{figure}[ht]
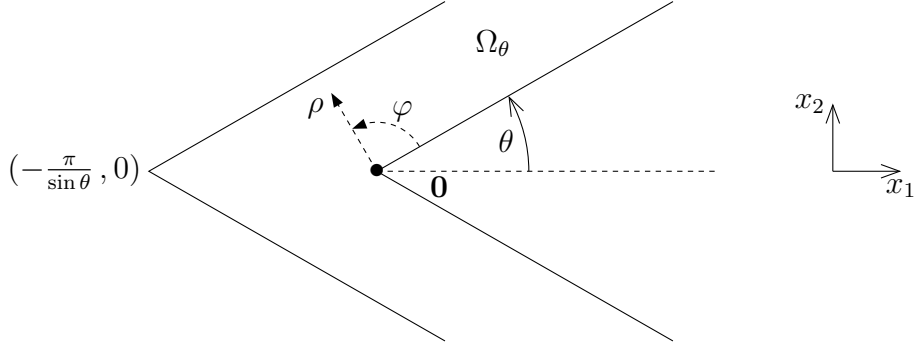

% 1. Definition of characteristic points
    \figinit{1mm}
    \figpt 1:(  0,  0)
    \figpt 2:( 45,  0)
    \figptrot 3: = 2 /1, 30/
    \figpt 4:(-30,  0)
    \figpt 5:(12, 0)
    \figptrot 6: = 5 /1, 120/
    \figvectP 101 [1,4]
    \figptstra 14 = 3/1, 101/
    \figptssym 23 = 3, 14 /1, 2/
    \figpt 10:(60,  0)

% 2. Creation of the postscript file
    \def\MyPSfile{}
    \psbeginfig{\MyPSfile}
    \psaxes 10(9)
    \figptsaxes 11 : 10(9)
    \psline[23,1,3]
    \psline[24,4,14]
    \psarrowcirc 1 ; 20 (0,30)
    \psset arrowhead (fillmode=yes, length=2)
    \psset (dash=4)
    \psarrow[1,6]
    \psarrowcirc 1 ; 6.5 (30,120)
    \psline[1,2]
  \psendfig

% 3. Writing text on the figure
    \figvisu{\figbox}{}{%
    \figinsert{\MyPSfile}
    \figwrites 11 :{$x_1$}(1)
    \figwritew 12 :{$x_2$}(1)
    \figwritegcw 4 :{$(-\frac{\pi}{\sin\theta}\,,0)$}(1,0)
    \figwritegce 1 :{$\Omega_{\theta}$}(13,17)
    \figwritegce 1 :{$\varphi$}(2,8)
    \figwritegce 1 :{$\theta$}(16,4)
    \figwritegcw 6 :{$\rho$}(1,-2)
    \figsetmark{$\bullet$}
    \figwritegce 1 :{${\bf 0}$}(7,-2.1)
    }
\centerline{\box\figbox}
\caption{The broken guide $\Omega_{\theta}$ (here $\theta=\frac\pi6$). Cartesian and polar coordinates.}
\label{F:1}
\end{figure}

Let us denote by $(x_1,x_2)$ the Cartesian coordinates of the plane and by ${\bf0}=(0,0)$ the origin. Let us define our so-called \enquote{broken waveguides}. For any angle $\theta\in\left(0,\frac{\pi}{2}\right)$ we introduce
\begin{equation}
\label{E:Omtheta}
   \Omega_{\theta}=\left\{(x_{1},x_{2})\in\mathbb{R}^2 : x_{1}\tan\theta<|x_{2}|<\left(x_{1}+\frac{\pi}{\sin\theta}\right)\tan\theta\right\}\,.
\end{equation}
Note that its width is independent from $\theta$, normalized to $\pi$, see Figure \ref{F:1}. The limit case where $\theta=\frac\pi2$ corresponds to the straight strip $(-\pi,0)\times\R$.

The operator $-\Delta^\Dir_{\Omega_{\theta}}$ is a positive unbounded self-adjoint operator with domain
\[
   \Dom(-\Delta^\Dir_{\Omega_{\theta}}) = \{\psi\in \sH^1_0(\Omega_\theta):\quad
   -\Delta\psi\in \sL^2(\Omega_\theta)\}\,.
\]
When $\theta\in\left(0,\frac{\pi}{2}\right)$, the boundary of $\Omega_\theta$ is not smooth, it is polygonal. The presence of the non-convex corner with vertex ${\bf0}$ is the reason for the space $\Dom(-\Delta^\Dir_{\Omega_{\theta}})$ to be distinct from $\sH^2\cap \sH^1_0(\Omega_\theta)$. 
We have the following description of the domain (see the classical references \cite{Kondratev67,Grisvard85}):
\begin{equation}
\label{eq:dom}
   \Dom(-\Delta^\Dir_{\Omega_{\theta}}) = \left(\sH^2\cap \sH^1_0(\Omega_\theta)\right) \oplus [\psi^\theta_\sing]
\end{equation}
where $[\psi^\theta_\sing]$ denotes the space generated by the singular function $\psi^\theta_\sing$ defined in the polar coordinates $(\rho,\varphi)$ near the origin by
\begin{equation}
\label{eq:sing}
   \psi^{\theta}_\sing(x_1,x_2) = \chi(\rho)\, \rho^{\pi/\omega} \sin\frac{\pi\varphi}{\omega}\quad
   \mbox{with}\quad \omega = 2(\pi-\theta)
\end{equation}
where where $\chi$ is a radial cutoff function near the origin.

We gather in the following statement several important preliminary properties for the spectrum of $-\Delta^\Dir_{\Omega_{\theta}}$. All these results are proved in the literature. 
\begin{prop}\label{P:ess}
We have:
\begin{enumerate}[(i)]

\item If $\theta=\frac\pi2$, $-\Delta^{\Dir}_{\Omega_{\theta}}$ has no discrete spectrum. Its essential spectrum is the closed  interval $[1,+\infty)$. 

\item For any $\theta$ in the open interval $(0,\frac\pi2)$ the essential spectrum of  $-\Delta^{\Dir}_{\Omega_{\theta}}$ coincides with $[1,+\infty)$. 

\item For any $\theta\in(0,\frac\pi2)$, the discrete spectrum of $-\Delta^{\Dir}_{\Omega_{\theta}}$ is {\em nonempty}.

\item\label{P:essiv} For any $\theta\in(0,\frac\pi2)$ and any eigenvalue in the discrete spectrum of $-\Delta^{\Dir}_{\Omega_{\theta}}$, the associated eigenvectors $\psi$ are even with respect to the horizontal axis: $\psi(x_1,-x_2)=\psi(x_1,x_2)$.

\item For any $\theta\in(0,\frac\pi2)$, let $\mu_{\Gui,n}(\theta)$, $n=1,\ldots$, be the $n$-th Rayleigh quotient of $-\Delta^{\Dir}_{\Omega_{\theta}}$. Then, for any $n\ge1$, the function $\theta\mapsto\mu_{\Gui,n}(\theta)$ is continuous and increasing.
\end{enumerate}
\end{prop}
It is also possible to prove that the number of eigenvalues below the essential spectrum is exactly $1$ as soon as $\theta$ is close enough to $\frac{\pi}{2}$ (see \cite{Naz13}). In this book we will provide an proof of the following proposition which is inspired by \cite[Theorem 2.1]{MoTr05} (see Chapter \ref{chapter-models}, Section \ref{sec.bound-number}).
\begin{prop}
\label{prop:5-1}
For any $\theta\in(0,\frac{\pi}{2})$, the number of eigenvalues of $-\Delta^\Dir_{\Omega_\theta}$ below $1$, denoted by $\mathsf{N}(-\Delta^\Dir_{\Omega_\theta},1)$, is finite.
\end{prop}
As a consequence of the parity properties of the eigenvectors of $-\Delta^{\Dir}_{\Omega_{\theta}}$, cf. point \eqref{P:essiv} of Proposition \ref{P:ess}, we can reduce the spectral problem to the  half-guide
\begin{equation}
\label{E:Omthetaplus}
   \Omega_{\theta}^+=\left\{(x_{1},x_{2})\in \Omega_\theta : \ 
   x_{2}>0\right\}\,.
\end{equation}
We define the Dirichlet part of the boundary by
$\partial_\Dir\Omega^+_{\theta} = 
\partial\Omega_{\theta} \cap \partial\Omega^+_{\theta}$,
and the corresponding form domain
$$\sH^1_\Mix(\Omega^+_{\theta}) = \big\{\psi\in \sH^1(\Omega^+_{\theta}):\quad 
   \psi=0 \ \mbox{ on } \ \partial_\Dir\Omega^+_{\theta} \big\}\,.$$
Then the new operator of interest, denoted by $-\Delta^\Mix_{\Omega_{\theta}^+}$, is the Laplacian with mixed Dirichlet-Neumann conditions on $\Omega_{\theta}^+$. Its domain is:
\[
  \Dom(-\Delta^\Mix_{\Omega_{\theta}^+}) =
  \big\{ \psi\in \sH^1_\Mix(\Omega_{\theta}^+) : \ \ \Delta\psi\in \sL^2(\Omega_{\theta}^+)
  \ \ \mbox{and}\ \  \partial_{2}\psi=0 \ \mbox{ on }\ x_{2}=0 \big\}.
\]
Then the operators $-\Delta^{\Dir}_{\Omega_{\theta}}$ and $-\Delta^\Mix_{\Omega_{\theta}^+}$ have the same eigenvalues below $1$ and the eigenvectors of the latter are the restriction to $\Omega_{\theta}^+$ of the former.

\begin{figure}[ht]
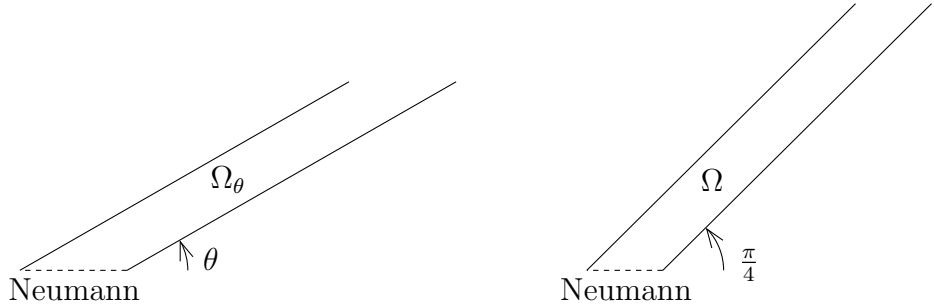

% 1. Definition of characteristic points
    \figinit{1mm}
    \figpt 1:(  0,  0)
    \figpt 2:( 50,  0)
    \figptrot 3: = 2 /1, 30/
    \figpt 4:(-14.1,  0)
    \figvectP 101 [1,4]
    \figptstra 14 = 3/1, 101/
    \figpt 31:(  0,  0)
    \figpt 32:( 50,  0)
    \figptrot 33: = 32 /31, 45/
    \figpt 34:(-10,  0)
    \figvectP 301 [31,34]
    \figptstra 44 = 33/1, 301/
    \figptstra 50 = 31, 33, 34, 44/-5, 101/

% 2. Creation of the postscript file
    \def\MyPSfile{F2.pf}
    \psbeginfig{\MyPSfile}
    \psline[1,3]
    \psline[4,14]
    \psline[50,51]
    \psline[52,53]
    \psarrowcirc 1 ; 8 (0,30)
    \psarrowcirc 50 ; 8 (0,45)
    \psset (dash=4)
    \psline[1,4]
    \psline[50,52]
  \psendfig

% 3. Writing text on the figure
    \figvisu{\figbox}{}{%
    \figinsert{\MyPSfile}
    \figwritebe 1 :{$\theta$}(10)
    \figwritegce 1 :{$\Omega_{\theta}$}(11,12)
    \figwritebe 50 :{$\frac\pi4$}(10)
    \figwritegce 50 :{$\Omega$}(5,12)
    \figwritegce 4 :{Neumann}(-1.5,-2.5)
    \figwritegce 52 :{Neumann}(-3.5,-2.5)
    }
\centerline{\box\figbox}
\caption{The half-guide $\Omega_{\theta}^+$ for $\theta=\frac\pi6$ and the reference domain $\Omega$.}
\label{F:2}
\end{figure}

In order to analyze the asymptotics $\theta\to0$, it is useful to rescale the integration domain and transfer the dependence on $\theta$ into the coefficients of the operator. For this reason, let us perform the following linear change of coordinates:
\begin{equation}
\label{E:xy}
   x=x_{1}\sqrt{2}\sin\theta\,, \quad y=x_{2}\sqrt{2}\cos\theta\,,
\end{equation}
which maps $\Omega^+_\theta$ onto the $\theta$-independent domain $\Omega^+_{\pi/4}$, see Fig.\ \ref{F:2}. That is why we set for simplicity
\begin{equation}
\label{E:Omega}
   \Omega := \Omega^+_{\pi/4}\,, \ \ 
   \partial_\Dir\Omega = \partial_\Dir \Omega^+_{\pi/4}\,, \ \ \mbox{and}\ \ 
   \sH^1_\Mix(\Omega) = \big\{\psi\in \sH^1(\Omega): 
   \psi=0 \ \mbox{ on } \ \partial_\Dir\Omega \big\}\,.
\end{equation}

Then, $\Delta^{\Mix}_{\Omega_{\theta}^+}$ is unitarily equivalent to the operator defined on $\Omega$ by:
\begin{equation}
\label{E:DGui}
 \mathcal{D}_\Gui(\theta) = -2\sin^2\!\theta\,\dr_{x}^2-2\cos^2\!\theta\,\dr_{y}^2\,,
\end{equation}
with Neumann condition on $y=0$ and Dirichlet everywhere else on the boundary of $\Omega$. 
We let $h=\tan\theta$ ; after a division by $2\cos^2\theta$, we get the new operator:
\begin{equation}
\label{E:LGui}
   \mathcal{L}_{\Gui}(h) = -h^2\dr_{x}^2-\dr_{y}^2\,,
\end{equation}
with domain:
\[
  \Dom(\mathcal{L}_{\Gui}(h))=\big\{\psi\in \sH^1_\Mix(\Omega) :  \ \ \mathcal{L}_{\Gui}(h)\psi\in \sL^2(\Omega)
  \ \ \mbox{and}\ \  
  \dr_{y}\psi=0 \ \mbox{ on } \ y=0 \big\}\,.
\]
The Born-Oppenheimer approximation of $\mathcal{L}_{\Gui}(h)$ (see Chapter \ref{chapter-BOE}) is
\begin{equation}\label{HG}
\mathcal{H}_{\BO, \Gui}(h)=-h^2\dr_{x}^2+V(x)\,,
\end{equation}
where 
\[V(x)=\left\{\begin{array}{cl}
\displaystyle\frac{\pi^2}{4(x+\pi\sqrt{2})^2}&\mbox{ when } x\in(-\pi\sqrt{2},0)\,,\\[2.5ex]
\displaystyle\frac{1}{2}&\mbox{ when } x\geq 0\,.
\end{array}\right.\]

\subsubsection{Eigenvalues induced by a strongly broken waveguide}
Let us now state the main result concerning the asymptotic expansion of the eigenvalues of the broken waveguide (see Chapter \ref{chapter-triangles} for the proof of the two terms asymptotic expansion).
\begin{theo}\label{spectrum-guide}
For all $N_{0}$, there exists $h_{0}>0$, such that for $h\in(0,h_{0})$ the $N_{0}$ first eigenvalues of $\mathcal{L}_{\Gui}(h)$ exist. These eigenvalues, denoted by $\lambda_{\Gui,n}(h)$, admit the expansions:
$$
   \lambda_{\Gui,n}(h)\underset{h\to 0}{\sim}\sum_{j\ge0}\gamma_{j,n}h^{\frac j 3}
   \quad
   \mbox{with} \ \ \gamma_{0,n}=\frac{1}{8}, \ \ \gamma_{1,n}=0\,,
   \ \ \mbox{and}\ \ \gamma_{2,n}=(4\pi\sqrt{2})^{-2/3}z_{\Ai}(n)\,.
$$
\end{theo}

\chapter{On some connected non linear problems}\label{intro-NL}
\begin{flushright}
\begin{minipage}{0.6\textwidth}
L'explication que nous devons juger satisfaisante est celle qui adh\`ere \`a son objet : point de vide entre eux, pas d'interstice o\`u une autre explication puisse aussi bien se loger ; elle ne convient qu'\`a lui, il ne se pr\^ete qu'\`a elle.
\begin{flushright}
\textit{La pens\'ee et le mouvant}, Bergson
\end{flushright}
\vspace*{0.5cm}
\end{minipage}
\end{flushright}

In this chapter we present two problems related to the non linear Schr\"odinger equation:
\begin{enumerate}[(i)]
\item the semiclassical limit for the $p$-eigenvalues of the magnetic Laplacian,
\item the dimensional reduction of the time dependent non linear Schr\"odinger equation.
\end{enumerate}

\section{Non linear magnetic eigenvalues}

\subsection{Definition of the non linear eigenvalue}
Let $\Omega$ be a bounded simply connected open set of $\R^2$. We introduce the following \enquote{nonlinear eigenvalue} (or optimal magnetic Sobolev constant):
\begin{equation}\label{inf}
\lambda(\Omega, \A, p,h)=\inf_{\psi\in \sH^1_{0}(\Omega), \psi\neq 0}\frac{\mathfrak{Q}_{h,\A}(\psi)}{\left(\int_{\Omega}|\psi|^p\dx x\right)^{\frac{2}{p}}}=\inf_{\underset{ \|\psi\|_{\sL^p(\Omega)}=1}{\psi\in \sH^1_{0}(\Omega),}}\mathfrak{Q}_{h,\A}(\psi)\,,
\end{equation}
where the magnetic quadratic form is defined by
$$\forall \psi\in\sH^1_{0}(\Omega),\quad\mathfrak{Q}_{h,\A}(\psi)=\int_{\Omega}|(-ih\nabla+\A)\psi|^2\dx\x\,.$$
\begin{lem}
The infimum in \eqref{inf} is attained.
\end{lem}
\begin{proof}
Consider a minimizing sequence $(\psi_{j})$ that is normalized in $\sL^p$-norm. 
Then, by a H\"{o}lder inequality and using that $\Omega$ has bounded measure, $(\psi_{j})$ is bounded in $\sL^2$.
Since $\A \in \sL^{\infty}(\Omega)$, we conclude that $(\psi_{j})$ is bounded in $\sH^1_{0}(\Omega)$.
By the Banach-Alaoglu Theorem there exists a subsequence (still denoted by $(\psi_{j})$) and $\psi_{\infty} \in \sH^1_{0}(\Omega)$ such that $\psi_j \rightharpoonup \psi_{\infty}$ weakly in $\sH^1_{0}(\Omega)$ and $\psi_j \rightarrow \psi_{\infty}$ in $\sL^q(\Omega)$ for all $q\in[2,+\infty)$.
This is enough to conclude.
\end{proof}
\begin{lem}
The minimizers (which belong to $\sH^1_{0}(\Omega)$) of the $\sL^p$-normalized version of \eqref{inf} satisfy the following equation in the sense of distributions:
\begin{equation}\label{NLS}
(-ih\nabla+\A)^2\psi=\lambda(\Omega, \A, p, h)|\psi|^{p-2}\psi\,,\qquad \|\psi\|_{\sL^p(\Omega)}=1\,.
\end{equation}
In particular (by Sobolev embedding), the minimizers belong to the domain of $\mathfrak{L}_{h,\A}$.
\end{lem}

\subsection{A result by Esteban and Lions}
By using the famous concentration-compactness method, Esteban and Lions proved the following proposition in \cite{EL89}.
\begin{prop}\label{prop.inf=min}
Let $\A\in\mathcal{L}(\R^d,\R^d)$ such that $\B\neq 0$ and $p\in\left(2,2^*\right)$, with $2^*=\frac{2d}{d-2}$. We let
\begin{equation}\label{lambdapA}
S=\inf_{\psi\in \Dom(\mathfrak Q_{\A}), \psi\neq 0}\frac{\mathfrak Q_{\A}(\psi)}{\|\psi\|^2_{\sL^p(\R^d)}}\,.
\end{equation}
Then, the infimum in \eqref{lambdapA} is attained.
\end{prop}
Note that $S>0$. Indeed, if $(\phi_{j})_{j\geq 1}$ is a minimizing sequence, normalized in $\sL^p$, such that $\mathfrak{Q}_{\A}(\phi_{j})\to 0$, we deduce that, by diamagnetism, $|\phi_{j}| \to 0$ in $\sH^1(\R^d)$. By the Sobolev embedding $\sH^1(\R^d)\subset \sL^p(\R^d)$, we get that $(\phi_{j})_{j\geq 1}$ goes to zero in $\sL^p(\R^d)$.
We prove this proposition in Chapter \ref{chapter-non-linear}, Section \ref{conc-comp} by using an alternative method to the concentration-compactness principle. Moreover, it is possible to prove that the minimizers of \eqref{lambdapA} have an exponential decay.
\begin{prop}\label{prop.expdecnl}
There exists $\alpha>0$ such that, for any minimizer $\psi$ of \eqref{lambdapA}, we have $e^{\alpha|\x|}\psi\in\sL^2(\R^d)$.
\end{prop}
We focus on the two dimensional case.
\begin{definition}
For $p\in(2,+\infty)$, we define
\begin{equation}\label{lambda0}
\lambda^{[0]}(p )=\lambda(\R^2, \A^{[0]}, p,1)=\inf_{\psi\in \Dom(\mathfrak{Q}_{\A^{[0]}}), \psi\neq 0}\frac{\mathfrak{Q}_{\A^{[0]}}(\psi)}{\|\psi\|^2_{\sL^p}}\,,
\end{equation}
where $\A^{[0]}(x,y)=\left(0, -x\right)$. Here 
\begin{align*}
\mathfrak{Q}_{\A^{[0]}}(\psi) = \int_{\R^2} |(-i\nabla +\A^{[0]})\psi|^2\dx \x\,,
\end{align*}
with domain
\begin{align*}
\Dom(\mathfrak{Q}_{\A^{[0]}})=\left\{ \psi \in \sL^2(\R^2)\,:\, (-i\nabla +\A^{[0]})\psi\in \sL^2(\R^2) \right\}\,.
\end{align*}
\end{definition}
Let us now state the main theorem of this section (the proof is given in Chapter \ref{chapter-non-linear}, Section \ref{sec.nla}).
\begin{theo}\label{theo1}
Let $p\geq 2$. Let us assume that $\A$ is smooth on $\overline{\Omega}$, that $\B=\nabla \times \A$ does not vanish on $\overline{\Omega}$ and that its minimum $b_{0}$ is attained in $\Omega$. Then there exist $C>0$ and $h_{0}>0$ such that, for all $h\in(0,h_{0})$,
$$(1-Ch^{\frac{1}{8}})\lambda^{[0]}(p )b^{\frac{2}{p}}_{0}h^2h^{-\frac{2}{p}}\leq\lambda(\Omega,\A, p, h)\leq (1+Ch^{1/2})\lambda^{[0]}(p)b_{0}^{\frac{2}{p}}h^2h^{-\frac{2}{p}}\,.$$
\end{theo}

\section{Non linear dynamics in waveguides}
Let us now discuss another non linear problem.

With the same formalism, we will consider the case of unbounded curves and the case of closed curves. Consider a smooth, simple curve $\Gamma$ in $\R^2$ defined by its normal parametrization $\gamma : x_{1}\mapsto \gamma(x_{1})$. For $\eps>0$ we introduce the map
\begin{equation}\label{Phi-NL}
\Phi_{\eps} : \mathcal{S}=\M\times (-1,1)\ni(x_{1},x_{2})\mapsto \gamma(x_{1})+\eps x_{2}\nu(x_{1})=\mathsf{x}\,,
\end{equation}
where $\nu(x_{1})$ denotes the unit normal vector at the point $\gamma(x_{1})$ such that $\det(\gamma'(x_{1}),\nu(x_{1}))=1$ and where
$$
\M=\left\{\begin{array}{ll}\R\quad&\mbox{for an unbounded curve,}\\
\T=\R/(2\pi\Z)\quad&\mbox{for a closed curve.}
\end{array}\right.
$$
We recall that the curvature at the point $\gamma(x_{1})$, denoted by $\kappa(x_{1})$, is defined by
$$\gamma''(x_{1})=\kappa(x_{1})\nu(x_{1})\,.$$
The waveguide is $\Omega_{\eps}=\Phi_{\eps}(\mathcal{S})$ and we will work under the following assumption which states that waveguide does not overlap itself and that $\Phi_{\eps}$ is a smooth diffeomorphism.
\begin{assumption}
\label{assumption1}
We assume that the function $\kappa$ is bounded, as well as its derivatives $\kappa'$ and $\kappa''$. Moreover, we assume that there exists $\eps_{0}\in (0,\frac{1}{\|\kappa\|_{\sL^\infty}})$ such that, for $\eps\in(0,\eps_{0})$, $\Phi_{\eps}$ is injective. 
\end{assumption}
We will denote by $-\Delta_{\Omega_{\eps}}^\Dir$ the Dirichlet Laplacian on $\Omega_{\eps}$. We are interested in the following equation:
\begin{equation}\label{CNLS}
i\dr_{t}\psi^\eps=-\Delta_{\Omega_{\eps}}^\Dir\psi^\eps+\lambda \eps^\alpha |\psi^\eps|^2\psi^\eps
\end{equation}
on $\Omega_\eps$ with a Cauchy condition $\psi^\eps(0;\cdot)=\psi^\eps_{0}$ and where $\alpha\geq 1$ and $\lambda\in \R$ are parameters.

By using the diffeomorphism $\Phi_{\eps}$, we may rewrite \eqref{CNLS} in the space coordinates $(x_{1},x_{2})$ given by \eqref{Phi-NL}. For that purpose, let us introduce $m_{\eps}(x_{1},x_{2})=1-\eps x_{2}\kappa(x_{1})$ and consider the function $\psi^\eps$ transported by $\Phi_{\eps}$, 
$$\mathcal{U}_{\eps}\psi^\eps(t; x_{1},x_{2})=\phi^\eps(t ; x_{1}, x_{2})=\eps^{1/2}m_{\eps}(x_{1},x_{2})^{1/2}\psi^\eps(t ; \Phi_{\eps}(x_{1},x_{2}))\,.$$
Note that $\mathcal{U}_{\eps}$ is unitary from $\sL^2(\Omega_{\eps}, \dx \mathsf{x})$ to $\sL^2(\mathcal{S}, \dx x_{1} \dx x_{2})$ and maps $\sH^1_0(\Omega_{\eps})$ (resp. $\sH^2(\Omega_\eps)$) to $\sH^1_0(\mathcal{S})$ (resp. to $\sH^2(\mathcal{S}))$. Moreover, the operator $-\Delta_{\Omega_{\eps}}^\Dir$ is unitarily equivalent to the self-adjoint operator on $\sL^2(\mathcal{S}, \dx x_{1} \dx x_{2})$,
\[\mathcal{U}_{\eps}(-\Delta^\Dir_{\Omega_{\eps}})\mathcal{U}^{-1}_{\eps}=\mathcal{H}_{\eps}+V_{\eps},\quad\mbox{ with } \mathcal{H}_{\eps}=\mathcal{P}^2_{\eps,1}+\mathcal{P}^2_{\eps,2}\,,\]
where $$\mathcal{P}_{\eps,1}=m_{\eps}^{-1/2}D_{x_{1}}m_{\eps}^{-1/2}, \qquad \mathcal{P}_{\eps,2}=\eps^{-1}D_{x_{2}}$$ and where the effective electric $V_\eps$ potential is defined by
\[V_{\eps}(x_{1},x_{2})=-\frac{\kappa(x_{1})^2}{4(1-\eps x_{2}\kappa(x_{1}))^{2}}\,.\]
Notice that, for all $\eps<\eps_0$, we have $m_\eps\geq 1-\eps_0\|\kappa\|_{\sL^\infty}>0$. The problem \eqref{CNLS} becomes
\begin{equation}\label{CNLS'}
i\dr_{t}\phi^\eps=\mathcal{H}_{\eps}\phi^\eps+V_{\eps}\phi^\eps+\lambda \eps^{\alpha-1}m_{\eps}^{-1} |\phi^\eps|^2\phi^\eps
\end{equation}
with Dirichlet boundary conditions $\phi^\eps(t;x_{1},\pm 1)=0$ and the Cauchy condition
$ \phi^\eps(\cdot ; 0)=\phi_{0}^\eps=\mathcal{U}_{\eps}\psi^\eps_0$. We notice that the domains of $\mathcal{H}_{\eps}$ and  $\mathcal{H}_{\eps}+V_{\eps}$ coincide with $\sH^2(\mathcal{S})\cap \sH^1_{0}(\mathcal{S})$.

In order to study \eqref{CNLS''}, it is natural to conjugate the equation by the unitary group $e^{it\mathcal{H}_{\eps}}$ so that the problem \eqref{CNLS''} becomes
\begin{equation}\label{CNLS''-conj}
i\dr_{t}\widetilde\phy^\eps=e^{it\mathcal{H}_{\eps}}(V_{\eps}-\eps^{-2}\mu_{1})e^{-it\mathcal{H}_{\eps}}\widetilde\phy^\eps+\lambda W_{\eps}(t;\widetilde\phy^\eps) ,\qquad \widetilde\phy^\eps(0;\cdot)=\phi_{0}^\eps\,,
\end{equation}
where
\begin{equation}\label{Weps}
W_{\eps}(t;\phy)=e^{it\mathcal{H}_{\eps}}m_{\eps}^{-1} |e^{-it\mathcal{H}_{\eps}}\phy|^2 e^{-it\mathcal{H}_{\eps}}\phy
\end{equation}
and where $\widetilde\phy^\eps=e^{it\mathcal{H}_{\eps}}\phy^\eps$ which satisfies $\widetilde\phy^\eps(t;x_{1},\pm 1)=0$.

We will analyze the critical case $\alpha=1$ where the nonlinear term is of the same order as the parallel kinetic energy associated to $D_{x_1}^2$. It is well-known that \eqref{CNLS} (thus \eqref{CNLS'} also) has two conserved quantities: the $\sL^2$ norm and the nonlinear energy. Let us introduce the first eigenvalue $\mu_1=\frac{\pi^2}{4}$ of $D_{x_2}^2$ on $(-1,1)$ with Dirichlet boundary conditions, associated to the eigenfunction $e_1(x_{2})=\cos\left(\frac{\pi}{2}x_{2}\right)$ and define the energy functional
\begin{multline}\label{Energy}
\mathcal{E}_{\eps}(\phi)=\frac{1}{2}\int_{\mathcal{S}}|\mathcal{P}_{\eps,1}\phi|^2 \dx x_{1} \dx x_{2}+\frac{1}{2}\int_{\mathcal{S}}|\mathcal{P}_{\eps,2}\phi|^2 \dx x_{1} \dx x_{2}+\frac{1}{2}\int_{\mathcal{S}}\left(V_{\eps}-\frac{\mu_1}{\eps^2}\right)|\phi|^2 \dx x_{1} \dx x_{2}\\
+\frac{\lambda}{4}\int_{\mathcal{S}}m_{\eps}^{-1}|\phi|^4 \dx x_{1} \dx x_{2}\,.
\end{multline}
Notice that we have substracted the conserved quantity $\frac{\mu_1}{2\eps^{2}}\|\phi\|_{\sL^2}^2$ to the usual nonlinear energy, in order to deal with bounded energies. Indeed, we will consider initial conditions with bounded mass and energy, which means more precisely the following assumption.
\begin{assumption}
\label{assumption2}
There exists two constants $M_0>0$ and $M_1>0$ such that the initial data $\phi_0^\eps$ satisfies, for all $\eps\in (0,\eps_0)$,
$$\|\phi^\eps_0\|_{\sL^2}\leq M_0 \quad \mbox{and}\quad \mathcal{E}_{\eps}(\phi_0^\eps)\leq M_1\,.$$
\end{assumption}
Let us define the projection $\Pi_1$ on $e_1$ by letting $\Pi_1 u=\langle u, e_1\rangle_{\sL^2((-1,1))} e_1$. A consequence of Assumption \ref{assumption2} is that $\phi_0^\eps$ has a bounded $\sH^1$ norm and is close to its projection $\Pi_1\phi_{0}^\eps$. Indeed, we will prove the following lemma (see Chapter \ref{NLWG}, Section \ref{sec.lbnl}).
\begin{lem}\label{cauchy-tensorial}
Assume that $\phi_0^\eps$ satisfies Assumption \ref{assumption2}. Then there exists $\eps_1(M_0)\in (0,\eps_0)$ and a constant $C>0$ independent of $\eps$ such that, for all $\eps\in(0,\eps_1(M_0))$,
\begin{equation}
\label{conf-NL}
\|\phi_0^\eps\|_{\sH^1(\mathcal S)}\leq C\quad\mbox{and}\quad \|\phi_0^\eps-\Pi_1\phi_{0}^\eps\|_{\sL^2(\M,\sH^1(-1,1))}\leq C\eps\,.
\end{equation}
\end{lem}
It will be convenient to consider the following change of temporal gauge $\phi^\eps(t; x_{1}, x_{2})=e^{-i\mu_{1}\eps^{-2} t}\phy^\eps(t; x_{1}, x_{2})$. This leads to the equation
\begin{equation}\label{CNLS''}
i\dr_{t}\phy^\eps=\mathcal{H}_{\eps}\phy^\eps+(V_{\eps}-\eps^{-2}\mu_{1})\phy^\eps+\lambda m_{\eps}^{-1} |\phy^\eps|^2\phy^\eps
\end{equation}
with conditions $\phy^\eps(t;x_{1},\pm 1)=0$, $\phy^\eps(0;\cdot)=\phi^\eps_{0}$.

We can now state the main theorem of this section (see Chapter \ref{NLWG}, Section \ref{sec.ge}).
\begin{theo}
\label{mainthmH2}
Assume that $\phi_0^\eps\in \sH^2\cap \sH^1_0(\mathcal S)$ and that there exist $M_0>0$, $M_2>0$ such that, for all $\eps\in (0,\eps_0)$, 
\begin{equation}
\label{ass3}
\|\phi_0^\eps\|_{\sL^2}\leq M_0,\qquad \left\|(\mathcal H_\eps-\frac{\mu_1}{\eps^2})\phi_0^\eps\right\|_{\sL^2}\leq M_2\,.
\end{equation}
Then $\phi_0^\eps$ satisfies Assumption \ref{assumption2} and, for all $\eps\in (0,\eps_1(M_0))$, \eqref{CNLS''} admits a unique solution $\phy^\eps\in \mathcal{C}(\R_+;\sH^2\cap \sH^1_0(\mathcal{S}))\cap \mathcal{C}^1(\R_+;\sL^2(\mathcal{S}))$. Moreover, there exists $C>0$ such that, for all $\eps\in (0,\eps_1(M_0))$ and $t\geq 0$, we have
\[\|\varphi^\eps(t)-\Pi_{1}\varphi^\eps(t)\|_{\sL^2}\leq C\eps \,.\]
\end{theo}

%\mainmatter
\part{Methods and examples}\label{Part.Models}

\chapter{Elements of spectral theory}\label{chapter-spectral-theory}

\begin{flushright}
\begin{minipage}{0.5\textwidth}
It will neither be necessary to deliberate nor to trouble ourselves, as if we shall do this thing, something definite will occur, but if we do not, it will not occur.
\begin{flushright}
\textit{Organon}, On Interpretation, Aristotle
\end{flushright}
\end{minipage}
\vspace*{0.5cm}
\end{flushright}

This chapter is devoted to recall basic tools in spectral analysis.

\section{Spectrum}
\subsection{Spectrum of an unbounded operator}
Let $\mathfrak{L}$ be an unbounded operator on a separable Hilbert space $(\sH,\langle\cdot,\cdot\rangle)$ with domain $\Dom(\mathfrak{L})$, dense in $\sH$. Let us first recall the following two definitions.
\begin{definition}
The operator $(\mathfrak{L},\Dom(\mathfrak{L}))$ is closed if and only if 
$$\Dom(\mathfrak{L}) \ni u_{n}\to u\in\sH,\qquad\mathfrak{L}u_{n}\to v\Longrightarrow u\in\Dom(\mathfrak{L}),\qquad \mathfrak{L}u=v\,.$$ 
\end{definition}

\begin{definition}
The adjoint of $(\mathfrak{L},\Dom(\mathfrak{L}))$ is defined as follows. We let 
$$\Dom(\mathfrak{L}^*):=\left\{u\in\Dom(\mathfrak{L}),\quad v\mapsto \langle \mathfrak{L}v,u\rangle\quad \mbox{is continuous on }\Dom(\mathfrak{L})\right\}$$
and, for $u\in\mathfrak{L}^*$, $\mathfrak{L}^*u$ is defined (thanks to the Riesz theorem) as the unique element in $\sH$ such that for all $v\in\sH$, $\langle \mathfrak{L}v,u\rangle=\langle v,\mathfrak{L}^*u\rangle$.

We say that $(\mathfrak{L},\Dom(\mathfrak{L}))$ is self-adjoint when $(\mathfrak{L}^*,\Dom(\mathfrak{L}^*))=(\mathfrak{L},\Dom(\mathfrak{L}))$.
\end{definition}

\begin{prop}
The operator $(\mathfrak{L}^*,\Dom(\mathfrak{L}^*))$ is always a closed operator (\textit{i.e.} with closed graph). If $(\mathfrak{L},\Dom(\mathfrak{L}))$ is closable, then $\Dom(\mathfrak{L}^*)$ is dense and $(\mathfrak{L}^*)^*=\overline{\mathfrak{L}}$, where $\overline{\mathfrak{L}}$ is the smallest closed extension of $\mathfrak{L}$.
\end{prop}

\begin{definition}
An operator is said to be Fredholm if its kernel is finite dimensional, its range is closed and with finite codimension.
\end{definition}

We now recall the following definitions of its spectrum $\sp(\mathfrak{L})$, its essential spectrum $\spe(\mathfrak{L})$ and its discrete spectrum $\spd(\mathfrak{L})$.
\begin{definition}
We define
\begin{enumerate}
\item the spectrum: $\lambda\in\sp(\mathfrak{L})$ if and only if $(\mathfrak{L}-\lambda \,\Id)$ is not invertible, with bounded inverse, from $\Dom(\mathfrak{L})$ onto $\sH$, 
\item the essential spectrum: $\lambda\in\spe(\mathfrak{L})$  if and only if  $(\mathfrak{L}-\lambda \,\Id)$ is not Fredholm from $\Dom(\mathfrak{L})$ into $\sH$,
\item the discrete spectrum: $\spd(\mathfrak{L}) = \sp(\mathfrak{L})\setminus\spe(\mathfrak{L})$.
\end{enumerate}
\end{definition}
Note that we have obviously $\spe(\mathfrak{L})\subset \sp(\mathfrak{L})$. For the convenience of the reader, let us recall the proof of some classical lemmas (see \cite[Chapter VI]{ReSi75} and \cite[Chapter 3]{L-B03}) which can also be treated as exercises.
\begin{lem}
If $\mathfrak{L}$ is self-adjoint, we have the equivalence: $\lambda\in\sp(\mathfrak{L})$ if and only if there exists a sequence $(u_{n})\in \Dom(\mathfrak{L})$ such that $\|u_{n}\|_{\sH}=1$, $(u_{n})$ and $(\mathfrak{L}-\lambda \,\Id)u_{n}\underset{n\to+\infty}{\to }0$ in $\sH$.
\end{lem}
\begin{proof}
Let us notice that if there exists a sequence $(u_{n})\in \Dom(\mathfrak{L})$ such that $\|u_{n}\|_{\sH}=1$, $(u_{n})$ and $(\mathfrak{L}-\lambda \,\Id)u_{n}\underset{n\to+\infty}{\to }0$ then $\lambda\in\sp(\mathfrak{L})$ (if not we could apply the bounded inverse and get a contradiction).

If $\lambda\notin\R$, since $\mathfrak{L}$ is self-adjoint, $\mathfrak{L}-\lambda$ is invertible (with bounded inverse since $\mathfrak{L}$ is closed). Now, for $\lambda\in\R$, if there is no sequence $(u_{n})\in \Dom(\mathfrak{L})$ such that $\|u_{n}\|_{\sH}=1$, $(u_{n})$ and $(\mathfrak{L}-\lambda \,\Id)u_{n}\underset{n\to+\infty}{\to }0$, then we can find $c>0$ such that
$$\|(\mathfrak{L}-\lambda)u\|\geq c\|u\|,\qquad \forall u\in\Dom(\mathfrak{L})\,.$$
Therefore $\mathfrak{L}-\lambda$ is injective with closed range. But, since $\mathfrak{L}-\lambda=(\mathfrak{L}-\lambda)^*$, the range of $\mathfrak{L}-\lambda$ is dense in $\sH$ and so $\mathfrak{L}-\lambda$ is surjective.

\end{proof}

\begin{lem}[Weyl criterion]\label{Weyl}
If $\mathfrak{L}$ is self-adjoint, we have the equivalence: $\lambda\in\spe(\mathfrak{L})$ if and only if there exists a sequence $(u_{n})\in \Dom(\mathfrak{L})$ such that $\|u_{n}\|_{\sH}=1$, $(u_{n})$ has no subsequence converging in $\sH$ and $(\mathfrak{L}-\lambda \,\Id)u_{n}\underset{n\to+\infty}{\to }0$ in $\sH$.
\end{lem}
\begin{proof}
If $\lambda\in\sp(\mathfrak{L})\setminus\sp_{\ess}(\mathfrak{L})$, the operator $\mathfrak{L}-\lambda$ is Fredholm. Let $(u_{n})\in \Dom(\mathfrak{L})$ such that $\|u_{n}\|_{\sH}=1$ and $(\mathfrak{L}-\lambda \,\Id)u_{n}\underset{n\to+\infty}{\to }0$. The operator $\mathfrak{L}-\lambda : \ker(\mathfrak{L}-\lambda)^\perp\to \mathrm{range}(\mathfrak{L}-\lambda)$ is injective with closed range. Therefore, there exists $c>0$ such that, for all $w\in\ker(\mathfrak{L}-\lambda)^\perp$, $\|(\mathfrak{L}-\lambda)w\|\geq c\|w\|$. We write $u_{n}=v_{n}+w_{n}$, with $v_{n}\in\ker(\mathfrak{L}-\lambda)$ and $w_{n}\in\ker(\mathfrak{L}-\lambda)^\perp$. We have $\|(\mathfrak{L}-\lambda)u_{n}\|^2=\|(\mathfrak{L}-\lambda)v_{n}\|^2+\|(\mathfrak{L}-\lambda)w_{n}\|^2$ and we deduce that $w_{n}\to 0$. Moreover $(v_{n})$ is bounded in finite dimension, thus there exists a converging subsequence of $(u_{n})$.

Conversely, let us assume that $\lambda\in\sp(\mathfrak{L})$ and that all sequence $(u_{n})\in \Dom(\mathfrak{L})$ such that $\|u_{n}\|_{\sH}=1$ and  $(\mathfrak{L}-\lambda)u_{n}\underset{n\to+\infty}{\to }0$ has a converging subsequence.The kernel $\ker(\mathfrak{L}-\lambda)$ is finite dimensional. Indeed, if it were of infinite dimension, one could construct a infinite orthonormal family $(u_{n})$ of $\ker(\mathfrak{L}-\lambda)$ and in particular we would get $u_{n}\rightharpoonup 0$ that is a contradiction. Let us now check that there exists $c>0$ such that, for all $u\in\ker(\mathfrak{L}-\lambda)^\perp$, $\|(\mathfrak{L}-\lambda)u\|\geq c\|u\|$. If not, there exists  a normalized sequence $(u_{n})$ in $\ker(\mathfrak{L}-\lambda)^\perp$ such that $\|(\mathfrak{L}-\lambda)u_{n}\|\to 0$. By assumption, we may assume that $(u_{n})$ converges towards some $u_{\infty}$ that necessarily belongs to $\ker(\mathfrak{L}-\lambda)^\perp$. But since $\mathfrak{L}-\lambda$ is closed (it is self-adjoint), we have $(\mathfrak{L}-\lambda)u_{\infty}=0$ so that $u_{\infty}=0$ and this is a contradiction. We deduce that the image of $\mathfrak{L}-\lambda$ is closed.

\end{proof}

\begin{lem}\label{isolated}
If $\mathfrak{L}$ is self-adjoint, the discrete spectrum is formed by isolated eigenvalues of finite multiplicity and conversely.
\end{lem}
\begin{proof}
Let us consider $\lambda\in\sp(\mathfrak{L})\setminus\spe(\mathfrak{L})$. There exists a Weyl sequence $(u_{n})$ of unit vectors such that $(\mathfrak{L}-\lambda)u_{n}\to 0$. We may assume that $(u_{n})$ converges to some $u$ and we get $(\mathfrak{L}-\lambda)u=0$. The eigenvalue $\lambda$ has finite multiplicity. Let us prove that it is isolated. If it were not the case, then one could consider a non stationnary sequence $\lambda_{n}$ tending to $\lambda$. Moreover, one could find a sequence $(u_{n})$ of unit vectors such that:
$$\|(\mathfrak{L}-\lambda_{n})u_{n}\|\leq\frac{|\lambda-\lambda_{n}|}{n}\,.$$
By assumption, we may assume that $(u_{n})$ converges towards some $u$ and thus one would get $(\mathfrak{L}-\lambda)u=0$ and so
$$\langle(\mathfrak{L}-\lambda_{n})u,u_{n}\rangle=(\lambda-\lambda_{n})\langle u,u_{n}\rangle\,.$$
By the Cauchy-Schwarz inequality, we find that $\langle u_{n},u\rangle\to 0$ and we get $u=0$ that is a contradiction.

For the converse, we have just to prove that the image of $\mathfrak{L}-\lambda$ is closed when $\lambda$ is an isolated eigenvalue of finite multiplicity. This fact is a consequence of the spectral theorem (see Theorem \ref{spectral-theorem0}).

\end{proof}

\subsection{The example of the magnetic Laplacian}

\subsubsection{Recalling the Lax-Migram theorem}
Let us recall the famous Lax-Milgram theorem that will allows the definition of many operators in this book. We refer to the book by Helffer \cite[Section 3.3]{Hel13} for a proof.
\begin{theo}[Lax-Milgram]\label{Lax-Milgram}
Let us consider two Hilbert spaces $\sV$ and $\sH$ such that $\sV\subset \sH$ with continuous injection and with $\sV$ dense in $\sH$. If $\mathfrak{B}$ is a continuous sesquilinear form on $\sV$ that is coercive, \textit{i.e.}
$$\exists \alpha>0,\quad\mathfrak{B}(u,u)\geq \alpha\|u\|^2_{\sV}\,,$$
then we may define an operator $(\mathfrak{L},\Dom(\mathfrak{L}))$ whose domain is
$$\Dom(\mathfrak{L}):=\left\{u\in\sV : v\mapsto \mathfrak{B}(u,v) \mbox{ is continuous on } \sV \mbox{ for the topology of $\sH$}\right\}$$
and such that, for $u\in\Dom(\mathfrak{L})$,
$$\mathfrak{B}(u,v)=\langle\mathfrak{L}u,v\rangle_{\sH},\quad\forall v\in\sV\,.$$
The operator $\mathfrak{L}:\Dom(\mathfrak{L})\to\sH$ is bijective and its inverse is continuous. Moreover $\Dom(\mathfrak{L})$ is dense in $\sH$.

If $\mathfrak{B}$ is also Hermitian, then $\mathfrak{L}$ is self-adjoint and its domain is dense in $\sV$.

\end{theo}

Note that this theorem is directly related to the Friedrichs procedure (see for instance \cite[p. 177]{ReSi75}).

\subsubsection{The Dirichlet realization}
Let us consider the following sesquilinear form, defined for $u,v\in\sV=\sH^1_{0}(\Omega)$, by:
$$\mathfrak{B}_{h,\A}(u,v)=\int_{\Omega} (-ih\nabla+\A)u\, \overline{(-ih\nabla+\A)v}\dx x\,.$$
We have obviously
$$\forall u\in\sV,\quad\mathfrak{B}_{h,\A}(u,u)+\|u\|^2_{\sH}\geq \|u\|_{\sH}^2$$
and this involves the coercivity on $\sV$. For this shifted sesquilinear form, $\sV$ is an Hilbert space.
Here the domain of $\mathfrak{L}$ is given by
$$\Dom(\mathfrak{L}^\Dir_{h,\A})=\left\{u\in \sH^1_{0}(\Omega) : \mathfrak{L}_{h,\A}u\in \sL^2(\Omega)   \right\}\,.$$
The self-adjoint operator $\mathfrak{L}=\mathfrak{L}^\Dir_{h,\A}$ satisfies
$$\langle \mathfrak{L}^\Dir_{h,\A}u,v\rangle=\mathfrak{B}_{h,\A}(u,v),\quad \forall u\in\Dom(\mathfrak{L}_{h,\A}^\Dir),\quad \forall v\in\sH^1_{0}(\Omega)\,.$$
When $\Omega$ is regular, we have the characterization:
$$\Dom(\mathfrak{L}^\Dir_{h,\A})=\sH^1_{0}(\Omega)\cap \sH^2(\Omega)\,.$$
Note that we could have defined the initial quadratic form on $\mathfrak{C}^\infty_{0}(\Omega)$ but this space is not complete for the $\sH^1_{0}(\Omega)$-norm. Completing $\mathfrak{C}^\infty_{0}(\Omega)$ for the norm induced by the quadratic form and then defining the self-adjoint operator $\mathfrak{L}$ is called the Friedrichs procedure.
\subsubsection{The Neumann realization}
We consider the other quadratic form defined by: 
$$\mathfrak{Q}_{h,\A}(u)=\int_{\Omega} |(-ih\nabla+\A)u|^2\dx x, \quad u\in \sH^1(\Omega)\,.$$
We can define a self-adjoint operator $\mathfrak{L}_{h,\A}^\Neu$ whose domain is given by:
$$\Dom(\mathfrak{L}^\Neu_{h,\A})=\left\{u\in \sH^1(\Omega) : \mathfrak{L}_{h,\A}u\in \sL^2(\Omega), (-ih\nabla+\A)u\cdot\bfn=0,\, \mbox{ on }\dr\Omega  \right\}\,.$$
When $\Omega$ is regular, this becomes:
$$\Dom(\mathfrak{L}^\Neu_{h,\A})=\left\{u\in \sH^1(\Omega) : u\in \sH^2(\Omega), (-ih\nabla+\A)u\cdot\bfn=0,\, \mbox{ on }\dr\Omega  \right\}\,.$$

\subsubsection{Riesz-Fr\'echet-Kolmogorov criterion and compact resolvent}
Let us recall a criterion of relative compactness in $\sL^p$ (see \cite{Brezis}). 
\begin{theo}[Riesz-Fr\'echet-Kolmogorov]\label{RFK}
Let $\Omega\subset\R^N$ be an open set and $\mathcal{F}$ a bounded subset of $\sL^p(\Omega)$, with $p\in[1,+\infty)$. We assume that
$$\forall \eps>0,\, \exists \omega\subset\subset\Omega,\quad \forall f\in\mathcal{F},\quad \|f\|_{\sL^p(\Omega\setminus\omega)}\leq\eps$$
and that
$$\forall\eps>0, \forall  \omega\subset\subset\Omega,\,\exists \delta>0,\quad \delta<\dist(\omega,\complement\Omega),\quad \forall |h|\leq\delta,\quad  \forall f\in\mathcal{F}, \quad\|\tau_{h}f-f\|_{\sL^p(\omega)}\leq\eps\,,$$
where $\tau_{h}f(x)=f(x+h)-f(x)$.
\end{theo}

By using a density argument and the Taylor formula, we can get the following proposition (see \cite[Proposition 9.3]{Brezis}).
\begin{prop}\label{quotient}
Let $p\in(1,+\infty)$ and $u\in\sL^p(\Omega)$. Then $u\in \sW^{1,p}(\Omega)$ if and only if, for all $\omega\subset\subset \Omega$ and $h\in(0,\dist(\omega,\complement\Omega))$, we have
$$\|\tau_{h}u\|_{\sL^p(\omega)}\leq C|h|\,.$$
In this case, we can take $C=\|\nabla u\|_{\sL^p(\Omega)}$.
\end{prop}

Let us provide a useful criterion for the compactness of a resolvent.
\begin{prop}
An operator $(\mathfrak{L},\Dom(\mathfrak{L}))$ has compact resolvent if and only if the injection $(\Dom(\mathfrak{L}),\|\cdot\|_{\mathfrak{L}})\hookrightarrow \sH$ is compact.
\end{prop}
\begin{proof}
Thanks to the closed graph theorem, for $z\notin\sp(\mathfrak{L})$, $(\mathfrak{L}-z)^{-1} : (\sH,\|\cdot\|_{\sH})\to (\Dom(\mathfrak{L}),\|\cdot\|_{\mathfrak{L}})$ is bounded.
\end{proof}
\begin{prop}\label{criterion-comp}
Let us consider two Hilbert spaces $\sV$ and $\sH$ such that $\sV\subset \sH$ with continuous injection and with $\sV$ dense in $\sH$. Assume that $\mathfrak{B}$ is a continuous, coercive and Hermitian sesquilinear form on $\sV$ and if $\mathfrak{L}$ denotes the self-adjoint operator associated with $\mathfrak{B}$. Let us denote by $\|\cdot\|_{\mathfrak{B}}$ the norm induced by $\mathfrak{B}$, \textit{i.e.} $\|u\|_{\mathfrak{B}}=\sqrt{\mathfrak{B}(u,u)}$, and by $\|\cdot\|_{\mathfrak{L}}$ the graph norm on $\Dom(\mathfrak{L})$.

If $(\Dom(\mathfrak{B}),\|\cdot\|_{\mathfrak{B}})\hookrightarrow \sH$ is compact, then $\mathfrak{L}$ has compact resolvent.
\end{prop}
\begin{proof}
By the Cauchy-Schwarz inequality, $(\Dom(\mathfrak{L}),\|\cdot\|_{\mathfrak{L}})\to(\Dom(\mathfrak{B}),\|\cdot\|_{\mathfrak{B}})$ is bounded. The conclusion follows since the compact operators form a ideal of the set of bounded operators.
\end{proof}

\subsubsection{Reminder about compact operators}
In the following theorem, we recall some fundamental facts about compact operators. In particular, we will notice that the non zero spectrum of a compact operator is discrete.
\begin{theo}[About the Fredholm alternative]
Let $\mathfrak{L}\in\mathfrak{L}(\sH)$ be a compact operator. Then, we have
\begin{enumerate}
\item If $\sH$ is of infinite dimension, then $0\in\sp(\mathfrak{L})$.
\item $\dim\left(\ker(\mathfrak{L}-\mathsf{Id})\right)$ is finite.
\item $\mathrm{range}(\mathfrak{L}-\mathsf{Id})$ is closed.
\item $\ker(\mathfrak{L}-\mathsf{Id})=\{0\}$ iff $\mathrm{range}(\mathfrak{L}-\mathsf{Id})=\sH$.
\item If $\lambda\in\sp(\mathfrak{L})\setminus\{0\}$, $\mathfrak{L}-\lambda$ is a Fredholm operator (and thus $\lambda$ belongs to the discrete spectrum).
\item The elements of $\sp(\mathfrak{L})\setminus\{0\}$ are isolated and the only accumulation point of the spectrum is $0$.
\end{enumerate}
\end{theo}

If $\Omega$ is bounded and Lipschitzian, the form domains $\sH^1_{0}(\Omega)$ and $\sH^{1}(\Omega)$ are compactly embedded in $\sL^2(\Omega)$ (their unit balls satisfy the assumptions of the Riesz-Fr\'echet-Kolmogorov criterion, see \cite{Brezis} for details). Therefore $\mathfrak{L}_{h,\A}^\Dir$ and $\mathfrak{L}_{h,\A}^\Neu$ have compact resolvents. Therefore these operators have discrete spectra. We can consider the non decreasing sequences of their eigenvalues.

\begin{exe}
We consider $\mathfrak{L}=\mathfrak{L}_{h,\A}^\Dir$ when $\Omega$ is bounded and regular. Let us take $\lambda$ an eigenvalue of $\mathfrak{L}$ ($\lambda\in\R$ since $\mathfrak{L}$ is self-adjoint). As we said $\ker(\mathfrak{L}-\lambda)$ has finite dimension. Since $P$ is self-adjoint, we can write:
$$\overline{\mathrm{range}(\mathfrak{L}-\lambda)}=\ker(\mathfrak{L}-\lambda)^\perp\,.$$
Prove that the image of $\mathfrak{L}-\lambda$ is closed by using that $K=(\mathfrak{L}-\lambda+i)^{-1}$ is compact. 
\end{exe}

\section{Min-max principle and spectral theorem}
\subsection{Statement of the theorems}
We state a theorem which will be one of the fundamental tools in this book.
\begin{theo}\label{spectral-theorem0}
Let us assume that $(\mathfrak{L},\Dom(\mathfrak{L}))$ is a self-adjoint operator. Then, if $\lambda\notin\sp(\mathfrak{L})$, we have:
\[\|(\mathfrak{L}-\lambda)^{-1}\|\leq \frac{1}{\dist(\lambda,\sp(\mathfrak{L}))}\,.\]
\end{theo}
\begin{rem}
A proof using the \enquote{spectral theorem} can be found in \cite{ReSi78} and \cite[Section VI.5]{Kato66} . An immediate consequence of this theorem is that, for all $\psi\in\Dom (\mathfrak{L})$:
\[\|\psi\|\dist(\lambda,\sp(\mathfrak{L}))\leq\|(\mathfrak{L}-\lambda)\psi\|\,.\]
In particular, if we find $\psi\in \Dom (\mathfrak{L})$ such that $\|\psi\|=1$ and $\|(\mathfrak{L}-\lambda)\psi\|\leq \eps$, we get: $\dist(\lambda,\sp(\mathfrak{L}))\leq \eps$.
\end{rem}

\begin{proof}
This result may be proved without the general spectral theorem. Let us provide the elements of the proof.
Let us first establish the result when $\mathfrak{L}$ is bounded and normal (\textit{i.e.} $[\mathfrak{L},\mathfrak{L}^*]=0$). For that purpose, we will use the results of the following exercises.

\begin{exe} 
If $P$ is a polynomial, we have $\lambda\in\sp(\mathfrak{L})$ iff $P(\lambda)\in\sp(\mathfrak{L})$. 
\end{exe}

\begin{exe}
We define the spectral radius as
\[\rho(\mathfrak{L})=\sup_{\lambda\in\sp(\mathfrak{L})}|\lambda|\,.\]
\begin{enumerate}
\item By using the convergence of a Neumann series, prove that
\[\rho(\mathfrak{L})=\limsup_{n\to+\infty}\|\mathfrak{L}^n\|^{\frac{1}{n}}\]
and then $\rho(\mathfrak{L})=\inf_{n} \|\mathfrak{L}^n\|^{\frac{1}{n}}$. 
\item By using $\|\mathfrak{L}^*\mathfrak{L}\|=\|\mathfrak{L}\|^2$, prove that $\rho(\mathfrak{L})=\|\mathfrak{L}\|$ and deduce $\|P(\mathfrak{L})\|=\|P\|_{\infty}$ where $\|\cdot\|_{\infty}$ is the uniform norm on the spectrum $K$ of $\mathfrak{L}$ that is compact. 
\item By using the Stone-Weierstrass theorem, extend this equality to continuous functions on $K$. If $f$ is a continuous function on $K$, explain how we may define $f(\mathfrak{L})$.
\end{enumerate}
\end{exe}
If $\lambda\notin\sp(\mathfrak{L})$, the function $r : K\ni z\mapsto (z-\lambda)^{-1}$ is continuous and the result follows when $\mathfrak{L}$ is bounded and normal as soon as we have noticed that $r(\mathfrak{L})=(\mathfrak{L}-\lambda)^{-1}$.

Now, let us assume that $\mathfrak{L}$ is self-adjoint with domain $\Dom(\mathfrak{L})$. This is not difficult to prove that $\mathfrak{L}\pm i\,\Id$ is invertible. We introduce the function, called Cayley transform,
\[g(x)=\frac{x-i}{x+i},\qquad x\in\R\]
and the bounded and unitary operator 
\[g\left(\mathfrak{L}\right):=U:=\left(\mathfrak{L}-i\mathsf{Id}\right)(\mathfrak{L}+i\mathsf{Id})^{-1}\,.\]
In particular $g(\mathfrak{L})$ is normal. Easy computations provide that $g: \sp\left(\mathfrak{L}\right)\mapsto\sp\left(g(\mathfrak{L})\right)$ is bijective. Then, for $\mu\notin\sp(\mathfrak{L})$, we define, on $\sp\left(g\left(\mathfrak{L}\right)\right)$,
\[f(y)=\frac{1}{g^{-1}(y)-\mu}\,.\]
From the case of bounded and normal operators, we infer that
\[\|f(g(\mathfrak{L}))\|\leq \|f\|_{\infty,\sp(g(\mathfrak{L}))}=\|(\cdot-\mu)^{-1}\|_{\infty,\sp(\mathfrak{L})}\,.\]
It remains to write that $f(g(\mathfrak{L}))=(\mathfrak{L}-\mu)^{-1}$ by noticing that $g^{-1}(U)(\mathsf{Id}-U)=i(U+\mathsf{Id})$ (that implies that $g^{-1}(U)=\mathfrak{L}$ on $\Dom(\mathfrak{L})$) and $(g^{-1}(U)-\mu)f(U)=\mathsf{Id}$, where $g^{-1}(U)$ is understood in the sense of functional calculus of bounded and normal operators.

\end{proof}
As a consequence of the proof of Theorem \ref{spectral-theorem0}, we may deduce the Stone theorem.
\begin{theo}\label{theo.Stone}
Let $(\mathfrak{L},\Dom(\mathfrak{L}))$ a self-adjoint operator. For all $\psi_{0}\in\Dom(\mathfrak{L})$, there exists a unique local $\mathcal{C}^1$-solution $t\mapsto S(t)\psi_{0}$ of the equation
\[\psi'(t)=i\mathfrak{L}\psi(t),\qquad \psi(0)=\psi_{0}\,.\]
This solution is global and, for all $t\in\R$, $\|S(t)\psi_{0}\|=\|\psi_{0}\|$. For all $t\in\R$ and for all $\psi_{0}\in\Dom(\mathfrak{L})$, we have $S(t)\psi_{0}\in\Dom(\mathfrak{L})$. We denote $S(t)=e^{it\mathfrak{L}}$ and $(e^{it\mathfrak{L}})_{t\in\R}$ is a semi-group.
\end{theo}
\begin{proof}
We let $S(t)=e^{itg^{-1}(U)}$, where $g^{-1}(U)$ is defined in the proof of Theorem \ref{spectral-theorem0}. We have $S'(t)=ig^{-1}(U)S(t)=iS(t)g^{-1}(U)$ so that, for $\psi\in\Dom(\mathfrak{L})$, $S'(t)\psi=iS(t)\mathfrak{L}\psi$. Let us prove that, for all $t\in\R$ and $\psi\in\Dom(\mathfrak{L})$, we have $S(t)\psi\in\Dom(\mathfrak{L})$ and that $\mathfrak{L}S(t)\psi=S(t)\mathfrak{L}\psi$.
We have, for all $\varphi\in\sH$ and $\psi\in\Dom(\mathfrak{L})$,
\[\langle g^{-1}(U)e^{itg^{-1}(U)}\varphi,\psi\rangle=\langle e^{itg^{-1}(U)}\varphi,\mathfrak{L}\psi\rangle\,.\]
This implies, by definition, that  $e^{itg^{-1}(U)}\varphi\in\Dom(\mathfrak{L}^*)=\Dom(\mathfrak{L})$ and that $\mathfrak{L} e^{itg^{-1}(U)}\varphi= g^{-1}(U)e^{itg^{-1}(U)}\varphi$. The proof of the uniqueness and of the group property is left to the reader.
\end{proof}

\begin{exe}\label{funct-calc} The aim of this exercise is to investigate the functional calculus of a simple self-adjoint operator on $\sL^2(\R)$ and provide an explicit functional calculus.
Let us recall the expression of the Fourier transform on $\R$. For $\psi\in\mathcal{S}(\R)$, we let, for all $\xi\in\R$,
\[\mathcal{F}\psi(\xi)=\frac{1}{\sqrt{2\pi}}\int_{\R}e^{-ix\xi}\psi(x)\dx x\,.\]
It is well-known that $\mathcal{F}$ extends to an isometry of $\sL^2(\R)$ and that, for all $\psi\in\mathcal{S}'(\R)$,
\[\mathcal{F}(D_{x}\psi)=\xi\mathcal{F}(\psi)\]
that may be written as $\mathcal{F}D_{x}\mathcal{F}^{-1}=\xi$. In other words, the self-adjoint operator $D_{x}$ with domain $\sH^1(\R)$ is diagonalized thanks to the Fourier transform. 

Let us now consider a smooth function on $\R$ denoted by $\delta$ bounded as well as its derivatives and such that there exists $\delta_{0}>0$ such that $\delta\geq\delta_{0}$.
\begin{enumerate}
\item Solve the equation $\delta D_{x}(\delta\psi)=\xi\psi$ for $\xi\in\R$.
\item For $\psi\in\mathcal{S}(\R)$, we let
\[\mathcal{F}_{\delta}(\psi)(\xi)=\frac{1}{\sqrt{2\pi}}\int_{\R} \delta(x)^{-1}e^{i\xi\int_{0}^x\delta^{-2}(y)\dx y}\psi(x)\dx x\,.\]
Prove that $\mathcal{F}_{\delta}$ is unitary in $\sL^2(\R)$.
\item Prove that it diagonalizes the operator $\delta D_{x}\delta$ and that $\mathcal{F}^{-1}_{\delta} D_{\xi}\mathcal{F}_{\delta}=\int_{0}^x \delta^{-2}(y) \dx y$.
\end{enumerate}

\end{exe}

We now give a standard method to estimate the discrete spectrum and the bottom of the essential spectrum of a self-adjoint operator $\mathfrak{L}$ on an Hilbert space $\sH$.
We recall first the definition of the Rayleigh quotients of a self-adjoint operator $\mathfrak{L}$. 

\begin{definition}
The Rayleigh quotients associated with the self-adjoint operator $\mathfrak{L}$ on $\sH$ of domain $\Dom(\mathfrak{L})$ are defined for all positive natural number $n$ by
$$\mu_n (\mathfrak{L})= \sup_{\substack {\psi_1,\ldots,\psi_{n-1}} } \ \ \inff_{\underset{u\in\Dom(\mathfrak{L}), u\neq 0}{u\in\spann(\psi_1,\ldots,\psi_{n-1})^\perp} }\frac{\langle \mathfrak{L}u,u\rangle_\sH} {\langle u,u\rangle_\sH} \,.$$
\end{definition}

\begin{lem}\label{positivity-spec}
If $\mathfrak{L}$ is self-adjoint with non negative spectrum, we have $\mu_{1}(\mathfrak{L})\geq 0$.
\end{lem}
\begin{proof}
Let us assume that $\mu_{1}(\mathfrak{L})<0$.
We may define the sesquilinear form $\mathcal{B}(u,v)=\langle(\mathfrak{L}-\mu_{1}(\mathfrak{L}))^{-1}u,v\rangle$ on $\sH$ and it is non negative. Thus, the Cauchy-Schwarz inequality provides, for $u,v\in\sH$,
$$|\langle(\mathfrak{L}-\mu_{1}(\mathfrak{L}))^{-1}u,v\rangle|\leq \langle(\mathfrak{L}-\mu_{1}(\mathfrak{L}))^{-1}u,u\rangle^{\frac{1}{2}} \langle(\mathfrak{L}-\mu_{1}(\mathfrak{L}))^{-1}v,v\rangle^{\frac{1}{2}}\,.$$
We take $v=(\mathfrak{L}-\mu_{1}(\mathfrak{L}))^{-1}u$ and deduce for all $u\in\sH$,
$$\|(\mathfrak{L}-\mu_{1}(\mathfrak{L}))^{-1}u\|\leq \|(\mathfrak{L}-\mu_{1}(\mathfrak{L}))^{-1}\|^{\frac{1}{2}}\langle(\mathfrak{L}-\mu_{1}(\mathfrak{L}))^{-1}u,u\rangle^{\frac{1}{2}}\,.$$
and thus, for all $v\in\Dom(\mathfrak{L})$,
$$\|v\|\leq \|(\mathfrak{L}-\mu_{1}(\mathfrak{L}))^{-1}\|^{\frac{1}{2}}\langle v,(\mathfrak{L}-\mu_{1}(\mathfrak{L}))v\rangle^{\frac{1}{2}}\,.$$
By definition of $\mu_{1}(\mathfrak{L})$ we may find a sequence $(v_{n})$ such that $\|v_{n}\|=1$ such that $\langle \mathfrak{L}v_{n},v_{n}\rangle\to \mu_{1}(\mathfrak{L})$ and we get a contradiction.
\end{proof}

The following statement gives the relation between Rayleigh quotients and eigenvalues.
\begin{theo}
\label{th:1-1}
Let $\mathfrak{L}$ be a self-adjoint operator of domain $\Dom(\mathfrak{L})$. We assume that $\mathfrak{L}$ is semi-bounded from below.
Then the Rayleigh quotients $\mu_n$ of $\mathfrak{L}$ form a non-decreasing sequence and one of the following holds
\begin{enumerate}
\item $\mu_n(\mathfrak{L})$ is the $n$-th eigenvalue (counted with mutliplicity) eigenvalue of $\mathfrak{L}$ and $\mathfrak{L}$ has only discrete spectrum in $(-\infty,\mu_{n}(\mathfrak{L})]$.
\item $\mu_{n}(\mathfrak{L})$ is the bottom of the essential spectrum and, for all $j\geq n$, $\mu_{j}(\mathfrak{L})=\mu_{n}(\mathfrak{L})$.
\end{enumerate}
\end{theo}
\begin{proof}
Let us provide an elementary proof which does not use the so-called spectral projections. First it is easy to see that the sequence $(\mu_{n})$ is non-decreasing. Then, we notice that
\begin{equation}\label{crit-dis}
a<\mu_{n}\Longrightarrow(-\infty, a)\cap\spe(\mathfrak{L})=\emptyset\,.
\end{equation}
Indeed, if $\lambda\in(-\infty,a)$ were in the essential spectrum, by Lemma \ref{isolated} and thanks to Weyl sequences, for all $N\geq 1$ and $\eps>0$, we could find an orthonormal family $(u_{j})_{j\in\{1,\ldots,N\}}$ such that $\|(\mathfrak{L}-\lambda)u_{j}\|\leq\frac{\eps}{\sqrt{N}}$. Then, given $n\geq 1$ and taking $N\geq n$, for all $(\psi_{1},\ldots,\psi_{n-1})\in\sH$, there exists a non zero $u$ in the intersection $\spann(u_{1},\ldots,u_{N})\cap\spann(\psi_{1},\ldots,\psi_{n-1})^\perp$. We write $u=\sum_{j=1}^N \alpha_{j} u_{j}$ and notice that
$$\frac{\langle \mathfrak{L}u,u\rangle_\sH} {\langle u,u\rangle_\sH}\leq \lambda+\frac{\|(\mathfrak{L}-\lambda)u\|}{\|u\|}\leq\lambda+ \left(\sum_{j=1}^N \|(\mathfrak{L}-\lambda)u_{j}\|^2\right)^{\frac{1}{2}}\leq\lambda+\eps$$
and thus $\mu_{n}\leq \lambda+\eps$. For $\eps$ small enough, we get $\mu_{n}\leq a$, that is a contradiction. If $\gamma$ is the infimum of the essential spectrum (suppose that it is not empty), we have $\mu_{n}\leq\gamma$. Note also that if $\mu_{n}=+\infty$ for some $n$, then the essential spectrum is empty. This implies the second point.

It remains to prove the first point. Thus, we assume that $\mu_{n}<\gamma$. By using the same considerations as above, if $a<\mu_{n}$, the number of eigenvalues (with multiplicity) lying in $(-\infty,a)$ is less than $n-1$. Let us finally show that, if $a\in(\mu_{n},\gamma)$, then the number of eigenvalues in $(-\infty, a)$ is at least $n$. If not the direct sum of eigenspaces associated with eigenvalues below $a$ would be generated by $\psi_{1},\ldots,\psi_{n-1}$ and 
$$\mu_{n}\geq  \inff_{\underset{u\in\Dom(\mathfrak{L}), u\neq 0}{u\in\spann(\psi_1,\ldots,\psi_{n-1})^\perp} }\frac{\langle \mathfrak{L}u,u\rangle_\sH} {\langle u,u\rangle_\sH}\geq a\,,$$
where we have used Lemma \ref{positivity-spec} and that $\sp(\mathfrak{L}_{| F})\subset[a,+\infty)$, with $F=\spann(\psi_{1},\ldots,\psi_{n-1})^\perp$.
\end{proof}

A consequence of this theorem (or of its proof) which is often used is the following proposition.
\begin{prop}
Suppose that there exists $a\in\R$ and an $n$-dimensional space $\mathsf{V}\subset \Dom \mathfrak{L}$ such that
\[\langle \mathfrak{L}\psi,\psi\rangle_{\sH}\leq a\|\psi\|^2\,.\]
Then, we have:
\[\lambda_{n}(\mathfrak{L})\leq a\,.\]
\end{prop}

\subsection{Examples of applications}
Let us provide some applications of the min-max principle. 
\subsubsection{Sturm-Liouville's theory}
We consider the following operator $D_{x} g(x) D_{x}+V(x)$, with $g, V\in\mathcal{C}^{\infty}([0,1])$, $g\geq c>0$ on $[0,1]$ and domain
\[\left\{\psi\in\sH^1_{0}((0,1)) : (D_{x} g(x) D_{x}+V(x))\psi\in\sL^2((0,1))\right\}\,.\]
It is clearly a self-adjoint operator, denoted by $\mathfrak{L}$, with compact resolvent. Therefore, we may consider the non decreasing sequence of its eigenvalues $(\lambda_{n})_{n\geq 1}$. By the Cauchy-Lipshitz theorem, we also notice that these eigenvalues are simple. For all $n\geq 1$, let us consider an eigenfunction $u_{n}$ associated with $\lambda_{n}$. Notice that $\langle u_{n}, u_{m}\rangle=0$ if $n\neq m$ and that the zeros of $u_{n}$ are simple and thus isolated.
\begin{prop}\label{ST}
For all $n\geq 1$, the function $u_{n}$ admits exactly $n-1$ zeros in $(0,1)$.
\end{prop}
\begin{proof}
Let us denote by $Z_{n}$ the number of zeros of $u_{n}$ in $(0,1)$. 

Let us prove that $Z_{n}\leq n-1$. If the eigenfunction $u_{n}$ admits at least $n$ zeros in $(0,1)$, denoted by $z_{1}, \ldots, z_{n+1}$ and we may define $(u_{n,j})_{j=0,\ldots, n}$ by $u_{n,j}(x)=u_{n}(x)$ for $x\in[z_{j}, z_{j+1}]$ and $u_{n,j}(x)=0$ elsewhere. It is clear that these functions belong to the form domain of $\mathcal{L}$ and that they form an orthogonal family. We may establish (by using an integration by parts) that
\[\forall v\in\spann_{j\in\{0,\ldots, n\}}u_{n,j}, \qquad\mathfrak{Q}(v)\leq \lambda_{n}\|v\|^2_{\sL^2((0,1))}\,.\]
By the min-max principle, we get $\lambda_{n+1}\leq \lambda_{n}$ and this contradicts the simplicity of the eigenvalues.

Let us now prove that $Z_{n}\geq Z_{n-1}+1$. It is sufficient to prove that if $u_{n-1}$ is zero in $z_{0}$ and $z_{1}$, then $u_{n}$ vanishes in $(z_{0}, z_{1})$. Indeed, this would imply that $u_{n}$ vanishes at least $Z_{n-1}+1$ times. For that purpose we introduce $W(f_{1},f_{2})=g\left(f_{1}'f_{2}-f_{1}f_{2}'\right)$ and compute 
\[W(u_{n-1}, u_{n})'=(\lambda_{n}-\lambda_{n-1})u_{n-1} u_{n}\,.\]
We have $W(u_{n-1}, u_{n})(z_{0})=W(u_{n-1}, u_{n})(z_{1})=0$, thus $W(u_{n-1}, u_{n})'$ vanishes somewhere in $(z_{0}, z_{1})$ and so does $u_{n}$.

The conclusion follows easily.
\end{proof}

\subsubsection{Another example coming from spheric coordinates}
\begin{notation}
For $\alpha\in\left(0,\pi\right)$, let us consider the operator on $\sL^2\left(\left(0,\frac{1}{2}\right), \sin(\alpha\varphi)\dx\varphi\right)$ 
defined by
\[\mathfrak{P}_{\alpha}=-\frac{1}{\sin(\alpha\varphi)}\dr_{\varphi}\sin(\alpha\varphi)\dr_{\varphi}\,,\]
with domain
\begin{multline*}
\Dom\left(\mathfrak{P}_{\alpha}\right)=\Big\{\psi\in \sL^2\left((0,\tfrac{1}{2}),\sin(\alpha\varphi)\dx\varphi\right)\,, \\
\frac{1}{\sin(\alpha\varphi)}\dr_{\varphi}\sin(\alpha\varphi)\dr_{\varphi}\psi\in  \sL^2\left((0,\tfrac{1}{2}),\sin(\alpha\varphi)\dx\varphi\right)\,, 
\dr_{\varphi}\psi\left(\tfrac{1}{2}\right)=0, \ \psi(0)=0\Big\}.
\end{multline*}
We denote by $\nu_{1}(\alpha)$ its first eigenvalue.
\end{notation}
The aim of this section is to establish the following lemma.
\begin{lem}\label{lowest-ev-dir}
There exists $c_{0}>0$ such that for all $\alpha\in(0,\pi)$:
\[\nu_{1}(\alpha)\geq c_{0}\,.\]
\end{lem}
\begin{proof}
We consider the associated quadratic form $\mathfrak{p}_{\alpha}$:
\[\mathfrak{p}_{\alpha}(\psi)=\int_{0}^{\frac{1}{2}} \sin(\alpha\varphi) |\dr_{\varphi}\psi|^2\dx\varphi\,.\]
We have the elementary lower bound:
\begin{equation*}
\mathfrak{p}_{\alpha}(\psi)
\geq\int_{0}^{\frac{1}{2}} \alpha\varphi\left(1-\frac{(\alpha\varphi)^2}{6}\right) |\dr_{\varphi}\psi|^2\dx\varphi
\geq\frac{1}{2}\int_{0}^{\frac{1}{2}} \alpha\varphi  |\dr_{\varphi}\psi|^2\dx\varphi\,,
\end{equation*}
since $0\leq \alpha\varphi\leq \frac\pi2$. 
We are led to analyze the lowest eigenvalue $\gamma\geq 0$ of the operator on $\sL^2\left(\left(0,\frac{1}{2}\right),\varphi \dx\varphi\right)$ defined by $-\frac 1\varphi\dr_{\varphi}\varphi\dr_{\varphi}$ with Dirichlet condition at $\varphi=0$ and Neumann condition at $\varphi=\frac{1}{2}$. Let us prove that $\gamma>0$. If it were not the case, the corresponding eigenvector $\psi$ would satisfy:
\[-\frac 1\varphi\dr_{\varphi}\varphi\dr_{\varphi}\psi=0\,,\]
so that
\[\psi(\varphi)=c\ln\varphi+d,\quad\mbox{ with }c,d\in\R\,.\]
The boundary conditions provide $c=d=0$ and thus $\psi=0$. By contradiction, we infer that $\gamma>0$. We deduce that
\[\mathfrak{p}_{\alpha}(\psi)\geq \frac{\gamma}{2}\int_{0}^{\frac{1}{2}} \alpha\varphi  |\psi|^2\dx\varphi
\geq \frac{\gamma}{2}\int_{0}^{\frac{1}{2}} \sin(\alpha\varphi)  |\psi|^2\dx\varphi\,.\]
By the min-max principle, we conclude that, for all $\alpha\in(0,\pi)$,
\[\nu_{1}(\alpha)\geq\frac{\gamma}{2}=:c_{0}>0\,.\]
\end{proof}

\subsubsection{An example with small magnetic field}
In this section, we let 
\[\Omega= \mathcal{B}(0,1)\subset\R^2\,,\qquad\A_{0}(x_{1}, x_{2})=\frac{1}{2}\left(x_{2},-x_{1}\right)\,,\]
and we consider the magnetic Neumann Laplacian $\mathfrak{L}^\Neu_{\alpha\A_{0}}$ on $\Omega$ with $\alpha>0$.
\begin{prop}\label{prop.sB}
If $\mu(\alpha)$ denotes the lowest eigenvalue of $\mathfrak{L}^\Neu_{\alpha\A_{0}}$, we have
\[\mu(\alpha)=\frac{\alpha^2}{|\Omega|}\int_{\Omega}|\A_{0}(\x)|^2\dx\x+\mathcal{O}(\alpha^{\frac{5}{2}})\,.\]
\end{prop}
\begin{proof}
Let us first notice that $\A_{0}(\x)\cdot \n(\x)=0$ on $\partial\Omega$ and that $\nabla\cdot \A_{0}=0$. Therefore the magnetic Neumann condition $(-i\nabla+\alpha\A_{0})\psi\cdot\n=0$ becomes $\nabla\psi\cdot\n=0$ on $\partial\Omega$. In particular, we notice that the domain is independent from $\alpha$ (due to our special choice of gauge). By using the test function $\psi=1$ and the min-max principle, we get
\[\mu(\alpha)\leq\frac{\alpha^2}{|\Omega|}\int_{\Omega}|\A_{0}(\x)|^2\dx\x\,.\]
Let us now consider a $\sL^2$-normalized eigenfunction $\psi_{\alpha}$ associated with $\mu(\alpha)$. We have
\[\int_{\Omega}|(-i\nabla+\alpha\A_{0})\psi_{\alpha}|^2\dx\x=\mu(\alpha)=\mathcal{O}(\alpha^2)\,.\]
By using a classical inequality, we get that, for all $\eps>0$, we have
\[\int_{\Omega}|(-i\nabla+\alpha\A_{0})\psi_{\alpha}|^2\dx\x\geq (1-\eps)\|\nabla\psi_{\alpha}\|^2_{\sL^2(\Omega)}-\eps^{-1}\alpha^2|\Omega|\max_{\x\in\Omega} |\A_{0}(\x)|^2\,.\]
Taking $\eps=\alpha$, we deduce that
\[\|\nabla\psi_{\alpha}\|^2_{\sL^2(\Omega)}=\mathcal{O}(\alpha)\,.\]
We have
\[\|\nabla\psi_{\alpha}\|^2_{\sL^2(\Omega)}=\left\|\nabla\left(\psi_{\alpha}-\frac{1}{|\Omega|}\int_{\Omega}\psi_{\alpha}\dx\x\right)\right\|^2_{\sL^2(\Omega)}\geq \lambda_{2}(-\Delta^\Neu, \Omega)\left\|\psi_{\alpha}-\frac{1}{|\Omega|}\int_{\Omega}\psi_{\alpha}\dx\x\right\|^2_{\sL^2(\Omega)}\,,\]
where $\lambda_{2}(-\Delta^\Neu, \Omega)$ is the second Rayleigh quotient associate with the Neumann Laplacian on $\Omega$ and we used the fact that $\psi_{\alpha}-\frac{1}{|\Omega|}\int_{\Omega}\psi_{\alpha}\dx\x$ is orthogonal to the constant functions and the min-max principle. We now use that the first eigenvalue of the Neumann Laplacian (that is $0$) on $\Omega$ is simple and associated with the constant functions. This fact will be explained in general in Section \ref{sec.Harnack} and is also known as the Poincar\'e inequality. We deduce that
\begin{equation}\label{eq.approx-alpha}
\left\|\psi_{\alpha}-\underline{\psi_{\alpha}}\right\|_{\sL^2(\Omega)}=\mathcal{O}(\alpha^{\frac{1}{2}})\,,\qquad \underline{\psi_{\alpha}}=\frac{1}{|\Omega|}\int_{\Omega}\psi_{\alpha}\dx\x\,.
\end{equation}
We have, for all $\eps>0$,
\begin{multline*}
\int_{\Omega}|(-i\nabla+\alpha\A_{0})\psi_{\alpha}|^2\dx\x\\
\geq (1-\eps)\int_{\Omega}|-i\nabla\psi_{\alpha}+\alpha\A_{0}\underline{\psi_{\alpha}}|^2\dx\x-\eps^{-1}\alpha^2|\Omega|\max_{\x\in\Omega} |\A_{0}(\x)|^2\left\|\psi_{\alpha}-\underline{\psi_{\alpha}}\right\|^2_{\sL^2(\Omega)}\,.
\end{multline*}
Then we notice that
\begin{multline*}
\int_{\Omega}|-i\nabla\psi_{\alpha}+\alpha\A_{0}\underline{\psi_{\alpha}}|^2\dx\x\\
=\|\nabla\psi_{\alpha}\|^2_{\sL^2(\Omega)}+\alpha^2|\underline{\psi_{\alpha}}|^2\int_{\Omega}|\A_{0}(\x)|^2\dx\x+2\alpha\Im\overline{\underline{\psi_{\alpha}}}\langle\nabla\psi_{\alpha},\A_{0}\rangle_{\sL^2(\Omega)}\,.
\end{multline*}
Thanks to the Green-Riemann formula and the fact that $\nabla\cdot\A_{0}=0$ and $\A_{0}\cdot\n=0$, we get
\[\langle\nabla\psi_{\alpha},\A_{0}\rangle_{\sL^2(\Omega)}=0\,,\]
so that we find
\[\int_{\Omega}|-i\nabla\psi_{\alpha}+\alpha\A_{0}\underline{\psi_{\alpha}}|^2\dx\x\geq\alpha^2|\underline{\psi_{\alpha}}|^2|\Omega|\int_{\Omega}|\A_{0}(\x)|^2\dx\x\,.\]
We take $\eps=\alpha^{\frac{1}{2}}$ and, with \eqref{eq.approx-alpha}, we deduce that
\[\mu(\alpha)=\int_{\Omega}|(-i\nabla+\alpha\A_{0})\psi_{\alpha}|^2\dx\x\geq\alpha^2(1-\alpha^{\frac{1}{2}})|\underline{\psi_{\alpha}}|^2\int_{\Omega}|\A_{0}(\x)|^2\dx\x-C\alpha^{\frac{5}{2}}\,.\]
By using again \eqref{eq.approx-alpha}, we get that
\[|\underline{\psi_{\alpha}}|=\frac{1}{|\Omega|^{\frac{1}{2}}}+\mathcal{O}(\alpha^{\frac{1}{2}})\,,\]
and the conclusion follows.
\end{proof}
\begin{rem}
The result of Proposition \ref{prop.sB} may be easily generalized to smooth domains by choosing a vector potential $\A_{0}$ such that
\[\nabla\cdot\A_{0}=0\,,\quad\mbox{ in }\Omega\,,\qquad \A_{0}\cdot\n=0\,,\quad\mbox{ on }\partial\Omega\,.\]
Such a vector potential may be found by minimizing $\int_{\Omega}|\A-\nabla\varphi|^2\dx\x$ for $\varphi\in\sH^1(\Omega)$ for the initial $\A$ such that $\nabla\times\A=B$. Moreover, for this particular choice of vector potential, the domain does not depend on the parameter $\alpha$ and we may apply the analytic perturbation theory (see Chapter \ref{chapter-examples}, Section \ref{Sec.analytic}) to get the analyticity of $\mu(\alpha)$.
\end{rem}

\subsection{Persson's theorem}
Let us give a characterization of the bottom of the essential spectrum in the Schr\"odinger case (see \cite{Persson60} and also \cite{FouHel10}).

\begin{theo}\label{Persson}
Let $V$ be real-valued, semi-bounded potential and $\A\in\mathcal{C}^1(\R^n)$ a magnetic potential. Let $\mathfrak{L}_{\A,V}$ be the corresponding self-adjoint, semi-bounded Schr\"odinger operator. Then, the bottom of the essential spectrum is given by:
$$\inf\spe(\mathfrak{L}_{\A,V})=\Sigma(\mathfrak{L}_{\A,V})\,,$$
where:
$$\Sigma(\mathfrak{L}_{\A,V})=\sup_{K\subset\R^n}\left[\inf_{\|\phi\|=1}\langle \mathfrak{L}_{\A,V}\phi,\phi\rangle_{\sL^2} \,|\,\phi\in\mathcal{C}^\infty_{0}(\R^n\setminus K)\right]\,.$$
\end{theo}
Let us notice that generalizations including the presence of a boundary are possible.

In fact, we will not really need this theorem in this book, but only the following criterion.
\begin{prop}\label{Persson-improved}
Let $\Omega\subset\R^d$ a non empty open set. Let us consider a quadratic form $\mathfrak{Q}$ defined on the dense subset $\Dom(\mathfrak{Q})\subset\sL^2(\Omega)$, bounded from below by $1$ and such that $(\Dom(\mathfrak{Q}),\sqrt{\mathfrak{Q}(\cdot)})$ is an Hilbert space. We denote by $(\mathfrak{L},\Dom(\mathfrak{L}))$ the corresponding self-adjoint operator. For all $R>0$, we let $\Omega_{R}=\Omega\cap B(0,R)$ and $\iota_{R} : \psi\to \psi_{|\Omega_{R}}$. 

We assume that
\begin{enumerate}
\item For all $M\geq 0$ and $R>0$, $\iota_{R}\left(\left\{\psi\in\Dom(\mathfrak{Q}) : \mathfrak{Q}(\psi)\leq M\right\}\right)$ is a precompact part of $\sL^2(\Omega_{R})$.
\item\label{PI-ii} For all smooth cutoff function $\chi$ supported in a neighborhood of $0$ and for all $\psi\in\Dom(\mathfrak{Q})$, $\chi\psi\in\Dom(\mathfrak{Q})$. Moreover, for all smooth cutoff function $0\leq \chi\leq 1$ being $0$ in $B(0,1)$ and $1$ on $\complement B(0,2)$ and for all $\eps>0$, there exists $R_{0}>0$ such that for all $R\geq R_{0}$ and all $\psi\in\Dom(\mathfrak{L})$,
$$\mathfrak{Q}(\chi_{R}\psi)\leq\langle\mathfrak{L}\psi,\chi_{R}^2\psi\rangle_{\sL^2(\Omega)}+\eps\|\psi\|^2_{\sL^2(\Omega)},\qquad\mbox{ with } \chi_{R}(x)=\chi\left(R^{-1}x\right)$$
\item There exist $\mu\in\R$ and $R_{0}>0$ such that for all $R\geq R_{0}$, all $\psi\in\Dom(\mathfrak{Q})$ and all $\chi$ supported in $\complement B(0,R)$,
$$\mathfrak{Q}(\chi\psi)\geq \mu\|\chi\psi\|^2_{\sL^2(\Omega)}\,.$$
\end{enumerate}
Then we have $\inf \sp_{\ess}\left(\mathfrak{L}\right)\geq\mu$.
\end{prop}
\begin{proof}
Let us consider $\lambda\in\sp(\mathfrak{L})$ with $\lambda<\mu$. We shall prove that $\lambda$ is in the discrete spectrum. Let us introduce a sequence $(\psi_{n})_{n\geq 0}\subset\Dom(\mathfrak{L})$ such that we have 
$$\|\psi_{n}\|_{\sL^2(\Omega)}=1\quad\mbox{ and }\quad\|(\mathfrak{L}-\lambda)\psi_{n}\|_{\sL^2(\Omega)}\to 0\,.$$ 
Let us show that $(\psi_{n})_{n\geq 0}$ is precompact in $\sL^2(\Omega)$. There exists $N\geq 0$ such that, for all $n\geq N$, $\|(\mathfrak{L}-\lambda)\psi_{n}\|_{\sL^2(\R)}\leq \eps$. Then we notice that that there exists $R_{0}>0$ such that for all $R\geq R_{0}$ and all $n\geq N$, we have
$$\mathfrak{Q}(\chi_{R}\psi_{n})\leq\langle\mathfrak{L}\psi_{n},\chi_{R}^2\psi_{n}\rangle_{\sL^2(\Omega)}+\eps\|\psi_{n}\|^2_{\sL^2(\Omega)}$$
so that
$$\mathfrak{Q}(\chi_{R}\psi_{n})\leq \lambda\|\chi_{R}\psi_{n}\|^2_{\sL^2(\Omega)}+2\eps\,.$$
We get 
$$(\mu-\lambda)\|\chi_{R}\psi_{n}\|^2_{\sL^2(\Omega)}\leq 2\eps\,.$$
Up to choosing $R_{0}$ larger, we have, for all $R\geq R_{0}$ and all $n\in\N$,
$$(\mu-\lambda)\|\chi_{R}\psi_{n}\|^2_{\sL^2(\Omega)}\leq 2\eps$$
that implies
$$(\mu-\lambda)\|\psi_{n}\|^2_{\sL^2(\Omega\cap\complement B(0,2R_{0}))}\leq 2\eps$$
Now, we use the precompactness of the sequence $\left(\iota_{2R_{0}}(\psi_{n})\right)_{n\in\N}$ and this enough to conclude that $(\psi_{n})$ is a precompact part of $\sL^2(\Omega)$. It remains to use Proposition \ref{Weyl}.
\end{proof}

\begin{exe}
Prove the lower bound of the infimum of the essential spectrum in Theorem \ref{Persson} by using Proposition \ref{Persson-improved}. 
\end{exe}

The following exercise may be done to prepare the understanding of the next proposition.
\begin{exe}
If $\A\in\sL^2_{\loc}$, we recall that $\sH^1_{\A}(\R^d)$ denotes
\[\{\psi\in\sL^2(\R^d) : (-i\nabla+\A)\psi\in\sL^2(\R^d)\}\,.\]
\begin{enumerate}
\item Prove that, equipped with the scalar product,
\[\langle\phi,\psi\rangle_{\sH^1_{\A}(\R^d)}=\int_{\R^d}(-i\nabla+\A)\phi\,\overline{(-i\nabla+\A)\psi}\dx\x+\int_{\R^d}\phi\,\overline{\psi}\dx\x\,,\]
it is a Hilbert space.
\item Prove that $\mathcal{C}^\infty_{0}(\R^d)$ is dense in $\sH^1_{\A}(\R^d)$.
\end{enumerate}
\end{exe}
Let us now provide an estimate of the essential spectrum of an electro-magnetic Laplacian when we assume that the electric potential is \enquote{small} (sufficiently integrable) at infinity.
\begin{prop}\label{prop.elecnl}
Let $d\geq 3$. Let us consider $\A\in\sL^2_{\loc}(\R^d,\R^d)$ and $V\in\sL^{\frac{p}{p-2}}(\R^d)$, where $p\in(2,2^*)$ and $2^*=\frac{2d}{d-2}$. Then, the quadratic form, defined for $\varphi\in\sH^1_{\A}(\R^d)$, by
\[\mathfrak{Q}_{\A,V}(\varphi)=\int_{\R^d}|(-i\nabla+\A)\varphi|^2\dx\x+\int_{\R^d} V|\varphi|^2\dx\x\,,\]
is well-defined, bounded from below and closed. Moreover, for all $\eps>0$, there exists $R>0$ such that, for all $\psi\in\sH^1_{\A}(\R^d)$ with $\supp\,\psi\subset\complement D(0,R)$, we have 
\begin{equation}\label{lbnlAV}
\mathfrak{Q}_{\A,V}(\psi)\geq (1-\eps)\mathfrak{Q}_{\A}(\psi)\,.
\end{equation}
If $\A$ is linear, we have $\inf\spe(\mathfrak{L}_{\A,V})\geq \sup_{1\leq k<\ell\leq d}|B_{k\ell}|$ where $\mathfrak{L}_{\A,V}$ is the operator associated with the (closed) form $\mathfrak{Q}_{\A,V}$.

When $d=2$, the same results hold if $V\in\sL^q(\R^d)$ for some $q>1$.
\end{prop}
\begin{proof}
First we notice that, for $\psi\in\sH^1_{\A}(\R^d)$,
\[\int_{\R^d} |V||\psi|^2\dx\x\leq\|V\|_{\sL^{\frac{p}{p-2}}(\R^d)}\|\psi\|^2_{\sL^p(\R^d)}\leq C\|V\|_{\sL^{\frac{p}{p-2}}(\R^d)}\||\psi|\|^2_{\sH^{1}(\R^d)} \,,\]
where we used the H¬\"older inequality and the Sobolev embedding $\sH^1(\R^d)\subset \sL^p(\R^d)$.
Then, the diamagnetic inequality implies that
\[\int_{\R^d} |V||\psi|^2\dx\x\leq C\|V\|_{\sL^{\frac{p}{p-2}}(\R^d)}\mathfrak{Q}_{\A}(\psi)\,.\]
Thus the quadratic form is well-defined on $\sH^1_{\A}(\R^d)$. Note that, this argument also show that, if $\psi$ is supported in $\complement D(0,R)$, we have
\[\int_{\R^d} |V||\psi|^2\dx\x\leq C\|V\|_{\sL^{\frac{p}{p-2}}(\complement D(0,R))}\mathfrak{Q}_{\A}(\psi)\,.\]
This implies \eqref{lbnlAV}. Let us now prove that the quadratic form is bounded from below. We have
\[\int_{\R^d} |V||\psi|^2\dx\x\leq C\|V\|_{\sL^{\frac{p}{p-2}}(\R^d)}\|\psi\|^2_{\sL^p(\R^d)}\,,\]
and by interpolation, we get
\[\|\psi\|^2_{\sL^p(\R^d)}\leq\|\psi\|^{2\theta}_{\sL^2(\R^d)}\|\psi\|^{2(1-\theta)}_{\sL^{2^*}(\R^d)}\,,\]
for $\theta\in(0,1)$ defined by $\frac{1}{p}=\frac{\theta}{2}+\frac{1-\theta}{2}$. With the Sobolev embedding and the diamagnetic inequality, we get
\[\|\psi\|^2_{\sL^p(\R^d)}\leq C\|\psi\|^{2\theta}_{\sL^2(\R^d)}\mathfrak{Q}_{\A}(\psi)^{1-\theta}\,.\]
We recall the convexity inequality
\[\forall a,b\geq 0,\quad \forall \theta\in(0,1),\qquad a^{\theta}b^{1-\theta}\leq \theta a+(1-\theta)b\,,\]
that implies
\[\forall\eps,\quad\forall a,b\geq 0,\quad \forall \theta\in(0,1),\qquad a^{\theta}b^{1-\theta}\leq \theta \eps^{-1}a+(1-\theta)\eps^{\frac{\theta}{1-\theta}} b\,,\]
It follows that
\[\|\psi\|^2_{\sL^p(\R^d)}\leq C\left(\theta \eps^{-1}\|\psi\|^2_{\sL^2(\R^d)}+(1-\theta)\eps^{\frac{\theta}{1-\theta}}\mathfrak{Q}_{\A}(\psi)\right)\]
and the lower bound follows by taking $\eps$ small enough.

The estimate of the essential spectrum comes from Proposition \ref{Persson-improved} (\eqref{PI-ii} comes from the formula \eqref{loc-chi} that will be proved later) and Proposition \ref{minoration-B}.

We leave the case $d=2$ as an exercise.
\end{proof}

\section{Simplicity and Harnack's inequality}\label{sec.Harnack}
This section is devoted to establish the simplicity of the lowest eigenvalue of operators in the form $-\Delta+V$. For that purpose, we will use the following version of the Harnack inequality.

\begin{prop}\label{Harnack}
Let $\Omega$ be an non empty open set of $\R^d$ and $V\in\mathcal{C}^\infty(\Omega)$. Let us fix a ball $D\subset\subset\Omega$. Then, there exists $C>0$ such that, for all positive function $u\in\mathcal{C}^\infty(\Omega)$ solution of 
$$(-\Delta+V)u=0,\qquad\mbox{ on }\Omega\,,$$
we have
$$\max_{D} u\leq C\min_{D} u\,.$$
\end{prop}

\begin{proof}
Let us provide an elementary proof inspired by the presentation by Evans (see \cite[p. 351]{Evans10}). Let $D\subset\subset\Omega$ and $\chi$ a smooth cutoff function supported in $\Omega$ and being $1$ in a neighborhood of $D$. For notation simplicity, in this proof, we will denote by $C$ all the constants that only depends on $\Omega$, $V$, $\chi$ and $D$.

We write $v=\ln u$ and notice that, on $\Omega$,
\begin{equation}\label{eq-Deltav}
-\Delta v-|\nabla v|^2+V=0\,.
\end{equation}
We let $w=|\nabla v|^2$. We want to get a bound on $w$ on $\overline{D}$ that only depends on $D$ and $V$. We consider $\x_{0}\in\Omega$ where the maximum of $z=\chi^4 w$ is attained. Note here that the presence of the cutoff function is due to the fact that we do not know if the maximum of $w$ is reached in $\Omega$ (it might be on the boundary). If $v$ is not constant on $D$,   we have $z(\x_{0})>0$ and thus $\chi(\x_{0})>0$, $w(\x_{0})>0$. Indeed, if $z(\x_{0})=0$, we get that, for all $\x\in D$, $w(\x)=0$. Therefore we assume that $z(\x_{0})>0$.

Since $z$ is maximal at $\x_{0}\in\Omega$, we have
\begin{equation}\label{z-max}
\nabla z(\x_{0})= 0,\qquad \Delta z(\x_{0})\leq 0\,.
\end{equation}
We deduce from the second inequality that
$$\chi^4(\x_{0})\Delta w(\x_{0})\leq -(\Delta \chi^4)(\x_{0}) w(\x_{0})-2\nabla \chi^4(\x_{0}) \cdot \nabla w(\x_{0})\,.$$
By using the first equality in \eqref{z-max} and $\chi(\x_{0})\neq 0$, we find
$$\chi(\x_{0})\nabla w(\x_{0})+4w(\x_{0})\nabla\chi(\x_{0})=0\,.$$
We find
\begin{equation}\label{Deltaw-w}
\chi^4(\x_{0})\Delta w(\x_{0})\leq Cw(\x_{0})\,.
\end{equation}
We obtain by a simple computation
$$\Delta w=2|\nabla^2 v|^2+2\sum_{k=1}^d \left(\partial_{k} V-\partial_{k}w\right)\partial_{k}v\,,$$
where $|\nabla^2 v|^2$ is the sum of the squares of the elements of the Hessian matrix of $v$.
In particular, on $D$, we have
$$\Delta w\geq 2|\nabla^2 v|^2-C|\nabla v|-2\nabla w\cdot\nabla v $$
and, at $\x_{0}$, we find, by using again the first equality in \eqref{z-max},
$$\chi^4(\x_{0})\Delta w(\x_{0})\geq  2\chi^4(\x_{0})|\nabla^2 v(\x_{0})|^2-Cw(\x_{0})^{\frac{1}{2}}-C\chi^3(\x_{0})w(\x_{0})^{\frac{3}{2}}\,.$$
With \eqref{Deltaw-w} we get
$$\chi^4(\x_{0}) |\nabla^2 v(\x_{0})|^2\leq C\chi^3(\x_{0})w(\x_{0})^{\frac{3}{2}}+Cw(\x_{0})+Cw(\x_{0})^{\frac{1}{2}}$$
Then, by using \eqref{eq-Deltav}, we find $w^2(\x_{0})\leq C+|\nabla^2 v(\x_{0})|^2$ and thus
$$\chi^4(\x_{0}) w(\x_{0})^2\leq C\chi^3(\x_{0})w(\x_{0})^{\frac{3}{2}}+Cw(\x_{0})+Cw(\x_{0})^{\frac{1}{2}}+C\chi^4(\x_{0})\,.$$
We infer
$$(\chi^4(\x_{0}) w(\x_{0})-C\chi^3(\x_{0})w(\x_{0})^{\frac{1}{2}})w(\x_{0})\leq C\chi^4(\x_{0})+Cw(\x_{0})\,.$$
If $\chi^4(\x_{0}) w(\x_{0})-C\chi^3(\x_{0})w(\x_{0})^{\frac{1}{2}}\leq 0$, then $\chi(\x_{0})w(\x_{0})^{\frac{1}{2}}\leq C$ and the reader can go to \eqref{bound-unif-z}. If not, we can write
$$(\chi^4(\x_{0}) w(\x_{0})-C\chi^3(\x_{0})w(\x_{0})^{\frac{1}{2}})\chi^4(\x_{0}) w(\x_{0})\leq C\chi^4(\x_{0})+Cw(\x_{0})\,.$$
so that
$$\left(\chi^4(\x_{0}) w(\x_{0})-\frac{C\chi^3(\x_{0})w(\x_{0})^{\frac{1}{2}}}{2}\right)^2\leq C\chi^4(\x_{0})+Cw(\x_{0})\,.$$
We find
$$\chi^4(\x_{0}) w(\x_{0})\leq C\chi^2(\x_{0})+Cw(\x_{0})^{\frac{1}{2}}$$
and we can play the same game to find
$$\chi^4(\x_{0}) w(\x_{0})^{\frac{1}{2}}\leq C\chi^2(\x_{0})\,.$$
In any case, we get
\begin{equation}\label{bound-unif-z}
\chi^4(\x_{0}) w(\x_{0})\leq C\,.
\end{equation}
In particular, since $\x_{0}$ is the maximum of $z$, we get
$$\forall \x\in D,\qquad |\nabla v(\x)|^2\leq C\,.$$
We infer that
$$\forall\x,\mathbf{y}\in D,\qquad |v(\x)-v(\mathbf{y})|\leq \sqrt{C}|\x-\mathbf{y}|\leq  \sqrt{C} \mathsf{diam}(D)$$
so that
$$\forall\x,\mathbf{y}\in D,\qquad \frac{u(\x)}{u(\mathbf{y})}\leq C$$
and the conclusion follows.
\end{proof}

\begin{cor}
Let $\Omega$ be an non empty open set of $\R^d$ and $V\in\mathcal{C}^\infty(\Omega)$. Let us fix a ball $D\subset\subset\Omega$. Then, there exists $C>0$ such that, for all function $u\geq 0$ belonging to $\mathcal{C}^\infty(\Omega)$ and solution of 
$$(-\Delta+V)u=0,\qquad\mbox{ on }\Omega\,,$$
we have
$$\max_{D} u\leq C\min_{D} u\,.$$
\end{cor}
\begin{proof}
We apply Proposition \ref{Harnack} to $u_{\eps}=u+\eps$ and make $\eps$ go to $0$.
\end{proof}

\begin{cor}\label{cor:Harnack}
Let $\Omega$ be an non empty open connected set of $\R^d$ and $V\in\mathcal{C}^\infty(\Omega)$. We also assume that $V\geq 1$. We define 
$$\left\{\psi\in\sH^1(\Omega) : \sqrt{V}\psi\in\sL^2(\Omega)\right\}\,.$$
and the quadratic form
$$\forall \psi\in\Dom(\mathfrak{Q}_{V}),\qquad\mathfrak{Q}_{V}(\psi)=\int_{\Omega} |\nabla\psi|^2+V(\x)|\psi|^2 \dx \x\,.$$
If $\mathfrak{L}_{V}$ denotes the associated self-adjoint operator, and if the infimum of its spectrum belongs to the discrete spectrum, then it is simple and there exists an associated eigenfunction that is positive on $\Omega$.
\end{cor}

\begin{proof}
We first notice that (see \cite[Proposition 2.1.2]{FouHel10}), if $\psi\in\sH^1(\Omega)$, then we have $|\psi|\in\sH^1(\Omega)$ and 
$$\|\nabla|\psi|\|_{\sL^2(\Omega)}\leq \|\nabla\psi\|_{\sL^2(\Omega)}\,.$$
Let $\psi$ be an eigenfunction associated with $\lambda_{1}$. We have $\mathfrak{Q}_{V}(|\psi|)\leq \mathfrak{Q}_{V}(\psi)$ so that, by the min-max principle and using that $\lambda_{1}$ is the smallest Rayleigh quotient, we find that $|\psi|$ is also an eigenfunction associated with $\lambda_{1}$. By an elliptic regularity argument, $u=|\psi|$ is smooth on $\Omega$ since it satisfies
$$-\Delta u+(V-\lambda_{1}) u=0\,.$$
If $u$ vanishes at $\x_{0}$, then, by the Harnack inequality, it must vanish in a neighborhood of $\x_{0}$. By a connexity argument, $u$ is identically zero if it vanishes at some point of $\Omega$.
Therefore, all the eigenfunctions associated with $\lambda_{1}$ do not vanish in $\Omega$. If $\lambda_{1}$ were of multiplicity at least two, we would consider $u_{1}$ and $u_{2}$ two orthogonal eigenfunctions. But this is impossible since they do not vanish in $\Omega$.

\end{proof}

\chapter{Examples}\label{chapter-examples}
\begin{flushright}
\begin{minipage}{0.6\textwidth}
Bene quidam dixit de amico suo : dimidium animae suae. Nam ego sensi animam meam et animam illius unam fuisse animam in duobus corporibus, 
et ideo mihi horrori erat vita, quia nolebam dimidius vivere et ideo forte mori metuebam, ne totus ille moreretur, quem multum amaveram.
\begin{flushright}
\textit{Confessiones}, Augustinus
\end{flushright}
\vspace*{0.5cm}
\end{minipage}
\end{flushright}

This chapter aims at exemplifying some questions discussed in Chapter \ref{chapter-spectral-theory}.

\section{Harmonic oscillator}\label{Harmonic}
Before going further we shall discuss the spectrum of the harmonic oscillator which we will encounter many times in this book. 
We are interested in the self-adjoint realization on $\sL^2(\R)$ of:
$$\mathfrak{H}_{\harm}=D_{x}^2+x^2\,.$$
This operator is defined as the self-adjoint operator associated with the quadratic form defined by
$$\mathfrak{Q}_{\harm}(\psi)=\|\psi'\|^2+\|x\psi\|^2,\quad \psi\in \sB^1(\R)\,,$$
where 
$$\sB^1(\R)=\{\psi\in \sL^2(\R) : \psi'\in \sL^2(\R),\quad x\psi\in \sL^2(\R)\}\,.$$
Note that $\sB^1(\R)$ is an Hilbert space for the scalar product
$$\langle u,v\rangle_{\sB^1(\R)}=\int_{\R} u'\overline{v}'\dx x+\int_{\R} x^2u\overline{v}\dx x\,.$$

\begin{exe} 
Prove that $\sB^1(\R)$ is dense in $\sL^2(\R)$ and that $\mathcal{C}^\infty_{0}(\R)$ is dense in $\sB^1(\R)$.
\end{exe}
The domain of $\mathfrak{H}_{\harm}$ is given by
$$\Dom(\mathfrak{H}_{\harm})=\{\psi\in\sB^1(\R),\quad (D_{x}^2+x^2)\psi\in\sL^2(\R)\}\,.$$
The domain of the operator can be characterized with the following proposition.
\begin{prop}
We have
$$\Dom(\mathfrak{H}_{\harm})=\{\psi\in \sL^2(\R) : \psi''\in \sL^2(\R),\quad x^2\psi\in \sL^2(\R)\}\,.$$
\end{prop}
\begin{proof}
Let us provide an instructive proof. We use the difference quotients method (see \cite[Theorem 9.25]{Brezis}). Let us consider $\psi\in\Dom(\mathfrak{H}_{\harm})$. It is sufficient to prove that $\psi''\in\sL^2(\R)$.
There exists $f\in\sL^2(\R)$ such that, in the sense of distributions, we have
$$\forall \varphi\in\mathcal{S}^(\R),\qquad\langle (D_{x}^2+x^2)\psi,\varphi\rangle=\langle f,\varphi\rangle$$
so that
$$\forall \varphi\in\mathcal{S}(\R),\qquad\langle \partial_{x}\psi,\partial_{x}\varphi\rangle+\langle x\psi, x\varphi\rangle=\langle f,\varphi\rangle\,,$$
where the bracket is now the $\sL^2$-bracket.

Since $\psi\in\sB^1(\R)$ and that $\mathcal{S}(\R)$ is dense in $\sB^1(\R)$, we can extend this equality:
$$\forall \varphi\in\sB^1(\R),\qquad\langle \partial_{x}\psi,\partial_{x}\varphi\rangle+\langle x\psi, x\varphi\rangle=\langle f,\varphi\rangle\,.$$
Let us define the difference quotient
$$Q_{h}\varphi(x)=\frac{\varphi(x+h)-\varphi(x)}{h},\quad x\in\R,\quad h\neq 0\,.$$
If $\varphi\in\sB^1(\R)$, then $Q_{h}\varphi\in\sB^1(\R)$. We get
$$\forall \varphi\in\sB^1(\R),\qquad\langle \partial_{x}\psi,\partial_{x}Q_{h}\varphi\rangle+\langle x\psi, xQ_{h}\varphi\rangle=\langle f,Q_{h}\varphi\rangle\,.$$
We find
$$\langle \partial_{x}\psi,\partial_{x}Q_{h}\varphi\rangle=-\langle \partial_{x} Q_{-h}\psi,\partial_{x}\varphi\rangle$$
and
$$\langle x\psi, xQ_{h}\varphi\rangle=-\langle xQ_{-h}\psi,x\varphi\rangle-\langle \psi(x-h),x\varphi\rangle-\langle x\psi,\varphi(x+h)\rangle\,.$$
We find, for all $\varphi\in\sB^1(\R)$ and $h\neq 0$,
$$\langle \partial_{x} Q_{-h}\psi,\partial_{x}\varphi\rangle+\langle xQ_{-h}\psi,x\varphi\rangle=-\langle f, Q_{h}\varphi\rangle-\langle \psi(x-h),x\varphi\rangle-\langle x\psi,\varphi(x+h)\rangle\,.$$
and we apply this equality to $\varphi=Q_{-h}\psi$. We deduce
\begin{align*}
&\langle \partial_{x} Q_{-h}\psi,\partial_{x}Q_{-h}\psi\rangle+\langle xQ_{-h}\psi,xQ_{-h}\psi\rangle\\
&=-\langle f, Q_{h}Q_{-h}\psi\rangle-\langle \psi(x-h),xQ_{-h}\psi\rangle-\langle x\psi,Q_{-h}\psi(x+h)\rangle.
\end{align*}
Then we notice that
\begin{align*}
|\langle f, Q_{h}Q_{-h}\psi\rangle|       &\leq \|f\|_{\sL^2(\R)}\| Q_{h}Q_{-h}\psi\|_{\sL^2(\R)}\\
							 &\leq\|f\|_{\sL^2(\R)}\|\partial_{x}Q_{-h}\psi\|_{\sL^2(\R)}\\
							 &\leq\frac{1}{2}\left(\|f\|^2_{\sL^2(\R)}+\|\partial_{x}Q_{-h}\psi\|^2_{\sL^2(\R)}\right),
\end{align*}
where we have used Proposition \ref{quotient}. We can deal with the other terms in the same way and we get
\begin{align*}
&\|\partial_{x} Q_{-h}\psi\|^2_{\sL^2(\R)}+\|x Q_{-h}\psi\|^2_{\sL^2(\R)}\\
&\leq\frac{1}{2}\left(\|f\|^2_{\sL^2(\R)}+\|\partial_{x}Q_{-h}\psi\|^2_{\sL^2(\R)}+\|\psi\|^2_{\sL^2(\R)}+\|xQ_{-h}\psi\|^2_{\sL^2(\R)}+\|\psi\|_{\sB^1(\R)}^2+|h|\|\psi\|_{\sH^1(\R)}^2\right).
\end{align*}
We deduce that
$$\|Q_{-h}\partial_{x}\psi\|^2_{\sL^2(\R)}+\|x Q_{-h}\psi\|^2_{\sL^2(\R)}\leq \|f\|^2_{\sL^2(\R)}+\|\psi\|^2_{\sL^2(\R)}+\|\psi\|_{\sB^1(\R)}^2+|h|\|\psi\|_{\sH^1(\R)}^2\,.$$
We may again use Proposition \ref{quotient} and we deduce that $\partial_{x}\psi\in\sH^1(\R)$ and $x\psi\in\sH^1(\R)$.
\end{proof}
The self-adjoint operator $\mathfrak{H}_{\harm}$ has compact resolvent since $\sB^1(\R)$ is compactly embedded in $\sL^2(\R)$. Its spectrum is a sequence of eigenvalues which tends to $+\infty$.
Let us explain how we can get the spectrum of $\mathfrak{H}_{\harm}$. We let:
$$a=\frac{1}{\sqrt{2}}\left(\frac{d}{dx}+x\right),\quad a^*=\frac{1}{\sqrt{2}}\left(-\frac{d}{dx}+x\right)\,.$$
We have: 
$$[a,a^*]=aa^*-a^*a=1\,.$$
We let:
$$f_{0}(x)=e^{-x^2/2}\,.$$
We investigate the spectrum of $a^* a$. We have: $af_{0}=0$. We let $f_{n}=(a^*)^n f_{0}$. This is easy to prove that $a^* a f_{n}=n f_{n}$ and that $af_{n}=n f_{n-1}$. 

\begin{exe} 
Prove that the $(f_{n})$ form a Hilbertian basis of $\sL^2(\R)$. These functions are called Hermite's functions.The eigenvalues of $\mathfrak{H}_{\harm}$ are the numbers $2n+1, n\in\N$. They are simple and associated with the normalized Hermite's functions.
\end{exe}

\begin{exe}
This exercise is an example of exact WKB expansions. We will recognize Laguerre's polynomials. We wish to study the $2D$ harmonic oscillator: $-\Delta+|x|^2$.
\begin{enumerate}
\item Write the operator in terms of radial coordinates.
\item Explain how the spectral analysis can be reduced to the study of:
$$-\dr_{\rho}^2-\rho^{-1}\dr_{\rho}+\rho^{-2}m^2+\rho^2\,,$$
on $\sL^2(\rho d \rho)$ with $m\in\mathbb{Z}$.
\item Perform the change of variable $t=\rho^2.$
\item For which $\alpha$ is $t\mapsto t^{\alpha} e^{-t/2}$ an eigenfunction ?
\item Conjugate the operator by $t^{-m/2}e^{t/2}$. On which space is the new operator $\mathfrak{L}_{m}$ acting ? Describe the new scalar product.
\item Find eigenvalues of $\mathfrak{L}_{m}$ by noticing that  $\R_{N}[X]$ is stable by $\mathfrak{L}_{m}$.
\item Conclude.
\end{enumerate}
\end{exe}

\section{A $\delta$-interaction}
In this section we discuss a model on the line related to the so-called $\delta$-interactions. The reader is referred to \cite[Chapter II.2]{AGHH88} and to \cite{BEKS94, EN01, EN03, BEW08, Exner08} where the spectral properties of $\delta$-interactions are analyzed.
Let us define, for $\psi\in\sH^1(\R)$,
\[\mathfrak{q}^\delta(\psi)=\int_{\R} |\psi'(y)|^2\dx y-|\psi(0)|^2\,.\]
\begin{prop}
The quadratic form $\mathfrak{q}^\delta$ is well defined and semi-bounded from below. Moreover, there exists $C>0$ such that $\sqrt{\mathfrak{q}^\delta(\cdot)+C\|\cdot\|_{\sL^2(\R)}^2}$ is a norm equivalent to $\|\cdot\|_{\sH^1(\R)}$.
\end{prop}
\begin{proof}
Let us recall the classical Sobolev embedding:
$$\exists C>0,\quad \forall u\in\sH^1(\R),\quad \|u\|^2_{\sL^\infty(\R)}\leq C\|u\|^2_{\sH^1(\R)}=C(\|u\|^2_{\sL^2(\R)}+\|u'\|^2_{\sL^2(\R)})\,.$$
We apply this inequality to $u(x)=v(\lambda x)$ for $\lambda>0$ and $v\in\sH^1(\R)$. Choosing the appropriate $\lambda$ we get
$$\|v\|^2_{\sL^\infty(\R)}\leq 2C\|v\|_{\sL^2(\R)}\|v'\|_{\sL^2(\R)}$$
and thus, for all $\eps\in(0,1)$,
$$\|v\|^2_{\sL^\infty(\R)}\leq C\left(\eps^{-1}\|v\|^2_{\sL^2(\R)}+\eps\|v'\|^2_{\sL^2(\R)}\right)\,.$$
We deduce that, for all $\psi\in\sH^1(\R)$,
\begin{equation}\label{delta-H1}
\mathfrak{q}^\delta(\psi)\geq -C\eps^{-1}\|\psi\|^2_{\sL^2(\R)}+(1-C\eps)\|\psi'\|^2_{\sL^2(\R)}\,.
\end{equation}
Choosing $\eps$ small enough, the conclusion follows.
\end{proof}

\begin{prop}
If $\mathfrak{L}^\delta$ denotes the self-adjoint operator associated with $\mathfrak{q}^\delta$, we have
$$\Dom(\mathfrak{L}^\delta)=\left\{u\in\sH^1(\R) : \sH^2(\R\setminus\{0\})\mbox{ and } u'(0^+)-u'(0^-)=-u(0)\right\}\,.$$
Moreover we have $\spe(\mathfrak{L}^\delta)=[0,+\infty)$ and $\sp_{\dis}(\mathfrak{L}^\delta)=\{-\frac{1}{4}\}$.
\end{prop}
\begin{proof}
By definition, we have, for all $u\in\Dom(\mathfrak{L}^\delta)$ and $v\in\sH^1(\R)$,
$$\langle\mathfrak{L}^\delta u, v\rangle=\mathfrak{b}^\delta(u,v)\,.$$
For, $v\in\mathcal{C}^\infty_{0}(\R\setminus\{0\})$, we get, in $\mathcal{D}'(\R\setminus\{0\})$, $\mathfrak{L}^{\delta}u=-u''\in\sL^2(\R\setminus\{0\})$ so that we deduce $u\in\sH^2(\R\setminus\{0\})$. We deduce that $u'(0^+)$ and $u'(0^-)$ are well defined by Sobolev embedding. Then, for all $v\in\mathcal{C}^\infty_{0}(\R)$, an integration by parts gives
$$\mathfrak{b}^\delta(u,v)=-\int_{\R}u'' \overline{v} \dx x+(u'(0^+)-u'(0^-)+u(0))\overline{v(0)}\,.$$
But, we have $\langle\mathfrak{L}^\delta u, v\rangle=-\int_{\R} u'' \overline{v}\dx x$ and thus $u'(0^+)-u'(0^-)+u(0)=0$. Conversely, if $u\in\sH^1(\R)\cap\sH^2(\R\setminus\{0\})$ satisfies this jump condition, it is in the domain.

Let us show that $0\in\sp_{\ess}(\mathfrak{L}^\delta)$. We consider $\chi_{R}(x)=\chi_{1}(R^{-1}(x-R))$ with $\chi_{1}$ a smooth cutoff function supported in $[0,+\infty)$. We get $\|\mathfrak{L}^\delta \chi_{R}\|_{\sL^2(\R)}$ tends to $0$ when $R\to+\infty$. We apply the Weyl criterion. If we use $\chi_{R,\xi}(x)=e^{ix\xi}\chi_{R}(x)$, for $\xi\in\R$, we find that $\xi^2\in\sp_{\ess}(\mathfrak{L}^\delta)$ so that $[0,+\infty)\subset \sp_{\ess}(\mathfrak{L}^\delta)$. Let us now establish the converse inclusion.

Let us consider $\lambda\in\sp(\mathfrak{L}^\delta)$ with $\lambda<0$. We shall prove that $\lambda$ is in the discrete spectrum. For that purpose, we use Proposition \ref{Persson-improved}: the first item comes from \eqref{delta-H1} and the fact that $\sH^1((a,b))$ is compactly embedded in $\sL^2((a,b))$, the second item from the formula
$$\langle\mathfrak{L}^\delta\psi,\chi_{R}^2\psi\rangle_{\sL^2(\R)}=\mathfrak{q}^\delta(\chi_{R}\psi)-\|\chi_{R}'\psi\|^2_{\sL^2(\R)}$$
and the third from the fact that the Laplacian is non negative.

Finally, if $\lambda\in\sp_{\dis}(\mathfrak{L}^\delta)$, we may easily solve the eigenvalue equation and we find that $\lambda=-\frac{1}{4}$ is associated with the eigenfunction $\psi(x)=e^{-|x|/2}$.
\end{proof}

\begin{exe} For $x\geq 0$, we introduce the quadratic form $\mathfrak{q}_{x}$ defined for $\psi\in \sH^1(\R)$ by
\begin{equation}\label{qx}
\mathfrak{q}_{x}(\psi)=\int_{\R} |\psi'(y)|^2\dx y-|\psi(-x)|^2-|\psi(x)|^2\,.
\end{equation}
\begin{enumerate}
\item Prove that $\mathfrak{q}_{x}$ is semi-bounded from below. 
\item We introduce the associated self-adjoint operator denoted by $\mathfrak{D}_{x}$. Prove that
$$\Dom(\mathfrak{D}_{x})=\left\{\psi\in \sH^1(\R)\cap \sH^2(\R\setminus\{\pm x\}) : \psi(\pm x^{+})-\psi(\pm x^{-})=-\psi(\pm x)\right\}\,.$$
\item Show that, for all $x\geq 0$, the essential spectrum of $\mathfrak{D}_{x}$ is given by
$$\spe(\mathfrak{D}_{x})=[0,+\infty)\,.$$
\item For $x\geq 0$, we denote by $\mu_{1}(x)$ the lowest eigenvalue of $\mathfrak{D}_{x}$ and by $u_{x}$ the corresponding positive and $\sL^2$-normalized eigenfunction. Establish the following properties
\begin{enumerate}
\item For $x\geq 0$, we have
$$\mu_{1}(x)=-\left(\frac{1}{2}+\frac{1}{2x}W(xe^{-x})\right)^2\,,$$
where $W : [-e^{-1},+\infty) \to [-1,+\infty)$ is the Lambert function, \textit{i.e.} the inverse of $[-1,+\infty)\ni w\mapsto we^w\in[-e^{-1},+\infty)$.
\item The second eigenvalue $\mu_{2}(x)$ only exists for $x>1$ and is given by
$$\mu_{2}(x)=-\left(\frac{1}{2}+\frac{1}{2x}W(-xe^{-x})\right)^2\,.$$
\item\label{1} $\mu_{1}(x)\underset{x\to 0}{=}-1+2x+\mathcal{O}(x^2)$,
\item\label{2} $\mu_{1}(x)\underset{x\to+\infty}{=}-\frac{1}{4}-\frac{e^{-x}}{2}+\mathcal{O}(xe^{-2x})$, $\mu_{2}(x)\underset{x\to+\infty}{=}-\frac{1}{4}+\frac{e^{-x}}{2}+\mathcal{O}(xe^{-2x})$,
\item\label{3} For all $x\geq 0$, $-1\leq \mu_{1}(x)<-\frac{1}{4}$ and for all $x>1$, $\mu_{2}(x)>-\frac{1}{4}$,
\item\label{4} $\mu_{1}$ admits a unique minimum at $0$,
\item\label{5} For all $x\geq 0$ and all $\psi\in \sH^1(\R)$, we have $\mathfrak{q}_{x}(\psi)\geq -\Vert\psi\Vert^2$,
\item\label{6} $R(x)=\Vert\partial_{x}u_{x}\Vert^2_{\sL^2(\R_{y})}$ defines a bounded function for $ x> 0$.
\item\label{7} $\Vert\partial_{y}u_{x}\Vert^2_{\sL^2(\R_{y})}$ defines a bounded function for $x\geq 0$. 
\end{enumerate}
\end{enumerate}
\end{exe}

\section{Robin Laplacian}
In this section, we discuss some properties of a model closely related to the $\delta$-interaction. 

Let us define, for $\psi\in\sH^1(\R_{+})$ and $\gamma>0$,
\[\mathfrak{q}_{\gamma}^{\mathsf{R}}(\psi)=\int_{0}^{+\infty} |\psi'(y)|^2\dx y-\gamma|\psi(0)|^2\,.\]
\begin{prop}\label{prop.Robin}
The quadratic form $\mathfrak{q}_{\gamma}^{\mathsf{R}}$ is well-defined on $\sH^1(\R_{+})$ and bounded from below. If $\mathfrak{L}_{\gamma}^{\mathsf{R}}$ denotes the self-adjoint operator associated with $\mathfrak{q}_{\gamma}^{\mathsf{R}}$, we have
\[\Dom(\mathfrak{L}_{\gamma}^\mathsf{R})=\left\{u\in\sH^1(\R_{+}) : \sH^2(\R_{+})\mbox{ and } \psi'(0)=-\gamma\psi(0)\right\}\,.\]
Moreover, we have $\spe(\mathfrak{L}_{\gamma}^\mathsf{R})=[0,+\infty)$ and $\sp_{\dis}(\mathfrak{L}_{\gamma}^\mathsf{R})=\{-\gamma^2\}$ and the $\sL^2$-normalized eigenfunction associated with $-\gamma^2$ is $(2\gamma)^{\frac{1}{2}}e^{-\gamma x}$.
\end{prop}
\begin{proof}
The definition of the operator and the characterization of the domain follows as for the $\delta$-interaction. The characterization of the essential spectrum also follows from the same arguments. Let us just determine the discrete spectrum. We want to solve
\[-\psi''=-\omega^2\psi\,,\qquad \psi'(0)=-\gamma\psi(0)\,,\]
with $\psi\in\sH^2(\R_{+})$ and $\omega>0$. Thus, we have $\psi(x)=Ae^{-\omega x}$ so that, with the boundary condition, $\omega=\gamma$.
\end{proof}

Let us now introduce a model that can be useful in practice (see for instance \cite{HelKac15}). For $\psi\in\sH^1((0,L))$ with $\psi(L)=0$ and $\gamma, L>0$, we let
\[\mathfrak{q}_{\gamma, L}^{\mathsf{R}}(\psi)=\int_{0}^{L} |\psi'(y)|^2\dx y-\gamma|\psi(0)|^2\,.\]
We would like to investigate the behavior of the lowest eigenvalue when $L\to+\infty$.
\begin{prop}
The quadratic form $\mathfrak{q}_{\gamma, L}^{\mathsf{R}}$ is well-defined on $\{\psi\in\sH^1((0,L))\,,\psi(L)=0\}$ and bounded from below. If $\mathfrak{L}_{\gamma, L}^{\mathsf{R}}$ denotes the self-adjoint operator associated with $\mathfrak{q}_{\gamma, L}^{\mathsf{R}}$, we have
\[\Dom(\mathfrak{L}_{\gamma, L}^\mathsf{R})=\left\{u\in\sH^1((0,L)) : \sH^2((0, L))\mbox{ and } \psi'(0)=-\gamma\psi(0)\,,\quad\psi(L)=0\right\}\,.\]
The operator $\mathfrak{L}_{\gamma, L}^\mathsf{R}$ has compact resolvent and there exists only one negative eigenvalue $\lambda_{1}(\gamma, L)$ as soon as $L$ is large enough and its satisfies, for all $\eps>0$,
\[\lambda_{1}(\gamma, L)\underset{L\to+\infty}{=}-\gamma^2+\mathcal{O}(e^{-2(1-\eps)\gamma L})\,.\]
A corresponding eigenfunction is
\[\psi_{\gamma, L}(x)=\sqrt{2\gamma}\left\{e^{-\omega_{\gamma, L}x}+e^{-2\omega_{\gamma, L}L}e^{\omega_{\gamma, L}x}\right\}\,,\quad \omega_{\gamma, L}=\sqrt{-\lambda_{1}(\gamma, L)}\,,\]
and we have, for all $\eps>0$,
\[\|e^{-\omega_{\gamma, L}x}-e^{-\gamma x}\|^2_{\sL^2((0, L))}=\mathcal{O}(L^3e^{-(1-\eps)4\gamma L})\,,\]
and  
\[\|e^{-2\omega_{\gamma, L}L}e^{\omega_{\gamma, L}x}\|^2_{\sL^2((0, L))}=\mathcal{O}\left(Le^{-(1-\eps)2\gamma L}\right)\,.\]
\end{prop}
\begin{proof}
Let us just describe the negative spectrum (the considerations of domain are left to the reader as an exercise). We want to solve
\[-\psi''=-\omega^2\psi\,,\quad \psi'(0)=-\gamma\psi(0)\,,\quad \psi(L)=0\,,\]
where $\omega>0$. We have
\[\psi(x)=Ae^{\omega x}+B e^{-\omega x}\,.\]
The boundary conditions lead to 
\[Ae^{\omega L}+B e^{-\omega L}=0\,,\qquad A(\omega+\gamma)+B(\gamma-\omega)=0\,.\]
This leads to the condition
\[F_{\gamma, L}(\omega)=\omega-\gamma+e^{-2\omega L}(\omega+\gamma)=0\,.\]
We consider the function $F_{\gamma}$ on $[0+\infty)$. We have $F_{\gamma}(0)=0$ and $\displaystyle{\lim_{\omega\to+\infty} F_{\gamma}(\omega)=+\infty}$. We get
\[F'_{\gamma, L}(\omega)=1+e^{-2\omega L}(1-2L\gamma-2L\omega)\,,\]
and
\[F''_{\gamma, L}(\omega)=-4Le^{-2\omega L}(1-L\gamma-L\omega)\,.\]
If $L>\frac{1}{\gamma}$, we have, for all $\omega>0$,  $F''_{\gamma, L}(\omega)>0$. Thus, $F'_{\gamma, L}$ is increasing from $2(1-L\gamma)$ to $1$ and $F'_{\gamma, L}$ has only one zero $z_{\gamma, L}$ in $(0, +\infty)$. We deduce that $F_{\gamma, L}$ decreases on $(0, z_{\gamma, L})$ and increases on $(z_{\gamma, L}, +\infty)$. For $L>\frac{1}{\gamma}$, $F_{\gamma, L}$ admits a unique zero $\omega_{\gamma, L}$ in $(0, \infty)$. Therefore there is a unique non negative eigenvalue that is $\lambda_{1}(\gamma, L)=-\omega^2_{\gamma, L}$.

By using Proposition \ref{prop.Robin} and the min-max principle, we have
\[\forall L>0\,,\qquad \lambda_{1}(\gamma, L)\geq -\gamma^2\,.\]
By using the test function $(2\gamma)^{\frac{1}{2}}\chi(L^{-1}x)e^{-\gamma x}$, with $\chi$ a smooth function being $1$ on $|x|\leq 1-\eps$ and $0$ for $x\geq 1$,  and the min-max principle, we get the wished upper bound.The estimates of the first eigenfunction easily follows.
\end{proof}

\section{De Gennes operator and applications}\label{Sec:deGennes}

\subsection{The de Gennes operator}
The analysis of the two dimensional magnetic Laplacian with Neumann condition on $\R^2_{+}$ makes the so-called de Gennes operator to appear. We refer to \cite{DauHel} where this model is studied in details (see also \cite{FouHel10}). This operator is defined as follows. For $\zeta\in\R$, we consider the Neumann realization $\mathfrak{L}_{\zeta}^{[0]}$ in $\sL^2(\R_+)$ associated with the operator $D_{t}^2+(\zeta-t)^2$ with domain
\begin{equation}\label{oh}
\Dom(\mathfrak{L}_{\zeta}^{[0]})=\left\{u\in \mathsf{B}^1(\R_+) : \left(D_{t}^2+(t-\zeta)^2\right) u\in\sL^2(\R_{+}),\quad u'(0)=0\right\}\,.
\end{equation}
\begin{rem}
Note that, by the difference quotient method, we may establish that
$$\left\{u\in \mathsf{B}^1(\R_+) : \left(D_{t}^2+(t-\zeta)^2\right) u\in\sL^2(\R_{+})\right\}\subset\sH^2(\R_{+})\,,$$
so that, with a Sobolev embedding, $u'(0)$ is well defined.
\end{rem}
The operator $\mathfrak{L}_{\zeta}^{[0]}$ has compact resolvent by standard arguments. By the Cauchy-Lipschitz theorem, all the eigenvalues are simple.
\begin{notation}
The lowest eigenvalue of $\mathfrak{L}_{\zeta}^{[0]}$ is denoted $\nu_{1}^{[0]}(\zeta)$.
\end{notation}
\begin{lem}
The function $\zeta\mapsto\nu^{[0]}_{n}(\zeta)$ is analytic.
\end{lem}
\begin{proof}
The family $(\mathfrak{L}_{\zeta}^{[0]})_{\zeta\in\R}$ is analytic of type $(A)$ in the sense of Kato (see \cite[p. 375]{Kato66}) and thus one might directly apply the analytic perturbation theory. Nevertheless, let us provide an elementary proof. Let us fix $\zeta_{1}\in\R$ and prove that $\nu_{n}^{[0]}$ is continuous at $\zeta_{1}$. We have, for all $\psi\in \sB^1(\R_{+})$,
$$\left|\mathfrak{Q}^{[0]}_{\zeta}(\psi)-\mathfrak{Q}^{[0]}_{\zeta_{1}}(\psi)\right|\leq |\zeta^2-\zeta_{1}^2|\|\psi\|^2+2|\zeta-\zeta_{1}| \|t^{\frac{1}{2}}\psi\|^2$$
so that
$$\left|\mathfrak{Q}^{[0]}_{\zeta}(\psi)-\mathfrak{Q}^{[0]}_{\zeta_{1}}(\psi)\right|\leq  |\zeta^2-\zeta_{1}^2|\|\psi\|^2+4|\zeta-\zeta_{1}|\mathfrak{Q}^{[0]}_{\zeta_{1}}(\psi)+4\zeta_{1}^2|\zeta-\zeta_{1}|\|\psi\|^2\,.$$
We deduce that
$$\mathfrak{Q}^{[0]}_{\zeta}(\psi)\leq\left(1+4|\zeta-\zeta_{1}|\right)\mathfrak{Q}^{[0]}_{\zeta_{1}}(\psi)+4\zeta_{1}^2|\zeta-\zeta_{1}|\|\psi\|^2+  |\zeta^2-\zeta_{1}^2|\|\psi\|^2$$
and
$$\mathfrak{Q}^{[0]}_{\zeta}(\psi)\geq\left(1-4|\zeta-\zeta_{1}|\right)\mathfrak{Q}^{[0]}_{\zeta_{1}}(\psi)-4\zeta_{1}^2|\zeta-\zeta_{1}|\|\psi\|^2  -|\zeta^2-\zeta_{1}^2|\|\psi\|^2\,.$$
It remains to apply the min-max principle and we get the comparisons between the eigenvalues.
We shall now prove the analyticity. Let us fix $\zeta_{1}\in\R$ and $z\in\mathbb{C}\setminus\sp(\mathfrak{L}_{\zeta_{1}}^{[0]})$. We observe that $t (\mathfrak{L}_{\zeta_{1}}^{[0]}-z)^{-1}$ is bounded with a uniform bound with respect to $z$ in a compact avoiding the spectrum so that for $\zeta$ close enough to $\zeta_{1}$, $\mathfrak{L}_{\zeta}^{[0]}-z$ is invertible. Indeed, we can write
\begin{align*}
\mathfrak{L}_{\zeta}^{[0]}-z&=\mathfrak{L}_{\zeta_{1}}^{[0]}-z+2(\zeta_{1}-\zeta)t+\zeta^2-\zeta_{1}^2\\
&=\left(\mathsf{Id}+2(\zeta_{1}-\zeta)t\left(\mathfrak{L}_{\zeta_{1}}^{[0]}-z\right)^{-1}+(\zeta^2-\zeta_{1}^2)\left(\mathfrak{L}_{\zeta_{1}}^{[0]}-z\right)^{-1}\right)\left(\mathfrak{L}_{\zeta_{1}}^{[0]}-z\right).
\end{align*}
By using the expression of the inverse and the fact that the domain of $\mathfrak{L}_{\zeta}^{[0]}$ does not depend on $\zeta$, we infer that  $\zeta\mapsto (\mathfrak{L}_{\zeta}^{[0]}-z)^{-1}\in\mathcal{L}\left(\sL^2(\R_{+}),\left(\Dom(\mathfrak{L}_{\zeta_{1}}^{[0]}),\|\cdot\|_{\mathfrak{L}_{\zeta_{1}}^{[0]}} \right)\right)$ is analytic near $\zeta_{1}$, uniformly for $z$ in a compact. Since $\mathfrak{L}_{\zeta}^{[0]}$ has compact resolvent and is self-adjoint, the application, defined as a Riemannian integral,
$$P_{\Gamma}(\zeta)=\frac{1}{2i\pi}\int_{\Gamma} (\mathfrak{L}_{\zeta}^{[0]}-z)^{-1}\dx z$$
is the projection on the space generated by the eigenfunctions associated with eigenvalues enclosed by the smooth contour $\Gamma$. It is possible to consider a contour which encloses only $\nu_{n}^{[0]}(\zeta_{1})$ and, thus, only $\nu_{n}^{[0]}(\zeta)$ as soon as $\zeta$ is close enough to $\zeta_{1}$. We leave the last details to the reader.
\end{proof}
We may now consider the $\sL^2$-normalized and positive eigenfunction $u^{[0]}_{\zeta}=u^{[0]}(\cdot,\zeta)$ associated with $\nu_{1}^{[0]}(\zeta)$ and that depends on $\zeta$ analytically.
\begin{prop}
The function $u^{[0]}_{\zeta}$ belongs to $\mathcal{S}(\overline{\R_{+}})$.
\end{prop}
\begin{proof}
This is a consequence of the following (by using difference quotients). If $u\in\sB^k(\R)$ is such that $\mathfrak{L}_{\zeta}^{[0]} u=f$ with $f\in\sB^{k-1}(\R_{+})$, then $u\in\sB^{k+1}(\R_{+})$. We have denoted
$$\sB^k(\R_{+})=\left\{u\in\sL^2(\R_{+}) : x^\alpha \partial_{x}^\beta u\in\sL^2(\R_{+}),\quad \alpha+\beta\leq k\right\}\,.$$
Then, we notice that $\displaystyle{u^{[0]}_{\zeta}\in\bigcap_{k\geq 0}\sB^k(\R_{+})}\subset \mathcal{S}(\overline{\R_{+}})$.
\end{proof}
We have used the notion of holomorphic functions valued in a Banach space. The aim of the following exercises is to prove that all the natural definitions of holomorphy coincide.

\begin{exe}\label{exo.13} Let $\sB$ be a Banach space and $\Omega\subset\C$ an open set. We say that $f:\Omega\to\sB$ is holomorphic if, for all $z_{0}\in\Omega$, $\frac{f(z)-f(z_{0})}{z-z_{0}}$ converges when $z$ goes to $z_{0}$. We say that $f$ is weakly  holomorphic on $\Omega$ if, for all $\ell\in\sB^*$, $\ell\circ f$ is holomorphic on $\Omega$.
\begin{enumerate}
\item Let us assume that $f$ is weakly holomorphic on $\Omega$ and consider, for $z_{0}\in\Omega$ and $r>0$ such that $D(z_{0},r)\subset \Omega$,
$$C:=\left\{\frac{f(z)-f(z_{0})}{z-z_{0}}, z\in D(z_{0}, r)\setminus\{z_{0}\}\right\}\,.$$
Prove that $\ell\left(C\right)$ is bounded for all $\ell\in \sB^*$.
\item Deduce that $C$ is bounded by using the Banach-Steinhaus theorem.
\item By using the Cauchy-formula and the Hahn-Banach theorem, prove that $f$ is holomorphic on $\Omega$.
\end{enumerate}
\end{exe}
\begin{exe}\label{exe.an.res} Let $\sB$ be a Banach space and $\sH$ be a Hilbert space. Let $\Omega\subset\C$ an open set.
\begin{enumerate}
\item By using the Banach-Steinhaus theorem, the Cauchy formula and the Hahn-Banach theorem show that if $f : \Omega\to\mathcal{L}(\sB,\sH)$ is such that $\Omega \ni z\mapsto \langle f(z)\psi,\varphi\rangle_{\sH}$ is holomorphic for all $\psi\in\sB$ and $\varphi\in\sH$, then $f$ is holomorphic.
\item If $(\mathfrak{L},\Dom(\mathfrak{L}))$ is a closed operator on a Hilbert space $\sH$, show that $\Omega\ni z\mapsto R(z)=(\mathfrak{L}-z)^{-1}$ is holomorphic on the resolvent set as well as if $R$ is valued $(\sH,\|\cdot\|_{\sH})$ or in $(\Dom(\mathfrak{L}),\|\cdot\|_{\mathfrak{L}})$ (where $\|\cdot\|_{\mathfrak{L}}$ is the graph norm).
\end{enumerate}
\end{exe}
\begin{lem} 
$\zeta\mapsto \nu_{1}^{[0]}(\zeta)$ admits a unique minimum and it is non degenerate.
 \end{lem}
\begin{proof}
An easy application of the min-max principle gives:
$$\lim_{\zeta\to-\infty}\nu_{1}^{[0]}(\zeta)=+\infty\,.$$
Let us now show that:
$$\lim_{\zeta\to+\infty}\nu_{1}^{[0]}(\zeta)=1\,.$$
The de Gennes operator is equivalent to the operator $-\dr_{t}^2+t^2$ on $(-\zeta,+\infty)$ with Neumann condition at $-\zeta$. Let us begin with upper bound.
An easy and explicit computation gives:
$$\nu_{1}^{[0]}(\zeta)\leq\langle(-\dr_{t}^2+t^2)e^{-t^2/2}, e^{-t^2/2} \rangle_{\sL^2((-\zeta,+\infty))}\underset{\zeta\to+\infty}{\to}1\,.$$
Let us investigate the converse inequality. Let us prove some concentration of $u_{\zeta}^{[0]}$ near $0$ when $\zeta$ increases (the reader can compare this with the estimates of Agmon of Section \ref{Agmon}). We have
$$\int_{0}^{{+\infty}}(t-\zeta)^2 |u_{\zeta}^{[0]}(t)|^2 \dx t\leq  \nu_{1}^{[0]}(\zeta)\,.$$
If $\lambda(\zeta)$ is the lowest Dirichlet eigenvalue, we have:
$$ \nu_{1}^{[0]}(\zeta)\leq\lambda(\zeta)\,.$$
By monotonicity of the Dirichlet eigenvalue with respect to the domain, we have, for $\zeta>0$:
$$\lambda(\zeta)\leq \lambda(0)=3\,.$$
It follows that:
$$\int_{0}^1 |u_{\zeta}^{[0]}(t)|^2\dx t\leq\frac{3}{(\zeta-1)^2},\quad\zeta\geq 2\,.$$
Let us introduce the test function: $\chi(t)u_{\zeta}^{[0]}(t)$ with $\chi$ supported in $(0,+\infty)$ and being $1$ for $t\geq 1$. We have
\begin{multline*}
\langle(-\dr_{t}^2+(t-\zeta)^2)\chi(t)u_{\zeta}^{[0]}(t),\chi(t)u_{\zeta}^{[0]}(t) \rangle_{\sL^2(\R)}\geq \|\chi(\cdot+\zeta)u_{\zeta}^{[0]}(\cdot+\zeta)\|^2_{\sL^2(\R)}=\|\chi u_{\zeta}^{[0]}\|^2_{\sL^2(\R)}\\
=1+\mathcal{O}(|\zeta|^{-2}).
\end{multline*}
Moreover, we get:
$$\langle(-\dr_{t}^2+(t-\zeta)^2)\chi(t)u_{\zeta}^{[0]}(t),\chi(t)u_{\zeta}^{[0]}(t) \rangle_{\sL^2(\R)}=\langle(-\dr_{t}^2+(t-\zeta)^2)\chi(t)u_{\zeta}^{[0]}(t),\chi(t)u_{\zeta}^{[0]}(t) \rangle_{\sL^2(\R_{+})}\,.$$
We have:
$$\langle(-\dr_{t}^2+(t-\zeta)^2)\chi(t)u_{\zeta}^{[0]}(t),\chi(t)u_{\zeta}^{[0]}(t) \rangle_{\sL^2(\R_{+})}= \nu_{1}^{[0]}(\zeta)\|\chi u_{\zeta}^{[0]}\|^2+\|\chi' u_{\zeta}^{[0]}\|^2$$
which can be controlled by the concentration result. We infer that, for $\zeta$ large enough,
$$ \nu_{1}^{[0]}(\zeta)\geq 1-C|\zeta|^{-1}\,.$$
From these limits, we deduce the existence of a minimum strictly less than $1$.

We now use the Feynman-Hellmann formula which will be established later (see Section \ref{Sec:FH}). We have:
$$(\nu_{1}^{[0]})'(\zeta)=-2\int_{t>0} (t-\zeta)|u_{\zeta}^{[0]}(t)|^2\dx t\,.$$
For $\zeta<0$, we get an increasing function. Moreover, we see that $\nu(0)=1$. The minima are obtained for $\zeta>0$. 

We can write that:
$$(\nu_{1}^{[0]})'(\zeta)=2\int_{t>0} (t-\zeta)^2 u_{\zeta}^{[0]} (u^{[0]}_{\zeta})'\dx t+\zeta^2u_{\zeta}^{[0]}(0)^2\,.$$
This implies:
$$(\nu_{1}^{[0]})'(\zeta)=(\zeta^2- \nu_{1}^{[0]}(\zeta))u_{\zeta}^{[0]}(0)^2\,.$$
Let $\zeta_{c}$ a critical point for $\nu_{1}^{[0]}$. We get:
$$(\nu_{1}^{[0]})''(\zeta_{c})=2\zeta_{c}u^{[0]}_{\zeta_{c}}(0)^2\,.$$
The critical points are all non degenerate. They correspond to local minima.We conclude that there is only one critical point and that is the minimum. We denote it $\zeta_{0}$ and we have $\nu_{1}^{[0]}(\zeta_{0})=\zeta_{0}^2.$
\end{proof}
We let: 
\begin{equation}\label{C1}
\Theta_0=\nu_{1}^{[0]}(\zeta_0),\qquad C_1=\frac{(u^{[0]}_{\zeta_0})^2(0)}{3}\,.
\end{equation}

\begin{figure}[h!t]\label{mutau14}
\includegraphics[height=5cm]{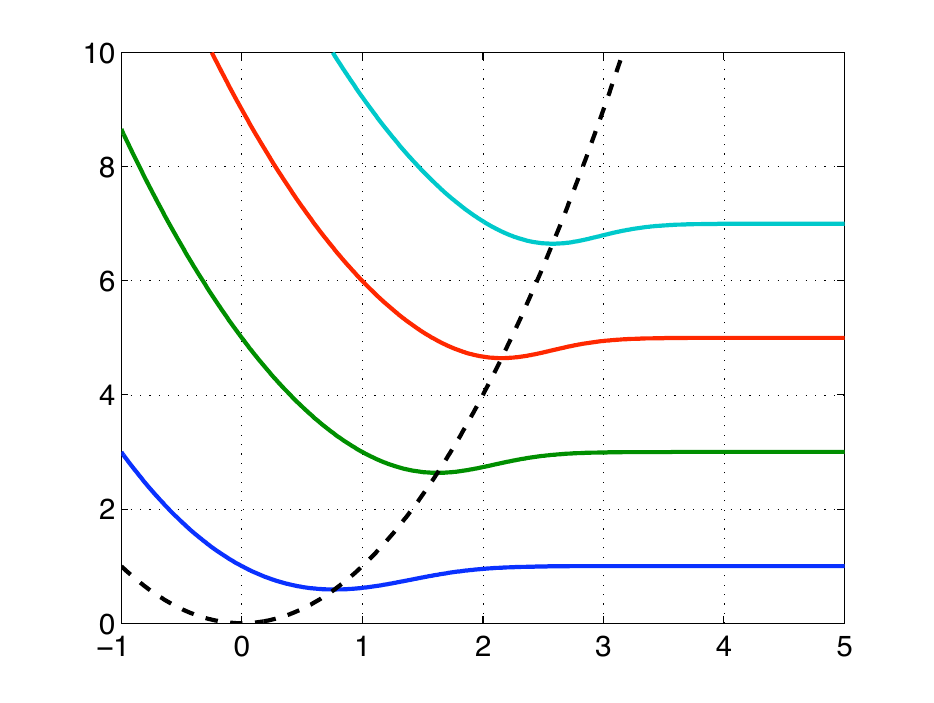}
\caption{$\zeta\mapsto\nu^{[0]}_{k}(\zeta)$, for $k=1, 2, 3,4$}
\end{figure}

\subsection{De Gennes operator and magnetic wall}

Let us now explain how we can investigate the spectral properties of a Hamiltonian with the following discontinuous magnetic field
$$\bB(x,y)=b_{1}\one_{\R_{-}}(x)+b_{2}\one_{\R_{+}}(x)\,,$$
where $\gb=(b_{1},b_{2})\in\R^2$. An associated vector potential is given by :
$$\bA(x,y)=(0,a_{b}(x)),\qquad a_{\gb}(x)=b_{1}x\one_{\R_{-}}(x)+b_{2}x\one_{\R_{+}}(x)\,.$$
The magnetic Hamiltonian is 
$$\mathfrak{L}_{\gb}=(-i\nabla-\bA)^2=D_{x}^2+(D_{y}-a_{\gb}(x))^2\,.$$ 
We will see that this example (inspired from \cite{HPPR14}) will give the flavor of many spectral methods related to the theory of ODE's. In particular, we will investigate the relation between the essential and the discrete spectrum (how many are the eigenvalues below the essential spectrum?) by using the so-called Sturm-Liouville theory, the min-max principle and some special functions related to the de Gennes operator.

We notice that $\overline{\mathfrak{L}_{b}}=\mathfrak{L}_{-b}$ so that we may assume that $b_{2}\geq 0$. If $S$ denotes the symmetry $(x,y)\mapsto (-x,y)$, we have $S\mathfrak{L}_{b_{1},b_{2}}S=\mathfrak{L}_{-b_{2},-b_{1}}=\overline{\mathfrak{L}_{b_{2},b_{1}}}$. For $B>0$, we introduce the $\sL^2$-unitary transform
$$U_{B}\psi(x,y)=B^{-1/2}\psi(B^{-1/2}x,B^{-1/2}y)$$
and we have
$$U_{B}^{-1}\mathfrak{L}_{\gb}U_{B}=B^{-1}\mathfrak{L}_{B \gb}\,.$$
These considerations allow the following reductions: 
\begin{enumerate}
\item If $b_{1}$ or $b_{2}$ is zero we may assume that $b_{1}=0$ and, if $b_{2}\neq 0$, we may assume that $b_{2}=1$. We call the case $(b_{1},b_{2})=(0,1)$ the \enquote{magnetic wall}.
\item If $b_{1}$ and $b_{2}$ have opposite signs and $|b_{1}|\neq |b_{2}|$, we may only consider the case  $|b_{1}|<|b_{2}|$ and then $b_{1}<0<b_{2}=1$. We call this case the \enquote{trapping magnetic step}.
\item If $b_{1}$ and $b_{2}$ are such that $|b_{1}|= |b_{2}|$, we may only consider the cases $(b_{1},b_{2})=(1,1)$ and $(b_{1},b_{2})=(-1,1)$.
\item If $b_{1}$ and $b_{2}$ have the same sign, we may assume that $0<b_{1}<b_{2}=1$. We call this case the \enquote{non-trapping magnetic step}.
\end{enumerate}
In order to perform the spectral analysis, we can use the translation invariance in the $y$-direction and thus the direct integral decomposition (see \cite[XIII.16]{ReSi78}) associated with the Fourier transform with respect to $y$, denoted by $\mathcal{F}_{y}$,
$$\mathcal{F}_{y}\mathfrak{L}_{\gb}\mathcal{F}_{y}^{-1}=\int_{k\in\R}^{\oplus} \gh_{\gb}(k) \dx k\,,$$
where 
$$\gh_{\gb}(k)=D_{x}^2+V_{\gb}(x,k),\qquad \mbox{ with }\qquad V_{\gb}(x,k)=(k-a_{\gb}(x))^2\,.$$
The domain of $\gh_{\gb}(k)$ is given by
$$\Dom(\gh_{\gb}(k))=\{\psi\in\Dom(\gq_{\gb}(k)) : (D_{x}^2+V_{\gb}(x,k))\psi\in\sL^2(\R)\}\,,$$
where the quadratic form $\gq_{\gb}(k)$ is defined by
$$\gq_{\gb}(k)(\psi)=\int_{x\in\R} |\psi'(x)|^2+|(k-V_{\gb}(x,k))\psi|^2\dx x\,.$$
We have:
$$\sp(\mathfrak{L}_{\gb})=\overline{\bigcup_{k\in\R}\sp(\gh_{\gb}(k))} \,.$$
\begin{notation}
We denote by $\lambda_{\gb,n}(k)$ the $n$-th Rayleigh quotient of $\gh_{\gb}(k)$. We recall that if $\lambda_{\gb,n}(k)$ is strictly less than the infimum of the essential spectrum, it coincides with the $n$-th eigenvalue of $\gh_{\gb}(k)$.
\end{notation}

We restrict ourselves to the case $\gb=(0,1)$. Since we have $V_{\gb}(x,k)=k^2$ for $x\leq 0$ and $\lim_{x\to+\infty}V_{\gb}(x,k)=+\infty$, we easily deduce the following proposition.
\begin{prop}
For $\gb=(0,1)$ and $k\in\R$, the essential spectrum of $\gh_{b}(k)$ is given by
$$\spe(\gh_{b}(k))=[k^2,+\infty)\,.$$
Moreover we have $\sp(\mathcal{L}_{\gb})=[0,+\infty)$.
\end{prop}
In fact, we can prove slightly more.
\begin{prop}
For $\gb=(0,1)$ and $k\in\R$,  the operator $\gh_{\gb}(k)$ has no embedded eigenvalues in its essential spectrum.
\end{prop}
\begin{proof}
Let us consider $\lambda\geq k^2$ and $\psi\in \Dom(\gh_{\gb}(k))$ such that:
\begin{equation}\label{eve}
-\psi''+(k-a_{\gb}(x))^2\psi=\lambda \psi\,.
\end{equation}
For $x<0$, we have $-\psi''=(\lambda-k^2) \psi$ whose only solution in $\sL^2(\R_{-})$ is zero. But since the solutions of \eqref{eve} belongs to $\sH^2_{\loc}$ and are in $\mathcal{C}^1(\R)$, this implies that $\psi(0)=\psi'(0)=0$ and thus $\psi=0$ by the Cauchy-Lipschitz theorem.
\end{proof}
Let us now describe the discrete spectrum, that is the eigenvalues $\lambda<k^2$. Since, for $k\leq 0$, we have $\gq_{\gb}(k)\geq k^2$, we deduce the following proposition by the min-max principle.
\begin{prop}
For $\gb=(0,1)$ and $k\leq 0$, we have
$$\sp(\gh_{\gb}(k))=\spe(\gh_{\gb}(k))=[k^2,+\infty)\,.$$
\end{prop}
Therefore we must only analyze the case when $k>0$. The following lemma is a reformulation of the eigenvalue problem.
\begin{lem}\label{lemma-trans}
The eigenvalues $\lambda<k^2$ of $\gh_{\gb}(k)$ are exactly the $\lambda$ such that there exists a non zero function $\psi\in\sL^2(\R_{+})$ satisfying
\begin{subnumcases}{}
-\psi''(x)+(x-k)^2\psi=\lambda \psi(x),\\
\psi'(0)-\sqrt{k^2-\lambda}\, \psi(0)=0\, .
\end{subnumcases}
Moreover, the eigenfunctions of $\gh_{\gb}(k)$ can only vanish on $\R_{+}$. The eigenvalues of $\gh_{\gb}(k)$ are simple.
\end{lem}
\begin{proof}
We consider $\gh_{\gb}(k)\psi=\lambda\psi$. Since $\psi\in\sL^2(\R_{-})$, there exists $A\in\R$ such that, for $x\leq 0$, $\psi(x)=Ae^{x\sqrt{k^2-\lambda}}$. Then we have to solve $-\psi''+V_{\gb}(x,k)\psi=\lambda\psi$ for $x\geq 0$ with the transmission conditions
$$\psi(0)=A \ \ \mbox{and} \ \ \psi'(0)=A\sqrt{k^2-\lambda} \, $$
or equivalently
$$\psi'(0)-\sqrt{k^2-\lambda}\,\psi(0)=0\,.$$
In particular, $A$ cannot be zero. The simplicity is a consequence of the Cauchy-Lipschitz theorem.
\end{proof}

\begin{lem}\label{non-decreasing}
The functions $\R_{+}\ni k\mapsto \lambda_{\gb,n}(k)$ are non decreasing.
\end{lem}
\begin{proof}
We use the translation $x=y+k$ to see that $\gh_{b}(k)$ is unitarily equivalent to $D_{y}^2+\tilde V(y,k)$ with $\tilde V(y,k)=\one_{(-\infty,-k)}(y)k^2+\one_{(-k,+\infty)}(y)y^2$. For $0<k_{1}<k_{2}$, we have:
\begin{align*}
\tilde V(y,k_{2})-\tilde V(y,k_{1})&=\one_{(-k_{2},-k_{1})}(y)y^2+\one_{(-\infty,-k_{2})}(y)k_{2}^2-\one_{(-\infty,-k_{1})}(y)k_{1}^2\\
&\geq \one_{(-k_{2},-k_{1})}(y)k_{1}^2+\one_{(-\infty,-k_{2})}(y)k_{2}^2-\one_{(-\infty,-k_{1})}(y)k_{1}^2\\
&=\one_{(-\infty,-k_{2})}(y)k_{2}^2-\one_{(-\infty,-k_{2})}(y)k_{1}^2\geq 0\,.
\end{align*}
By the min-max principle, we infer the desired monotonicity.
\end{proof}
The next lemma is a consequence of the Sturm-Liouville theory.
\begin{lem}
Let $n\in\N$ be such that $\lambda_{\gb,n}(k)<k^2$. Then, the corresponding eigenspace is one dimensional and is generated by a normalized function $\psi_{\gb,n}(k)$, depending analytically on $k$, which has exactly $n-1$ zeros which are positive.
\end{lem}
\begin{proof}
We have only to explain the part of the statement concerned with the zeros. Thanks to Lemma \ref{lemma-trans}, one knows that the zeros are necessarily positive. Then, we apply the strategy of the proof of Proposition \ref{ST} (the integrability at infinity replaces the cancellation of the eigenfunction).
\end{proof}

\begin{notation}
We let $E_{0}=0$ and for $n\geq 1$, $E_{n}=2n-1$.
\end{notation}

By using the harmonic approximation in the semiclassical limit (see Chapter \ref{chapter-spectral-theory}, Section \ref{Harmonic} and, in this chapter, Sections \ref{Harmonic-2} and \ref{Harmonic-3}; see also \cite{DiSj99}), we can prove the following lemma.
\begin{lem}\label{limit}
For all $n\geq 1$, we have
$$\lim_{k\to+\infty}\lambda_{\gb,n}(k)=E_{n}\,.$$
In particular, for $k$ large enough, we have $\lambda_{\gb,n}(k)\leq E_{n}<k^2$.
\end{lem}
Let us now prove that the $n$-th band function lies between the two consecutive Landau levels $E_{n-1}$ and $E_{n}$.
\begin{prop}\label{strip}
For all $n\geq 1$ and for all $k>0$ such that $\lambda_{\gb,n}(k)<k^2$ we have $\lambda_{\gb,n}(k)\in\left(E_{n-1},E_{n}\right)$.
\end{prop}
\begin{proof}
By Lemmas \ref{non-decreasing} and \ref{limit}, we have $\lambda_{\gb,n}(k)<E_{n}$ (the strict inequality comes from the analyticity). It remains to prove that $\lambda_{\gb,n}(k)>E_{n-1}$. We have clearly $\lambda_{\gb,1}(k)>E_{0}$. Let us introduce $\gh^{D}(\zeta)$ the Dirichlet realization on $\R_{+}$ of $D_{t}^2+(t-\zeta)^2$ and its eigenvalues $(\mu^D_{\ell}(\zeta))_{\ell\geq 1}$. For $n\geq 2$, we consider the function $\varphi_{n}(t)=\psi_{\gb, n}(t+z_{n,1}(k))$ which satisfies $\gh^D(k-z_{n,1}(k))\varphi_{n}=\lambda_{\gb,n}(k)\varphi_{n}$ and has exactly $n-2$ zeros on $\R_{+}$. By the Sturm's oscillation theorem, $\varphi_{n}$ is the $(n-1)$-th eigenfunction of $\gh^D(k-z_{n,1}(k))$. Therefore we have $\lambda_{\gb,n}(k)=\mu^D_{n-1}(k-z_{n,1}(k))$. Moreover for all $\ell\geq 1$ and $\zeta\in\R$, $\mu^D_{\ell}(\zeta)>E_{\ell}$ (see \cite{dBiePu99}). This provides the desired conclusion.
\end{proof}

\begin{notation}
We recall (modulo a slight adaptation of the last section) that, for all $\ell\geq 1$, the function $\nu^{[0]}_{\ell}$ admits a unique and non-degenerate minimum at $\zeta=\zeta^{[0]}_{\ell-1}$, denoted by $\Theta_{\ell-1}$ that belongs to $(E_{\ell-1},E_{\ell})$. Moreover, $\zeta^{[0]}_{\ell-1}$ is also the unique solution of the equation $\nu^{[0]}_{\ell}(\zeta)=\zeta^2$ (see Figure \ref{mutau14}).
\end{notation}

\begin{prop}
For all $n\geq 1$, the equation $\lambda_{\gb,n}(k)=k^2$ has a unique non negative solution,  $k=\zeta^{[0]}_{n-1}$, such that, locally, for $k>k_{n}$, $\lambda_{\gb,n}(k)<k^2$. Moreover we have $\lambda_{\gb,n}(\zeta^{[0]}_{n-1})=\Theta_{n-1}$.
\end{prop}
\begin{proof}
Thanks to Lemma \ref{limit}, we can define $k_{n}=\max\{k\geq0 : \lambda_{\gb,n}(k)=k^2\}$. By continuity, we have, for all $k>k_{n}$, $\lambda_{\gb,n}(k)<k^2$. Let us now prove the uniqueness. Let us consider a solution $\tilde k_{n}\geq 0$. For all integer $p\geq p_{0}$ with $p_{0}$ large enough, we have $\lambda_{\gb,n}\left(\tilde k_{n}+\frac 1 p\right)<\left(\tilde k_{n}+\frac 1 p\right)^2$. 
Let us now consider the eigenvalue equation
\begin{equation}\label{eve'}
D_{x}^2\varphi_{n,p}+(\tilde k_{n,p}-a_{\gb}(x))^2\varphi_{n,p}=\lambda_{n,p}\varphi_{n,p}\,,
\end{equation}
where $\varphi_{n,p}=\psi_{\gb,n}(\tilde k_{n,p})$, $\tilde k_{n,p}=\tilde k_{n}+\frac{1}{p}$ and $\lambda_{n,p}=\lambda_{\gb,n}(\tilde k_{n,p})$. Let us investigate the limit $p\to+\infty$. As seen in the proof of Lemma \ref{lemma-trans}, we know that there exists $\alpha\in \R^{*}$ such that, for $x\leq 0$,
$$\varphi_{n,p}(x)=\alpha e^{x\sqrt{\tilde k_{n,p}^2-\lambda_{n,p}}}\,.$$
We can relate the de Gennes eigenfunctions to the Weber functions (see for instance \cite{AB64}).
\begin{notation}
We denote by $U(a,x)$ the first Weber parabolic special function which is solution of the linear ODE:
$$-y''(x)+\frac{1}{4}x^2y(x)=-ay(x) \,.$$
It decays exponentially for $x\to+\infty$. We let $\hat{U}=\Re(U)$.
\end{notation}
By solving \eqref{eve'} on $x\geq0$ and using the parabolic cylinder $\hat{U}$ function, we find that there exits $\beta$ such that, for $x\geq 0$,
$$\varphi_{n,p}(x)=\beta\hat U\left(-\frac{\lambda_{n,p}}{2} ; \sqrt{2}(x-\tilde k_{n,p})  \right)\,.$$
Since $\varphi_{n,p}\neq 0$, we have $(\alpha,\beta)\neq (0,0)$ and $\varphi_{n,p}$ is $\mathcal{C}^1$ at $x=0$, we get the transmission condition:
$$\sqrt{\tilde k_{n,p}^2-\lambda_{n,p}}\hat U\left(-\frac{\lambda_{n,p}}{2};-\tilde k_{n,p}\sqrt{2}\right)-\sqrt{2}\hat U'\left(-\frac{\lambda_{n,p}}{2} ;-\tilde k_{n,p}\sqrt{2} \right)=0\,.$$
 By continuity and taking the limit $p\to+\infty$, we get
$$\hat U'\left(-\frac{\tilde k_{n}^2}{2} ;-\tilde k_{n}\sqrt{2} \right)=0\,.$$
Notice that the function $x\mapsto \hat U\left(-\frac{\tilde k_{n}^2}{2}; \sqrt{2}(x-\tilde k_{n})\right)$ solves the differential equation
\begin{equation*}
\left\{ 
\begin{aligned}
& -y''(x)+(x-\tilde k_{n})^2y(x)=\tilde k_{n}^2 y(x)
\\
& y'(0)=0 \ \ \text{and} \ \ y(0) \neq 0
\end{aligned}
\right.
\end{equation*}
Moreover it belongs to $\mathcal{S}(\overline{\R_{+}})$.

Therefore, there exists $\ell\geq 1$ such that $\nu^{[0]}_{\ell}(\tilde k_{n})=\tilde k_{n}^2$, and therefore $\tilde k_{n}=\zeta^{[0]}_{\ell-1}$ and $\tilde k_{n}^2=\Theta_{\ell-1}$. By Proposition \ref{strip}, we know that $\tilde k_{n}^2=\lambda_{\gb,n}(\tilde k_{n})\in[E_{n-1},E_{n}]$. Moreover, we recall that $\Theta_{\ell-1}\in\left(E_{\ell-1},E_{\ell}\right)$. This implies that $\ell=n$.
\end{proof}

\begin{cor}
For $n\geq 1$ and $\zeta^{[0]}_{n-1}<k<\zeta^{[0]}_{n}$, the operator $\gh_{\gb}(k)$ admits $n$ simple eigenvalues below the threshold of its essential spectrum.
\end{cor}

\section{Towards analytic families}\label{Sec.analytic}

\subsection{Kato-Rellich's theorem}
In Section \ref{Sec:deGennes}, we proved that $(\mathfrak{L}^{[0]}_{\zeta})_{\zeta\in\R}$ is an \enquote{analytic family}. In fact, this comes from the general theory of Kato (see \cite[Chapter 7]{Kato66} and also the older reference \cite{Rellich42}).

\begin{theo}\label{KT}
Let us consider a family of self-adjoint operators with compact resolvents $(\mathfrak{L}_{\zeta})_{\zeta\in\R}$. We assume that
\begin{enumerate}
\item\label{KT1} the domain $\Dom(\mathfrak{L}_{\zeta})$ does not depend on $\zeta$,
\item\label{KT2} for all $\zeta_{1}\in\R$, there exists $r>0$, such that for all $u\in\Dom(\mathfrak{L}_{\zeta})$, $\zeta\mapsto\mathfrak{L}_{\zeta}u$ is analytic in $\mathcal{B}(\zeta_{1}, r)$.
\end{enumerate}
Let $\zeta_{1}\in\R$ and $\Gamma$ a smooth contour avoiding the spectrum of $\mathfrak{L}_{\zeta_{1}}$. There exists $r>0$ such that if $|\zeta-\zeta_{1}|\leq r$, $\Gamma$ avoids the spectrum of $\mathfrak{L}_{\zeta}$ and
\[P_{\Gamma}(\zeta)=\frac{1}{2i\pi}\int_{\Gamma} (\mathfrak{L}_{\zeta}-z)^{-1}\dx z\]
is analytic near $\zeta_{1}$ and coincides with the orthogonal projection on the direct sum of the eigenspaces associated with the eigenvalues of $\mathfrak{L}_{\zeta}$.

Moreover, if $\mu$ is an eigenvalue of $\mathfrak{L}_{\zeta_{1}}$, with multiplicity $m$, then, in a neighborhood of $\zeta_{1}$, the eigenvalues of $\mathfrak{L}_{\zeta}$ can be represented as the union of $m$ analytic curves $(\nu_{k}(\zeta))_{k\in\{1,\ldots,m\}}$ crossing at $\mu$ and there exists a corresponding analytic family of eigenfunctions $(w_{k}(\zeta))_{k\in\{1,\ldots,m\}}$.
\end{theo}

\begin{proof}
Let $\zeta_{1}\in\R$ and $K$ be a compact set avoiding the spectrum of $\mathfrak{L}_{\zeta_{1}}$. For $z\in K$, we write
\[\mathfrak{L}_{\zeta}-z=\mathfrak{L}_{\zeta_{1}}-z+\mathfrak{L}_{\zeta}-\mathfrak{L}_{\zeta_{1}}=\left(\mathsf{Id}+\left(\mathfrak{L}_{\zeta}-\mathfrak{L}_{\zeta_{1}}\right)\left(\mathfrak{L}_{\zeta_{1}}-z\right)^{-1}\right)\left(\mathfrak{L}_{\zeta_{1}}-z\right)\,\,.\]
We let 
\[\mathcal{D}^z_{\zeta}=\left(\mathfrak{L}_{\zeta}-\mathfrak{L}_{\zeta_{1}}\right)\left(\mathfrak{L}_{\zeta_{1}}-z\right)^{-1}=\left(\mathfrak{L}_{\zeta}-\mathfrak{L}_{\zeta_{1}}\right)\left(\mathfrak{L}_{\zeta_{1}}-i\right)^{-1}\left(\mathfrak{L}_{\zeta_{1}}-i\right)\left(\mathfrak{L}_{\zeta_{1}}-z\right)^{-1}\]
and 
\[A(\zeta)=\left(\mathfrak{L}_{\zeta}-\mathfrak{L}_{\zeta_{1}}\right)\left(\mathfrak{L}_{\zeta_{1}}-i\right)^{-1},\qquad B(z)=\left(\mathfrak{L}_{\zeta_{1}}-i\right)\left(\mathfrak{L}_{\zeta_{1}}-z\right)^{-1}\]
so that
\[\mathcal{D}^z_{\zeta}=A(\zeta)B(z)\,.\]
We have already seen in Exercise \ref{exe.an.res} that $B$ is analytic. Let us show that $A$ is analytic. In order to see this, we have just to notice that it is pointwise analytic, i.e. $\zeta\mapsto A(\zeta)\psi$ is analytic for all $\psi\in\Dom(\mathfrak{L}_{\zeta_{1}})$. Indeed, if this is the case, for all $v\in\sH$, we may find a sequence $A_{n}$ of linear applications such that
\[A(\zeta)v=\sum_{n=0}^\infty (\zeta-\zeta_{1})^n A_{n}(v)\,,\qquad \zeta\in\mathcal{B}(\zeta_{1}, r)\,.\]
By using the Cauchy formula (in the spirit of Exercise \ref{exo.13}) or the Cauchy inequalities (with the Banach-Steinhaus theorem), we infer that $\zeta\mapsto A(\zeta)$ is analytic and
\[A(\zeta)=\sum_{n=0}^\infty (\zeta-\zeta_{1})^n A_{n},\qquad\mbox{ with }\qquad A_{0}=0\,.\]
We will denote by $R_{A}>0$ the convergence radius of this series. 

Then we have $\mathsf{Id}+\mathcal{D}^z_{\zeta}=\mathsf{Id}+A(\zeta)B(z)$ and we notice that it is invertible for $\zeta$ close enough to $\zeta_{1}$, uniformly in $z\in K$. Let us write
\[\mathsf{Id}+A(\zeta)B(z)=\sum_{k=1}^\infty (\zeta-\zeta_{1})^k a_{k}(z)\qquad a_{k}(z)=A_{k}B(z)\,\,,\]
where the $a_{k}(z)$ satisfy, for $r\in(0,R_{A})$,
\[r^k\|a_{k}(z)\|\leq \|B\|_{\sL^\infty(K)}M=\tilde M\,.\]
We may consider the formal inverse of the power series $\sum_{k=1}^\infty (\zeta-\zeta_{1})^k a_{k}(z)$, denoted by $\sum_{k\geq 0} (\zeta-\zeta_{1})^k b_{k}(z)$ where the sequence $(b_{k}(z))_{k\geq 1}$ is defined through the Cauchy product. It is a classical exercise to see that
\[b_{0}(z)=1,\qquad |b_{k}(z)|\leq\frac{\tilde M (\tilde M+1)^{k-1}}{r^k},\quad k\geq 1\,\,,\]
so that the convergence radius of $\sum_{k\geq 0} (\zeta-\zeta_{1})^k b_{k}(z)$ is at least $r'=\frac{r}{\tilde M+1}>0$. Moreover the convergence is uniform for $z\in K$ and on $D\left(\zeta_{1},\frac{r'}{2}\right)$. We have
\[\left(\mathsf{Id}+A(\zeta)B(z)\right)^{-1}=\sum_{k\geq 0}^\infty (\zeta-\zeta_{1})^k b_{k}(z)\,.\]
We get
\[\left(\mathfrak{L}_{\zeta}-z\right)^{-1}=\left(\mathfrak{L}_{\zeta_{1}}-z\right)^{-1}\sum_{k\geq 0}^\infty (\zeta-\zeta_{1})^k b_{k}(z)\,\,,\]
uniformly with respect to $z\in K$. It remains to integrate with respect to $z$ on $\Gamma$.

Let us consider an eigenvalue $\mu_{n}(\zeta_{1})$ of multiplicity $m$ and a contour $\Gamma$ enclosing only $\mu_{n}(\zeta_{1})$. Let us show that $P_{\Gamma}(\zeta)$ has constant rank $m$ as soon as $\zeta$ is close to $\zeta_{1}$. We use an argument of Kato (see \cite[I.4.6]{Kato66}). We choose $r>0$ such that, for $\zeta\in D(\zeta_{1},r)$, we get $\|P_{\Gamma}(\zeta)-P_{\Gamma}(\zeta_{1})\|<1$. We let $P=P_{\Gamma}(\zeta)$ and $Q=P_{\Gamma}(\zeta_{1})$. We let
\begin{align*}
U&=QP+(\mathsf{Id}-Q)(\mathsf{Id}-P)\in\mathcal{L}\left(\mathrm{range}(P),\mathrm{range}(Q)\right)\\
V&=PQ+(\mathsf{Id}-P)(\mathsf{Id}-Q)\in\mathcal{L}\left(\mathrm{range}(Q),\mathrm{range}(P)\right)\,
\end{align*}
and notice that $UV=VU=\mathsf{Id}-(P-Q)^2$. Thus $UV$ and $VU$ are invertible, so are $U$ and $V$ and then $\mathrm{range}(P)=\mathrm{range}(Q)$.

If $u_{1},\ldots, u_{m}$ is an eigenbasis associated with $\mu_{n}$, the family $(P_{\Gamma}(\zeta)u_{k})_{k\in\{1,\ldots,m\}}$ is a basis of the range of $P_{\Gamma}(\zeta)$ (for $\zeta$ close enough to $\zeta_{1}$).

If we let $v_{k}(\zeta)=P_{\Gamma}(\zeta)u_{k}$, we notice that $v_{k}$ is analytic. Since $\mathrm{range}(P_{\Gamma}(\zeta))$ is stable by $\mathfrak{L}_{\zeta}$, the spectrum of $\mathfrak{L}_{\zeta}$ enclosed in $\Gamma$ is nothing but the spectrum of the restriction of $P_{\Gamma}(\zeta)$ to this finite dimensional subspace. We may also orthonormalize the family $(P_{\Gamma}(\zeta)u_{k})_{k\in\{1,\ldots,m\}}$ to get an orthonormal basis depending on $\zeta$ analytically and the investigation is reduced to a finite dimensional matrix depending on $\zeta$ analytically. In this case the analytic representation of the eigenvalues and eigenfunctions is well-known (see \cite[Chapter II, Theorem 6.1]{Kato66}). 

\end{proof}

\subsection{An application to the Lu-Pan operator}\label{HLP-op}
Let us recall that $\mathfrak{L}^\LP_\theta$ is defined by:
\[\mathfrak{L}^\LP_\theta=-\Delta+V_{\theta}=D_{s}^2+D_{t}^2+V_{\theta}\,,\] 
where $V_{\theta}$ is defined for any $\theta\in (0,\tfrac\pi2)$ by
\[V_\theta \colon (s,t)\in \R^2_+ \longmapsto  (t\cos\theta-s\sin\theta)^2\,.\]
We can notice that $V_\theta$ reaches its minimum $0$ all along the line $t\cos\theta=s\sin\theta$, which makes the angle $\theta$ with  $\partial \R^2_+$.
We denote by $\Dom(\mathfrak{L}^\LP_\theta)$ the domain of $\mathcal{L}_\theta$ and we consider the associated quadratic form $\mathfrak{Q}^\LP_\theta$ defined by:
\[\mathfrak{Q}^\LP_\theta(u)=\int_{ \R^2_+} \big(|\nabla u|^2+V_\theta|u|^2\big) \dx s\dx t\,,\]
whose domain $\Dom(\mathfrak{Q}^\LP_{\theta})$ is
\[\Dom(\mathfrak{Q}^\LP_{\theta}) = \{ u\in \sL^2( \R^2_+),\ \nabla u \in \sL^2( \R^2_+),\  \sqrt{V_{\theta}} \,u \in \sL^2( \R^2_+) \}.\]
Let $\mathfrak{s}_{n}(\theta)$ denote the $n$-th Rayleigh quotient of $\mathfrak{L}^\LP_\theta$.
Let us recall some fundamental spectral properties of $\mathfrak{L}^\LP_\theta$ when $\theta\in\left(0,\frac{\pi}{2}\right)$.

It is proved in \cite{HelMo02} that $\spe(\mathfrak{L}^\LP_\theta)=[1,+\infty)$ and that $\theta\mapsto\mathfrak{s}_{1}(\theta)$ is non decreasing. Thanks to Corollary \ref{cor:Harnack}, we know that $\mathfrak{s}_{1}(\theta)$ is a simple eigenvalue for$\left(0,\frac\pi2\right)$. It is possible to prove that, modulo a rotation and a rescaling, depending on $\theta$ analytically, that $\mathfrak{L}^\LP_\theta$ is an analytic family (it satisfies \eqref{KT1} and \eqref{KT2} in Theorem \ref{KT}). As a consequence (we do not really need to care about the essential spectrum since $\mathfrak{s}_{1}$ is strictly below), we deduce that $\theta\mapsto \mathfrak{s}_{1}(\theta)$ is analytic. Then, we can show that the function $(0,\frac\pi2)\ni\theta\mapsto\mathfrak{s}_{1}(\theta)$ is increasing (see \cite[Lemma 3.6]{LuPan00a} and Chapter \ref{chapter-models}, Section \ref{Sec.Lu-Pan} where a close problem is analyzed).

\section{Examples of Feynman-Hellmann formulas}\label{Sec:FH}

In this section, we give examples of the so-called Feynman-Hellmann formulas (that we used in Section \ref{Sec:deGennes}).

\subsection{De Gennes operator}
Let us prove propositions which are often used in the study of the magnetic Laplacian.

For $\rho>0$ and $\zeta\in\R$, let us introduce the Neumann realization on $\R_{+}$ of:
\[\mathfrak{L}^{[0]}_{\rho,\zeta}=-\rho^{-1}\dr_{\tau}^2+(\rho^{1/2}\tau-\zeta)^2\,.\]
By scaling, we observe that $\mathfrak{L}^{[0]}_{\rho,\zeta}$ is unitarily equivalent to $\mathfrak{L}^{[0]}_{\zeta}$ and that $\mathfrak{L}^{[0]}_{1,\zeta}=\mathfrak{L}^{[0]}_{\zeta}$ (the corresponding eigenfunction is $u^{[0]}_{1,\zeta}=u^{[0]}_{\zeta}$). 
\begin{rem}
The introduction of the scaling parameter $\rho$ is related to the Virial theorem (see \cite{Weidmann}) which was used by physicists in the theory of superconductivity (see \cite{DAS96} and also \cite{ABG10,CDG95}). We also refer to the papers \cite{Ray10c} and \cite{Ray11b} where it is used many times.
\end{rem}

The domain of $\mathfrak{L}^{[0]}_{\rho,\zeta}$ is $\left\{u\in\mathsf{B}^2(\R_{+}), u'(0)=0\right\}$ and is independent from $\rho$ and $\zeta$ so that the family $\left(\mathfrak{L}^{[0]}_{\rho,\zeta}\right)_{\rho>0,\zeta\in\R}$ is an analytic family of type (A). The lowest eigenvalue of $H_{\rho,\zeta}$ is $ \nu_{1}^{[0]}(\zeta)$ and we will denote by $u_{\rho,\zeta}$ the corresponding normalized eigenfunction:
\[u^{[0]}_{\rho,\zeta}(\tau)=\rho^{1/4}u_{\zeta}^{[0]}(\rho^{1/2}\tau)\,.\]
In order to lighten the notation and when it is not ambiguous we will write $\mathfrak{L}$ for $\mathfrak{L}^{[0]}_{\rho,\zeta}$, $u$ for $u^{[0]}_{\rho,\zeta}$ and $\nu$ for $ \nu_{1}^{[0]}(\zeta)$.

The main idea is now to take derivatives of:
\begin{equation}\label{Hu}
\mathfrak{L} u=\nu u
\end{equation}
with respect to $\rho$ and $\zeta$.
Taking the derivative with respect to $\rho$ and $\zeta$, we get the following proposition.
\begin{prop}\label{S}
We have:
\begin{equation}\label{dxiu}
(\mathfrak{L}-\nu)\dr_{\zeta}u=2(\rho^{1/2}\tau-\zeta)u+\nu'(\zeta)u
\end{equation}
and
\begin{equation}\label{drhou}
(\mathfrak{L}-\nu)\dr_{\rho}u=\left(-\rho^{-2}\dr^2_{\tau}-\zeta\rho^{-1}(\rho^{1/2}\tau-\zeta)-\rho^{-1}\tau(\rho^{1/2}\tau-\zeta)^2\right)u\,.
\end{equation}
Moreover, we get:
\begin{equation}\label{Su}
(\mathfrak{L}-\nu)(Su)=Xu\,,
\end{equation}
where 
\[X=-\frac{\zeta}{2}\nu'(\zeta)+\rho^{-1}\dr_{\tau}^2+(\rho^{1/2}\tau-\zeta)^2\]
and
\[S=-\frac{\zeta}{2}\dr_{\zeta}-\rho\dr_{\rho}\,.\]
\end{prop}
\begin{proof}
Taking the derivatives with respect to $\zeta$ and $\rho$ of \eqref{Hu}, we get:
\[(\mathfrak{L}-\nu)\dr_{\zeta}u=\nu'(\zeta)u-\dr_{\zeta}\mathfrak{L} u\]
and
\[(\mathfrak{L}-\nu)\dr_{\rho}u=-\dr_{\rho}\mathfrak{L}\,.\]
We have: $\dr_{\zeta}\mathfrak{L}=-2(\rho^{1/2}\tau-\zeta)$ and $\dr_{\rho}\mathfrak{L}=\rho^{-2}\dr_{\tau}^2+\rho^{-1/2}\tau(\rho^{1/2}\tau-\zeta)$.
\end{proof}
Taking $\rho=1$ and $\zeta=\zeta_{0}$ in \eqref{dxiu}, we deduce, with the Fredholm alternative:
\begin{cor}
We have
\[(\mathfrak{L}^{[0]}_{\zeta_{0}}-\nu(\zeta_{0}))v^{[0]}_{\zeta_{0}}=2(t-\zeta_{0})u^{[0]}_{\zeta_{0}}\,,\]
with
\[v^{[0]}_{\zeta_{0}}=\left(\dr_{\zeta}u_{\zeta}^{[0]}\right)_{|\zeta=\zeta_{0}}\,.\]
Moreover, we have
\[\int_{\tau>0}(\tau-\zeta_{0})(u^{[0]}_{\zeta_{0}})^2\dx\tau=0\,.\]
\end{cor}
\begin{cor}\label{d-rho-u-u}
We have, for all $\rho>0$:
$$\int_{\tau>0} (\rho^{1/2}\tau-\zeta_{0})(u^{[0]}_{\rho,\zeta_{0}})^2\dx \tau=0$$
and:
$$\int_{\tau>0}(\tau-\zeta_{0})\left(\dr_{\rho}u\right)_{\rho=1,\zeta=\zeta_{0}} u\dx \tau=-\frac{\zeta_{0}}{4}\,.$$
\end{cor}
\begin{cor}\label{viriel}
We have:
$$(\mathfrak{L}^{[0]}_{\zeta_{0}}-\nu(\zeta_{0}))S_{0}u=\left(\dr_{\tau}^2+(\tau-\zeta_{0})^2\right)u^{[0]}_{\zeta_{0}}\,,$$
where: 
$$S_{0}u=-\left(\dr_{\rho}u^{[0]}_{\rho,\zeta}\right)_{|\rho=1,\zeta=\zeta_{0}}-\frac{\zeta_{0}}{2}v^{[0]}_{\zeta_{0}}\,.$$
Moreover, we have:
$$\|\dr_{\tau}u^{[0]}_{\zeta_{0}}\|^2=\|(\tau-\zeta_{0})u^{[0]}_{\zeta_{0}}\|^2=\frac{\Theta_{0}}{2}\,.$$
\end{cor}
The next proposition deals with the second derivative of \eqref{Hu} with respect to $\zeta$.
\begin{prop}\label{mu''}
We have:
$$(\mathfrak{L}^{[0]}_{\zeta}- \nu_{1}^{[0]}(\zeta))w^{[0]}_{\zeta_{0}}=4(\tau-\zeta_{0})v^{[0]}_{\zeta_{0}}+((\nu_{1}^{[0]})''(\zeta_{0})-2)u^{[0]}_{\zeta_{0}}\,,$$
with 
$$w^{[0]}_{\zeta_{0}}=\left(\dr_{\zeta}^2 u_{\zeta}^{[0]}\right)_{|\zeta=\zeta_{0}}\,.$$
Moreover, we have:
$$\int_{\tau>0}(\tau-\zeta_{0}) v^{[0]}_{\zeta_{0}} u^{[0]}_{\zeta_{0}}\dx \tau=\frac{2-(\nu_{1}^{[0]})''(\zeta_{0})}{4}\,.$$
\end{prop}
\begin{proof}
Taking the derivative of \eqref{dxiu} with respect to $\zeta$ (with $\rho=1$), we get:
$$(\mathfrak{L}^{[0]}_{\zeta}- \nu_{1}^{[0]}(\zeta))\dr_{\zeta}^2 u_{\zeta}^{[0]}=2\nu'(\zeta)\dr_{\zeta}u_{\zeta}^{[0]}+4(\tau-\zeta)\dr_{\zeta}u_{\zeta}^{[0]}+(\nu''(\zeta)-2)u_{\zeta}^{[0]}\,.$$
It remains to take $\zeta=\zeta_{0}$ and to write the Fredholm alternative.
\end{proof}

\subsection{Lu-Pan operator (bis)}\label{Sec.Lu-Pan}
The following result is obtained in \cite{BDPR11}.
\begin{prop}\label{Ctheta>0}
For all $\theta\in\left(0,\frac{\pi}{2}\right)$, we have:
$$\mathfrak{s}_{1}(\theta)\cos\theta - \mathfrak{s}_{1}'(\theta)\sin\theta >0\,.$$
Moreover, we have:
$$\lim_{\substack{\theta\to\frac{\pi}{2}\\ \theta<\frac{\pi}{2} }} \mathfrak{s}'_{1}(\theta)=0\,.$$
\end{prop}

\begin{proof}
For $\gamma\geq 0$, we introduce the operator (see \cite{Ray12}):
$$\mathfrak{L}^\LP_{\theta,\gamma}=D_{s}^2+D_{t}^2+(t(\cos\theta +\gamma)-s\sin\theta)^2$$
and we denote by $\mathfrak{s}_{1}(\theta,\gamma)$ the bottom of its spectrum.
Let $\rho>0$ and $\alpha\in(0,\frac\pi2)$ satisfy
\[
   \cos\theta+\gamma=\rho\cos\alpha \quad\mbox{and}\quad \sin\theta=\rho\sin\alpha.
\]
We perform the rescaling $t=\rho^{-1/2}\hat{t}$, $s=\rho^{-1/2}\hat{s}$ and obtain that $\mathfrak{L}^\LP_{\theta,\gamma}$ is unitarily equivalent to:
$$\rho(D_{\hat{s}}^2+D_{\hat{t}}^2+(\hat{t}\cos\alpha-\hat{s} \sin\alpha)^2)=\rho \mathfrak{L}^\LP_{\alpha}\,.$$
In particular, we observe that $\mathfrak{s}_{1}(\theta,\gamma) = \rho\mathfrak{s}_1(\alpha)$ is a simple eigenvalue: there holds
\begin{equation}\label{sigmathetagamma}
\mathfrak{s}_{1}(\theta,\gamma)=\sqrt{(\cos\theta+\gamma)^2+\sin^2\theta}\,\,\,\, \mathfrak{s}_{1}\left(\arctan \left(\frac{\sin\theta}{\cos\theta+\gamma}\right)\right)\,.
\end{equation}
Performing the rescaling  $\tilde{t}=(\cos\theta +\gamma)t$, we get the operator $\tilde{\mathfrak{L}}^\LP_{\theta,\gamma}$ which is unitarily equivalent to $\mathfrak{L}^\LP_{\theta,\gamma}$ :
$$\tilde{\mathfrak{L}}^\LP_{\theta,\gamma}=D_{s}^2+(\cos\theta+\gamma)^2 D_{\tilde{t}}^2+(\tilde{t}-s \sin\theta)^2\,.$$
We observe that the domain of $\tilde{\mathfrak{L}}^\LP_{\theta,\gamma}$ does not depend on $\gamma\geq 0$. Denoting by $\tilde{u}_{\theta,\gamma}$ the $\sL^2$-normalized and positive eigenfunction of $\tilde{\mathfrak{L}}^\LP_{\theta,\gamma}$ associated with $\mathfrak{s}_{1}(\theta,\gamma)$, we write:
$$\tilde{\mathfrak{L}}^\LP_{\theta,\gamma}\tilde{u}^\LP_{\theta,\gamma}=\mathfrak{s}_{1}(\theta,\gamma)\tilde{u}^\LP_{\theta,\gamma}\,.$$
Taking the derivative with respect to $\gamma$, multiplying by $\tilde{u}^\LP_{\theta,\gamma}$ and integrating, we get the Feynman-Hellmann formula:
$$\dr_{\gamma}\mathfrak{s}_{1}(\theta,\gamma)=2(\cos\theta+\gamma)\int_{ \R^2_+} |D_{t} \tilde{u}^\LP_{\theta,\gamma}|^2 \dx\x\geq 0\,.$$
We deduce that, if $\dr_{\gamma}\mathfrak{s}_{1}(\theta,\gamma)=0$, then $D_{t} \tilde{u}^\LP_{\theta,\gamma}=0$ and $\tilde{u}^\LP_{\theta,\gamma}$ only depends on $s$, which is a contradiction with $\tilde{u}^\LP_{\theta,\gamma}\in \sL^2( \R^2_+)$.
Consequently, we have  $\dr_{\gamma}\mathfrak{s}_{1}(\theta,\gamma)>0$ for any $\gamma\geq 0$.
An easy computation using formula (\ref{sigmathetagamma}) provides: 
$$\dr_{\gamma}\mathfrak{s}_{1}(\theta,0)= \mathfrak{s}_{1}(\theta)\cos\theta-\mathfrak{s}_{1}'(\theta)\sin\theta\,.$$
The function $\mathfrak{s}_{1}$ is analytic and increasing. 
Thus we deduce:
$$\forall\theta\in\left(0,\frac{\pi}{2}\right),\qquad 0 \leq \mathfrak{s}'_{1}(\theta)<\frac{\cos\theta}{\sin\theta}\mathfrak{s}_{1}(\theta)\,.$$
We get:
$$0\leq 
\liminf_{\substack{\theta\to\frac{\pi}{2}\\ \theta<\frac{\pi}{2} }}\mathfrak{s}'_{1}(\theta) \leq \limsup_{\substack{\theta\to\frac{\pi}{2}\\ \theta<\frac{\pi}{2} }}\mathfrak{s}'_{1}(\theta)\leq 0,
$$
which ends the proof.
\end{proof}

\chapter{First semiclassical examples}\label{chapter-semi-ex}
\begin{flushright}
\begin{minipage}{0.5\textwidth}
Nous appelons ici intuition la sympathie par laquelle on se transporte \`a l'int\'erieur d'un objet pour co\"{\i}ncider avec ce qu'il a d'unique et par cons\'equent d'inexprimable.
\begin{flushright}
\textit{ La pens\'ee et le mouvant}, Bergson
\end{flushright}
\end{minipage}
\vspace*{0.5cm}
\end{flushright}

In this chapter, we give the first semiclassical examples of this book. In particular, we essentially deal with the electric Laplacian in dimension one:
\begin{enumerate}[(i)]
\item we prove a version of the Weyl's law,
\item we start the discussion about the harmonic approximation.
\end{enumerate}

\section{Semiclassical estimate of the number of eigenvalues}
In this subsection we explain how we can estimate the number of eigenvalues of $h^2D_{x}^2+V(x)$ by using the spirit of partitions of unity and reduction to local models. We propose to prove the following version of the Weyl's law in dimension one (see Remark \ref{rem-counting}).
\begin{prop}\label{number-1D}
Let us consider $V : \R\to \R$ a piecewise Lipschitzian with a finite number of discontinuities which satisfies:
\begin{enumerate}
\item $V$ tends to $\ell_{\pm\infty}$ when $x\to\pm\infty$ with $\ell_{+\infty}\leq\ell_{-\infty}$,
\item  $\sqrt{(\ell_{+\infty}-V)_{+}}$ belongs to $L^1(\R)$.
\end{enumerate}
We consider the operator $\mathfrak{h}_{h}=h^2D_{x}^2+V(x)$ and we assume that the function $(0,1)\ni h\mapsto E(h)\in(-\infty,\ell_{+\infty})$ satisfies
\begin{enumerate}
\item for any $h\in(0,1)$, $\{x\in\R : V(x)\leq E(h)\}=[x_{\min}(E(h)),x_{\max}(E(h))]$,
\item  $h^{1/3}(x_{\max}(E(h))-x_{\min}(E(h)))\underset{h\to 0}{\to} 0$,
\item $E(h)\underset{h\to 0}{\to} E_{0}\leq \ell_{+\infty}$.
\end{enumerate}
Then we have:
$$\mathsf{N}(\mathfrak{h}_{h},E(h))\underset{h\to 0}{\sim} \frac{1}{\pi h}\int_{\R} \sqrt{(E_{0}-V)_{+}}\dx x\,.$$
\end{prop}

\begin{proof}
The strategy of the proof is well-known but we recall it since the usual result does not deal with a moving threshold $E(h)$. We consider a subdivision of the real axis $(s_{j}(h^{\alpha}))_{j\in\Z}$, which contains the discontinuities of $V$, such that there exists $c>0$, $C>0$ such that, for all $j\in\Z$ and $h>0$, $ch^\alpha\leq s_{j+1}(h^\alpha)-s_{j}(h^\alpha)\leq Ch^\alpha$, where $\alpha>0$ is to be determined. We introduce 
$$J_{\min}(h^{\alpha})=\min\{j\in\Z :  s_{j}(h^\alpha)\geq x_{\min}(E(h))\}\,,$$
$$J_{\max}(h^{\alpha})=\max\{j\in\Z :  s_{j}(h^\alpha)\leq x_{\max}(E(h))\}\,.$$
For $j\in\Z$ we may introduce the Dirichlet (resp. Neumann) realization on $(s_{j}(h^{\alpha}), s_{j+1}(h^\alpha))$ of $h^2D_{x}^2+V(x)$ denoted by $\mathfrak{h}_{h,j}^\Dir$ (resp. $\mathfrak{h}_{h,j}^\Neu$). The so-called Dirichlet-Neumann bracketing (\textit{i.e.} the application of the min-max principle and easy domain inclusions, see~\cite[Chapter XIII, Section 15]{ReSi78}) implies:
$$\sum_{j=J_{\min}(h^\alpha)}^{J_{\max}(h^\alpha)} \mathsf{N}(\mathfrak{h}^\Dir_{h,j},E(h)) \leq\mathsf{N}(\mathfrak{h}_{h},E(h))\leq \sum_{j=J_{\min}(h^\alpha)-1}^{J_{\max}(h^\alpha)+1} \mathsf{N}(\mathfrak{h}^\Neu_{h,j},E(h))\,.$$
Let us estimate $\mathsf{N}(\mathfrak{h}^\Dir_{h,j},E(h))$. If $\mathfrak{q}_{h,j}^\Dir$ denotes the quadratic form of $\mathfrak{h}^\Dir_{h,j}$, we have:
$$\mathfrak{q}_{h,j}^\Dir(\psi)\leq \int_{s_{j}(h^\alpha)}^{s_{j+1}(h^\alpha)} h^2|\psi'(x)|^2+V_{j,\sup,h}|\psi(x)|^2\dx x,\quad\forall\psi\in\mathcal{C}^\infty_{0}((s_{j}(h^{\alpha}), s_{j+1}(h^\alpha)))\,,$$
where 
$$V_{j,\sup,h}=\sup_{x\in(s_{j}(h^{\alpha}), s_{j+1}(h^\alpha))} V(x)\,.$$
We infer that
$$\mathsf{N}(\mathfrak{h}^\Dir_{h,j},E(h))\geq\#\left\{n\geq 1 : n\leq \frac{1}{\pi h}(s_{j+1}(h^{\alpha})-s_{j}(h^\alpha))\sqrt{\left(E(h)-V_{j,\sup,h}\right)_{+}}\right\}$$
so that:
$$\mathsf{N}(\mathfrak{h}^\Dir_{h,j},E(h))\geq  \frac{1}{\pi h}(s_{j+1}(h^{\alpha})-s_{j}(h^\alpha))\sqrt{\left(E(h)-V_{j,\sup,h}\right)_{+}}-1$$
and thus:
\begin{multline*}
\sum_{j=J_{\min}(h^\alpha)}^{J_{\max}(h^\alpha)} \mathsf{N}(\mathfrak{h}^\Dir_{h,j},E(h))\geq \\
\frac{1}{\pi h}\sum_{j=J_{\min}(h^\alpha)}^{J_{\max}(h^\alpha)}(s_{j+1}(h^{\alpha})-s_{j}(h^\alpha))\sqrt{\left(E(h)-V_{j,\sup,h}\right)_{+}} -(J_{\max}(h^\alpha)-J_{\min}(h^\alpha)+1)\,.
\end{multline*}
Let us consider the function 
\[f_{h}(x)=\sqrt{\left(E(h)-V(x)\right)_{+}}\]
and analyze
\begin{multline*}
\left|\sum_{j=J_{\min}(h^\alpha)}^{J_{\max}(h^\alpha)}(s_{j+1}(h^{\alpha})-s_{j}(h^\alpha))\sqrt{\left(E(h)-V_{j,\sup,h}\right)_{+}}       -\int_{\R} f_{h}(x)\dx x\right|\\
\leq \left|\sum_{j=J_{\min}(h^\alpha)}^{J_{\max}(h^\alpha)} \int_{s_{j}(h^\alpha)}^{s_{j+1}(h^\alpha)} \sqrt{\left(E(h)-V_{j,\sup,h}\right)_{+}}-f_{h}(x)\dx x\right|\\
+\int_{s_{J_{\max}}(h^\alpha)}^{x_{\max}(E(h))} f_{h}(x)\dx x+\int_{x_{\min}(E(h))}^{s_{J_{\min}(h^\alpha)}} f_{h}(x)\dx x\\
\leq  \left|\sum_{j=J_{\min}(h^\alpha)}^{J_{\max}(h^\alpha)} \int_{s_{j}(h^\alpha)}^{s_{j+1}(h^\alpha)} \sqrt{\left(E(h)-V_{j,\sup,h}\right)_{+}}-f_{h}(x)\dx x\right|+\tilde C h^\alpha\,.
\end{multline*}
Using the trivial inequality $|\sqrt{a_{+}}-\sqrt{b_{+}}|\leq \sqrt{|a-b|}$, we notice that 
$$|f_{h}(x)-\sqrt{\left(E(h)-V_{j,\sup,h}\right)_{+}}|\leq \sqrt{|V(x)-V_{j,\sup,h}|}\,.$$
Since $V$ is Lipschitzian on $(s_{j}(h^{\alpha}), s_{j+1}(h^\alpha))$, we get:
$$\left|\sum_{j=J_{\min}(h^\alpha)}^{J_{\max}(h^\alpha)} \int_{s_{j}(h^\alpha)}^{s_{j+1}(h^\alpha)}\sqrt{\left(E(h)-V_{j,\sup,h}\right)_{+}}-f_{h}(x)\dx x\right|\leq (J_{\max}(h^\alpha)-J_{\min}(h^\alpha)+1) \tilde Ch^\alpha h^{\alpha/2}\,.$$
This leads to the optimal choice $\alpha=\frac{2}{3}$ and we get the lower bound:
$$\sum_{j=J_{\min}(h^{2/3})}^{J_{\max}(h^{2/3})} \mathsf{N}(\mathfrak{h}^\Dir_{h,j},E(h))\geq \frac{1}{\pi h}\left( \int_{\R} f_{h}(x)\dx x-\tilde Ch (J_{\max}(h^{2/3})-J_{\min}(h^{2/3})+1)\right)\,.$$
Therefore we infer
$$\mathsf{N}(\mathfrak{h}_{h},E(h))\geq \frac{1}{\pi h}\left(\int_{\R} f_{h}(x)\dx x-\tilde Ch^{1/3}(x_{\max}(E(h))-x_{\min}(E(h))-\tilde C h\right)\,.$$
We notice that: $f_{h}(x)\leq \sqrt{(\ell_{+\infty}-V(x))_{+}}$ so that we can apply the dominate convergence theorem.
We can deal with the Neumann realizations in the same way.
\end{proof}
\begin{rem}\label{rem-counting}
Classical results (see~\cite{ReSi78, Ro87, DiSj99, Z13}) impose a fixed security distance below the edge of the essential spectrum ($E(h)=E_0<l_{+\infty}$) or deal with non-negative potentials, $V$, with compact support. Both these cases are recovered by Proposition~\ref{number-1D}. In our result, the maximal threshold for which one can ensure that the semiclassical behavior of the counting function holds is dictated by the convergence rate of the potential towards its limit at infinity, through the assumption
\[ h^{1/3}(x_{\max}(E(h))-x_{\min}(E(h)))\underset{h\to 0}{\to} 0. \]
More precisely, assume that $l_{-\infty}>l_{+\infty}$ so that $x_{\min}(E(h))\geq x_{\min}(l_{+\infty})$ is uniformly bounded for $E(h)$ in a neighborhood of $l_{+\infty}$. Then
\begin{enumerate}[(i)]
\item If $l_{+\infty}-V(x)\leq C  x^{-\gamma}$ for any $x\geq x_0$ and given $x_0,C>0$ and $\gamma>2$, then one can choose $E(h)=l_{+\infty}-C h^\rho$ and $x_{\max}(E(h))\leq h^{-\rho/\gamma}$, provided $\rho<\gamma/3$.
\item If $l_{+\infty}-V(x)\leq C_1 \exp(-C_2  x)$ for any $x\geq x_0$ and given $x_0,C_1,C_2>0$, then one can choose $E(h)=l_{+\infty}-C_1\exp(C_2 h^{-1/3}\times o(h))$ and the assumption is satisfied.
\end{enumerate}
\end{rem}

\section{Harmonic approximation in dimension one}\label{Harmonic-one}

We illustrate the application of the spectral theorem in the case of the electric Laplacian $\mathfrak{L}_{h,V}=-h^2\Delta+V(x)$. We assume that $V\in\mathcal{C}^{\infty}(\R,\R)$, that $V(x)\to+\infty$ when $|x|\to+\infty$ and that it admits a unique and non degenerate minimum at $0$. This example is also the occasion to understand more in details how we construct quasimodes in general. From a heuristic point of view, we guess that the lowest eigenvalues correspond to functions localized near the minimum of the potential (this intuition comes from the classical mechanics). Therefore we can use a Taylor expansion of $V$ near $0$:
$$V(x)=\frac{V''(0)}{2} x^2+\mathcal{O}(|x|^3)\,.$$
We can then try to compare $h^2D_{x}^2+V(x)$ with $\displaystyle{h^2D_{x}^2+\frac{V''(0)}{2} x^2}$.

\begin{prop}\label{prop-harmonic-1}
For all $n\geq 1$, there exists a sequence $(\mu_{n,j})$ such that, for all $J\geq 1$, there exists $h_{0},C>0$ such that, for all $h\in(0,h_{0})$,
$$\dist\left(h\sum_{j=0}^J \mu_{n,j}h^{\frac{j}{2}},\sp(\mathfrak{L}_{h,V})\right)\leq Ch^{\frac{J+1}{2}}\,.$$
Moreover we have $\mu_{n,0}=\sqrt{\frac{V''(0)}{2}}$.
\end{prop}
\begin{proof}
For an homogeneity reason, we try the rescaling $x=h^{1/2}y$. The electric operator becomes:
$$\mathcal{L}_{h,V}=hD_{y}^2+V(h^{1/2}y)\,.$$
Let us use the Taylor formula:
$$V(h^{1/2}y)\sim\frac{V''(0)}{2}hy^2+\sum_{j\geq 3} h^{j/2}\frac{V^{(j)}(0)}{j!} y^{j}\,.$$
This provides the formal expansion:
$$\mathcal{L}_{h,V}\sim h\left(L_{0}+\sum_{j\geq 1}h^{j/2}L_{j}\right)\,,$$
where 
$$L_{0}=-\dr_{y}^2+\frac{V''(0)}{2}y^2\,.$$
We look for a quasimode and an eigenvalue in the form
$$u\sim\sum_{j\geq 0} u_{j}(y) h^{j/2},\quad\mu\sim h\sum_{j\geq 0} \mu_{j} h^{j/2}\,.$$
Let us investigate the system of PDE that we get when solving in the formal series:
$$\mathcal{L}_{h,V} u\sim \mu u\,.$$
We get the equation:
$$L_{0}u_{0}=\mu_{0}u_{0}\,.$$
Therefore we can take for $(\mu_{0}, u_{0})$ a $\sL^2$-normalized eigenpair of the harmonic oscillator.
Then we solve:
$$(L_{0}-\mu_{0})u_{1}=(\mu_{1}-L_{1})u_{0}\,.$$
We want to determine $\mu_{1}$ and $u_{1}$. We can verify that $H_{0}-\mu_{0}$ is a Fredholm operator so that a necessary and sufficient condition to solve this equation is given by:
$$\langle(\mu_{1}-L_{1})u_{0},u_{0}\rangle_{\sL^2}=0\,.$$
\begin{lem}
Let us consider the equation:
\begin{equation}\label{lem-L0}
(L_{0}-\mu_{0})u=f\,,
\end{equation}
with $f\in\mathcal{S}(\R)$ such that $\langle f,u_{0}\rangle_{\sL^2}=0$. The \eqref{lem-L0} admits a unique solution which is orthogonal to $u_{0}$ and this solution is in the Schwartz class.
\end{lem}
\begin{proof}
Let us just sketch the proof to enlighten the general idea. We know that we can find $u\in\Dom(H_{0})$ and that $u$ is determined modulo $u_{0}$ which is in the Schwartz class. Therefore, we have: $y^2 u\in \sL^2(\R)$ and $u\in \sH^2(\R)$. Let us introduce a smooth cutoff function $\chi_{R}(y)=\chi\left(R^{-1}y\right)$. $\chi_{R}y^2 u$ is in the form domain of $H_{0}$ as well as in the domain of $H_{0}$ so that we can write:
$$\langle L_{0}(\chi_{R}y^2 u), \chi_{R} y^2u \rangle_{\sL^2}=\langle[L_{0},\chi_{R}y^2]u, \chi_{R} y^2u\rangle_{\sL^2}+\langle \chi_{R} y^2u( \mu_{0}u+f),  \chi_{R} y^2u\rangle_{\sL^2}\,.$$
The commutator can easily be estimated and, by dominate convergence, we find the existence of $C>0$ such that for $R$ large enough we have:
$$\|\chi_{R} y^3 u\|^2\leq C\,.$$
The Fatou lemma involves:
$$y^3 u\in \sL^2(\R)\,.$$
This is then a standard iteration procedure which gives that $\dr_{y}^l(y^k u)\in L^2(\R)$. The Sobolev injection ($\sH^s(\R)\hookrightarrow \mathcal{C}^{s-\frac{1}{2}}(\R)$ for $s>\frac{1}{2}$) gives the conclusion.

\end{proof}

This determines a unique value of $\mu_{1}=\langle L_{1}u_{0},u_{0}\rangle_{\sL^2}.$ For this value we can find a unique $u_{1}\in\mathcal{S}(\R)$ orthogonal to $u_{0}$.

This is easy to see that this procedure can be continued at any order.

Let us consider the $(\mu_{j}, u_{j})$ that we have constructed and let us introduce:
$$U_{J,h}=\sum_{j=0}^J u_{j}(y)h^{j/2},\quad \mu_{J,h}=h\sum_{j=0}^J \mu_{j}h^{j/2}\,.$$
We estimate:
$$\|(\mathcal{L}_{h,V}-\mu_{J,h})U_{J,h}\|\,.$$
By using the Taylor formula and the definition of the $\mu_{j}$ and $u_{j}$, we have:
$$\|(\mathcal{L}_{h,V}-\mu_{J,h})U_{J,h}\|\leq C_{J}h^{(J+1)/2}\,,$$
since $h^{(J+1)/2}\|y^{(J+1)/2}U_{J,h}\|\leq C_{J} h^{(J+1)/2}$ due to the fact that $u_{j}\in\mathcal{S}(\R).$
The spectral theorem implies:
$$\dist\left(\mu_{J,h},\spd(\mathcal{L}_{h,V})\right)\leq C_{J} h^{(J+1)/2}\,.$$
\end{proof}

\section{Helffer-Kordyukov's toy operator}
Let us now give an explicit example of construction of quasimodes for the magnetic Laplacian in $\R^2$. We investigate the operator:
\[\mathfrak{L}_{h,\A}=(hD_{1}+A_{1})^2+(hD_{2}+A_{2})^2\,,\]
with domain
\[\Dom \mathfrak{L}_{h,\A}=\{\psi\in \sL^2(\R^2) : \left((hD_{1}+A_{1})^2+(hD_{1}+A_{2})^2\right)\psi\in \sL^2(\R^2)\}\,.\]

\subsection{Compact resolvent ?}
Let us state an easy lemma.
\begin{lem}\label{lb-magnetic}
We have:
$$\mathfrak{Q}_{h,\A}(\psi)\geq \left|\int_{\R^2} h\B(x) |\psi|^2\,\dx x\right|, \quad \forall \psi\in\mathcal{C}^{\infty}_{0}(\R^2)\,.$$
\end{lem}
\begin{proof}
This is a consequence of Proposition \ref{minoration-B}.
\end{proof}
\begin{prop}\label{prop.compact2D}
Suppose that $\A\in\mathcal{C}^\infty(\R^2,\R^2)$ and that $\B=\nabla\times\A\geq 0$ and $\B(x)\underset{|x|\to+\infty}{\to}+\infty$. Then, $\mathfrak{L}_{h,\A}$ has compact resolvent.
\end{prop}
\begin{proof}
This is an application of Theorem \ref{RFK} and Proposition \ref{criterion-comp}.
\end{proof}

\subsection{Quasimodes}\label{sec.quasimodes.ex}
Let us give a simple example inspired by \cite{HelKo11}. Let us choose $\A$ such that $\B=1+x^2+y^2$. We take $A_{1}=0$ and $A_{2}=x+\frac{x^3}{3}+y^2 x$.
We study the Friedrichs extension of
\[\mathfrak{L}^\ex_{h,\A}=h^2 D^2_{x}+\left(hD_{y}+x+\frac{x^3}{3}+y^2 x\right)^2\,.\]
\begin{prop}
There exists $c\in\R$ such that for all $m\in\N$, there exists $C_{m}>0$ and $h_{0}>0$ such that, for $h\in(0,h_{0})$,
\[\dist\left(h+(2m+1+c)h^2,\spd(\mathfrak{L}^\ex_{h,\A})\right)\leq C_{m} h^3\,.\]

\end{prop}

\begin{proof}

Let us try the rescaling $x=h^{1/2}u$, $y=h^{1/2}v$. We get a new operator:
$$\mathcal{L}_{h,\A}=h D_{u}^2+h\left(D_{v}+u+h\frac{u^3}{3}+hv^2 u\right)^2\,.$$
Let us conjugate by the partial Fourier transform with respect to $v$ ; we get the unitarily equivalent operator:
$$\hat  {\mathcal{L}}_{h,\A}=h D_{u}^2+h\left(\xi+u+h\frac{u^3}{3}+huD_{\xi}^2\right)^2\,.$$
Let us now use the transvection: $u=\check u-\check\xi, \xi=\check\xi$. We have:
$$D_{u}=D_{\check u}, \quad D_{\xi}=D_{\check u}+D_{\check \xi}\,.$$
We are reduced to the study of:
$$\check  {\mathcal{L}}_{h,\A}=h D_{\check u}^2+h\left(\check u+h\frac{(\check u-\check\xi)^3}{3}+h(\check u-\check \xi)(D_{\xi}+D_{\check u})^2\right)^2$$
We can expand $\check  {\mathcal{L}}_{h,\A}$ in formal power series:
$$\check  {\mathcal{L}}_{h,\A}=hP_{0}+h^2 P_{1}+\ldots\,,$$
where $P_{0}=D_{\check u}^2+\check u^2$ and $P_{1}=\frac{2}{3}\check u(\check u-\check\xi)^3+(\check u-\check \xi)(D_{\check\xi}+D_{\check u})^2\check u+\check u(\check u-\check \xi)(D_{\check\xi}+D_{\check u})^2$.

Let us look for quasimodes in the form
$$\lambda\sim h\lambda_{0}+h^2\lambda_{1}+\ldots,\quad \psi\sim \psi_{0}+h \psi_{1}+\ldots$$
We solve the equation:
$$P_{0}\psi_{0}=\lambda_{0}\psi_{0}\,.$$
We take $\lambda_{0}=1$ and $\psi_{0}(\check u,\check\xi)=g_{0}(\check u)f_{0}(\check\xi)$ where $g_{0}$ is the first normalized eigenfunction of the harmonic oscillator. $f_{0}$ is a function to be determined. The second equation of the formal system is:
$$(P_{0}-\lambda_{0})\psi_{1}=(\lambda_{1}-P_{1})\psi_{0}\,.$$
The Fredholm condition gives, for all $\check \xi$:
$$\langle(\lambda_{1}-P_{1})\psi_{0},g_{0}\rangle_{\sL^2(\R_{\check u})}=0\,.$$
Let us analyze the different terms which appear in this differential equation. There should be a term in $\check\xi^3$. Its coefficient is:
$$\int_{\R} \check ug_{0}(\check u)^2\dx\check u=0\,.$$
For the same parity reason, there is no term in $\check\xi$. Let us now analyze the term in $D_{\check \xi}$. Its coefficient is:
$$\langle (D_{\check u}\check u+\check u D_{\check u})g_{0},\check u g_{0}\rangle_{\sL^2(\R_{\check u})}=0\,,$$
for a parity reason. In the same way, there is no term in $\check\xi D^2_{\check \xi}$. The coefficient of $\check \xi D_{\check \xi}$ is:
$$2\int_{\R} (\check u D_{\check u}-D_{\check u}\check u)g_{0} g_{0}\dx\check u=0\,.$$
The compatibility equation is in the form:
$$(a D_{\check \xi}^2+b \check\xi^2+c)f_{0}=\lambda_{1}f_{0}\,.$$
It turns out that (exercise):
$$a=b=2\int_{\R} \check u^2 g_{0}^2\dx\check u=1\,.$$
In the same way $c$ can be explicitly found. This leads to a family of choices for $(\lambda_{1}, f_{0})$: We can take $\lambda_{1}=c+(2m+1)$ and $f_{0}=g_{m}$ the corresponding Hermite function.

This construction provides us a family of quasimodes (which are in the Schwartz class) and we can apply the spectral theorem.
\end{proof}
\begin{rem}
One could continue the expansion at any order and one could also consider the other possible values of $\lambda_{0}$ (next eigenvalues of the harmonic oscillator).
\end{rem}

\begin{rem}
The fact that the construction can be continued as much as the appearance of the harmonic oscillator is a clue that our initial scaling is actually the good one. We can also guess that the lowest eigenfunctions are concentrated near zero at the scale $h^{1/2}$ if the quasimodes approximate the true eigenfunctions.
\end{rem}

\chapter{From local models to global estimates}\label{chapter-models}

\begin{flushright}
\begin{minipage}{0.55\textwidth}
Zeno's reasoning, however, is fallacious, when he says that if everything when it occupies an equal space is at rest, and if that which is in locomotion is always occupying such a space at any moment, the flying arrow is therefore motionless. This is false, for time is not composed of indivisible moments any more than any other magnitude is composed of indivisibles. 
\begin{flushright}
\textit{Physics}, Aristotle
\end{flushright}
\vspace*{0.5cm}
\end{minipage}
\end{flushright}

In this chapter we introduce the notions of partition of unity and of localization and provide some examples.

\section{A localization formula}\label{Sec.Agmon0}

We explain in this section how we can perform a reduction of the magnetic Laplacian to local models. 

\subsection{Partition of unity and localization formula}
The presentation is inspired by \cite{CFKS87}. 

\begin{lem}
There exists $C>0$ such that for all $R>0$, there exists a family of smooth cutoff functions $(\chi_{j,R})_{j\in\Z^2}$ on $\R^d$ such that
$$\sum_{j} \chi_{j,R}^2=1,\qquad \sum_{j}\|\nabla\chi_{j,R}\|^2\leq CR^{-2}\,.$$
Moreover, the support of $\chi_{j,R}$ is a ball of center $\x_{j}$ and radius $R$.
\end{lem}

\begin{proof}
We may consider a cutoff function $\chi$ being $1$ on $\mathcal{B}(0,1)$ and $0$ away from $\mathcal{B}(0,1)$. We let
$$S_{R}(\x)=\sum_{j\in\Z^2} \chi^2\left(\frac{\x-Rj}{R}\right)\,.$$
There exists $N>0$ such that for all $R>0$ and all $\x\in\R^d$, $S_{R}(\x)\leq N$. Moreover, we have $S_{R}(\x)\geq 1$ for all $\x\in\R^d$ and thus we may define
$$\chi_{j,R}(\x)=\frac{ \chi\left(\frac{\x-Rj}{R}\right)}{S_{R}(\x)}\,.$$
It remains to notice that
$$\nabla S_{R}(\x)=2R^{-1}\sum_{j} \chi_{j,R}\left(\frac{\x-Rj}{R}\right)\nabla \chi_{j,R}\left(\frac{\x-Rj}{R}\right)$$
so that
$$\|\nabla S_{R}(\x)\|\leq 2D\sum_{j} \mathds{1}_{\mathcal{B}(Rj,R)}(\x)\,.$$
By using support considerations, we get $\sum_{j} \mathds{1}_{\mathcal{B}(Rj,R)}(\x)\leq N$ and $\|\nabla S_{R}(\x)\|\leq \tilde DR^{-1}$ and easy arguments provide the control of the gradients.
\end{proof}

The following formula is sometimes called, by a slight abuse, \enquote{IMS formula} and allows to localize the electro-magnetic Laplacian.
\begin{prop}
We have
\begin{equation}\label{loc-chi}
\forall\psi\in\Dom(\mathfrak{L}_{h,\A, V})\,,\quad\forall\chi\in\mathcal{C}_{0}^\infty(\R^d)\,,\qquad  \mathfrak{Q}_{h,\A,V}(\chi\psi)=\langle\mathfrak{L}_{h,\A,V}\psi,\chi^2\psi\rangle_{\sL^2}+\|(\nabla\chi)\psi\|_{\sL^2}^2\,,
\end{equation}
and
\begin{equation}\label{loc-chi-sum}
\forall\psi\in \Dom(\mathfrak{Q}_{h,\A,V})\,,\qquad\mathfrak{Q}_{h,\A,V}(\psi)=\sum_{j} \mathfrak{Q}_{h,\A,V}(\chi_{j,R}\psi)-h^2\sum_{j} \|\nabla \chi_{j,R}\psi\|^2\,.
\end{equation}
\end{prop}
\begin{proof}
The proof is easy and instructive. By a density argument, it is enough to prove the formulas for $\psi\in\Dom (\mathfrak{L}_{h,\A,V})$. 

We let $P=hD_{k}+A_{k}$ and $\chi=\chi_{j,R}$. We estimate
\begin{align*}
\langle P \psi,P\chi^2 \psi\rangle_{\sL^2}&=\langle \chi P\psi,[P,\chi]\psi\rangle_{\sL^2}+\langle \chi P\psi, P \chi \psi\rangle_{\sL^2}\\
&=\langle \chi P\psi,[P,\chi]\psi\rangle_{\sL^2}+\langle P\chi \psi,P \chi \psi\rangle_{\sL^2}+\langle [\chi,P]\psi,P\chi \psi\rangle_{\sL^2}\\
&=\langle P\chi \psi,P \chi \psi\rangle_{\sL^2}-\|[P, \chi]\psi\|^2+\langle \chi P\psi,[P,\chi]\psi\rangle_{\sL^2}-\langle [P,\chi]\psi,\chi P\psi\rangle_{\sL^2}.
\end{align*}
Taking the real part, we find
\begin{equation*}
\langle P\psi,P\chi^2 \psi\rangle_{\sL^2}= \|P \chi \psi\|^2-\|[P,\chi]\psi\|^2\,.
\end{equation*}
We have $[P,\chi]=-i h\dr_{k}\chi$ and it remains to sum over $k$ and integrate by parts.

To get \eqref{loc-chi-sum}, we write
\[\langle \mathfrak{L}_{h,\A,V}\psi,\psi \rangle_{\sL^2}=\sum_{j}\langle \mathfrak{L}_{h,\A,V}\psi,\chi_{j,R}^2\psi \rangle_{\sL^2}\,,\]
and we apply \eqref{loc-chi}.
\end{proof}
Let us illustrate a possible use of \eqref{loc-chi}.
\begin{exe} 
Consider $\left(-\Delta_{x}+V(x), \mathcal{C}^\infty_{0}(\R^N)\right)$ where $V\in\mathcal{C}^0(\R^N,\R)$.
\begin{enumerate}
\item Give the domain of the adjoint.
\item Prove that this operator is symmetric.
\item We recall that a symmetric operator (with dense domain) is closable and that, by definition, it is essentially self-adjoint if its closure is self-adjoint. We also recall the characterization: $(\mathfrak{L}, \Dom(\mathfrak{L}))$ is self-adjoint iff $\ker(\mathfrak{L}^*\pm i)=\{0\}$. Prove that $\left(-\Delta_{x}+V(x), \mathcal{C}^\infty_{0}(\R^N)\right)$ is essentially self-adjoint. For that purpose, we will notice that the elements of the above kernels are in $\sH^2_{\loc}(\R^N)$. One will use a cutoff function $\chi_{R}(\x)=\chi(R^{-1}\x)$ with $\chi\in\mathcal{C}^\infty_{0}(\R^d)$ being $1$ near $0$. 
\end{enumerate}
\end{exe}

\subsection{Harmonic approximation in dimension one (bis)}\label{Harmonic-2}

In this section, we continue the analysis started in Chapter \ref{chapter-spectral-theory}, Section \ref{Harmonic}. We recall that the operator is expressed as $\mathfrak{L}_{h,V}=h^2D_{x}^2+V$.

\begin{prop}
We have
$$\lambda_{1}(\mathfrak{L}_{h,V})=h\sqrt{\frac{V''(0)}{2}}+\mathcal{O}(h^{6/5})\,.$$
\end{prop}

\begin{proof}
There exist $\delta_{0}>0$, $\eps_{0}>0$, $C>0$ such that:
$$V(x)\geq \delta_{0}\quad\mbox{ for } |x|\geq \eps_{0}$$
and 
$$\left|V(x)-\frac{V''(0)}{2}x^2\right|\leq C|x|^3\quad \mbox{ for } |x|\leq \eps_{0}\,.$$
We introduce a partition of unity on $\R$ with balls of size $r>0$ and centers $x_{j}$ and such that:
\begin{equation}\label{partition-R}
\sum_{j} \chi_{j,r}^2=1,\quad \sum_{j} \chi_{j,r}'^2\leq Cr^{-2}.
\end{equation}
We may assume that $x_{0}=0$ and that there exists $c>0$ such that, for all $j\neq 0$, $|x_{j}|\geq cr$. This localization formula gives 
$$\mathfrak{Q}_{h,V}(\psi)=\sum_{j}\mathfrak{Q}_{h,V}(\chi_{j,r}\psi)-h^2\sum_{j}\|\chi_{j,r}'\psi\|^2\geq \sum_{j}\mathfrak{Q}_{h,V}(\chi_{j,r}\psi)-Ch^2r^{-2}\|\psi\|^2\,.$$
There exists $\tilde c>0$ such that for $j\neq 0$, we have
$$\mathfrak{Q}_{h,V}(\chi_{j,r}\psi)\geq \min(\delta_{0},\tilde cr^2)\|\chi_{j,r}\psi\|^2\,.$$
Moreover, by using a Taylor expansion and then the min-max principle for the harmonic oscillator, we get
\begin{multline*}
\mathfrak{Q}_{h,V}(\chi_{0,r}\psi)\geq \int_{\R} |hD_{x}(\chi_{0,r}\psi)|^2+\frac{V''(0)}{2}x^2|\chi_{0,r}\psi|^2\dx x-Cr^3\|\chi_{0,r}\psi\|^2\\
\geq\left(h\sqrt{\frac{V''(0)}{2}}-Cr^3\right) \|\chi_{0,r}\psi\|^2 .
\end{multline*}
We choose $r=h^{\rho}$ for some $\rho>0$ and we optimize the remainders by taking $2-2\rho=3\rho$ and thus $\rho=\frac{2}{5}$.
\end{proof}

\subsection{Magnetic example}\label{Mag-Ex}
As we are going to see, the localization formula is very convenient to prove lower bounds for the spectrum. We consider an open bounded set $\Omega\subset\R^3$ and the Dirichlet realization of the magnetic Laplacian $\mathfrak{L}_{h,\A}^{\Dir}$. Then we have the lower bound for the lowest eigenvalues.
\begin{prop}\label{models-IMS}
For all $n\in\N^*$, there exist $h_{0}>0$ and $C>0$ such that for $h\in(0,h_{0})$:
$$\lambda_{n}(h)\geq \min_{\Omega}\|\B\|\, h-Ch^{5/4}\,.$$
\end{prop}
\begin{proof}
We introduce a partition of unity with radius $R>0$ denoted by $(\chi_{j,R})_{j}$. Let us consider an eigenpair $(\lambda,\psi)$. We have:
$$\mathfrak{Q}_{h,\A}(\psi)=\sum_{j} \mathfrak{Q}_{h,\A}(\chi_{j,R}\psi)-h^2\sum_{j} \|\nabla \chi_{j,R}\psi\|^2$$
so that:
$$\mathfrak{Q}_{h,\A}(\psi)\geq\sum_{j} \mathfrak{Q}_{h,\A}(\chi_{j,R}\psi)-CR^{-2}h^2\|\psi\|^2$$
and:
$$\lambda\|\psi\|^2\geq\sum_{j} \mathfrak{Q}_{h,\A}(\chi_{j,R}\psi)-CR^{-2}h^2\|\psi\|^2\,.$$
It remains to provide a lower bound for $\mathfrak{Q}_{h,\A}(\chi_{j,R}\psi)$. We choose $R=h^{\rho}$ with $\rho>0$, to be chosen. We approximate the magnetic field in each ball by the constant magnetic field $\B_{j}$:
$$\|\B-\B_{j}\|\leq C\|x-x_{j}\|\,.$$
In a suitable gauge (using Lemma \ref{lem.Poincare}), we have:
$$\|\A-\A_{j}^\lin \|\leq C\|x-x_{j}\|^2\,,$$
where $C>0$ does not depend on $j$ but only on the magnetic field. Then, we have, for all $\eps\in(0,1)$:
$$\mathfrak{Q}_{h,\A}(\chi_{j,R}\psi)\geq (1-\eps)\mathfrak{Q}_{h,\A^\lin_{j}}(\chi_{j,R}\psi)-C^2\eps^{-1}R^{4}\|\chi_{j,R}\psi\|^2\,.$$
From the min-max principle, we deduce:
$$\mathfrak{Q}_{h,\A}(\chi_{j,R}\psi)\geq \left((1-\eps)\|\B_{j}\|h-C^2\eps^{-1}h^{4\rho}\right)\|\chi_{j,R}\psi\|^2\,.$$
Optimizing $\eps$, we take $\eps=h^{2\rho-1/2}$ and it follows:
$$\mathfrak{Q}_{h,\A}(\chi_{j,R}\psi)\geq \left(\|\B_{j}\|h-Ch^{2\rho+1/2}\right)\|\chi_{j,R}\psi\|^2\,.$$
We now choose $\rho$ such that $2\rho+1/2=2-2\rho.$ We are led to take $\rho=\frac{3}{8}$ and the conclusion follows.
\end{proof}

\begin{exe}\label{exo-minmag}
This exercise aims at proving \eqref{eq.helmo}.
\begin{enumerate}
\item Let $\Omega$ be a bounded subset of $\R^2$ with $0\in\Omega$. Assume that the magnetic field has a positive minimum at $0$ and consider the Dirichlet magnetic Laplacian.
By using a test function $\psi$ in the form $\psi(\x)=\chi(\x)e^{-\rho |\x|^2/h}$ with $\rho>0$ to be determined and $\chi$ a smooth cutoff function near $0$, prove that 
\[\lambda_{1}(h)=h\min_{\Omega} B+o(h)\,.\]
\item Prove the same kind of asymptotic expansion in dimension three.
\end{enumerate}
\end{exe}

\subsection{Using a partition of unity to bound a number of eigenvalues}\label{sec.bound-number}

\begin{figure}[ht]
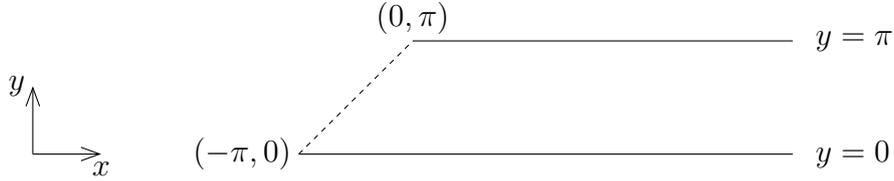

% 1. Definition of characteristic points
    \figinit{1mm}
    \figpt 1:(  0,  0)
    \figpt 2:( 50,  0)
    \figpt 3: (50, 15)
    \figpt 4: (0,  15)
    \figpt 5:(-15,  0)
    \figpt 10:(-50,  0)

% 2. Creation of the postscript file
    \def\MyPSfile{F3.pf}
    \psbeginfig{\MyPSfile}
    \psaxes 10(9)
    \figptsaxes 11 : 10(9)
    \psline[5,2]
    \psline[3,4]
    \psset (dash=4)
    \psline[5,4]
  \psendfig

% 3. Writing text on the figure
    \figvisu{\figbox}{}{%
    \figinsert{\MyPSfile}
    \figwritee  2 :$y=0$(3)
    \figwritee  3 :$y=\pi$(3)
    \figwriten  4 :{$(0,\pi)$}(1)
    \figwritew  5 :{$(-\pi,0)$}(1)
    \figwrites 11 :{$x$}(1)
    \figwritew 12 :{$y$}(1)
    }
\centerline{\box\figbox}
\caption{The reference half-guide $\widetilde\Omega:=\widetilde{\Omega}_{\pi/4}$.}
\label{F:3}
\end{figure}

This section is devoted to the proof of Proposition \ref{prop:5-1} stated in Chapter \ref{intro-wg}. We introduce the open set $\widetilde{\Omega}_{\theta}$ isometric to $\Omega_{\theta}^+$, see Figure \ref{F:3},
$$
   \widetilde{\Omega}_{\theta} = 
   \left\{(\tilde x,\tilde y)\in \left(-\frac{\pi}{\tan\theta},+\infty\right)\times\left(0,\pi\right) : \quad
   \tilde y<\tilde x\tan\theta + \pi\ 
   \mbox{ if }\ \tilde x\in\left(-\frac{\pi}{\tan\theta},0\right)\right\}.
$$

The part $\partial_\Dir\widetilde{\Omega}_{\theta}$ of the boundary carrying the Dirichlet condition is the union of its horizontal parts. 
Let us now perform the change of variable: 
\[
   x=\tilde x\tan\theta, \quad y=\tilde y,
\]
so that the new integration domain $\widetilde\Omega:=\widetilde{\Omega}_{\pi/4}$ is independent of $\theta$. The bilinear form $b$ on $\widetilde{\Omega}_{\theta}$ is transformed into the form $\mathfrak{B}_{\theta}$ on the fixed set $\widetilde\Omega$:
\begin{equation}
\label{eq:3-2}
   \mathfrak{B}_{\theta}(\psi,\psi') = \int_{\widetilde\Omega}
   \tan^2\!\theta \,(\partial_x\psi\,\partial_x\psi') + (\partial_y\psi\,\partial_y\psi') \dx x\dx y,
\end{equation}
with associated form domain 
\begin{equation}
\label{eq:3-3}
   \sV:=\{\psi\in \sH^1(\widetilde\Omega) :  \psi=0 \mbox{ on } \partial_{\Dir}\widetilde\Omega \}
\end{equation}
independent from $\theta$. We let
\[
 \mathfrak{Q}(\psi)=\mathfrak{Q}_{\theta}(\psi) = \mathfrak{B}_{\theta}(\psi,\psi) = \int_{\widetilde\Omega} 
   \tan^2\!\theta\,|\partial_{x}\phi|^2+|\partial_{y}\phi|^2\dx x \dx y.
\]
We recall that the form domain $\sV$ is the subspace of $\psi\in \sH^1(\widetilde\Omega)$ which satisfy the Dirichlet condition  on $\partial_{\Dir}\widetilde\Omega$. We want to prove that
\[
   \mathsf{N}(\mathfrak{Q},1)\quad\mbox{is finite}.
\]
We consider a partition of unity $(\chi_{0},\chi_{1})$ such that 
\[
   \chi_{0}(x)^2+\chi_{1}(x)^2=1
\]
with $\chi_{0}(x)=1$ for $x<1$ and $\chi_{0}(x)=0$ for $x>2$. 
For $R>0$ and $\ell\in\{0,1\}$, we introduce:
$$
   \chi_{\ell,R}(x)=\chi_{\ell}(R^{-1}x).
$$
Thanks to the localization formula, we can split the quadratic form as:
\begin{equation}
\label{eq:IMS}
   \mathfrak{Q}(\psi)=\mathfrak{Q}(\chi_{0,R}\psi)+\mathfrak{Q}(\chi_{1,R}\psi)
   -\|\chi'_{0,R}\psi\|^2_{\widetilde\Omega}-\|\chi'_{1,R}\psi\|^2_{\widetilde\Omega}\,.
\end{equation}
We can write
\[
   |\chi'_{0,R}(x)|^2+|\chi'_{1,R}(x)|^2 = R^{-2}W_R(x)\quad
   \mbox{with}\quad W_R(x) = |\chi'_{0}(R^{-1}x)|^2+|\chi'_{1}(R^{-1}x)|^2 \,.
\]
Then
\begin{align}
\label{eq:5-1}
   \|\chi'_{0,R}\psi\|^2_{\widetilde\Omega}+\|\chi'_{1,R}\psi\|^2_{\widetilde\Omega} &=
   \int_{\widetilde\Omega} R^{-2} W_{R}(x)|\psi|^2\dx x \dx y 
   \nonumber\\ &= 
   \int_{\widetilde\Omega} R^{-2} W_{R}(x) 
   \big(|\chi_{0,R}\psi|^2 + |\chi_{1,R}\psi|^2 \big)\dx x \dx y.
\end{align}
Let us introduce the subsets of $\widetilde\Omega$:
$$
   \mathcal{U}_{0,R} = \{(x,y)\in\widetilde\Omega : x<2R\}
   \quad\mbox{ and }\quad
   \mathcal{U}_{1,R}=\{(x,y)\in\widetilde\Omega : x>R\}
$$
and the associated form domains
\[
\begin{gathered}
   \sV_0 = \left\{\phi\in \sH^1(\mathcal{U}_{0,R}) :  \quad
   \phi=0 \ \mbox{ on }\ \partial_{\Dir}\widetilde\Omega \cap \partial \mathcal{U}_{0,R}
   \ \mbox{ and on }\ \{2R\}\times(0,\pi)\right\} \\
   \sV_1 = \sH^1_0(\mathcal{U}_{1,R}).
\end{gathered}
\]
We define the two quadratic forms $\mathfrak{Q}_{0,R}$ and $\mathfrak{Q}_{1,R}$ by
\begin{equation}
\label{eq:Qell}
   \mathfrak{Q}_{\ell,R}(\phi) = \int_{\mathcal{O}_{\ell,R}} 
   \tan^2\theta|\partial_{x}\phi|^2+|\partial_{y}\phi|^2-R^{-2}W_{R}(x)|\phi|^2\dx x \dx y
   \quad\mbox{for}\quad \psi\in \sV_\ell,\ \ \ell=0,1.
\end{equation}
As a consequence of \eqref{eq:IMS} and \eqref{eq:5-1} we find
\begin{equation}
\label{Q=Q0+Q1}
   \mathfrak{Q}(\psi) = \mathfrak{Q}_{0,R}(\chi_{0,R}\psi) + \mathfrak{Q}_{1,R}(\chi_{1,R}\psi)\quad \forall\psi\in \sV.
\end{equation}
Let us prove

\begin{lem}
\label{lem:5-1}
We have: 
\[
   \mathsf{N}(\mathfrak{Q},1) \leq \mathsf{N}(\mathfrak{Q}_{0,R},1) + \mathsf{N}(\mathfrak{Q}_{1,R},1).
\]
\end{lem}

\begin{proof}
We recall the formula for the $j$-th Rayleigh quotient of $\mathfrak{Q}$:
\[
   \lambda_j =
   \inff_{\substack{E\subset \sV\\ \dim E=j}} \ \sup_{\substack{
    \psi\in E}}   \frac{\mathfrak{Q}(\psi)}{\|\psi\|^2_{\widetilde\Omega}} \,.
\]
The idea is now to give a lower bound for $\lambda_j$. Let us introduce:
\[
\mathcal{J} : \left\{\begin{array}{ccc}
\sV &\to& \sV_0 \times \sV_1 \\
\psi&\mapsto&(\chi_{0,R}\psi\ ,\,\chi_{1,R}\psi) \,.
\end{array}\right.
\]
As $(\chi_{0,R},\chi_{1,R})$ is a partition of the unity, $\mathcal{J}$ is injective. In particular, we notice that $\mathcal{J} : V \to\mathcal{J}(V)$ is bijective so that we have:
\begin{align*}
   \lambda_j &=
   \inf_{\substack{F\subset \mathcal{J}(\sV)\\ \dim F=j}} \ \sup_{\substack{
    \psi\in \mathcal{J}^{-1}(F)}}   \frac{\mathfrak{Q}(\psi)}{\|\psi\|^2_{\widetilde\Omega}}\\
    &= \inf_{\substack{F\subset \mathcal{J}(\sV)\\ \dim F=j}} \ \sup_{\substack{
   \psi\in \mathcal{J}^{-1}(F)}}   \frac{\mathfrak{Q}_{0,R}(\chi_{0,R}\psi)+\mathfrak{Q}_{1,R}(\chi_{1,R}\psi)}
   {\|\chi_{0,R}\psi\|^2_{\widetilde\Omega} + \|\chi_{1,R}\psi\|^2_{\widetilde\Omega}}\\
    &= \inf_{\substack{F\subset \mathcal{J}(\sV)\\ \dim F=j}} \ \sup_{\substack{
    (\psi_{0},\psi_{1})\in F}}   \frac{\mathfrak{Q}_{0,R}(\psi_{0})+\mathfrak{Q}_{1,R}(\psi_{1})}
    {\|\psi_{0}\|^2_{\mathcal{U}_{0,R}}+\|\psi_{1}\|^2_{\mathcal{U}_{1,R}}}.
\end{align*}
As  $\mathcal{J}(V) \subset \sV_0 \times \sV_1$, we deduce
\begin{align}\label{eq:5-2}
   \lambda_j &\geq \inf_{\substack{F\subset \sV_0\times \sV_1\\ \dim F=j}} \ \sup_{\substack{
    (\psi_{0},\psi_{1})\in F}}   \frac{\mathfrak{Q}_{0,R}(\psi_{0})+\mathfrak{Q}_{1,R}(\psi_{1})}
    {\|\psi_{0}\|^2_{\mathcal{U}_{0,R}} + \|\psi_{1}\|^2_{\mathcal{U}_{1,R}}}
    =: \nu_{j},
\end{align}
Let $A_{\ell,R}$ be the self-adjoint operator with domain $\Dom(A_{\ell,R})$ associated with the coercive bilinear form corresponding to the quadratic form $\mathfrak{Q}_{\ell,R}$ on $V_\ell$.
We see that $\nu_{j}$ in \eqref{eq:5-2} is the $j$-th Rayleigh quotient of the diagonal self-adjoint operator $A_R$
\[
   \begin{pmatrix}
   A_{0,R} & 0\\
   0 & A_{1,R}
   \end{pmatrix} \quad\mbox{with domain}\quad \Dom(A_{0,R}) \times  \Dom(A_{1,R})\,.
\]
The Rayleigh quotients of $A_{\ell,R}$ are associated with the quadratic form $\mathfrak{Q}_{\ell,R}$ for $\ell=0,1$.
Thus $\nu_j$ is the $j$-th element of the ordered set
\[
   \{\lambda_k(\mathfrak{Q}_{0,R}), \ k\ge1\} \cup \{\lambda_k(\mathfrak{Q}_{1,R}), \ k\ge1\}.
\]
Lemma \ref{lem:5-1} follows.
\end{proof}

Since the operator $A_{0,R}$ has a compact resolvent, we get the following lemma.

\begin{lem}
For all $R>0$, $\mathsf{N}(\mathfrak{Q}_{0,R},1)$ is finite.
\end{lem}

To achieve the proof of Proposition \ref{prop:5-1}, it remains to establish the following lemma.

\begin{lem}
There exists $R_{0}>0$  such that, for $R\geq R_{0}$, $\mathsf{N}(\mathfrak{Q}_{1,R},1)$ is finite.
\end{lem}

\begin{proof}
For all $\phi\in V_1$, we write: 
$$\phi=\Pi_{0}\phi+\Pi_{1}\phi\,,$$
where 
\begin{equation}
\label{eq:Pi0}
   \Pi_{0}\phi(x,y) = \Phi(x) \sin y\quad\mbox{with}\quad
   \Phi(x) = \int_0^\pi \phi(x,y)\sin{y} \dx y
\end{equation}
is the projection on the first eigenfunction of $-\partial^2_y$ on $\sH^1_0(0,\pi)$, and $\Pi_{1}=\Id-\Pi_{0}$.
We have, for all $\eps>0$:
\begin{align}\mbox{}\hskip-1ex
   \mathfrak{Q}_{1,R}(\phi)
   &=\mathfrak{Q}_{1,R}(\Pi_{0}\phi)+\mathfrak{Q}_{1,R}(\Pi_{1}\phi)
    -2\int_{\mathcal{U}_{1,R}} R^{-2}W_{R}(x)\Pi_{0}\phi\, \Pi_{1}\phi\dx x \dx y
    \nonumber\\
   &\geq \mathfrak{Q}_{1,R}(\Pi_{0}\phi)+\mathfrak{Q}_{1,R}(\Pi_{1}\phi)
    -\eps^{-1}\int_{\mathcal{U}_{1,R}} R^{-2}W_{R}(x)|\Pi_{0}\phi|^2\dx x \dx y\nonumber\\
    &-\eps\int_{\mathcal{U}_{1,R}} R^{-2}W_{R}(x)|\Pi_{1}\phi|^2\dx x \dx y.
    \label{eq:Q1R}
\end{align}
Since the second eigenvalue of $-\partial^2_y$ on $\sH^1_0(0,\pi)$ is $4$, we have:
\[
   \int_{\mathcal{U}_{1,R}} 
   |\partial_{y}\Pi_{1}\phi|^2\dx x \dx y \ge 4 \|\Pi_{1}\phi\|^2_{\mathcal{U}_{1,R}}\,.
\]
Denoting by $M$ the maximum of $W_R$ (which is independent of $R$), and using \eqref{eq:Qell} we deduce
$$
   \mathfrak{Q}_{1,R}(\Pi_{1}\phi)\geq (4-MR^{-2})\|\Pi_{1}\phi\|^2_{\mathcal{U}_{1,R}}\,.
$$
Combining this with \eqref{eq:Q1R} where we take $\varepsilon=1$, and with the definition \eqref{eq:Pi0} of $\Pi_0$, we find
$$
   \mathfrak{Q}_{1,R}(\phi) \geq q_{R}(\Phi) +(4-2MR^{-2})\|\Pi_{1}\phi\|^2_{\mathcal{U}_{1,R}},
$$
where 
\begin{align*}
   q_{R}(\Phi) &= \int_{R}^\infty \tan^2\theta|\partial_{x}\Phi|^2+|\Phi|^2
  -R^{-2}W_{R}(x)|\Phi|^2\dx x \\
  &\ge \int_{R}^\infty \tan^2\theta|\partial_{x}\Phi|^2+|\Phi|^2
  -R^{-2}M\mathds{1}_{[R,2R]}|\Phi|^2\dx x.
\end{align*}
We choose $R=\sqrt{M}$ so that $(4-2MR^{-2})=2$, and then
\begin{equation}
\label{eq:5-3}
   \mathfrak{Q}_{1,R}(\phi) \geq \tilde q_{R}(\Phi) + 2\|\Pi_{1}\phi\|^2_{\mathcal{U}_{1,R}},
\end{equation}
where now
\begin{equation}
\label{eq:5-4}
   \tilde q_{R}(\Phi) = \int_{R}^\infty \tan^2\theta|\partial_{x}\Phi|^2+
   (1-\mathds{1}_{[R,2R]})|\Phi|^2\dx x.
\end{equation}
Let $\tilde a_R$ denote the one dimensional operator associated with the quadratic form $\tilde q_R$.
From \eqref{eq:5-3}-\eqref{eq:5-4}, we deduce that the $j$-th Rayleigh quotient of $A_{1,R}$ admits as lower bound the $j$-th Rayleigh quotient of the diagonal operator:
\[
  \begin{pmatrix}
  \tilde a_R & 0\\
  0 & 2\,\Id
  \end{pmatrix}
\]
so that we find
\[
   \mathsf{N}(\mathfrak{Q}_{1,R},1)\leq \mathsf{N}(\tilde q_{R},1).
\]
Finally, the eigenvalues $<1$ of $\tilde a_R$ can be computed explicitly and this is an elementary exercise to deduce that $\mathsf{N}(\tilde q_{R},1)$ is finite.
\end{proof}

This concludes the proof of Proposition \ref{prop:5-1}.

\section{Agmon-Persson estimates}\label{Agmon}

\subsection{Agmon formula}
This section is devoted to the Agmon formula in the semiclassical framework. We refer to the classical references \cite{Agmon82, Agmon85, Hel88, HelSj84, HelSj85}.

\begin{prop}\label{models-Agmon}
Let $ \Omega$ be an open domain in $\R^m$ with Lipschitzian boundary. Let $V\in\mathcal{C}^0(\overline{ \Omega},\R)$, $\A\in\mathcal{C}^{0}(\overline{ \Omega}, \R^m)$ and $\Phi$ a real valued Lipschitzian and bounded function on $\overline{\Omega}$. Then, for $u\in\Dom(\mathfrak{L}_{h,\A,V})$ (with Dirichlet or magnetic Neumann  condition), we have:
\begin{align*}
&\int_{ \Omega} |(-ih\nabla+\A)e^{\Phi} u|^2\,\dx x+\int_{ \Omega} \left(V-h^2|\nabla\Phi|^2 e^{2\Phi}\right)|u|^2\,\dx x=\Re\langle \mathfrak{L}_{h,\A,V}u,e^{2\Phi}u\rangle_{\sL^2(\Omega)}\,.
\end{align*}
\end{prop}
\begin{proof}
We give the proof when $\Phi$ is smooth. Let us use the Green-Riemann formula:
$$\sum_{k=1}^m\langle (-ih\dr_{k}+A_{k})^2 u, e^{2\Phi} u\rangle_{\sL^2}=\sum_{k=1}^m\langle (-ih\dr_{k}+A_{k}) u,(-ih\dr_{k}+A_{k}) e^{2\Phi} u\rangle_{\sL^2}\,,$$
where the boundary term has disappeared thanks to the boundary condition. In order to lighten the notation, we let $P=-ih\dr_{k}+A_{k}$.
\begin{align*}
\langle P u,Pe^{2\Phi} u\rangle_{\sL^2}&=\langle e^{\Phi}Pu,[P,e^{\Phi}]u\rangle_{\sL^2}+\langle e^{\Phi} Pu, P e^{\Phi} u\rangle_{\sL^2}\\
&=\langle e^{\Phi}Pu,[P,e^{\Phi}]u\rangle_{\sL^2}+\langle Pe^{\Phi}u,P e^{\Phi}u\rangle_{\sL^2}+\langle [e^{\Phi},P]u,Pe^{\Phi}u\rangle_{\sL^2}\\
&=\langle Pe^{\Phi}u,P e^{\Phi}u\rangle_{\sL^2}-\|[P, e^{\Phi}]u\|^2+\langle e^{\Phi}Pu,[P,e^{\Phi}]u\rangle_{\sL^2}-\langle [P,e^{\Phi}]u,e^{\Phi}Pu\rangle_{\sL^2}\,.
\end{align*}
We deduce:
$$\Re\left(\langle P u,Pe^{2\Phi} u\rangle_{\sL^2}\right)=\langle Pe^{\Phi}u,P e^{\Phi}u\rangle_{\sL^2}-\|[P, e^{\Phi}u]\|^2\,.$$
This is then enough to conclude.

\end{proof}
In fact we can prove a more general localization formula (which generalizes Propositions \ref{models-IMS} and \ref{models-Agmon}).
\begin{prop}[\enquote{Localization} of $P^2$ with respect to $A$]\label{P2A}
Let $(\mathsf{H},\langle\cdot,\cdot\rangle)$ be a Hilbert space and two unbounded operators $P$ and $\gA$ defined on a domain $\mathsf{D}\subset \mathsf{H}$. We assume that $P$ is symmetric and that $P(\mathsf{D})\subset \mathsf{D}$, $\gA(\mathsf{D})\subset \mathsf{D}$ and $\gA^*(\mathsf{D})\subset \mathsf{D}$.
Then, for $\psi\in \mathsf{D}$, we have:
\begin{multline}\label{IMS-formula}
\Re\langle P^2\psi, \gA \gA ^*\psi \rangle=\|P(\gA ^*\psi)\|^2-\|[\gA ^*,P]\psi\|^2+\Re\langle P\psi, [[P, \gA], \gA ^*]\psi\rangle\\
+\Re\left(\langle P\psi, \gA ^*[P, \gA]\psi \rangle-\overline{\langle P\psi, \gA[P, \gA ^*]\psi \rangle}\right)\,.
\end{multline}
\end{prop}

\subsection{Agmon-Persson estimates}
It turns out that the estimates of Agmon are closely related to the estimates of Persson. These estimates state that, if an eigenfunction of the electro-magnetic Laplacian is associated with a discrete eigenvalue less than the bottom of the essential spectrum, then it has an exponential decay. The following proposition is very convenient in concrete situations.
\begin{prop}\label{prop.APE}
Let $V\in\mathcal{C}^0(\overline{ \Omega},\R)$ bounded from below and $\A\in\mathcal{C}^{1}(\overline{ \Omega}, \R^m)$. Let us also assume that there exists $R_{0}>0, \mu^*\in\R, h_{0}\in(0,1)$ such that, for all $h\in(0,h_{0})$ and for all $\psi\in\Dom(\mathfrak{L}_{h,\A,V})$ with support in $\complement D(0,R_{0})$, we have
\[\mathfrak{Q}_{h,\A,V}(\psi)\geq \mu^*\|\psi\|^2\,.\]
Then, for $h\in(0,h_{0})$, we have $\inf\spe(\mathcal{L}_{h,\A,V})\geq \mu^*$. Moreover, if $\psi$ is an eigenfunction associated with $\mu<\mu^*$, then for all $\eps\in(0,\sqrt{\mu^*-\mu})$, we have $e^{\eps |\x|}\psi\in\sL^2(\Omega)$ and even $e^{\eps |\x|}\psi\in\Dom(\mathfrak{Q}_{h,\A,V})$.
\end{prop}

\begin{proof}
The first part of the statement is a consequence of Proposition \ref{Persson-improved}.

Let $\eps\in(0,\sqrt{\mu^*-\mu})$. We introduce $\chi_{m}(y)=\chi(m^{-1}y)$, with $\chi$ a smooth cutoff function being $1$ in a (suitable) neighborhood of $0$. With Proposition \ref{models-Agmon}, we deduce that
\[\mathfrak{Q}_{h,\A,V}(e^{\eps\chi_{m}(|\x|)|\x|}\psi)\leq \mu\|e^{\eps\chi_{m}(|\x|)|\x|}\psi\|^2+\|(e^{\eps\chi_{m}(|\x|)|\x|})'\psi\|^2\,.\]
But we have
\[\|(e^{\eps\chi_{m}(|\x|)|\x|})'\psi\|^2=\eps^2\|\left( m^{-1}\chi'(m^{-1}|\x|)|\x|\nabla |\x|+\chi_{m}(|\x|)\nabla |\x|\right)\tilde\psi\|^2\,,\]
where $\tilde\psi=e^{\eps\chi_{m}(|\x|)|\x|}\psi$. We deduce, for all $\gamma\in(0,1)$, 
\[\|(e^{\eps\chi_{m}(|\x|)|\x|})'\psi\|^2\leq\eps^2\left((1+\gamma^{-1})\|\chi'\|_{\infty}^2+(1+\gamma)\right)\|e^{\eps\chi_{m}(|\x|)|\x|}\psi\|^2\,.\]
We choose $\gamma=\|\chi'\|_{\infty}$ so that
\[\|(e^{\eps\chi_{m}(|\x|)|\x|})'\psi\|^2\leq\eps^2\left(1+\|\chi'\|_{\infty}\right)^2\|e^{\eps\chi_{m}(|\x|)|\x|}\psi\|^2\,.\]
We get
\begin{equation}\label{eq.Qleq}
\mathfrak{Q}_{h,\A,V}(e^{\eps\chi_{m}(|\x|)|\x|}\psi)\leq \left(\mu+\eps^2\left(1+\|\chi'\|_{\infty}\right)^2\right)\|e^{\eps\chi_{m}(|\x|)|\x|}\psi\|^2\,.
\end{equation}
We consider a partition of the unity $\chi_{R,1}^2+\chi_{R,2}^2=1$ with $\sum_{j=1}^2 |\nabla\chi_{R,j}|\leq CR^{-2}$ and $\supp(\chi_{R,2})\subset\complement \mathcal{B}(0,R)$ (with $R\geq R_{0}$). With the localization formula, we find
\begin{multline*}
\mathfrak{Q}_{h,\A,V}(e^{\eps\chi_{m}(|\x|)|\x|}\psi)\geq \mathfrak{Q}_{h,\A,V}(\chi_{R,1}e^{\eps\chi_{m}(|\x|)|\x|}\psi)+\mathfrak{Q}_{h,\A,V}(\chi_{R,2}e^{\eps\chi_{m}(|\x|)|\x|}\psi)\\
-CR^{-2}\|e^{\eps\chi_{m}(|\x|)|\x|}\psi\|^2\,,
\end{multline*}
so that
\begin{multline*}
\mathfrak{Q}_{h,\A,V}(e^{\eps\chi_{m}(|\x|)|\x|}\psi)\geq \mathfrak{Q}_{h,\A,V}(\chi_{R,1}e^{\eps\chi_{m}(|\x|)|\x|}\psi)+\mu^*\|\chi_{R,2}e^{\eps\chi_{m}(|\x|)|\x|}\psi\|^2\\
-CR^{-2}\|e^{\eps\chi_{m}(|\x|)|\x|}\psi\|^2\,.
\end{multline*}
We deduce the existence of $C(R, \eps)>0$ such that, for all $m\geq 1$,
\[\left(\mu^*-\mu-\eps^2\left(1+\|\chi'\|_{\infty}\right)^2-CR^{-2}\right)\|\chi_{R,2}e^{\eps\chi_{m}(|\x|)|\x|}\psi\|^2\leq C(R, \eps)\|\psi\|^2\,.\]
We choose $\chi$ such that $\mu^*-\mu-\eps^2\left(1+\|\chi'\|_{\infty}\right)^2\geq \frac{\mu^*-\mu}{2}>0$. Then, for $R$ large enough, we find $c(R,\eps)>0$ such that, for all $m\geq 1$,
\[c(R,\eps)\|\chi_{R,2}e^{\eps\chi_{m}(|\x|)|\x|}\psi\|^2\leq C(R, \eps)\|\psi\|^2\,.\]
We get the existence of $\tilde C(R,\eps)>0$ such that, for all $m\geq 1$,
\begin{equation}\label{eq.APE-R}
\|e^{\eps\chi_{m}(|\x|)|\x|}\psi\|^2\leq \tilde C(R, \eps)\|\psi\|^2\,.
\end{equation}
Then, we take the limit $m\to+\infty$ and use the Fatou lemma. To get the control of $e^{\eps|\x|}\psi$ in the norm of the quadratic form we use \eqref{eq.Qleq}.
\end{proof}

\section{Applications}
\subsection{Harmonic approximation in dimension one (ter)}\label{Harmonic-3}
We continue the analysis of Section \ref{Harmonic-2}. With Proposition \ref{prop-harmonic-1}, we have $\lambda_{n}\left(\mathfrak{L}_{h,V}\right)=\mathcal{O}(h)$.
\begin{prop}\label{Agmon-1D}
For all $\eps\in(0,1)$, there exists $C>0$ and $h_{0}>0$ such that, for all $h\in(0,h_{0})$,
\begin{equation}\label{Agmon000}
\|e^{\eps\,\Phi_{0}/h}\psi\|^2\leq C\|\psi\|^2,\qquad \mathfrak{Q}_{h,V}(e^{\eps\,\Phi_{0}/h}\psi)\leq Ch\|\psi\|^2,
\end{equation}
where $\Phi_{0}=\left|\int_{0}^x \sqrt{V(y)}\dx y\right|$.
\end{prop}
\begin{proof}
The proof follows from the same strategy as the one of Proposition \ref{prop.APE}. For $\eps\in(0,1)$, we introduce $\Phi=\eps\Phi_{0}$  and $\chi_{m}(y)=\chi(m^{-1}y)$, with $\chi$ a smooth cutoff function being $1$ in a neighborhood of $0$. Let us consider an eigenvalue $\lambda$ ($=\mathcal{O}(h)$) and an associated eigenfunction $\psi$. We have
$$\mathfrak{Q}_{h,V}(e^{\eps\chi_{m}(\Phi_{0})\Phi_{0}/h}\psi)\leq \lambda\|e^{\eps\chi_{m}(\Phi_{0})\Phi_{0}/h}\psi\|^2+h^2\|(e^{\eps\chi_{m}(\Phi_{0})\Phi_{0}/h})'\psi\|^2\,.$$
We have, for all $\gamma\in(0,1)$,
\begin{align*}
\|h(e^{\eps\chi_{m}(\Phi_{0})\Phi_{0}/h})'\psi\|^2&=\|\chi_{m}'(\Phi_{0})\Phi_{0}'\Phi_{0}\tilde\psi+\chi_{m}(\Phi_{0})\Phi_{0}'\tilde\psi\|^2\\
&\leq\eps^{2} (1+\gamma^{-1})\|\chi_{m}'(\Phi_{0})\Phi_{0}\sqrt{V}\tilde\psi\|^2+\eps^2(1+\gamma)\|\chi_{m}(\Phi_{0})\sqrt{V}\tilde\psi\|^2\\
&\leq  \eps^{2}\left((1+\gamma^{-1})\|\chi'\|^2_{\infty}+(1+\gamma)\right)\|\sqrt{V}\tilde\psi\|^2,
\end{align*}
with $\tilde\psi=e^{\eps\chi_{m}(\Phi_{0})\Phi_{0}/h}\psi$.
We choose $\gamma=\|\chi'\|_{\infty}$ and we get
\[\|h(e^{\eps\chi_{m}(\Phi_{0})\Phi_{0}/h})'\psi\|^2\leq  \eps^{2}\left(1+\|\chi'\|_{\infty}\right)^2\|\sqrt{V}\tilde\psi\|^2\,.\]
Given $\eps\in(0,1)$, we may find $\chi$ such that $\|\chi'\|_{\infty}$ small enough so that there exists $\tilde\eta>0$ such that $1- \eps^{2}\left(1+\|\chi'\|_{\infty}\right)^2\geq \tilde\eta$. We get
$$\mathfrak{Q}_{h,\tilde\eta V}(e^{\eps\chi_{m}(\Phi_{0})\Phi_{0}/h}\psi)\leq Ch\|e^{\eps\chi_{m}(\Phi_{0})\Phi_{0}/h}\psi\|^2$$
and thus
$$\int_{\R} \tilde\eta Ve^{2\eps\chi_{m}(\Phi_{0})\Phi_{0}/h}|\psi|^2\dx x\leq Ch\|e^{\eps\chi_{m}(\Phi_{0})\Phi_{0}/h}\psi\|^2\,.$$
Given $C_{0}>0$, we write
$$\int_{\R}  Ve^{2\eps\chi_{m}(\Phi_{0})\Phi_{0}/h}|\psi|^2\dx x=\int_{|x|\geq C_{0}h^{1/2}}\!\!\!\!\!  V e^{2\eps\chi_{m}(\Phi_{0})\Phi_{0}/h}|\psi|^2\dx x+\int_{|x|\leq C_{0}h^{1/2}}\!\!\!\! \!\!\! Ve^{2\eps\chi_{m}(\Phi_{0})\Phi_{0}/h}|\psi|^2\dx x\,.$$
Using the quadratic approximation of $V$ at $0$ and the fact that $V$ admits a unique and non degenerate minimum, we deduce that there exists $c>0$ such that for all $C_{0}>0$, there exist $C,h_{0}>0$ such that, for $h\in(0,h_{0})$ and all $m\geq 1$,
$$\int_{|x|\leq C_{0}h^{1/2}}  V e^{2\eps\chi_{m}(\Phi_{0})\Phi_{0}/h}|\psi|^2\dx x\leq Ch\|\psi\|^2$$
and
$$\int_{|x|\geq C_{0}h^{1/2}}  V e^{2\eps\chi_{m}(\Phi_{0})\Phi_{0}/h}|\psi|^2\dx x\geq cC_{0}^2h \int_{|x|\geq C_{0}h^{1/2}} e^{2\eps\chi_{m}(\Phi_{0})\Phi_{0}/h}|\psi|^2\dx x\,.$$
Taking $C_{0}$ large enough, we deduce that
$$\int_{|x|\geq C_{0}h^{1/2}} e^{2\eps\chi_{m}(\Phi_{0})\Phi_{0}/h}|\psi|^2\dx x\leq C\|\psi\|^2\,.$$
We deduce that there exist $C>0, h_{0}>0$ such that, for all $m\geq1$ and $h\in(0,h_{0})$,
$$\|e^{\eps\chi_{m}(\Phi_{0})\Phi_{0}/h}\psi\|^2\leq C\|\psi\|^2\,.$$
Then we consider the limit $m\to+\infty$ and use the Fatou lemma. We deduce the first estimate in \eqref{Agmon000}. We easily deduce that
\begin{equation}\label{Agmon000'}
\mathfrak{Q}_{h,V}(e^{\eps\Phi_{0}/h}\psi)\leq Ch\|\psi\|^2.
\end{equation}
\end{proof}

\begin{exe}
Prove that for all $\zeta\in\R$, we have
$$\int_{\R_{+}} e^{2t} |u_{\zeta}^{[0]}(t)|^2 \dx t<+\infty,\qquad\mbox{ and }\qquad \int_{\R_{+}} e^{2t} |(u_{\zeta}^{[0]})'(t)|^2 \dx t<+\infty\,.$$
\end{exe}

\begin{prop}\label{el-expansion}
For all $n\geq 1$, there exists a sequence $(\mu_{n,j})$ such that, for all $J\geq 1$, there exists $h_{0},C>0$ such that, for all $h\in(0,h_{0})$,
$$\lambda_{n}\left(\mathfrak{L}_{h,V}\right)\sim \sum_{j\geq 0} \mu_{n,j}h^{\frac{j}{2}}\,.$$
\end{prop}
\begin{proof}
For $N\geq 1$, we may define a family of eigenpairs $((\lambda_{n}(\mathfrak{L}_{h,V}),\psi_{n,h}))_{n=1,\ldots, N}$ such that $(\psi_{n,h})_{n=1,\ldots,N}$ is an orthonormal family. We let $$\mathcal{E}_{N}(h)=\underset{n=1,\ldots,N}{\spann}\psi_{n,h}\,.$$
We leave as an exercise to check that the elements of $\mathcal{E}_{N}(h)$ still satisfy the estimates of Agmon \eqref{Agmon000}.
By using these estimates of Agmon, we easily get that, for all $\psi\in\mathcal{E}_{N}(h)$,
$$\mathfrak{Q}_{h,V}(\psi)\geq\int_{\R} h^2|D_{x}\psi|^2+\frac{V''(0)}{2}x^2|\psi|^2\dx x-Ch^{3/2}\|\psi\|^2\,.$$
Then, the min-max principle implies
$$\lambda_{N}(\mathcal{L}_{h,V})\geq (2N-1)h\sqrt{\frac{V''(0)}{2}}-Ch^{3/2}\,.$$
Then, the expansion at any order of the $n$-th eigenvalue follows from Proposition \ref{prop-harmonic-1}.
\end{proof}
It turns out that the estimates of Agmon are related to the so-called WKB constructions. We provide an example of such a construction in the following proposition (see \cite[Chapter 3]{DiSj99} for further details and generalizations).
\begin{prop}\label{el-WKB}
For all $n\geq 1$, there exist a neighborhood of $0$ denoted by $\mathcal{V}$ and a smooth function $a_{n,0}$ defined on $\mathcal{V}$ and $h_{0},C>0$ such that, for all $h\in(0,h_{0})$,
$$\left\|\left(\mathfrak{L}_{h,V}-(2n-1)h\sqrt{\frac{V''(0)}{2}}\right)\chi a_{n,0}e^{-\Phi_{0}/h}\right\|\leq Ch^2\,,$$
with $\Phi_{0}=\left|\int_{0}^x \sqrt{V(y)}\dx y\right|$ and $\chi$ a smooth cutoff function being $1$ near $0$.
\end{prop}
\begin{proof}
Let us compute 
\begin{equation}\label{Lae}
\mathfrak{L}_{h,V}\left(a_{0}e^{-\Phi_{0}/h}\right)=\left(h^2D_{x}^2a_{0}-2h D_{x}a_{0}D_{x}\Phi_{0}+ha_{0}\Phi''_{0}-(\Phi'_{0})^2a_{0}+Va_{0}\right)e^{-\Phi_{0}/h}
\end{equation}
and we solve
$$\mathfrak{L}_{h,V}\left(a_{0}e^{-\Phi_{0}/h}\right)=\lambda h a_{0}e^{-\Phi_{0}/h}\,.$$
We have $(\Phi'_{0})^2=V$. Then, we consider
$$\Phi'_{0} \partial_{x}a_{0}+\partial_{x}(\Phi'_{0} a_{0})=\lambda a_{0}\,.$$
We have to solve this equation in a neighborhood of $0$ (so that $\Phi'_{0}$ only vanishes at $0$). It can be explicitly solved on $x>0$ and $x<0$. Since we look for a smooth function $a_{0}$ we can check that this implies $\lambda=(2n-1)\sqrt{\frac{V''(0)}{2}}$, for $n\geq 1$. Moreover $a_{n,0}$ behaves like $s^n$ near $0$.
Finally, we write
\begin{multline*}
\left(\mathfrak{L}_{h,V}-(2n-1)h\sqrt{\frac{V''(0)}{2}}\right)\chi a_{n,0}e^{-\Phi_{0}/h}\\
=\chi\left(\mathcal{L}_{h,V}-(2n-1)h\sqrt{\frac{V''(0)}{2}}\right)a_{n,0}e^{-\Phi_{0}/h}+[\mathcal{L}_{h,V},\chi] a_{n,0}e^{-\Phi_{0}/h}.
\end{multline*}
With support considerations, the second term in the r.h.s. is $\mathcal{O}(h^\infty)$. By using \eqref{Lae}, the first term in the l.h.s. is $\mathcal{O}(h^2)$.
\end{proof}
Proposition \ref{el-WKB} can be used to prove that there are no odd powers of $h^{\frac{1}{2}}$ in the expansion given in Proposition \ref{el-expansion}.

\subsection{A model with parameter}

The estimates of Agmon may be useful to analyze the dependence of eigenvalues with respect to some parameters, especially when the dependence of the quadratic form on the parameters is not clear. In this section, we deal with a simple example of such a situation. For $a\in[0,1]$, we consider the Friedrichs extension $\mathfrak{L}_{a}$ of the differential operator, acting on $\mathcal{C}^\infty_{0}(\R^2)$,
\[\left(D_{x}+ay+y(x^2+y^2)\right)^2+D^2_{y}\,.\]
We recall that the domain of the associated quadratic form $\mathfrak{Q}_{a}$ is
\[\Dom(\mathfrak{Q}_{a})=\left\{\psi\in\sL^2(\R^2) : \qquad D_{y}\psi\in\sL^2(\R^2),\qquad (D_{x}+ay+y(x^2+y^2))^2\psi\in\sL^2(\R^2)\right\}\,.\]
The magnetic field is $B_{a}(\x)=a+|\x|^2$ and tends to $+\infty$ when $|\x|\to+\infty$. In particular, with Proposition \ref{prop.compact2D}, $\mathfrak{L}_{a}$ has compact resolvent. We recall that is comes from the lower bound
\begin{equation}\label{eq.LbLa}
\mathfrak{Q}_{a}(\psi)\geq\int_{\R^2} (a+|\x|^2)|\psi|^2\dx\x\geq \int_{\R^2} |\x|^2|\psi|^2\dx\x\,.
\end{equation}
We consider its lowest eigenvalue $\lambda(a)$.
\begin{prop}
There exists $C>0$ such that, for all $a\in[0,1]$, 
\[|\lambda(a)-\lambda(0)|\leq Ca\,.\]
\end{prop}
\begin{proof}
Let us consider a normalized eigenfunction $u_{a}$ associated with $\lambda(a)$. From the lower bound \eqref{eq.LbLa} and Proposition \ref{prop.APE}, we have $e^{|\x|}u_{0}\in\sL^2(\R^2)$.Then we have
\[\mathfrak{Q}_{a}(u_{0})=\mathfrak{Q}_{0}(u_{0})+2a\Re\langle(D_{x}+y(x^2+y^2))u_{0},yu_{0}\rangle_{\sL^2(\R^2)}+a^2\|yu_{0}\|^2_{\sL^2(\R^2)}\,,\]
so that
\[\lambda(a)\leq\mathfrak{Q}_{a}(u_{0})\leq \lambda(0)+2a\sqrt{\mathfrak{Q}_{0}(u_{0})}\|yu_{0}\|_{\sL^2(\R^2)}+a^2\|yu_{0}\|^2_{\sL^2(\R^2)}\,,\]
and we deduce the upper bound. Now, we know that there exists $C_{0}>0$ such that for all $a\in[0,1]$,
\begin{equation}\label{eq.uub}
\lambda(a)\leq \lambda(0)+C_{0}a\leq \lambda(0)+C_{0}\,.
\end{equation}
From the $a$-independent bounds \eqref{eq.LbLa} and \eqref{eq.uub} and from the proof of Proposition \ref{prop.APE}, we deduce that there exists $C>0$ such that, for all $a\in[0,1]$,
\begin{equation}\label{unif-AP}
\int_{\R^2}e^{2|\x|}|u_{a}|^2\dx\x\leq C\,.
\end{equation}
More precisely, it comes from the fact that, for all $\mu^*\geq 2+\lambda_{0}+C_{0}$, there exists $R_{0}>0$ such that, for all $\psi$ supported in $\complement \mathcal{B}(0,R_{0})$ and all $a\in [0,1]$, we have $\mathfrak{Q}_{a}(\psi)\geq\mu^*\|\psi\|^2_{\sL^2(\R^2)}$. We also notice that the constant in \eqref{eq.APE-R} does not depend on $a\in[0,1]$.

In the same way as previously, we have
\[\mathfrak{Q}_{0}(u_{a})=\lambda(a)-2\Re\langle\langle(D_{x}+ay+y(x^2+y^2))u_{a},yu_{a}\rangle_{\sL^2(\R^2)}+a^2\|yu_{a}\|^2_{\sL^2(\R^2)}\,,\]
and thus
\[\lambda(0)\leq\mathfrak{Q}_{0}(u_{a})\leq\lambda(a)+2a\sqrt{\mathfrak{Q}_{a}(u_{a})}\|yu_{a}\|_{\sL^2(\R^2)}+a^2\|yu_{a}\|^2_{\sL^2(\R^2)}\,.\]
The conclusion easily follows since $\mathfrak{Q}_{a}(u_{a})=\lambda(a)\leq \lambda(0)+C_{0}$ and with the uniform estimate \eqref{unif-AP}.
\end{proof}

\subsection{Pan-Kwek's operator}\label{sec.theo-Rp}
We prove Theorem~\ref{theo-Rp}.

\subsubsection{Changing the parameters}
To analyze the family of operators $\mathcal{M}^\Neu_{x,\xi}$ depending on parameters $(x,\xi)$, we introduce the new parameters $(x,\eta)$ using a change of variables. 
Let us introduce the following change of parameters: 
$$\mathcal{P}(x,\xi)=(x,\eta)=\left(x,\xi+\frac{x^2}{2}\right)\,.$$
A straight forward computation provides that $\mathcal{P} : \R^2\to\R^2$ is a $\mathcal{C}^\infty$-diffeomorphism such that:
$$|x|+|\xi|\to+\infty\Leftrightarrow |\mathcal{P}(x,\xi)|\to+\infty\,.$$
We have $\mathcal{M}^\Neu_{x,\xi}=\mathcal{N}^\Neu_{x,\eta}$, where:
$$\mathcal{N}^\Neu_{x,\eta}=D_{t}^2+\left(\frac{(t-x)^2}{2}-\eta\right)^2\,,$$
with Neumann condition on $t=0$. Let us denote by $\nu_{1}^\Neu(x,\eta)$ the lowest eigenvalue of $\mathcal{N}^\Neu_{x,\eta}$, so that:
$$\mu^\Neu_{1}(x,\xi)=\nu^\Neu_{1}(x,\eta)=\nu^\Neu_{1}\left(\mathcal{P}(x,\xi)\right)\,.$$
We denote by $\Dom(\mathcal{Q}^\Neu_{x,\eta})$ the form domain of the operator and by 
$\mathcal{Q}^\Neu_{x,\eta}$ the associated quadratic form. 

\subsubsection{Existence of a minimum for $\mu_{1}^\Neu(x,\xi)$}
To prove Theorem \ref{theo-Rp}, we establish the following result:
\begin{theo}\label{theo-Rp2}
The function $\R\times\R\ni (x,\eta)\mapsto \nu_{1}^\Neu(x,\eta)$ admits a minimum. Moreover we have:
$$\liminf_{|x|+|\eta|\to+\infty}\nu_{1}^\Neu(x,\eta)\geq\nu_{\Mont}>\min_{(x,\eta)\in\R^2}\nu_{1}^\Neu(x,\eta)\,.$$
\end{theo}
To prove this result, we decompose the plane in subdomains and analyze in each part. 

\begin{lem}\label{derivees-parabole}
For all $(x,\eta)\in\R^2$ such that $\eta\geq\frac{x^2}{2}$, we have:
$$-\dr_{x}\nu_{1}^\Neu(x,\eta)+\sqrt{2\eta}\dr_{\eta}\nu_{1}^\Neu(x,\eta)>0\,.$$
Thus there is no critical point in the area $\{\eta\geq\frac{x^2}{2}\}$.
\end{lem}

\begin{proof}
The Feynman-Hellmann formulas provide:
\begin{equation*}
\dr_{x}\nu_{1}^\Neu(x,\eta)=-2\int_{0}^{+\infty} \left(\frac{(t-x)^2}{2}-\eta\right)(t-x)u_{x,\eta}^2(t)\,\dx t,
\end{equation*}
\begin{equation*}
\dr_{\eta}\nu_{1}^\Neu(x,\eta)=-2\int_{0}^{+\infty} \left(\frac{(t-x)^2}{2}-\eta\right)u_{x,\eta}^2(t)\,\dx t.
\end{equation*}
We infer:
$$-\dr_{x}\nu_{1}^\Neu(x,\eta)+\sqrt{2\eta}\dr_{\eta}\nu_{1}^\Neu(x,\eta)=\int_{0}^{+\infty}(t-x-\sqrt{2\eta})(t-x+\sqrt{2\eta})(t-x-\sqrt{2\eta})u_{x,\eta}^2(t)\,\dx t\,.$$
We have:
$$\int_{0}^{+\infty}(t-x-\sqrt{2\eta})^2(t-x+\sqrt{2\eta})u_{x,\eta}^2(t)\,\dx t>0\,.$$
\end{proof}

\begin{lem}\label{lem.minnuad}
We have:
$$\inf_{(x,\eta)\in\R^2}\nu_{1}^\Neu(x,\eta)<\nu_{\Mont}\,.$$
\end{lem}
\begin{proof}
We apply Lemma \ref{derivees-parabole} at $x=0$ and $\eta=\eta_{\Mont}$ to deduce that:
$$\dr_{x}\nu_{1}^\Neu(0,\eta_\Mont)<0\,.$$
\end{proof}

The following lemma is obvious:
\begin{lem}\label{lem.deltaneg}
For all $\eta\leq 0$, we have:
$$\nu_{1}^\Neu(x,\eta)\geq \eta^2\,.$$
In particular, we have
$$\nu_{1}^\Neu(x,\eta)> \nu_{\Mont},\qquad \forall \eta<-\sqrt{\nu_{\Mont}}\,.$$
\end{lem}

\begin{lem}\label{lem.parab}
For $x\leq 0$ and $\eta\leq\frac{x^2}{2}$, we have:
$$\nu_{1}^\Neu(x,\eta)\geq\nu_{1}^{[1]}({0}) >\nu_{\Mont}\,.$$
\end{lem}
\begin{proof}
We have, for all $\psi\in\Dom(\mathcal{Q}^\Neu_{x,\eta})$:
$$\mathcal{Q}^\Neu_{x,\eta}(\psi)=\int_{\R_{+}} |D_{t}\psi|^2+\left(\frac{(t-x)^2}{2}-\eta\right)^2|\psi|^2\dx t$$
and
$$\left(\frac{(t-x)^2}{2}-\eta\right)^2=\left(\frac{t^2}{2}-x t+\frac{x^2}{2}-\eta\right)^2\geq \frac{t^4}{4}\,.$$
The min-max principle provides:
$$\nu_{1}^\Neu(x,\eta)\geq\nu_{1}^{[1]}({0})\,.$$
Moreover, thanks to the Feynman-Hellmann theorem, we get:
$$\left(\dr_{\eta}\nu_{1}^{[1]}({\eta})\right)_{\eta=0}=-\int_{\R_{+}}t^2u_{0}(t)^2\dx t<0\,.$$
\end{proof}

\begin{lem}\label{lem.3.6}
There exist $C,D>0$ such that for all $x\in\R$ and $\eta\geq D$ such that $\frac{x}{\sqrt{\eta}}\geq -1$:
$$\nu_{1}^\Neu(x,\eta)\geq C\eta^{1/2}\,.$$
\end{lem}

\begin{proof}
For $x\in\R$ and $\eta>0$, we can perform the change of variable:
$$\tau=\frac{t-x}{\sqrt{\eta}}\,.$$ 
The operator $\eta^{-2}\mathcal{N}^\Neu_{x,\eta}$ is unitarily equivalent to:
$$\hat{\mathcal{N}}^\Neu_{\hat{x},h}=h^2D_{\tau}^2+\left(\frac{\tau^2}{2}-1\right)^2\,,$$
on $\sL^2\left((-\hat{x},+\infty)\right)$, with $\hat{x}=\frac{x}{\sqrt{\eta}}$ and $h=\eta^{-3/2}$. 

By using the harmonic approximation (see Section \ref{Harmonic-2}), we deduce
$$\nu_{1}^\Neu(x,\eta) \geq c\eta^{-3/2}\,,$$
for $\eta$ large enough.
\end{proof}

\begin{lem}\label{Agmon-Mont}
Let $u_{\eta}$ be an eigenfunction associated with the first eigenvalue of ${\mathfrak L}_{\eta}^{\Mont,+}$. 
Let $D>0$. There exist $\eps_{0},C>0$ such that, for all $\eta$ such that $|\eta|\leq D$, we have:
$$\int_{0}^{+\infty} e^{2\eps_{0} t^3}|u_{\eta}|^2\dx t\leq C\|u_{\eta}\|^2\,.$$
\end{lem}

\begin{proof}
We leave the proof to the reader as an exercise: it is sufficient to apply Proposition \ref{prop.APE}.
\end{proof}

\begin{lem}\label{lem.3.8}
For all $D>0$, there exist $A>0$ and $C>0$ such that for all $|\eta|\leq D$ and $x\geq A$, we have:
$$\left|\nu_{1}(x,\eta)-\nu_{1}^{[1]}({\eta})\right|\leq Cx^{-2}\,.$$
\end{lem}

\begin{proof}
We perform the translation $\tau=t-x$, so that $\mathcal{N}^\Neu_{x,\eta}$ is unitarily equivalent to:
$$\tilde{\mathcal{N}}^\Neu_{x,\eta}=D_{\tau}^2+\left(\frac{\tau^2}{2}-\eta\right)^2\,,$$
on $\sL^2(-x,+\infty)$. The corresponding quadratic form writes:
$$\tilde{\mathcal{Q}}^\Neu_{x,\eta}(\psi)=\int_{-x}^{+\infty} |D_{\tau}\psi|^2+\left(\frac{\tau^2}{2}-\eta\right)^2|\psi|^2\,\dx\tau\,.$$

Let us first prove the upper bound.
We take $\psi(\tau)=\chi_{0}(x^{-1}\tau)u_{\eta}(\tau)$. The \enquote{IMS} formula provides:
$$\tilde{\mathcal{Q}}^\Neu_{x,\eta}(\chi_{0}(x^{-1}\tau)u_{\eta}(\tau))=\nu_{1}^{[1]}({\eta})\|\chi_{0}(x^{-1}\tau)u_{\eta}(\tau)\|^2+\|(\chi_{0}(x^{-1}\tau))'u_{\eta}(\tau)\|^2\,.$$
Jointly min-max principle with Lemma \ref{Agmon-Mont}, we infer that:
\begin{eqnarray*}
\nu_{1}(x,\eta)
&\leq& \nu_{1}^{[1]}({\eta})+\frac{\|(\chi_{0}(x^{-1}\tau))'u_{\eta}(\tau)\|^2}{\|\chi_{0}(x^{-1}\tau)u_{\eta}(\tau)\|^2}\\
&\leq& \nu_{1}^{[1]}({\eta})+\frac{Cx^{-2}}{ e^{2c\eps_{0} x^3}}. 
\end{eqnarray*}
Let us now prove the lower bound.
Let us now prove the converse inequality. We denote by $\tilde{u}_{x,\eta}$ the positive and $\sL^2$-normalized groundstate of $\tilde{\mathcal{N}}^\Neu_{x,\eta}$. On the one hand, with the localization formula \eqref{loc-chi}, we have:
$$\tilde{\mathcal{Q}}^\Neu_{x,\eta}(\chi_{0}(x^{-1}\tau)\tilde{u}_{x,\eta})\leq\nu_{1}(x,\eta)\|\chi_{0}(x^{-1}\tau)\tilde{u}_{x,\eta}\|^2+Cx^{-2}\,.$$
On the other hand, we notice that:
$$\int_{-x}^{+\infty} t^4|\tilde{u}_{x,\eta}|^2\,\dx \tau\leq C,\qquad
\int_{-x}^{-\frac{x}{2}} t^4|\tilde{u}_{x,\eta}|^2\,\dx \tau\leq C\,,$$
and thus:
$$\int_{-x}^{-\frac{x}{2}} |\tilde{u}_{x,\eta}|^2\,\dx \tau\leq \tilde Cx^{-4}\,.$$
We infer that:
$$\tilde{\mathcal{Q}}^\Neu_{x,\eta}(\chi_{0}(x^{-1}\tau)\tilde{u}_{x,\eta})\leq (\nu_{1}(x,\eta)+Cx^{-2})\|\chi_{0}(x^{-1}\tau)\tilde{u}_{x,\eta}\|^2\,.$$
We deduce that:
$$\nu_{1}^{[1]}({\eta})\leq \nu_{1}(x,\eta)+Cx^{-2}\,.$$
\end{proof}
We have proved in Lemmas~\ref{lem.deltaneg}-\ref{lem.3.6} and \ref{lem.3.8} that the  limit inferior of $\nu_{1}(x,\eta)$ in these areas are not less than $\nu_{\Mont}$.
Then, we deduce the existence of a minimum with Lemma~\ref{lem.minnuad}.

\subsection{Helffer-Kordyukov's operator}
Let us now apply the estimates of Agmon to the model introduced in Chapter \ref{chapter-models}, Section \ref{sec.quasimodes.ex}.
\begin{prop}
There exist $\tilde C>0, h_{0}>0$, $\eps>0$ such that, for $h\in(0,h_{0})$ and $(\lambda,\psi)$ an eigenpair of ${\mathfrak{L}}^{\ex}_{h,\A}$ satisfying $\lambda\leq h+Ch^2$, we have:
$$\int_{\R^2}e^{\eps h^{-1/4}|x|}|\psi|^2\,\dx x\leq \tilde C\|\psi\|^2\,.$$
\end{prop}
\begin{proof}
We consider an eigenpair $(\lambda,\psi)$ as in the proposition and we use the Agmon identity, jointly with the localization formula (with balls of size $h^{3/8}$):
$$\mathfrak{Q}_{h,\A}^{\ex}(e^{\Phi/h^{\delta}}\psi)-h^{2-2\delta}\|\nabla\Phi e^{\Phi/h^{\delta}}\psi\|^2=\lambda\|e^{\Phi/h^{\delta}}\psi\|\,,$$
where $\delta>0$ and $\Phi$ are to be determined. For simplicity we choose $\Phi(\x)=\eps\|\x\|$.
We infer that:
$$\int_{\R^2}(h\B(x,y)-h-Ch^2-2\eps^2 h^{2-2\delta})|e^{\Phi/h^{\delta}}\psi|^2\dx x\dx y\leq 0\,.$$
We recall that $\B(x,y)=1+x^2+y^2$. We choose $\delta$ so that $hh^{2\delta}=h^{2-2\delta}$ and we get $\delta=\frac{1}{4}$. We now split the integral into two parts: $\|\x\|\geq C_{0}h^{1/4}$ and $\|\x\|\leq C_{0}h^{1/4}$.
If $\eps$ is small enough, we infer that:
$$\|e^{\Phi/{h^{1/4}}}\psi\|\leq \tilde C\|\psi\|\,.$$
\end{proof}

\chapter{Birkhoff normal form in dimension one}\label{chapter-appendix}

\begin{flushright}
\begin{minipage}{0.5\textwidth}
Cut away all that is excessive, straighten all that is crooked, bring light to all that is overcast, labour to make all one glow of beauty.
\begin{flushright}
\textit{Enneads}, I. 6. 9, Plotinus
\end{flushright}
\end{minipage}
\vspace*{0.5cm}
\end{flushright}

This chapter is an invitation to symplectic geometry and pseudo-differential calculus. Therefore we do not try to be the most general as possible and focus on an elementary application (the Birkhoff normal form in dimension one) that will be very helpful in Chapter \ref{chapter-birk}. Since we only wish to highlight the main aspects of the proofs, we will often keep some details in the shadow and refer to the nice introductions to semiclassical analysis \cite{DiSj99, Martinez02, Z13}.

\section{Symplectic geometry and pseudo-differential calculus}

\subsection{Differential geometry in action: a Darboux-Moser-Weinstein result}

\subsubsection{Some definitions}
Let us recall basic concepts related to differential forms. We mainly refer to \cite[Appendix B]{Z13} for a concise introduction and to \cite[Chapter 7]{Arnold} for more details. We present the concepts when the dimension is even (and equals to $2d$), even if most of them do not depend on the parity of the dimension. If $\kappa : \R^{2d}\to\R^{2d}$ is smooth mapping, the pull-back by $\kappa$ of a $m$-differential form $\omega$ in $\R^{2d}$, denoted by $\kappa^*\omega$, is the $m$-differential form defined by
$$\forall (u_{1},\ldots, u_{m})\in(\R^{2d})^m,\qquad (\kappa^*\omega)_{x}(u_{1},\ldots, u_{m})=\omega_{\kappa(x)}(d\kappa_{x}(u_{1}),\ldots, d\kappa_{x}(u_{m}))\,,$$
where $d\kappa_{x}$ is the usual differential of $\kappa$ at the point $x$.

We say that $\kappa$ is symplectic when
$$\kappa^*\omega_{0}=\omega_{0}\,,\qquad \mbox{ with }\qquad \omega_{0}=\dx \xi\wedge \dx x\,.$$
In other words, $\kappa$ is symplectic if it preserves the canonical symplectic form $\omega_{0}$ in $\R^{2d}$. 

If $X$ is a vector field on $\R^{2d}$ and $\phi_{s}$ the associated flow, that is $\phi'=X(\phi)$, the Lie derivative of a $m$-differential form $\omega$ is by definition
$$\mathcal{L}_{X}\omega=(\partial_{s}\phi_{s}^{*}\omega)_{s=0}\,.$$
The Lie derivative may be expressed thanks to the Cartan formula:
$$\mathcal{L}_{X}\omega=\dx (\iota_{X}\omega)+\iota_{X}\dx\omega\,,$$
where $\iota_{X}$ associate to a $m$-differential form $\omega$ the $m-1$-differential form obtained by replacing the first entry of $\omega$ by $X$.

Let us provide an abstract and fundamental example of symplectic mapping. Let us consider a smooth function $H$ (the Hamiltonian) and the vector field defined $X_{H}$ by $dH(\cdot)=\omega_{0}(\cdot, X_{H})$. The flow associated with $X_{H}$, denoted by $\phi_{s}=e^{sX_{H}}$, is called the Hamiltonian flow and, for all $s$, we have $\phi_{s}^*\omega_{0}=\omega_{0}$. In other words, for all $s$, the application $\phi_{s} : (x,\xi)\mapsto e^{sX_{H}}(x,\xi)$ is symplectic. This can be seen from the Cartan formula. Finally, we will use the standard definition of the Poisson bracket of smooth functions:
$$\{f,g\}=\omega_{0}(\nabla f,\nabla g)=\partial_{\xi}f\cdot \partial_{x}g-\partial_{x}f\cdot\partial_{\xi}g\,.$$

\subsubsection{A lemma}
The aim of this section is to prove the following classical lemma.
\begin{lem}\label{symplectize}
  Let us consider $\omega_{0}$ and $\omega_{1}$ two $2$-forms on
  $\R^{2d}$ which are closed and non degenerate. Let us assume that
  ${\omega_{1}}={\omega_{0}}$ on $\{0\}\times\Omega$ where $\Omega$ is a bounded open set of $\R^{2d-1}$. In a neighborhood of $\{0\}\times\Omega$ there exists a change of coordinates
  $\psi_{1}$ such that:
$$\psi_{1}^*\omega_{1}=\omega_{0}\quad \mbox{ and }\quad
{\psi_{1}}={\Id} + \mathcal{O}(x_{1}^2)\,.$$
\end{lem}

\begin{proof}
The reader is referred to \cite[p. 92]{McSa98}.

Let us begin to explain how we can find a $1$-form $\sigma$ on $\R^{2d}$ such that,
in a neighborhood of $\{0\}\times\Omega$,
$$\tau=\omega_{1}-\omega_{0}=\dx\sigma\quad \mbox{ and }\quad\sigma=\mathcal{O}(x_{1}^2)\,.$$ 
In other words, we want to establish an explicit Poincar\'e lemma. For that purpose we introduce the family of
diffeomorphisms $(\phi_{t})_{0<t\leq 1}$ defined by
$$\phi_{t}( x_{1}, \tilde x)=(t x_{1}, \tilde x)$$
and we let
$$\phi_{0}(x_{1},\tilde x)=(0,\tilde x)\,.$$
We have
\begin{equation}
  \phi_{0}^*\tau=0\quad\mbox{ and }\quad \phi_{1}^*\tau=\tau\,.
  \label{equ:phi_t}
\end{equation}
Let us denote by $X_{t}$ the vector field associated with $\phi_{t}$. We have
$$ X_{t}=\frac{d\phi_{t}}{dt}(\phi_{t}^{-1})=(t^{-1}x_{1}, 0)= t^{-1} x_1e_1\,.$$
Let us compute the Lie derivative of $\tau$ along $X_{t}$,
$$\frac{d}{dt}\phi_{t}^*\tau=\phi_{t}^*\mathcal{L}_{X_{t}}\tau\,.$$ 
From the Cartan formula, we have, 
$$\mathcal{L}_{X_{t}}=\iota(X_{t})\dx\tau+\dx(\iota(X_{t})\tau)\,.$$ 
Since $\tau$ is closed on $\R^{2d}$, we have $\dx\tau=0$. Therefore it follows that
\begin{equation}
  \frac{d}{dt}\phi_{t}^*\tau=\dx(\phi_{t}^*\iota(X_{t})\tau)\,.
  \label{equ:ddt_phi_t}
\end{equation}
We consider the $1$-form 
$$\sigma_{t}=\phi_{t}^*\iota(X_{t})\tau=x_1 \tau_{\phi_t(x_{1}, x_{2}, \xi_{1},  \xi_{2})}(e_1,\nabla\phi_t(\cdot))=\mathcal{O}(x_1^2)\,.$$  
We see from~\eqref{equ:ddt_phi_t} that the map $t\mapsto \phi_t^* \tau$ is
smooth on $[0,1]$.  To conclude, let $\sigma$ be the $1$-form  defined on a neighborhood of $\{0\}\times\Omega$ by $\sigma=\int_{0}^1
\sigma_{t}\,\dx t$; it follows from~\eqref{equ:phi_t}
and~\eqref{equ:ddt_phi_t} that:
$$\frac{d}{dt}\phi_{t}^*\tau=\dx\sigma_{t}\quad \mbox{ and }\quad \tau=\dx\sigma\,.$$
Finally we use a standard deformation argument due to Moser. For $t\in [0,1]$,
we let 
$$\omega_{t}=\omega_{0}+t(\omega_{1}-\omega_{0})\,.$$ 
The $2$-form $\omega_{t}$ is closed and non degenerate (up to choosing a neighborhood of $x_{1}=0$ small enough). We look for $\psi_{t}$
such that
$$\psi_{t}^*\omega_{t}=\omega_{0}\,.$$
For that purpose, let us determine a vector field $Y_{t}$ such that
$$\frac{d}{dt}\psi_{t}=Y_{t}(\psi_{t}).$$
By using again the Cartan formula, we get
$$0=\frac{d}{dt}\psi_{t}^*\omega_{t}=\psi_{t}^*\left(\frac{d}{dt}\omega_{t}+\iota(Y_{t})\dx\omega_{t}+\dx(\iota(Y_{t})\omega_{t})\right)\,.$$
This becomes
$$\omega_{0}-\omega_{1}=\dx(\iota(Y_{t})\omega_{t})\,.$$
We are led to solve:
$$\iota(Y_{t})\omega_{t}=-\sigma\,.$$
By non degeneracy of $\omega_{t}$, this determines $Y_{t}$.  Since $Y_{t}$ vanishes on $\{0\}\times\Omega$, there exists a neighborhood of $\{0\}\times\Omega$ small enough in the $x_{1}$-direction such that $\psi_{t}$
exists until the time $t=1$ and satisfies
$\psi_{t}^*\omega_{t}=\omega_{0}$. Since $\sigma=\mathcal{O}( x_1^2)$, we
get $\psi_{1}=\Id+\mathcal{O}(x_{1}^2)$.
\end{proof}

\subsection{Pseudo-differential calculus}

\subsubsection{Symbols}
Here we refer to \cite[Chapter 7]{DiSj99} and \cite[Chapter 4]{Z13}.

A function $m:\R^{2d}\to
[0,\infty)$ is an order function if there exist constants $N_0, C_0>0$
such that 
\[
m(X)\leq C_0\langle X-Y\rangle^{N_0} m(Y)
\]
for any $X,Y\in\R^{2d}$. For $\delta\in\left(0,\frac{1}{2}\right)$, the symbol class $S_{\delta}(m)$ is the space of smooth
$h$-dependent functions $a_h:\R^{2d}\to\C$ such that
\[
\forall
\alpha\in\N^{2d}, \qquad \abs{\partial_x^\alpha a_h(x)} \leq C_\alpha h^{-|\alpha|\delta}m(x), \qquad  \forall h\in(0,1].
\]
We let $S(m)=S_{0}(m)$. For a classical symbol
$a_{ h}=a(x,\xi;h)\in S_{\delta}(m)$,  the Weyl quantization of $a$ is defined as:
\begin{equation}\label{a-Weyl}
\Op_{h}^w a\, (\psi)(x)=\frac{1}{(2\pi h)^d}\int_{\R^{2d}} e^{i\pscal{x-y}{\xi}/h} a\left(\frac{x+y}{2}, \xi\right)\psi(y)\dx y\dx \xi,\quad \forall \psi\in\mathcal{S}(\R^d)\,.
\end{equation}
It can be proved that the integral in \eqref{Weyl} is actually convergent thanks to a succession of integrations by parts and that $\Op_{h}^w(a)$ sends $\mathcal{S}(\R^d)$ into $\mathcal{S}(\R^d)$.

If $m_{1}$ and $m_{2}$ are order functions and $a\in S_{\delta}(m_{1})$, $b\in S_{\delta}(m_{2})$, we may define the Moyal product of $a$ and $b$ by letting
$$a\star b(x,\xi)=e^{\frac{ih}{2}\omega_{0}(D_{x}, D_{\xi}, D_{y}, D_{\eta})}a(x,\xi)b(y,\eta)_{|(y,\eta)=(x,\xi)}\,$$
and
$$a\star b\in S_{\delta}(m_{1}m_{2})\,,\qquad \Op_{h}^w\left(a\star b\right)=\Op_{h}^w\left(a\right)\Op_{h}^w\left(b\right)\,,$$
as operators defined on $\mathcal{S}(\R^d)$. Note that the verification is just a computation using the inverse Fourier transform when $a$ and $b$ belong to $\mathcal{S}(\R^d)$.

Moreover, thanks to the exponential expression and by the Taylor formula, the Moyal product can be expanded in the sense of the $S(m_{1}m_{2})$-topology as
$$a\star b=ab+\frac{h}{2i}\{a,b\}+\mathcal{O}_{S(m_{1}m_{2})}(h^{1-2\delta})\,.$$

We recall the so-called Borel's theorem: If $(a_{j})_{j\geq 0}$ is a sequence of symbols in $S_{\delta}(m)$, there exists a symbol in $S_{\delta}(m)$ such that
$$a\sim\sum_{j=0}^{+\infty} h^ja_{j},\qquad \mbox{ in } S_{\delta}(m)\,.$$

We will sometimes use the \textit{Calderon-Vaillancourt theorem}: If $a\in S(1)$, then $\Op_{h}^w(a)$ is a bounded operator from $\sL^2(\R^d)$ to $\sL^2(\R^d)$ and 
\[\|\Op_{h}^w a\|\leq\sum_{|\alpha|\leq M d}\sup_{\R^d}\|\partial^\alpha a\|\,.\]
Another important and classical theorem in the pseudo-differential theory is the \textit{G\aa rding inequality}: If $a\in S(1)$ is a real symbol such that $a\geq 0$, then there exists $C>0$, $h_{0}>0$ such that, for all $\psi\in\sL^2(\R^d)$ and $h\in(0,h_{0})$,
\[\langle\Op_{h}^w a\,\psi,\psi\rangle\geq -Ch\|\psi\|^2\,.\]

\subsubsection{Egorov theorems}\label{app.egorov}
We now recall the classical result (see for instance \cite[Theorem
11.1]{Z13} and~\cite[Th\'eor\`eme IV.10]{Ro87}).
\begin{theo}[\mbox{\cite[Theorem 11.1, Remark (ii)]{Z13}}]
\label{theo:egorov1}
  Let $P$ and $Q$ be $h$-pseudo-differential operators on $\R^d$, with
  $P\in S(1)$ and $Q\in S(1)$.  Then the operator $e^{\frac{i}{h}Q} P
  e^{-\frac{i}{h}Q}$ is a pseudo-differential operator in $S(1)$, and
  \[
e^{\frac{i}{h}Q} P e^{-\frac{i}{h}Q} - \Op^w_h(p\circ \kappa) \in
  h S(1).
\] 
Here $p$ is the Weyl symbol of $P$, and the canonical
transformation $\kappa$ is the time-1 Hamiltonian flow associated to
principal symbol of $Q$.
\end{theo}
From this classical version of Egorov's theorem, one can deduce the
following refinement that is useful when $P$ does not belong to $S(1)$ (see \cite[Appendix]{HKRVN14}).
\begin{theo}
\label{theo:egorov2}
  Let $P$ and $Q$ be $h$-pseudo-differential operators on $\R^d$, with
  $P\in S(m)$ and $Q\in S(m')$, where $m$ and $m'$ are order functions
  such that:
  \begin{equation}
    m' = \mathcal{O}(1); \quad m m' = \mathcal{O}(1)\,.\label{equ:egorov0}
\end{equation}

Then the operator $e^{\frac{i}{h}Q} P e^{-\frac{i}{h}Q}$ is a
pseudo-differential operator in $S(m)$, and we have $e^{\frac{i}{h}Q} P
e^{-\frac{i}{h}Q} - \Op^w_h(p\circ \kappa) \in h S(1)$.
\end{theo}

\section{Birkhoff normal form}\label{sec.BFelec}
This section provide some insights concerning the semiclassical Birkhoff normal form in the simple case of $h^2D_{x}^2+V(x)$. 
We will consider 
$$\Op_{h}^w\left(H\right)=h^2D_{x}^2+V(x),\qquad H(x,\xi)=\xi^2+V(x)\,,$$
where $V$ is a standard symbol so that, for some order function $m$, $H\in S(m)$. We recall in Appendix \ref{chapter-appendix} some elements of symplectic geometry as well as standard facts coming from the pseudo-differential theory. If the reader wishes to go further in the proofs of the results recalled there, he is referred to the books \cite{DiSj99, Martinez02, Z13}. In this section we follow the spirit of \cite{VuCha08, Vu09} (see also \cite{Vu06}).

The aim of this section is to prove the following eigenvalue estimate (which improves the result of Proposition \ref{el-expansion}).
\begin{theo}\label{theo.elec.birk}
Let $\eta\in(0,1)$. There exists a smooth function $f^\star$ with compact support arbitrarily small and satisfying $|f^\star(Z,h)|=\mathcal{O}((Z+h)^{\frac{3}{2}})$ as $(Z,h)\to 0$ such that the eigenvalues of the operators $\Op_{h}^w\left(H\right)$ and $\Op_{h}^w\left(|z|^2\right)+f^\star\left(\Op_{h}^w\left(|z|^2\right),h\right)$ below $h^\eta$ coincide modulo $\mathcal{O}(h^\infty)$. Moreover, if we let 
$$\mathsf{N}_{h}=\left\{n\in\N^* : (2n-1)h\leq h^{\frac{1}{2}}\right\}\,,$$
and if $\lambda_{n}(h)$ is the $n$-th eigenvalue of $\Op_{h}^w(H)$ we have
$$\lambda_{n}(h)=(2n-1)h+\mathcal{O}(h^{\frac{3}{2}})\,,$$
uniformly for $n\in\mathsf{N}_{h}$ and $h\in(0,h_{0})$.

\end{theo}

\subsection{Formal series and homological equations}
We introduce the space of formal series $\mathcal{E}=\R[[x,\xi, h]]$. We endow
$\mathcal{E}$ with the Moyal product (compatible with the Weyl
quantization) denoted by $\star$.
\begin{notation}
  The degree of $x^\alpha\xi^\beta h^l$ is
  $\alpha+\beta+2l$. $\mathcal{D}_{N}$ denotes the space of the
  monomials of degree $N$. $\mathcal{O}_{N}$ is the space of formal
  series with valuation at least $N$.  For $\tau,\gamma\in\mathcal{E}$, we denote
  $\ad_{\tau}\gamma = [\tau,\gamma]=\tau\star\gamma-\gamma\star\tau$. We notice that $[
  \mathcal{O}_{N_1}, \mathcal{O}_{N_2}] \subset
  \mathcal{O}_{N_1+N_2}$.
\end{notation}

\begin{lem}\label{cohom}
We let $z=x+i\xi$. We have
$$\mathcal{E}=\ker\{|z|^2,\cdot\}\oplus\mathrm{Im}\{|z|^2,\cdot\},$$
where the Poisson bracket is given by 
$$\{f,g\}=\frac{\partial f}{\partial \xi}\frac{\partial g}{\partial x}-\frac{\partial f}{\partial x}\frac{\partial g}{\partial\xi}=\frac{1}{i}\left(\frac{\partial f}{\partial \overline{z}}\frac{\partial g}{\partial z}-\frac{\partial f}{\partial z}\frac{\partial g}{\partial \overline{z}}\right),$$
where 
$$\frac{\partial}{\partial z}=\frac{1}{2}\left(\frac{\partial}{\partial x}-i\frac{\partial}{\partial y}\right),\qquad\frac{\partial}{\partial z}=\frac{1}{2}\left(\frac{\partial}{\partial x}+i\frac{\partial}{\partial y}\right).$$
\end{lem}
\begin{proof}
The family $\left(z^\alpha  \overline{z}^\beta h^\gamma\right)_{(\alpha,\beta,\gamma)\in\N^3}$ is a basis of $\mathcal{E}$. Then it is sufficient to notice that $\{|z|^2,\mathcal{D}_{N}\}\subset\mathcal{D}_{N}$ and
$$\{|z|^2, z^\alpha \overline{z}^\beta\}=\frac{2}{i}(\alpha-\beta)z^\alpha \overline{z}^\beta.$$
\end{proof}

\begin{prop}\label{prop:formal-normal-form0}
  Given $\gamma\in\mathcal{O}_{3}$, there exist formal power series
  $\tau,\kappa\in\mathcal{O}_{3}$ such that
$$e^{ih^{-1}\ad_{\tau}}(|z|^2+\gamma)=|z|^2+\kappa\,,$$
with $[\kappa,|z|^2]=0.$
\end{prop}
\begin{proof}
First, we notice $ih^{-1}\ad_{\tau}$ sends $\mathcal{O}_{N}$ into $\mathcal{O}_{N+1}$ so that the exponential is well defined in the formal series. Then, we proceed by induction. Let $N\geq 1$. Assume that we have, for $N\geq 1$ and $\tau_{N}\in\mathcal{O}_{3}$:
$$e^{ih^{-1}\ad_{\tau_{N}}}(|z|^2+\gamma)=|z|^2+K_{3}+\cdots+K_{N+1}+R_{N+2}+\mathcal{O}_{N+3}\,,$$
where $K_{i}\in\mathcal{D}_{i}$ commutes with $|z|^2$ and
where $R_{N+2}\in\mathcal{D}_{N+2}$.

Let $\tau'\in \mathcal{D}_{N+2}$. A computation provides:
\begin{align*}
  &e^{ih^{-1}\ad_{\tau_{N}+\tau'}}(|z|^2+\gamma)=H^0+K_{3}+\cdots+K_{N+1}+K_{N+2}+\mathcal{O}_{N+3},
\end{align*}
with:
$$K_{N+2}=R_{N+2}+ih^{-1}\ad_{\tau'} |z|^2=R_{N+2}-ih^{-1}\ad_{|z|^2} \tau'\,,$$
We can write
$$R_{N+2}=K_{N+2}+ih^{-1}\ad_{|z|^2} \tau'\,.$$
Note that $ih^{-1}\ad_{|z|^2}=\{|z|^2,\cdot\}$. With Lemma \ref{cohom}, we deduce the existence of
$\tau'$ and $K_{N+2}$ such that $K_{N+2}$ commutes with $|z|^2$. 
\end{proof}

\subsection{Quantizing}\label{sec.quantizing.elec}
Let us now quantize the formal procedure.
\begin{prop}\label{prop.quant.elec}
There exists a real and compactly supported symbol $c(x,\xi,h)$ and a smooth function $f^\star$ with arbitrarily small compact support such that:
$$e^{ih^{-1}\Op_{h}^w( c)}\Op_{h}^w\left(H\right)e^{-ih^{-1}\Op_{h}^w( c)}$$
is a pseudo-differential operator in $S(m)$ and 
$$e^{ih^{-1}\Op_{h}^w( c)}\Op_{h}^w\left(H\right)e^{-ih^{-1}\Op_{h}^w( c)}=\mathcal{N}_{h}+\Op_{h}^w\left(s_{h}\right)+h^\infty S(1)\,,$$
\begin{enumerate}[(i)]
\item with $\mathcal{N}_{h}=\Op_{h}^w\left(|z|^2\right)+f^\star\left(\Op_{h}^w\left(|z|^2\right),h\right)$,
\item where $s_{h}$ is a symbol in $S(m)$ whose Taylor series at $(0, 0, 0)$ is zero.
\end{enumerate}
\end{prop}
\begin{proof}
Thanks to the Borel lemma, we may find a smooth function with compact support $c(x,\xi,h)$ whose Taylor series at $(0,0,0)$ is the series $\tau$ given in Proposition \ref{prop:formal-normal-form0}. In particular, $\Op_{h}^w( c)$ is a bounded self-adjoint operator (by, for instance, the Calderon-Vaillancourt theorem). Then, we consider the conjugate operator
$$e^{ih^{-1}\Op_{h}^w( c)}\Op_{h}^w\left(H\right)e^{-ih^{-1}\Op_{h}^w( c)}$$
that is a pseudo-differential operator, with symbol $N_{h}$, by the Egorov theorem. By the Taylor formula, we can write
\begin{multline*}
e^{ih^{-1}\Op_{h}^w( c)}\Op_{h}^w\left(H\right)e^{-ih^{-1}\Op_{h}^w( c)}=\sum_{n=0}^N \frac{1}{n!}\ad^n_{ih^{-1}\Op_{h}^w( c)}\Op_{h}^w\left(H\right)\\
+\frac{h^{-N-1}}{N!}\int_{0}^1 (1-t)^Ne^{ith^{-1}\Op_{h}^w( c)}\ad^{N+1}_{ih^{-1}\Op_{h}^w( c)}\Op_{h}^w\left(H\right)e^{-ith^{-1}\Op_{h}^w( c)}\dx t\,.
\end{multline*}
By the Egorov theorem, the integral remainder
$$\frac{h^{-N-1}}{N!}\int_{0}^1 (1-t)^Ne^{ith^{-1}\Op_{h}^w( c)}\ad^{N+1}_{ih^{-1}\Op_{h}^w( c)}\Op_{h}^w\left(H\right)e^{-ith^{-1}\Op_{h}^w( c)}\dx t$$
is a pseudo-differential operator whose symbol admits a Taylor expansion in $\mathcal{O}_{N+1}$. Moreover, the symbol of
$$\sum_{n=0}^N \frac{1}{n!}\ad^n_{ih^{-1}\Op_{h}^w( c)}\Op_{h}^w\left(H\right)$$
admits a Taylor expansion that coincides with $|z|^2+\kappa$ modulo $\mathcal{O}_{N+1}$. In other words, the Taylor series of $N_{h}$ is $|z|^2+\kappa$ where $\kappa$ is in the form $\sum_{k+\ell\geq 1} d_{k,\ell}|z|^{2k} h^\ell$. Using again the Borel lemma, we may find a smooth function $f(I,h)$ with compact support (as small as we want) such that its Taylor series at $(0,0)$ is
$$\sum_{k+\ell\geq 1} d_{k,\ell}I^{2k} h^\ell\,.$$
We have :
$$e^{ih^{-1}\Op_{h}^w( c)}\Op_{h}^w\left(H\right)e^{-ih^{-1}\Op_{h}^w( c)}=\Op_{h}^w\left(|z|^2\right)+\Op_{h}^w\left(f(|z|^2,h)\right)+\Op_{h}^w\left(R_{h}\right)\,.$$
with $R_{h}=\Op_{h}^w(r_{h})$ where the Taylor series of $r_{h}$ at $(0,0,0)$ is $0$.

Note that  $\kappa$ can also be written in the form $\sum_{k+\ell\geq 1} d^\star_{k,\ell}\left(|z|^{2}\right)^{\star k} h^\ell$ and we may also find a smooth function, with support arbitrarily small, $f^\star(I,h)$ with Taylor series
$$\sum_{k+\ell\geq 1} d^\star_{k,\ell}I^{2k} h^\ell\,,$$
and we have, by using the Taylor formula and  the functional calculus of pseudo-differential operators (see \cite[Theorem 8.7]{DiSj99} for a detailed presentation) to estimate the Taylor remainder,
$$\Op_{h}^w\left(|z|^2\right)+\Op_{h}^w\left(f(|z|^2,h)\right)=\Op_{h}^w\left(|z|^2\right)+f^\star(\Op_{h}^w\left(|z|^2\right),h)+\Op_{h}^w\left(\tilde R_{h}\right)+h^\infty S(1)\,,$$
where $\tilde R_{h}=\Op_{h}^w(\tilde r_{h})$ where the Taylor series of $\tilde r_{h}$ at $(0,0,0)$ is $0$.
\end{proof}

\subsection{Microlocalizing}
First, we get the following.
\begin{lem}\label{lem.nb.vp1}
We have:
$$\mathsf{N}\left(\mathcal{N}_{h},\beta \right)=\mathcal{O}(h^{-1})\,.$$
\end{lem}
\begin{proof}
If the support of $f^\star$ is small enough, for all $\eps\in(0,1)$, we have, for all $\psi\in\mathcal{C}^\infty_{0}(\R)$ and $h$ small enough,
\begin{equation}\label{relative-lb-1D}
\langle\mathcal{N}_{h}\psi,\psi\rangle\geq (1-\eps)\langle\Op_{h}^w(|z|^2)\psi,\psi\rangle\,.
\end{equation}
Thus, by the min-max principle, we infer that
$$\mathsf{N}\left(\mathcal{N}_{h},\beta \right)\leq \mathsf{N}\left(\Op_{h}^w(|z|^2),(1-\eps)^{-1}\beta \right)\,$$
and the result follows by counting the eigenvalues of the harmonic oscillator.
\end{proof}
Then, we may use the Weyl's asymptotic estimate (see for instance Chapter \ref{chapter-semi-ex}, Proposition \ref{number-1D}).
\begin{lem}\label{lem.nb.vp2}
If $\beta\in\left(0, \underset{|x|\to+\infty}{\liminf}V\right)$, we have:
$$\mathsf{N}\left(\Op_{h}^w(H),\beta \right)=\mathcal{O}(h^{-1})\,.$$
\end{lem}
The following proposition is devoted to microlocalization estimates of the eigenfunctions of $\Op_{h}^w(H)$ and $\mathcal{N}_{h}$.
\begin{prop}\label{prop.micro.elec}
Let $\eta\in(0,1)$, $\delta\in\left(0,\frac{\eta}{2}\right)$ and $\chi$ a smooth cutoff function supported away from a compact neighborhood of $0$. Then, there exists $h_{0}>0$ such that for all $h\in(0,h_{0})$, all eigenvalue $\lambda$ of $\Op_{h}^w(H)$ or of $\mathcal{N}_{h}$ such that $\lambda\leq h^{\eta}$ and all associated eigenfunction $\psi$, we have
$$\|\Op_{h}^w(\chi(h^{-\delta}(x,\xi)))\psi\|_{\sL^2(\R)}=\mathcal{O}(h^\infty)\|\psi\|_{\sL^2(\R)}\,.$$
\end{prop}
\begin{proof}
Let us prove this for the eigenfunctions of $\mathcal{N}_{h}$. We write the eigenvalue equation
$$\mathcal{N}_{h}\psi=\lambda \psi\,.$$
We have
$$\mathcal{N}_{h}\Op_{h}^w(\chi(h^{-\delta}(x,\xi))))\psi=\lambda \Op_{h}^w(\chi(h^{-\delta}x))\psi+[\mathcal{N}_{h},\Op_{h}^w(\chi(h^{-\delta}(x,\xi)))]\psi\,.$$
Taking the scalar product with $\Op_{h}^w(\chi(h^{-\delta}(x,\xi)))\psi$, we infer that
\begin{multline*}
\langle\mathcal{N}_{h}\Op_{h}^w(\chi(h^{-\delta}(x,\xi)))\psi, \Op_{h}^w(\chi(h^{-\delta}(x,\xi)))\psi\rangle_{\sL^2(\R)}\leq h^\eta \|\Op_{h}^w(\chi(h^{-\delta}(x,\xi)))\psi\|_{\sL^2(\R)}^2\\
+\langle[\mathcal{N}_{h},\Op_{h}^w(\chi(h^{-\delta}(x,\xi)))]\psi, \Op_{h}^w(\chi(h^{-\delta}(x,\xi)))\psi\rangle_{\sL^2(\R)}\,.
\end{multline*}
We use again the lower bound \eqref{relative-lb-1D} to get
\begin{multline*}
\langle\mathcal{N}_{h}\Op_{h}^w(\chi(h^{-\delta}(x,\xi)))\psi, \Op_{h}^w(\chi(h^{-\delta}(x,\xi)))\psi\rangle_{\sL^2(\R)}\\
\geq (1-\eps)\langle\Op_{h}^w\left(|z|^2\right)\Op_{h}^w(\chi(h^{-\delta}(x,\xi)))\psi,\Op_{h}^w(\chi(h^{-\delta}(x,\xi)))\psi \rangle\,.
\end{multline*}
By a rescaling $(x,\xi)=h^\delta(\tilde x,\tilde\xi)$, a support consideration and the G\aa rding inequality with semiclassical parameter $h^{1-2\delta}$, we get
\begin{multline*}
\langle\Op_{h}^w\left(|z|^2\right)\Op_{h}^w(\chi(h^{-\delta}(x,\xi)))\psi,\Op_{h}^w(\chi(h^{-\delta}(x,\xi)))\psi \rangle\\
\geq h^{2\delta}(1-Ch^{1-2\delta})\|\Op_{h}^w(\chi(h^{-\delta}(x,\xi)))\psi\|^2_{\sL^2(\R)}\,,
\end{multline*}
so that we deduce 
\begin{multline*}
\langle\mathcal{N}_{h}\Op_{h}^w(\chi(h^{-\delta}(x,\xi)))\psi, \Op_{h}^w(\chi(h^{-\delta}(x,\xi)))\psi\rangle_{\sL^2(\R)}\\
\geq (1-\eps)(h^{2\delta}-Ch^{1-2\delta})\|\Op_{h}^w(\chi(h^{-\delta}(x,\xi)))\psi\|^2_{\sL^2(\R)}\,,
\end{multline*}
and thus
\begin{multline*}
((1-\eps)(h^{2\delta}-Ch)-h^\eta)\|\Op_{h}^w(\chi(h^{-\delta}(x,\xi)))\psi\|^2_{\sL^2(\R)}\\
\leq \langle[\mathcal{N}_{h},\Op_{h}^w(\chi(h^{-\delta}(x,\xi)))]\psi, \Op_{h}^w(\chi(h^{-\delta}(x,\xi)))\psi\rangle_{\sL^2(\R)}\,.
\end{multline*}
The pseudo-differential operator $[\mathcal{N}_{h},\Op_{h}^w(\chi(h^{-\delta}(x,\xi)))]$ has a symbol in the standard class $S_{\delta}(m)$ (here we have $\delta\in\left(0,\frac{1}{2}\right)$). Its symbol is supported in $\supp (\chi(h^{-\delta}(x,\xi)))$ modulo $h^\infty S_{\delta}(1)$ and its main term is of order $h^{1-2\delta}$. Therefore, if we consider a cutoff function $\underline{\chi}$ supported on a slightly bigger support than $\chi$, we deduce
$$((1-\eps)(h^{2\delta}-Ch)-h^\eta)\|\Op_{h}^w(\chi(h^{-\delta}(x,\xi)))\psi\|^2_{\sL^2(\R)}\leq Ch^{1-2\delta}\|\Op_{h}^w(\underline{\chi}(h^{-\delta}(x,\xi)))\psi\|^2_{\sL^2(\R)}\,.$$
This implies the existence of $\tilde\delta>0$ such that
$$\|\Op_{h}^w(\chi(h^{-\delta}(x,\xi)))\psi\|^2_{\sL^2(\R)}\leq Ch^{\tilde\delta}\|\Op_{h}^w(\underline{\chi}(h^{-\delta}(x,\xi)))\psi\|^2_{\sL^2(\R)}\,.$$
Then the result follows by a iterative argument by replacing $\chi$ by $\underline{\chi}$.
\end{proof}
It is now easy to deduce the following corollary.
\begin{cor}\label{cor.micro.elec}
Let $\eta\in(0,1)$, $\delta\in\left(0,\frac{\eta}{2}\right)$ and $\chi$ a smooth cutoff function supported away from a compact neighborhood of $0$. Then, there exists $h_{0}>0$ such that for all $h\in(0,h_{0})$, for all $\psi\in\mathds{1}_{[-\infty,h^\eta)}(\mathcal{N}_{h})$ or $\psi\in\mathds{1}_{[-\infty,h^\eta)}(\Op^w_{h}(H))$, we have
$$\|\Op_{h}^w(\chi(h^{-\delta}(x,\xi)))\psi\|_{\sL^2(\R)}=\mathcal{O}(h^\infty)\|\psi\|_{\sL^2(\R)}\,.$$
\end{cor}
\begin{proof}
By Proposition \ref{prop.micro.elec}, the estimate is clear when $\psi$ is an eigenfunction. Thanks to Lemmas \ref{lem.nb.vp1} and \ref{lem.nb.vp2}, we have
$$\dim\mathrm{range}\left(\mathds{1}_{[-\infty,h^\eta)}(\mathcal{N}_{h})\right)=\mathcal{O}(h^{-1}),\qquad\dim\mathrm{range}\left(\mathds{1}_{[-\infty,h^\eta)}(\Op^w_{h}(H))\right)=\mathcal{O}(h^{-1})\,.$$
In particular, these numbers of eigenvalues below $h^\eta$ do not increase more than polynomially in $h^{-1}$. Then, the result follows by considering orthonormal bases of the spaces $\mathrm{range}\left(\mathds{1}_{[-\infty,h^\eta)}(\mathcal{N}_{h})\right)$ and $\mathrm{range}\left(\mathds{1}_{[-\infty,h^\eta)}(\Op^w_{h}(H))\right)$ and by applying Proposition \ref{prop.micro.elec} to the elements of these bases.
\end{proof}

\subsection{Proof of Theorem \ref{theo.elec.birk}}
We have now all the elements to deduce Theorem \ref{theo.elec.birk}. It essentially follows from an application of the min-max principle. Let us consider the sequence of the eigenvalues of $\mathcal{N}_{h}$ denoted by $(\lambda_{j}(\mathcal{N}_{h}))_{j\geq 1}$. We may consider an associated orthonormal family of eigenfunctions $(\psi_{j,h})_{\geq 1}$. Let us consider a positive integer $M$ less than $\dim\mathrm{range}\left(\mathds{1}_{[-\infty,h^\eta)}(\mathcal{N}_{h})\right) $. With the notations of Proposition \ref{prop.quant.elec}, we let
$$\varphi_{j,h}=e^{-ih^{-1}\Op_{h}^w(c)}\psi_{j,h}$$
and we introduce
$$V_{h}=\underset{1\leq j\leq M}{\spann} \varphi_{j,h}\,.$$
Then, with Proposition \ref{prop.quant.elec}, for all $\varphi\in V_{h}$, we have
$$\langle\Op_{h}^w(H)\varphi,\varphi\rangle\leq \lambda_{M}(\mathcal{N}_{h})\|\psi\|^2_{\sL^2(\R)}+\langle\Op_{h}^w(s_{h})\psi,\psi\rangle+\mathcal{O}(h^{\infty})\|\psi\|^2_{\sL^2(\R)}\,.$$
with $\psi=e^{ih^{-1}\Op_{h}^w(c)}\varphi$. Thanks to Corollary \ref{cor.micro.elec} and the fact that the Taylor series of $s_{h}$ with respect to $(x,\xi,h)$ is zero at $(0,0,0)$, we deduce, by symbolic calculus for pseudo-differential operators, that
$$|\langle\Op_{h}^w(s_{h})\psi,\psi\rangle|=\mathcal{O}(h^\infty)\|\psi\|^2_{\sL^2(\R)}\,.$$
From the min-max principle, we infer that the $M$-th eigenvalue $\lambda_{M}(h)$ of $\Op_{h}^w(H)$ satisfies:
$$\lambda_{M}(h)\leq  \lambda_{M}(\mathcal{N}_{h})+\mathcal{O}(h^\infty)\,.$$
We leave the proof of the reverse inequality to the reader. The rest of the proof of the theorem easily follows by the functional calculus for self-adjoint operators.

\chapter{Semiclassical non linear magnetic eigenvalues}\label{chapter-non-linear}
\begin{flushright}
\begin{minipage}{0.4\textwidth}
Je pr\'ef\'erais t\^atonner dans le noir sans le secours de faibles lampes.
\begin{flushright}
\textit{M\'emoires d'Hadrien}, Yourcenar
\end{flushright}
\vspace*{0.5cm}
\end{minipage}
\end{flushright}

In this chapter, we illustrate the methods of Chapter \ref{chapter-models} by analyzing a non linear eigenvalue problem.

\section{About the concentration-compactness principle}\label{conc-comp}
This section is devoted to recall the famous concentration-compactness lemma.

\subsection{Concentration-compactness lemma}
Before stating the famous concentration-compactness lemma, let us establish an elementary lemma.
\begin{lem}[Helly's theorem]\label{lem-monotone}
Let $M>0$ and let us consider a sequence of non decreasing functions $(f_{n})_{n\in\N}\in[0,M]^\R$. Then, there exists a subsequence $(f_{n_{k}})_{k\in\N}$ such that  for all $x\in\R$, $(f_{n_{k}}(x))_{k\in\N}$ converges.
\end{lem}
\begin{proof}
Thanks to the Tychonov theorem, one knows that $[0,M]^\Q$ is compact and, since $\Q$ is countable, one also knows that the topology of $[0,M]^\Q$ is given by a distance. Therefore, by the Borel-Lebesgue theorem, the sequence $(f_{n})_{n\in\N}\in[0,M]^\Q$ (in fact its restriction to $\Q$) admits a converging subsequence $(f_{n_{k}})_{k\in\N}$. For $x\in\Q$, we let $\displaystyle{f(x)=\lim_{k\to+\infty} f_{n_{k}}(x)}$. Of course, $f : \Q\to\R$ is non decreasing. We let 
$$\mathcal{E}=\{x\in\R : \lim_{\Q\ni q\to x^-} f(q)= \lim_{\Q\ni q\to x^+} f(q)=\ell(x)\}\,.$$
We notice that $\complement\mathcal{E}$ is at most countable. Indeed, if $x\in\complement\mathcal{E}$, the exists $q_{x}\in\Q\cap \left(\lim_{\Q\ni q\to x^-} f(q), \lim_{\Q\ni q\to x^+} f(q)\right)$ and the application $\complement\mathcal{E}\ni x\mapsto q_{x}$ is injective (since $f$ is non decreasing). Thus, up to another subsequence extraction, we may assume that $(f_{n_{k}}(x))_{k\geq 0}$ converges for $x\in\Q\cup\complement\mathcal{E}$. Let us now analyze the convergence for $x\in\mathcal{E}$. For all $\eps>0$, there exists $\eta>0$ such that for all $t\in[x-\eta,x+\eta]\cap\Q$, we have $|f(y)-\ell(x)|\leq\eps$. The, if $(\alpha,\beta)\in\left([x-\eta,x]\times[x,x+\eta]\right)\bigcap\Q^2$, we have
$$f_{n_{k}}(\alpha)\leq f_{n_{k}}(x)\leq f_{n_{k}}(\beta)$$
so that, for $k$ large enough,
$$\ell(x)-2\eps\leq f(\alpha)-\eps\leq f_{n_{k}}(\alpha)\leq f_{n_{k}}(x)\leq f_{n_{k}}(\beta)\leq f(\beta)+\eps\leq \ell(x)+2\eps\,.$$
and thus $(f_{n_{k}}(x))_{k\in\N}$ converges to $\ell(x)$.
\end{proof}

\begin{lem}\label{conc-comp-lem}
Let $\lambda>0$ and a sequence of non negative and integrable functions $(\rho_{n})_{n\in\N}$ such that 
\begin{equation}\label{massn}
\int_{\R^N} \rho_{n}(x) \dx x=\lambda\,.
\end{equation}
We denote by $\mu_{n}$ the measure associated with the density $\rho_{n}$.
Then, there exists a subsequence such that one of the following holds:
\begin{enumerate}[(i)]
\item (vanishing) $\forall t>0$, $\displaystyle{\lim_{k\to+\infty} \sup_{y\in\R^N} \mu_{n_{k}}(\mathcal{B}(y,t))=0}$.
\item (compactness) $\exists (y_{k})_{k\in\N},\quad\forall \eps>0,\quad \exists R>0,\quad \mu_{n_{k}}(\mathcal{B}(y_{k},R))\geq \lambda-\eps.$
\item (dichotomy) $\exists\alpha\in(0,\lambda),\quad\forall\eps>0,\quad \exists k_{0}\geq 1,\quad \exists (\rho_{k}^1)_{k\in\N}, (\rho_{k}^2)_{k\in\N},\quad\forall k\geq k_{0}:$
$$\|\rho_{n_{k}}-\rho_{k}^1-\rho_{k}^2\|_{\sL^1}\leq \eps,\quad |\|\rho_{k}^1\|_{\sL^1}-\alpha|\leq \eps\,,$$
and with $\dist(\supp(\rho_{k}^1),\supp(\rho_{k}^2))=+\infty$.
\end{enumerate}
\end{lem}
\begin{proof}
Let us introduce the \enquote{concentration} functions:
$$Q_{n}(t)=\sup_{y\in\R^N} \mu_{n}(\mathcal{B}(y,t))\,.$$
The functions $Q_{n}$ are non negative and non decreasing and, $\forall t\geq 0$, $Q_{n}(t)\leq \mu_{n}(\R^N)=\lambda$. Note that $\lim_{t\to+\infty} Q_{n}(t)=\lambda$. Thus, $Q_{n}(t)$ goes from $0$ to $\lambda$ when t goes from $0$ to $+\infty$. We may use Lemma \ref{lem-monotone} and find a subsequence such that $Q_{n_{k}}(t)$ converges to $Q(t)$ when $k\to+\infty$. The function $Q$ is still non negative, non decreasing and bounded by $\lambda$. Therefore we may define
$$\alpha=\lim_{k\to+\infty} Q(t)\in[0,\lambda]\,.$$
\begin{enumerate}[(i)]
\item If $\alpha=0$, then $Q=0$.
\item Assume that $\alpha=\lambda$. For all $\mu\in\left[\frac{\lambda}{2},\lambda\right)$, there exists $t_{\mu}>0$, such that, for $k\geq 1$, we have
$$Q_{n_{k}}(t_{\mu})>\mu\,.$$
Indeed, we have the existence of $\tilde t_{\mu}>0$ such that $Q(\tilde t_{\mu})>\mu$ so that there exists $k_{0}\geq 1$ such that for $k\geq k_{0}$, $Q_{n_{k}}(\tilde t_{\mu})>\mu$. Furthermore, there exists $t'>0$ such that, for $k\in\{1,\ldots, k_{0}\}$, $Q_{n_{k}}(t')>\mu$. We take $t_{\mu}=\max(\tilde t_{\mu},t')$.

We get the existence of $(y_{k}(\mu))$ such that 
$$\mu_{n_{k}}(\mathcal{B}(y_{k}(\mu),t_{\mu}))>\mu\,.$$
Now, for $\mu>\frac{\lambda}{2}$, we notice that 
$$\mathcal{B}(y_{k}(\mu),t_{\mu})\cap \mathcal{B}\left(y_{k}\left(\frac{\lambda}{2}\right),t_{\frac{\lambda}{2}}\right)\neq \emptyset\,.$$
Indeed, if these balls were disjoint, the total mass of $\mu_{n_{k}}$ would exceed $\lambda$. This implies that, for all $\mu>\frac{\lambda}{2}$,
$$\mu_{n_{k}}\left( \mathcal{B}\left(y_{k}\left(\frac{\lambda}{2}\right),t_{\frac{\lambda}{2}}+2t_{\mu}\right)\right)>\mu\,.$$
\item Let us finally assume that $\alpha\in(0,\lambda)$. Let $\eps>0$ and $t>0$ such that $\alpha-\eps<Q(t)\leq\alpha$. We get, for $k\geq k_{0}$, 
$$\alpha-\eps<Q_{n_{k}}(t)<\alpha+\eps$$
and thus a sequence $(y_{k})$ such that 
$$\alpha-\eps<\mu_{n_{k}}\left(\mathcal{B}(y_{k},t)\right)<\alpha+\eps\,.$$
We may find a sequence $(T_{k})$ tending to $+\infty$ and such that
$$\mu_{n_{k}}\left(\mathcal{B}(y_{k},T_{k})\right)\leq Q_{n_{k}}(T_{k})\leq \alpha+\eps\,.$$
Indeed, we may define
$$T_{k}=\sup\{t\geq 0 : Q_{n_{k}}(t)\leq \alpha+\eps\}$$
and, if $(T_{k})$ has a converging subsequence, it is bounded by $T$ and for $t\geq T$, and $k\geq 1$, $Q_{n_{k}}(t)>\alpha+\eps$. This is a contradiction when $k$ goes to $+\infty$.

We define
$$\rho_{k}^1=\rho_{n_{k}}\mathds{1}_{\mathcal{B}(y_{k},t)},\quad \rho_{k}^2=\rho_{n_{k}}\mathds{1}_{\complement\mathcal{B}(y_{k},T_{k})}$$
and a straightforward computation gives
$$\|\rho_{n_{k}}-\rho_{k}^1-\rho_{k}^2\|_{\sL^1}=\mu_{n_{k}}\left(\mathcal{B}(y_{k},T_{k})\right)-\mu_{n_{k}}\left(\mathcal{B}(y_{k},t)\right)\leq 2\eps\,.$$
\end{enumerate}

\end{proof}

\begin{rem}
In Lemma \ref{conc-comp-lem}, we can replace \eqref{massn} by
$$\int_{\R^N} \rho_{n}(x) \dx x\underset{n\to+\infty}{\to}\lambda\,.$$
\end{rem}

\subsection{Application of the principle}
Let us now prove Proposition \ref{prop.inf=min} (we leave the case $p=2$ as an exercise). 

In order to prove the proposition, we could use, as in \cite{EL89}, the concentration compactness lemma. Nevertheless, we will use here a slightly more elementary point of view (even if we will recognize the concentration-compactness alternative in the proof!) that was suggested to the author by L. Le Treust.

\subsubsection{Excluding vanishing}
Let us start with a useful lemma. Let us introduce
\[M(\psi)=\sup_{\kb\in\Z^d}\|\psi\|_{\sL^2(\Omega_{\kb})} \mbox{ with }\Omega_{\kb}=[0,1]^2+\kb\,,\]
that is well-defined for $\psi\in\sL^2(\R^d)$.

In what follows, we will assume that $d\geq 3$ (and we leave to the reader the easy adapations to deal with the case $d=2$). Let $C_{0}>0$ be the optimal Sobolev constant for the critical embedding
\[\|\psi\|_{\sL^{2^*}([0,1]^d)}\leq C_{0}\|\nabla\psi\|_{\sL^2([0,1]^d)}\,.\]

\begin{lem}\label{lem.vanM}
We let $q_{0}=2+\frac{4}{d}\in\left(2,2^*\right)$. We have
\begin{equation}\label{eq.q0}
\forall\psi\in\sH^1(\R^d)\,,\qquad\|\psi\|^{q_{0}}_{\sL^{q_{0}}(\R^d)}\leq C_{0}^2M(\psi)^{\frac{4}{d}}\|\nabla\psi\|^2_{\sL^2(\R^d)}\,.
\end{equation}
Then, we have the following estimates.
\begin{enumerate}[(i)]
\item For $q\in(2,q_{0})$, we have,
\begin{equation}
\|\psi\|_{\sL^q(\R^d)}\leq C_{0}^{\frac{2(1-\theta)}{q_{0}}}\|\psi\|^\theta_{\sL^2(\R^d)}M(\psi)^{\frac{4(1-\theta)}{q_{0}d}}\|\nabla\psi\|^{\frac{2(1-\theta)}{q_{0}}}_{\sL^2(\R^d)}\,,\mbox{ with } \frac{1}{q}=\frac{\theta}{2}+\frac{1-\theta}{q_{0}}\,.
\end{equation}
\item For $q\in(q_{0}, 2^*)$, we have
\begin{equation}
\|\psi\|_{\sL^q(\R^d)}\leq C_{0}^{1-\theta+\frac{2\theta}{q_{0}}}M(\psi)^{\frac{4\theta}{q_{0}d}}\|\nabla\psi\|^{1-\theta+\frac{2\theta}{q_{0}}}_{\sL^2(\R^d)}\,,\mbox{ with } \frac{1}{q}=\frac{\theta}{q_{0}}+\frac{1-\theta}{2^*}\,.
\end{equation}
\end{enumerate}
\end{lem}

\begin{proof}
Let $q\in(2,2^*)$. We have
\[\|\psi\|^q_{\sL^q(\R^d)}=\sum_{\kb\in\Z^2}\int_{\Omega_{\kb}} |\psi|^q \dx x\,.\]
We notice that, by interpolation,
\[ \|\psi\|_{\sL^q(\Omega_{\kb})}^q\leq \|\psi\|^{\theta q}_{\sL^2(\Omega_{\kb})}\|\psi\|^{(1-\theta)q}_{\sL^{2^*}(\Omega_{\kb})}\,,\]
where $\frac{1}{q}=\frac{\theta}{2}+\frac{1-\theta}{2^*}$ or $\theta=1-d\left(\frac{1}{2}-\frac{1}{q}\right)\in(0,1)$. By Sobolev embedding, we get
\[\|\psi\|_{\sL^{2^*}(\Omega_{\kb})}\leq C_{0}\|\nabla\psi\|_{\sL^2(\Omega_{\kb})}\,,\]
where $C_{0}$ is the Sobolev constant associated with $k=0$. We deduce
\[ \|\psi\|_{\sL^q(\Omega_{\kb})}^q\leq C_{0}^{(1-\theta)q}\|\psi\|^{\theta q}_{\sL^2(\Omega_{\kb})}\|\nabla\psi\|^{(1-\theta)q}_{\sL^{2}(\Omega_{\kb})}\,.\]
Then, we look for $q\in(2,2^*)$ such that $(1-\theta)q=2$. We get $q=q_{0}$. The corresponding $\theta$ is $\theta_{0}=\frac{2}{d+2}$. We find \eqref{eq.q0}. The last two estimates follow from an interpolation argument.
\end{proof}
Let us consider $(\psi_{j})_{j\geq 1}$ a minimizing sequence of \eqref{lambdapA} such that $\|\psi_{j}\|_{\sL^p(\R^d)}=1$. Thanks to the diamagnetic inequality, we find that $(|\psi_{j}|)_{j\geq 1}$ is bounded in $\sH^1(\R^d)$. Taking $q=p$ in Lemma \ref{lem.vanM}, we find that
\[\liminf_{j\to+\infty} M(\psi_{j})=m>0\,.\]
Indeed, if not, the normalization of $(\psi_{j})$ in $\sL^p$ would lead to a contradiction. Therefore we may assume that $(\psi_{j})$ is such that $(M(\psi_{j}))$ is larger than $\frac{m}{2}>0$. Thus, there exists $(\tau_{j})\subset\Z^d$ such that, for all $j\geq 1$,
\[\|\psi_{j}(\cdot-\tau_{j})\|_{\sL^p([0,1]^d)}\geq \frac{m}{2}>0\,.\]
We let 
\[\varphi_{j}(\x)=e^{-i\A(\tau_{j})\cdot\x}\psi_{j}(\x-\tau_{j})\,,\]
so that ($\A$ is linear):
\[(-i\nabla+\A)\varphi_{j}=e^{-i\A(\tau_{j})\cdot\x}(-i\nabla+\A(\x-\tau_{j}))\psi_{j}(x-\tau_{j})\,.\]
Thus $(\varphi_{j})_{j\geq 1}$ is also a minimizing sequence and it satisfies
\[\|\varphi_{j}\|_{\sL^2([0,1]^d)}\geq\frac{m}{2}>0\,.\]
Up to another subsequence extraction, we may assume that $\varphi_{j}$ weakly converges to $\varphi$ in $\sH^1_{\A}(\R^d)$ (and also pointwise). Therefore, since $\sH^1_{\A}([0,1]^d)$ is compactly embedded in $\sL^2([0,1]^d)$, we have strong convergence in $\sL^2([0,1]^d)$ and
\[\|\varphi\|_{\sL^2([0,1]^d)}\geq\frac{m}{2}>0\,.\]
In particular, the function $\varphi\in\sH^1_{\A}(\R^d)$ is not zero. By the Fatou lemma, we have also $\|\varphi\|_{\sL^p(\R^d)}\leq 1$.

\subsubsection{Excluding dichotomy}
We introduce $\psi_{j}=\varphi_{j}-\varphi$ that weakly converges to $0$ in $\sH^1_{\A}(\R^d)$. We have
\[\mathfrak{Q}_{\A}(\varphi_{j})=\mathfrak{Q}_{\A}(\psi_{j})+\mathfrak{Q}_{\A}(\varphi)+2\Re \mathfrak{B}_{\A}(\psi_{j},\varphi)\,,\]
where $\mathfrak{B}_{\A}$ is the sesquilinear form associated with $\mathfrak{Q}_{\A}$. Since $\psi_{j}$ weakly converges to $0$ in $\sH^1_{\A}(\R^d)$, we deduce that $\mathfrak{B}_{\A}(\psi_{j},\varphi)\to 0$. In other words, we can write
\begin{equation}\label{eq.splitQA}
\mathfrak{Q}_{\A}(\varphi_{j})=\mathfrak{Q}_{\A}(\psi_{j})+\mathfrak{Q}_{\A}(\varphi)+\eps_{j}\,,
\end{equation}
with $\eps_{j}\to 0$.

We must prove that the $\sL^p$ norm also splits into two parts:
\begin{equation}\label{eq.splitLp}
\|\varphi_{j}-\varphi\|^p_{\sL^p(\R^d)}+\|\varphi\|^p_{\sL^p(\R^d)}-\|\varphi_{j}\|^p_{\sL^p(\R^d)}=\tilde\eps_{j}\to 0\,.
\end{equation}
Let us temporarily assume that \eqref{eq.splitLp} holds. Thanks to \eqref{eq.splitQA}, we have
\[\mathfrak{Q}_{\A}(\varphi_{j})\geq S\left(\|\psi_{j}\|^2_{\sL^p(\R^d)}+\|\varphi\|^2_{\sL^p(\R^d)}\right)+\eps_{j}\,,\]
and with \eqref{eq.splitLp}, we deduce that
\[\mathfrak{Q}_{\A}(\varphi_{j})\geq S\left(\left(1-\alpha+\tilde\eps_{j}\right)^{\frac{2}{p}}+\alpha^{\frac{2}{p}}\right)+\eps_{j}\,,\mbox{ with } \alpha=\|\varphi\|^p_{\sL^p(\R^d)}\in(0,1]\,.\]
Since $(\varphi_{j})_{j\geq 1}$ is a minimizing sequence, we get
\[S\geq S\left((1-\alpha)^{\frac{2}{p}}+\alpha^{\frac{2}{p}}\right)\,.\]
But we have $S>0$ so that
\[(1-\alpha)^{\frac{2}{p}}+\alpha^{\frac{2}{p}}\leq 1\,,\mbox{ with }\alpha\in(0,1]\,.\]
Since $p>2$ and by strict convexity, we must have $\alpha=1$. Therefore we conclude that $\|\varphi\|_{\sL^p(\R^d)}=1$. Finally it remains to notice that
\[S=\liminf_{j\to+\infty}\mathfrak{Q}_{\A}(\varphi_{j})\geq \mathfrak{Q}_{\A}(\varphi)\geq S\|\varphi\|^2_{\sL^p(\R^d)}=S\,,\]
and thus $\varphi$ is a minimizer. This achieves the proof of Proposition \ref{prop.inf=min}, modulo the proof of \eqref{eq.splitLp}. For that purpose we write
\[\tilde\eps_{j}:=\int_{\R^d} |\varphi_{j}-\varphi|^p-|\varphi_{j}|^p+|\varphi|^p\dx\x\,.\]
To see that $(\tilde\eps_{j})_{j\geq 1}$ tends to zero, we provide an argument slightly more elementary than the one by Struwe in \cite[p. 38]{Struwe08}.

Let us prove that the family $(|\varphi_{j}-\varphi|^p-|\varphi_{j}|^p)_{j\geq 1}$ is equi-integrable on $\R^d$. There exists $C(p)>0$ such that,
\[\left||\varphi_{j}-\varphi|^p-|\varphi_{j}|^p\right|\leq C(p)(|\varphi_{j}|^{p-1}+|\varphi|^{p-1})|\varphi|\,.\]
For $R>0$, by the H\"older inequality, we get
\[\int_{|\x|\geq R} |\varphi_{j}|^{p-1}|\varphi|\dx\x\leq\left(\int_{|\x|\geq R}|\varphi_{j}|^p\dx\x\right)^{\frac{p-1}{p}}\left(\int_{|\x|\geq R}|\varphi|^p\right)^{\frac{1}{p}}\leq\left(\int_{|x|\geq R}|\varphi|^p\right)^{\frac{1}{p}} \,.\]
Thus, for all $\eps>0$, there exists $R>0$, such that for all $j\geq 1$, we have
\[\left|\int_{|\x|\geq R} |\varphi_{j}-\varphi|^p-|\varphi_{j}|^p+|\varphi|^p\dx\x\right|\leq\frac{\eps}{2}\,.\]
Moreover, the embedding $\sH^1(\mathcal{B}(0,R))\subset \sL^p(\mathcal{B}(0,R))$ is compact so that $(\varphi_{j})_{j\geq 1}$ strongly converges to $\varphi$ in $\sL^p(\mathcal{B}(0,R))$ and thus, for $j\geq j(R,\eps)$,
\[\left|\int_{|\x|\leq R} |\varphi_{j}-\varphi|^p-|\varphi_{j}|^p+|\varphi|^p\dx\x\right|\leq\frac{\eps}{2},.\]
This implies that $|\tilde\eps_{j}|\leq \eps$.

\subsection{Exponential decay}
Let us now give the proof of Proposition \ref{prop.expdecnl}. This is a consequence of the following proposition.
\begin{prop}
For all $p\in(2,2^*)$,  there exists $\alpha>0$ such that for any minimizer $\psi$ of \eqref{lambdapA}, we have $e^{\alpha|\x|}\psi\in\sL^2(\R^d)$.
\end{prop}
\begin{proof}
If $\psi$ is an $\sL^p$-normalized minimizer, it satisfies the Euler-Lagrange equation:
\[(-ih\nabla+\A)^2\psi=S|\psi|^{p-2}\psi\,,\]
that can be rewritten as
\[(-ih\nabla+\A)^2\psi+V\psi=0\,,\mbox{ with }V=-S|\psi|^{p-2}\,,\]
or
\[\mathfrak{L}_{h,\A,V}\psi=0\,.\]
It remains to apply Proposition \ref{prop.elecnl} and \ref{prop.APE}.
\end{proof}

\section{Proof of the non linear semiclassical asymptotics}\label{sec.nla}
This section is devoted to the proof of Theorem \ref{theo1}.
\subsection{Upper bound}
Let us consider $v$ a minimizer associated with \eqref{lambda0} for $k=0$ and let
$$\psi(\x)=h^{-\frac{1}{p}}e^{i\frac{\phi(\x)}{h}}\chi(\x)v\left(\frac{\x-\x_{0}}{h^{\frac{1}{2}}}\right)\,,$$
where $\x_{0}$ denotes a point in $\Omega$ where the minimum of the magnetic field is obtained and where $\phi$ is a real function such that $\tilde \A=\A+\nabla \phi$
satisfies in a fixed neighborhood of $\x_{0}$:
\[\left|\tilde \A(\x)-b_{0}\tilde\A^{[0]}(\x)\right|\leq C |\x-\x_{0}|^2,\qquad \tilde\A^{[0]}(\x)=\frac{1}{2} e_{3}\times (\x-\x_{0})\,.\]
The existence of such a $\phi$ comes from Lemma \ref{lem.Poincare}. Note that $C$ only depend on the magnetic field. We have 
$$\int_{\Omega} |\psi(\x)|^p\dx \x=\int_{\R^2} |v(\y)|^p\dx \y$$
and
$$\mathfrak{Q}_{h,\A}(\psi)=h^{-\frac{2}{p}}\int_{\Omega} \left|\left(-ih\nabla+\tilde\A\right)v\left(\frac{\x-\x_{0}}{h^{\frac{1}{2}}}\right)\right|^2\dx \x$$
so that, for all $\eps>0$,
\begin{multline*}
 h^{\frac{2}{p}}\mathfrak{Q}_{h,\A}(\psi)\leq(1+\eps)\int_{\Omega} \left|\left(-ih\nabla+b_{0}\tilde\A^{[0]}\right)v\left(\frac{\x-\x_{0}}{h^{\frac{1}{2}}}\right)\right|^2\dx \x\\
+(1+\eps^{-1})\int_{\Omega} \left|\left(\tilde\A-b_{0}\tilde\A^{[0]}\right)v\left(\frac{\x-\x_{0}}{h^{\frac{1}{2}}}\right)\right|^2\dx \x\,.
\end{multline*}
Due to the exponential decay of $v$, 
we have 
$$\int_{\R^2} |\y|^4 |v(\y)|^2\dx \y<+\infty\,,$$
and thus
\begin{multline*}
 h^{\frac{2}{p}}\mathfrak{Q}_{h,\A}(\psi)\leq(1+\eps)h^2\int_{\R^2} \left|\left(-i\nabla+b_{0}\tilde\A^{[0]}\right)v(\y)\right|^2\dx \y\\
+C^2(1+\eps^{-1})h^3\int_{\R^2} |v(\y)|^2\dx \x.
\end{multline*}
We have:
$$\int_{\R^2} \left|\left(-i\nabla+b_{0}\tilde\A^{[0]}\right)v(\y)\right|^2\dx \y\geq b_{0}\int_{\R^2} |v(\y)|^2 \dx \y\,.$$
We deduce the upper bound:
\begin{multline*}
 h^{\frac{2}{p}}\mathfrak{Q}_{h,\A}(\psi)\leq\left((1+\eps)h^2+b_{0}^{-1}C^2(1+\eps^{-1})h^3\right)\int_{\R^2} \left|\left(-i\nabla+b_{0}\tilde\A^{[0]}\right)v(\y)\right|^2\dx \y\,.
\end{multline*}
We take $\eps=h^{1/2}$ so that,
$$ h^{\frac{2}{p}}\lambda(\Omega,\A,p,h)\leq\left(h^2+Ch^{5/2}\right)\frac{\int_{\R^2} \left|\left(-i\nabla+b_{0}\tilde\A^{[0]}\right)v(\y)\right|^2\dx \y}{\left(\int_{\R^2} |v(\y)|^p \dx \y\right)^{\frac{2}{p}}}\,.$$
We get
$$\lambda(\Omega,\A,p,h)\leq h^{-\frac{2}{p}}\left(h^2+Ch^{5/2}\right)\lambda(1,b_{0}\tilde\A^{[0]},p)\,.$$
By homogeneity and gauge invariance, we have
$$\lambda(\R^2, b_{0}\tilde\A^{[0]},p,1)=b_{0}^{\frac{2}{p}}\lambda(\R^2,\A^{[0]},p,1)\,.$$
We infer the upper bound
$$\lambda(\Omega,\A,p,h)\leq h^{-\frac{2}{p}} \left(b_{0}^{\frac{2}{p}}h^2\lambda(\R^2,\A^{[0]},p,1)+Ch^{\frac{5}{2}}\right)\,.$$

\subsection{Lower bound}

\subsubsection{Semiclassical localization formula adapted to $\sL^p$-norms}
Let us introduce a quadratic partition of unity \enquote{with small interaction supports}.
\begin{lem}\label{lemma-partition}
Let us consider $E=\{(\alpha,\rho,h,\ell)\in(\R_{+})^3\times\Z^2 : \alpha\geq \rho\}$. There exists a family of smooth cutoff functions $(\chi^{[\ell]}_{\alpha,\rho,h})_{(\alpha,\rho,h,\ell)\in E}$ on $\R^2$ such that $0\leq \chi^{[\ell]}_{\alpha,\rho,h}\leq 1$, 
\begin{align*}
\chi^{[\ell]}_{\alpha,\rho,h}&=1, \quad \text{  on }\quad |\x-(2h^{\rho}+h^{\alpha})\ell|_{\infty}\leq h^{\rho}\,,\\
\chi_{\alpha,\rho,h}&=0, \quad\text{ on }\quad |\x-h^{\rho}\ell|_{\infty}\geq h^{\rho}+h^{\alpha}\,,
\end{align*} and such that
$$\sum_{\ell\in\Z^2}\left(\chi_{\alpha,\rho,h}^{[\ell]}\right)^2=1\,.$$
Moreover there exists $D>0$ such that, for all $h>0$,
\begin{equation}\label{partition-remainder}
\sum_{\ell\in\Z^2} |\nabla\chi_{\alpha,\rho,h}^{[\ell]}|^2\leq Dh^{-2\alpha}\,.
\end{equation}
\end{lem}

\begin{proof}
Let us consider $F=\{(\alpha,\rho,h)\in(\R_{+})^3 : \alpha\geq \rho\}$.
There exists a family of smooth cutoff functions of one real variable $(\chi_{\alpha,\rho,h})_{(\alpha,\rho,h)\in F}$ such that $0\leq \chi_{\alpha,\rho,h}\leq 1$, $\chi_{\alpha,\rho,h}=1$ on $|x|\leq h^{\rho} + \frac{1}{2} h^{\alpha}$ and $\chi_{\alpha,\rho,h}=0$ on $|x|\geq h^{\rho}+h^{\alpha}$, and such that for all $(\alpha,\rho)$ with $\alpha\geq\rho>0$, there exists $C>0$ such that for all $h>0$, $|\nabla\chi_{\alpha,\rho,h}|\leq Ch^{-\alpha}$.
Then, we define :
$$S_{\alpha,\rho,h}(x)=\sum_{\ell\in\Z^2}\chi^2_{\alpha,\rho, h}\big(x_{1}-(2h^{\rho}+h^{\alpha})\ell_{1}\big)\chi^2_{\alpha,\rho, h}\big(x_{2}-(2h^{\rho}+h^{\alpha})\ell_{2}\big)\,,$$
and we have 
$$\forall x\in\R^2,\qquad 1\leq S_{\alpha,\rho,h}(x)\leq 4\,.$$
We let
$$\chi_{\alpha,\rho,h}^{[\ell]}(x)=\frac{\chi_{\alpha,\rho, h}(x_{1}-(2h^{\rho}+h^{\alpha})\ell_{1})\chi_{\alpha,\rho, h}(x_{2}-(2h^{\rho}+h^{\alpha})\ell_{2})}{\sqrt{S_{\alpha,\rho,h}(x)}}\,,$$
which satisfies the wished estimates by standard arguments.
\end{proof}

Given a "grid" and a non negative and integrable function $f$, the following lemma states that, up to a translation of the net, the mass of $f$ carried by a slightly thickened grid is controlled by a slight fraction of the total mass of $f$.
\begin{lem}\label{translation}
For $r>0$ and $\delta>0$, we define the net $\Lambda_{r}=((r\Z)\times\R)\cup(\R\times (r\Z))$ and the thickened net
$$\Lambda_{r,\delta}=\{x\in\R^2 : \dist(x,\Lambda_{r})\leq \delta\}\,.$$
Let us consider a non negative function $f$ belonging to $\sL^1(\R^2)$. Then there exists $\tau(r,\delta,f)=\tau\in\R^2$ such that :
$$\int_{\Lambda_{r,\delta}+\tau} f(x)\dx x\leq\frac{3\delta}{r+2\delta}\int_{\R^2} f(x)\dx x\,.$$
\end{lem}

\begin{proof}
We let $\e=\frac{1}{\sqrt{2}}(1,1)$. We notice that
$$\sum_{j=0}^{\lfloor \frac{r}{2\delta}\rfloor+1}\int_{\Lambda_{r,\delta}+j\delta \e} f(x) \dx x=\int_{\R^2} g_{r,\delta}(x) f(x)\dx x,\qquad\mbox{ with } g_{r,\delta}(x)=\sum_{j=0}^{\lfloor \frac{r}{2\delta}\rfloor+1} \mathds{1}_{\Lambda_{\delta}+j\delta \e}(x)\,.$$
We have, for almost all $x$, $g_{r,\delta}(x)\leq 3$, so that we get
$$\sum_{j=0}^{\lfloor \frac{r}{2\delta}\rfloor+1}\int_{\Lambda_{r,\delta}+j\delta \e} f(x) \dx x\leq 3\int_{\R^2} f(x)\dx x\,.$$
Therefore, there exists $j\in\left\{0,\ldots,\lfloor \frac{r}{\delta}\rfloor+1\right\}$, such that
$$\int_{\Lambda_{r,\delta}+j\delta \e} f(x) \dx x\leq \frac{3}{\lfloor \frac{r}{2\delta}\rfloor+2}\int_{\R^2} f(x)$$
and the conclusion easily follows.
\end{proof}
We can now establish the following lemma which permits to recover the total $\sL^p$-norm from the local contributions defined by the quadratic partition of unity.
\begin{lem}\label{Lp-partition}
Let us consider the partition of unity $(\chi_{\alpha,\rho,h}^{[\ell]})$ defined in Lemma \ref{lemma-partition}, with $\alpha>\rho>0$. There exist $C>0$ and $h_{0}>0$ such that for all $\psi\in \sL^p(\Omega)$ and $h\in(0,h_{0})$, there exists $\tau_{\alpha,\rho,h,\psi}=\tau\in\R^2$ such that
$$\sum_{\ell} \int_{\Omega} |\tilde\chi_{\alpha,\rho,h}^{[\ell]}\psi(\x)|^p \dx \x\leq\int_{\Omega}|\psi(\x)|^p\dx \x\leq (1+Ch^{\alpha-\rho})\sum_{\ell} \int_{\Omega} |\tilde\chi_{\alpha,\rho,h}^{[\ell]}\psi(\x)|^p \dx \x\,,$$
with $\tilde\chi_{\alpha,\rho,h}^{[\ell]}(\x)=\tilde\chi_{\alpha,\rho,h}^{[\ell]}(\x-\tau)$.
Moreover, the translated partition $(\tilde\chi_{\alpha,\rho,h}^{[\ell]})$ still satisfies \eqref{partition-remainder}.
\end{lem}

\begin{proof}
The first inequality is obvious since the cutoff functions are bounded by $1$ and their squares sum to unity. For the second one, we write, for any translation $\tau$,
$$\int_{\Omega}|\psi(\x)|^p\dx \x=\sum_{\ell}\int_{\Omega}\left(\tilde\chi_{\alpha,\rho,\ell}^{[\ell]}\right)^p|\psi(\x)|^p\dx \x+\int_{\Omega}\varphi_{\alpha,\rho}(\x)|\psi(\x)|^p\dx \x\,,$$
where
$$\varphi_{\alpha,\rho}=\sum_{\ell}\left(\left(\tilde\chi_{\alpha,\rho,\ell}^{[\ell]}\right)^2-\left(\tilde\chi_{\alpha,\rho,\ell}^{[\ell]}\right)^p\right)\,.$$
The smooth function $\varphi_{\alpha,\rho}$ is supported on $\tau+\Lambda_{h^\rho+\frac{h^\alpha}{2},2h^{\alpha}}$ and 
$$\int_{\Omega}\varphi_{\alpha,\rho}(\x)|\psi(\x)|^p\dx \x\leq \int_{\tau+\Lambda_{h^\rho+\frac{h^\alpha}{2},2h^{\alpha}}}f(\x)\dx \x\,,$$
where $f(\x)=|\psi(\x)|^p$ for $\x\in\Omega$ and $f(\x)=0$ elsewhere. Thus, by Lemma \ref{translation}, we find $\tau$ such that
$$\int_{\Omega}\varphi_{\alpha,\rho}(\x)|\psi(\x)|^p\dx \x\leq Ch^{\alpha-\rho}\int_{\R^2} f(\x)\dx\x$$
and the conclusion easily follows. 
\end{proof}

\subsubsection{Lower bound}

Let us consider $\psi\in\Dom(\mathfrak{Q}_{h,\A})$. With the localization formula associated with the partition $(\tilde\chi_{\alpha,\rho,h}^{[\ell]})$ that is adapted to $\psi$, we infer
$$\mathfrak{Q}_{h,\A}(\psi)=\sum_{\ell}\mathfrak{Q}_{h,\A}(\tilde\chi_{\alpha,\rho,h}^{[\ell]}\psi) -h^2\sum_{\ell}\|\nabla\tilde\chi_{\alpha,\rho,h}^{[\ell]}\psi\|_{\sL^2(\Omega)}^2\,.$$
We have
\begin{equation}\label{IMS-lb}
\mathfrak{Q}_{h,\A}(\psi)\geq\sum_{\ell}\left(\mathfrak{Q}_{h,\A}(\tilde\chi_{\alpha,\rho,h}^{[\ell]}\psi) -Dh^{2-2\alpha}\|\tilde\chi_{\alpha,\rho,h}^{[\ell]}\psi\|_{\sL^2(\Omega)}^2\right)\,.
\end{equation}
By the min-max principle, we get
\begin{equation}\label{relative-control}
\lambda(\Omega,\A, 2, h)\|\tilde\chi_{\alpha,\rho,h}^{[\ell]}\psi\|_{\sL^2(\Omega)}^2\leq \mathfrak{Q}_{h,\A}(\tilde\chi_{\alpha,\rho,h}^{[\ell]}\psi)
\end{equation}
and we recall that (see \eqref{eq.helmo} and Exercise \ref{exo-minmag})
\begin{equation}\label{HM}
\lambda(\Omega,\A, 2, h)=b_{0}h+\mathcal{O}(h^{\frac{3}{2}})
\end{equation}
so that
$$\mathfrak{Q}_{h,\A}(\psi)\geq(1-Dh^{1-2\alpha})\sum_{\ell}\mathfrak{Q}_{h,\A}(\tilde\chi_{\alpha,\rho,h}^{[\ell]}\psi)\,.$$
Then, we bound the local energies from below. Thanks to support considerations, we have, modulo a local change of gauge,
$$\mathfrak{Q}_{h,\A}(\tilde\chi_{\alpha,\rho,h}^{[\ell]}\psi) \geq (1-\eps)\mathfrak{Q}_{h,b_{j}\A^{[1]}}(\tilde\chi_{\alpha,\rho,h}^{[\ell]}\psi) -C\eps^{-1}h^{4\rho}\|\tilde\chi_{\alpha,\rho,h}^{[\ell]}\psi\|_{\sL^2(\Omega)}^2$$
so that it follows, by using again \eqref{relative-control},
$$\mathfrak{Q}_{h,\A}(\tilde\chi_{\alpha,\rho,h}^{[\ell]}\psi) \geq (1-\eps-C\eps^{-1}h^{4\rho-1})\mathfrak{Q}_{h,b_{j}\A^{[1]}}(\tilde\chi_{\alpha,\rho,h}^{[\ell]}\psi)\,.$$
We take $\eps= h^{2\rho-\frac{1}{2}}$ and we deduce
\begin{equation}\label{lb}
\mathfrak{Q}_{h,\A}(\psi)\geq (1-Dh^{1-2\alpha}-Ch^{2\rho-\frac{1}{2}})\sum_{\ell}b^{2/p}_{\ell}h^2h^{-2/p}\lambda^{[0]}( p)\|\tilde\chi_{\alpha,\rho,h}^{[\ell]}\psi\|_{\sL^p(\Omega)}^2
\end{equation}
so that
$$\mathfrak{Q}_{h,\A}(\psi)\geq (1-Dh^{1-2\alpha}-Ch^{2\rho-\frac{1}{2}})b^{2/p}_{0}h^2h^{-2/p}\lambda^{[0]}( p)\sum_{\ell}\|\tilde\chi_{\alpha,\rho,h}^{[\ell]}\psi\|_{\sL^p(\Omega)}^2$$
Since $p\geq 2$, we have
$$\sum_{\ell}\|\tilde\chi_{\alpha,\rho,h}^{[\ell]}\psi\|_{\sL^p(\Omega)}^2\geq \left(\sum_{\ell}\int_{\Omega} |\tilde\chi_{\alpha,\rho,h}^{[\ell]}\psi|^p \dx \x\right)^{\frac{2}{p}}\,.$$
Using Lemma \ref{Lp-partition}, we infer
\begin{equation}\label{recol-Lp}
\sum_{\ell}\|\tilde\chi_{\alpha,\rho,h}^{[\ell]}\psi\|_{\sL^p(\Omega)}^2 \geq (1-\tilde Ch^{\alpha-\rho})\|\psi\|^2_{\sL^p(\Omega)}\,.
\end{equation}
Finally, we get
$$\mathfrak{Q}_{h,\A}(\psi)\geq (1-Dh^{1-2\alpha}-Ch^{2\rho-\frac{1}{2}})(1-\tilde Ch^{\alpha-\rho})b^{2/p}_{0}h^2h^{-2/p}\lambda^{[0]}(p )\|\psi\|^2_{\sL^p(\Omega)}\,.$$
Optimizing the remainders, we choose $1-2\alpha=2\rho-\frac{1}{2}=\alpha-\rho$ so that $\rho=\frac{5}{16}$ and $\alpha=\frac{7}{16}$ and
$$\mathfrak{Q}_{h,\A}(\psi)\geq (1-Ch^{\frac{1}{8}})b^{2/p}_{0}h^2h^{-2/p}\lambda^{[0]}( p)\|\psi\|^2_{\sL^p(\Omega)}\,.$$

\part{Spectral reductions}\label{Part.Spec.Red}

\chapter{Electric Born-Oppenheimer approximation}\label{chapter-BOE}

\begin{flushright}
\begin{minipage}{0.5\textwidth}
Le \textit{cogito} d'un r\^eveur cr\'ee son propre cosmos, un cosmos singulier, un cosmos bien \`a lui. Sa r\^everie est d\'erang\'ee, son cosmos est troubl\'e si le r\^eveur a la certitude que la r\^everie d'un autre oppose un monde \`a son propre monde.
\begin{flushright}
\textit{La flamme d'une chandelle}, Bachelard
\end{flushright}
\vspace*{0.5cm}
\end{minipage}
\end{flushright}

This chapter presents the main idea behind the electric Born-Oppenheimer approximation (see \cite{CDS81, Martinez89}). We prove Theorem \ref{BOelec}.

\section{Quasimodes}
Let us explain the main steps in the construction of quasimodes behind Theorem \ref{BOelec}.
We recall that
$$\mathcal{V}(s)u_{s}=\nu(s)u_{s}\,.$$
By using Feynman-Hellmann formulas (see Chapter \ref{chapter-models}, Section \ref{Sec:FH}), this is easy to prove that 
$$\langle \mathcal{V}'(s_{0})u_{s_{0}}, u_{s_{0}}\rangle=0\,,$$
$$\left(\mathcal{V}(s_{0})-\nu(s_{0})\right)\left(\frac{\dx}{\dx s}u_{s}\right)_{|s=s_{0}}=-\mathcal{V}'(s_{0})u_{s_{0}}$$
and
$$\left\langle \mathcal{V}'(s_{0})\left(\frac{\dx}{\dx s}u_{s}\right)_{|s=s_{0}}+\frac{\mathcal{V}''(s_{0})}{2}u_{s_{0}},u_{s_{0}} \right\rangle=\frac{\nu''(s_{0})}{2}\,.$$
\begin{notation}\label{notation-FH}
We let
$$v_{s_{0}}(\tau)=\left(\frac{d}{d s}u_{s}\right)_{|s=s_{0}},\quad w_{s_{0}}(\tau)=\left(\frac{d^2}{d s^2}u_{s}\right)_{|s=s_{0}}\,.$$
\end{notation}
As usual we begin with the construction of suitable quasimodes.
We perform the change of variables $s=s_{0}+h^{\frac{1}{2}}\sigma,\,\, t=\tau$ and, instead of $\mathfrak{H}_{h}$, we study
$$\mathcal{H}_{h}=hD_{\sigma}^2+\mathcal{V}(s_{0}+h^{\frac{1}{2}}\sigma)\,.$$
In terms of formal power series, we have:
$$\mathcal{H}_{h}=\mathcal{V}(s_{0})+h^{\frac{1}{2}}\sigma\mathcal{V}'(s_{0})+h\left(\sigma^2\frac{\mathcal{V}''(s_{0})}{2}+D_{\sigma}^2\right)+\ldots$$
We look for quasi-eigenpairs in the form
$$\lambda\sim \lambda_{0}+h^{\frac{1}{2}}\lambda_{1}+h\lambda_{2}+\ldots, \quad \psi\sim \psi_{0}+h^{\frac{1}{2}}\psi_{1}+h\psi_{2}+\ldots$$
We must solve:
$$\mathcal{V}(s_{0})\psi_{0}=\lambda_{0}\psi_{0}\,.$$
Therefore, we choose $\lambda_{0}=\nu(s_{0})$ and $\psi_{0}(\sigma,\tau)=u_{s_{0}}(\tau) f_{0}(\sigma).$

We now meet the following equation
$$(\mathcal{V}(s_{0})-\lambda_{0})\psi_{1}=(\lambda_{1}-\sigma\mathcal{V}'(s_{0}))\psi_{0}\,.$$
The Feynman-Hellmann formula jointly with the Fredholm alternative implies that $\lambda_{1}=0$ and that we can take
$$\psi_{1}(\sigma,\tau)=\sigma f_{0}(\sigma)v_{s_{0}}(\tau)+f_{1}(\sigma)u_{s_{0}}(\tau)\,.$$
The crucial equation is given by
\begin{equation}\label{crucial}
(\mathcal{V}(s_{0})-\nu(s_{0}))\psi_{2}=\lambda_{2}\psi_{0}-\sigma \mathcal{V}'(s_{0})\psi_{1}-\left(\sigma^2\frac{\mathcal{V}''(s_{0})}{2}+D_{\sigma}^2\right)\psi_{0}.
\end{equation}
The Fredholm alternative jointly with the Feynman-Hellmann formula provides
$$\left(D_{\sigma}^2+\frac{\nu''(s_{0})}{2}\sigma^2\right)f_{0}=\lambda_{2}f_{0}\,.$$
This obliges to choose 
$$\lambda_{2}\in\left\{(2n-1)h\sqrt{\frac{\nu''(s_{0})}{2}}\quad n\geq 1\right\}$$
and for $f_{0}$ the corresponding rescaled Hermite function. With these choices, we may find a unique solution $\psi_{2}^\perp(\sigma,\cdot)\in\Dom(\mathcal{V}(s_{0}))$ of \eqref{crucial} that is orthogonal to $u_{s_{0}}$ for each $\sigma$. Thus the solutions of \eqref{crucial} can written in the form
$$\psi_{2}(\sigma,\tau)=\psi^\perp_{2}(\sigma,\tau)+f_{2}(\sigma)u_{s_{0}}(\tau)\,.$$

\begin{exe}
This exercise aims at proving some properties of the quasimodes and to conclude the proof.
\begin{enumerate}
\item Prove that the construction can be continued at any order, at least formally.
\item Prove that there exists $\eps_{0}>0$ such that
$$\int_{\Omega} e^{2\eps_{0}|\tau|}|u_{s_{0}}(\tau)|^2\dx \tau,\quad \int_{\Omega} e^{2\eps_{0}|\tau|}|v_{s_{0}}(\tau)|^2\dx \tau\,,$$
are finite. One will use estimates of Agmon.
\item Show that the $f_{j}$ belong to $\mathcal{S}(\R)$ and that for all $j$ there exists $\eta_{j}>0$ such that
$$\int_{\R} e^{\eta_{j}|\sigma|}|f_{j}| \dx \sigma<+\infty\,.$$
Establish that, for all $j\geq 0$, there exists $\eps_{j}>0$ such that
$$\int_{\R\times\Omega} e^{\eps_{j}(|\sigma|+|\tau|)}|\psi_{j}|\dx \sigma \dx \tau<+\infty\,.$$
Show that the $\psi_{j}$ belong to the domain of $\mathcal{H}_{h}$. One will proceed by induction.
\item Conclude that, for all $J\geq 0$, the exist $h_{0}>0$ and $C>0$ such that, for $h\in(0, h_{0})$,
$$\left\|\left(\mathcal{H}_{h}-\sum_{j=0}^J h^{\frac{j}{2}}\lambda_{j}\right)\sum_{j=0}^J h^{\frac{j}{2}}\psi_{j}\right\|\leq Ch^{\frac{J+1}{2}}\,.$$

\end{enumerate}

\end{exe}

\section{Essential spectrum and Agmon estimates} 
Let us first state a localization estimate.
\begin{prop}
Under Assumption \ref{essential}, there exists $h_{0}>0, C>0, \eps_{0}>0$ such that,  for $h\in(0,h_{0})$, for all eigenpair $(\lambda,\psi)$ such that $\lambda\leq \nu(s_{0})+C_{0}h$, we have:
$$\int_{\R\times\Omega} e^{2\eps_{0}(|s|+|\tau|)}|\psi|^2\dx s\dx \tau\leq C\|\psi\|^2\,.$$
\end{prop}
\begin{proof}
It is a straightforward application of Proposition \ref{prop.APE}.
\end{proof}
We are now led to prove some localization behavior of the eigenfunctions associated with eigenvalues $\lambda$ such that $|\lambda-\nu(s_{0})|\leq C_{0}h$.
\begin{prop}
There exist $\eps_{0}, h_{0}, C>0$ such that for all eigenpair $(\lambda,\psi)$ such that $|\lambda-\nu(s_{0})|\leq C_{0}h$, we have:
$$\int_{\R\times\Omega} e^{2\eps_{0}h^{-1/2}|s|}|\psi|^2 \dx x \leq C\|\psi\|^2\,.$$
and:
$$\int_{\R\times\Omega} \left|h\dr_{s}\left(e^{\eps_{0}h^{-1/2}|s|}\psi\right)\right|^2\dx x\leq Ch\|\psi\|^2\,.$$
\end{prop}
\begin{proof}
Let us write an estimate of Agmon
$$\mathcal{Q}_{h}(e^{h^{-1/2}\eps_{0}|s|}\psi)-h\eps_{0}^2\|e^{h^{-1/2}\eps_{0}|s|}\psi\|^2=\lambda\|e^{h^{-1/2}\eps_{0}|s|}\psi\|^2\leq (\nu(s_{0})+C_{0}h)\|e^{h^{-1/2}\eps_{0}|s|}\psi\|^2\,.$$
But we notice that
$$\mathcal{Q}_{h}(e^{h^{-1/2}\eps_{0}|s|}\psi)\geq \int_{\R\times\Omega} h^2 \left|\dr_{s} \left(e^{h^{-1/2}\eps_{0}|s|}\psi\right)\right|^2+\nu(s)\left|\left(e^{h^{-1/2}\eps_{0}|s|}\psi\right)\right|^2\dx x$$
and this implies
$$\int_{\R\times\Omega} (\nu(s)-\nu(s_{0})-C_{0}h-\eps_{0}^2h)\left|\left(e^{h^{-1/2}\eps_{0}|s|}\psi\right)\right|^2\dx x\leq 0\,.$$
The conclusion follows from a slight adaptation of the proof of Proposition \ref{Agmon-1D}.
\end{proof}

\section{Projection argument}
As we have observed, it can be more convenient to study $\mathcal{H}_{h}$ instead of $\mathfrak{H}_{h}$. Let us introduce the Feshbach-Grushin projection (see \cite{Gru72}) on $u_{s_{0}}$:
$$\Pi_{0}\psi=\langle\psi,u_{s_{0}}\rangle_{\sL^2(\Omega)} u_{s_{0}}(\tau)\,.$$
We want to estimate the projection of the eigenfunctions associated with eigenvalues $\lambda$ such that $|\lambda-\nu(s_{0})|\leq C_{0}h$. For that purpose, let us introduce the quadratic form:
$$q_{s_{0}}(\psi)=\int_{\R\times\Omega} |\dr_{\tau}\psi|^2+V(s_{0},\tau)|\psi|^2\dx \sigma \dx \tau\,.$$
This quadratic form is associated with the operator: $\Id_{\sigma} \otimes \mathcal{V}(s_{0})$ whereas $\Pi_{0}$ is the projection on its first eigenspace.

\begin{prop}\label{BOE-projection}
There exist $C,h_{0}>0$ such that, for $h\in(0,h_{0})$, for all eigenfunction $\psi$ of $\mathcal{H}_{h}$ associated with $\lambda$ such that $\lambda\leq\nu(s_{0})+C_{0}h$,
\begin{equation}\label{ub-quadratic}
0\leq q_{s_{0}}(\psi)-\nu(s_{0})\|\psi\|^2\leq Ch^{\frac{1}{2}}\|\psi\|^2.
\end{equation}
Moreover, we have:
\begin{equation}\label{ub-qs0}
\|\psi-\Pi_{0}\psi\|+\|\dr_{\tau}(\psi-\Pi_{0}\psi)\|\leq Ch^{\frac{1}{4}}\|\psi\|\,.
\end{equation}
\end{prop}

\begin{proof}
The proof uses the spectral gap $\nu_{2}(s_{0})-\nu_{1}(s_{0})>0$. We write
\begin{equation}\label{ev-eq-quadra}
h\|\dr_{\sigma}\psi\|^2+\|\dr_{\tau}\psi\|^2+\int_{\R\times\Omega} V(s_{0}+h^{\frac{1}{2}}\sigma,\tau)|\psi|^2\dx s\dx \tau \leq (\lambda+C_{0}h)\|\psi\|^2\,.
\end{equation}
Using the fact that $V$ is a polynomial and the fact that, for $k,n\in\N$:
$$\int |\tau |^n |\sigma|^k |\psi|^2\dx \sigma\dx \tau\leq C\|\psi\|^2\,,$$
we get \eqref{ub-quadratic}.

We notice that:
$$q_{s_{0}}(\psi)-\nu(s_{0})\|\psi\|^2=q_{s_{0}}(\psi-\Pi_{0}\psi)-\nu(s_{0})\|\psi-\Pi_{0}\psi\|^2\,,$$
due to the fact that $\Pi_{0}\psi$ belongs to the kernel of $\Id_{\sigma} \otimes \mathcal{V}(s_{0})-\nu(s_{0})\Id$.
We observe then that:
$$q_{s_{0}}(\psi-\Pi_{0}\psi)-\nu(s_{0})\|\psi-\Pi_{0}\psi\|^2\geq\int_{\R} \int_{\Omega}|\dr_{\tau}(\psi-\Pi_{0}\psi)|^2+V(s_{0},\tau)|(\psi-\Pi_{0}\psi)|^2\dx \tau\dx \sigma\,.$$
Since for each $u$, we have $\langle\psi-\Pi_{0}\psi,u_{s_{0}}\rangle_{\sL^2(\Omega)}=0$, we have the lower bound (by using the min-max principle):
$$q_{s_{0}}(\psi-\Pi_{0}\psi)-\nu(s_{0})\|\psi-\Pi_{0}\psi\|^2\geq\int_{\R} (\nu_{2}(s_{0})-\nu(s_{0}))\int_{\Omega}|\psi-\Pi_{0}\psi|^2\dx \tau\dx \sigma\,.$$
The estimate \eqref{ub-qs0} follows.
\end{proof}

\begin{prop}
There exist $C,h_{0}>0$ such that, for $h\in(0,h_{0})$, for eigenfunction $\psi$ of $\mathcal{H}_{h}$ associated with $\lambda$ such that $\lambda\leq\nu(s_{0})+C_{0}h$,
\begin{equation}\label{BOE-estim1}
0\leq q_{s_{0}}(\sigma\psi)-\nu(s_{0})\|\sigma\psi\|^2\leq Ch^{\frac{1}{2}}\|\psi\|^2
\end{equation}
and
\begin{equation}\label{BOE-estim2}
0\leq q_{s_{0}}(\dr_{\sigma}\psi)-\nu(s_{0})\|\dr_{\sigma}\psi\|^2\leq Ch^{\frac{1}{4}}\|\psi\|^2\,.
\end{equation}
Moreover, we have:
$$\|\sigma(\psi-\Pi_{0})\psi\|+\|\sigma\dr_{\tau}(\psi- \Pi_{0}\psi)\|\leq Ch^{\frac{1}{4}}\|\psi\|$$
and
$$\|\dr_{\sigma}(\psi-\Pi_{0}\psi)\|+\|\dr_{\sigma}(\dr_{t}(\psi-\Pi_{0}\psi))\|\leq Ch^{\frac{1}{8}}\|\psi\|\,.$$
\end{prop}

\begin{proof}
Using the \enquote{IMS} formula, we get:
$$\mathcal{Q}_{h}(\sigma\psi)=\lambda\|\sigma\psi\|^2+h\|\psi\|^2\leq (\nu(s_{0})+C_{0}h)\|\sigma\psi\|^2+h\|\psi\|^2\,.$$
Using the estimates of Agmon, we find \eqref{BOE-estim1}.
Let us analyze the estimate with $\dr_{\sigma}$. We take the derivative with respect to $\sigma$ in the eigenvalue equation:
\begin{equation}
\left(hD_{\sigma}^2+D_{\tau}^2+V(s_{0}+h^{\frac{1}{2}}\sigma,\tau)\right)\dr_{\sigma}\psi=\lambda\dr_{\sigma}\psi+\left[V(s_{0}+h^{\frac{1}{2}}\sigma,\tau),\dr_{\sigma}\right]\psi\,.
\end{equation}
Taking the scalar product with $\dr_{\sigma}\psi$, using $\|\partial_{\sigma}\psi\|\leq C\|\psi\|$ (that comes from \eqref{ev-eq-quadra}) and the estimates of Agmon, we find
\begin{equation}\label{du-eigen-equation}
\mathcal{Q}_{h}(\dr_{\sigma}\psi)\leq (\nu(s_{0})+C_{0}h)\|\dr_{\sigma}\psi\|^2+Ch^{1/2}\|\psi\|^2
\end{equation}
and we deduce 
\begin{equation}\label{BOE-ds2}
\|\dr_{\sigma}^2\psi\|\leq Ch^{-1/4}\|\psi\|+C\|\dr_{\sigma}\psi\|\,. 
\end{equation}
Then we must estimate
$$\int_{\R\times\Omega} \left(V(s_{0}+h^{\frac{1}{2}}\sigma,\tau)-V(s_{0},\tau)\right)|\partial_{\sigma}\psi|^2\dx \sigma \dx\tau$$
and thus only terms in the form
$$\int_{\R\times\Omega} h^{\frac{k}{2}}\sigma^k \tau^\ell|\partial_{\sigma}\psi|^2\dx \sigma \dx\tau=h^{\frac{k}{2}}\langle \sigma^k \tau^\ell\partial_{\sigma}\psi,\partial_{\sigma}\psi\rangle,\qquad k\geq 1\,.$$
By integration by parts, we have 
\begin{align*}
h^{\frac{k}{2}}\langle \sigma^k \tau^\ell\partial_{\sigma}\psi,\partial_{\sigma}\psi\rangle&=-h^{\frac{k}{2}}\langle \partial_{\sigma}(\sigma^k \tau^\ell\partial_{\sigma}\psi),\psi\rangle\\
																    &=-h^{\frac{k}{2}}\langle \partial^2_{\sigma}\psi, \sigma^k \tau^\ell\psi\rangle-kh^{\frac{k}{2}}\langle\partial_{\sigma}\psi,\sigma^{k-1}\tau\psi\rangle\,.
\end{align*}
Thanks to the Cauchy-Schwarz inequality and the estimates of Agmon, we get
$$kh^{\frac{k}{2}}|\langle\partial_{\sigma}\psi,\sigma^{k-1}\tau\psi\rangle|\leq Ch^{\frac{1}{2}}\,.$$
Moreover, using \eqref{BOE-ds2}, we get in the sameway
$$h^{\frac{k}{2}}|\langle \partial^2_{\sigma}\psi, \sigma^k \tau^\ell\psi\rangle|\leq Ch^{\frac{1}{4}}\,.$$
This is enough to get \eqref{BOE-estim2}. The approximation results easily follow.
\end{proof}
We can now use our approximation results to reduce the investigation to a model operator in dimension one.
\section{Accurate lower bound}
For all $N\geq 1$, let us consider an orthonormal family of eigenfunctions $(\psi_{n,h})_{1\leq n\leq N}$ of $\mathcal{H}_{h}$ such that $\psi_{n,h}$ is associated with $\lambda_{n}(h)$. We consider the $N$ dimensional space defined by
$$\mathcal{E}_{N}(h)=\underset{1\leq n\leq N}{\spann} \psi_{n,h}\,.$$ 
It is rather easy to observe that, for $\psi\in\mathcal{E}_{N}(h)$:
$$\mathcal{Q}_{h}(\psi)\leq \lambda_{N}(h)\|\psi\|^2\,.$$
We are going to prove a lower bound of $\mathcal{Q}_{h}$ on $\mathcal{E}_{N}(h)$. 
We notice that:
$$\mathcal{Q}_{h}(\psi)\geq \int_{\R\times\Omega} h|\dr_{\sigma}\psi|^2+\nu(s_{0}+h^{\frac{1}{2}}\sigma)|\psi|^2\dx \sigma \dx \tau\,.$$
We have:
\begin{multline*}
\int h|\dr_{\sigma}\psi|^2+\nu(s_{0}+h^{\frac{1}{2}}\sigma)|\psi|^2\dx \sigma\dx t=\int_{|\sigma h^{1/2}|\leq\eps_{0}} h|\dr_{\sigma}\psi|^2+\nu(s_{0}+h^{\frac{1}{2}}\sigma)|\psi|^2\dx \sigma\dx \tau\\
+ \int_{|\sigma h^{\frac{1}{2}}|\geq \eps_{0}} h|\dr_{\sigma}\psi|^2+\nu(s_{0}+h^{\frac{1}{2}}\sigma)|\psi|^2\dx \sigma\dx \tau\,.
\end{multline*}
With the Taylor formula, we can write:
\begin{multline*}
\int_{|\sigma h^{1/2}|\leq\eps_{0}} h|\dr_{\sigma}\psi|^2+\nu(s_{0}+h^{\frac{1}{2}}\sigma)|\psi|^2\dx \sigma \dx \tau\geq\\ 
\int_{|\sigma h^{1/2}|\leq\eps_{0}} h|\dr_{\sigma}\psi|^2+\nu(s_{0})+h\frac{\nu''(s_{0})}{2}\sigma^2|\psi|^2\dx \sigma\dx \tau
-Ch^{\frac{3}{2}}\int_{|\sigma h^{\frac{1}{2}}|\leq\eps_{0}} |\sigma|^3|\psi|^2\dx \sigma \dx \tau\,.
\end{multline*}
Thus, the estimates of Agmon imply that
\begin{align*}
&\int_{|\sigma h^{\frac{1}{2}}|\leq\eps_{0}} h|\dr_{\sigma}\psi|^2+\nu(s_{0}+h^{\frac{1}{2}}\sigma)|\psi|^2\dx \sigma\dx t\\
&\geq \int_{|\sigma h^{\frac{1}{2}}|\leq\eps_{0}} h|\dr_{\sigma }\psi|^2+\nu(s_{0})|\psi|^2+h\frac{\nu''(s_{0})}{2}\sigma^2|\psi|^2\dx \sigma \dx \tau-Ch^{\frac{3}{2}}\|\psi\|^2\,.
\end{align*}
Using again the estimates of Agmon, we notice that
$$\int_{|\sigma h^{1/2}|\geq\eps_{0}} h|\dr_{\sigma}\psi|^2+\nu(s_{0})|\psi|^2+h\frac{\nu''(s_{0})}{2}\sigma^2|\psi|^2\dx \sigma\dx \tau=\mathcal{O}(h^{\infty})\|\psi\|^2\,.$$
It follows that
$$\mathcal{Q}_{h}(\psi)\geq \int_{\R\times\Omega} h|\dr_{\sigma}\psi|^2+\nu(s_{0})|\psi|^2+h\frac{\nu''(s_{0})}{2}\sigma^2|\psi|^2\dx \sigma\dx \tau-Ch^{\frac{3}{2}}\|\psi\|^2\,.$$
\begin{exe}
By using the approximation results, prove that
$$\mathcal{Q}_{h}(\psi)\geq \nu(s_{0})\|\psi\|^2+\int_{\R\times\Omega} h|\dr_{\sigma}\Pi_{0}\psi|^2+h\frac{\nu''(s_{0})}{2}\sigma^2|\Pi_{0}\psi|^2\dx \sigma\dx \tau+o(h)\|\psi\|^2\,.$$
\end{exe}
Thanks to the orthogonality of the $\psi_{n,h}$ with respect to the bilinear form associated with $\mathcal{Q}_{h}$, we get
$$\lambda_{N}(h)\|\psi\|^2\geq \mathcal{Q}_{h}(\psi)\geq \nu(s_{0})\|\psi\|^2+\int_{\R\times\Omega} h|\dr_{\sigma}\Pi_{0}\psi|^2+h\frac{\nu''(s_{0})}{2}\sigma^2|\Pi_{0}\psi|^2\dx \sigma\dx \tau+o(h)\|\psi\|^2\,.$$
This becomes
$$\int_{\R} h|\dr_{\sigma}\langle\psi,u_{s_{0}}\rangle|^2+h\frac{\nu''(s_{0})}{2}\sigma^2|\langle\psi,u_{s_{0}}\rangle|^2\dx \sigma \leq (\lambda_{N}(h)-\nu(s_{0})+o(h))\|\langle\psi,u_{s_{0}}\rangle\|^2_{\sL^2(\R_{\sigma})}\,.$$
Due to Proposition \ref{BOE-projection}, the space $\left\{\langle\psi,u_{s_{0}}\rangle,\quad\psi\in\mathcal{E}_{N}(h)\right\}$ is of dimension $N$. Thus, by the min-max principle, we deduce
$$\lambda_{N}(h)\geq \nu(s_{0})+(2N-1)h\left(\frac{\nu''(s_{0})}{2}\right)^{1/2}+o(h)\,.$$

\section{Examples}
Let us now give examples which can be treated as exercises.

\subsection{Lu-Pan/de Gennes operator}
Our first example (which comes from \cite{BDPR11} and \cite{Ray11b}) is the Neumann realization of the operator acting on $\sL^2(\R^2_{+}, \dx \zeta \dx \tau)$:
$$h^2 D_{\zeta}^2+D_{\tau}^2+(\zeta-\tau)^2\,,$$
where $\R^2_{+}=\{(\zeta,\tau)\in\R^2 : \tau>0\}.$
\subsection{Montgomery operator}
The second example (which is the core of \cite{DomRay12}) is the self-adjoint realization on $\sL^2(\dx \zeta \dx \tau)$ of
$$h^2 D_{\zeta}^2+D_{\tau}^2+\left(\zeta-\frac{\tau^2}{2}\right)^2\,.$$

\subsection{Popoff operator}
Our last example (which comes from \cite{PoRay12}) corresponds to the Neumann realization on $\sL^2(\mathcal{E}_{\alpha},\dx \zeta \dx z \dx \tau)$ of
$$h^{2}D_{\zeta}^2+D_{\tau}^2+D_{z}^2+(\zeta-\tau)^2\,.$$

\section{An alternative point of view}\label{sec.chapter-BOE-number}
This section is devoted to the proof of Theorem \ref{number}. We recall that we consider $\mathfrak{L}_{h}=h^2D_{s}^2+D_{t}^2+V(s,t)$ whose associated quadratic form is
$$\mathfrak{Q}_{h}(\psi)=\int_{\R^2} h^2|D_{s}\psi|^2+|D_{t}\psi|^2+V(s,t)|\psi|^2 \dx s \dx t\,.$$
We denote by $u_{s}$ the first normalized eigenfunction of $D_{t}^2+V(s,t)$ and we introduce the projections defined for $\psi\in \sL^2(\R^2)$ by
$$\Pi_{s}\psi(s,t)=\langle\psi,u_{s}\rangle_{\sL^2(\R_{t})} u_{s}(t),\quad \Pi^\perp_{s}\psi(s,t)=\psi(s,t)-\Pi_{s}\psi(s,t)\,.$$
\begin{lem}\label{op-red}
For all $\psi\in\Dom(\mathfrak{Q}_{h})$, the function $\Pi_{s}\psi$ belongs to $\Dom(\mathfrak{Q}_{h})$ and we have
$$\mathfrak{Q}_{h}(\Pi_{s}\psi)=\int_{\R_{s}} h^2|f'(s)|^2+(\nu_{1}(s)+h^2 R(s))|f(s)|^2  \dx s,\quad \mbox{ with } f(s)=\langle\psi, u_{s}\rangle_{\sL^2(\R_{t})}\,.$$
\end{lem}
\begin{proof}
It follows immediately that, for any $\psi\in \Dom(\mathfrak{Q}_{h})$,
\[ \partial_s \big(\Pi_{s}\psi) \ = \ f(s)\partial_s u_{s}(t) +f'(s)u_{s}(t)\in\sL^2(\R^2),\]
since $\sup_{s\in\R}  |f'(s)|\leq\sup_{s\in\R}  |\langle\psi, \partial_s  u_{s}\rangle_{\sL^2(\R_{t})}|+\sup_{s\in\R}  |\langle \partial_s \psi, u_{s}\rangle_{\sL^2(\R_{t})}|<\infty$, and
\[ \partial_t \big(\Pi_{s}\psi) \ = \ \langle\psi,u_{s}\rangle_{\sL^2(\R_{t})}\partial_t u_{s}(t) \in\sL^2(\R^2).\]
Thus one has $\Pi_{s}\psi\in \Dom(\mathfrak{Q}_{h})$, and the calculations thereafter are valid.

By definition, one has
\begin{align*}
&\mathfrak{Q}_{h}(\Pi_{s}\psi)\\
&=\int_{\R^2} h^2 |f(s)\partial_s u_{s}(t) +f'(s)u_{s}(t)|^2 +|f(s)|^2|\partial_t  u_{s}(t)|^2\dx s\dx t +\!\!\int_{\R^2}V(s,t)|f(s)|^2|u_{s}(t)|^2 \dx s,
\end{align*}
and thus
\begin{multline*}
\mathfrak{Q}_{h}(\Pi_{s}\psi)=\int_{\R_s} h^2 |f'(s)|^2+h^2|f(s)|^2\Vert \partial_s u_{s}(t)\Vert_{\sL^2(\R_t)}^2\dx s \\
+\int_{\R_s}|f(s)|^2\left(\int_{\R_t}|\partial_t u_{s}(t)|^2+V(s,t)|u_{s}(t)|^2 \dx t\right) \dx s\\
  =\int_{\R_{s}} h^2|f'(s)|^2\dx s+\int_{\R_{s}} h^2  |f(s)|^2R(s) \dx s+\int_{\R_{s}} \nu_{1}(s)|f(s)|^2  \dx s.
\end{multline*}
where we used Fubini's theorem, and the following properties on $u_{s}$:
\begin{enumerate}[(a)]
\item $\forall s\in \R$, $u_{s}$ is normalized in $\sL^2(\R_t)$, and in particular, 
\[ \langle u_{s},\partial_s u_{s}\rangle_{L^2(\R_{t})}=\frac{\dx}{\dx s}\langle u_{s},u_{s}\rangle_{\sL^2(\R_{t})} =0.\]
\item $\forall s\in\R$, one has $\mathfrak{q}_{s}( u_{s})=\mu_1(s)$.
\end{enumerate}
\end{proof}

\begin{exe}\label{BOE-Bil}
Prove that, for all $\psi_{1},\psi_{2}\in\Dom(\mathfrak{Q}_{h})$, we have
$$\mathfrak{B}_{h}(\Pi_{s}\psi_{1},\Pi_{s}\psi_{2})=\int_{\R_{s}} h^2f_{1}'(s)\overline{f'_{2}(s)}+(\nu_{1}(s)+h^2 R(s))f_{1}'(s)\overline{f'_{2}(s)}  \dx s\,,$$
where $f_{j}(s)=\langle\psi_{j}, u_{s}\rangle_{\sL^2(\R_{t})}$.
\end{exe}

\begin{prop}\label{dimensional-reduction-lb}
For all $\psi\in\Dom(\mathfrak{Q}_{h})$ and all $\eps\in(0,1)$, we have
\begin{multline*}
\mathfrak{Q}_{h}(\psi)\geq \int_{\R_s} (1-\eps)h^2|f'(s)|^2+\big(\nu_1(s)-4\eps^{-1} h^2 R(s)\big)|f(s)|^2 \dx s\\
+\int_{\R_s} (1-\eps)h^2\Vert\partial_{s}\Pi_{s}^\perp\psi\Vert_{\sL^2(\R_{t})}^2+\big(\nu_2(s)-4\eps^{-1} h^2 R(s)\big)\Vert\Pi_{s}^\perp\psi\Vert_{\sL^2(\R_t)}^2 \dx s\,.
\end{multline*}
\end{prop}
\begin{proof}
We write
\begin{equation}\label{dimensional-reduction-lb-1}
\mathfrak{Q}_{h}(\psi)=\int_{\R^2} h^2|D_{s}\psi|^2\dx s \dx t+\int_{\R_{s}}\mathfrak{q}_{s}(\psi_{s})\dx s\,.
\end{equation}
On the one hand, we have
\begin{equation}\label{dimensional-reduction-lb-2}
\mathfrak{q}_{s}(\psi_{s})=\mathfrak{q}_{s}(\Pi_{s}\psi)+\mathfrak{q}_{s}(\Pi_{s}^\perp\psi)\geq\nu_{1}(s)|f(s)|^2+\nu_{2}(s)\|\Pi_{s}^\perp\psi\|^2_{\sL^2(\R_{t})}\,.
\end{equation}
On the other hand, we get
$$\int_{\R^2} |\partial_{s}\psi|^2\dx s \dx t=\int_{\R^2} |\Pi_{s}\partial_{s}\psi|^2\dx s\dx t+\int_{\R^2} |\Pi^\perp_{s}\partial_{s}\psi|^2\dx s\dx t$$
But we have
$$[\Pi_{s},\partial_{s}]\psi=-\langle\psi,\partial_{s}u_{s}\rangle_{\sL^2(\R_{t})}u_{s}-\langle\psi,u_{s}\rangle_{\sL^2(\R_{t})}\partial_{s} u_{s}=-[\Pi_{s}^\perp,\partial_{s}]$$
and
$$\|[\Pi_{s},\partial_{s}]\psi\|^2=\|[\Pi^\perp_{s},\partial_{s}]\psi\|^2=\|\langle\psi,\partial_{s}u_{s}\rangle_{\sL^2(\R_{t})}u_{s}\|^2+\|\langle\psi,u_{s}\rangle_{\sL^2(\R_{t})}\partial_{s} u_{s}\|^2$$
so that
$$\|[\Pi_{s},\partial_{s}]\psi\|^2=\|[\Pi^\perp_{s},\partial_{s}]\psi\|^2\leq \int_{\R^2} R(s)|\psi|^2\dx s \dx t\,.$$
Writing $\Pi_{s}\partial_{s}\psi=\partial_{s}\Pi_{s}\psi+[\Pi_{s},\partial_{s}]\psi$, we get
\begin{multline*}
\int_{\R^2} |\partial_{s}\psi|^2\dx s \dx t\geq (1-\eps)\int_{\R_{s}} |f'(s)|^2\dx s+(1-\eps)\int_{\R_{s}} \|\partial_{s}\Pi_{s}^\perp\psi\|^2_{\sL^2(\R_{t})} \dx s\\
-4\eps^{-1}\int _{\R_{s}} R(s)|f(s)|^2\dx s-4\eps^{-1}\int _{\R_{s}} R(s)\|\Pi_{s}^\perp\psi\|_{\sL^2(\R_{t})}^2\dx s\,.
\end{multline*}
Combining this last estimate with \eqref{dimensional-reduction-lb-1} and \eqref{dimensional-reduction-lb-2}, the result follows.

\end{proof}

\begin{prop}\label{P.QvsQtens}
Let us consider the following quadratic form, defined on $\Dom(\mathfrak{q}^\mode_{h})\times\Dom(\mathfrak{Q}_{h})$, by
\begin{multline*}
\mathfrak{Q}_{h}^\tens(f,\varphi)=\\
(1-h)\int_{\R_{s}} h^2|f'(s)|^2+\big(\nu_1(s)-4M h\big)|f(s)|^2 \dx s+(1-h)\int_{\R^2} h^2|\partial_{s}\varphi|^2+\big(\nu_2(s)-4M h\big)|\varphi|^2 \dx s\dx t,\\
\quad \forall (f,\varphi)\in \Dom(\mathfrak{q}^\mode_{h})\times\Dom(\mathfrak{Q}_{h})\,.
\end{multline*}
If $\mathfrak{H}^\tens_{h}$ denotes the associated operator, then we have, for all $n\geq 1$
$$\lambda_{n}(h)\geq \lambda^\tens_{n}(h)\,.$$
\end{prop}

\begin{proof}
We use Proposition~\ref{dimensional-reduction-lb} with $\eps=h$ and we get, for all $\psi\in\Dom(\mathfrak{Q}_{h})$, 
\begin{multline*}
\mathfrak{Q}_{h}(\psi)\geq \int_{\R_{s}} (1-h)h^2|f'|^2+\big(\nu_1(x)-4M h\big)|f|^2 \dx s\\
+\int_{\R^2} (1-h)h^2|\partial_{s}\Pi_{s}^\perp\psi|^2+\big(\nu_2(s)-4M h\big)|\Pi_{s}^\perp\psi|^2 \dx s\dx t\,.
\end{multline*}
Thus we have
\begin{equation}\label{tensorisation}
\mathfrak{Q}_{h}(\psi)\geq \mathfrak{Q}^\tens_{h}(\langle\psi,u_{s}\rangle,\Pi_{s}^\perp\psi),\quad \Vert\psi\Vert^2=\Vert f\Vert^2+\Vert\Pi_{s}^\perp\psi\Vert^2\,.
\end{equation}
With \eqref{tensorisation} we infer
\[
   \lambda_n(h)\geq 
   \inf_{\substack{G\subset \sH^1(\R^2)\\ \dim G=n}} \ \sup_{\substack{
    \psi\in G}}   \frac{\mathfrak{Q}^\tens_{h}(\langle\psi, u_{s}\rangle,\Pi_{s}^\perp\psi)}{\Vert\Pi_{s}\psi\Vert^2+\Vert\Pi_{s}^\perp\psi\Vert^2} \,.
\]
Now, we define the linear injection
$$
\mathcal{J} : \left\{\begin{array}{ccc}
\Dom(\mathfrak{Q}_{h}) &\to& \Dom(\mathfrak{q}^\mode_{h})\times\Dom(\mathfrak{Q}_{h}) \\
\psi&\mapsto&(\langle\psi,u_{s}\rangle\ ,\,\Pi_{s}^\perp\psi)
\end{array}\right..
$$
so that we have
$$ \inf_{\substack{G\subset \Dom(\mathfrak{Q}_{h})\\ \dim G=n}} \ \sup_{\substack{
    \psi\in G}}   \frac{\mathfrak{Q}^\tens_{h}(\Pi_{s}\psi,\Pi_{s}^\perp\psi)}{\Vert\Pi_{s}\psi\Vert^2+\Vert\Pi_{s}^\perp\psi\Vert^2}=\inf_{\substack{\tilde G\subset \mathcal{J}(\Dom(\mathfrak{Q}_{h}))\\ \dim \tilde G=n}} \ \sup_{\substack{(f,\varphi)\in \tilde G}}   \frac{\mathfrak{Q}^\tens_{h}(f,\varphi)}{\Vert f\Vert^2+\Vert\varphi\Vert^2} \\,,$$
and
$$\inf_{\substack{\tilde G\subset \mathcal{J}(\Dom(\mathfrak{Q}_{h}))\\ \dim \tilde G=n}} \ \sup_{\substack{(f,\varphi)\in \tilde G}}   \frac{\mathfrak{Q}^\tens_{h}(f,\varphi)}{\Vert f\Vert^2+\Vert\varphi\Vert^2}\geq \inf_{\substack{\tilde G\subset \Dom(\mathfrak{q}^\mode_{h})\times\Dom(\mathfrak{Q}_{h}) \\ \dim \tilde G=n}} \ \sup_{\substack{(f,\varphi)\in \tilde G}}   \frac{\mathfrak{Q}^\tens_{h}(f,\varphi)}{\Vert f\Vert^2+\Vert\varphi\Vert^2}\,.$$
We recognize the $n$-th Rayleigh quotient of $\mathfrak{H}^\tens_{h}$ and the conclusion follows.
\end{proof}
We can now prove Theorem \ref{number}. Let us introduce the model quadratic forms
$$\mathfrak{q}^\mode_{h}(f)=\int_{\R} h^2|f'(s)|^2+\nu_{1}(s)|f(s)|^2  \dx s\,.$$
Thanks to Exercise \ref{BOE-Bil} and by using the eigenfunctions of the operator associated with $\mathfrak{q}^\mode_{h}$, we get
$$\mathsf{N}\left(\mathfrak{Q}_{h},E\right)\geq \mathsf{N}\left(\mathfrak{q}^\mode_{h},E-Mh^2\right)\,.$$
Conversely, we use the result of Proposition \ref{P.QvsQtens} to get 
$$\mathsf{N}\left(\mathfrak{Q}_{h},E\right)\leq\mathsf{N}\left(\mathfrak{Q}_{h}^\tens, E\right)\leq  \mathsf{N}\left(\mathfrak{q}^\mode_{h},(E+4Mh)(1-h)^{-1}\right)\,,$$
the last inequality coming from the fact that, when $h$ is small enough, there are no eigenvalues of $\mathfrak{H}_{h}^\tens$ below the threshold (by assumption in Theorem \ref{number}).

Therefore we are reduced to the estimate of the counting function of $\mathfrak{q}^\mode_{h}$ in one dimension and we apply Theorem \ref{number-1D}.

\chapter{Magnetic Born-Oppenheimer approximation}\label{chapter-BOM}

\begin{flushright}
\begin{minipage}{0.5\textwidth}
Pour l'ach\`evement de la science, il faut passer en revue une \`a une toutes les choses qui se rattachent \`a notre but par un mouvement de pens\'ee continu et sans nulle interruption, et il faut les embrasser dans une \'enum\'eration suffisante et m\'ethodique.
\begin{flushright}
\textit{R\`egles pour la direction de l'esprit}, Descartes
\end{flushright}
\vspace*{0.5cm}
\end{minipage}
\end{flushright}

We explain in this chapter the main steps to the proof of Theorem \ref{main-theorem-BOM}. In particular the reader is supposed to be familiar with the basics of pseudo-differential calculus. We establish general Feynman-Hellmann formulas and we also recall the fundamental properties of coherent states.

\section{Quasimodes}\label{Sec.quasimodes}
This section is devoted to the proof of the following proposition.
\begin{prop}\label{quasimodes-BOM}
Let us assume Assumption \ref{hyp-gen}. For all $n\geq 1$, there exist a sequence $(\gamma_{j,n})_{j\geq 0}$ such that for all $J\geq 0$ there exist $C>0$ and $h_{0}>0$ such that for $h\in(0,h_{0})$:
$$\dist\left(\sum_{j=0}^J \gamma_{j,n}h^{j/2},\sp(\mathfrak{L}_{h})\right)\leq Ch^{(J+1)/2}\,,$$
where:
$$\gamma_{0,n}=\mu_{0},\quad \gamma_{1,n}=0,\quad \gamma_{2,n}=\nu_{n}\left(\frac{1}{2}\Hess_{x_{0},\xi_{0}}\,\mu(\sigma,D_{\sigma})\right)\,.$$
\end{prop}
In order to perform the investigation we use the following rescaling:
$$s=h^{1/2}\sigma$$
so that $\mathfrak{L}_{h}$ becomes:
\begin{equation}\label{Lch}
\mathcal{L}_{h}=(-i\nabla_{\tau}+A_{2}(x_{0}+h^{1/2}\sigma,\tau))^2+(\xi_{0}-ih^{1/2}\nabla_{\sigma}+A_{1}(x_{0}+h^{1/2}\sigma,\tau))^2\,.
\end{equation}
We will also need generalizations of the Feynman-Hellmann formulas which are obtained by taking the derivative of the eigenvalue equation
$$\mathcal{M}_{x,\xi}u_{x,\xi}=\mu(x,\xi)u_{x,\xi}$$
with respect to $x_{j}$ and $\xi_{k}$.
\begin{prop}\label{FH}
We have:
\begin{equation}\label{FH1}
(\mathcal{M}_{x,\xi}-\mu(x,\xi))(\dr_{\eta} u)_{x,\xi}=(\dr_{\eta}\mu(x,\xi)-\dr_{\eta}\mathcal{M}_{x,\xi}) u_{x,\xi}
\end{equation}
and:
\begin{multline}\label{FH2}
(\mathcal{M}_{x_{0},\xi_{0}}-\mu_0)(\dr_{\eta}\dr_{\theta}u)_{x_{0},\xi_{0}}\\
=\dr_{\eta}\dr_{\theta}\mu(x_{0},\xi_{0})u_{x_{0},\xi_{0}}- \dr_{\eta}\mathcal{M}_{x_{0},\xi_{0}}(\dr_{\theta} u)_{x_{0},\xi_{0}}
-\dr_{\theta}\mathcal{M}_{x_{0},\xi_{0}}(\dr_{\eta} u)_{x_{0},\xi_{0}}
-\dr_{\eta}\dr_{\theta}\mathcal{M}_{x_{0},\xi_{0}}u_{x_{0},\xi_{0}},
\end{multline}
where $\eta$ and $\theta$ denote one of the $x_{j}$ or $\xi_{k}$. Moreover we have
\begin{equation}\label{FH2bis}
\dr_{\eta}\mu(x,\xi)=\int_{\R^n} \dr_{\eta}{\mathcal{M}_{x,\xi}}\, u_{x,\xi}(\tau) u_{x,\xi}(\tau)\dx\tau\,.
\end{equation}
\end{prop}
We can now prove Proposition \ref{quasimodes-BOM}. Since $A_{1}$ and $A_{2}$ are polynomials, we can write, for some $M\in\N$:
$$\mathcal{L}_{h}=\sum_{j=0}^M h^{j/2}\mathcal{L}_{j}$$
with:
$$\mathcal{L}_{0}=\mathcal{M}_{x_{0},\xi_{0}},\quad \mathcal{L}_{1}=\sum_{j=1}^m   (\dr_{x_{j}}\mathcal{M})_{x_{0},\xi_{0}}\sigma_{j}+\sum_{j=1}^m (\dr_{\xi_{j}}\mathcal{M})_{x_{0},\xi_{0}} D_{\sigma_{j}}\,,$$
\begin{multline*}
\mathcal{L}_{2}=\frac{1}{2}\sum_{k,j=1}^m\Big( (\dr_{x_{j}}\dr_{x_{k}}\mathcal{M})_{x_{0},\xi_{0}} \sigma_{j}\sigma_{k}+(\dr_{\xi_{j}}\dr_{\xi_{k}}\mathcal{M})_{x_{0},\xi_{0}} D_{\sigma_{j}}D_{\sigma_{k}}+(\dr_{\xi_{j}}\dr_{x_{k}}\mathcal{M})_{x_{0},\xi_{0}} D_{\sigma_{j}}\sigma_{k}\\ +(\dr_{x_{k}}\dr_{\xi_{j}}\mathcal{M})_{x_{0},\xi_{0}}\sigma_{k}D_{\sigma_{j}}\Big).
\end{multline*}
We look for quasimodes in the form:
$$\psi\sim\sum_{j\geq 0} h^{j/2}\psi_{j}$$
and quasi-eigenvalues in the form:
$$\gamma\sim\sum_{j\geq 0} h^{j/2}\gamma_{j}$$
so that they solve in the sense of formal series:
$$\mathcal{L}_{h}\psi\sim\gamma \psi\,.$$
By collecting the terms of order $h^0$, we get the equation:
$$\mathcal{M}_{x_{0},\xi_{0}}\psi_{0}=\gamma_{0}\psi_{0}\,.$$
This leads to take $\gamma_{0}=\mu_{0}$ and :
$$\psi_{0}(\sigma,\tau)=f_{0}(\sigma)u_{0}(\tau)\,,$$
where $u_{0}=u_{x_{0},\xi_{0}}$ and $f_{0}$ is a function to be determined in the Schwartz class.
By collecting the terms of order $h^{1/2}$, we find:
$$(\mathcal{M}_{x_{0},\xi_{0}}-\mu(x_{0},\xi_{0}))\psi_{1}=(\gamma_{1}-\mathcal{L}_{1})\psi_{0}\,.$$
By using \eqref{FH1} and the Fredholm alternative (applied for $\sigma$ fixed) we get $\gamma_{1}=0$
and the solution:
\begin{equation}\label{psi1}
\psi_{1}(\sigma,\tau)= \sum_{j=1}^m (\dr_{x_{j}}u)_{x_{0},\xi_{0}}\, \sigma_{j} f_{0}+\sum_{j=1}^m (\dr_{\xi_{j}}u)_{x_{0},\xi_{0}}\, D_{\sigma_{j}} f_{0} +f_{1}(\sigma) u_{0}(\tau)\,,
\end{equation}
where $f_{1}$ is a function to be determined in the Schwartz class.
The next equation reads:
$$(\mathcal{M}_{x_{0},\xi_{0}}-\mu(x_{0},\xi_{0}))\psi_{2}=(\gamma_{2}-\mathcal{L}_{2})\psi_{0}-\mathcal{L}_{1}\psi_{1}\,.$$
The Fredholm condition is:
\begin{equation}\label{Fredholm}
\langle\mathcal{L}_{2}\psi_{0}+\mathcal{L}_{1}\psi_{1},  u_{0}\rangle_{L^2(\R^n,\dx \tau)}=\gamma_{2}f_{0}\,.
\end{equation}
We obtain (exercise):
$$\frac{1}{2}\Hess\, \mu(x_{0},\xi_{0}) (\sigma,D_{\sigma})f_{0}=\gamma_{2} f_{0}\,.$$
We take $\gamma_{2}$ in the spectrum of $\frac{1}{2}\Hess\, \mu(x_{0},\xi_{0}) (\sigma,D_{\sigma})$ and we choose $f_{0}$ a corresponding normalized eigenfunction. The construction can be continued at any order.

We deduce from Propositions \ref{essential-BOM} and \ref{quasimodes-BOM}:
\begin{cor}\label{rub}
For all $n\geq 1$ there exist $h_{0}>0$ and $C>0$ such that for all $h\in(0,h_{0})$ the $n$-th eigenvalue of $\mathfrak{L}_{h}$ exists and satisfies:
$$\lambda_{n}(h)\leq \mu_{0}+Ch\,.$$
\end{cor}
\section{Rough estimates of the eigenfunctions}\label{micro}
This section is devoted to recall the basic and rough localization and microlocalization estimates satisfied by the eigenfunctions resulting from Assumptions  \ref{hyp-gen} and \ref{confining} and Corollary \ref{rub}. The following two propositions are applications of Proposition \ref{prop.APE}.
\begin{prop}\label{Agmon-tau}
Let $C_{0}>0$. There exist $h_{0},C, \eps_{0}>0$ such that for all eigenpairs $(\lambda,\psi)$ of $\mathfrak{L}_{h}$ such that $\lambda\leq\mu_0+C_{0}h$ we have:
$$\left\|e^{\eps_{0}|\tau|}\psi\right\|^2\leq C\|\psi\|^2,\quad\mathfrak{Q}_{h}\left(e^{\eps_{0}|\tau|}\psi\right)\leq C\|\psi\|^2\,.$$
\end{prop}

\begin{prop}\label{Agmon-s-BOM}
Let $C_{0}>0$. There exist $h_{0},C, \eps_{0}>0$ such that for all eigenpairs $(\lambda,\psi)$ of $\mathfrak{L}_{h}$ such that $\lambda\leq\mu_0+C_{0}h$, we have:
$$\left\|e^{\eps_{0}|s|}\psi\right\|^2\leq C\|\psi\|^2,\quad \mathfrak{Q}_{h}\left(e^{\eps_{0}|s|}\psi\right)\leq C\|\psi\|^2\,.$$
\end{prop}

We deduce from Propositions \ref{Agmon-tau} and \ref{Agmon-s-BOM} the following corollary.
\begin{cor}
Let $C_{0}>0$ and $k,l\in\N$. There exist $h_{0},C, \eps_{0}>0$ such that for all eigenpairs $(\lambda,\psi)$ of $\mathfrak{L}_{h}$ such that $\lambda\leq\mu_0+C_{0}h$, we have:
$$\|\tau^k s^l \psi\|\leq C\|\psi\|,\quad \mathfrak{Q}_{h}(\tau^k s^l \psi)\leq C\|\psi\|^2\,,$$
$$\|-i\nabla_{\tau} s^l \tau^k \psi \|\leq C\|\psi\|^2,\quad \|-ih\nabla_{s} s^l \tau^k \psi \|\leq C\|\psi\|^2\,.$$
\end{cor}
Taking successive derivatives of the eigenvalue equation we deduce by induction:
\begin{cor}\label{poly-bound}
Let $C_{0}>0$ and $k,l,p\in\N$. There exist $h_{0},C, \eps_{0}>0$ such that for all eigenpairs $(\lambda,\psi)$ of $\mathfrak{L}_{h}$ such that $\lambda\leq\mu_0+C_{0}h$  and all $h\in(0,h_{0})$, we have:
$$\|(-i\nabla_{\tau})^p s^l \tau^k \psi \|\leq C\|\psi\|^2,\quad \|(-ih\nabla_{s})^p s^l \tau^k \psi \|\leq C\|\psi\|^2\,.$$
\end{cor}
Using again Propositions \ref{Agmon-tau} and \ref{Agmon-s-BOM} and an induction argument we get:
\begin{prop}\label{loc-space}
Let $k\in\N$. Let $\eta>0$ and $\chi$ a smooth cutoff function being zero in a neighborhood of $0$.  There exists $h_{0}>0$ such that for all eigenpairs $(\lambda,\psi)$ of $\mathfrak{L}_{h}$ such that $\lambda\leq\mu_0+C_{0}h$ and all $h\in(0,h_{0})$, we have:
$$\|\chi(h^{\eta}s)\psi\|_{\sB^k(\R^{m+n})}\leq \mathcal{O}(h^{\infty})\|\psi\|, \quad \|\chi(h^{\eta}\tau)\psi\|_{\sB^k(\R^{m+n})}\leq \mathcal{O}(h^{\infty})\|\psi\|\,,$$
where $\|\cdot\|_{\sB^k(\R^{n+m})}$ is the standard norm on: 
$$\sB^k(\R^{m+n})=\{\psi\in L^2(\R^{m+n}) :  y_{j}^q \dr_{y_{l}}^p \psi\in \sL^2(\R^{n+m}),\forall j,k\in\{1,\cdots,m+n\},\, p+q\leq k\}\,.$$
\end{prop}

By using a rough pseudo-differential calculus jointly with the space localization of Proposition \ref{loc-space} and standard elliptic estimates, we get: 
\begin{prop}\label{loc-phase}
Let $k\in\N$. Let $\eta>0$ and $\chi$ a smooth cutoff function being zero in a neighborhood of $0$. There exists $h_{0}>0$ such that for all eigenpairs $(\lambda,\psi)$ of $\mathfrak{L}_{h}$ such that $\lambda\leq\mu_0+C_{0}h$, we have:
$$\|\chi(h^\eta hD_{s})\psi\|_{\sB^k(\R^{m+n})}\leq \mathcal{O}(h^{\infty})\|\psi\|,\quad \|\chi(h^{\eta}D_{\tau})\psi\|_{\sB^k(\R^{m+n})}\leq \mathcal{O}(h^{\infty})\|\psi\|\,.$$
\end{prop}

\section{Coherent states and microlocalization}\label{coherent}
\subsection{A first lower bound}
By using the formalism introduced in Chapter \ref{intro-models}, Section \ref{intro-coherent}, we get the following proposition.
\begin{prop}
 There exist $h_{0},C>0$ such that for all eigenpairs $(\lambda,\psi)$ of $\mathcal{L}_{h}$ such that $\lambda\leq\mu_0+C_{0}h$  and all $h\in(0,h_{0})$, we have
\begin{equation}\label{rlb-cs}
\mathcal{Q}_{h}(\psi)\geq \int_{\R^{2m}} Q_{h,u,p}(\psi_{u,p})\dx u \dx p-Ch\|\psi\|^2\geq (\mu(x_{0},\xi_{0})-Ch)\|\psi\|^2\,,
\end{equation}
where $Q_{h,u,p}$ is the quadratic form associated with the operator $\mathcal{M}_{x_{0}+h^{1/2}u,\xi_{0}+h^{1/2}p}$.
\end{prop}
\begin{proof}
We use \eqref{ordering}. Then the terms of $\mathcal{R}_{h}$ (see \eqref{remainders}) are in the form $h h^{p/2} \sigma ^l D_{\sigma}^q \tau ^\alpha D^\beta_{\tau}$ with $l+q\leq p$ and $\beta=0,1$. With Corollary \ref{poly-bound} and the rescaling \eqref{rescaling}, we have:
$$\|h^{p/2} \sigma ^l D_{\sigma}^q \tau ^\alpha D^\beta_{\tau}\psi\|\leq C\|\psi\|$$
and the conclusion follows.
\end{proof}

\subsection{Localization in the phase space}
This section is devoted to elliptic regularity properties (both in space and frequency) satisfied by the eigenfunctions.
We will use the generalization of the localization formula given in Chapter \ref{chapter-models}, Formula \eqref{IMS-formula}.
The following lemma is a straightforward consequence of Assumption \ref{hyp-gen}.
\begin{lem}\label{min-non-deg}
Under Assumption \ref{hyp-gen}, there exist $\eps_{0}>0$ and $c>0$ such that 
%for $|x|+|\xi|\leq \eps_{0}$:
$$\mu(x_{0}+x,\xi_{0}+\xi)-\mu(x_{0},\xi_{0})\geq c(|x|^2+|\xi|^2),\qquad \forall (x,\xi)\in{\mathcal{B}}(\varepsilon_{0})\,,$$
and %for $|x|+|\xi|\geq \eps_{0}$
$$\mu(x_{0}+x,\xi_{0}+\xi)-\mu(x_{0},\xi_{0})\geq c,\qquad \forall (x,\xi)\in\complement{\mathcal{B}}(\varepsilon_{0})\,,$$
where ${\mathcal{B}}(\varepsilon_{0})=\{(x,\xi),  |x|+|\xi|\leq \eps_{0}\}$ and $\complement {\mathcal{B}}(\varepsilon_{0})$ is its complement.
\end{lem}

\begin{notation}
In what follows we will denote by $\tilde\eta>0$ all the quantities which are multiples of $\eta>0$,i.e. in the form $p\eta$ for $p\in\N\setminus\{0\}$. We recall that $\eta>0$ can be chosen arbitrarily small. 
\end{notation}

\begin{prop}\label{sigma-dsigma}
There exist $h_{0},C, \eps_{0}>0$ such that for all eigenpairs $(\lambda,\psi)$ of $\mathcal{L}_{h}$ such that $\lambda\leq\mu_0+C_{0}h$, we have:
$$\|\sigma\psi\|^2+\|\nabla_{\sigma}\psi\|^2\leq C\|\psi\|^2\,.$$
\end{prop}
\begin{proof}
Let  $(\lambda,\psi)$ be an eigenpair such that $\lambda\leq\mu_0+C_{0}h$. 
We recall that \eqref{rlb-cs} holds. We have 
$$\mathcal{Q}_{h}(\psi)=\lambda\|\psi\|^2\leq (\mu_{0}+C_{0}h)\|\psi\|^2\,.$$ 
We deduce that
$$\int_{\R^{2m}} Q_{h,u,p}(\psi_{u,p})-\mu_{0}|\psi_{u,p}|^2\dx u \dx p\leq Ch\|\psi\|^2$$
and thus by the min-max principle
$$\int_{\R^{2m}} \left(\mu(x_{0}+h^{1/2}u,\xi_{0}+h^{1/2}p)-\mu_{0}\right)|\psi_{u,p}|^2\dx u \dx p\leq Ch\|\psi\|^2\,.$$
We use the $\eps_{0}>0$ given in Lemma \ref{min-non-deg} and we split the integral into two parts.
Therefore, we find:
\begin{eqnarray}\label{eq.decoup}
\int_{\Bhe} (|u|^2+|p|^2)|\psi_{u,p}|^2\dx u\dx p\leq C\|\psi\|^2,\\ 
\int_{\CBhe} |\psi_{u,p}|^2\dx u\dx p\leq Ch\|\psi\|^2\,. 
\end{eqnarray}
The first inequality is not enough to get the conclusion. We also need a control of momenta in the region $\CBhe$. 
For that purpose, we write:
\begin{equation}\label{Qa*}
\mathcal{Q}_{h}(\ga_{j}^*\psi)=\int_{\R^{2m}} Q_{h,u,p}\left(\frac{u_{j}-ip_{j}}{\sqrt{2}}\psi_{u,p}\right)\dx u \dx p+\langle\mathcal{R}_{h}\ga_{j}^*\psi,\ga_{j}^*\psi\rangle\,.
\end{equation}
Up to lower order terms we must estimate terms in the form:
$$h\langle  h^{p/2} \sigma ^l D_{\sigma}^q \tau ^\alpha D^\beta_{\tau}\ga_{j}^*\psi  ,\ga_{j}^*\psi\rangle\,,$$
with $l+q=p$, $\alpha\in\N$ and $\beta=0,1$.
By using the \textit{a priori} estimates of Propositions \ref{loc-space} and \ref{loc-phase}, we have:
$$\| h^{p/2} \sigma ^l D_{\sigma}^q  \tau ^\alpha D^\beta_{\tau} \ga_{j}^*\psi\|\leq Ch^{-\tilde\eta}\|\ga_{j}^*\psi\|\,.$$
The remainder is controlled by:
$$|\langle\mathcal{R}_{h}\ga_{j}^*\psi,\ga_{j}^*\psi\rangle|\leq Ch^{1-\tilde \eta}(\|\nabla_{\sigma}\psi\|^2+\|\sigma\psi\|^2)\,.$$
Then we analyze $\mathcal{Q}_{h}(\ga_{j}^*\psi)$ by using \eqref{IMS-formula} (Chapter \ref{chapter-models}) with $\gA =\ga_{j}$. We need to estimate the different remainder terms. We notice that:
\begin{eqnarray*}
\|[\ga_{j}^*, P_{k, r, h}]\psi\|&\leq& C h^{1/2}\|\psi\|,\\ 
|\langle P_{k, r, h}\psi, \ga_{j}^*[ P_{k, r, h},\ga_{j}]\psi\rangle|&\leq& \| P_{k, r , h}\psi\|\ \| \ga_{j}^*[ P_{k, r, h},\ga_{j}]\psi\|\,,\\
 |\langle P_{k, r, h}\psi, \ga_{j}[ P_{k, r, h},\ga_{j}^*]\psi\rangle|&\leq& \| P_{k, r , h}\psi\|\ \|\ga_{j}[ P_{k, r, h},\ga_{j}^*]\psi\|\,,\\
 |\langle P_{k, r, h}\psi, [[ P_{k, r, h},\ga_{j}],\ga_{j}^*]\psi\rangle|&\leq& \| P_{k, r , h}\psi\|\ \|[[ P_{k, r, h},\ga_{j}],\ga_{j}^*]\psi\|\,,
\end{eqnarray*}
where $P_{1, r, h}$ denotes  the magnetic momentum $h^{1/2}D_{\sigma_{r}}+A_{1,r}(x_{0}+h^{1/2}\sigma, \tau)$ and $P_{2, r, h}$ denotes $D_{\tau_{r}}+A_{2,r}(x_{0}+h^{1/2}\sigma, \tau)$.
We have:
$$\| P_{k, r, h}\psi\|\leq C\|\psi\|$$
and:
$$\| \ga_{j}^*[ P_{k, r, h},\ga_{j}]\psi\|\leq Ch^{1/2}\|\ga^*_{j} Q(h^{1/2}\sigma,\tau)\psi\|\,,$$
where $Q$ is polynomial. The other terms can be bounded in the same way.
We apply the estimates of Propositions \ref{loc-space} and \ref{loc-phase} to get:
$$\|\ga^*_{j} Q(h^{1/2}\sigma,\tau)\psi\|\leq Ch^{-\tilde \eta}\|\ga_{j}^*\psi\|\,.$$
We have:
$$\mathcal{Q}_{h}(\ga_{j}^*\psi)=\lambda\|\ga_{j}^*\psi\|^2+\mathcal{O}(h)\|\psi\|^2+\mathcal{O}(h^{\frac{1}{2}-\tilde \eta})(\|\nabla_{\sigma}\psi\|^2+\|\sigma\psi\|^2)\,.$$
so that:
$$\mathcal{Q}_{h}(\ga_{j}^*\psi)\leq \mu(x_{0},\xi_{0})\|\ga_{j}^*\psi\|^2+Ch\|\psi\|^2+\mathcal{O}(h^{\frac{1}{2}-\tilde \eta})(\|\nabla_{\sigma}\psi\|^2+\|\sigma\psi\|^2)\,.$$
By using \eqref{Qa*} and splitting again the integral into two parts, it follows:
\begin{eqnarray*}
\int_{\Bhe} (|u|^2+|p|^2)|(u_{j}-ip_{j})\psi_{u,p}|^2\dx u\dx p&\leq& C\|\psi\|^2+Ch^{-\frac{1}{2}-\tilde \eta}(\|\nabla_{\sigma}\psi\|^2+\|\sigma\psi\|^2)\,,\\
\int_{\CBhe} |(u_{j}-ip_{j})\psi_{u,p}|^2\dx u\dx p&\leq &Ch\|\psi\|^2+Ch^{\frac{1}{2}-\tilde \eta}(\|\nabla_{\sigma}\psi\|^2+\|\sigma\psi\|^2)\,.
\end{eqnarray*}
Combining the last inequality with the first one of \eqref{eq.decoup} and the Parseval formula we get the conclusion.

\end{proof}
By using the same ideas, we can establish the following proposition.
\begin{prop}\label{sigma2-dsigma2}
Let $P\in\mathbb{C}_{2}[X_{1},\ldots,X_{2n}]$. There exist $h_{0},C, \eps_{0}>0$ such that for all eigenpairs $(\lambda,\psi)$ of $\mathfrak{L}_{h}$ such that $\lambda\leq\mu_0+C_{0}h$, we have:
$$\|P(\sigma,D_{\sigma})\psi\|^2\leq Ch^{-\frac{1}{2}-\tilde \eta}\|\psi\|^2\,.$$
\end{prop}

\subsection{Approximation lemmas}
We introduce the projection
$$\Psi_{0}=\Pi_{0}\psi=\langle\psi,u_{x_{0},\xi_{0}}\rangle_{\sL^2(\R^n,\dx \tau)} u_{x_{0},\xi_{0}}$$
and, inspired by \eqref{psi1} where $f_{0}$ is replaced by $\langle\psi,u_{x_{0},\xi_{0}}\rangle_{\sL^2(\R^n,\dx \tau)}$ and $f_{1}$ by $0$,
\begin{equation}
\Psi_{1}=\sum_{j=1}^m (\dr_{x_{j}}u)_{x_{0},\xi_{0}}\, \sigma_{j} \langle\psi,u_{x_{0},\xi_{0}}\rangle_{\sL^2(\R^n,\dx \tau)}+\sum_{j=1}^m (\dr_{\xi_{j}}u)_{x_{0},\xi_{0}}\, D_{\sigma_{j}} \langle\psi,u_{x_{0},\xi_{0}}\rangle_{\sL^2(\R^n,\dx \tau)}.
\end{equation}
This leads to defined the corrected Feshbach projection
\begin{equation}
\Pi_{h}\psi=\Psi_{0}+h^{1/2}\Psi_{1}
\end{equation}
and
\[R_{h}=\psi-\Pi_{h}\psi\,.\]
We may notice that  this corrected Feshbach correction shares some features with the \enquote{super-adiabatic} projections \textit{\`a la Panati-Spohn-Teufel-Wachsmuth} (see for instance \cite{Teufel03, GST07, TW13}).

Note that the functions $\Psi_{0}$ and $\Psi_{1}$ will be {\it a priori} $h$-dependent. By the $\sL^2$-normalization of $u_{x,\xi}$ (when $\xi\in\R^m$), $\Psi_{1}$ and $R_{h}$ are orthogonal (with respect to the $\tau$-variable) to $u_{0}$ (and $\Psi_{0}$). Furthermore, we have by construction and Proposition \ref{FH},
\begin{equation}\label{L1Psi0}
(\mathcal{L}_{0}-\mu_{0})\Psi_{1}=-\mathcal{L}_{1}\Psi_{0}
\end{equation}
and, by the Fredholm alternative,
$$\langle\mathcal{L}_{1}\Psi_{0},\Psi_{0}\rangle_{\sL^2(\R^n,\dx \tau)}=0.$$
We can prove a first approximation.
\begin{prop}
There exist $h_{0},C>0$ such that for all eigenpairs $(\lambda,\psi)$ of $\mathcal{L}_{h}$ such that $\lambda\leq\mu_0+C_{0}h$, we have
$$\|\psi-\Pi_{0}\psi\|\leq Ch^{1/2-\tilde\eta}\|\psi\|$$
\end{prop}
\begin{proof}
We can write:
$$(\mathcal{L}_{0}-\mu_{0})\psi=(\lambda-\mu_{0})\psi-h^{1/2}\mathcal{L}_{1}\psi-h\mathcal{L}_{2}\psi+\ldots-h^{M/2}\mathcal{L}_{M}\psi\,.$$
By using the rough microlocalization given in Propositions \ref{loc-space} and \ref{loc-phase} and Proposition \ref{sigma2-dsigma2}, we infer that for $p\geq 2$:
\begin{equation}\label{control-reste}
h^{p/2}\|\tau^\alpha D_{\tau} ^\beta \sigma^l D_{\sigma}^q \psi\|\leq C h^{\frac p2-\frac{p-2}2-\frac14-\tilde \eta}\|\psi\|=Ch^{\frac{3}{4}-\tilde\eta}\|\psi\|\,,
\end{equation}
and thanks to Proposition \ref{sigma-dsigma}:
$$\|\mathcal{L}_{1}\psi\|\leq Ch^{-\tilde\eta}\|\psi\|\,,$$
so that:
$$\|(\mathcal{L}_{0}-\mu_{0})\psi\|\leq Ch^{\frac{1}{2}-\tilde\eta}\|\psi\|\,,$$
and the conclusion follows.
\end{proof}

\begin{cor}\label{cor.approx0}
There exist $h_{0},C>0$ such that for all eigenpairs $(\lambda,\psi)$ of $\mathcal{L}_{h}$ such that $\lambda\leq\mu_0+C_{0}h$, we have:
$$\|\sigma(\psi-\Pi_{0}\psi)\|\leq Ch^{1/4-\tilde\eta}\|\psi\|,\quad \|D_{\sigma}(\psi-\Pi_{0}\psi)\|\leq Ch^{1/4-\tilde\eta}\|\psi\|$$
\end{cor}
We can now estimate $\psi-\Pi_{h}\psi$.

\begin{prop}\label{control-Rh}
There exist $h_{0},C>0$ such that for all eigenpairs $(\lambda,\psi)$ of $\mathcal{L}_{h}$ such that $\lambda\leq\mu_0+C_{0}h$, we have:
$$\|\psi-\Pi_{h}\psi\|\leq Ch^{3/4-\tilde \eta}\|\psi\|\,.$$
\end{prop}

\begin{proof}
Let us write:
$$\mathcal{L}_{h}\psi=\lambda\psi\,.$$
We have:
$$(\mathcal{L}_{0}+h^{1/2}\mathcal{L}_{1})\psi=(\mu_{0}+\mathcal{O}(h))\psi-h\mathcal{L}_{2}\psi-\ldots-h^{M/2}\mathcal{L}_{M}\psi\,.$$
Let us notice that, as in \eqref{control-reste}, for $p\geq 2$:
$$h^{p/2}\|\mathcal{L}_{p}\psi\|\leq C h^{\frac 3 4-\tilde\eta}\|\psi\|\,.$$
We get:
$$(\mathcal{L}_{0}-\mu_{0})R_{h}=-h^{1/2}\mathcal{L}_{1}(\psi-\Psi_{0})+\mathcal{O}(h)\psi-h\mathcal{L}_{2}\psi-\ldots-h^{M/2}\mathcal{L}_{M}\psi$$
It remains to apply Corollary \ref{cor.approx0} to get:
$$h^{1/2}\|\mathcal{L}_{1}(\psi-\Psi_{0})\|\leq  \tilde C h^{\frac34-\tilde\eta}\|\psi\|\,.$$
\end{proof}

Let us introduce a subspace of dimension $P\geq 1$. For $j\in\{1,\cdots,P\}$ we can consider a $\sL^2$-normalized eigenfunction of $\mathcal{L}_{h}$ denoted by $\psi_{j,h}$ and so that the family $(\psi_{j,h})_{j\in\{1,\cdots,P\}}$ is orthogonal. We let:
$$\mathcal{E}_{P}(h)=\underset{j\in\{1,\cdots,P\}}\spann\psi_{j,h}\,.$$
\begin{rem}
We can extend all the local and microlocal estimates as well as our approximations to $\psi\in\mathcal{E}_{P}(h)$.
\end{rem}
Then we can prove a lower bound for the quadratic form on $\mathcal{E}_{P}(h)$ by replacing $\psi\in\mathcal{E}_{P}(h)$ by $\Pi_{h}\psi$, in the spirit of Chapter \ref{chapter-BOE}.

We have:
\[\mathcal{Q}_{h}(\psi)=\langle\mathcal{L}_{0}\psi,\psi\rangle+h^{1/2}\langle\mathcal{L}_{1}\psi,\psi\rangle+h\langle\mathcal{L}_{2}\psi,\psi\rangle+\ldots+h^{p/2}\langle\mathcal{L}_{p}\psi,\psi\rangle+\ldots+h^{M/2}\langle\mathcal{L}_{M}\psi,\psi\rangle\,.\]
Using Propositions \ref{sigma-dsigma}, \ref{sigma2-dsigma2}, \ref{loc-space} and \ref{loc-phase}, we have, for $\ell\geq 3$
\[|h^{\ell/2}\langle\mathcal{L}_{\ell}\psi,\psi\rangle|\leq Ch^{\frac \ell2-\frac{\ell-3}2-\tilde\eta-\frac14}\|\psi\|^2=Ch^{\frac 5 4-\tilde\eta}\|\psi\|^2\,.\]
We infer
\[\mathcal{Q}_{h}(\psi)\geq\langle\mathcal{L}_{0}\psi,\psi\rangle+h^{1/2}\langle\mathcal{L}_{1}\psi,\psi\rangle+h\langle\mathcal{L}_{2}\psi,\psi\rangle-Ch^{\frac 5 4-\tilde\eta}\|\psi\|^2\,.\]
It remains to analyze the different terms.
We have
\[\langle\mathcal{L}_{0}\psi,\psi\rangle=\langle\mathcal{L}_{0}(\Psi_{0}+h^{1/2}\Psi_{1}+R_{h}),\Psi_{0}+h^{1/2}\Psi_{1}+R_{h}\rangle\,.\]
The orthogonality (with respect to $\tau$) cancels the terms $\langle\mathcal{L}_{0}\Psi_{1},\Psi_{0}\rangle$ and $\langle R_{h},\Psi_{0}\rangle$. Moreover, we have, with Propositions \ref{loc-space} and \ref{loc-phase},
\[h^{1/2}|\langle\mathcal{L}_{0}R_{h},\Psi_{1}\rangle|\leq h^{1/2-\tilde\eta}\|R_{h}\|\|\Psi_{1}\|\,,\]
and we use Proposition \ref{sigma-dsigma} to get
\[\|\Psi_{1}\|\leq C\|\psi\|\,,\]
so that, with Proposition \ref{control-Rh},
\[\langle\mathcal{L}_{0}\psi,\psi\rangle=\mu_{0}\|\Psi_{0}\|^2+h\langle\mathcal{L}_{0}\Psi_{1},\Psi_{1}\rangle+\mathcal{O}(h^{\frac 5 4-\tilde\eta})\|\psi\|^2\,.\]
We have
\[\langle\mathcal{L}_{1}\psi,\psi\rangle=\langle\mathcal{L}_{1}\Psi_{0},\Psi_{0}\rangle+2h^{1/2}\langle\mathcal{L}_{1}\Psi_{0},\Psi_{1}\rangle+h\langle\mathcal{L}_{1}\Psi_{1},\Psi_{1}\rangle+2\langle\mathcal{L}_{1}\psi,R_{h}\rangle\,.\]
Then, a Feynman-Hellmann formula provides $\langle\mathcal{L}_{1}\Psi_{0},\Psi_{0}\rangle=0$. Using again Propositions \ref{loc-space}, \ref{loc-phase}, \ref{sigma-dsigma}, \ref{sigma2-dsigma2} and \ref{control-Rh}, we notice that
\[\langle\mathcal{L}_{1}\psi,\psi\rangle=2h^{1/2}\langle\mathcal{L}_{1}\Psi_{0},\Psi_{1}\rangle+\mathcal{O}(h^{\frac 34-\tilde\eta})\|\psi\|^2\,.\]
We notice
\[\langle\mathcal{L}_{2}\psi,\psi\rangle=\langle\mathcal{L}_{2}\Psi_{0},\Psi_{0}\rangle+\langle\mathcal{L}_{2}(\psi-\Psi_{0}),\psi\rangle+\langle\mathcal{L}_{2}\psi,\psi-\Psi_{0}\rangle\,.\]
Writing $\psi-\Psi_{0}=h^{1/2} \Psi_{1}+R_{h}$, we have the estimate
\[|\langle\mathcal{L}_{2}(\psi-\Psi_{0}),\psi\rangle+\langle\mathcal{L}_{2}\psi,\psi-\Psi_{0}\rangle|\leq Ch^{-\frac14-\tilde\eta}h^{\frac{1}{2}-\tilde\eta}\|\psi\|^2\,.\]
We infer
\[\mathcal{Q}_{h}(\psi)\geq \mu_{0}\|\Psi_{0}\|^2+h\langle\mathcal{L}_{0}\Psi_{1},\Psi_{1}\rangle+h\langle\mathcal{L}_{1}\Psi_{0},\Psi_{1}\rangle+h\langle\mathcal{L}_{1}\Psi_{1},\Psi_{0}\rangle+h\langle\mathcal{L}_{2}\Psi_{0},\Psi_{0}\rangle-Ch^{\frac 5 4-\tilde\eta}\|\psi\|^2\,.\]
Using \eqref{L1Psi0}, we get
\[h\langle\mathcal{L}_{0}\Psi_{1},\Psi_{1}\rangle+h\langle\mathcal{L}_{1}\Psi_{0},\Psi_{1}\rangle=h\mu_{0}\|\Psi_{1}\|^2\,,\]
so that, by orthogonality,
\[\mathcal{Q}_{h}(\psi)\geq \mu_{0}\|\Psi_{0}+h^{1/2}\Psi_{1}\|^2+h\langle\mathcal{L}_{1}\Psi_{1},\Psi_{0}\rangle+h\langle\mathcal{L}_{2}\Psi_{0},\Psi_{0}\rangle-Ch^{\frac 5 4-\tilde\eta}\|\psi\|^2\,.\]
Since $\langle R_{h},\Psi_{0}\rangle=0$ we deduce that
\[\|\Psi_{0}+h^{1/2}\Psi_{1}\|^2=\|\Psi_{0}+h^{1/2}\Psi_{1}+R_{h}\|^2+\mathcal{O}(h^{\frac 5 4-\tilde\eta})\|\psi\|^2\,.\]
It follows that
\[\mathcal{Q}_{h}(\psi)- \mu_{0}\|\psi\|^2\geq h\langle\mathcal{L}_{1}\Psi_{1},\Psi_{0}\rangle+h\langle\mathcal{L}_{2}\Psi_{0},\Psi_{0}\rangle+\mathcal{O}(h^{\frac 5 4-\tilde\eta})\|\psi\|^2\,,\]
and, since $\mathcal{Q}_{h}(\psi)\leq \lambda_{P}(h)\|\psi\|^2$, we have
\[(\lambda_{P}(h)-\mu_{0})\|\psi\|^2\geq h\langle\mathcal{L}_{1}\Psi_{1},\Psi_{0}\rangle+h\langle\mathcal{L}_{2}\Psi_{0},\Psi_{0}\rangle+\mathcal{O}(h^{\frac 5 4-\tilde\eta})\|\psi\|^2\,.\]
Thus we get
\[(\lambda_{P}(h)-\mu_{0})\|\Psi_{0}\|^2\geq h\langle\mathcal{L}_{1}\Psi_{1},\Psi_{0}\rangle+h\langle\mathcal{L}_{2}\Psi_{0},\Psi_{0}\rangle+\mathcal{O}(h^{\frac 5 4-\tilde\eta})\|\psi\|^2\,.\]
We recall that (see \eqref{Fredholm} and below)
\begin{multline*}
\langle\mathcal{L}_{1}\Psi_{1},\Psi_{0}\rangle+\langle\mathcal{L}_{2}\Psi_{0},\Psi_{0}\rangle\\
=\left\langle\tfrac{1}{2}\Hess\, \mu(x_{0},\xi_{0}) (\sigma,D_{\sigma})(\langle\psi, u_{0}\rangle_{\sL^2(\R^n,\dx \tau)}),\langle\psi, u_{0}\rangle_{\sL^2(\R^n,\dx \tau)}\right\rangle_{\sL^2(\R^m,\dx \sigma)}.
\end{multline*}
Finally we apply the min-max principle to the $P$-dimensional space $\left\langle\mathfrak{E}_{P}(h),u_{0}\right\rangle_{\sL^2(\R^n,\dx \tau)}$ to get the spectral gap.

\chapter{Examples of magnetic WKB constructions}\label{chapter-WKB}
\begin{flushright}
\begin{minipage}{0.4\textwidth}
Mais la vision la plus belle qui nous reste d'une \oe{}uvre est souvent celle qui s'\'eleva au-dessus des sons faux tir\'es par des doigts malhabiles, d'un piano d\'esaccord\'e.
\begin{flushright}
\textit{Du c\^ot\'e de chez Swann}, Proust
\end{flushright}
\vspace*{0.5cm}
\end{minipage}
\end{flushright}
In this chapter we give some examples of magnetic WKB constructions. Let us underline that these examples are the first known results in the direction of WKB constructions in presence of a pure magnetic field.
\section{Vanishing magnetic fields}\label{WKB-vanishing}
This section in devoted to the proof of Theorem \ref{WKB-van}. The fundamental ingredients to succeed are a normal form procedure, an operator valued WKB construction (see Proposition \ref{el-WKB} for an electric example) and a complex extension of the standard model operators.
\begin{lem}\label{eikonal0}
For $r>0$, let us consider a holomorphic function $\nu : D(0,r)\to\C$ such that $\nu(0)=\nu'(0)=0$ and $\nu''(0)\in\R_{+}$. Let us also introduce a smooth $F$ defined in a real neighborhood of $\sigma=0$ such that $\sigma=0$ is a non degenerate maximum. Then, there exists a neighborhood of $\sigma=0$ such that the equation
\begin{equation}\label{eikonal}
\nu(i\varphi(\sigma))=F(\sigma)
\end{equation}
admits a smooth solution $\varphi$ solution such that $\varphi(0)=0$ and $\varphi'(0)>0$.
\end{lem}
\begin{proof}
We can apply the Morse lemma to deduce that \eqref{eikonal} is equivalent to
$$\tilde\nu(i\varphi(\sigma))^2=-f(\sigma)^2\,,$$
where $f$ is a non negative function such that $f'(0)=\sqrt{-\frac{F''(0)}{2}}$ and $F(\sigma)=-f(\sigma)^2$ and $\tilde\nu$ is a holomorphic function in a neighborhood of $0$ such that $\tilde\nu^2=\nu$ and $\tilde\nu'(0)=\sqrt{\frac{\nu''(0)}{2}}$. This provides the equations
$$\tilde\nu(i\varphi(\sigma))=if(\sigma),\quad \tilde\nu(i\varphi(\sigma))=-if(\sigma)\,.$$
Since $\tilde\nu$ is a local biholomorphism and $f(0)=0$, we can write the equivalent equations
$$\varphi(\sigma)=-i\tilde\nu^{-1}(if(\sigma)),\quad \varphi(\sigma)=-i\tilde\nu^{-1}(-if(\sigma))\,.$$
The function $\varphi(s)=-i\tilde\nu^{-1}(if(s))$ satisfies our requirements since $\varphi'(0)=\sqrt{-\frac{F''(0)}{\nu''(0)}}$.
\end{proof}
\subsection{Renormalization}
We use the canonical transformation associated with the change of variables:
\begin{equation}\label{can}
t=(\gamma(\sigma))^{-\frac{1}{k+2}}\tau,\quad s=\sigma\,,
\end{equation}
we deduce that  $\mathfrak{L}_{h}^{[k]}$ is unitarily equivalent to the operator on $\sL^2(\dx \sigma\dx \tau)$:
$$\mathfrak{L}_{h}^{[k],\neww}=\gamma(\sigma)^{\frac{2}{k+2}}D_{\tau}^2+\left(hD_{\sigma}-\gamma(\sigma)^{\frac{1}{k+2}}\frac{\tau^{k+1}}{k+1}+\frac{h}{2(k+2)}\frac{\gamma'(\sigma)}{\gamma(\sigma)}(\tau D_{\tau}+D_{\tau}\tau)\right)^2\,.$$
We may change the gauge
\begin{align*}
&e^{-ig(\sigma)/h}\mathfrak{L}_{h}^{[k],\neww}e^{ig(\sigma)/h}\\
&=\gamma(\sigma)^{\frac{2}{k+2}}D_{\tau}^2+\left(hD_{\sigma}+\zeta^{[k]}_{0}\gamma(\sigma)^{\frac{1}{k+2}}-\gamma(\sigma)^{\frac{1}{k+2}}\frac{\tau^{k+1}}{k+1}+\frac{h}{2(k+2)}\frac{\gamma'(\sigma)}{\gamma(\sigma)}(\tau D_{\tau}+D_{\tau}\tau)\right)^2.
\end{align*}
with
$$g(\sigma)=\zeta_{0}^{[k]}\int_{0}^{\sigma}\gamma(\tilde\sigma)^{\frac{1}{k+2}}\dx\tilde\sigma\,.$$
For some function $\Phi=\Phi(\sigma)$ to be determined, we consider
$$\mathfrak{L}^{[k],\wgt}_{h}=e^{\Phi/h}e^{-ig(\sigma)/h}\mathfrak{L}_{h}^{[k],\neww}e^{ig(\sigma)/h}e^{-\Phi/h}=\mathfrak{L}^{[k],\wgt,0}+h\mathfrak{L}^{[k],\wgt,1}+h^2\mathfrak{L}^{[k],\wgt,2}\,,$$
with
\begin{align*}
&\mathfrak{L}^{[k],\wgt,0}=\gamma(\sigma)^{\frac{2}{k+2}}\mathfrak{L}^{[k]}_{w(\sigma)}\,,\\
&\mathfrak{L}^{[k],\wgt,1}=\tfrac{1}{2}\left(\gamma(\sigma)^{\frac{1}{k+2}}\partial_{\zeta}\mathfrak{L}_{\zeta}^{[k]}D_{\sigma}+D_{\sigma}\gamma(\sigma)^{\frac{1}{k+2}}\partial_{\zeta}\mathfrak{L}_{\zeta}^{[k]}\right)+\mathfrak{R}_{1}(\sigma, \tau; D_{\tau})\,,\\
&\mathfrak{L}^{[k],\wgt,2}=D_{\sigma}^2+\mathfrak{R}_{2}(\sigma,\tau; D_{\sigma}, D_{\tau})\,,
\end{align*}
where
$$w(\sigma)=\zeta^{[k]}_{0}+i\gamma(\sigma)^{-\frac{1}{k+2}}\Phi'\,.$$
and 
where the $\mathfrak{R}_{1}(\sigma, \tau; D_{\tau})$ is of order zero in $D_{\sigma}$ and cancels for $\sigma=0$ whereas $\mathfrak{R}_{2}(\sigma,\tau; D_{\sigma}, D_{\tau})$ is of order one with respect to $D_{\sigma}$.

Now, let us try to solve, as usual, the eigenvalue equation 
$$\mathfrak{L}^{[k],\wgt}_{h}\an=\lambda\an$$
in the sense of formal series in $h$:
$$\an\sim\sum_{j\geq 0}h^j\an_{j},\quad \lambda\sim\sum_{j\geq 0}h^j\lambda_{j}\,.$$

\subsection{Solving the operator valued eikonal equation}
The first equation is
$$\mathfrak{L}^{[k],\wgt,0}\an_{0}=\lambda_{0}\an_{0}\,.$$
We must choose 
$$\lambda_{0}=\gamma_{0}^{\frac{2}{k+2}}\nu_{1}(\zeta_{0}^{[k]})$$
and we are led to take
\begin{equation}\label{psi0}
\an_{0}(\sigma,\tau)=f_{0}(\sigma)u^{[k]}_{w(\sigma)}(\tau)
\end{equation}
so that the equation becomes
$$\nu_{1}^{[k]}\left(w(\sigma)\right)-\nu_{1}(\zeta_{0}^{[k]})=\left(\gamma_{0}^{\frac{2}{k+2}}\gamma(\sigma)^{-\frac{2}{k+2}}-1\right)\nu^{[k]}_{1}(\zeta_{0}^{[k]})\,.$$
Therefore we are in the framework of Lemma \ref{eikonal0}.
We use the lemma with $F(\sigma)=\left(\gamma_{0}^{\frac{2}{k+2}}\gamma(\sigma)^{-\frac{2}{k+2}}-1\right)\nu^{[k]}_{1}(\zeta_{0}^{[k]})$ and, for the function $\varphi$ given by the lemma, we have
$$\Phi'(\sigma)=\gamma(\sigma)^{\frac{1}{k+2}}\varphi(\sigma)$$
and we take
$$\Phi(\sigma)=\int_{0}^\sigma \gamma(\tilde\sigma)^{\frac{1}{k+2}}\varphi(\tilde\sigma)\dx \tilde\sigma\,,$$
which is defined in a fixed neighborhood of $0$ and satisfies $\Phi(0)=\Phi'(0)=0$ and 
\begin{equation}\label{Phi''}
\Phi''(0)=\gamma_{0}^{\frac{1}{k+2}}\sqrt{\frac{2}{k+2}\frac{\gamma''(0)\nu_{1}^{[k]}(\zeta_{0}^{[k]})}{\left(\nu_{1}^{[k]}\right)''(\zeta_{0}^{[k]})\gamma(0)}}>0\,.
\end{equation}
Therefore \eqref{psi0} is well defined in a neighborhood of $\sigma=0$.
\subsection{Solving the transport equation}
We can now deal with the operator valued transport equation
$$(\mathfrak{L}^{[k],\wgt,0}-\lambda_{0})\an_{1}=(\lambda_{1}-\mathfrak{L}^{[k],\wgt,1})\an_{0}\,.$$
For each $\sigma$ the Fredholm condition is
$$\left\langle(\lambda_{1}-\mathfrak{L}^{[k],\wgt,1})\an_{0}, \overline{u^{[k]}_{w(\sigma)}}\right\rangle_{\sL^2(\R_{\tau})}=0\,,$$
where the complex conjugation is needed since $\mathfrak{L}^{[k],\wgt,1}$ is not necessarily self-adjoint.
Let us examine
$$\left\langle\mathfrak{L}^{[k],\wgt,1}\an_{0}, \overline{u^{[k]}_{w(\sigma)}}\right\rangle_{\sL^2(\R_{\tau})}\,.$$
We recall the Feynman-Hellmann formula
$$\frac{1}{2}\left(\nu_{1}^{[k]}\right)'(\zeta)=\int_{\R}\left(\zeta-\frac{\tau^{k+1}}{k+1}\right)u^{[k]}_{\zeta}u^{[k]}_{\zeta} \dx\tau\,,$$
and the formula
$$\int_{\R}u^{[k]}_{\zeta}u^{[k]}_{\zeta} \dx\tau=1$$
which are valid for $\zeta\in\mathbb{C}$ close to $\zeta^{[k]}_{0}$ by holomorphic extension of the formulas valid for $\zeta\in\R$.
We get an equation in the form
\begin{multline*}
\left\langle\mathfrak{L}^{[k],\wgt,1}\an_{0}, \overline{u^{[k]}_{w(\sigma)}}\right\rangle_{\sL^2(\R_{\tau})}\\
=\tfrac{1}{2}\left\{\gamma(\sigma)^{\frac{1}{k+2}}\left(\nu_{1}^{[k]}\right)'(w(\sigma))D_{\sigma}+D_{\sigma}\gamma(\sigma)^{\frac{1}{k+2}}\left(\nu_{1}^{[k]}\right)'(w(\sigma))\right\}\an_{0}+R^{[k]}(\sigma)\an_{0}\,,
\end{multline*}
where $R^{[k]}$ is smooth an vanishes at $\sigma=0$.
Thus we are reduced to solve the transport equation
\begin{multline*}
\frac{1}{2}\left\{\gamma(\sigma)^{\frac{1}{k+2}}\left(\nu_{1}^{[k]}\right)'(w(\sigma))D_{\sigma}+D_{\sigma}\gamma(\sigma)^{\frac{1}{k+2}}\left(\nu_{1}^{[k]}\right)'(w(\sigma))\right\}\an_{0}+R^{[k]}(\sigma)\an_{0}=\lambda_{1}\an_{0}\,.
\end{multline*}
The only point that we should verify is that the linearized transport equation near $\sigma=0$ is indeed a transport equation in the sense of \cite[Chapter 3]{DiSj99} so that we have just to consider the linearization of the first part of the equation. The linearized operator is
$$\frac{\left(\nu_{1}^{[k]}\right)''(\zeta_{0}^{[k]})\Phi''(0)}{2}(\sigma\dr_{\sigma}+\dr_{\sigma}\sigma)\,.$$
The eigenvalues of this operator are
\begin{equation}\label{evset}
\left\{\frac{\left(\nu_{1}^{[k]}\right)''(\zeta_{0}^{[k]})\Phi''(0)}{2}(2j+1),\quad j\in\N\right\}\,.
\end{equation}
Let us notice that
$$
\frac{\left(\nu_{1}^{[k]}\right)''(\zeta_{0}^{[k]})\Phi''(0)}{2}=\frac{\gamma_{0}^{\frac{1}{k+2}}}{2}\sqrt{\frac{2}{k+2}\frac{\gamma''(0)\nu_{1}^{[k]}(\zeta_{0}^{[k]})\left(\nu_{1}^{[k]}\right)''(\zeta_{0}^{[k]})}{\gamma(0)}}.
$$
This is exactly the expected expression for the second term in the asymptotic expansion of the eigenvalues (see Theorem \ref{main-theorem-BOM}).
Therefore $\lambda_{1}$ has to be chosen in the set \eqref{evset}, the transport equation can be solved in a neighborhood of $\sigma=0$ and the construction can be continued at any order (see \cite[Chapter 3]{DiSj99}). Since the first eigenvalues are simple, the spectral theorem implies that the constructed functions $f_{0}(\sigma)u^{[k]}_{\zeta^{[k]}_{0}+i\gamma(\sigma)^{-\frac{1}{k+2}}\Phi'}(\tau) e^{-\frac{\Phi(\sigma)}{h}}$ are approximations of the true eigenfunctions of $e^{-ig(\sigma)}\mathfrak{L}_{h}^{[k],\neww}e^{ig(\sigma)}$. This is the content of Theorem \ref{WKB-van}.

\section{Curvature induced magnetic bound states}\label{WKB-FouHel}
Let us prove Theorem \ref{WKB-FH}. Let us introduce a phase function $\Phi=\Phi(\sigma)$ defined in a neighborhood of $\sigma=0$ the unique and non-degenerate maximum of the curvature $\kappa$. We consider the conjugate operator
$$\mathfrak{L}^{\FH,\wgt}_{h}=e^{\Phi(\sigma)/h^{\frac{1}{4}}}\mathfrak{L}^{\FH}_{h}e^{-\Phi(\sigma)/h^{\frac{1}{4}}}\,.$$
As usual, we look for
$$\an\sim \sum_{j\geq 0} h^{\frac j 4}\an_{j},\quad \lambda\sim\sum_{j\geq 0} \lambda_{j}h^{\frac j 4}$$
such that, in the sense of formal series we have
$$\mathfrak{L}^{\FH,\wgt}_{h}\an \sim \lambda\an\,.$$
We may write
$$\mathfrak{L}^{\FH,\wgt}_{h}\sim \mathfrak{L}_{0}+ h^{\frac 1 4}\mathfrak{L}_{1}+ h^{\frac 1 2}\mathfrak{L}_{2}+ h^{\frac 3 4}\mathfrak{L}_{3}+\ldots\,,$$
where
\begin{align*}
&\mathfrak{L}_{0}=D_{\tau}^2+(\zeta_{0}-\tau)^2\,, \\
&\mathfrak{L}_{1}=2(\zeta_{0}-\tau)i\Phi'(\sigma)\,,\\
&\mathfrak{L}_{2}=\kappa(\sigma)\dr_{\tau}+2\left(D_{\sigma}+\kappa(\sigma)\frac{\tau^2}{2}\right)(\zeta_{0}-\tau)-\Phi'(\sigma)^2+2\kappa(\sigma)(\zeta_{0}-\tau)^2\tau\,,\\
&\mathfrak{L}_{3}=\left(D_{\sigma}+\kappa(\sigma)\frac{\tau^2}{2}\right)(i\Phi'(\sigma))+(i\Phi'(\sigma))\left(D_{\sigma}+\kappa(\sigma)\frac{\tau^2}{2}\right)+4i\Phi'(\sigma)\tau \kappa(\sigma)(\zeta_{0}-\tau)\,.
\end{align*}
Let us now solve the formal system. The first equation is
$$\mathfrak{L}_{0}\an_{0}=\lambda_{0}\an_{0}$$
and leads to take
$$\lambda_{0}=\Theta_{0},\quad \an_{0}(\sigma,\tau)=f_{0}(\sigma)u_{\zeta_{0}}(\tau)\,,$$
where $f_{0}$ has to be determined. The second equation is
$$(\mathfrak{L}_{0}-\lambda_{0})\an_{1}=(\lambda_{1}-\mathfrak{L}_{1})\an_{0}=(\lambda_{1}-2(\zeta_{0}-\tau))u_{\zeta_{0}}(\tau) i\Phi'(\sigma) f_{0}$$
and, due to the Fredholm alternative, we must take $\lambda_{1}=0$ and we take
$$\an_{1}(\sigma,\tau)=i\Phi'(\sigma) f_{0}(\sigma)\left(\dr_{\zeta}u\right)_{\zeta_{0}}(\tau)+f_{1}(\sigma)u_{\zeta_{0}}(\tau)\,,$$
where $f_{1}$ is to be determined in a next step. Then the third equation is
$$(\mathfrak{L}_{0}-\lambda_{0})\an_{2}=(\lambda_{2}-\mathfrak{L}_{2})\an_{0}-\mathfrak{L}_{1}\an_{1}\,.$$
Let us explicitly write the r.h.s. It equals
\begin{multline*}
\lambda_{2}u_{\zeta_{0}} f_{0}+\Phi'^2(u_{\zeta_{0}}+2(\zeta_{0}-\tau)(\dr_{\zeta}u)_{\zeta_{0}})f_{0}-2(\zeta_{0}-\tau)u_{\zeta_{0}}(i\Phi' f_{1}-i\dr_{\sigma}f_{0})\\
+\kappa(\sigma)f_{0}(\dr_{\tau}u_{\zeta_{0}}-2(\zeta_{0}-\tau)^2\tau u_{\zeta_{0}}-\tau^2(\zeta_{0}-\tau)u_{\zeta_{0}})\,.
\end{multline*}
Therefore the equation becomes
$$(\mathfrak{L}_{0}-\lambda_{0})\tilde\an_{2}=\lambda_{2}u_{\zeta_{0}} f_{0}+\frac{\nu_{1}''(\zeta_{0})}{2}\Phi'^2 u_{\zeta_{0}}f_{0}+\kappa(\sigma)f_{0}(-\dr_{\tau}u_{\zeta_{0}}-2(\zeta_{0}-\tau)^2\tau u_{\zeta_{0}}-\tau^2(\zeta_{0}-\tau)u_{\zeta_{0}})\,,$$
where $$\tilde\an_{2}=\an_{2}-v_{\zeta_{0}}(i\Phi' f_{1}-i\dr_{\sigma}f_{0})+\tfrac{1}{2}(\dr_{\zeta}^2 u)_{\zeta_{0}}\Phi'^2 f_{0}\,.$$
Let us now use the Fredholm alternative (with respect to $\tau$). We will need the following lemma the proof of which relies on Feynman-Hellmann formulas (like in Proposition \ref{FH}) and on \cite[p. 19]{FouHel06a} (for the last one).
\begin{lem}
We have:
\begin{equation*}
\begin{gathered}
\int_{\R_{+}} (\zeta_{0}-\tau)u^2_{\zeta_{0}}(\tau)\dx \tau=0,\qquad \int_{\R_{+}} (\dr_{\zeta} u)_{\zeta_{0}}(\tau) u_{\zeta_{0}}(\tau)\dx \tau=0\,,\\
2\int_{\R_{+}} (\zeta_{0}-\tau)(\dr_{\zeta} u)_{\zeta_{0}}(\tau) u_{\zeta_{0}}(\tau)\dx \tau=\frac{\nu_{1}''(\zeta_{0})}{2}-1\,,\\
\int_{\R_{+}} \left(2\tau(\zeta_{0}-\tau)^2+\tau^2(\zeta_{0}-\tau)\right)u^2_{\zeta_{0}}+u_{\zeta_{0}}\dr_{\tau}u_{\zeta_{0}} \dx\tau=-C_{1}\,.
\end{gathered}
\end{equation*}
\end{lem}
We get the equation
$$\lambda_{2}+\frac{\nu_{1}''(\zeta_{0})}{2}\Phi'^2(\sigma)+C_{1}\kappa(\sigma)=0,\qquad C_{1}=\frac{u^2_{\zeta_{0}}(0)}{3}\,.$$
This eikonal equation is the eikonal equation of a pure electric problem in dimension one whose potential is given by the curvature. Thus we take
$$\lambda_{2}=-C_{1}\kappa(0)\,,$$
and
$$\Phi(\sigma)=\left(\frac{2C_{1}}{\nu_{1}''(\zeta_{0})}\right)^{1/2}\left|\int_{0}^\sigma (\kappa(0)-\kappa(s))^{1/2}\dx s\right|\,.$$
In particular we have:
$$\Phi''(0)=\left(\frac{k_{2}C_{1}}{\nu_{1}''(\zeta_{0})}\right)^{1/2}\,,$$
where $k_{2}=-\kappa''(0)>0$.\\
This leads to take
$$\an_{2}=f_{0}\hat\an_{2}+(\dr_{\zeta}u)_{\zeta_{0}}(i\Phi' f_{1}-i\dr_{\sigma}f_{0})-\tfrac{1}{2}(\dr^2_{\eta}u)_{\zeta_{0}}\Phi'^2 f_{0}+f_{2}u_{\zeta_{0}}\,,$$
where $\hat\an_{2}$ is the unique solution, orthogonal to $u_{\zeta_{0}}$ for all $\sigma$, of
$$(\mathfrak{L}_{0}-\nu_{0})\hat\an_{2}=\nu_{2}u_{\zeta_{0}} +\frac{\nu_{1}''(\zeta_{0})}{2}\Phi'^2 u_{\zeta_{0}}+\kappa(\sigma)\left(-\dr_{\tau}u_{\zeta_{0}}-2(\zeta_{0}-\tau)^2\tau u_{\zeta_{0}}-\tau^2(\zeta_{0}-\tau)u_{\zeta_{0}}\right)\,,$$
and $f_{2}$ has to be determined.\\
Finally we must solve the fourth equation given by
$$(\mathfrak{L}_{0}-\lambda_{0})\an_{3}=(\lambda_{3}-\mathfrak{L}_{3})\an_{0}+(\lambda_{2}-\mathfrak{L}_{2})\an_{1}-\mathfrak{L}_{1}\an_{2}\,.$$
The Fredholm condition provides the following equation in the variable $\sigma$:
$$\langle \mathfrak{L}_{3}\an_{0}+(\mathfrak{L}_{2}-\lambda_{2})\an_{1}+\mathfrak{L}_{1}\an_{2} ,u_{\zeta_{0}}\rangle_{\sL^2(\R_{+},\dx\tau)}=\lambda_{3}f_{0}\,.$$
Using the previous steps of the construction, it is not very difficult to see that this equation does not involve $f_{1}$ and $f_{2}$ (due to the choice of $\Phi$ and $\lambda_{2}$ and Feynman-Hellmann formulas). Using the same formulas, we may write it in the form
\begin{equation}\label{transport-FH}
\frac{\nu_{1}''(\zeta_{0})}{2}\left(\Phi'(\sigma)\dr_{\sigma}+\dr_{\sigma}\Phi'(\sigma)\right)f_{0}+F(\sigma)f_{0}=\lambda_{3}f_{0}\,,
\end{equation}
where $F$ is a smooth function which vanishes at $\sigma=0$. Therefore the linearized equation at $\sigma=0$ is given by
$$\Phi''(0)\frac{\nu_{1}''(\zeta_{0})}{2}\left(\sigma\dr_{\sigma}+\dr_{\sigma}\sigma\right)f_{0}=\lambda_{3}f_{0}\,.$$
We recall that
$$\frac{\nu_{1}''(\zeta_{0})}{2}=3C_{1}\Theta_{0}^{1/2}$$
so that the linearized equation becomes
$$C_{1}\Theta_{0}^{1/4}\sqrt{\frac{3k_{2}}{2}}\left(\sigma\dr_{\sigma}+\dr_{\sigma}\sigma\right)f_{0}=\lambda_{3}f_{0}\,.$$
We have to choose $\lambda_{3}$ in the spectrum of this transport equation, which is given by the set
$$\left\{(2n-1)C_{1}\Theta_{0}^{1/4}\sqrt{\frac{3k_{2}}{2}},\quad n\geq 1\right\}\,.$$
If $\lambda_{3}$ belongs to this set, we may solve locally the transport equation \eqref{transport-FH} and thus find $f_{0}$. This procedure can be continued at any order.

%\mainmatter
\part{Magnetic wells in dimension two}\label{Part.MW}

%Commands
\newcommand{\new}{\mathsf{new}}

\chapter[Vanishing magnetic fields in dimension two]{Vanishing magnetic fields in dimension two}\label{chapter-vanishing}

\begin{flushright}
\begin{minipage}{0.5\textwidth}
For it is not from any sureness in myself that I cause others to doubt: it is from being in more doubt than anyone else that I cause doubt in others.
\begin{flushright}
\textit{Meno}, Plato
\end{flushright}
\vspace*{0.5cm}
\end{minipage}
\end{flushright}

This chapter presents the main elements of the proof of Theorem \ref{main-result-vanishing}. We provide a flexible and \enquote{elementary} proof which can be adapted to other situations, especially less regular situations as in Chapter \ref{chapter-edge}. A more conceptual proof, using a WKB method, is possible by using the material introduced in Chapter \ref{chapter-BOM}, Section \ref{BOM-mag}. Nevertheless, the approach chosen for this chapter has the interest to reduce explicitly the spectral analysis to an electric Laplacian in the Born-Oppenheimer form. In particular, we do not need the notions of coherent states or of microlocalization. 

\section{Normal form}

\subsection{Toward a normal form}
Let us start with an exercise.
\begin{exe}
We recall that $\Phi : (s,t)\mapsto c(s)+t\n(s)$ defines a local diffeomorphism near $(s,t)=(0,0)$. We let $m(s,t)=1-t\kappa(s)$ and we use tildes to indicate that we consider a function in the variables $(s,t)$.
\begin{enumerate}[(i)]
\item Prove that, for all smooth function supported near $(0,0)$, the quadratic form $\mathfrak{Q}_{h,\A}$ becomes
\[\tilde{\mathfrak{Q}}_{h,\A}(\psi)=\int \left(|hD_{t}\psi|^2+(1-t\kappa(s))^{-2}|\tilde P\psi|^2\right)\,m(s,t)\dx s\dx t\,,\]
where (read Chapter \ref{intro}, Section \ref{sec.MBtoB})
\begin{equation}\label{eq.tildeA}
\tilde P=hD_{s}-\tilde A (s,t),\qquad \tilde A(s,t)=\int_{0}^t (1-\kappa(s)t')\tilde \B(s,t')\dx t'\,.
\end{equation}
\item Prove that, near $(0,0)$, the operators become 
\[\widetilde{\mathfrak{L}}_{h,\A} = h^2(1-t\kappa(s))^{-1}D_{t}(1-t\kappa(s))D_{t}+(1-t\kappa(s))^{-1}\tilde P(1-t\kappa(s))^{-1}\tilde P\,.\]
\end{enumerate}
\end{exe}

By a change of function (see \cite[Theorem 18.5.9 and below]{Hor07}), we are led to the following operator on $\sL^2(\R^2)$ that is unitarily equivalent to $\widetilde{\mathfrak{L}}_{h,\A}$:
$$\mathfrak{L}^{\new}_{h,\A}=m^{1/2}\widetilde{ \mathfrak{L}}_{h,\A} m^{-1/2}=P_{1}^2+P_{2}^2-\frac{h^2 \kappa(s)^2}{4m^2}\,,$$
with $P_{1}=m^{-1/2}(hD_{s}-\tilde A(s,t))m^{-1/2}$ and $P_{2}=hD_{t}$.

We wish to use a system of coordinates more adapted to the magnetic situation.
Let us perform a Taylor expansion near $t=0$. We have:
$$\tilde \B(s,t)=\gamma(s)t+\dr^2_{t}\tilde\B(s,0)\frac{t^2}{2}+\mathcal O(t^3)\,.$$
This provides:
\begin{equation}\label{eq.dltildeA}
\tilde A(s,t)=\frac{\gamma(s)}{2}t^2+k(s)t^3+\mathcal O(t^4)\,,
\end{equation}
with
$$k(s)=\frac{1}{6}\dr^2_{t}\tilde\B(s,0)-\frac{\kappa(s)}{3}\gamma(s)\,.$$
This suggests, as for the model operator, to introduce the new magnetic coordinates in a fixed neighborhood of $(0,0)$:
$$\check t=\gamma(s)^{1/3}t, \quad\check s=s\,.$$
This change of variable is fundamental in the analysis of the models introduced in Chapter \ref{chapter-BOM}, Section \ref{BOM-mag}.
The change of coordinates for the derivatives is given by:
$$D_{t}= \gamma(\check{s})^{1/3}D_{\check{t}}, \quad D_{s}=D_{\check{s}}+\frac 1 3 \gamma'\gamma^{-1}\check{t} D_{\check{t}}\,.$$
The space $\sL^2(\dx s\dx t)$ becomes $\sL^2(\gamma(\check{s})^{-1/3}\dx\check{s} \dx\check{t})$.
In the same way as previously, we shall conjugate $\mathfrak{L}^{\new}_{h,\A}$. We introduce the self-adjoint operator on $\sL^2(\R^2)$:
$$\check {\mathfrak{L}}_{h,\A}=\gamma^{-1/6}\mathcal{L}^{\new}_{h,\A}\gamma^{1/6}\,.$$
We deduce:
$$\check{\mathfrak{L}}_{h,\A}=h^2 \gamma(\check{s})^{2/3}D_{\check{t}}^2+\check P^2\,,$$
where
$$\check P=\gamma^{-1/6}\check m^{-1/2}\left(hD_{\check{s}}-\check A(\check{s},\check{t})+h\frac 1 3 \gamma'\gamma^{-1}\check{t} D_{\check{t}}\right)\check m^{-1/2}\gamma^{1/6}\,,$$
with:
$$\check A(\check{s},\check{t})=\tilde A(\check{s},\gamma(\check{s})^{-1/3}\check{t})\,.$$
A straight forward computation provides
$$\check P= \check m^{-1/2}\left(hD_{\check{s}}-\check A(\check{s},\check{t})+h\frac 1 6 \gamma'\gamma^{-1}(\check{t} D_{\check{t}}+D_{\check{t}}\check{t})\right)\check m^{-1/2}\,,$$
where we make the generator of dilations $\check{t} D_{\check{t}}+D_{\check{t}}\check{t}$ to appear (and which is related to the virial theorem, see \cite{Ray10c, Ray11b} where this theorem is often used).
Up to a change of gauge, we can replace $\check P$ by
$$\check m^{-1/2}\left(hD_{\check{s}}+\zeta^{[1]}_{0}(\gamma(\check{s}))^{1/3}h^{2/3}-\check A(\check{s},\check{t})+h\frac 1 6 \gamma'\gamma^{-1}(\check{t} D_{\check{t}}+D_{\check{t}}\check{t})\right)\check m^{-1/2}\,.$$
\subsection{Normal form $\check{\mathfrak{L}}_{h,\A}$}
Therefore, the operator takes the form \enquote{\`a la H¬\"ormander}:
\begin{equation}\label{normal-form}
\check{\mathfrak{L}}_{h,\A}=P_{1}(h)^2+P_{2}(h)^2-\frac{h^2 \kappa(\check{s})^2}{4m(\check{s},\gamma(\check{s})^{1/3}\check{t})^2}\,,
\end{equation}
where
\begin{align*}
&P_{1}(h)=\check m^{-1/2}\left(hD_{\check{s}}+\zeta^{[1]}_{0}(\gamma(\check{s}))^{1/3}h^{2/3}-\check A(\check{s},\check{t})+h\frac 1 6 \gamma'\gamma^{-1}(\check{t} D_{\check{t}}+D_{\check{t}}\check{t})\right)\check m^{-1/2},\\
&P_{2}(h)=h \gamma(\check{s})^{1/3}D_{\check{t}}.
\end{align*}
Computing a commutator, we can rewrite $P_{1}(h)$ as
\begin{align}\label{P1}
P_{1}(h)=\check m^{-1}\left(hD_{\check{s}}+\zeta^{[1]}_{0}(\gamma(\check{s}))^{1/3}h^{2/3}-\check A(\check{s},\check{t})+h\frac 1 6 \gamma'\gamma^{-1}(\check{t} D_{\check{t}}+D_{\check{t}}\check{t})\right)+C_{h},              
\end{align}
where:
$$C_{h}=h\check m^{-1/2}(D_{\check{s}}\check m^{-1/2})+\frac{h\gamma'\gamma^{-1}}{3}\check{t} \check m^{-1/2} (D_{\check{t}} \check m^{-1/2})\,.$$
\begin{notation}
The quadratic form corresponding to $\check{\mathcal{L}}_{h}$ will be denoted by $\check{\mathcal{Q}}_{h}$.
\end{notation}

\subsection{Quasimodes}
We can construct quasimodes using the classical recipe (see Chapter \ref{chapter-BOM}) involving the scaling 
\begin{align}\label{scaling}
\check{t}=h^{1/3}\tau, \quad \check{s}= h^{1/6}\sigma,
\end{align}
and the Feynman-Hellmann formulas.
\begin{notation}
The operator $h^{-4/3}\check{\mathfrak{L}}_{h,\A}$ will be denoted by $\mathcal{L}_{h}$ in these rescaled coordinates.
\end{notation}
This provides the following proposition.
\begin{prop}\label{theo-quasimodes}
We assume \eqref{Assumption-vanishing}. For all $n\geq 1$, there exist a sequence $(\theta_{j}^n)_{j\geq 0}$ such that, for all $J\geq 0$, there exists  $h_{0}>0$ such that, for $h\in(0,h_{0})$, we have:
$$\dist\left(h^{4/3}\sum_{j=0}^J \theta^n_{j} h^{j/6}, \sp(\mathfrak{L}_{h,\A})\right)\leq Ch^{4/3}h^{(J+1)/6}\,.$$
Moreover, we have:
$$\theta_{0}^{n}=\gamma_{0}^{2/3}\nu^{[1]}_{1}(\zeta^{[1]}_{0}), \quad \theta_{1}^n=0, \quad \theta_{2}^n=\gamma_{0}^{2/3}C_{0}+\gamma_{0}^{2/3}(2n-1)\left(\frac{\alpha\nu^{[1]}_{1}(\zeta^{[1]}_{0})(\nu^{[1]}_{1})''(\zeta^{[1]}_{0})}{3}\right)^{1/2}\,.$$
\end{prop}
Thanks to the localization formula and a partition of unity, we may prove the following proposition.
\begin{prop}\label{prop.reflbvan}
For all $n\geq 1$, there exist $h_{0}>0$ and $C>0$ such that, for $h\in(0,h_{0})$,
$$\lambda_{n}(h)\geq \gamma_{0}^{2/3}\nu^{[1]}_{1}(\zeta^{[1]}_{0})h^{\frac{4}{3}}-Ch^{\frac{4}{3}+\frac{2}{15}}\,.$$
\end{prop}
\begin{proof}
We use a partition of unity (see Chapter \ref{chapter-models}, Section \ref{Sec.Agmon0}) with balls of size $h^{\rho}$:
$$\sum_{j} \chi_{j,h}^2=1$$
and such that:
$$\sum_{j} |\nabla\chi_{j,h}|^2\leq C h^{-2\rho}\,.$$
We let
$$\mathcal{B}_{j,h}=\supp \chi_{j,h}\,.$$
If $\lambda$ is an eigenvalue and $\psi$ a corresponding eigenfunction, we have the localization formula
$$\sum_{j} \mathfrak Q_{h, \A}(\chi_{j,h}\psi)-h^{2}\|\nabla\chi_{j,h} \psi\|^2=\lambda\sum_{j}\|\chi_{j,h}\psi\|^2\,.$$
We distinguish between the balls which intersect $t=0$ and the others so that we introduce:
\[J_{1}(h)=\{j : \mathcal{B}_{j,h}\cap\mathcal{C}\neq \emptyset\},\qquad J_{2}(h)=\{j : \mathcal{B}_{j,h}\cap\mathcal{C}= \emptyset\}\,.\]
If $j\in J_{2}(h)$ , we use the inequality of Lemma \ref{lb-magnetic} combined with the non-degeneracy of the cancellation of $\B$ and Assumption \ref{limit-beta}. We deduce the existence of $h_{0}>0$ and $c>0$ such that, for $h\in(0,h_{0})$,
\[\mathfrak Q_{h,\A}(\chi_{j,h}\psi)\geq h\left|\int \B(\x)|\chi_{j,h}\psi|^2\,\dx\x\right|\geq ch^{1+\rho}\|\chi_{j,h}\psi\|^2\,.\]
If $j\in J_{1}(h)$, we write:
\[\mathfrak Q_{h,\A}(\chi_{j,h}\psi)\geq (1-Ch^{\rho})\int |h\dr_{t}(\chi_{j,h}\psi)|^2+|(ih\dr_{s}+\tilde A)(\chi_{j,h}\psi)|^2\,\dx s\dx t\,,\]
where $\tilde A$ is defined in \eqref{eq.tildeA}. Thanks to a Taylor expansion (see \eqref{eq.dltildeA}), we infer, for all $\eps\in(0,1)$,
\begin{align*}
&\mathfrak Q_{h,\A}(\chi_{j,h}\psi)\geq\\
&(1-Ch^{\rho})\left((1-\eps)\int |h\dr_{t}(\chi_{j,h}\psi)|^2+|(ih\dr_{s}+\frac{\gamma(s_{j})t^2}{2})(\chi_{j,h}\psi)|^2\,\dx s\dx t-\frac{Ch^{6\rho}}{\eps}\|\chi_{j,h}\psi\|^2\right)\,,
\end{align*}
and we deduce (see Section \ref{subsec.Mont}):
\[\mathfrak Q_{h,\A}(\chi_{j,h}\psi)\geq (1-Ch^{\rho})\left((1-\eps)h^{4/3}\nu^{[1]}_{1}(\zeta^{[1]}_{0})\gamma_{j}^{2/3}\|\chi_{j,h}\psi\|^2-\eps^{-1}Ch^{6\rho}\|\chi_{j,h}\psi\|^2\right)\,.\]
Optimizing with respect to $\eps$, we choose: $\eps=h^{3\rho-\frac{2}{3}}.$ Then, we take $\rho$ such that $2-2\rho=3\rho+\frac{2}{3}$ and we deduce $\rho=\frac{4}{15}$.
\end{proof}

\section{Agmon estimates}
Two kinds of Agmon's estimates can be proved by using standard partition of unity arguments. 
\begin{prop}\label{Agmon-normal}
Let $(\lambda,\psi)$ be an eigenpair of $\mathfrak{L}_{h,\A}$. There exist $h_{0}>0$, $C>0$ and $\eps_{0}>0$ such that, for $h\in(0,h_{0})$:
\begin{equation}\label{L2-normal}
\int e^{2\eps_{0}|t(\x)|h^{-1/3}}|\psi|^2\dx \x\leq C\|\psi\|^2
\end{equation}
and
\begin{equation}\label{H1-normal}
\mathfrak{Q}_{h,\A}( e^{\eps_{0}|t(\x)|h^{-1/3}}\psi)\leq Ch^{4/3}\|\psi\|^2.
\end{equation}
\end{prop}
\begin{proof}
Let us consider an eigenpair $(\lambda,\psi)$ of $\mathfrak{L}_{h,\A}$. We begin to write the localization formula:
\begin{equation}\label{IMS-normal}
\mathfrak Q_{h, \A}(e^{\Phi}\psi)=\lambda\|e^\Phi\psi\|^2+h^2\|\nabla\Phi  e^{\Phi}\psi\|^2.
\end{equation}
We use a partition of unity with balls of size $Rh^{1/3}$:
$$\sum_{j} \chi_{j,h}^2=1$$
and such that:
$$\sum_{j} |\nabla\chi_{j,h}|^2\leq C R^{-2} h^{-2/3}.$$
We may assume that the balls which intersect the line $t=0$ have their centers on it.
Using again the localisation formula, we get the decomposition into local "energies":
\[\sum_{j} \mathfrak Q_{h, \A}(\chi_{j,h}e^{\Phi}\psi)-\lambda\|\chi_{j,h}e^{\Phi}\psi\|^2-h^2\|\chi_{j,h}\nabla\Phi e^{\Phi}\psi\|^2-h^{2}\|\nabla\chi_{j,h} e^{\Phi}\psi\|^2=0\,.\]
We distinguish between the balls which intersect $t=0$ and the others:
\[J_{1}(h)=\{j : \mathcal{B}_{j,h}\cap\mathcal{C}\neq \emptyset\},\quad J_{2}(h)=\{j : \mathcal{B}_{j,h}\cap\mathcal{C}= \emptyset\}\,.\]
If $j\in J_{2}(h)$, we get the existence of $c>0$ (independent from $R$) and $h_{0}>0$ such that, for $h\in(0,h_{0})$,
$$\mathfrak Q_{h,\A}(\chi_{j,h}e^{\Phi}\psi)\geq h\left|\int \B(\x)|\chi_{j,h}e^{\Phi}\psi|^2\dx\x\right|\geq cRh^{4/3}\|\chi_{j,h}e^{\Phi}\psi\|^2\,.$$
If $j\in J_{1}(h)$, we write
$$\mathfrak Q_{h,\A}(\chi_{j,h}e^{\Phi}\psi)\geq (1-CRh^{1/3})\left((1-\eps)h^{4/3}\nu^{[1]}_{1}(\zeta^{[1]}_{0})\gamma_{j}^{2/3}-\eps^{-1}Ch^{2}\||\chi_{j,h}e^{\Phi}\psi\|^2\right)\,.$$
We take $\eps=h^{1/3}$. 
We use Proposition \ref{theo-quasimodes} to get that $\lambda_{n}(h)\leq \gamma_{0}^{2/3}\nu^{[1]}_{1}(\zeta^{[1]}_{0})h^{\frac{4}{3}}+Ch^{\frac{5}{3}}$. We are led to choose $\Phi(\x)=\eps_{0}|t(\x)|h^{-1/3}$ so that
$$h^2|\nabla\Phi|^2\leq h^{4/3}\eps_{0}^2\,.$$
Taking $\eps_{0}$ small enough and $R$ large enough, we infer the existence of $\tilde c>0, C>0$ and $h_{0}>0$ such that, for $h\in(0,h_{0})$,
$$\tilde c h^{4/3} \sum_{j\in J_{1}(h)}\int e^{2\Phi}|\chi_{j,h}\psi|^2\dx\x\leq Ch^{4/3}\sum_{j\in J_{2}(h)} \int e^{2\Phi}|\chi_{j,h}\psi|^2\dx\x\,.$$
Then, due to the support of $\chi_{j,h}$ when $j\in J_{2}(h)$, we infer:
$$\sum_{j\in J_{2}(h)} \int e^{2\Phi}|\chi_{j,h}\psi|^2\,\dx\x\leq \tilde C\sum_{j\in J_{2}(h)}\int |\chi_{j,h}\psi|^2\dx\x\,.$$
We deduce \eqref{L2-normal}. Finally, \eqref{H1-normal} follows from \eqref{L2-normal} and \eqref{IMS-normal}.
\end{proof}
By using the same method, we can prove the following localization with respect to the tangential variable $s$ (here we use the fact that $\gamma$ is non degenerately minimal at $s=0$).
\begin{prop}\label{rl}
Let $(\lambda,\psi)$ be an eigenpair of $\mathfrak{L}_{h,\A}$. There exist $h_{0}>0$, $C>0$ and $\eps_{0}>0$ such that, for $h\in(0,h_{0})$:
\begin{equation}\label{rl-L2}
\int e^{2\chi(t(\x))|s(\x)|h^{-1/15}}|\psi|^2\dx \x\leq C\|\psi\|^2
\end{equation}
and
\begin{equation}\label{rl-H1}
\mathfrak{Q}_{h,\A}( e^{\chi(t(\x))|s(\x)|h^{-1/15}}\psi)\leq Ch^{4/3}\|\psi\|^2,
\end{equation}
where $\chi$ is a fixed smooth cutoff function being $1$ near $0$.
\end{prop}
From Propositions \ref{Agmon-normal} and \ref{rl}, we are led to introduce a cutoff function living near $x_{0}$. We take $\eps>0$ and we let:
$$\chi_{h,\eps}(\x)=\chi\left(h^{-1/3+\eps}t(\x)\right)\chi\left(h^{-1/15+\eps}s(\x)\right)\,.$$
where $\chi$ is a fixed smooth cutoff function supported near $0$.
\begin{notation}
We will denote by $\check\psi$ the function $\chi_{h,\eps}(\x)\psi(\x)$ in the coordinates $(\check{s},\check{t})$. 
\end{notation}
The following exercise aims at proving some a priori estimates on the truncated eigenfunctions in the coordinates $(\check s,\check t)$. They will be quite convenient in the rest of the proof.
\begin{exe}\label{exo-Agmon-t}
Let $\psi_{n,h}$ be a $\sL^2$-normalized eigenfunction associated with $\lambda_{n}(h)$. 
\begin{enumerate}
\item By using the estimates of Agmon, show that we have 
$$\check{\mathfrak{Q}}_{h}(\check\psi_{n,h})=\lambda_{n}(h)\| \check\psi_{n,h}\|^2+\mathcal{O}(h^{\infty})\,.$$
\item By applying the usual localization procedure to $\langle P_{j}(h)^2\check\psi_{n,h},\check t^{2k} \check\psi_{n,h}\rangle$, prove that, for all $k\geq 1$,
$$\check{\mathfrak{Q}}_{h}(\check t^k\check\psi_{n,h})\leq\lambda_{n}(h)\|\check t^k\check\psi_{n,h}\|^2+Ch^2\|\check t^{k-1}\check\psi_{n,h}\|^2+Ch^2\|\check t^k\check\psi_{n,h}\|^2+\mathcal{O}(h^\infty)\,.$$
\item By using the estimates of Agmon, deduce that, for all $k\geq 1$, 
$$\check{\mathfrak{Q}}_{h}(\check t^k\check\psi_{n,h})=\mathcal{O}(h^{\frac{4}{3}}h^{\frac{2k}{3}})\,.$$
\item Prove that for all $\check\psi$ in the domain of $\check{\mathfrak{Q}}_{h}$ and supported in a region $|\check t|\leq Ch^{\frac{1}{3}-\eps}$, we have
$$\check{\mathfrak{Q}}_{h}(\check\psi)\geq\frac{1}{2}\|hD_{\check s}\check\psi\|^2+(\gamma_{0}^{\frac{2}{3}}-Ch^{\frac{2}{3}-2\eps})\|hD_{\check t}\check\psi\|^2-Ch^{\frac{4}{3}}\|\check\psi\|^2-C\|\check t^2\check\psi\|^2-Ch^2\|\check t\check\psi\|^2\,.$$
\item Deduce that, for all $k\geq 1$, we have
$$\|hD_{\check t}(\check t^{k}\check\psi_{n,h})\|^2=\mathcal{O}(h^{\frac{4}{3}}h^{\frac{2k}{3}}),\qquad \|hD_{\check s}(\check t^{k}\check\psi_{n,h})\|^2=\mathcal{O}(h^{\frac{4}{3}}h^{\frac{2k}{3}})\,.$$
\end{enumerate}
\end{exe}
Let us now establish the following proposition.
\begin{prop}\label{rlb}
For all $n\geq 1$, there exist $h_{0}>0$ and $C>0$ s. t., for $h\in(0,h_{0})$:
$$\lambda_{n}(h)\geq \gamma_{0}^{2/3}\nu^{[1]}_{1}(\zeta^{[1]}_{0})h^{4/3}-Ch^{5/3}\,.$$
Moreover, we have
$$\|\check s\check\psi_{n,h}\|\leq Ch^{\frac{1}{6}}\|\check\psi_{n,h}\|\,.$$
\end{prop}
\begin{proof}
We use the notations and the results of Exercise \ref{exo-Agmon-t}. We write
$$\check{\mathfrak{Q}}_{h}(\check \psi_{n,h})=\lambda_{n}(h)\|\check \psi_{n,h}\|^2+\mathcal{O}(h^{\infty})\|\check \psi_{n,h}\|^2\,.$$
Then, we have
\begin{align*}
&\check{\mathfrak{Q}}_{h}(\check \psi_{n,h})\geq\\
&\int \check m^{-2}\left|\left(hD_{\check{s}}+\zeta^{[1]}_{0}\gamma^{1/3}h^{2/3}-\check A+\frac{h}{6}\gamma'\gamma^{-1}(\check{t} D_{\check{t}}+D_{\check{t}}\check{t})+C_{h}\right)\check \psi_{n,h}\right|^2\dx\check{s}\dx\check{t}\\
                                                         &+h^2 \|\gamma^{1/3}D_{\check{t}}\check \psi_{n,h}\|^2-Ch^2\|\check \psi_{n,h}\|^2.
\end{align*}
Let us now use a Taylor expansion the get rid of the metrics $\check m$. The remainder can be controlled with the results of Exercise \ref{exo-Agmon-t} and we get
\begin{align}\label{rlb-metrics}
&\check{\mathfrak{Q}}_{h}(\check \psi_{n,h})\geq\\
\nonumber&\int \left|\left(hD_{\check{s}}+\zeta^{[1]}_{0}\gamma^{1/3}h^{2/3}-\check A+\frac{h}{6}\gamma'\gamma^{-1}(\check{t} D_{\check{t}}+D_{\check{t}}\check{t})+C_{h}\right)\check \psi_{n,h}\right|^2\dx\check{s}\dx\check{t}\\
\nonumber                                                         &+h^2 \|\gamma^{1/3}D_{\check{t}}\check \psi_{n,h}\|^2-Ch^{\frac{5}{3}}\|\check \psi_{n,h}\|^2.
\end{align}
Expanding the square, we get
\begin{multline}\label{rlb-A}
\check{\mathfrak{Q}}_{h}(\check \psi_{n,h})\geq(1-\eta)\int \left|\left(hD_{\check{s}}+\zeta^{[1]}_{0}\gamma^{1/3}h^{2/3}-\gamma^{\frac{1}{3}}\frac{\check t^2}{2}\right)\check \psi_{n,h}\right|^2\dx\check{s}\dx\check{t}\\
-C\eta^{-1}\left(\|\check t^3\check\psi_{n,h}\|^2+h^{2+\frac{2}{15}}\|\check\psi_{n,h}\|^2\right)+h^2 \|\gamma^{1/3}D_{\check{t}}\check \psi_{n,h}\|^2-Ch^{\frac{5}{3}}\|\check \psi_{n,h}\|^2,
\end{multline}
where we have used that $0$ is a critical point of $\gamma$ as well as the size of the support in $\check s$.
We choose $\eta=h^{\frac{1}{3}}$ and we find
\begin{multline*}
\check{\mathfrak{Q}}_{h}(\check \psi_{n,h})\geq(1-h^{\frac{1}{3}})\int \left|\left(hD_{\check{s}}\gamma^{-\frac{1}{3}}+\zeta^{[1]}_{0}h^{2/3}-\frac{\check t^2}{2}\right)\gamma^{\frac{1}{3}}\check \psi_{n,h}\right|^2\dx\check{s}\dx\check{t}\\
+h^2 \|\gamma^{1/3}D_{\check{t}}\check \psi_{n,h}\|^2-Ch^{\frac{5}{3}}\|\check \psi_{n,h}\|^2.
\end{multline*}
Then, we write the symmetrization 
$$D_{\check{s}}\gamma^{-1/3}=\gamma^{-1/6}D_{\check{s}}\gamma^{-1/6}-i\gamma^{-1/6}(\gamma^{-1/6})'\,.$$
Then we estimate the double product involved by $i\gamma^{-1/6}(\gamma^{-1/6})'$ to get
\begin{multline*}
\int \left|\left(hD_{\check{s}}\gamma^{-\frac{1}{3}}+\zeta^{[1]}_{0}h^{2/3}-\frac{\check t^2}{2}\right)\gamma^{\frac{1}{3}}\check \psi_{n,h}\right|^2\dx\check{s}\dx\check{t}\\
\geq \int\left|\left(h\gamma^{-1/6}D_{\check{s}}\gamma^{-1/6}+\zeta^{[1]}_{0}h^{2/3}-\frac{\check{t}^2}{2}\right)\gamma^{\frac{1}{3}}\check \psi_{n,h}\right|^2\dx\check{s}\dx\check{t}-Ch^2\|\psi_{n,h}\|^2.
\end{multline*}
We deduce that
\begin{align}\label{lower-bound-optimal}
\nonumber\check{\mathfrak{Q}}_{h}(\check \psi_{n,h})\geq& h^2 \|\gamma^{\frac{1}{3}}D_{\check{t}}\check \psi_{n,h}\|^2+\int\left|\left(h\gamma^{-1/6}D_{\check{s}}\gamma^{-1/6}+\zeta^{[1]}_{0}h^{2/3}-\frac{\check{t}^2}{2}\right)\gamma^{\frac{1}{3}}\check \psi_{n,h}\right|^2\dx\check{s}\dx\check{t}\\
                                                         &-Ch^{\frac{5}{3}}\|\check \psi_{n,h}\|^2.
\end{align}
We can apply the functional calculus to the self-adjoint operator $\gamma^{-1/6}D_{\check{s}}\gamma^{-1/6}$ (see Exercise \ref{funct-calc}) and the following lower bound follows
\begin{align*}
\check{\mathfrak{Q}}_{h}(\check \psi_{n,h})\geq& h^{4/3}\nu^{[1]}_{1}(\zeta^{[1]}_{0})\|\gamma^{\frac{1}{3}}\check\psi_{n,h}\|^2-Ch^{5/3}\|\check \psi_{n,h}\|^2.
\end{align*}
This implies the lower bound for $\lambda_{n}(h)$. Since $\lambda_{n}(h)\leq \gamma_{0}^{\frac{1}{3}}\nu^{[1]}_{1}(\zeta^{[1]}_{0})h^{\frac{4}{3}}+Ch^{\frac{5}{3}}$, we get
$$\int \left(\gamma(\check s)^{\frac{1}{3}}-\gamma_{0}^{\frac{2}{3}}\right)|\check\psi_{n,h}|^2\dx\check s\dx\check t\leq Ch^{\frac{1}{3}}$$
and it remains to use the non degeneracy of the minimum of $\gamma$ at $0$.
\end{proof}
For all $N\geq 1$, let us consider a $\sL^2$-orthonormalized family $(\psi_{n,h})_{1\leq n\leq N}$ where $\psi_{n,h}$ is an eigenfunction associated with $\lambda_{n}(h)$. We consider the $N$ dimensional space defined by:
$$\mathfrak{E}_{N}(h)=\underset{1\leq n\leq N}{\spann} \check\psi_{n,h}\,.$$ 
An easy consequence of Proposition \ref{rlb} gives the following.
\begin{prop}\label{loc-sigma}
There exist $h_{0}>0$, $C>0$ such that, for $h\in(0,h_{0})$ and for all $\check\psi\in\mathfrak{E}_{N}(h)$:
$$\|\check{s}\check\psi\|\leq Ch^{1/6}\|\check\psi\|\,.$$
\end{prop}
With Proposition \ref{loc-sigma}, we have a better lower bound for the quadratic form.
\begin{prop}\label{improved-lb}
There exists $h_{0}>0$ such that for $h\in(0,h_{0})$ and $\check\psi\in\mathfrak{E}_{N}(h)$:
\begin{align*}
\check{\mathcal{Q}}_{h}(\check\psi)\geq&\gamma_{0}^{2/3}\int(1+2 \kappa_{0}\check{t}\gamma_{0}^{-1/3})|(\gamma^{-1/6}hD_{\check{s}}\gamma^{-1/6}+\zeta^{[1]}_{0}h^{2/3}-\frac{\check{t}^2}{2}-\gamma_{0}^{-4/3}k(0)\check{t}^3)\check\psi|^2\dx\check{s}\dx\check{t}\\
&+\int \gamma_{0}^{2/3}|hD_{\check{t}}\check\psi|^2\dx\check{s} \dx\check{t}+\frac{2}{3}\gamma_{0}^{2/3}\alpha \nu^{[1]}_{1}(\zeta^{[1]}_{0})h^{4/3}\|\check{s}\check\psi\|^2+o(h^{5/3})\|\check\psi\|^2,
\end{align*}
where $\alpha$ is defined in \eqref{alpha-vanishing}.
\end{prop}
\begin{proof}
Let us only indicate the changes that have to be made in the proof of Proposition \ref{rlb}. We shall keep the next term in the expansion of the metrics in \eqref{rlb-metrics}. In \eqref{rlb-A} we also keep one more term in the expansion of $\check A$ and we may choose a slightly smaller $\eta$.
\end{proof}
\section{Projection argument}
In this section, we establish a dimensional reduction. For that purpose, one needs a localization result for $D_{\check s}$.
\begin{prop}\label{cor-ds}
There exist $h_{0}>0$, $C>0$ such that, for $h\in(0,h_{0})$ and for all $\check\psi\in\mathfrak{E}_{N}(h)$:
$$\|D_{\check{s}}\check\psi\|\leq Ch^{-1/6}\|\check\psi\|\,.$$
\end{prop}
\begin{proof}
We only give some hints for the proof. We recall \eqref{lower-bound-optimal} and Exercise \ref{funct-calc} and we get
\begin{equation}\label{loc-van}
h^{\frac{4}{3}}\int \left(\nu^{[1]}_{1}(\zeta^{[1]}_{0}+h^{\frac{1}{3}}\zeta)-\nu^{[1]}_{1}(\zeta^{[1]}_{0})\right)|\check\phi|^2 \dx\zeta \dx\check t\leq Ch^{\frac{5}{3}}\|\check\phi\|^2
\end{equation}
where
$$\check\phi=\mathcal{F}_{\gamma^{-\frac{1}{6}}}(\gamma^{\frac{1}{3}}\check\psi)\,.$$
Choosing $\eps_{0}>0$ small enough and using the uniqueness and non-degeneracy of the minimum of $\nu^{[1]}_{1}$, we get
$$\int_{|h^\frac{1}{3}\zeta|\leq\eps_{0}} |h^{\frac{1}{3}}\zeta|^2\left|\check\phi\right|^2\dx\zeta \dx\check t \leq Ch^{\frac{1}{3}}\|\check\phi\|^2$$
and
$$\int_{|h^\frac{1}{3}\zeta|\geq\eps_{0}} \left|\check\phi\right|^2\dx\zeta \dx\check t \leq Ch^{\frac{5}{3}}\|\check\phi\|^2\,.$$
By using the localization formula of Proposition \ref{P2A} and estimating 
$$\langle \check{\mathfrak{L}}_{h}\check\psi, \left(\gamma^{-\frac{1}{6}}D_{\check s}\gamma^{-\frac{1}{6}}\right)^2\check\psi\rangle$$
we may essentially replace $\check\psi$ by $\gamma^{-\frac{1}{6}}D_{\check s}\gamma^{-\frac{1}{6}}\check\psi$ in \eqref{loc-van} and deduce that
$$\int |\zeta|^2\left|\check\phi\right|^2\dx\zeta \dx\check t \leq Ch^{-\frac{1}{3}}\|\check\phi\|^2\,.$$
\end{proof}
We can now prove an approximation result for the eigenfunctions. Let us recall the rescaled coordinates (see \eqref{scaling}):
\begin{equation}
\check{s}=h^{1/6}\sigma, \quad \check{t}=h^{1/3} \tau.
\end{equation}
\begin{notation}
$\mathcal{L}_{h}$ denotes $h^{-4/3}\check{\mathfrak{L}}_{h}$ in the coordinates $(\sigma, \tau)$. The corresponding quadratic form will be denoted by $\mathcal{Q}_{h}$. We will use the notation $\mathcal{E}_{N}(h)$ to denote $\mathfrak{E}_{N}(h)$ after rescaling. 
\end{notation}
We introduce the Feshbach-Grushin projection:
$$\Pi_{0}\phi=\langle\phi,u^{[1]}_{\zeta^{[1]}_{0}}\rangle_{\sL(\R_{\tau})} u^{[1]}_{\zeta^{[1]}_{0}}(\tau)\,.$$
We will need to consider the quadratic form:
$$\hat{\mathcal{Q}}_{0} (\phi)=\gamma_{0}^{2/3}\int |D_{\tau}\phi|^2+\left|\left(-\zeta^{[1]}_{0}+\frac{\tau^2}{2}\right)\phi\right|^2\dx\sigma \dx\tau\,.$$
The fundamental approximation result is given in the following proposition.
\begin{prop}
There exist $h_{0}>0$ and $C>0$ such that for $h\in(0,h_{0})$ and $\hat\psi\in\mathcal{E}_{N}(h)$:
\begin{align}\label{approx-Dombrowski}
0\leq \mathcal{Q}_{0}(\hat\psi)-\gamma_{0}^{2/3}\nu^{[1]}_{1}(\zeta^{[1]}_{0})\|\hat\psi\|^2\leq Ch^{1/6}\|\hat\psi\|^2
\end{align}
and
\begin{align}
\label{Pi0-approx}&\|\Pi_{0}\hat\psi-\hat\psi\|\leq Ch^{1/12}\|\hat\psi\|\\
\nonumber&\|D_{\tau}(\Pi_{0}\hat\psi-\hat\psi)\|\leq Ch^{1/12}\|\hat\psi\|,\\
\nonumber&\|\tau^2(\Pi_{0}\hat\psi-\hat\psi)\|\leq Ch^{1/12}\|\hat\psi\|.
\end{align}
\end{prop}
This permits to simplify the lower bound.
\begin{prop}
There exist $h_{0}>0$, $C>0$ such that, for $h\in(0,h_{0})$ and  $\check\psi\in\mathfrak{E}_{N}(h)$:
\begin{align*}
\check{\mathfrak{Q}}_{h}(\check\psi)&\geq \int \gamma_{0}^{2/3}\left(|hD_{\check{t}}\check\psi|^2+|(\gamma^{-1/6}hD_{\check{s}}\gamma^{-1/6}+\zeta^{[1]}_{0}h^{2/3}-\frac{\check{t}^2}{2})\check\psi|^2\right)\dx\check{s} \dx\check{t}\\ 
&+\frac{2}{3}\gamma_{0}^{2/3}\alpha \nu^{[1]}_{1}(\zeta^{[1]}_{0})h^{4/3}\|\check{s}\check\psi\|^2+C_{0}h^{5/3}\|\check\psi\|^2+o(h^{5/3})\|\check\psi\|^2,
\end{align*}
where $C_{0}$ is defined in \eqref{C0-vanishing}.
\end{prop}
\begin{proof}
We leave the proof to the reader, the main idea being to approximate the \enquote{curvature terms} by their averages in the quantum state $u^{[1]}_{\zeta^{[1]}_{0}}$.
\end{proof}
It remains to diagonalize $\gamma^{-1/6}D_{\check{s}}\gamma^{-1/6}$.
\begin{cor}\label{last-cor}
There exist $h_{0}>0$, $C>0$ such that, for $h\in(0,h_{0})$ and  $\check\psi\in\mathfrak{E}_{N}(h)$:
\begin{align*}
\check{\mathcal{Q}}_{h}(\check\psi)&\geq \int \gamma_{0}^{2/3}\left(|hD_{\check{t}}\check\phi|^2+|(h\zeta+\zeta^{[1]}_{0}h^{2/3}-\frac{\check{t}^2}{2})\check\phi|^2\right)\dx\mu \dx\check{t}\\ 
&+\frac{2}{3}\gamma_{0}^{2/3}\alpha \nu_{1}(\zeta^{[1]}_{0})h^{4/3}\|D_{\zeta}\check\phi\|^2+C_{0}h^{5/3}\|\check\phi\|^2+o(h^{5/3})\|\check\phi\|^2,
\end{align*}
with $\check\phi=\mathcal{F}_{\gamma}\check\psi.$
\end{cor}
Let us introduce the operator on $\sL^2(\R^2, \dx\mu \dx\check{t})$:
\begin{equation}\label{BO-approx}
\frac{2}{3}\gamma_{0}^{2/3}\alpha \nu^{[1]}_{1}(\zeta^{[1]}_{0})h^{4/3}D_{\zeta}^2+\gamma_{0}^{2/3}\left(h^2D_{\check{t}}^2+\left(h\zeta+\zeta^{[1]}_{0}h^{2/3}-\frac{\check{t}^2}{2}\right)^2\right)+C_{0}h^{5/3}.
\end{equation}
\begin{exe} 
Determine the asymptotic expansion of the lowest eigenvalues of this operator thanks to the Born-Oppenheimer theory and prove the following theorem.
\end{exe}
\begin{theo}\label{spectral-gap-vanishing}
We assume \eqref{Assumption-vanishing}. For all $n\geq 1$, there exists $h_{0}>0$ such that for $h\in(0,h_{0})$, we have:
$$\lambda_{n}(h)\geq\theta_{0}^n h^{4/3}+\theta_{2}^n h^{5/3}+o(h^{5/3})\,.$$
\end{theo}
This implies Theorem \ref{main-result-vanishing}.

\chapter{Non vanishing magnetic fields}\label{chapter-birk}

\begin{flushright}
\textgreek{Mhde'is >agewm'etrhtos e>is'itw mou t`hn st'eghn.}
\vspace*{0.5cm}
\end{flushright}

This chapter is devoted to the elements of the proofs of Theorems \ref{main-theo-Bir} and \ref{spectrum} announced in Chapter \ref{intro-van-birk}, Section \ref{intro-birk}. Many ideas involved in this chapter may be found in Chapter \ref{chapter-appendix}.

\section{Magnetic Birkhoff normal form}\label{BNF}
In this section we prove Theorem \ref{main-theo-Bir}.
\subsection{Symplectic normal bundle of $\Sigma$}
We introduce the submanifold of all particles at rest:
$$\Sigma := H^{-1}(0) = \{(q,p); \qquad p = \A(q)\}\,.$$
Since it is a graph, it is an embedded submanifold of $\R^4$,
parameterized by $q\in\R^2$.
\begin{lem}\label{lem:Sigma}
$\Sigma$ is a symplectic submanifold of $\R^4$, in the sense that the restriction of $\omega_{0}$ to $\Sigma$ is a non degenerate $2$-form. In fact,
$$j^*\omega_{\upharpoonright\Sigma} = \dx\A \simeq B\,,$$
where $j:\R^2\to \Sigma$ is the embedding $j(q)=(q,\A(q))$.
\end{lem}
\begin{proof}
  We compute 
  $$j^*\omega = j^*(\dx p_1\wedge \dx q_1+\dx p_2\wedge \dx q_2) =(-\deriv{A_1}{q_2}+\deriv{A_2}{q_1}) \dx q_1\wedge \dx q_2\neq 0\,.$$
\end{proof}

Since we are interested in the low energy regime, we wish to describe
a small neighborhood of $\Sigma$ in $\R^4$, which amounts to
understanding the normal symplectic bundle of $\Sigma$. 
\begin{notation}
To avoid a confusion with the exterior derivative $\dx$, for any $X$ and differentiable function $f$, we denote by $T_X f$ the tangent map of f at $X$. 
\end{notation}
The vectors $(Q,T_q\A(Q))$, with $Q\in T_q\Omega=\R^2$, span the tangent space
$T_{j(q)}\Sigma$. It is interesting to notice that the symplectic
orthogonal $T_{j(q)}\Sigma^\perp$ is very easy to describe as well.
\begin{lem}
  \label{lem:basis}
  For any $q\in \Omega$, the vectors
  \[
  u_1 := \frac{1}{\sqrt{\abs{B}}}(e_1, (T_q\A)^{\mathsf{T}}(e_1));
  \quad v_1 := \frac{\sqrt{\abs{B}}}{B}(e_2, (T_q\A)^{\mathsf{T}}(e_2))
  \]
  form a symplectic basis of $T_{j(q)}\Sigma^\perp$.
\end{lem}
\begin{proof}
  Let $(Q_1,P_1)\in T_{j(q)}\Sigma$ and $(Q_2,P_2)$ with $P_2=(T_q\A)^{\mathsf{T}}(Q_2)$.  Then from~\eqref{equ:omega} we get
  \begin{align*}
    \omega_{0}((Q_1,P_1), (Q_2,P_2)) & = \pscal{T_q\A(Q_1)}{Q_2} -
    \pscal{(T_q\A)^{\mathsf{T}}(Q_2)}{Q_1}\\
    & = 0\,.
  \end{align*}
  This shows that $u_1$ and $v_1$ belong to
  $T_{j(q)}\Sigma^\perp$. Finally
  \begin{align*}
    \omega_{0}(u_1,v_1) & = \frac{1}{B}\left( \pscal{(T_q\A)^{\mathsf{T}}(e_1)}{e_2} -\pscal{(T_q\A)^{\mathsf{T}}(e_2)}{e_1}\right)\\
    & = \frac{1}{B} \pscal{e_1}{(T_q\A-(T_q\A)^{\mathsf{T}}(e_2)}\\
    & = \frac{1}{B}\pscal{e_1}{\vec B \wedge e_2} =
    -\frac{B}{B}\pscal{e_1}{e_1} = -1\,.
  \end{align*}
\end{proof}

Thanks to this lemma, we are able to give a simple formula for the
transversal Hessian of $H$, which governs the linearized (fast)
motion.
\begin{lem}
  The transversal Hessian of $H$, as a quadratic form on
  $T_{j(q)}\Sigma^\perp$, is given by
  \[
  \forall q\in \Omega, \forall (Q,P)\in T_{j(q)}\Sigma^\perp, \quad
  T^2_qH ((Q,P)^2) = 2 \|Q\wedge \vec B\|^2.
  \]
\end{lem}
\begin{proof}
  Let $(q,p)=j(q)$.  From~\eqref{equ:hamiltonian} we get
  \[
  T_{(q,p)}H = 2 \pscal{p-\A}{\dx p - T_q\A\circ \dx q}\,.
  \]
  Thus
  \[
  T^2H_{(q,p)}((Q,P)^2) = 2 \|(\dx p - T_q\A\circ \dx q)(Q,P)\|^2 +
  \pscal{p-\A}{M((Q,P)^2)}\,,
  \]
  and it is not necessary to compute the quadratic form $M$, since
  $p-\A=0$. We obtain
  \begin{align*}
    T^2H_{(q,p)}((Q,P)^2) & = 2 \| P - T_q\A(Q) \| ^2\\
    & = 2 \| ((T_q\A)^{\mathsf{T}} - T_q\A)(Q) \| ^2 = 2 \| Q
    \wedge \vec B \|^2\,.
  \end{align*}
\end{proof}
We may express this Hessian in the symplectic basis $(u_1,v_1)$ given
by Lemma~\ref{lem:basis}:
\begin{equation}
  \label{equ:hessian2}
  T^2H _{\upharpoonright  T_{j(q)}\Sigma^\perp} = 
  \begin{pmatrix}
    2\abs{B} & 0 \\
    0 & 2\abs{B}
  \end{pmatrix}\,.
\end{equation}
Indeed, $\|e_1\wedge \vec B\|^2= B^2$, and the off-diagonal term is
$\frac{1}{B}\pscal{e_1\wedge \vec B}{e_2 \wedge \vec B} = 0$.

\subsection{Proof of Theorem~\ref{main-theo-Bir}}

We use the notation of the previous section. We endow $\C_{z_1}\times
\R^2_{z_2}$ with canonical variables $z_1=x_1+i\xi_1$,
$z_2=(x_2,\xi_2)$, and the symplectic structure induced by $\omega_0=\dx \xi\wedge \dx x$. 

Let us notice that there exists a diffeomorphism $g:\Omega\to g(\Omega)\subset\R^2_{z_2}$ such that $g(q_0)=0$
and
\[
g^*(\dx\xi_2\wedge \dx x_2) = j^*\omega\,.
\]
(We identify $g$ with $\varphi$ in the statement of the theorem.)  In
other words, the new embedding $\tilde{\jmath}:=j\circ
g^{-1}:\R^2\to\Sigma$ is symplectic. In fact we can give an explicit
choice for $g$ by introducing the global change of variables:
$$x_{2}=q_{1},\quad \xi_{2}=\int_{0}^{q_{2}} B(q_{1},s)\dx s\,.$$
Consider the following map $\tilde\Phi$ (where we identify $\Omega$ and $g(\Omega)$):
\begin{align}
  \label{equ:phi}
  \C\times\Omega & \stackrel{\tilde\Phi}\longrightarrow
  N\Sigma \\
  (x_1+i\xi_1, z_2) & \mapsto x_1 u_1(z_2) + \xi_1 v_1(z_2)\,,
\end{align}
where $u_1(z_2)$ and $v_1(z_2)$ are the vectors defined in
Lemma~\ref{lem:basis} with $q=g^{-1}(z_2)$. This is an isomorphism
between the normal symplectic bundle of $\{0\}\times \Omega$ and
$N\Sigma$, the normal symplectic bundle of $\Sigma$. Indeed,
Lemma~\ref{lem:basis} says that for fixed $z_2$, the map $z_1\mapsto
\tilde\Phi(z_1,z_2)$ is a linear symplectic map. This implies, by a
general result of Weinstein~\cite{weinstein-symplectic}, that there exists a symplectomorphism $\Phi$ from a neighborhood of
$\{0\}\times\Omega$ to a neighborhood of
$\tilde\jmath(\Omega)\subset\Sigma$ whose differential at
$\{0\}\times \Omega$ is equal to $\tilde\Phi$. Let us recall how to
prove this.

First, we may identify $\tilde\Phi$ with a map into $\R^4$ by
\[
\tilde{\Phi}(z_1,z_2) = \tilde\jmath(z_2) + x_1 u_1(z_2) + \xi_1
v_1(z_2)\,.
\]
Its Jacobian at $z_1=0$ in the canonical basis of $T_{z_1}\C\times
T_{z_2}\Omega=\R^4$ is a matrix with column vectors
$\left[u_1,v_1,T_{z_2}\tilde\jmath(e_1),T_{z_2}\tilde\jmath(e_2)\right]$,
which by Lemma~\ref{lem:basis} is a basis of $\R^4$. Thus $\tilde{\Phi}$ is a local diffeomorphism at every $(0,z_2)$. Therefore
if $\epsilon>0$ is small enough, $\tilde\Phi$ is a diffeomorphism of
$D(\epsilon)\times\Omega$ into its image ($D(\epsilon)\subset\C$ is the open ball of radius $\epsilon$).

Since $\tilde{\jmath}$ is symplectic, Lemma~\ref{lem:basis} implies
that the basis
$\left[u_1,v_1,T_{z_2}\tilde\jmath(e_1),T_{z_2}\tilde\jmath(e_2)\right]$
is symplectic in $\R^4$; thus the Jacobian of $\tilde\Phi$ on
$\{0\}\times\Omega$ is symplectic. This can be expressed by
saying that the $2$-form
\[
\omega_0 - \tilde\Phi^*\omega_{0}
\]
vanishes on $\{0\}\times\Omega$.

\begin{lem}
  There exists a smooth and injective map $S:B(\epsilon)\times\Omega\to
  B(\epsilon)\times\Omega$, which is tangent to the identity along
  $\{0\}\times\Omega$, such that
  \[
  S^* \tilde\Phi^*\omega = \omega_0.
  \]
\end{lem}
  
\begin{proof}
  It is sufficient to apply Lemma \ref{symplectize} to
  $\omega_{1}=\tilde\Phi^*\omega_{0}$.
\end{proof}
  
We let $\Phi:=\tilde\Phi\circ S$; this is the claimed symplectic map. We let $(z_{1},z_{2})=\Phi(\hat z_{1},\hat z_{2})$. Let us now analyze how the Hamiltonian $H$ is transformed under
$\Phi$. The zero-set $\Sigma=H^{-1}(0)$ is now $\{0\}\times\Omega$,
and the symplectic orthogonal
$T_{\tilde\jmath(0,\hat{z}_2)}\Sigma^\perp$ is canonically equal to
$\C\times\{\hat{z}_2\}$. By~\eqref{equ:hessian2}, the matrix of the
transversal Hessian of $H\circ\Phi$ in the canonical basis of $\C$ is
simply
\begin{equation}
  \label{equ:hessian3}
  T^2(H\circ\Phi)_{\upharpoonright  \C\times\{\hat z_2\}} = T^2_{\Phi(0,\hat z_2)}H\circ (T\Phi)^2 =  
  \begin{pmatrix}
    2\abs{B(g^{-1}(\hat z_2))} & 0 \\
    0 & 2\abs{B(g^{-1}(\hat z_2))}
  \end{pmatrix}\,.
\end{equation}
Therefore, by Taylor's formula in the $\hat z_1$ variable (locally
uniformly with respect to the $\hat z_2$ variable seen as a
parameter), we get
\begin{align*}
  H\circ\Phi(\hat z_1,\hat z_2) & = H\circ\Phi_{\upharpoonright  \hat z_1=0} +
  TH\circ\Phi_{\upharpoonright  \hat z_1=0}(\hat z_1) + \frac{1}{2}
  T^2(H\circ\Phi)_{\upharpoonright  \hat z_1=0}(\hat z_1^2) + \mathcal{O}(\abs{\hat z_1}^3)\\
  & = 0 + 0 +\abs{B(g^{-1}(\hat z_2))}\abs{\hat z_1}^2 +
  \mathcal{O}(\abs{\hat z_1}^3).
\end{align*}
In order to obtain the result claimed in the theorem, it remains to
apply a formal Birkhoff normal form in the $\hat z_1$ variable, to
simplify the remainder $\mathcal{O}(\hat z_1^3)$.  This classical
normal form is a particular case of the semiclassical normal form that
we prove below (Proposition~\ref{prop:formal-normal-form}). Therefore we simply refer to this proposition, and this finishes the proof of
the theorem, where, for simplicity of notation, the variables $(z_1,z_2)$ actually
refer to $(\hat{z}_1,\hat{z}_2)$.

\subsection{Semiclassical Birkhoff normal form}
In the coordinates $\hat
x_{1}, \hat\xi_{1}, \hat x_{2}, \hat\xi_{2}$ (which are defined in a neighborhood of $\{0\}\times\Omega$), the Hamiltonian takes
the form:
\begin{equation}\label{DL3}
  \hat{H}(\hat z_{1}, \hat z_{2})=H^0+\mathcal{O}(|\hat z_{1}|^3),\quad\mbox{ where } H^0=B(g^{-1}(\hat z_{2}))|\hat z_{1}|^2\,.
\end{equation}
Let us now consider the space of the formal power series in $\hat
x_{1}, \hat \xi_{1}, h$ with coefficients smoothly depending on
$(\hat x_{2},\hat \xi_{2})$ : $\mathcal{E}=\mathcal{C}^\infty_{\hat
  x_{2}, \hat\xi_{2}}[[\hat x_{1}, \hat\xi_{1},h]]$. We endow
$\mathcal{E}$ with the Moyal product (compatible with the Weyl
quantization) denoted by $\star$ and the commutator of two series
$\kappa_{1}$ and $\kappa_{2}$ (in all variables $(\hat x_{1},
\hat\xi_{1},\hat x_{2}, \hat\xi_{2})$) is defined as
$$[\kappa_{1},\kappa_{2}]=\kappa_{1}\star \kappa_{2}-\kappa_{2}\star\kappa_{1}.$$
Explicitly, we have
\[
[ \kappa_1,\kappa_2 ] (\hat{x}, \hat{\xi} , h ) = 2 \sinh \Bigl(
\frac{h}{2i} \square \Bigr)\bigl( f ( x, {\xi}, h ) g (
{y}, {\eta}, h ) \bigr) \Bigr|_{x=y=\hat x,\atop \xi = \eta = \hat{\xi}}
\]
where
$$  \square = \sum_{j=1}^2 \partial_{\xi_j} \partial_{y_j} -
\partial_{x_j} \partial_{\eta_j}. $$
\begin{prop}\label{prop:formal-normal-form}
  Given $\gamma\in\mathcal{O}_{3}$, there exist formal power series
  $\tau,\kappa\in\mathcal{O}_{3}$ such that:
$$e^{ih^{-1}\ad_{\tau}}(H^0+\gamma)=H^0+\kappa\,,$$
with $[\kappa,|\hat z_{1}|^2]=0$.
\end{prop}
\begin{proof}
The proof is essentially the same as the proof of Proposition \ref{prop:formal-normal-form0}. The only point to notice is that
\[
ih^{-1}\ad_{\tau'}H_0 = B(g^{-1}(\hat z_{2}))ih^{-1}\ad_{\tau'}|\hat
z_{1}|^2 + \mathcal{O}_{N+4}.
\]
\end{proof}

\subsection{Proof of Theorem \ref{main-theo-Bir}}

\subsubsection{First Egorov theorem}
Using \eqref{DL3} and applying the Egorov theorem (see \cite[Theorems 5.5.5 and
5.5.9]{Martinez02}, \cite{Ro87} or \cite[Theorem 11.5]{Z13}), we can find a unitary operator $V_{h}$ (a \enquote{Fourier Integral Operator}) such that
$$V_{h}^* \mathfrak{L}_{h,\A}V_{h}=C_{0} h+ \mathcal{H}^0_{ h}+\Op_{h}^w(r_{h})\,,$$
so that
$\sigma^{\T,w}\left(\Op_{h}^w(r_{h})\right)=\gamma\in\mathcal{O}_{3}$, where $\sigma^{\T,w}$ means that we consider the formal Taylor series
of the Weyl symbol with respect to $(h,\hat z_{1})$. In fact, one can choose $V_h$ such that the subprincipal symbol is
preserved by conjugation (see for instance \cite[Appendix A]{HelSj89}), which implies that $C_0=0$. Note that this version of the Egorov theorem is more general than the one recalled in Chapter \ref{chapter-appendix}.

\subsubsection{Second Egorov theorem}
Let us now quantize the formal result of Proposition \ref{prop:formal-normal-form}, as in Chapter \ref{chapter-appendix}, Section \ref{sec.quantizing.elec}. Since the formal series $\kappa$ given by Proposition
\ref{prop:formal-normal-form} commutes with $|\hat z_{1}|^2$, we can
write it as a formal series in $|\hat z_{1}|^2$, that is
$$\kappa=\sum_{k\geq 0} \sum_{l+m=k} h^l c_{l,m}(\hat z_{2}) |\hat z_{1}|^{2m}.$$
This formal series can be reordered by using the monomials $(|\hat
z_{1}|^2)^{\star m}$:
$$\kappa=\sum_{k\geq 0} \sum_{l+m=k} h^l c^\star_{l,m}(\hat z_{2})(|\hat z_{1}|^2)^{\star m}.$$
Thanks to the Borel lemma, there exists a smooth function $f^\star(h,|\hat z_{1}|^2,\hat z_{2})$, compactly supported, with a support in $\hat z_{1}$ arbitrarily small, such that the Taylor
expansion with respect to $(h,|\hat z_{1}|^2)$ of $f^\star(h,|\hat
z_{1}|^2,\hat z_{2})$ is given by $\kappa$ and, locally in $\hat z_{2}$,
\begin{equation}\label{f-star}
  \sigma^{\T,w}\left(\Op_{h}^w \left(f^{\star}(h, \mathcal{I}_h, \hat z_{2})\right)\right) = \kappa\,.
\end{equation}
Here, the
operator $\Op_{h}^w \left(f^{\star}(h,\mathcal{I}_h, \hat z_{2})\right)$
has to be understood as the Weyl quantization with respect to $\hat
z_{2}$ of an operator valued symbol.  We can write it in the form:
$$\Op_{h}^w f^\star(h,\mathcal{I}_{h},\hat z_{2}) =
C_0 h + \mathcal{H}^0 + \Op_{h}^w \tilde
f^\star(h,\mathcal{I}_{h},\hat z_{2})\,,$$ where
$\mathcal{H}^0_{h}=\Op_{h}^w (H^0)$ and $\sigma^{\T,w}(\Op_{h}^w (\tilde{f}^{\star}(h,
    \mathcal{I}_h, \hat z_{2})))$ is in $\mathcal{O}_4$.
Thus, by using the Calderon-Vaillancourt theorem, given any $\eta>0$, we may choose the support of $f^{\star}$
small enough (with respect to $\hat z_{1}$) in order to have, for all $\psi\in\mathcal{C}^\infty_0(\R^2)$,
\begin{equation}
  |\langle\Op_{h}^w \tilde{f}^\star(h,\mathcal{I}_{h}, \hat z_{2})\psi,
  \psi\rangle|\leq \eta \|\mathcal{I}_{h}^{1/2} \psi\|^2\,.
  \label{equ:relative}
\end{equation}
Now we introduce a smooth symbol $a_{h}$ with compact support such that, locally in $\hat z_{2}$, $\sigma^{\T,w}(a_{h})=\tau$. 

It remains to use Proposition \ref{prop:formal-normal-form} and again the Egorov theorem (see Chapter \ref{chapter-appendix}, Section \ref{app.egorov}) to notice
that $e^{ih^{-1}\Op_{h}^w(a_{h})} \Op_{h}^w(r_{h})
e^{-ih^{-1}\Op_{h}^w(a_{h})}$ is a pseudo-differential operator
such that the formal Taylor series of its symbol is $\kappa$.
Therefore, recalling \eqref{f-star}, we have found a unitary Fourier Integral Operator $U_{h}$ such that
\begin{equation}\label{Egorov}
  U_{h}^*\mathfrak{L}_{h,\A} U_{h}= \mathcal{H}^0_{h}+\Op_{h}^w
  \left(\tilde{f}^{\star}(h,\mathcal{I}_h,\hat z_{2})\right)+R_{h}+S_{h}\,,
\end{equation}
where $R_{h}$ and $S_{h}$ are like in Theorem \ref{main-theo-Bir}.

This ends the proof of Theorem \ref{main-theo-Bir}.

\section{Microlocalization}\label{spec}
This section is devoted to the proof of Theorem \ref{spectrum}. The strategy is presented in Chapter \ref{chapter-appendix}, Section \ref{sec.BFelec}. The main idea is to use the eigenfunctions of $\mathfrak{L}_{h,\A}$ and
$\mathcal{L}^{\nor}_{h}$ as test functions in the pseudo-differential
identity \eqref{Egorov} given in Theorem \ref{main-theo-Bir} and to apply
the variational characterization of the eigenvalues given by the
min-max principle. In order to control the remainders we shall just prove
the microlocalization of the eigenfunctions of $\mathfrak{L}_{h,\A}$
and $\mathcal{L}^{\nor}_{h}$ thanks to the confinement assumption
\eqref{conf}. This is the aim of the next sections.

\subsection{Counting the eigenvalues}
Let us first roughly estimate the numbers of eigenvalues.
\begin{lem}\label{Nh}
There exists $C>0$ such that for all $h>0$, we have
$$\mathsf{N}(\mathfrak{L}_{h,\A},C_{1}h)=\mathcal{O}(h^{-1})\,.$$
\end{lem}
\begin{proof}
We notice that:
$$\mathsf{N}(\mathfrak{L}_{h,\A},C_{1}h)=\mathsf{N}(\mathfrak{L}_{1,h^{-1}A},C_{1}h^{-1})$$
and that, for all $\eps\in\left(0,1\right)$:
$$\mathfrak{Q}_{1,h^{-1}\A}(\psi)\geq (1-\eps)\mathfrak{Q}_{1,h^{-1}\A}(\psi)+\eps\int_{\R^2} \frac{B(x)}{h}|\psi|^2 \dx x$$
so that we infer:
$$\mathsf{N}(\mathcal{L}_{h,\A},C_{1}h)\leq \mathsf{N}(\mathfrak{L}_{1,h^{-1}\A}+\eps(1-\eps)^{-1}h^{-1}B,(1-\eps)^{-1}C_{1}h^{-1}).$$
Then, the diamagnetic inequality \footnote{See \cite[Theorem 1.13]{CFKS87} and the link with the control of the resolvent kernel in \cite{Kato72, Simon79}.} jointly with a Lieb-Thirring estimate (see the original paper \cite{ LT76})
provides for all $\gamma>0$ the existence of $L_{\gamma,2}>0$ such
that, for all $h>0$ and $\lambda>0$,
$$\sum_{j=1}^{\mathsf{N}(\mathcal{H}_{1,h^{-1}\A}+\eps(1-\eps)^{-1}h^{-1}B,\lambda)}  \left|\tilde\lambda_{j}(h)-\lambda\right|^\gamma\leq L_{\gamma,2}\int_{\R^2} (\eps(1-\eps)^{-1}h^{-1}B(x)-\lambda)_{-}^{1+\gamma}\dx x.$$
We apply this inequality with
$\lambda=(1+\eta)(1-\eps)^{-1}C_{1}h^{-1}$, for some $\eta>0$. This implies that:
$$\sum_{j=1}^{N_{\epsilon,h,\eta}}
\left|\tilde\lambda_{j}(h)-\lambda\right|^\gamma\leq
L_{\gamma,2}\int_{B(x)\leq(1+\eta)C_{1}/\epsilon}
(\lambda-\eps(1-\eps)^{-1}h^{-1}B(x))^{1+\gamma}\dx x$$
with $N_{\epsilon,h,\eta}:= \mathsf{N}(\mathfrak{L}_{1,h^{-1}\A}+\eps(1-\eps)^{-1}h^{-1}B,(1-\eps)^{-1}C_{1}h^{-1})$,
so that:
\begin{multline*}
  (\eta(1-\eps)^{-1}C_{1}h^{-1})^\gamma N_{\epsilon,h,\eta}
  \leq L_{\gamma,2} (h(1-\eps))^{-1-\gamma}\int_{B(x)\leq
    \frac{(1+\eta)C_{1}}{\eps}} ((1+\eta)C_{1}-\eps B(x))^{1+\gamma}\dx x.
\end{multline*}
For $\eta$ small enough and $\eps$ is close to 1, we have $
(1+\eta)\eps^{-1}C_{1}< \tilde C_{1}$ so that the integral is finite,
which gives the required estimate.
\end{proof}

\begin{lem}\label{H0}
There exists $C>0$ and $h_{0}>0$ such that for all $h\in(0,h_{0})$, we have:
$$\mathsf{N}(\mathcal{L}^{\nor}_{h}, C_{1}h)=\mathcal{O}(h^{-1})\,.$$
\end{lem}
\begin{proof}
Let $\eps\in(0,1)$. By point~\eqref{item:Q} of Theorem~\ref{main-theo-Bir}, it is enough to prove that $\mathsf{N}(\mathcal{H}^0_h,\frac{C_1h}{1-\eps}) =\mathcal{O}(h^{-1})$ since
\begin{equation}\label{(1-eps)H0}
\forall\psi\in\mathcal{C}^\infty_{0}(\R^2),\qquad\langle\mathcal{L}^{\nor}_{h}\psi,\psi \rangle\geq (1-\eps) \langle\mathcal{H}^0_h\psi,\psi\rangle\,.
\end{equation}

The eigenvalues and eigenfunctions of $\mathcal{H}_h^0$ can be found by separation of variables: $\mathcal{H}_h^0 = \mathcal{I}_h\otimes\Op_h^w(B\circ\varphi^{-1})$, where $\mathcal{I}_h$ acts on $\sL^2(\R_{x_1})$ and $\hat B_h:=\Op_h^w(B\circ\varphi^{-1})$ acts on $\sL^2(\R_{x_2})$. Thus,
$$
\mathsf{N}(\mathcal{H}_h^0, hC_{1,\eps}) = \# \{(n,m)\in (\N^*)^2; \quad
(2n-1)h\gamma_m(h) \leq  hC_{1,\eps}\}\,,
$$
where $C_{1,\eps}:=\frac{C_1}{1-\eps}$, and
$\gamma_1(h)\leq \gamma_2(h) \leq \cdots$ are the eigenvalues of
$\hat B_h$. A simple estimate gives
$$
\mathsf{N}(\mathcal{H}_h^0, C_{1,\eps}) \leq
\left(1+\left\lfloor\dfrac12+\frac{C_{1,\eps}}{2\gamma_1(h)}\right\rfloor\right) \cdot \#\{m\in\N^*;
\quad \gamma_m(h)\leq C_{1,\eps}\}\,.
$$
If $\eps$ is small enough, $C_{1,\eps}< \tilde{C_1}$, and then
Weyl asymptotics (see for instance \cite[Chapter 9]{DiSj99}) for $\hat{B}_h$ gives
$$
\mathsf{N}(\hat{B}_h,C_{1,\eps}) \sim \frac{1}{2\pi h}\textup{vol}\{B\circ\varphi^{-1}\leq C_{1,\eps}\}\,,
$$
and G\aa{}rding's inequality implies $\displaystyle{\gamma_1(h)\geq \min_{q\in\R^2} B - \mathcal{O}(h)}$, which
finishes the proof.
\end{proof}
In the same spirit, if we consider the eigenvalues of $\mathcal{L}^{\nor}_{h}$ lying below the threshold $C_{1}h$, only a finite number of components of $\mathcal{L}^{\nor}_{h}$ in the Hilbertian decomposition 
$$\mathcal{L}^{\nor}_{h}=\bigoplus_{n\geq 1}\mathcal{L}^{\nor,(n)}_{h}$$ 
has to contribute to the spectrum.
\begin{lem}\label{lem.dec.hilb}
There exists $h_{0}>0$ such that for all $h\in(0,h_{0})$, and all $n\geq 1$, the lowest eigenvalue of $\mathcal{L}^{\nor,(n)}_{h}$ satisfies
$$\lambda_{1}\left(\mathcal{L}^{\nor,(n)}_{h}\right)\geq (1-2\eps)(2n-1)h\min B\,.$$
In particular, there exists $h_{0}>0$ and $K\geq 1$ such that for all $h\in(0,h_{0})$,
$$\bigcup_{n\geq 1}\sp\left(\mathcal{L}^{\nor}_{h}\right)\cap\{\lambda\leq C_{1}h\}\subset\bigcup_{n=1}^K \sp\left(\mathcal{L}^{\nor, (n)}_{h}\right)\cap\{\lambda\leq C_{1}h\}\,.$$
Moreover, for all eigenvalue $\lambda$ of $\mathcal{L}^{\nor}_{h}$ such that $\lambda\leq C_{1}h$, we may find an basis of $\ker\left(\mathcal{L}^{\nor}_{h}-\lambda\right)$ in the form $(e_{k,h}(x_{1})f_{j,h}(x_{2}))_{\substack{1\leq k\leq K\\ 1\leq j\leq J(h)}}$ where $e_{k,h}$ is the $k$-th rescaled Hermite function (associated with $h^2D_{x_{1}}^2+x_{1}^2$) and $J(h)=\mathcal{O}(h^{-1})$.
\end{lem}

\begin{proof}
It is sufficient to apply the relative bound \eqref{(1-eps)H0} to functions in the form $e_{n,h}(x_{1})f(x_{2})$ and then to use the G\aa rding inequality to see that $\Op_{h}^w\left(B\right)\geq \min B-Ch$. The rest of the proof is standard and the bound on $J(h)$ comes from Lemma \ref{H0}.
\end{proof}

\subsection{Localization and microlocalization of the eigenfunctions of $\mathfrak{L}_{h,\A}$ and $\mathcal{L}^{\nor}_{h}$}
\label{sec:local-micr-eigenf}
The space localization of the eigenfunctions of $\mathfrak{L}_{h,\A}$,
which is the quantum analog of Theorem \ref{theo:confining}, is a
consequence of the Agmon estimates (see Chapter \ref{chapter-models}, Section \ref{Agmon}).
\begin{prop}\label{Agmon-Bir}
  Let us assume \eqref{conf}.
  Let us fix $0<C_{1}<\tilde C_{1}$ and $\alpha\in(0,\frac12)$. There
  exist $C,h_{0},\eps_0>0$ such that for all $0<h\leq h_0$ and for all eigenpair $(\lambda, \psi)$
  of $\mathfrak{L}_{h,\A}$ such that $\lambda\leq C_{1}h$, we have:
$$\int |e^{\chi(q)h^{-\alpha}|q|} \psi|^2\dx q\leq C\|\psi\|^2\,,$$
where $\chi$ is zero for $|q|\leq M_{0}$ and $1$ for $|q|\geq M_0 +
\eps_{0}$. Moreover, we also have the weighted $\sH^1$ estimate
$$\int |e^{\chi(q)h^{-\alpha}|q|} (-ih\nabla+\A)\psi|^2 \dx q\leq Ch\|\psi\|^2.$$
\end{prop}
\begin{rem}
  This estimate is interesting when $|x|\geq M_0+ \eps_{0}$. In this
  region, we deduce by standard elliptic estimates that
  $\psi=\mathcal{O}(h^{\infty})$ in suitable norms (see for instance
  \cite[Proposition 3.3.4]{Hel88} or more recently~\cite[Proposition
  2.6]{Ray12}). Therefore, the eigenfunctions are localized in space
  in the ball of center $(0,0)$ and radius $M_0+\eps_0$.
\end{rem}
We shall now prove the microlocalization of the eigenfunctions near
the zero set of the magnetic Hamiltonian $\Sigma$. For the sake of simplicity, we express this microlocalization result in terms of functional calculus.
\begin{prop}\label{micro-loc-L}
  Let us assume \eqref{conf}. Let us fix $0<C_{1}<\tilde C_{1}$ and
  consider $\delta\in\left(0,\frac{1}{2}\right)$. Let $(\lambda,\psi)$ be
  an eigenpair of $\mathfrak{L}_{h,\A}$ with $\lambda\leq C_{1}h$. Then,
  we have:
$$\psi=\chi_{1}\left(h^{-2\delta}\mathfrak{L}_{h,\A}\right)\chi_{0}(q)\psi+\mathcal{O}(h^{\infty})\,,$$
where $\chi_{0}$ is smooth cutoff function supported in a compact set
in the ball of center $(0,0)$ and radius $M_{0}+\eps_{0}$ and where $\chi_{1}$ a smooth cutoff function being $1$ near
$0$.
\end{prop}
\begin{proof}
In view of Proposition~\ref{Agmon-Bir}, it is enough to prove that
  \begin{equation}\label{chi-chi0}
    \left(1-\chi_{1}\left(h^{-2\delta}\mathfrak{L}_{h,\A}\right)\right) (\chi_{0}(q)\psi)=\mathcal{O}(h^{\infty})\,.
  \end{equation}
By the space localization, we have
$$\mathfrak{L}_{h,\A} (\chi_{0}(q)\psi)=\lambda \chi_{0}(q)\psi+\mathcal{O}(h^{\infty}).$$
Then, we get
$$   \left(1-\chi_{1}\left(h^{-2\delta}\mathfrak{L}_{h,\A}\right)\right)\mathfrak{L}_{h,\A} (\chi_{0}(q)\psi)=\lambda   \left(1-\chi_{1}\left(h^{-2\delta}\mathfrak{L}_{h,\A}\right)\right)\left( \chi_{0}(x)\psi\right)+\mathcal{O}(h^{\infty})\,.$$
This implies
\begin{multline*}
  h^{2\delta}\| \left(1-\chi_{1}\left(h^{-2\delta}\mathfrak{L}_{h,\A}\right)\right)(\chi_{0}(q)\psi)\|^2\leq q_{h\,A}\left( \left(1-\chi_{1}\left(h^{-2\delta}\mathfrak{L}_{h,\A}\right)\right)\mathfrak{L}_{h,\A} (\chi_{0}(q)\psi)\right)\\
  \leq C_{1}h\|
  \left(1-\chi_{1}\left(h^{-2\delta}\mathfrak{L}_{h,\A}\right)\right)(\chi_{0}(q)\psi)\|^2+\mathcal{O}(h^{\infty})\|\psi\|^2\,.
\end{multline*}
Since $\delta\in\left(0,\frac{1}{2}\right)$, we deduce
\eqref{chi-chi0}.
\end{proof}
\begin{rem}
The operator $\chi_{1}\left(h^{-2\delta}\mathfrak{L}_{h,\A}\right)\chi_{0}(q)$ is a pseudo-differential operator whose principal symbol is given by $\chi_{1}(h^{-2\delta}H(q,p))\chi_{0}(q)$ whereas the subprincipal terms are supported away from the region where the principal symbol is $1$. To see this, the reader can adapt \cite[Theorem 8.7]{DiSj99}. Due to the localization of the eigenfunctions induced by $\mathfrak{L}_{h,\A}$ in a compact $K$, we may also replace $\mathfrak{L}_{h,\A}$ by $\mathfrak{L}_{h,\A}+V$ where $V$ is a confining electric potential supported away from $K$ and apply \cite[Theorem 8.7]{DiSj99}.
\end{rem}

The next two propositions state the microlocalization of the eigenfunctions of the normal form $\mathcal{L}^{\nor}_{h}$.
\begin{prop}\label{loc-z2}
  Let us consider the pseudo-differential operator:
$$\mathcal{L}^{\nor}_{h}=\mathcal{H}_{h}^0+\Op_{h}^w \tilde f^\star(h,\mathcal{I}_{h},\hat z_{2})\,.$$
We assume the confinement assumption \eqref{conf}. We can consider $\tilde M_{0}>0$ such that $B\circ\varphi^{-1}(\hat z_{2})\geq \tilde C_{1}$ for $|\hat
z_{2}|\geq \tilde M_{0}$. Let us consider $C_{1}<\tilde C_{1}$ and an
eigenpair $(\lambda,\psi)$ of $\mathcal{L}^{\nor}_{h}$ such that $\lambda\leq
C_{1}h$. Then, for all $\eps_{0}>0$ and for all smooth cutoff function $\chi$ supported in
$|\hat z_{2}|\geq \tilde M_{0}+\eps_{0}$, we have:
$$\Op_{h}^w\left(\chi(\hat z_{2})\right)\psi=\mathcal{O}(h^{\infty})\,.$$
\end{prop}
\begin{proof}
Thanks to Lemma \ref{lem.dec.hilb}, it is sufficient to establish the lemma when $\psi$ is in the form $\psi(x_{1},x_{2})=e_{n,h}(x_{1})f(x_{2})$ (with $1\leq n\leq K$). But, we can write
$$\mathcal{L}^{\nor,(n)}_{h} f=\lambda f\,.$$
and we can apply the same kind of microlocal estimates as in the proof of Proposition \ref{prop.micro.elec}, the remainders being uniformly bounded with respect to $n$.
 \end{proof}

\begin{prop}\label{loc-z1}
  Keeping the assumptions and the notation of
  Proposition~\ref{loc-z2}, we consider
  $\delta\in\left(0,\frac{1}{2}\right)$ and an eigenpair $(\lambda,\psi)$
  of $\mathcal{L}^{\nor}_{h}$ with $\lambda\leq C_{1}h$. Then, we have:
$$\psi=\chi_{1}\left(h^{-2\delta}\mathcal{I}_{h}\right)\Op_{h}^w\left(\chi_{0}(\hat z_{2})\right)\psi+\mathcal{O}(h^{\infty})\,,$$
for all smooth cutoff function $\chi_{1}$ supported in a
neighborhood of zero and all smooth cutoff function $\chi_{0}$ being $1$ near zero and supported in the ball of center $0$ and radius $\tilde M_{0}+\eps_{0}$.
\end{prop}
\begin{proof}
The proof follows the same lines as for Proposition \ref{micro-loc-L}.
\end{proof}

%\mainmatter
\part{Boundary magnetic wells in dimension three}\label{Part.BMW}

\chapter{Magnetic half-space}\label{chapter-variable3D}
\begin{flushright}
\begin{minipage}{0.51\textwidth}
Sedulo curavi, humanas actiones non ridere, non lugere, neque detestari, sed intelligere.
\begin{flushright}
\textit{Tractatus politicus}, Spinoza
\end{flushright}
\vspace*{0.5cm}
\end{minipage}
\end{flushright}

This chapter is devoted to the proof of Theorem \ref{maintheo-variable3D}. We keep the notation of Chapter \ref{intro-semi}, Section \ref{intro-variable3D}. We analyze here how a smooth boundary combines with the magnetic field to generate a magnetic harmonic approximation.

\section{Quasimodes}

\begin{theo}\label{theo-ub}
For all $\alpha>0$, $\theta\in\left(0,\frac{\pi}{2}\right)$, there exists a sequence $(\mu_{j,n})_{j\geq 0}$ and there exist positive constants $C, h_{0}$ such that for $h\in(0,h_{0})$:
\[\dist\left(\sp(\mathfrak{L}_{h}),h\sum_{j=0}^J \mu_{j,n}h^j\right)\leq Ch^{J+2}\]
and we have $\mu_{0,n}=\mathfrak{s}(\theta)$ and $\mu_{1,n}$ is the $n$-th eigenvalue of $\alpha\mathfrak{S}_{\theta}(D_{\rho},\rho)$.
\end{theo}

\begin{proof}
 We perform the scaling (\ref{scaling-variable3D}) and, after division by $h$, $\mathfrak{L}_{h,\alpha,\theta}$ becomes
\[\mathcal{L}_{h}=D_{s}^2+D_{t}^2+(D_{r}+t\cos\theta-s\sin\theta+h\alpha t(r^2+s^2))^2\,.\]
Using the partial Fourier transform $\mathcal{F}_{r\to\eta}$ (see (\ref{Fourier})) and the translation $U_{\theta}$ (see (\ref{Trans})), we get the new expression of the operator
\[U_{\theta}\mathcal{F}_{r\to\eta}\mathcal{L}_{h}\mathcal{F}_{r\to\eta}^{-1}U_{\theta}^{-1}=D_{\sigma}^2+D_{\tau}^2+\left(V_{\theta}(\sigma,\tau)+h\alpha \tau\left(D_{\rho}-\frac{D_{\sigma}}{\sin\theta}\right)^2+\left(\sigma+\frac{\rho}{\sin\theta}\right)^2\right)^2\,.\]
This normal form will be denoted by $\mathcal{L}^{\nor}_{h}$ and the corresponding quadratic form by $\mathcal{Q}_{h}^{\nor}$. 
By expanding the square, we may write
\[\mathcal{L}^{\nor}_{h}=\mathfrak{L}^{\LP}_{\theta}+hL_{1}+h^2L_{2}\,,\]
where
\[L_{1}=\alpha \tau \left\{\left(D_{\rho}-\frac{D_{\sigma}}{\sin\theta}\right)^2 V_\theta+V_\theta\left(D_{\rho}-\frac{D_{\sigma}}{\sin\theta}\right)^2+2V_\theta\left(\sigma+\frac{\rho}{\sin\theta}\right)^2\right\}\,,\]
\[L_{2}=\alpha^2 \tau^2\left\{\left(D_{\rho}-\frac{D_{\sigma}}{\sin\theta}\right)^2+\left(\sigma+\frac{\rho}{\sin\theta}\right)^2\right\}^2\geq 0\,.\]
We look for formal eigenvalues and eigenfunctions in the form:
\[\mu\sim\sum_{j\geq 0} \mu_{j}h^j,\qquad \psi\sim\sum_{j\geq 0} \psi_{j}h^j\,.\]
In other words; we solve the following problem in the sense of formal series:
\[\mathcal{L}^{\nor}_{h}\psi\sim\mu\psi\,.\]
The term in $h^0$ leads to solve
\[\mathfrak{H}^{\Neu}_{\theta}\psi_{0}=\mu_{0}\psi_{0}\,.\]
We take $\mu_{0}=\mathfrak{s}(\theta)$ and
\[\psi_{0}(\rho,\sigma,\tau)=u^\LP_{\theta}(\sigma,\tau)f_{0}(\rho)\,,\]
$f_{0}$ being to be determined. Then, we must solve
\[(\mathfrak{H}^{\Neu}_{\theta}-\mathfrak{s}(\theta))\psi_{1}=(\mu_{1}-L_{1})\psi_{0}\,.\]
We apply the Fredholm alternative and we write
\[\langle(\mu_{1}-L_{1})\psi_{0},u^\LP_{\theta}\rangle_{\sL^2(\R_{+,\hat s,\hat t}^2)}=0\,.\]
The compatibility equation becomes
\[\alpha\mathfrak{S}_{\theta}(D_{\rho},\rho)f_{0}=\mu_{1}f_{0}\]
and we take  $\mu_{1}$ in the spectrum of $\alpha\mathfrak{S}_{\theta}(D_{\rho},\rho)$ and for $f_{0}$ the corresponding $L^2$-normalized eigenfunction. Then, we can write the solution $\psi_{1}$ in the form:
\[\psi_{1}=\psi_{1}^\perp+f_{1}(\rho)u_{\theta}({\sigma,\tau})\]
where $\psi_{1}^\perp$ is the unique solution orthogonal to $u^\LP_{\theta}$. We notice that it is the the Schwartz class. This construction can be continued at any order and we apply the spectral theorem.
\end{proof}

\section{Agmon estimates}
In this section we only state standard Agmon's estimates with respect to $(x,y)$ satisfied by an eigenfunction $u_{h}$ associated with $\lambda_{n}(h)$. The reader may consider them as an exercise.
They are related to the following lower bound (which can be proved by using the techniques of Chapter \ref{chapter-models}, Section \ref{Sec.Agmon0}, see also  \cite{LuPan00a} and \cite[Theorem 9.1.1]{FouHel10}).
\begin{prop}\label{roughlb}
There exist $C>0$ and $h_{0}>0$ such that, for $h\in(0,h_{0})$,
\[\lambda_n(h)\geq \mathfrak{s}(\theta) h-Ch^{5/4}\,.\]
\end{prop}

\subsection{Agmon estimates of first order}
We recall that $\B_{\mathfrak{s}}$ admits a unique and non degenerate minimum (as stated in Assumption \eqref{A2}), $\mathfrak{s}(\theta)$, at $(0,0)$. Thus, thanks to the computations leading to Proposition \ref{roughlb} and by using the techniques of Chapter \ref{chapter-models}, Section \ref{Sec.Agmon0}, we deduce the following estimates of Agmon.
\begin{prop}\label{tangential-agmon-v3D}
For all $\delta>0$, there exist $C>0$ and $h_0>0$ such that, for all $h\in(0,h_0)$,
\[\int_{\Omega_0} e^{\delta(x^2+y^2)/h^{1/4}} |u_h|^2\dx\x\leq C\|u_h\|^2\,,\quad\int_{\Omega_0} e^{\delta(x^2+y^2)/h^{1/4}} |\nabla u_h|^2\dx\x\leq Ch^{-1}\|u_h\|^2\,.\]
\end{prop}
Combining Proposition \ref{roughlb} and Theorem \ref{theo-ub}, we get that 
\[\lambda_{n}(h)=\mathfrak{s}(\theta) h+\mathcal{O}(h^{5/4})\,.\]
Thanks to Assumption \ref{A1} (the interior energy is higher than the boundary energy), this is standard to deduce the following normal Agmon estimates.
\begin{prop}\label{normal-agmon-v3D}
There exist $\delta>0$, $C>0$ and $h_0>0$ such that for all $h\in(0,h_{0})$, we have
\[\int_{\Omega_0} e^{\delta h^{-1/2}z}(|u_h|^2+h^{-1}|(-ih\nabla+\A)u_h|^2)\dx\x\leq C \|u_h\|^2\,.\]
\end{prop}
These last two propositions imply the following estimates.
\begin{cor}
For all $\gamma>0$ and $\ell\in\N$, we have
\[\int_{|x|+|y|\geq h^{1/8-\gamma}} |\x|^\ell (|u_{h}|^2+|\nabla u_{h}|^2)\dx\x+\int_{z\geq h^{1/2-\gamma}} |\x|^\ell (|u_{h}|^2+|\nabla u_{h}|^2)\dx\x=\mathcal{O}(h^{\infty})\|u_{h}\|^2\,.\]
\end{cor}
Thanks to this a priori localization of the eigenfunction near $(0, 0, 0)$, we may cutoff the eigenfunctions modulo a very small remainder. For that purpose, let us consider $\gamma>0$ small enough and introduce the cutoff function defined by
\[\chi_h(x,y, z)=\chi_0\left(h^{-1/8+\gamma}x, h^{-1/8+\gamma}y, h^{-1/2+\gamma}z\right)\,,\]
where $\chi_{0}$ is a smooth cutoff function being $1$ near $(0,0,0)$. We can notice, by elliptic regularity, that $\chi_{h}u_{h}$ is smooth (as it is supported away from the vertices).

Let us also consider $N\geq 1$. For $n=1,\cdots,N$, let us consider $u_{n,h}$ a $\sL^2$-normalized associated with $\lambda_{n}(h)$ so that $\langle u_{n,h},u_{m,h}\rangle=0$ for $n\neq m$. We let
\[\mathfrak{E}_{N}(h)=\underset{1\leq n\leq N}{\spann} u_{n,h}\,.\]
We notice that Propositions \ref{normal-agmon-v3D} and \ref{tangential-agmon-v3D} hold for the elements of $\mathfrak{E}_{N}(h)$.
As a consequence of Propositions \ref{normal-agmon-v3D} and \ref{tangential-agmon-v3D}, we get the following corollary.	
\begin{cor}\label{firstlb}
We have
\[\mathfrak{Q}_{h}(\tilde{u}_{h})\leq \lambda_{N}(h)+\mathcal{O}(h^{\infty}),\qquad\mbox{ with } \qquad\tilde{u}_{h}=\chi_{h}u_{h}\,,\]
where $u_{h}\in \mathfrak{E}_{N}(h)$ and where $\mathfrak{Q}_{h}$ denotes the quadratic form associated with $\mathfrak{L}_{h}$.
\end{cor}

\subsection{Agmon estimates of higher order}
In the last section we stated estimates of Agmon for $u_{h}$ and its first derivatives. We will also need estimates for the higher order derivatives.
The main idea to obtain such estimates can be found for instance in \cite{Hel88}. The basic idea to obtain them is to consider derivatives of the eigenvalue equation and use standard energy estimates. 
\begin{prop}
For all $\nu\in\N^3$, there exist $\delta>0$, $\gamma\geq 0$, $h_{0}>0$ and $C>0$ such that, for $h\in(0,h_{0})$,
\[\int_{\Omega_{0}} e^{\delta h^{-1/2}z} |D^{\nu}\tilde{u}_{h}|^2\dx\x+\int_{\Omega_{0}} e^{\delta h^{-1/4}(x^2+y^2)} |D^{\nu}\tilde{u}_{h}|^2\dx\x\leq Ch^{-\gamma}\|\tilde{u}_{h}\|^2\,,\]
where $u_{h}\in \mathfrak{E}_{N}(h).$
\end{prop}
These estimates only mean that the eigenfunctions and all their derivatives only live close to $(0, 0, 0)$. As usual, we immediately deduce the following.
\begin{cor}\label{exp-decay}
For all $\gamma>0$, we have, for all $\nu\in\N^3$ and $\ell\in\N$,
\[\int_{|x|+|y|\geq h^{1/8-\gamma}} |\x|^\ell |D^{\nu} \tilde{u}_{h}|^2\dx\x+\int_{z\geq h^{1/2-\gamma}}|\x|^\ell|D^{\nu} \tilde{u}_{h}|^2\dx\x=\mathcal{O}(h^{\infty})\|\tilde{u}_{h}\|^2\,,\]
where $u_{h}\in \mathfrak{E}_{N}(h).$
\end{cor}

\subsection{Normal form}
Let us now transfer initial eigenvalue problem onto the side of the normal form $\mathcal{L}^{\nor}_{h}$. For $u_{h}\in \mathfrak{E}_{N}(h)$, we introduce the rescaled and truncated function
\begin{equation}\label{wh}
w_{h}(r,s,t)=\chi_{h}^{\resc}(r,s,t)u^{\resc}_{h}(r,s,t)=\chi_{0}(h^{3/8+\gamma}r,h^{3/8+\gamma}s,h^{\gamma}t)u_{h}(h^{1/2}r,h^{1/2}s,h^{1/2}t)
\end{equation}
and its version on the side of normal coordinates
\[v_{h}(\rho,\sigma,\tau)=U_{\theta}\mathcal{F}_{r\to\eta}w_{h}\,.\]
We consider $\mathcal{F}_{N}(h)$ the image of  $\mathfrak{E}_{N}(h)$ by these transformations. We can reformulate Corollary \ref{firstlb}.
\begin{cor}
With the previous notation, we have, for $v_{h}\in\mathcal{F}_{N}(h)$,
\[\mathcal{Q}^{\nor}_{h}(v_{h})\leq \lambda^{\resc}_{N}(h)+\mathcal{O}(h^{\infty})\,,\]
where $\lambda^{\resc}_{N}(h)=h^{-1}\lambda_{N}(h).$
\end{cor}
We can also notice that, when $u_{h}$ is an eigenfunction associated with $\lambda_{p}(h)$, we have
\begin{equation}\label{eq-wh}
\mathcal{L}^{\nor}_{h}v_{h}=\lambda^{\resc}_{p}(h) v_{h}+r_{h}\,,
\end{equation}
where the remainder $r_{h}$ is $\mathcal{O}(h^{\infty})$ in the sense of Corollary \ref{exp-decay}.

In the following, we aim at proving localization and approximation estimates for $v_{h}$ rather than $u_{h}$.  Moreover, these approximations will allow us to estimate the energy $\mathcal{Q}^{\nor}_{h}(v_{h})$.

\section{Relative polynomial localizations in the phase space}
This section aims at estimating momenta of $v_{h}$ with respect to polynomials in the phase space.
Before starting the analysis, let us recall the link (cf. (\ref{Trans})) between the variables $(\eta,s,t)$ and $(\rho,\sigma,\tau)$:
\begin{equation}\label{corresp}
D_{\rho}=D_{\eta}+\frac{1}{\sin\theta}D_{s},\quad D_{\sigma}=D_{s},\quad D_{\tau}=D_{t}\,.
\end{equation}
We will use the following obvious remark.
\begin{rem}\label{compare-constant}
If $\phi$ is supported in $\supp(\chi_{h})$, we have, for all $\eps\in(0,1)$,
\[\mathcal{Q}_{h}(\phi)\geq (1-\eps)\mathfrak{Q}_{1,0,\theta}(\phi)-Ch^{1/2-6\gamma}\eps^{-1}\|\phi\|^2\,.\]
Optimizing in $\eps$, we have:
\[\mathcal{Q}_{h}(\phi)\geq (1-h^{1/4-3\gamma})\mathfrak{Q}_{1,0,\theta}(\phi)-Ch^{1/4-3\gamma}\|\phi\|^2\,.\]
Moreover, when the support of $\phi$ avoids the boundary, we have
\[\mathfrak{Q}_{1,0,\theta}(\phi)\geq\|\phi\|^2\,.\]
\end{rem}

\subsection{Localizations in $\sigma$ and $\tau$}
This section is concerned with many localizations lemmas  with respect to $\sigma$ and $\tau$.
\subsubsection{Estimates with respect to $\sigma$ and $\tau$}
We begin to prove estimates depending only on the variables $\sigma$ and $\tau$.
\begin{lem}\label{st}
Let $N\geq 1$. For all $k, n$, there exist $h_{0}>0$ and $C(k,n)>0$ such that, for all $h\in(0,h_{0})$:
\begin{align}
\label{control1}&\|\tau^k\sigma^{n+1} v_{h}\|\leq C(k,n)\|v_{h}\|\,,\\
\label{control2}&\|\tau^k D_{\sigma}(\sigma^{n} v_{h})\|\leq C(k,n)\|v_{h}\|\,,\\
\label{control3}&\|\tau^k D_{\tau}(\sigma^{n} v_{h})\|\leq C(k,n)\|v_{h}\|\,,
\end{align}
for $v_{h}\in\mathcal{F}_{N}(h)$.
\end{lem}
\begin{proof}
We prove the estimates when $v_{h}$ is the image of an eigenfunction associated to $\lambda_{p}(h)$ with $p=1,\ldots,N$.

Let us analyze the case $n=0$. The estimate (\ref{control3}) follows from the normal Agmon estimates.
By multiplying \eqref{eq-wh} by $\tau^k$ and taking the scalar product with $\tau^kv_{h}$, we get
\[\mathcal{Q}^{\nor}_{h}(\tau^k v_{h})\leq \lambda^{\resc}_{p}\|\tau^k v_{h}\|^2+|\langle[D_{\tau}^2,\tau^k]w_{h}, \tau^k v_{h}\rangle|+\mathcal{O}(h^{\infty})\|v_{h}\|^2\,.\]
The normal Agmon estimates provide
\[|\langle[D_{\tau}^2,\tau^k]v_{h}, \tau^k v_{h}\rangle|\leq C\|v_{h}\|^2\]
and thus
\[\mathcal{Q}^{\nor}_{h}(\tau^k v_{h})\leq C\|v_{h}\|^2\,.\]
We deduce (\ref{control2}).
We also have
\[\|\tau^k(-\sigma\sin\theta+\tau\cos\theta+R_{h})v_{h}\|^2\leq C\|v_{h}\|^2\,,\]
where
\begin{equation}\label{Rh}
R_{h}=h\alpha\tau\left\{\left(D_{\rho}-(\sin\theta)^{-1}D_{\sigma}\right)^2+\left(\sigma+(\sin\theta)^{-1}\rho\right)^2\right\}\,.
\end{equation}
We use the basic lower bound
\[\|\tau^k(-\sigma\sin\theta+\tau\cos\theta+R_{h})v_{h}\|^2\geq \frac{1}{2}\|\tau^k\sigma\sin\theta v_{h}\|^2-2\|(\tau^{k+1}\cos\theta+\tau^k R_{h})v_{h}\|^2\,.\]
Moreover, we have (using the support of $\chi^{\resc}_{h}$):
\[\|\tau^k R_{h}v_{h}\|\leq Ch(h^{-3/8-\gamma})^2\|\tau^{k+1}v_{h}\|\leq Ch(h^{-3/8-\gamma})^2\|v_{h}\|\,,\]
the last inequality coming from the normal Agmon estimates.
Thus, we get
\[\|\tau^k\sigma v_{h}\|^2\leq C\|v_{h}\|^2\,.\]
We now proceed by induction. We multiply (\ref{eq-wh}) by $\tau^k\sigma^{n+1}$, take the scalar product with ${\tau^k\sigma^{n+1} v_{h}}$ and it follows:
\begin{align*}
\mathcal{Q}^{\nor}_{h}(\tau^k\sigma^{n+1} v_{h})\leq& \lambda^{\resc}_{p}(h)\|\tau^k\sigma^{n+1} v_{h}\|^2+C\|\tau^{k-2}\sigma^{n+1} v_{h}\|\|\tau^k \sigma^{n+1} v_{h}\|\\
&+C\|\tau^{k-1}D_{\tau}\sigma^n v_{h}\|\|\tau^k\sigma^{n+1} v_{h}\|+C\|\tau^kD_{\sigma}\sigma^{n} w_{h}\|\|\tau^k\sigma^{n+1} v_{h}\|\\
&+C\|\tau^k\sigma^{n-1} v_{h}\|\|\tau^k\sigma^{n+1} v_{h}\|\\
&+|\langle \tau^k[\sigma^{n+1},(-\sigma\sin\theta+\tau\cos\theta+R_{h})^2]v_{h},\tau^k\sigma^{n+1}\rangle|\,.
\end{align*}
We have
\begin{align*}
&[\sigma^{n+1},(-\sigma\sin\theta+\tau\cos\theta+R_{h})^2]\\
&=[\sigma^{n+1},R_{h}](-\sigma\sin\theta+\tau\cos\theta+R_{h})+(-\sigma\sin\theta+\tau\cos\theta+R_{h})[\sigma^{n+1},R_{h}]\,.
\end{align*}
Let us analyze the commutator $[\sigma^{n+1},R_{h}]$. We can write
\[[\sigma^{n+1},R_{h}]=\alpha h\tau[\sigma^{n+1},\left(D_{\rho}-(\sin\theta)^{-1}D_{\sigma}\right)^2]\]
and
\begin{align*}
[\left(D_{\rho}-(\sin\theta)^{-1}D_{\sigma}\right)^2,\sigma^{n+1}]&=(\sin\theta)^{-2}n(n+1)\sigma^{n-1}\\
&+2i(\sin\theta)^{-1}(n+1)(D_{\rho}-(\sin\theta)^{-1}D_{\sigma})\sigma^n\,.\\
\end{align*}
We infer
\begin{align*}
&[\sigma^{n+1},(-\sigma\sin\theta+\tau\cos\theta+R_{h})^2]\\
&=\left(\alpha h\tau (\sin\theta)^{-2}n(n+1)\sigma^{n-1}+2i\alpha h\tau (\sin\theta)^{-1}(n+1)(D_{\rho}-(\sin\theta)^{-1}D_{\sigma})\sigma^n\right)(V_{\theta}+R_{h})\\
&+(V_{\theta}+R_{h})\left(\alpha h\tau (\sin\theta)^{-2}n(n+1)\sigma^{n-1}+2i\alpha h\tau (\sin\theta)^{-1}(n+1)(D_{\rho}-(\sin\theta)^{-1}D_{\sigma})\sigma^n\right)\,.
\end{align*}
After having computed a few more commutators, the terms of ${[\sigma^{n+1},(-\sigma\sin\theta+\tau\cos\theta+R_{h})^2]}$ are in the form: 
\[\tau^l \sigma^m\,,\]
\[h\tau^l(D_{\rho}-(\sin\theta)^{-1}D_{\sigma}) \sigma^m\,,\]
\[h^2\tau^l(D_{\rho}-(\sin\theta)^{-1}D_{\sigma})^3 \sigma^m\,,\]
\[h^2\tau^l(\sigma+(\sin\theta)^{-1}\rho)^2 (D_{\rho}+(\sin\theta)^{-1}D_{\sigma}) \sigma^m\,,\]
with $m\leq n+1$ and $l=0,1,2$. 

Let us examine for instance the term $h^2\tau^l(\sigma+(\sin\theta)^{-1}\rho)^2 (D_{\rho}+(\sin\theta)^{-1}D_{\sigma}) \sigma^m$. We have, after the inverse Fourier transform and translation:
$$h^2\|\tau^l(\sigma+(\sin\theta)^{-1}\rho)^2 (D_{\rho}+(\sin\theta)^{-1}D_{\sigma}) \sigma^m v_{h}\|\leq Ch^2 (h^{-3/8-\gamma})^3\|\tau^l \sigma^m v_{h}\|$$
where we have used the support of $\chi^{\resc}_{h}$ (see (\ref{wh})).
We get:
\begin{align*}
&|\langle \tau^k[\sigma{n+1},(-\sigma\sin\theta+\tau\cos\theta+R_{h})^2]v_{h},\tau^k\sigma^{n+1}v_{h}\rangle|\leq C\|\tau^k\sigma^{n+1}v_{h}\|\sum_{j=0}^{n+1}\sum_{l=0}^{k+2}\|\tau^l\sigma^{j}v_{h}\|\,.
\end{align*}
We deduce by the induction assumption:
$$\mathcal{Q}^{\nor}_{h}(\tau^k\sigma^{n+1} v_{h})\leq C\|v_{h}\|^2\,.$$
We infer that, for all $k$:
$$\|D_{\tau} (\tau^{k}\sigma^{n+1})v_{h}\|\leq C\|v_{h}\|\mbox{ and } \|D_{\sigma} (\tau^{k}\sigma^{n+1})v_{h}\|\leq C\|v_{h}\|\,.$$
Moreover, we also deduce:
$$\|(V_{\theta}+R_{h})\tau^{k}\sigma^{n+1} v_{h}\|\leq C\|v_{h}\|\,,$$
from which we find:
$$\|\tau^k \sigma^{n+2} v_{h}\|\leq C\|v_{h}\|\,.$$
\end{proof}
We also need a control of the derivatives with respect to $\sigma$. The next lemma is left to the reader as an exercise (take successive derivatives of the eigenvalue equation and estimate commutators by induction). Roughly speaking, it states that $\sigma$, $\tau$, $D_{\sigma}$ and $D_{\tau}$ are bounded.
\begin{lem}\label{Ds-v3D}
For all $m,n, k$, there exist $h_{0}>0$ and $C(m,n,k)>0$ such that for all ${h\in(0,h_{0})}$:
\begin{align}
&\|\tau^k D_{\sigma}^{m+1}\sigma^{n}v_{h}\|\leq C(k,m,n)\|v_{h}\|\,,\\
&\|\tau^k D_{\sigma}^{m}D_{\tau}\sigma^{n}v_{h}\|\leq C(k,m,n)\|v_{h}\|\,,
\end{align}
for $v_{h}\in\mathcal{F}_{N}(h)$.
\end{lem}
We now establish partial Agmon estimates with respect to $\sigma$ and $\tau$. Roughly speaking, we can write the previous lemmas with $\rho v_{h}$ and $D_{\rho} v_{h}$ instead of $v_{h}$.
\subsubsection{Partial estimates involving $\rho$}
Let us begin to prove the following lemma.
\begin{lem}\label{agmonttau}
For all $k\geq 0$, there exist $h_{0}>0$ and $C(k)>0$ such that, for all $h\in(0,h_{0})$,
\begin{eqnarray*}
&\|\tau^k \rho v_{h}\|&\leq C (\|\rho v_{h}\|+\|v_{h}\|)\,,\\ 
&\|\tau^{k} D_{\tau}\rho v_{h}\|&\leq C (\|\rho v_{h}\|+\|v_{h}\|)\,, \\
&\|\tau^k D_{\sigma}\rho v_{h}\|&\leq C(\|\rho v_{h}\|+\|v_{h}\|)\,,
\end{eqnarray*}
for $v_{h}\in\mathcal{F}_{N}(h)$.
\end{lem}
\begin{proof}
For $k=0$, we multiply \eqref{eq-wh} by $\rho$ and take the scalar product with $\rho v_{h}$. There is only one commutator to analyze:
\[[(V_{\theta}+R_{h})^2,\rho]=[(V_{\theta}+R_{h}),\rho](V_{\theta}+R_{h})+(V_{\theta}+R_{h})[(V_{\theta}+R_{h}),\rho]\]
so that
\[[(V_{\theta}+R_{h})^2,\rho]=[R_{h},\rho](V_{\theta}+R_{h})+(V_{\theta}+R_{h})[R_{h},\rho]\,.\]
We deduce, thanks to the support of $w_{h}$:
\[|\langle[(V_{\theta}+R_{h})^2,\rho]v_{h},\rho v_{h}\rangle|\leq C\|v_{h}\|\|\rho v_{h}\|\leq C(\|\rho v_{h}\|^2+\|v_{h}\|^2)\]
and we infer
\[\mathcal{Q}^{\nor}_{h}(\rho v_{h})\leq  C(\|\rho v_{h}\|^2+\|v_{h}\|^2)\,.\]
We get
\[\|D_{\tau}\rho v_{h}\|\leq C(\|\rho v_{h}\|+\|v_{h}\|) \mbox{ and } \|D_{\sigma}\rho v_{h}\|\leq C(\|\rho v_{h}\|+\|v_{h}\|)\,.\]
Then it remains to prove the case $k\geq 1$ by induction (use Remark \ref{compare-constant} and that $\mathfrak{s}(\theta)<1$).
\end{proof}
As an easy consequence of the proof of Lemma \ref{agmonttau}, we have the following.
\begin{lem}\label{agmonttau'}
For all $k\geq 0$, there exist $h_{0}>0$ and $C(k)>0$ such that, for all $h\in(0,h_{0})$:
$$\|\tau^k \sigma\rho v_{h}\|\leq C(k)(\|\rho v_{h}\|+\|v_{h}\|)\,,$$
for $v_{h}\in\mathcal{F}_{N}(h)$.
\end{lem}
We can now deduce the following lemma (exercise).
\begin{lem}
For all $k, n$, there exist $h_{0}>0$ and $C(k,n)>0$ such that, for all $h\in(0,h_{0})$:
\begin{align}
&\|\rho \tau^k\sigma^{n+1} v_{h}\|\leq C(k,n)(\|\rho v_{h}\|+\|v_{h}\|)\,,\\
&\| \rho \tau^k D_{\sigma}(\sigma^{n} v_{h})\|\leq C(k,n)(\|\rho v_{h}\|+\|v_{h}\|)\,,\\
&\| \rho \tau^k D_{\tau}(\sigma^{n} v_{h})\|\leq C(k,n)(\|\rho v_{h}\|+\|v_{h}\|)\,,
\end{align}
for $v_{h}\in\mathcal{F}_{N}(h)$.
\end{lem}
From this lemma, we deduce a stronger control with respect to the derivative with respect to $\sigma$.
\begin{lem}\label{partial-tau}
For all $m,n, k$, there exist $h_{0}>0$ and $C(m,n,k)>0$ such that for all ${h\in(0,h_{0})}$:
\begin{align}
&\|\rho \tau^k D_{\sigma}^{m+1}\sigma^{n}v_{h}\|\leq C(k,m,n)(\|\rho v_{h}\|+\|v_{h}\|)\,,\\
&\|\rho\tau^k D_{\sigma}^{m}D_{\tau}\sigma^{n}v_{h}\|\leq C(k,m,n)(\|\rho v_{h}\|+\|v_{h}\|)\,,
\end{align}
for $v_{h}\in\mathcal{F}_{N}(h)$.
\end{lem}
\begin{proof}
The proof can be done by induction. The case $m=0$ comes from the previous lemma. Then, the recursion is the same as for the proof of Lemma \ref{Ds-v3D} and uses Lemma \ref{Ds-v3D} to control the additional commutators.
\end{proof}
By using the symmetry between $\rho$ and $D_{\rho}$, we have finally the following important lemma.
\begin{lem}\label{partial-Dtau}
For all $m,n, k$, there exist $h_{0}>0$ and $C(m,n,k)>0$ such that for all ${h\in(0,h_{0})}$:
\begin{align}
&\|D_{\rho} \tau^k D_{\sigma}^{m+1}\sigma^{n}v_{h}\|\leq C(k,m,n)(\|D_{\rho} v_{h}\|+\|v_{h}\|)\,,\\
&\|D_{\rho} \tau^k D_{\sigma}^{m}D_{\tau}\sigma^{n}v_{h}\|\leq C(k,m,n)(\|D_{\rho} v_{h}\|+\|v_{h}\|)\,,
\end{align}
for $v_{h}\in\mathcal{F}_{N}(h)$.
\end{lem}
\subsection{Approximation of $v_{h}$}
In this section, we prove that $v_{h}$ behaves like $u^\LP_{\theta}(\sigma,\tau)$ with respect to $\sigma$ and $\tau$. Let us state the approximation result of this section.
\begin{prop}\label{approxvh}
There exists $C>0$ and $h_0>0$ such that, for $h\in(0, h_0)$,
\[\|v_h-\Pi v_{h}\|+\|V_{\theta}v_h-V_{\theta}\Pi v_{h}\|+\|\nabla_{\sigma,t}(v_h-\Pi v_{h})\|\leq C h^{1/4-2\gamma}\|v_h\|\,,\]
where $\Pi$ is the projection on $u^\LP_{\theta}$ and $v_{h}\in\mathcal{F}_{N}(h)$.
\end{prop}
\begin{proof}
As usual, we start to prove the inequality when $v_{h}$ is the image of an eigenfunction associated with $\lambda_{p}(h),$ the extension to $v_{h}\in\mathcal{F}_{N}(h)$ being standard.
We want to estimate 
\[\|(\mathfrak{H}^{\Neu}_{\theta}-\mathfrak{s}(\theta))v_{h}\|\,.\]
We have
\[\|(\mathfrak{H}^{\Neu}_{\theta}-\mathfrak{s}(\theta))v_{h}\|\leq \|(\mathfrak{H}^{\Neu}(\theta)-\lambda_{p}(h))v_{h}\|+Ch^{1/4}\|v_{h}\|\,.\]
With the definition of $v_{h}$ and with Corollary \ref{exp-decay}, we have:
\[\|(\mathfrak{H}^{\Neu}_{\theta}-\lambda_{p}(h))v_{h}\|\leq h\|L_{1}v_{h}\|+h^2\|L_{2}v_{h}\|+\mathcal{O}(h^{\infty})\|v_{h}\|\,.\]
Then, we can write
\[\|L_{1}v_{h}\|\leq C\left\|\tau V_{\theta}\left(D_{\rho}-\frac{D_{\sigma}}{\sin\theta}\right)^2 v_{h}\right\|+C\left\|\tau\left(D_{\rho}-\frac{D_{\sigma}}{\sin\theta}\right)^2V_{\theta} v_{h}\right\|+C\left\|\tau V_{\theta}\left(\sigma+\frac{\rho}{\sin\theta}\right)^2 v_h\right\|\,.\]
With Lemma \ref{st} and the support of $u_{h}$, we infer
\[h\|L_{1}v_{h}\|\leq Ch^{1/4-2\gamma}\|v_{h}\|\,.\]
In the same way, we get
\[h^2\|L_{2}v_{h}\|\leq Ch^{1/2-4\gamma}\|v_{h}\|\,.\]
We deduce
\[\|(\mathfrak{H}^{\Neu}_{\theta}-\mathfrak{s}(\theta))v_{h}\|\leq Ch^{1/4-2\gamma}\|v_{h}\|\,.\]
We have
\[\|(\mathfrak{H}^{\Neu}_{\theta}-\mathfrak{s}(\theta))v^{\perp}_{h}\|\leq Ch^{1/4-2\gamma}\|v_{h}\|\,,\qquad v_h=v^{\perp}_h+\Pi v_{h}\,.\]
The resolvent, valued in the form domain, being bounded, the result follows.
\end{proof}

\section{Localization induced by the effective harmonic oscillator}
In this section, we prove Theorem \ref{maintheo-variable3D}. In order to do that, we first prove a localization with respect to $\rho$ and then use it to improve the approximation of Proposition \ref{approxvh}.
\subsection{Control of $v_{h}$ with respect to $\rho$}
Let us prove an optimal localization estimate of the eigenfunctions with respect to $\rho$. Thanks to our relative boundedness lemmas (Lemmas \ref{partial-tau} and \ref{partial-Dtau}) we can compare the initial quadratic form with the model quadratic form.    
\begin{prop}\label{estimateC0}
There exist $h_{0}>0$ and $C>0$ such that, for all $C_{0}>0$ and $h\in(0,h_{0})$,
\begin{eqnarray*}
&\mathcal{Q}^\nor_{h}(v_{h})\geq&(1-C_{0}h)\left(\|D_{\tau} v_{h}\|^2+\|D_{\sigma} v_{h}\|^2+\left\|\left(V_{\theta}(\sigma,\tau)+\alpha h\tau \mathcal{H}_{\harm}\right)v_{h}\right\|^2\right)\\
&&-\frac{C}{C_{0}}h\langle \mathcal{H}_{\harm}v_{h},v_{h}\rangle-Ch\|v_{h}\|^2\,,
\end{eqnarray*}
for $v_{h}\in\mathcal{F}_{N}(h)$.
\end{prop}
\begin{proof}
Let us consider 
\[\mathcal{Q}^\nor_{h}(v_{h})=\|D_{\tau} v_{h}\|^2+\|D_{\sigma} v_{h}\|^2+\left\|\left(V_{\theta}(\sigma,\tau)+\alpha h\tau\left\{\mathcal{H}_{\harm}+L(\rho,D_{\rho},\sigma,D_{\sigma}) \right\}\right)v_{h}\right\|^2\,.\]
where
\[L(\rho,D_{\rho},\sigma,D_{\sigma})=(\sin\theta)^{-2}(-2\sin\theta D_{\sigma} D_{\rho}+2\sin\theta \sigma\rho+D_{\sigma}^2+\sigma^2)\,.\]
For all $\eps>0$, we have:
\begin{eqnarray*}
&\mathcal{Q}^\nor_{h}(v_{h})\geq&(1-\eps)\left(\|D_{\tau} v_{h}\|^2+\|D_{\sigma} v_{h}\|^2+\left\|\left(V_{\theta}(\sigma,\tau)+\alpha h\tau \mathcal{H}_{\harm}\right)v_{h}\right\|^2\right)\\
&&-\eps^{-1}\alpha^2 h^2\|\tau L(\rho,D_{\rho},\sigma,D_{\sigma}) v_{h}\|^2\,.
\end{eqnarray*}
We take $\eps=C_{0}h$. We apply Lemmas \ref{Ds-v3D}, \ref{partial-tau} and \ref{partial-Dtau} to get
\[\|\tau L(\rho,D_{\rho},\sigma,D_{\sigma}) v_{h}\|^2\leq C(\|D_{\rho}v_{h}\|^2+\|\rho v_{h}\|^2+\|v_{h}\|^2)\,.\]
\end{proof}
From the last proposition, we are led to study the model operator:
\[\mathcal{H}_{h}=D_{\sigma}^2+D_{\tau}^2+(V_{\theta}(\sigma,\tau)+\alpha h \tau \mathcal{H}_{\harm})^2\,.\]
We can write $\mathcal{H}_{h}$ as a direct sum:
\[\mathcal{H}_{h}=\bigoplus_{n\geq 1} \mathcal{H}_h^n\,,\]
with 
\[\mathcal{H}_h^n=D_{\sigma}^2+D_{\tau}^2+(V_{\theta}(\sigma,\tau)+\alpha h \tau \mu_{n})^2\,,\]
where $\mu_{n}$ is the $n$-the eigenvalue of $\mathcal{H}_{\harm}$.
Therefore we shall analyze (see Chapter \ref{chapter-models}, Section \ref{Sec.Lu-Pan}):
\[\mathfrak{L}^\LP_{\theta,g}=D_{\sigma}^2+D_{\tau}^2+(V_{\theta}(\sigma,\tau)+g\tau)^2\,.\]
We deduce the existence of $c>0$ such that, for all $g\geq 0$:
\[\mathfrak{s}(\theta,g)\geq\mathfrak{s}(\theta)+cg\,.\]
Taking $C_{0}$ large enough in Proposition \ref{estimateC0}, we deduce the following proposition.
\begin{prop}\label{control-tau}
There exist $C>0$ and $h_{0}>0$ such that, for all $h\in(0, h_{0})$,
\[\langle \mathcal{H}_{\harm}v_{h},v_{h}\rangle\leq C\|v_{h}\|^2, \mbox{ for } v_{h}\in\mathcal{F}_{N}(h)\]
and
\[\lambda^{\resc}_{N}(h)\geq \mathfrak{s}(\theta)-Ch\,.\]
\end{prop}

\subsection{Refined approximation of $v_{h}$}
The control of $v_{h}$ with respect to $\rho$ provided by Proposition \ref{control-tau} permits to improve the approximation of $v_{h}$.
\begin{prop}\label{approxvh2}
There exist $C>0$, $h_0>0$ and $\gamma>0$ such that, if $h\in(0, h_0)$ :
\begin{align*}
&\|V_{\theta}D_{\rho}v_h-V_{\theta} D_{\rho}\Pi v_{h}\|+\|D_{\rho}v_h-D_{\rho}\Pi v_{h}\|+\|\nabla_{\sigma,\tau}(D_{\rho}v_h-D_{\rho}\Pi v_{h})\|\leq C h^{\gamma}\|v_h\|\,,\\
&\|V_{\theta}\rho v_h-V_{\theta} \rho \Pi v_{h}\|+\|\rho v_h-\rho \Pi v_{h}\|+\|\nabla_{\sigma,\tau}(\rho v_h-\rho\Pi v_{h})\|\leq C h^{\gamma}\|v_h\|\,,
\end{align*}
for $v_{h}\in\mathcal{F}_{N}(h)$.
\end{prop}
\begin{proof}
Let us apply $D_{\rho}$ to \eqref{eq-wh}. We have the existence of $\gamma>0$ such that:
\[\|[\mathcal{L}^{\nor}_{h},D_{\rho}]v_{h}\|\leq Ch^\gamma\|v_{h}\|\,.\]
We can write
\[\|(\mathfrak{H}^{\Neu}_{\theta}-\sigma(\theta))D_{\rho}v_{h}\|\leq \|(\mathfrak{H}^{\Neu}_{\theta}-\lambda_{p}^\resc(h))D_{\rho}v_{h}\|+Ch^{1/4}\|D_{\rho}v_{h}\|\,.\]
Proposition \ref{control-tau} provides
\[\|(\mathfrak{H}^{\Neu}_{\theta}-\sigma(\theta))D_{\rho}v_{h}\|\leq \|(\mathfrak{H}^{\Neu}_{\theta}-\lambda_{p}^\resc(h))D_{\rho}v_{h}\|+Ch^{1/4}\|v_{h}\|\,.\]
Then, we get
\[\|hL_{1}D_{\rho}v_{h}\|\leq Ch^{1/4-2\gamma}\|v_{h}\|\]
and
\[\|h^2L_{2}D_{\rho}v_{h}\|\leq Ch^{1/2-4\gamma}\|v_{h}\|\,.\]
We deduce
\[\|(\mathfrak{H}^{\Neu}_{\theta}-\mathfrak{s}(\theta))D_{\rho}v_{h}\|\leq Ch^{1/4-\gamma}\|v_{h}\|\,.\]
The conclusion is the same as for the proof of Proposition \ref{approxvh}. The analysis for $\rho$ can be done exactly in the same way.
\end{proof}

\subsection{Conclusion: proof of Theorem \ref{maintheo-variable3D}}
We recall that
\[\mathcal{Q}^{\nor}_{h}(v_{h})=\|D_{\tau} v_{h}\|^2+\|D_{\sigma} v_{h}\|^2+\left\|\left(V_{\theta}(\sigma,\tau)+\alpha h\tau\left\{\mathcal{H}_{\harm}+L(\rho,D_{\rho},\sigma,D_{\sigma}) \right\}\right)v_{h}\right\|^2\]
so that we get
\begin{align*}
&\mathcal{Q}^{\nor}_{h}(v_{h})\geq\mathfrak{s}(\theta)\|v_{h}\|^2\\ 
&+\alpha h\langle 2\tau V_{\theta}(\sigma,\tau)\mathcal{H}_{\harm}+\tau V_{\theta}L(\rho,D_{\rho},\sigma,D_{\sigma})+\tau L(\rho,D_{\rho},\sigma,D_{\sigma})V_{\theta}(\sigma,\tau)v_{h},v_{h}\rangle\,.
\end{align*}
It remains to approximate $v_{h}$ by $\Pi v_{h}$ modulo lower order remainders (exercise!). This implies:
\[\mathcal{Q}^{\nor}_{h}(v_{h})\geq\mathfrak{s}(\theta)\|v_{h}\|^2+\alpha h\langle\mathfrak{S}_{\theta}(D_{\rho},\rho)\phi_{h},\phi_{h}\rangle_{\sL^2(\R_{\rho})}+o(h)\|v_{h}\|^2\,,\]
where $\phi_{h}=\langle v_{h},u_{\theta}\rangle_{\sL^2(\R_{\sigma,\tau})}$ and $v_{h}\in\mathcal{F}_{N}(h)$. 
With the min-max principle, we deduce the spectral gap between the lowest eigenvalues and it remains to use Proposition \ref{approxvh}.

\chapter[Magnetic wedge]{Magnetic wedge}\label{chapter-edge}
\begin{flushright}

\begin{minipage}{0.5\textwidth}
On oublie vite du reste ce qu'on n'a pas pens\'e avec profondeur, ce qui vous a \'et\'e dict\'e par l'imitation, par les passions environnantes.
\begin{flushright}
\textit{\`A la recherche du temps perdu},\\ 
\textit{La Prisonni\`ere}, Proust
\end{flushright}
\vspace*{0.5cm}
\end{minipage}
\end{flushright}

This chapter is devoted to the proof of Theorem \ref{main-result-edge} announced in Chapter \ref{intro-semi}, Section \ref{intro-edge}. We focus on the specific features induced by the presence of a non smooth boundary.

\section{Quasimodes}
Before starting the analysis, we use the following scaling:
\begin{equation}\label{scaling-Popoff}
\check s=h^{1/4}\sigma ,\quad \check t=h^{1/2}\tau,\quad \check z=h^{1/2}\cz
\end{equation}
so that we denote by $\mathcal{L}_{h}$ and $\mathcal{C}_{h}$ the operators $h^{-1}\check{\mathcal{L}}_{h}$ and $h^{-1/2}\check{\mathcal{C}}_{h}$  in the coordinates $(\sigma , \tau, \cz)$. 

Using Taylor expansions, we can write in the sense of formal power series the magnetic Laplacian near the edge and the associated magnetic Neumann boundary condition: 
$$\mathcal{L}_{h}\underset{h\to 0}\sim \sum_{j\geq 0} \mathcal{L}_{j} h^{j/4}$$
and
$$\mathcal{C}_{h}\underset{h\to 0}\sim \sum_{j\geq 0} \mathcal{C}_{j} h^{j/4}\,,$$
where the first $\mathcal{L}_{j}$ and $\mathcal{T}_{j}$ are given by (see Conjecture \ref{min-nu-edge}):
\begin{align}
\label{L0-Popoff}&\mathcal{L}_{0}=D_{\tau}^2+D_{\cz}^2+(\tau-\zeta^\Po_{0})^2\,,\\
\label{L1-Popoff}&\mathcal{L}_{1}=-2(\tau-\zeta^\Po_{0})D_{\sigma }\,,\\
\label{L2-Popoff}&\mathcal{L}_{2}=D_{\sigma }^2+2\kappa \mathcal{T}_{0}^{-1}\sigma ^2 D_{\cz}^2\,,
\end{align}
where 
\begin{align*}
&\mathcal{C}_{0}=(-\tau+\zeta^\Po_{0},D_{\tau},D_{\cz}),\\
&\mathcal{C}_{1}=(D_{\sigma },0,0),\\
&\mathcal{C}_{2}=(0,0,\kappa \mathcal{T}_{0}^{-1}\sigma ^2 D_{\cz}),
\end{align*}
where $\kappa$ is defined in \eqref{kappa}.

We will also use an asymptotic expansion of the normal $\hat \n(h)$. We recall that we have $\check{\n}=(-\mathcal{T}'(\check{s})\check{t},-\mathcal{T}(\check{s}),\pm1)$ so that we get:
$$\hat\n(h) \underset{h\to 0}\sim \sum_{j\geq 0} \n_{j} h^{j/4}\,,$$
with
\begin{equation}
\n_{0}=(0,-\mathcal{T}_{0},\pm 1), \quad \n_{1}=(0,0,0),\quad \n_{2}=(0,\kappa \sigma ^2,0)\,.
\end{equation}
We look for $(\hat\lambda(h),\hat\psi(h))$ in the form:
$$\hat\lambda(h)\underset{h\to 0}\sim \sum_{j\geq 0} \mu_{j} h^{j/4}\,,$$
$$\hat\psi(h)\underset{h\to 0}\sim \sum_{j\geq 0} \psi_{j} h^{j/4}\,,$$
which satisfies, in the sense of formal series, the following boundary value problem:
\begin{equation}
\label{E:Pbformelwedge}
\left\{
\begin{aligned}
\mathcal{L}_{h}\hat\psi(h)&\underset{h\to 0}\sim \hat\lambda(h)\hat\psi(h),
\\
 \hat\n\cdot\mathcal{C}_{h}\hat\psi(h)&\underset{h\to 0}\sim 0 \quad \mbox{ on }\quad \dr_{\Neu}\mathcal{W}_{\alpha_{0}}\,.
 \end{aligned}
 \right.
 \end{equation}
This provides an infinite system of PDE's. We will use Notation \ref{notation-FH} introduced in Chapter \ref{chapter-BOE}.

\subsection{Terms in $h^0$}
We solve the equation:
$$\mathcal{L}_{0}\psi_{0}=\mu_{0}\psi_{0}, \mbox{ in }\mathcal{W}_{\alpha_{0}}, \quad \n_{0}\cdot \mathcal{C}_{0}\psi_{0}=0, \mbox{ on }\dr_{\Neu}\mathcal{W}_{\alpha_{0}}\,.$$
We notice that the boundary condition is exactly the Neumann condition. We are led to choose $\mu_{0}=\nu_{1}^\Po(\alpha_{0},\zeta^\Po_{0})$ and $\psi_{0}(\sigma ,\tau, \cz)=u^\Po_{\zeta^\Po_{0}}(\tau,\cz)f_{0}(\sigma )$ where $f_{0}$ will be chosen (in the Schwartz class) in a next step. 

\subsection{Terms in $h^{1/4}$}
Collecting the terms of size $h^{1/4}$, we find the equation:
$$(\mathcal{L}_{0}-\mu_{0})\psi_{1}=(\mu_{1}-\mathcal{L}_{1})\psi_{0}, \quad \n_{0}\cdot \mathcal{C}_{0}\psi_{1}=0, \mbox{ on } \dr_{\Neu}\mathcal{W}_{\alpha_{0}}\,.$$
As in the previous step, the boundary condition is just the Neumann condition. We use the Feynman-Hellmann formulas to deduce:
$$(\mathcal{L}_{0}-\mu_{0})(\psi_{1}+v^\Po_{\zeta^\Po_{0}}(\tau,\cz)D_{\sigma }f_{0}(\sigma ))=\mu_{1}\psi_{0}, \quad \n_{0}\cdot \mathcal{C}_{0}\psi_{1}=0, \mbox{ on } \dr_{\Neu}\mathcal{W}_{\alpha_{0}}\,.$$
Taking the scalar product of the r.h.s. of the first equation with $u^\Po_{\zeta^\Po_{0}}$ with respect to $(\tau,\cz)$ and using the Neumann boundary condition for $v^\Po_{\zeta^\Po_{0}}$ and $\psi_{1}$ when integrating by parts, we find $\mu_{1}=0$. This leads to choose:
$$\psi_{1}(\sigma , \tau, \cz)=v^\Po_{\zeta^\Po_{0}}(\tau,\cz)D_{\sigma }f_{0}(\sigma )+f_{1}(\sigma )u^\Po_{\zeta^\Po_{0}}(\tau,\cz)\,,$$
where $f_{1}$ will be determined in a next step.
\subsection{Terms in $h^{1/2}$}
Let us now deal with the terms of order $h^{1/2}$:
$$(\mathcal{L}_{0}-\mu_{0})\psi_{2}=(\mu_{2}-\mathcal{L}_{2})\psi_{0}-\mathcal{L}_{1}\psi_{1}, \quad \n_{0}\cdot \mathcal{C}_{0}\psi_{2}=-\n_{0}\cdot \mathcal{C}_{2}\psi_{0}-\n_{2}\cdot \mathcal{C}_{0}\psi_{0}, \mbox{ on } \dr_{\Neu}\mathcal{W}_{\alpha_{0}}\,.$$
We analyze the boundary condition:
\begin{align*}
\n_{0}\cdot \mathcal{C}_{2}\psi_{0}+\n_{2}\cdot \mathcal{C}_{0}\psi_{0}&=\pm \kappa \mathcal{T}_{0}^{-1}\sigma ^2 D_{\cz}\psi_{0}+\kappa \sigma ^2 D_{\tau}\psi_{0}\\
&=\kappa \mathcal{T}_{0}^{-1}\sigma ^2(\pm D_{\cz}+\mathcal{T}_{0}D_{\tau})\psi_{0}\\
&=\pm 2\kappa\mathcal{T}_{0}^{-1} \sigma ^2D_{\cz}\psi_{0}\,.
\end{align*}
where we have used the Neumann boundary condition of $\psi_{0}$. Then, we use  the Feynman-Hellmann formulas together with \eqref{L1-Popoff} and \eqref{L2-Popoff} to get:
\begin{equation}\label{equation-2}
(\mathcal{L}_{0}-\mu_{0})(\psi_{2}-v^\Po_{\zeta^\Po_{0}} D_{\sigma }f_{1}-\frac{w^\Po_{\zeta^\Po_{0}}}{2}D^2_{\sigma } f_{0} )=\mu_{2}\psi_{0}-\tfrac{\dr_{\zeta}^2\nu_{1}^\Po(\alpha_{0},\zeta^\Po_{0})}{2}D_{\sigma }^2\psi_{0}-2\kappa \mathcal{T}_{0}^{-1}\sigma ^2 D_{\cz}^2\psi_{0}\,,
\end{equation}
with boundary condition:
$$\n_{0}\cdot \mathcal{C}_{0}\psi_{2}=\mp 2\kappa\sigma ^2\mathcal{T}_{0}^{-1}D_{\cz}\psi_{0}, \mbox{ on } \dr_{\Neu}\mathcal{W}_{\alpha_{0}}\,.$$
We use the Fredholm condition by taking the scalar product of the r.h.s. of \eqref{equation-2} with $u^\Po_{\alpha_{0},\zeta^\Po_{0}}$ with respect to $(\tau,\cz)$. Integrating by parts and using the Green-Riemann formula (the boundary terms cancel), this provides the equation:
$$\mathcal{H}^\Po_{\harm} f_{0}=\mu_{2} f_{0}\,,$$
with
$$\mathcal{H}^\Po_{\harm}=\tfrac{\dr_{\eta}^2\nu_{1}^\Po(\alpha_{0},\zeta^\Po_{0})}{2}D_{\sigma }^2+2\kappa \mathcal{T}_{0}^{-1}\|D_{\cz}u^\Po_{ \zeta^\Po_{0}}\|_{\sL^2(\mathcal{S}_{\alpha_{0}})}^2\sigma ^2\,.$$
Up to a scaling, the 1D-operator $\mathcal{H}^\Po_{\harm}$ is the harmonic oscillator on the line (we have used that Conjecture \ref{min-nu-edge} is true). Its spectrum is given by:
$$\left\{(2n-1)\sqrt{\kappa \mathcal{T}_{0}^{-1}\|D_{\cz}u_{\zeta^\Po_{0}}\|^2\dr_{\zeta}^2\nu_{1}^\Po(\alpha_{0},\zeta^\Po_{0})},\quad n\geq 1    \right\}\,.$$
Therefore for $\mu_{2}$ we take:
\begin{equation}
\label{E:choixla2}
\mu_{2}=(2n-1)\sqrt{\kappa \mathcal{T}_{0}^{-1}\|D_{\cz}u^\Po_{\zeta^\Po_{0}}\|_{\sL^2(\mathcal{S}_{\alpha_{0}})}^2\dr_{\zeta}^2\nu_{1}^\Po(\alpha_{0},\zeta^\Po_{0})}
\end{equation}
with $n\in \N^*$ and for $f_{0}$ the corresponding normalized eigenfunction.
With this choice we deduce the existence of $\psi_{2}^\perp$ such that:
\begin{equation}
(\mathcal{L}_{0}-\mu_{0})\psi_{2}^\perp =\mu_{2}\psi_{0}-\tfrac{\dr_{\zeta}^2\nu_{1}^\Po(\alpha_{0},\zeta^\Po_{0})}{2}D_{\sigma }^2\psi_{0}-2\kappa \mathcal{T}_{0}^{-1}\sigma ^2 D_{\cz}^2\psi_{0}, \mbox{ and } \langle \psi_{2}^\perp ,u^\Po_{\zeta^\Po_{0}}\rangle_{\tau, \cz}=0.
\end{equation}
We can write $\psi_{2}$ in the form:
$$\psi_{2}=\psi_{2}^\perp+v^\Po_{\zeta^\Po_{0}} D_{\sigma }f_{1}+D^2_{\sigma } f_{0}\frac{w^\Po_{\zeta^\Po_{0}}}{2}+f_{2}(\sigma )u^\Po_{\zeta^\Po_{0}}\,,$$
where $f_{2}$ has to be determined in a next step.

The construction can be continued (exercise).

By using the spectral theorem, we infer that
\begin{equation}\label{rub-wedge}
\lambda_{n}(h)\leq \nu(\alpha_{0})h+Ch^{\frac{3}{2}}\,.
\end{equation}
\section{Agmon estimates}
Thanks to a standard partition of unity, we can establish the following estimate for the eigenvalues (use the strategy in the proof of Proposition \ref{prop.reflbvan}).
\begin{prop}\label{rlb-edge}
There exist $C$ and $h_{0}>0$ such that, for $h\in(0,h_{0}):$
$$\lambda_{n}(h)\geq \nu(\alpha_{0})h-Ch^{5/4}\,.$$
\end{prop}
From \eqref{rub-wedge} and Proposition \ref{rlb-edge}, we infer that the main term in the asymptotic expansion of $\lambda_{n}(h)$ is $\nu(\alpha_{0})h$. Then, due to the difference of energy between the smooth boundary and the wedge (see Assumption \ref{H:comparaisonspectre}), this implies, with the estimates of Agmon (see the proof of Proposition \ref{Agmon-normal} where the same ideas are used; here we choose balls of size $Rh^{\frac{1}{2}}$), a localization of the lowest eigenfunctions near $E$.
\begin{prop}\label{normal-agmon}
There exist $\eps_{0}>0, h_{0}>0$ and $C>0$ such that for all $h\in(0,h_{0})$:
\begin{align*}
&\int_{\Omega} e^{2\eps_{0}h^{-1/2}d(\x,E)}|\psi|^2\dx\x\leq C\|\psi\|^2,\\
&\mathfrak{Q}_{h}(e^{\eps_{0}h^{-1/2}d(\x,E)}\psi)\leq Ch\|\psi\|^2.
\end{align*}
\end{prop}
As a consequence, we can refine the lower bound.
\begin{prop}\label{Oh32}
For all $n\geq 1$, there exists $h_{0}>0$ such that for $h\in(0,h_{0})$, we have:
$$\lambda_{n}(h)=\nu(\alpha_{0},\zeta^\Po_{0})h+\mathcal{O}(h^{3/2})\,.$$
\end{prop}
\begin{proof}
We have:
$$\check{\mathfrak{Q}}_{h}(\check\psi)=\langle \check\nabla_{h}\check \psi,\check\nabla_{h}\check \psi\rangle_{\sL^2(\dx\check s \dx\check t \dx\check z)}\,.$$
With the estimates of Agmon with respect to $\check t$ and $\check z$, we infer:
$$\check{\mathfrak{Q}}_{h}(\check\psi)\geq \mathcal{Q}_{h}^{\fla}(\check \psi)-Ch^{3/2}\|\check\psi\|^2\,.$$
where:
\begin{align*}
&\check{\mathfrak{Q}}_{h}^{\fla}(\check \psi)= \|hD_{\check t}\check \psi\|^2+ \|h\mathcal{T}_{0}\mathcal{T}(\check s)^{-1}D_{\check z}\check \psi\|^2+\|(hD_{\check s}+\zeta^\Po_{0}h^{1/2}-\check t)\check \psi\|^2.
\end{align*}
Moreover, we have:
$$\check{\mathfrak{Q}}_{h}^{\fla}(\check \psi)\geq  \|hD_{\check t}\check \psi\|^2+ \|hD_{\check z}\check \psi\|^2+\|(hD_{\check s}+\zeta^\Po_{0}h^{1/2}-\check t)\check \psi\|^2\geq \nu(\alpha_{0},\zeta^\Po_{0})h\,.$$
\end{proof}
A rough localization estimate is given by the following proposition (that follows again by the estimates of Agmon related to Proposition \ref{rlb-edge}, see also Proposition \ref{rl}).
\begin{prop}\label{tangential-agmon}
There exist $\eps_{0}>0, h_{0}>0$ and $C>0$ such that for all $h\in(0,h_{0})$:
\begin{align*}
&\int_{\Omega} e^{\chi(\x)h^{-1/8}|s(\x)|}|\psi|^2\dx\x\leq C\|\psi\|^2,\\
&\mathfrak{Q}_{h}( e^{\chi(\x)h^{-1/8}|s(\x)|}\psi)\leq Ch\|\psi\|^2,
\end{align*}
where $\chi$ is a smooth cutoff function supported in a fixed neighborhood of $E$.
\end{prop}

We use a cutoff function $\chi_{h}(\x)$ near $\x_{0}$ such that:
$$\chi_{h}(\x)=\chi_{0}(h^{1/8-\gamma}\check s(\x))\chi_{0}(h^{1/2-\gamma}\check t(\x))\chi_{0}(h^{1/2-\gamma}\check z(\x))\,.$$
For all $N\geq 1$, let us consider $\sL^2$-normalized eigenpairs $(\lambda_{n}(h),\psi_{n,h})_{1\leq n\leq N}$ such that $\langle\psi_{n,h},\psi_{m,h}\rangle=0$ when $n \neq m$. We consider the $N$ dimensional space defined by:
$$\mathfrak{E}_{N}(h)=\underset{1\leq n\leq N}{\mathrm{span}} \tilde\psi_{n,h},\quad\mbox{ where }\quad\tilde\psi_{n,h}=\chi_{h}\psi_{n,h}\,.$$ 

\begin{notation}
We will denote by $\tilde \psi(=\chi_{h}\psi)$ the elements of $\mathfrak{E}_{N}(h)$.
\end{notation}

Let us state a proposition providing the localization of the eigenfunctions with respect to $D_{\check s}$ (the proof is left to the reader as an exercise, see Chapter \ref{chapter-vanishing} for a similar estimate).
\begin{prop}\label{Ds}
There exist $h_{0}>0$ and $C>0$ such that, for $h\in(0,h_{0})$ and $\check\psi\in\check{\mathfrak{E}}_{N}(h)$, we have:
$$\|D_{\check s}\check\psi\|\leq Ch^{-1/4}\|\check\psi\|\,.$$
\end{prop}

\section{Projection method}
The result of Proposition \ref{Ds} implies an approximation result for the eigenfunctions. 
Let us recall the scaling defined in \eqref{scaling-Popoff}:
\begin{equation}
\check s=h^{1/4}\sigma ,\quad \check t=h^{1/2}\tau,\quad \check z=h^{1/2}\cz.
\end{equation}

\begin{notation}
We will denote by $\mathcal{E}_{N}(h)$ the set of the rescaled elements of $\check{\mathfrak{E}}_{N}(h)$. The elements of $\mathcal{E}_{N}(h)$ will be denoted by $\hat\psi$. Moreover we will denote by $\mathcal{L}_{h}$ the operator $h^{-1}\check{\mathcal{L}}_{h}$ in the rescaled coordinates. The corresponding quadratic form will be denoted by $\mathcal{Q}_{h}$.
\end{notation}

\begin{lem}\label{approx}
There exist $h_{0}>0$ and $C>0$ such that, for $h\in(0,h_{0})$ and $\hat\psi\in\mathcal{E}_{N}(h)$, we have:
\begin{align}
\label{approx1}&\|\hat\psi-\Pi_{0}\hat\psi\|+\|D_{\tau}(\hat\psi-\Pi_{0}\hat\psi)\|+\|D_{\cz}(\hat\psi-\Pi_{0}\hat\psi)\|\leq Ch^{1/8}\|\hat\psi\|\\
\label{approx2}&\|\sigma (\hat\psi-\Pi_{0}\hat\psi)\|+\|\sigma  D_{\tau}(\hat\psi-\Pi_{0}\hat\psi)\|+\|\sigma D_{\cz}(\hat\psi-\Pi_{0}\hat\psi)\|\leq Ch^{1/8-\gamma}(\|\hat\psi\|+(\|\sigma \hat\psi\|)\,,
\end{align}
where $\Pi_{0}$ is the projection on $u_{\zeta^\Po_{0}}$:
$$\Pi_{0}\hat\psi=\langle \hat\psi,u^\Po_{\zeta^\Po_{0}}\rangle_{\sL^2(\mathcal{S}_{\alpha_{0}})}u^\Po_{\zeta^\Po_{0}}\,.$$
\end{lem}

This approximation result allows us to catch the behavior of the eigenfunction with respect to $\check s$. In fact, this is the core of the dimension reduction process of the next proposition. Indeed $\sigma ^2 D_{\cz}^2$ is not an elliptic operator, but, once projected on $u_{\zeta^\Po_{0}}$, it becomes elliptic.

\begin{prop}
There exist $h_{0}>0$ and $C>0$ such that, for $h\in(0,h_{0})$ and $\check\psi\in\check{\mathfrak{E}}_{N}(h)$, we have:
$$\|\check s\check\psi\|\leq Ch^{1/4}\|\check\psi\|\,.$$
\end{prop}
\begin{proof}
It is equivalent to prove that:
$$\|\sigma \hat\psi\|\leq C\|\hat\psi\|\,.$$
The proof of Proposition \ref{Oh32} provides the inequality:
$$\|D_{\tau}\hat \psi\|^2+ \|\mathcal{T}_{0}\mathcal{T}(h^{1/4}\sigma )^{-1}D_{\cz}\hat \psi\|^2+\|(h^{1/4}D_{\sigma }+\zeta^\Po_{0}-\tau)\hat \psi\|^2\leq (\nu^\Po_{1}(\alpha_{0},\zeta^\Po_{0})+Ch^{1/2})\|\hat\psi\|^2\,.$$
From the non-degeneracy of the maximum of $\alpha$, we deduce the existence of $c>0$ such that:
$$\|\mathcal{T}_{0}\mathcal{T}(h^{1/4}\sigma )^{-1}D_{\cz}\hat \psi\|^2\geq \|D_{\cz}\hat\psi\|^2+ch^{1/2}\|\sigma  D_{\cz}\hat\psi\|^2$$
so that we have:
$$ch^{1/2}\|\sigma  D_{\cz}\hat\psi\|^2\leq Ch^{1/2}\|\hat\psi\|^2$$
and:
$$\|\sigma  D_{\cz}\hat\psi\|\leq \tilde C\|\hat\psi\|\,.$$
It remains to use Lemma \ref{approx} and especially \eqref{approx2}. In particular, we have:
$$\|\sigma D_{\cz}(\hat\psi-\Pi_{0}\hat\psi)\|\leq Ch^{1/8-\gamma}(\|\hat\psi\|+(\|\sigma \hat\psi\|)\,.$$
We infer:
$$\|\sigma  D_{\cz}\Pi_{0}\hat\psi\|\leq \tilde C\|\hat\psi\|+Ch^{1/8-\gamma}(\|\hat\psi\|+(\|\sigma \hat\psi\|)\,.$$
Let us write 
$$\Pi_{0}\hat\psi=f_{h}(\sigma )u^\Po_{\zeta^\Po_{0}}(\tau,\cz)\,.$$
We have:
$$\|\sigma  D_{\cz}\Pi_{0}\hat\psi\|=\|D_{\cz} u^\Po_{\zeta^\Po_{0}}\|_{\sL^2(\mathcal{S}_{\alpha_{0}})}\|\sigma  f_{h}\|_{\sL^2(d\sigma )}=\|D_{\cz} u^\Po_{\zeta^\Po_{0}}\|_{\sL^2(\mathcal{S}_{\alpha_{0}})}\|\sigma  f_{h} u^\Po_{\zeta^\Po_{0}}\|=\|D_{\cz} u^\Po_{\zeta^\Po_{0}}\|_{\sL^2(\mathcal{S}_{\alpha_{0}})}\|\sigma  \Pi_{0}\hat\psi\|\,.$$
We use again Lemma \ref{approx} to get:
$$\|\sigma  D_{\cz}\Pi_{0}\hat\psi\|=\|D_{\cz} u_{\zeta^\Po_{0}}\|_{\sL^2(\mathcal{S}_{\alpha_{0}})}\|\sigma  \hat\psi\|+\mathcal{O}(h^{1/8-\gamma})(\|\hat\psi\|+\|\sigma \hat\psi\|)\,.$$
We deduce:
$$\|D_{\cz} u^\Po_{\zeta^\Po_{0}}\|_{\sL^2(\mathcal{S}_{\alpha_{0}})}\|\sigma  \hat\psi\|\leq \tilde C\|\hat\psi\|+2Ch^{1/8-\gamma}(\|\hat\psi\|+(\|\sigma \hat\psi\|)$$
and the conclusion follows.
\end{proof}
\begin{prop}\label{last-lb}
There exists $h_{0}>0$ such that for $h\in(0,h_{0})$ and $\hat\psi\in\hat{\mathfrak{E}}_{N}(h)$, we have:
\begin{align*}
\hat{\mathcal{Q}}_{h}(\hat\psi)\geq& \|D_{\tau}\hat\psi\|^2+\|D_{\cz}\hat\psi\|^2+\|(h^{1/4}D_{\sigma }-\tau+\zeta^\Po_{0})\hat\psi\|^2+h^{1/2}\mathcal{T}_{0}^{-1}\kappa\|D_{\cz}u^\Po_{\zeta^\Po_{0}}\|_{\sL^2(\mathcal{S}_{\alpha_{0}})}^2\sigma ^2\\
                                          &+o(h^{1/2})\|\hat\psi\|^2.
\end{align*}
\end{prop}
Let us introduce the operator:
\begin{equation}\label{pre-BO1}
D_{\tau}^2+D_{\cz}^2+(h^{1/4}D_{\sigma }-\tau+\zeta^\Po_{0})^2+h^{1/2}\mathcal{T}_{0}^{-1}\kappa\|D_{\cz}u^\Po_{\zeta^\Po_{0}}\|^2\sigma ^2.
\end{equation}
After Fourier transform with respect to $\sigma $, the operator \eqref{pre-BO1} becomes:
\begin{equation}\label{pre-BO2}
D_{\tau}^2+D_{\cz}^2+(h^{1/4}\xi-\tau+\zeta^\Po_{0})^2+h^{1/2}\mathcal{T}_{0}^{-1}\kappa\|D_{\cz}u^\Po_{\zeta^\Po_{0}}\|_{\sL^2(\mathcal{S}_{\alpha_{0}})}^2 D_{\xi}^2\,.
\end{equation}

\begin{exe}
Use the Born-Oppenheimer approximation to estimate the lowest eigenvalues of this last operator and deduce Theorem \ref{main-result-edge}.
\end{exe}

\chapter[Magnetic cone]{Magnetic cone}\label{chapter-cones}
\begin{flushright}
\begin{minipage}{0.5\textwidth}
Ignarus enim praeterquam quod a causis externis multis modis agitatur nec unquam vera animi acquiescentia potitur, vivit paeterea sui et Dei et rerum quasi inscius et simulac pati desinit, simul etiam esse desinit.
\begin{flushright}
\textit{Ethica}, Pars V, Spinoza
\end{flushright}
\vspace*{0.5cm}
\end{minipage}
\end{flushright}

This chapter deals with the proof of Theorem \ref{main-theo-cone}.

\section{Quasimodes in the axisymmetric case\label{Sec.QM}}
This section deals with the proof of the following proposition.
\begin{prop}\label{theo-quasimodes-cone}
For all $N\geq1$ and $J\geq1$, there exist $C_{N,J}$ and $\alpha_{0}$ such that for all $1\leq n\leq N$, and $0<\alpha<\alpha_{0}$, we have:
\[\dist\left(\spd(\mathfrak{L}_{\alpha,0}),\sum_{j=0}^J\gamma_{j,n}\alpha^{2j+1}\right)\leq C_{N,J}\ \alpha^{2J+3}\,,\]
where $\gamma_{0,n}=\mathfrak{l}_{N}=2^{-5/2}(4n-1)$.
\end{prop}

\begin{proof}
We construct quasimodes which do not depend on $\theta$. In other words, we look for quasimodes for:
\[\mathcal{L}_{\alpha,0}=-\frac{1}{\tau^{2}}\dr_{\tau}\tau^2\dr_{\tau}
+\frac{\sin^2(\alpha\varphi)}{4\alpha^2}\tau^2
-\frac{1}{\alpha^2\ \tau^2\sin(\alpha\varphi)}\dr_{\varphi}\sin(\alpha\varphi)\dr_{\varphi}\,.\]
We write a formal Taylor expansion of $\mathcal{L}_{\alpha,0}$ in powers of $\alpha^2$:
\[\mathcal{L}_{\alpha,0}\sim \alpha^{-2}\mathcal{M}_{-1}+\mathcal{M}_{0}+\sum_{j\geq 1} \alpha^{2j}\mathcal{M}_{j}\,,\]
where
\[\mathcal{M}_{-1}=-\frac{1}{\tau^2\varphi}\dr_{\varphi}\varphi\dr_{\varphi}, 
\qquad \mathcal{M}_{0}=-\frac{1}{\tau^{2}}\dr_{\tau}\tau^2\dr_{\tau}+\frac{\varphi^2 \tau^2}{4}+\frac{1}{3\tau^2}\varphi\dr_{\varphi}\,.\]
We look for quasimodes expressed as formal series:
\[\psi\sim \sum_{j\geq 0} \alpha^{2j}\psi_{j}, \qquad \lambda\sim \alpha^{-2}\lambda_{-1}+\lambda_{0}+\sum_{j\geq 1} \alpha^{2j}\lambda_{j}\,,\]
so that, formally, we have
\[\mathcal{L}_{\alpha,0}\psi\sim \lambda\psi\,.\]
We are led to solve the equation:
\[\mathcal{M}_{-1}\psi_{0}= -\frac{1}{\tau^2\varphi}\dr_{\varphi}\varphi\dr_{\varphi}\psi_{0}=\lambda_{-1}\psi_{0}\,.\]
We choose $\lambda_{-1}=0$ and $\psi_{0}(\tau,\varphi)=f_{0}(\tau)$, with $f_{0}$ to be chosen in the next step. We shall now solve the equation
\[\mathcal{M}_{-1}\psi_{1}=(\lambda_{0}-\mathcal{M}_{0})\psi_{0}\,.\]
We look for $\psi_{1}$ in the form: $\psi_{1}(\tau,\varphi)=t^2\tilde\psi_{1}(\tau,\varphi)+f_{1}(\tau)$. The equation provides
\begin{equation}\label{alpha0-normal}
-\frac 1\varphi\dr_{\varphi}\varphi\dr_{\varphi}\tilde\psi_{1}=(\lambda_{0}-\mathcal{M}_{0})\psi_{0}\,.
\end{equation}
For each $\tau>0$, the Fredholm condition is $\langle (\lambda_{0}-\mathcal{M}_{0})\psi_{0},1\rangle_{\sL^2((0,\frac 12),\varphi\dx\varphi)}=0$, that becomes
\[\int_{0}^{\frac 12} (\mathcal{M}_{0}\psi_{0})(\tau,\varphi)\,\varphi\dx \varphi=\frac{\lambda_{0}}{2^3}f_{0}(\tau)\,.\]
Moreover we have
\[\int_{0}^{\frac 12} (\mathcal{M}_{0}\psi_{0})(\tau,\varphi)\,\varphi\dx \varphi=-\frac{1}{2^3 \tau^2}\dr_{\tau}\tau^2\dr_{\tau}f_{0}(\tau)+\frac{1}{2^8}\tau^2f_{0}(\tau)\,,\]
so that we get
\[\left(-\frac{1}{\tau^{2}}\dr_{\tau}\tau^2\dr_{\tau}+\frac{1}{2^5}\tau^2\right)f_{0}=\lambda_{0}f_{0}\,.\]
We are led to take
\[\lambda_{0}=\mathfrak{l}_{N}\qquad\mbox{ and }\qquad  f_{0}(\tau)=\mathfrak{f}_{n}(\tau)\,.\]
For this choice of $f_{0}$, we infer the existence of a unique function denoted by $\tilde\psi_{1}^\perp$ (in the Schwartz class with respect to $t$) orthogonal to $1$ in $\sL^2((0,\tfrac12),\varphi \dx\varphi)$ which satisfies \eqref{alpha0-normal}. Using the decomposition of $\psi_{1}$, we have
\[\psi_{1}(\tau,\varphi)=\tau^2\tilde\psi_{1}^\perp(\tau,\varphi)+f_{1}(\tau)\,,\]
where $f_{1}$ has to be determined in the next step.

We leave the construction of the next terms to the reader.

We define
\begin{eqnarray}
\label{quasimode}\Psi_{n}^J(\alpha)(\tau,\theta,\varphi)&=&\sum_{j=0}^J\alpha^{2j}\psi_{j}(\tau,\varphi),\quad\forall (\tau,\theta,\varphi)\in\mathcal{P}\,,\\
\Lambda^J_{n}(\alpha)&=&\sum_{j=0}^J\alpha^{2j}\lambda_{j}\,.
\end{eqnarray}
Due to the exponential decay of the $\psi_{j}$ and thanks to Taylor expansions, there exists $C_{n,J}$ such that:
\[\|\left(\mathcal{L}_{\alpha}-\Lambda^J_{n}(\alpha)\right)\Psi_{n}^J(\alpha)\|_{\sL^2(\mathcal{P},\dx\tilde\mu)}
\leq C_{n,J}\alpha^{2J+2}\|\Psi_{n}^J(\alpha)\|_{\sL^2(\mathcal{P},\dx\tilde\mu)}\,.\]
Using the spectral theorem and going back to the operator $\mathfrak{L}_{\alpha}$ by change of variables, we conclude the proof of Proposition~\ref{theo-quasimodes-cone} with $\gamma_{j,n}=\lambda_{j}$.
\end{proof}
Considering the main term of the asymptotic expansion, we deduce the three following corollaries.
\begin{cor}
For all $N\geq 1$, there exist $C$ and $\alpha_{0}$ and for all $1\leq n\leq N$ and $0\leq\alpha\leq\alpha_{0}$, there exists an eigenvalue $\tilde\lambda_{k(n,\alpha)}$ of $\mathcal{L}_{\alpha}$ such that
\[|\tilde\lambda_{k(n,\alpha)}-\mathfrak{l}_{N}| \leq C\alpha^2\,.\]
\end{cor}
\begin{cor}
We observe that for $1\leq n\leq N$ and $\alpha\in(0,\alpha_{0})$:
\[0\leq\tilde\lambda_{n}(\alpha)\leq\tilde\lambda_{k(n,\alpha)}\leq \mathfrak{l}_{N}+C\alpha^2\,.\]
\end{cor}
\begin{cor}\label{rough-bound-on-la-n}
For all $n\geq 1$, there exist $\alpha_{0}(n)>0$ and $C_{n}>0$ such that, for all $\alpha\in(0,\alpha_{0}(n))$, the $n$-th eigenvalue exists and satisfies:
\[\lambda_{n}(\alpha)\leq C_{n}\alpha\,,\]
or equivalently $\tilde\lambda_{n}(\alpha)\leq C_{n}$.
\end{cor}
\section{Agmon estimates}\label{Sec.Agmon}
Let us first state the following convenient lemma.
\begin{lem}\label{lem.murho}
Let us consider $\rho>0$ and $\mu(\rho)$ is the lowest eigenvalue of the magnetic Neumann Laplacian on the disk of center $(0,0)$ and radius $\rho$. There exists $c>0$ such that, for all $\rho\geq 0$,
\[\mu(\rho)\geq c\min(\rho^2,1)\,.\]
\end{lem}
\begin{proof}
The magnetic Laplacian is in the form $\mathfrak{L}_{\A_{0}, \rho}=(-i\nabla+\A_{0})^2$ with 
\[\A_{0}(\x)=\frac{1}{2}(x_{2}, -x_{1})\,.\]
In Proposition \ref{prop.sB}, we noticed that the magnetic Neumann condition is just the classical Neumann condition. By using the rescaling $\x=\rho\mathsf{y}$, we get that $\mathfrak{L}_{\A_{0}, \rho}=(-i\nabla+\A_{0})^2$ is unitarily equivalent to $\rho^{-2}\mathfrak{L}^\Neu_{\rho^{2}\A_{0}}$ acting on $\sL^2(\mathcal{B}(0,1))$. Then it is easy to see that $\mu$ is a continuous  and positive function on $(0,+\infty)$. By Proposition \ref{prop.sB}, we get 
\[\mu(\rho)\underset{\rho\to 0}{=}\frac{\rho^2}{|\Omega|}\int_{\Omega}|\A_{0}(\x)|^2\dx\x+o(\rho^2)\,.\]
Moreover $\mathfrak{L}_{\A_{0}, \rho}$ is also equivalent to $\rho^2(-i\rho^{-2}\nabla+\A_{0})^2$ acting on $\sL^2(\mathcal{B}(0,1))$. The limit $\rho\to+\infty$ is a semiclassical limit ($h=\rho^{-2}$) and we deduce (see for instance \eqref{eq.BPT-result} and \cite[Theorem 8.1.1]{FouHel10}) that 
\[\mu(\rho)\underset{\rho\to+\infty}{\to}\Theta_{0}\,.\]

\end{proof}
Let us now prove the following fine estimate when $\beta\in\left[0,\frac{\pi}{2}\right)$.
\begin{prop}\label{optimal-Agmon-cone}
Let $C_{0}>0$ and $\eta\in\left(0,\frac{1}{2}\right)$. For all $\beta\in\left[0,\frac{\pi}{2}\right)$, there exist $\alpha_{0}>0$, $\eps_{0}$ and $C>0$ such that for any $\alpha\in(0,\alpha_{0})$ and for all eigenpair $(\lambda,\psi)$ of $\mathcal{L}_{\alpha,\beta}$ satisfying $\lambda\leq C_{0}\alpha$:
\begin{equation}
\int_{\Ca} e^{2\eps_{0}\alpha^{1/2}|z|}|\psi|^2\dx {\bf{x}}\leq C\|\psi\|^2.
\end{equation}
\end{prop}
\begin{proof}
Thanks to a change of gauge $\mathfrak{L}_{\A}$ is unitarily equivalent to the Neumann realization of:
\[\mathfrak{L}_{\hat\A}=D_{z}^2+(D_{x}+z\sin\beta)^2+(D_{y}+x\cos\beta)^2\,.\]
The associated quadratic form is
\[\mathfrak{Q}_{\hat\A}(\psi)=\int |D_{z}\psi|^2+|(D_{x}+z\sin\beta)\psi|^2+|(D_{y}+x\cos\beta)\psi|^2\dx x \dx y \dx z\,.\]
Let us introduce a smooth cut-off function $\chi$ such that $\chi=1$ near $0$ and let us also consider, for $R\geq 1$ and $\eps_{0}>0$,
\[\Phi_{R}(z)=\eps_{0}\alpha^{1/2} \chi\left(\frac{z}{R}\right) |z|\,.\]
The Agmon formula gives
\[\mathfrak{Q}_{\hat\A}(e^{\Phi_{R}}\psi)=\lambda \|e^{\Phi_{R}}\psi\|^2-\|\nabla\Phi_{R} e^{\Phi_{R}}\psi\|^2\,.\]
There exists $\alpha_{0}>0$ and $\tilde C_{0}$ such that for $\alpha\in(0,\alpha_{0})$, $R\geq 1$ and $\eps_{0}\in(0,1)$, we have:
\[\mathfrak{Q}_{\hat\A}(e^{\Phi_{R}}\psi)\leq \tilde C_{0}\alpha \|e^{\Phi_{R}}\psi\|^2\,.\]
We introduce a partition of unity with respect to $z$:
\[\chi_{1}^2(z)+\chi_{2}^2(z)=1\,,\]
where $\chi_{1}(z)=1$ for $0\leq z\leq 1$ and $\chi_{1}(z)=0$ for $z\geq 2.$ 
For $j=1,2$ and $\gamma>0$, we let
\[\chi_{j,\gamma}(z)=\chi_{j}(\gamma^{-1}z)\,,\]
so that
\[\|\chi'_{j,\gamma}\|\leq C\gamma^{-1}\,.\]
The localization formula provides
\begin{equation}\label{IMS-inequality}
\mathfrak{Q}_{\hat\A}(e^{\Phi_{R}}\chi_{1,\gamma}\psi)+\widehat{\mathfrak{Q}_{\A}}(e^{\Phi_{R}}\chi_{2,\gamma}\psi)-C^2\gamma^{-2}\|e^{\Phi_{R}}\psi\|^2
    \leq  \tilde C_{0}\alpha \|e^{\Phi_{R}}\psi\|^2.
\end{equation}
We want to write a lower bound for $\widehat{\mathfrak{Q}_{\A}}(e^{\Phi_{R}}\chi_{2,\gamma}\psi)$. Integrating by slices we have:
\[\mathfrak{Q}_{\hat\A}(\psi)\geq \cos\beta\int \mu(\sqrt{\cos\beta}\,z\tan(\alpha/2))\|\psi\|^2\dx z\]
where $\mu$ is defined in Lemma \ref{lem.murho}. From this lemma, we deduce
\[\mathfrak{Q}_{\hat\A}(e^{\Phi_{R}}\chi_{2,\gamma}\psi)\geq\int c \cos\beta \min(z^2\alpha^2\cos\beta,1) \|e^{\Phi_{R}}\chi_{2,\gamma}\psi\|^2\dx z\,.\]
We choose $\gamma=\eps_{0}^{-1}\alpha^{-1/2}(\cos\beta)^{-1/2}$. On the support of $\chi_{2,\gamma}$ we have $z\geq \gamma$. It follows
\[\mathfrak{Q}_{\hat\A}(e^{\Phi_{R}}\chi_{2,\gamma}\psi)\geq\int  c \cos\beta \min(\eps_{0}^{-2}\alpha,1) \|e^{\Phi_{R}}\chi_{2,\gamma}\psi\|^2\dx z\,.\]
For $\alpha$ such that $\alpha \leq \eps_{0}^2$, we have:
\[\mathfrak{Q}_{\hat\A}(e^{\Phi_{R}}\chi_{2,\gamma}\psi)\geq\int  c\alpha\eps_{0}^{-2} \cos\beta \|e^{\Phi_{R}}\chi_{2,\gamma}\psi\|^2\dx z\,.\]
We deduce that there exists $c>0$, $C>0$ and $\tilde C_{0}>0$ such that for all $\eps_{0}\in(0,1)$ there exists $\alpha_{0}>0$ such that for all  $R\geq 1$ and $\alpha\in(0,\alpha_{0})$:
\[( c \eps_{0}^{-2}\cos\beta \alpha-C\alpha )\|\chi_{2,\gamma}e^{\Phi_{R}}\psi\|^2\leq \tilde C_{0}\alpha \|\chi_{1,\gamma}e^{\Phi_{R}}\psi\|^2\,.\]
Since $\cos\beta>0$ and $\eta>0$, if we choose $\eps_{0}$ small enough, this implies
\[\|\chi_{2,\gamma}e^{\Phi_{R}}\psi\|^2\leq \tilde C \|\chi_{1,\gamma}e^{\Phi_{R}}\psi\|^2\leq \hat C \|\psi\|^2\,.\]
It remains to take the limit $R\to+\infty$.
\end{proof}

\begin{rem}
It turns out that Proposition \ref{optimal-Agmon-cone} is still true for $\beta=\frac{\pi}{2}$. In this case the argument must be changed as follows. Instead of decomposing the integration domain with respect to $z>0$ one should integrate by slices along a fixed direction which is not parallel to the axis of the cone. Therefore we are reduced to analyze the bottom of the spectrum of the Neumann Laplacian on ellipses instead of circles. We leave the details to the reader.
\end{rem}

\section{Axisymmetry of the first eigenfunctions\label{Sec.m0}}

\begin{notation}\label{notation-cone}
From Propositions \ref{inf-ess-sp} and \ref{theo-quasimodes-cone}, we infer that, for all $n\geq 1$, there exists $\alpha_{n}>0$ such that if $\alpha\in(0,\alpha_{n})$, the $n$-th eigenvalue $\tilde\lambda_{n}(\alpha)$ of $\mathcal{L}_{\alpha}$ exists. Due to the fact that $-i\dr_{\theta}$ commutes with the operator, one deduces that for each $n\geq 1$, we can consider a basis $(\psi_{n,j}(\alpha))_{j=1,\cdots J(n,\alpha)}$ of the eigenspace of $\mathcal{L}_{\alpha}$ associated with $\tilde\lambda_{n}(\alpha)$ such that 
$$\psi_{n,j}(\alpha)(\tau,\theta,\varphi)=e^{im_{n,j}(\alpha)\theta}\Psi_{n,j}(\tau,\varphi)\,.$$
\end{notation}
As an application of the localization estimates of Section \ref{Sec.Agmon}, we prove the following proposition.
\begin{prop}\label{m=0}
For all $n\geq 1$, there exists $\alpha_{n}>0$ such that if $\alpha\in(0,\alpha_{n})$, we have:
$$m_{n,j}(\alpha)=0, \quad \forall j=1,\ldots, J(n,\alpha)\,.$$
In other words, the functions of the $n$-th eigenspace are independent from $\theta$ as soon as $\alpha$ is small enough.
\end{prop}

 In order to succeed, we use a contradiction argument: We consider an $\sL^2$-normalized eigenfunction of $\mathcal{L}_{\alpha}$  associated to $\lambda_{n}(\alpha)$ in the form $e^{im(\alpha)\theta} \Psi_{\alpha}(\tau,\varphi)$ and we assume that there exists $\alpha>0$ as small as we want such that $m(\alpha)\neq 0$ or equivalently $|m(\alpha)|\geq 1$.

\subsection{Dirichlet condition on the axis $\varphi=0$}
Let us prove the following lemma.
\begin{lem}\label{Dir-cond}
For all $t>0$, we have $\Psi_{\alpha}(t,0)=0.$
\end{lem}
\begin{proof}
We recall the eigenvalue equation:
$$\mathcal{L}_{\alpha,0,m(\alpha)}\Psi_{\alpha}=\tilde\lambda_{n}(\alpha)\Psi_{\alpha}\,.$$
We deduce:
$$\mathcal{Q}_{\alpha,0,m(\alpha)}(\Psi_{\alpha})\leq C\|\Psi_{\alpha}\|^2_{\sL^2(\mathcal{R},\dx\mu)}\,.$$
This implies:
$$\int_{\mathcal{R}}\frac{1}{\tau^2\sin^2(\alpha \varphi)}\left(m(\alpha)+\frac{\sin^2(\alpha\varphi)}{2\alpha}\tau^2\right)^2|\Psi_{\alpha}(\tau,\varphi)|^2\dx\mu\leq C\|\Psi_{\alpha}\|^2_{\sL^2(\mathcal{R},\dx\mu)}<+\infty\,.$$
Using the inequality $(a+b)^2\geq \frac{1}{2}a^2-2b^2$, it follows:
$$\frac{m(\alpha)^2}{2}\int_{\mathcal{R}}\frac{1}{\tau^2\sin^2(\alpha \varphi)}|\Psi_{\alpha}(\tau,\varphi)|^2\dx\mu
-2\int_{\mathcal{R}}\frac{\tau^2\sin^2(\alpha \varphi)}{4\alpha^2}|\Psi_{\alpha}(\tau,\varphi)|^2\dx\mu<+\infty\,,$$
so that:
$$m(\alpha)^2\int_{\mathcal{R}}\frac{1}{\tau^2\sin^2(\alpha \varphi)}|\Psi_{\alpha}(\tau,\varphi)|^2\dx\mu<+\infty\,,$$
and:
\begin{equation}\label{integrability1}
\int_{\mathcal{R}}\frac{1}{\tau^2\sin^2(\alpha \varphi)}|\Psi_{\alpha}(\tau,\varphi)|^2\dx\mu<+\infty\,.
\end{equation}
Therefore, for almost all $\tau>0$, we have:
\begin{equation}\label{integrability2}
\int_{0}^{\frac 12}\frac{1}{\sin^2(\alpha \varphi)}|\Psi_{\alpha}(\tau,\varphi)|^2\,\sin(\alpha\varphi)\dx \varphi<+\infty\,.
\end{equation}
The function $\mathcal{R}\ni (\tau,\varphi)\mapsto\Psi_{\alpha}(\tau,\varphi)$ is smooth by elliptic regularity inside $\Ca$ (thus $\mathcal{R}$). In particular, it is continuous at $\varphi=0$. By the integrability property \eqref{integrability2}, this imposes that, for all $\tau>0$, we have $\Psi_{\alpha}(\tau,0)=0.$
\end{proof}

\subsection{End of the proof of Proposition $\ref{m=0}$}
We have
\begin{equation}\label{t-eigen-eq}
\mathcal{L}_{\alpha,0,m(\alpha)}(\tau\Psi_{\alpha})=\tilde\lambda_{n}(\alpha) \tau\Psi_{\alpha}+[\mathcal{L}_{\alpha,0,m(\alpha)},\tau]\Psi_{\alpha}\,.
\end{equation}
We have:
$$[\mathcal{L}_{\alpha,0,m(\alpha)},\tau]=[-\tau^{-2}\dr_{\tau}\tau^2\dr_{\tau},\tau]=-2\dr_{\tau}-\frac{2}{\tau}\,.$$
We take the scalar product of the equation \eqref{t-eigen-eq} with $t\Psi_{\alpha}$. We notice that:
 $$\langle [\mathcal{L}_{\alpha,0,m(\alpha)},\tau]\Psi_{\alpha},\tau\Psi_{\alpha}\rangle_{\sL^2(\mathcal{R},\dx\mu)}=-2\|\Psi_{\alpha}\|^2_{\sL^2(\mathcal{R},\dx\mu)}+3\|\Psi_{\alpha}\|^2_{\sL^2(\mathcal{R},\dx\mu)}=\|\Psi_{\alpha}\|^2_{\sL^2(\mathcal{R},\dx\mu)}\,.$$
The Agmon estimates provide:
$$|\langle \tau[\mathcal{L}_{\alpha,0,m(\alpha)},\chi_{\alpha,\eta}]\Psi_{\alpha},\tau\Psi_{\alpha}\rangle_{\sL^2(\mathcal{R},\dx\mu)}|={\mathcal O}(\alpha^\infty)\|\Psi_{\alpha}\|^2_{\sL^2(\mathcal{R},\dx\mu)}\,.$$
We infer:
$$\mathcal{Q}_{\alpha,0,m(\alpha)}(\tau\Psi_{\alpha})\leq C(\|\tau\Psi_{\alpha}\|^2_{\sL^2(\mathcal{R},\dx\mu)}+\|\Psi_{\alpha}\|^2_{\sL^2(\mathcal{R},\dx\mu)})\,,$$
 and especially:
$$\alpha^{-2}\int_{\mathcal{R}} |\dr_{\varphi}\Psi_{\alpha}|^2\dx\mu 
  \leq C\left(\|t\Psi_{\alpha}\|^2_{\sL^2(\mathcal{R},\dx\mu)}+\|\Psi_{\alpha}\|^2_{\sL^2(\mathcal{R},\dx\mu)}\right)\,.$$
Lemmas \ref{Dir-cond} and \ref{lowest-ev-dir} imply that:
$$c_{0}\alpha^{-2}\int_{\mathcal{R}}|\Psi_{\alpha}|^2\dx\mu
  \leq C\left(\|\tau\Psi_{\alpha}\|^2_{\sL^2(\mathcal{R},\dx\mu)}+\|\Psi^\cut\|^2_{\sL^2(\mathcal{R},\dx\mu)}\right)\,.$$
With the estimates of Agmon, we have:
$$c_{0}\alpha^{-2} \|\Psi_{\alpha}\|^2_{\sL^2(\mathcal{R},\dx\mu)} 
  \leq \tilde C\|\Psi_{\alpha}\|^2_{\sL^2(\mathcal{R},\dx\mu)}\,.$$
We infer that, for $\alpha$ small enough, 
$\Psi_{\alpha}=0$ 
and this is a contradiction. This ends the proof of Proposition \ref{m=0}.

\section{Spectral gap in the axisymmetric case \label{Sec.Synth}}
This section is devoted to the proof of the following proposition. 
\begin{prop}\label{spectral-gap}
For all $n\geq 1$, there exists $\alpha_{0}(n)>0$ such that, for all $\alpha\in(0,\alpha_{0}(n))$, the $n$-th eigenvalue exists and satisfies:
$$\lambda_{n}(\alpha,0)\geq \gamma_{0,n}\alpha +{o}(\alpha)\,,$$
or equivalently $\tilde\lambda_{n}(\alpha,0)\geq \gamma_{0,n} +{o}(1)$.
\end{prop}
We first establish approximation results satisfied by the eigenfunctions in order to catch their behavior with respect to the $t$-variable. Then, we can apply a reduction of dimension and we are reduced to a family of 1D model operators.

\subsection{Approximation of the eigenfunctions \label{Sec.Moy}}
Let us consider $N\geq 1$ and let us introduce:
$${\mathfrak{E}}_{N}(\alpha)=\spann\{\psi_{n,1}(\alpha),{1\leq n\leq N}\}\,,$$
where $\psi_{n,1}(\alpha)(t,\theta,\psi)=\Psi_{n,1}(t,\varphi)$ are considered as functions defined in $\mathcal{P}$.

\begin{prop}\label{approximations}
For all $N\geq 1$, there exist $\alpha_{0}(N)>0$ and $C_{N}>0$ such that, for all $\psi\in{\mathfrak{E}}_{N}(\alpha)$:
\begin{eqnarray}
\label{approx1-cone}\|\tau^{-1}(\psi-\underline{\psi})\|^2_{\sL^2(\mathcal{P},\dx\tilde\mu)}
  &\leq& C_{N}\alpha^2\|\psi\|^2_{\sL^2(\mathcal{P},\dx\tilde\mu)}\,,\\
\label{approx2-cone}\|\psi-\underline{\psi}\|^2_{\sL^2(\mathcal{P},\dx\tilde\mu)}
  &\leq& C_{N}\alpha^2\|\psi\|^2_{\sL^2(\mathcal{P},\dx\tilde\mu)}\,,\\
\label{approx3-cone}\|\tau(\psi-\underline{\psi})\|^2_{\sL^2(\mathcal{P},\dx\tilde\mu)}
  &\leq& C_{N}\alpha^2\|\psi\|^2_{\sL^2(\mathcal{P},\dx\tilde\mu)}\,,\quad\label{eq.t2moy}
\end{eqnarray}
where:
\begin{equation}\label{eq.underpsi}
\underline{\psi}(\tau)=\frac{1}{\int_{0}^{\frac12}\varphi\dx\varphi}\int_{0}^{\frac{1}{2}}\psi(\tau,\varphi)\,\varphi \dx\varphi\,.
\end{equation}
\end{prop}
\begin{proof}
It is sufficient to prove the proposition for $\psi=\psi_{n,1}(\alpha)$ and $1\leq n\leq N$. We have: 
\begin{equation}\label{eq-cone}
\mathcal{L}_{\alpha}\Psi_{n,1}(\alpha)=\tilde\lambda_{n}(\alpha)\Psi_{n,1}(\alpha)\,.
\end{equation}
We have:
$$\mathcal{Q}_{\alpha}(\psi)\leq C\|\psi\|^2_{\sL^2(\mathcal{P},\dx\tilde\mu)}\,,$$
and thus, seeing $\psi$ as a function on $\mathcal{P}$:
$$\frac{1}{\alpha^2}\int_{\mathcal{P}} \tau^{-2}|\dr_{\varphi}\psi|^2\dx\tilde\mu\leq C\|\psi\|^2_{\sL^2(\mathcal{P},\dx\tilde\mu)}\,.$$
We get:
$$\int_{\mathcal{P}} |\dr_{\varphi}\psi|^2\, \sin\alpha\varphi \dx \tau\dx\theta \dx\varphi
    \leq C\alpha^2\|\psi\|^2_{\sL^2(\mathcal{P},\dx\tilde\mu)}\,,$$
so that (using the inequality $\sin(\alpha\varphi)\geq\frac{\alpha\varphi}2$):
$$\int_{\mathcal{P}}  \frac{\alpha\varphi}{2} |\dr_{\varphi}\psi|^2 \dx \tau \dx\theta \dx\varphi
    \leq C\alpha^2\|\psi\|^2_{\sL^2(\mathcal{P},\dx\tilde\mu)}\,.$$
We infer:
$$\int_{\mathcal{P}} \alpha\varphi |\dr_{\varphi}(\psi-\underline{\psi})|^2 \dx \tau \dx\theta \dx\varphi
    \leq C\alpha^2\|\psi\|^2_{\sL^2(\mathcal{P},\dx\tilde\mu)}\,.$$
Let us consider the Neumann realization of the operator $-\frac1\varphi\partial_{\varphi}\varphi\partial_{\varphi}$ on $\sL^2((0,\frac 12),\varphi\dx\varphi)$. The first eigenvalue is simple, equal to $0$ and associated to constant functions. Let $\delta>0$ be the second eigenvalue.
The function $\psi-\underline\psi$ is orthogonal to constant functions in $\sL^2((0,\frac 12)\varphi\dx\varphi)$ by definition \eqref{eq.underpsi}. Then, we apply the min-max principle to $\psi-\underline\psi$ and deduce:
$$\int_{\mathcal{P}} \delta \alpha\varphi |\psi-\underline{\psi}|^2 \dx \tau \dx\theta \dx\varphi\leq C\alpha^2\|\psi\|^2_{\sL^2(\mathcal{P},\dx\tilde\mu)}\,,$$
and:
$$\int_{\mathcal{P}} \tau^{-2}|\psi-\underline{\psi}|^2 \dx\tilde\mu\leq\tilde C\alpha^2\|\psi\|^2_{\sL^2(\mathcal{P},\dx\tilde\mu)}\,,$$
which ends the proof of \eqref{approx1-cone}.
We multiply \eqref{eq-cone} by $t$ and we take the scalar product with $\tau\psi$ to get:
$$\mathcal{Q}_{\alpha}(\tau\psi)\leq \tilde\lambda_{n}(\alpha)\|\tau\psi\|^2_{\sL^2(\mathcal{P},\dx\tilde\mu)}+\left|\langle[-\tau^{-2}\dr_{\tau}\tau^2\dr_{\tau},\tau]\psi ,\tau\psi \rangle_{\sL^2(\mathcal{P},\dx\tilde\mu)}\right|\,.$$
We recall that:
$$[-\tau^{-2}\dr_{\tau}\tau^2\dr_{\tau},\tau]=-2\dr_{\tau}-\frac{2}{\tau}\,.$$
We get:
$$\mathcal{Q}_{\alpha,0}(t\psi)\leq C\|\psi\|^2_{\sL^2(\mathcal{P},\dx\tilde\mu)}\,.$$
We deduce \eqref{approx2-cone} in the same way as \eqref{approx1-cone}.\\
Finally, we easily get:
$$\mathcal{Q}_{\alpha,0}(\tau^2\psi)
\leq \tilde\lambda_{n}(\alpha)\|\tau^2\psi\|^2_{\sL^2(\mathcal{P},\dx\tilde\mu)}+\left|\langle[-\tau^{-2}\dr_{\tau}\tau^2\dr_{\tau},\tau^2]\psi ,\tau^2\psi \rangle_{\sL^2(\mathcal{P},\dx\tilde\mu)}\right|
\,.$$
The commutator is:
$$[-\tau^{-2}\dr_{\tau}\tau^2\dr_{\tau},\tau^2]=-6-4\tau\dr_{\tau}\,.$$
This implies:
$$\mathcal{Q}_{\alpha,0}(\tau^2\psi)\leq C\|\psi\|^2_{\sL^2(\mathcal{P},\dx\tilde\mu)}\,.$$
The approximation \eqref{approx3-cone} follows.
\end{proof}

\subsection{Proof of Proposition \ref{spectral-gap}}
We have now the elements to prove Proposition \ref{spectral-gap}. The main idea is to apply the min-max principle to the quadratic form $\mathcal{Q}_{\alpha,0}$ and to the space ${\mathfrak{E}}_{N}(\alpha)$.
\begin{lem}\label{lemma}
For all $N\geq 1$, there exist $\alpha_N>0$ and $C_{N}>0$ such that, for all $\alpha\in (0,\alpha_N)$ and for all $\psi\in{\mathfrak{E}}_{N}(\alpha)$:
\begin{equation*}
\int_{\mathcal{P}} \left(|\dr_{\tau}\psi|^2+ 2^{-5} |\tau\psi|^2+\frac{1}{\alpha^2 \tau^2}|\dr_{\varphi}\psi|^2\right) \dx\tilde\mu
    \leq \tilde\lambda_{n}(\alpha)\|\psi\|^2_{\sL^2(\mathcal{P},\dx\tilde\mu)}+C_{N}\alpha\|\psi\|^2_{\sL^2(\mathcal{P},\dx\tilde\mu)}\,.
\end{equation*}

\end{lem}
\begin{proof}
We recall that, for all $\psi\in{\mathfrak{E}}_{n}(\alpha)$, we have:
$$\mathcal{Q}_{\alpha,0}(\psi)\leq \tilde\lambda_{n}(\alpha)\|\psi\|^2_{\sL^2(\mathcal{P},\dx\tilde\mu)}\,.$$
We infer that:
\begin{equation*}
\int_{\mathcal{P}}\left( |\dr_{\tau}\psi|^2+\frac{\sin^2(\alpha\varphi)}{4\alpha^2} |\tau\psi|^2+\frac{1}{\alpha^2 \tau^2}|\dr_{\varphi}\psi|^2 \right)\dx\tilde\mu
\leq \tilde\lambda_{n}(\alpha)\|\psi\|^2_{\sL^2(\mathcal{P},\dx\tilde\mu)}\,.
\end{equation*}
We shall analyze the term $\int_{\mathcal{P}}\frac{\sin^2(\alpha\varphi)}{4\alpha^2} |\tau\psi|^2\dx\tilde\mu$. We get:
$$\left|\int_{\mathcal{P}} \frac{\sin^2(\alpha\varphi)}{4\alpha^2}\tau^2|\psi|^2\dx\tilde\mu-\int_{\mathcal{P}} \frac{\sin^2(\alpha\varphi)}{4\alpha^2}\tau^2|\underline{\psi}|^2\dx\tilde\mu\right|
\leq C\|\tau\psi-\tau\underline{\psi}\|_{\sL^2(\mathcal{P},\dx\tilde\mu)} \|\psi\|_{\sL^2(\mathcal{P},\dx\tilde\mu)}\,,$$
and thus:
$$\int_{\mathcal{P}} \frac{\sin^2(\alpha\varphi)}{4\alpha^2}\tau^2|\psi|^2\dx\tilde\mu\geq \int_{\mathcal{P}} \frac{\sin^2(\alpha\varphi)}{4\alpha^2}\tau^2|\underline{\psi}|^2\dx\tilde\mu-C\|\tau\psi-\tau\underline{\psi}\|_{\sL^2(\mathcal{P},\dx\tilde\mu)} \|\psi\|_{\sL^2(\mathcal{P},\dx\tilde\mu)}\,.$$
Proposition \ref{approximations} provides:
\begin{equation}\label{improved-approx}
\|\tau\psi-\tau\underline{\psi}\|_{\sL^2(\mathcal{P},\dx\tilde\mu)}\leq C\alpha\|\psi\|_{\sL^2(\mathcal{P},\dx\tilde\mu)}\,,
\end{equation}
so that:
$$\int_{\mathcal{P}} \frac{\sin^2(\alpha\varphi)}{4\alpha^2}\tau^2|\psi|^2\dx\tilde\mu
\geq \int_{\mathcal{P}} \frac{\sin^2(\alpha\varphi)}{4\alpha^2}\tau^2|\underline{\psi}|^2\dx\tilde\mu-C\alpha^{1/2-\eta}\|\psi\|^2_{\sL^2(\mathcal{P},\dx\tilde\mu)}\,.$$
We deduce:
\begin{equation}\label{lb-interm}
\int_{\mathcal{P}} \frac{\sin^2(\alpha\varphi)}{4\alpha^2}\tau^2|\psi|^2\dx\tilde\mu
\geq (2^{-5}-C\alpha^2)\int_{\mathcal{P}}|\tau\underline{\psi}|^2\dx\tilde\mu-C\alpha\|\psi\|^2_{\sL^2(\mathcal{P},\dx\tilde\mu)}\,.
\end{equation}
Proposition \ref{optimal-Agmon-cone} and \eqref{lb-interm} provide:
$$\int_{\mathcal{P}} \frac{\sin^2(\alpha\varphi)}{4\alpha^2}\tau^2|\psi|^2\dx\tilde\mu
\geq 2^{-5}\int_{\mathcal{P}}|\tau\psi|^2\dx\tilde\mu-C\alpha\|\psi\|^2_{\sL^2(\mathcal{P},\dx\tilde\mu)}\,.$$
\end{proof}
An straightforward consequence of Lemma \ref{lemma} is:
\begin{lem}\label{lemma2}
For all $N\geq 1$, there exist $\alpha_{N}>0$ and $C_{N}>0$ such that, for all $\alpha\in(0,\alpha_{N})$ and for all $\psi\in{\mathfrak{E}}_{N}(\alpha)$:
$$\int_{\mathcal{P}} \left(|\dr_{\tau}\psi|^2+ 2^{-5} |\tau\psi|^2+\frac{1}{\alpha^2 \tau^2}|\dr_{\varphi}\psi|^2\right) \dx{\Breve\mu}
\leq \left(\tilde\lambda_{n}(\alpha)+C_{N}\alpha\right)\|\psi\|^2_{\sL^2(\mathcal{P},\dx{\Breve\mu})}\,,$$
with $\dx{\Breve\mu}=t^2  \varphi \dx \tau \dx\varphi \dx\theta$.
\end{lem}
\begin{proof}
It is sufficient to write for any $\varphi\in(0,\frac 12)$:
$$\varphi=\frac{1}{\alpha}\sin(\alpha\varphi)\frac{\alpha\varphi}{\sin(\alpha\varphi)}=\frac{1}{\alpha}\sin(\alpha\varphi)
(1+{\mathcal O}(\alpha^2))\qquad\mbox{ as }\alpha\to 0\,.$$
\end{proof}
With Lemma~\ref{lemma2}, we deduce (from the min-max principle) that there exists $\alpha_{N}$ such that 
$$\forall \alpha\in(0,\alpha_{N}),\qquad \tilde\lambda_{n}(\alpha)\geq \mathfrak{l}_{N}-C\alpha\,.$$
This achieves the proof of Proposition~\ref{spectral-gap}.
\section{Dimensional reduction for a general orientation}
By using commutator formulas in the spirit of Proposition \ref{P2A} jointly with the estimates of Agmon, one can prove that:
\begin{lem}\label{app-t}
Let $k\geq 0$ and $C_{0}>0$. There exist $\alpha_{0}>0$ and $C>0$ such that for all $\alpha\in(0,\alpha_{0})$ and all eigenpair $(\lambda,\psi)$ of $\mathcal{L}_{\alpha,\beta}$ such that $\lambda\leq C_{0}$:
$$\|\tau^{k} \psi-\tau^{k}\underline{\psi}_{\theta}\|\leq C\alpha^{1/2}\|\psi\|\,,$$
with $$\underline{\psi}_{\theta}(\tau,\varphi)=\frac{1}{2\pi}\int_{0}^{2\pi}\psi(\tau,\theta,\varphi)\dx\theta\,.$$
\end{lem}
We also get an approximation of $D_{t}\psi$.
\begin{lem}\label{app-Dt}
Let $C_{0}>0$. There exist $\alpha_{0}>0$ and $C>0$ such that for all $\alpha\in(0,\alpha_{0})$ and all eigenpair $(\lambda,\psi)$ of $\mathcal{L}_{\alpha,\beta}$ such that $\lambda\leq C_{0}$, we have:
$$\|D_{\tau}\psi-D_{\tau}\underline{\psi}_{\theta}\|\leq C\alpha^{1/2}\|\psi\|\,.$$
\end{lem}
The last two lemmas imply the following proposition:
\begin{prop}
There exist $C>0$ and $\alpha_{0}>0$ such that for any $\alpha\in(0,\alpha_{0})$ and all $\psi\in\mathfrak{E}_{N}(\alpha)$, we have 
\begin{equation}
\mathcal{Q}_{\alpha,\beta}(\psi)\geq(1-\alpha) \mathcal{Q}_{\alpha,\beta}^\model(\psi)-C\alpha^{1/2}\|\psi\|^2,
\end{equation}
where:
\begin{multline*}
\mathcal{Q}_{\alpha,\beta}^\model(\psi)=\\
\int_{\mathcal{P}} |D_{\tau}\psi|^2\dx\tilde\mu+\frac{1}{2^4}\int_{\mathcal{P}} \cos^2(\alpha\varphi) \tau^2\sin^2\beta |\psi|^2\dx\tilde\mu+\int_{\mathcal{P}} \frac{1}{\tau^2\sin^2(\alpha\varphi)}|(D_{\theta}+A_{\theta,1})\psi|^2\dx\tilde\mu+\|P_{3}\psi\|^2.
\end{multline*}
\end{prop}
The spectral analysis is then reduced to an axisymmetric case.

%\mainmatter
\part{Waveguides}\label{Part.Wave}

\chapter{Magnetic effects in curved waveguides}\label{chapter-mwg}
\begin{flushright}
\begin{minipage}{0.65\textwidth}
Hic, ne deficeret, metuens avidusque videndi \\
Flexit amans oculos, et protinus illa relapsa est.\\ 
Bracchiaque intendens prendique et prendere certans\\ 
Nil nisi cedentes infelix arripit auras.\\
Jamque iterum moriens non est de coniuge quicquam\\
Questa suo (quid enim nisi se quereretur amatam?)\\
Supremumque vale, quod iam vix auribus ille\\
Acciperet, dixit revolutaque rursus eodem est.
\begin{flushright}
\textit{Metamorphoses}, Liber X, Ovidius
\end{flushright}
\vspace*{0.5cm}
\end{minipage}
\end{flushright}

In this chapter we prove Theorem  \ref{Thm-2D} and we give the main steps in the proof of Theorem \ref{Thm-3D} which is much more technically involved. In particular we show on this non trivial example how to establish the norm resolvent convergence (see Lemma \ref{NRC}).

\section{Two dimensional waveguides}\label{2D}

\subsection{Proof of Theorem \ref{Thm-2D}}
Let us consider $\delta\leq 1$ and $K\geq 2\sup\frac{\kappa^2}{4}$.
\paragraph{A first approximation}
We let:
$$\mathcal{L}^{[2]}_{\eps,\delta}=\mathcal{L}^{[2]}_{\eps,\eps^{-\delta}\mathcal{A}_{\eps}}-\eps^{-2}\lambda_{1}^\Dir(\omega)+K$$
and
$$\mathcal{L}^{\app,[2]}_{\eps,\delta}=(i\dr_{s}+\eps^{1-\delta}\B(s,0)\tau)^2-\frac{\kappa^2}{4}-\eps^{-2}\dr_{\tau}^2-\eps^{-2}\lambda_{1}^\Dir(\omega)+K\,.$$
The corresponding quadratic forms, defined on $\sH^1_{0}(\Omega)$, are denoted by $\mathcal{Q}^{[2]}_{\eps,\delta}$ and $\mathcal{Q}^{\app, [2]}_{\eps,\delta}$ whereas the sesquilinear forms are denoted by $\mathcal{B}^{[2]}_{\eps,\delta}$ and $\mathcal{B}^{\app, [2]}_{\eps,\delta}$.
We can notice that:
$$\left|V_{\eps}(s,\tau)-\left(-\frac{\kappa(s)^2}{4}\right)\right|\leq C\eps$$
so that the operators $\mathcal{L}^{[2]}_{\eps,\delta}$ and $\mathcal{L}^{\app,[2]}_{\eps,\delta}$ are invertible for $\eps$ small enough. Moreover there exists $c>0$ such that for all $\varphi\in \sH^1_{0}(\Omega)$:
$$\mathcal{Q}^{[2]}_{\eps,\delta}(\varphi)\geq c\|\varphi\|^2,\quad \mathcal{Q}^{\app,[2]}_{\eps,\delta}(\varphi)\geq c\|\varphi\|^2\,.$$
Let $\phi,\psi\in \sH^1_{0}(\Omega)$.
We have to analyse the difference of the sesquilinear forms:
$$\mathcal{B}^{[2]}_{\eps,\delta}(\phi,\psi)-\mathcal{B}^{\app,[2]}_{\eps,\delta}(\phi,\psi)\,.$$
We easily get:
$$\left|\langle V_{\eps}\phi,\psi\rangle-\langle -\frac{\kappa^2}{4}\phi,\psi\rangle\right|\leq C\eps\|\phi\|\|\psi\|\leq \tilde C\eps\sqrt{\mathcal{Q}^{[2]}_{\eps,\delta}(\psi)}\sqrt{\mathcal{Q}^{\app,[2]}_{\eps,\delta}(\phi)} \,.$$
We must investigate:
$$\langle m_{\eps}^{-1} (i\dr_{s}+b\mathcal{A}_{1}(s,\eps\tau))m^{-1/2}_{\eps}\phi,(i\dr_{s}+b\mathcal{A}_{1}(s,\eps\tau))m^{-1/2}_{\eps}\psi \rangle\,.$$
We notice that:
$$|\dr_{s}m^{-1/2}_{\eps}|\leq C\eps,\quad |m^{-1/2}_{\eps}-1|\leq C\eps\,.$$
We have:
\begin{multline*}
|\langle m_{\eps}^{-1} (i\dr_{s}+b\mathcal{A}_{1}(s,\eps\tau))m^{-1/2}_{\eps}\phi,(i\dr_{s}+b\mathcal{A}_{1}(s,\eps\tau))(m^{-1/2}_{\eps}-1)\psi \rangle|\\
\leq C\eps \|m^{-1/2}_{\eps} (i\dr_{s}+b\mathcal{A}_{1}(s,\eps\tau))m^{-1/2}_{\eps}\phi\|(\|\psi\|+\| m^{-1/2}_{\eps}(i\dr_{s}+b\mathcal{A}_{1}(s,\eps\tau))\psi\|)\\
\leq C\eps  (\|(i\dr_{s}+b\mathcal{A}_{1}(s,\eps\tau))\phi\|+\|\phi\|)(\|\psi\|+\| m^{-1/2}_{\eps}(i\dr_{s}+b\mathcal{A}_{1}(s,\eps\tau))\psi\|)\,.
\end{multline*}
By the Taylor formula, we get (since $\delta\leq 1$):
\begin{equation}\label{Taylor-A1}
|\mathcal{A}_{1}(s,\eps\tau)-\eps b\B(s,0)\tau|\leq Cb\eps^2\leq C\eps\,.
\end{equation}
so that:
$$\|(i\dr_{s}+b\mathcal{A}_{1}(s,\eps\tau))\phi\|\leq \|(i\dr_{s}+\eps b\B(s,0)\tau)\phi\|+Cb\eps^2\|\phi\|\,.$$
We infer that:
\begin{multline*}
|\langle m_{\eps}^{-1} (i\dr_{s}+b\mathcal{A}_{1}(s,\eps\tau))m^{-1/2}_{\eps}\phi,(i\dr_{s}+b\mathcal{A}_{1}(s,\eps\tau))(m^{-1/2}_{\eps}-1)\psi \rangle|\\
\leq C\eps\left(\|\phi\|\|\psi\|+\|\phi\|\sqrt{\mathcal{Q}^{[2]}_{\eps,\delta}(\psi)}+\|\psi\|\sqrt{\mathcal{Q}^{\app,[2]}_{\eps,\delta}(\phi)}+\sqrt{\mathcal{Q}^{[2]}_{\eps,\delta}(\psi)}\sqrt{\mathcal{Q}^{\app,[2]}_{\eps,\delta}(\phi)}\right)\\
\leq \tilde C\eps\sqrt{\mathcal{Q}^{[2]}_{\eps,\delta}(\psi)}\sqrt{\mathcal{Q}^{\app,[2]}_{\eps,\delta}(\phi)}\,.
\end{multline*}
It remains to analyse:
$$\langle m_{\eps}^{-1} (i\dr_{s}+b\mathcal{A}_{1}(s,\eps\tau))m^{-1/2}_{\eps}\phi,(i\dr_{s}+b\mathcal{A}_{1}(s,\eps\tau))\psi \rangle\,.$$
With the same kind of arguments, we deduce:
\begin{multline*}
|\langle m_{\eps}^{-1} (i\dr_{s}+b\mathcal{A}_{1}(s,\eps\tau))m^{-1/2}_{\eps}\phi,(i\dr_{s}+b\mathcal{A}_{1}(s,\eps\tau))\psi \rangle-\langle (i\dr_{s}+b\mathcal{A}_{1}(s,\eps\tau))\phi,(i\dr_{s}+b\mathcal{A}_{1}(s,\eps\tau))\psi \rangle|\\
\leq  \tilde C\eps\sqrt{\mathcal{Q}^{[2]}_{\eps,\delta}(\psi)}\sqrt{\mathcal{Q}^{\app,[2]}_{\eps,\delta}(\phi)}\,.
\end{multline*}
We again use \eqref{Taylor-A1} to infer:
\begin{multline*}
\langle (i\dr_{s}+b\mathcal{A}_{1}(s,\eps\tau))\phi,(i\dr_{s}+b\mathcal{A}_{1}(s,\eps\tau))\psi \rangle-\langle (i\dr_{s}+b\mathcal{A}_{1}(s,\eps\tau))\phi,(i\dr_{s}+b\eps\B(s,0)\tau)\psi \rangle|\\
\leq C\eps\|(i\dr_{s}+b\mathcal{A}_{1}(s,\eps\tau))\phi\|\|\psi\|.
\leq \tilde C\eps \sqrt{\mathcal{Q}^{[2]}_{\eps,\delta}(\psi)}\sqrt{\mathcal{Q}^{\app,[2]}_{\eps,\delta}(\phi)}\,.
\end{multline*}
In the same way, we deduce:
\begin{multline*}
\langle (i\dr_{s}+b\mathcal{A}_{1}(s,\eps\tau))\phi,(i\dr_{s}+b\mathcal{A}_{1}(s,\eps\tau))\psi \rangle-\langle (i\dr_{s}+b\eps\B(s,0)\tau)\phi,(i\dr_{s}+b\eps\B(s,0)\tau)\psi \rangle|\\
\leq \tilde C\eps \sqrt{\mathcal{Q}^{[2]}_{\eps,\delta}(\psi)}\sqrt{\mathcal{Q}^{\app,[2]}_{\eps,\delta}(\phi)}\,.
\end{multline*}
We get:
$$\left|\mathcal{B}^{[2]}_{\eps,\delta}(\phi,\psi)-\mathcal{B}^{\app,[2]}_{\eps,\delta}(\phi,\psi)\right|\leq C\eps \sqrt{\mathcal{Q}^{[2]}_{\eps,\delta}(\psi)}\sqrt{\mathcal{Q}^{\app,[2]}_{\eps,\delta}(\phi)}\,.$$
By Lemma \ref{NRC}, we infer that:
\begin{equation}\label{first-app}
\left\|\left(\mathcal{L}^{[2]}_{\eps,\delta}\right)^{-1}-\left(\mathcal{L}^{\app,[2]}_{\eps,\delta}\right)^{-1}\right\|\leq \tilde C\eps.
\end{equation}
\paragraph{Case $\delta<1$.} The same kind of arguments provides:
$$\left|\mathcal{B}^{\app,[2]}_{\eps,\delta}(\phi,\psi)-\mathcal{B}^{\eff,[2]}_{\eps,\delta}(\phi,\psi)\right|\leq C\eps^{1-\delta} \sqrt{\mathcal{Q}^{\app,[2]}_{\eps,\delta}(\psi)}\sqrt{\mathcal{Q}^{\eff,[2]}_{\eps,\delta}(\phi)}$$
By Lemma \ref{NRC}, we get that:
$$\left\|\left(\mathcal{L}^{\app,[2]}_{\eps,\delta}\right)^{-1}-\left(\mathcal{L}^{\eff,[2]}_{\eps,\delta}\right)^{-1}\right\|\leq \tilde C\eps^{1-\delta}\,.$$
\paragraph{Case $\delta=1$.}
This case is slightly more complicated to analyse. We must estimate the difference the sesquilinear forms:
$$\mathcal{D}_{\eps}(\phi,\psi)=\mathcal{B}^{\app,[2]}_{\eps,1}(\phi,\psi)-\mathcal{B}^{\eff, [2]}_{\eps,1}(\phi,\psi)\,.$$
We have:
$$\mathcal{D}_{\eps}(\phi,\psi)=\langle i\dr_{s}\phi, \B(s,0)\tau\psi \rangle+\langle \B(s,0)\tau\phi,i\dr_{s}\psi \rangle+\langle \B(s,0)^2\tau^2\phi,\psi\rangle-\|\tau J_{1}\|_{\omega}^2\langle \B(s,0)^2\phi,\psi\rangle\,.$$
We introduce the projection defined for $\varphi\in \sH^1_{0}(\Omega)$:
$$\Pi_{0}\varphi=\langle\varphi,J_{1} \rangle_{\omega}\, J_{1} $$
and we let, for all $\varphi\in \sH^1_{0}(\Omega)$:
$$\varphi^{\parallel}=\Pi_{0}\varphi,\quad\varphi^{\perp}=(\Id-\Pi_{0})\varphi\,.$$
We can write:
$$\mathcal{D}_{\eps}(\phi,\psi)=\mathcal{D}_{\eps}(\phi^\para,\psi^\para)+\mathcal{D}_{\eps}(\phi^\para,\psi^\perp)+\mathcal{D}_{\eps}(\phi^\perp,\psi^\para)+\mathcal{D}_{\eps}(\phi^\perp,\psi^\perp)\,.$$
By using that $\langle \tau J_{1}, J_{1}\rangle_{\omega}=0$, we get:
$$\mathcal{D}_{\eps}(\phi^\para,\psi^\para)=0\,.$$
Then we have:
\begin{equation}\label{phi-pa-psi-pe1}
\|\tau J_{1}\|_{\omega}^2\langle \B(s,0)^2\phi^\para,\psi^\perp\rangle=0\,,\quad |\langle \B(s,0)^2\tau^2\phi^\para,\psi^\perp\rangle|\leq C\|\phi^\para\|\|\psi^\perp\|\,.
\end{equation}
Thanks to the min-max principle, we deduce:
\begin{equation}\label{gap}
\mathcal{Q}^{\app,[2]}_{\eps,1}(\psi^\perp)\geq\frac{\lambda_{2}^\Dir(\omega)-\lambda_{1}^\Dir(\omega)}{\eps^2}\|\psi^\perp\|^2,\quad \mathcal{Q}^{\eff,[2]}_{\eps,1}(\phi^\perp)\geq\frac{\lambda_{2}^\Dir(\omega)-\lambda_{1}^\Dir(\omega)}{\eps^2}\|\phi^\perp\|^2.
\end{equation}
Therefore we get:
$$|\langle \B(s,0)^2\tau^2\phi^\para,\psi^\perp\rangle|\leq C\eps\|\phi\|\sqrt{\mathcal{Q}^{\app, [2]}_{\eps,1}(\psi^\perp)}\,.$$
We have:
$$\mathcal{Q}^{\app,[2]}_{\eps,1}(\psi)=\mathcal{Q}^{\app, [2]}_{\eps,1}(\psi^\para)+\mathcal{Q}^{\app,[2]}_{\eps,1}(\psi^\perp)+\mathcal{B}^{\app, [2]}_{\eps,1}(\psi^\para,\psi^\perp)+\mathcal{B}^{\app,[2]}_{\eps,1}(\psi^\perp,\psi^\para)\,.$$
We can write:
$$\mathcal{B}^{\app, [2]}_{\eps,1}(\psi^\para,\psi^\perp)=\langle (i\dr_{s}+\B(s,0)\tau)\psi^\para,(i\dr_{s}+\B(s,0)\tau)\psi^\perp\rangle\,.$$
We notice that:
\begin{equation}\label{phi-pa-psi-pe2}
\langle (i\dr_{s})\psi^\para,(i\dr_{s})\psi^\perp\rangle=0,\quad |\langle\B(s,0)\tau\psi^\para,\B(s,0)\tau\psi^\perp\rangle|\leq C\|\psi^\para\|\|\psi^\perp\|\leq C\|\psi\|^2.
\end{equation}
Moreover we have:
$$|\langle (i\dr_{s})\psi^\para, \B(s,0)\tau\psi^\perp\rangle|\leq C\|(i\dr_{s}\psi)^\para\|\|\psi^\perp\|\leq C\|i\dr_{s}\psi\|\|\psi\|\leq \tilde C\|\psi\|^2+\tilde C\|\psi\|\sqrt{\mathcal{Q}^{\app,[2]}_{\eps,1}(\psi)}\,.$$
The term $\mathcal{B}^{\app,[2]}_{\eps,1}(\psi^\perp,\psi^\para)$ can be analysed in the same way so that:
$$\mathcal{Q}^{\app,[2]}_{\eps,1}(\psi^\perp)\leq \mathcal{Q}^{\app,[2]}_{\eps,1}(\psi)+C\|\psi\|^2+C\|\psi\|\sqrt{\mathcal{Q}^{\app,[2]}_{\eps,1}(\psi)}\leq \tilde C(\|\psi\|^2+\mathcal{Q}^{\app,[2]}_{\eps,1}(\psi))\,.$$
We infer:
\begin{equation}\label{phi-pa-psi-pe3}
|\langle \B(s,0)^2\tau^2\phi^\para,\psi^\perp\rangle|\leq C\eps\|\phi\|\left(\|\psi\|+\sqrt{\mathcal{Q}^{\app,[2]}_{\eps,1}(\psi)}\right).
\end{equation}
We must now deal with the term
$$\langle i\dr_{s}\phi^\para, \B(s,0)\tau\psi^\perp \rangle\,.$$
We have:
$$|\langle i\dr_{s}\phi^\para, \B(s,0)\tau\psi^\perp \rangle|\leq C\|i\dr_{s}\phi\|\|\psi^\perp\|$$
and we easily deduce that:
\begin{equation}\label{phi-pa-psi-pe4}
|\langle i\dr_{s}\phi^\para, \B(s,0)\tau\psi^\perp \rangle|\leq C\eps\sqrt{\mathcal{Q}^{\eff,[2]}_{\eps,1}(\phi)}\left(\|\psi\|+\sqrt{\mathcal{Q}^{\app,[2]}_{\eps,1}(\psi)}\right),
\end{equation}
We also get the same kind of estimate by exchanging $\psi$ and $\phi$. Gathering \eqref{phi-pa-psi-pe1}, \eqref{phi-pa-psi-pe2}, \eqref{phi-pa-psi-pe3} and \eqref{phi-pa-psi-pe4}, we get the estimate:
$$|\mathcal{D}_{\eps}(\phi^\para,\psi^\perp)|\leq C\eps\sqrt{\mathcal{Q}^{\app,[2]}_{\eps,1}(\psi)}\sqrt{\mathcal{Q}^{\eff,[2]}_{\eps,1}(\phi)}\,.$$
By exchanging the roles of $\psi$ and $\phi$, we can also prove:
$$|\mathcal{D}_{\eps}(\phi^\perp,\psi^\para)|\leq C\eps\sqrt{\mathcal{Q}^{\app,[2]}_{\eps,1}(\psi)}\sqrt{\mathcal{Q}^{\eff,[2]}_{\eps,1}(\phi)}\,.$$
We must estimate $\mathcal{D}_{\eps}(\phi^\perp,\psi^\perp)$. With \eqref{gap}, we immediately deduce that:
$$|\langle \B(s,0)^2\tau^2\phi^\perp,\psi^\perp\rangle-\|\tau J_{1}\|_{\omega}^2\langle \B(s,0)^2\phi^\perp,\psi^\perp\rangle|\leq C\eps^2 \|\phi\|\|\psi\|\,.$$
We find that:
$$|\langle i\dr_{s}\phi^\perp, \B(s,0)\tau\psi^\perp \rangle|\leq C\|\psi^\perp\|\|i\dr_{s}\phi\|$$
and this term can treated as the others.
Finally we deduce the estimate:
$$|\mathcal{D}_{\eps}(\phi,\psi)|\leq C\eps \sqrt{\mathcal{Q}^{\app,[2]}_{\eps,1}(\psi)}\sqrt{\mathcal{Q}^{\eff,[2]}_{\eps,1}(\phi)}\,.$$
We apply Lemma~\ref{NRC} and the estimate \eqref{first-app} 
to obtain Theorem~\ref{Thm-2D}.

\subsection{Proof of Corollary~\ref{expansion-eigenvalues-2D}}
Let us expand the operator $\mathcal{L}^{[2]}_{\eps,b\mathcal{A}_{\eps}}$ in formal power series:
$$\mathcal{L}^{[2]}_{\eps,b\mathcal{A}_{\eps}}\sim \sum_{j=0} \eps^{j-2} L_{j}\,,$$
where 
$$L_{0}=-\dr_{\tau}^2,\quad L_{1}=0,\quad L_{2}=(i\dr_{s}+\tau\B(s,0))^2-\frac{\kappa(s)^2}{4}\,.$$
We look for a quasimode in the form of a formal power series:
$$\psi\sim\sum_{j\geq 0}\eps ^j \psi_{j}$$
and a quasi-eigenvalue:
$$\gamma\sim\sum_{j\geq 0}\gamma_{j}\eps^{j-2}\,.$$
We must solve:
$$(L_{0}-\gamma_{0}) u_{0}=0\,.$$
We choose $\gamma_{0}=\frac{\pi^2}{4}$ and we take:
$$\psi_{0}(s,t)=f_{0}(s)J_{1}(\tau)\,,$$
with $J_{1}(\tau)=\cos\left(\frac{\pi\tau}{2}\right)$.
Then, we must solve:
$$(L_{0}-\gamma_{0}) \psi_{1}=\gamma_{1}\psi_{0}\,.$$
We have $\gamma_{1}=0$ and $\psi_{1}=f_{1}(s)J_{1}(\tau).$
Then, we solve:
\begin{equation}\label{Eq-2}
(L_{0}-\gamma_{0}) \psi_{2}=\gamma_{2}u_{0}-L_{2}u_{0}\,.
\end{equation}
The Fredholm condition implies the equation:
$$-\dr_{s}^2 f+\left(\left(\frac{1}{3}+\frac{2}{\pi^2}\right)\B(s,0)^2-\frac{\kappa(s)^2}{4}\right)f_{0}=\mathcal{T}^{[2]}f_{0}=\gamma_{2}f_{0}$$
and we take for $\gamma_{2}=\gamma_{2,n}=\mu_{n}$ a negative eigenvalue of $\mathcal{T}^{[2]}$ and for $f_{0}$ a corresponding normalized eigenfunction (which has an exponential decay).

This leads to the choice:
$$\psi_{2}=\psi_{2}^\perp(s,\tau)+f_{2}(s)J_{1}(\tau)\,,$$
where $\psi_{2}^\perp$ is the unique solution of (\ref{Eq-2}) which satisfies $\langle \psi_{2}^\perp, J_{1}\rangle_{\tau}=0.$
We can continue the construction at any order where this formal series method is used in a semiclassical context). We write $(\gamma_{j,n}, \psi_{j,n})$ instead of $(\gamma_{j}, \psi_{j})$ to emphasize the dependence on $n$ (determined in the choice of $\gamma_{2}$). We let:
\begin{equation}\label{quasi}
\Psi_{J,n}(\eps)=\sum_{j=0}^J \eps^{j}\psi_{j,n}, \mbox{ and }\Gamma_{J,n}(\eps)=\sum_{j=0}^J \eps^{-2+j} \gamma_{j,n}\,.
\end{equation}
A computation provides:
$$\|(\mathcal{L}^{[2]}_{\eps,b\mathcal{A}_{\eps}}-\Gamma_{J,n}(\eps))\Psi_{J,n}(\eps)\|\leq C\eps^{J+1}\,.$$
The spectral theorem implies that:
$$\dist(\Gamma_{J,n}(\eps),\spd(\mathcal{L}^{[2]}_{\eps,b\mathcal{A}_{\eps}}))\leq C\eps^{J+1}\,.$$
It remains to use the spectral gap given by the approximation of the resolvent in Theorem~\ref{Thm-2D} and Corollary~\ref{expansion-eigenvalues-2D} follows.

\section{Three dimensional waveguides}\label{3D}

\subsection{Expression of the operator in curvilinear coordinates}
We will adopt the following notation.
\begin{notation}\label{notation}
Given an open set $U \subset \R^d$ 
and a vector field $\F=\F(y_{1},\cdots, y_{d}) : U\to\R^d$ 
in dimension $d=2,3$, 
we will use in our computations the following notation:
\[
  \curl\F = 
\begin{cases}
  \dr_{y_{1}}\F_{2}-\dr_{y_{2}}\F_{1}
  &\mbox{if}\quad d=2,
  \\
  (\dr_{y_{2}}\F_{3}-\dr_{y_{3}}\F_{2},
  \dr_{y_{3}}\F_{1}-\dr_{y_{1}}\F_{3},
  \dr_{y_{1}}\F_{2}-\dr_{y_{2}}\F_{1})
  &\mbox{if}\quad d=3.
\end{cases}
\]	
\end{notation}
We recall the relations between $\mathcal{A}$, $\mathcal{B}$ and $\A$, $\B$. This can be done in terms of differential forms.
Let us consider the $1$-form:
\[\xi_{\A}=\A_{1}\dx x_{1}+\A_{2}\dx x_{2}+\A_{3}\dx x_{3}\,.\]
We consider $\Phi$ the diffeomorphism defined in \eqref{Phi}. 
The pull-back of $\xi_{\A}$ by $\Phi$ is given by:
\[\Phi^*\xi_{\A}=\mathcal{A}_{1}\dx t_{1}+\mathcal{A}_{2}\dx t_{2}+\mathcal{A}_{3}\dx t_{3}\,.\]
where $\mathcal{A}= (d\Phi)^{\mathsf{T}} \A(\Phi)$ since we have $x=\Phi(t)$. Then, thanks to Chapter \ref{intro}, Section \ref{sec.MBtoB}, we get
\[\mathcal{B}=\widetilde{d\Phi}\B=\det (d\Phi) (d\Phi)^{-1}\B\,,\]
where $\widetilde{d\Phi}$ denotes the adjugate matrix of $d\Phi$. Let us give an interpretation of the components of $\mathcal{B}$.

A straightforward computation provides the following expression for $d\Phi$:
\[
[hT(s)+h_{2}(\sin\theta M_{2}-\cos\theta M_{3})+h_{3}(-\cos\theta M_{2}-\sin\theta M_{3}), \cos\theta M_{2}+\sin\theta M_{3}, -\sin\theta M_{2}+\cos\theta M_{3}]
\]
so that $\det d\Phi=h$ and
\begin{align*}
\mathcal{B}_{23}&=h(h^2+h_{2}^2+h_{3}^2)^{-1/2}\B\cdot T(s),\, \\
\mathcal{B}_{13}&=-h\B\cdot(-\cos\theta M_{2}-\sin\theta M_{3}),\,\\ 
\mathcal{B}_{12}&=h\B\cdot (-\sin\theta M_{2}+\cos\theta M_{3})\,.
\end{align*}
Let us check that $\mathfrak{L}^{[3]}_{\eps,b\A}$ (whose quadratic form is denoted by $\mathfrak{Q}^{[3]}_{\eps,b\A}$) is unitarily equivalent to $\mathfrak{L}^{[3]}_{\eps,b\mathcal{A}}$ given in \eqref{L-frak3}. For that purpose we let
$$G=(d\Phi)^{\mathsf{T}} d\Phi$$
and a computation provides:
$$G=\begin{pmatrix}
h^2+h_{2}^2+h_{3}^2&-h_{3} &-h_{2} \\
-h_{3}& 1& 0\\
-h_{2}& 0&1
\end{pmatrix}$$
and:
$$G^{-1}=\begin{pmatrix}
0&0&0 \\
0& 1& 0\\
0& 0&1
\end{pmatrix}+h^{-2}\begin{pmatrix} 1\\ h_{3}\\ h_{2} \end{pmatrix}\begin{pmatrix} 1&h_{3} &h_{2}\end{pmatrix}\,.
$$
We notice that $|G|=h^2$. In terms of quadratic form we write:
$$\mathfrak{Q}^{[3]}_{\eps,b\A}(\psi)=\int_{\R\times(\eps\omega)}|(d\Phi^{-1})^{\mathsf{T}}(-i\nabla_{t}+(d\Phi)^{\mathsf{T}} \A(\Phi))|^2 \,h\dx t$$
and
\begin{eqnarray*}
&\mathfrak{Q}^{[3]}_{\eps,b\A}(\psi)=&\int_{\R\times(\eps\omega)} \left(|(-i\dr_{t_{2}}+b\mathcal{A}_{2})\psi|^2+ |(-i\dr_{t_{3}}+b\mathcal{A}_{3})\psi|^2\right)\,h \dx t\\
&                                                               &+\int_{\R\times(\eps\omega)}h^{-2}|\left(-i\dr_{s}+b\mathcal{A}_{1}+h_{3}(-i\dr_{t_{2}}+b\mathcal{A}_{2})+h_{2}(-i\dr_{t_{3}}+b\mathcal{A}_{3})\right)\psi|^2\,h\dx t
\end{eqnarray*}
so that
\begin{multline*}
\mathfrak{Q}^{[3]}_{\eps,b\A}(\psi)\\
=\int_{\R\times(\eps\omega)}\!\!\!\!\!\!\! \left(|(-i\dr_{t_{2}}+b\mathcal{A}_{2})\psi|^2+ |(-i\dr_{t_{3}}+b\mathcal{A}_{3})\psi|^2+h^{-2}|(-i\dr_{s}+b\mathcal{A}_{1}-i\theta'\dr_{\alpha}+\mathcal{R})\psi|^2\right)\,h\dx t\,.
\end{multline*}
Since $\omega$ is simply connected (and so is $\Omega_{\eps}$) we may change the gauge and assume that the vector potential is given by:
\begin{eqnarray}\label{explicit-A}
\nonumber\quad&&\mathcal{A}_{1}(s,t_{2},t_{3})=-\frac{t_{2} t_{3}\dr_{s}\mathcal{B}_{23}(s,0,0)}{2}-\int_{0}^{t_{2}} \mathcal{B}_{12}(s,\tilde t_{2},t_{3})\dx \tilde t_{2}-\int_{0}^{t_{3}}\mathcal{B}_{13}(s,0,\tilde t_{3})\dx \tilde t_{3}\,,\\
&&\mathcal{A}_{2}(s,t_{2},t_{3})=-\frac{ t_{3}\mathcal{B}_{23}(s,0,0)}{2}\,,\\
\nonumber&&\mathcal{A}_{3}(s,t_{2},t_{3})=-\frac{ t_{2}\mathcal{B}_{23}(s,0,0)}{2}+\int_{0}^{t_{2}}\mathcal{B}_{23}(s,\tilde t_{2},t_{3})\dx \tilde t_{2}\,.
\end{eqnarray}
In other words, thanks to the Poincar\'e lemma, there exists a (smooth) phase function $\rho$ such that $(d\Phi)^{\mathsf{T}} \A(\Phi)+\nabla_{t}\rho=\mathcal{A}$. In particular, we have: $\mathcal{A}_{j}(s,0)=0, \dr_{j}\mathcal{A}_{j}(s,0)=0$ for $j\in\{1,2,3\}$.

\subsection{Proof of Theorem \ref{Thm-3D}}
Let us consider $\delta\leq 1$ and $K\geq 2\sup\frac{\kappa^2}{4}$.
\paragraph{A first approximation}
We let:
$$\mathcal{L}^{[3]}_{\eps,\delta}=\mathcal{L}^{[3]}_{\eps,\eps^{-\delta}\mathcal{A}_{\eps}}-\eps^{-2}\lambda_{1}^\Dir(\omega)+K$$
and
$$\mathcal{L}^{\app,[3]}_{\eps,\delta}=\sum_{j=2,3} (-i\eps^{-1}\dr_{\tau_{j}}+b\mathcal{A}^\lin_{j,\eps})^2+(-i\dr_{s}+b\mathcal{A}_{1,\eps}^\lin-i\theta'\dr_{\alpha})^2-\frac{\kappa^2}{4}-\eps^{-2}\dr_{\tau}^2-\eps^{-2}\lambda_{1}^\Dir(\omega)+K\,,$$
where:
$$\mathcal{A}^\lin_{j,\eps}(s,\tau)=\mathcal{A}_{j}(s,0)+\eps\tau_{2}\dr_{2}\mathcal{A}_{j}(s,0)+\eps\tau_{3}\dr_{3}\mathcal{A}_{j}(s,0)\,.$$
We recall that $\mathcal{A}$ is given by \eqref{explicit-A} and that $\mathcal{L}^{[3]}_{\eps,\eps^{-\delta}\mathcal{A}_{\eps}}$ is defined in \eqref{L3}. We have to analyse the difference of the corresponding sesquilinear forms:
$$\mathcal{B}^{[3]}_{\eps,\delta}(\phi,\psi)-\mathcal{B}^{\app, [3]}_{\eps,\delta}(\phi,\psi)\,.$$
We leave as an exercise the following estimate:
\begin{equation}\label{first-approx}
\left\|(\mathcal{L}^{[3]}_{\eps,\delta})^{-1}-(\mathcal{L}^{\app,[3]}_{\eps,\delta})^{-1}\right\|\leq \tilde C\eps\,.
\end{equation}
\subsubsection{Case $\delta<1$}
This case is similar to the case in dimension $2$ since $|b\mathcal{A}^\lin_{j,\eps}|\leq C\eps^{1-\delta}$. If we let:
$$\mathcal{L}^{\appp,[3]}_{\eps,\delta}=\sum_{j=2,3} (-i\eps^{-1}\dr_{\tau_{j}})^2+(-i\dr_{s}-i\theta'\dr_{\alpha})^2-\frac{\kappa^2}{4}-\eps^{-2}\dr_{\tau}^2-\eps^{-2}\lambda_{1}^\Dir(\omega)+K\,,$$
we easily get:
$$\left\|(\mathcal{L}^{\appp,[3]}_{\eps,\delta})^{-1}-(\mathcal{L}^{\app,[3]}_{\eps,\delta})^{-1}\right\|\leq \tilde C\eps^{1-\delta}\,.$$
It remains to decompose the sesquilinear form associated with $\mathcal{L}^{\appp,[3]}_{\eps,\delta}$ by using the orthogonal projection $\Pi_{0}$ and the analysis follows the same lines as in dimension $2$.
\subsubsection{Case $\delta=1$} This case cannot be analysed in the same way as in dimension 2. Using the explicit expression of the vector potential \eqref{explicit-A}, we can write our approximated operator in the form:
\begin{eqnarray*}
\mathcal{L}^{\appp,[3]}_{\eps,1}=&\left(-\eps^{-1}i\dr_{\tau_{2}}-\frac{\mathcal{B}_{23}(s,0,0)}{2}\tau_{3}\right)^2+\left(-\eps^{-1}i\dr_{\tau_{3}}+\frac{\mathcal{B}_{23}(s,0,0)}{2}\tau_{2}\right)^2\\
                                                           &+(-i\dr_{s}-i\theta'\dr_{\alpha}-\tau_{2}\mathcal{B}_{12}(s,0,0)-\tau_{3}\mathcal{B}_{13}(s,0,0))^2-\eps^{-2}\lambda_{1}^\Dir(\omega)+K\,.
\end{eqnarray*}

\subsubsection{Perturbation theory}
Let us introduce the operator on $\sL^2(\omega)$ (with Dirichlet boundary condition) and depending on $s$:
$$\mathcal{P}_{\eps}^2=\left(-\eps^{-1}i\dr_{\tau_{2}}-\frac{\mathcal{B}_{23}(s,0,0)}{2}\tau_{3}\right)^2+\left(-\eps^{-1}i\dr_{\tau_{3}}+\frac{\mathcal{B}_{23}(s,0,0)}{2}\tau_{2}\right)^2\,.$$
Thanks to perturbation theory the lowest eigenvalue $\nu_{1,\eps}(s)$ of $\mathcal{P}_{\eps}^2$ is simple and we may consider an associated $\sL^2$ normalized eigenfunction $u_{\eps}(s)$. Let us provide a estimate for the eigenpair $(\nu_{1,\eps}(s),u_{\eps}(s))$. We have to be careful with the dependence on $s$ in the estimates. Firstly, we notice that there exist $\eps_{0}>0$ and $C>0$ such that for all $s$, $\eps\in(0,\eps_{0})$ and all $\psi\in \sH^1_{0}(\omega)$:
\begin{multline}
\int_{\omega} \left|\left(-\eps^{-1}i\dr_{\tau_{2}}-\frac{\mathcal{B}_{23}(s,0,0)}{2}\tau_{3}\right)\psi\right|^2+\left|\left(-\eps^{-1}i\dr_{\tau_{3}}+\frac{\mathcal{B}_{23}(s,0,0)}{2}\tau_{2}\right)\psi\right|^2 \dx \tau\\
\geq \eps^{-2} \int_{\omega} |\dr_{\tau_{2}}\psi|^2+|\dr_{\tau_{3}}\psi|^2 \dx\tau-C\eps^{-1}\|\psi\|^2.
\end{multline}
From the min-max principle we infer that:
\begin{equation}\label{sgap}
\nu_{n,\eps}(s)\geq \eps^{-2}\lambda^\Dir_{n}(\omega)-C\eps^{-1}\,.
\end{equation}
Let us analyse the corresponding upper bound. Thanks to the Fredholm alternative, we may introduce $R_{\omega}$ the unique function such that:
\begin{equation}\label{J1}
(-\Delta^\Dir_{\omega}-\lambda_{1}^{\Dir}(\omega)) R_{\omega}=D_{\alpha} J_{1},\quad \langle R_{\omega}, J_{1}\rangle_{\omega}=0\,.
\end{equation}
We use $v_{\eps}=J_{1}+\eps\mathcal{B}_{23}(s,0,0)R_{\omega}$ as test function for $\mathcal{P}_{\eps}^2$ and an easy computation provides that there exist $\eps_{0}>0$ and $C>0$ such that for all $s$, $\eps\in(0,\eps_{0})$:
$$\left\|\left(\mathcal{P}_{\eps}^2-\left(\eps^{-2}\lambda^\Dir_{1}(\omega)+\mathcal{B}_{23}^2(s,0,0)\left(\frac{\|\tau J_{1}\|_{\omega}^2}{4}-\langle D_{\alpha}R_{\omega},J_{1}\rangle_{\omega}\right)\right)\right)v_{\eps}\right\|_{\omega}\leq C\eps\,.$$
The spectral theorem implies that there exists $n(\eps,s)\geq 1$ such that:
$$\left|\nu_{n(\eps,s),\eps}(s)-\eps^{-2}\lambda^\Dir_{1}(\omega)-\mathcal{B}_{23}^2(s,0,0)\left(\frac{\|\tau J_{1}\|_{\omega}^2}{4}-\langle D_{\alpha}R_{\omega},J_{1}\rangle_{\omega}\right)\right|\leq C\eps\,.$$
Due to the spectral gap uniform in $s$ given by \eqref{sgap} we deduce that  there exist $\eps_{0}>0$ and $C>0$ such that for all $s$, $\eps\in(0,\eps_{0})$:
$$\left|\nu_{1,\eps}(s)-\eps^{-2}\lambda^\Dir_{1}(\omega)-\mathcal{B}_{23}^2(s,0,0)\left(\frac{\|\tau J_{1}\|^2}{4}-\langle D_{\alpha}R_{\omega},J_{1}\rangle_{\omega}\right)\right|\leq C\eps\,.$$
This new information provides:
$$\left\|\left(\mathcal{P}_{\eps}^2-\nu_{1,\eps}(s)\right)v_{\eps}\right\|_{\omega}\leq \tilde C\eps$$
and thus:
$$\left\|\left(\mathcal{P}_{\eps}^2-\nu_{1,\eps}(s)\right)(v_{\eps}-\langle v_{\eps},u_{\eps}\rangle_{\omega} u_{\eps})\right\|_{\omega}\leq \tilde C\eps\,.$$
so that, with the spectral theorem and the uniform gap between the eigenvalues:
$$\left\|v_{\eps}-\langle v_{\eps},u_{\eps}\rangle_{\omega} u_{\eps}\right\|_{\omega}\leq C\eps^3\,.$$
Up to changing $u_{\eps}$ in $-u_{\eps}$, we infer that :
$$||\langle v_{\eps},u_{\eps}\rangle_{\omega}|-\|v_{\eps}\|_{\omega}|\leq C\eps^3,\quad \left\|v_{\eps}-\|v_{\eps}\|_{\omega} u_{\eps}\right\|_{\omega}\leq \tilde C\eps^3\,.$$
Therefore we get:
$$\left\|u_{\eps}-\tilde v_{\eps}\right\|_{\omega}\leq C\eps^3,\quad \tilde v_{\eps}=\frac{v_{\eps}}{\|v_{\eps}\|_{\omega}}$$
and this is easy to deduce:
\begin{equation}\label{approx-H1}
\left\|\nabla_{\tau_{2},\tau_{3}}\left(u_{\eps}-\tilde v_{\eps}\right)\right\|_{\omega}\leq C\eps^3\,.
\end{equation}
\subsubsection{Projection arguments} 
We shall analyse the difference of the sesquilinear forms:
$$\mathcal{D}_{\eps}(\phi,\psi)=\mathcal{L}^{\appp,[3]}_{\eps,1}(\phi,\psi)-\mathcal{L}^{\eff,[3]}_{\eps,1}(\phi,\psi)\,.$$
We write:
$$\mathcal{D}_{\eps}(\phi,\psi)=\mathcal{D}_{\eps,1}(\phi,\psi)+\mathcal{D}_{\eps,2}(\phi,\psi)\,,$$
where
$$\mathcal{D}_{\eps,1}(\phi,\psi)=\langle\mathcal{P}_{\eps}\phi,\mathcal{P}_{\eps}\psi\rangle-\left\langle\left(-\eps^{-2}\Delta_{\omega}^\Dir+\mathcal{B}_{23}^2(s,0,0)\left(\frac{\|\tau J_{1}\|_{\omega}^2}{4}-\langle D_{\alpha}R_{\omega},J_{1}\rangle_{\omega}\right)\right)\phi,\psi\right\rangle$$
and
$$\mathcal{D}_{\eps,2}(\phi,\psi)=\langle\mathcal{M}\phi,\psi\rangle-\langle\mathcal{M}^\eff\phi,\psi\rangle\,,$$
with:
$$\mathcal{M}=\left(-i\dr_{s}-i\theta'\dr_{\alpha}-\tau_{2}\mathcal{B}_{12}(s,0,0)-\tau_{3}\mathcal{B}_{13}(s,0,0)\right)^2\,,$$
$$\mathcal{M}^\eff=\langle(-i\dr_{s}-i\theta'\dr_{\alpha}-\mathcal{B}_{12}(s,0,0)\tau_{2}-\mathcal{B}_{13}(s,0,0)\tau_{3})^2 \Id(s)\otimes J_{1}, \Id(s)\otimes J_{1}\rangle_{\omega}\,.$$
We introduce the projection on $u_{\eps}(s)$:
$$\Pi_{\eps,s}\varphi=\langle\varphi ,u_{\eps}\rangle_{\omega}\, u_{\eps}(s)$$
and, for $\varphi\in \sH^1_{0}(\Omega)$, we let:
$$\varphi^{\para_{\eps}}=\Pi_{\eps,s}\varphi,\quad \varphi^{\perp_{\eps}}=\varphi-\Pi_{\eps,s}\varphi\,.$$
We can write the formula:
$$\mathcal{D}_{\eps,1}(\phi,\psi)=\mathcal{D}_{\eps,1}(\phi^{\para_{\eps}},\psi^\para)+\mathcal{D}_{\eps,1}(\phi^{\para_{\eps}},\psi^\perp)+\mathcal{D}_{\eps,1}(\phi^{\perp_{\eps}},\psi^\para)+\mathcal{D}_{\eps,1}(\phi^{\perp_{\eps}},\psi^\perp)\,,$$
where $\psi^\para=\Pi_{0}\psi=\langle\psi, J_{1}\rangle_{\omega}\, J_{1}$ and $\psi^\perp=\psi-\psi^\para$. Using our mixed decomposition, we can get the following bound on $\mathcal{D}_{\eps,1}(\phi,\psi)$:
\begin{equation}\label{D1}
|\mathcal{D}_{\eps,1}(\phi,\psi)|\leq C\eps\sqrt{\mathcal{Q}^{\app2,[3]}_{\eps,1}(\psi)}\sqrt{\mathcal{Q}^{\eff,[3]}_{\eps,1}(\phi)}\,.
\end{equation}
Moreover we easily get:
\begin{equation}\label{D2}
|\mathcal{D}_{\eps,2}(\phi,\psi)|\leq C\eps\sqrt{\mathcal{Q}^{\app2,[3]}_{\eps,1}(\psi)}\sqrt{\mathcal{Q}^{\eff,[3]}_{\eps,1}(\phi)}\,.
\end{equation}
Combining \eqref{D1} and \eqref{D2}, we infer that:
$$|\mathcal{D}_{\eps}(\phi,\psi)|\leq C\eps\sqrt{\mathcal{Q}^{\app2,[3]}_{\eps,1}(\psi)}\sqrt{\mathcal{Q}^{\eff,[3]}_{\eps,1}(\phi)}\,.$$
With Lemma \ref{NRC} we infer:
\begin{equation}\label{second-approx}
\left\| \left(\mathcal{L}^{\appp,[3]}_{\eps,1}\right)^{-1}-\left(\mathcal{L}^{\eff,[3]}_{\eps,1}\right)^{-1}\right\|\leq C\eps\,.
\end{equation}
Finally we deduce Theorem \ref{Thm-3D} from \eqref{first-approx} and \eqref{second-approx}.

\subsection{Proof of Corollary~\ref{expansion-eigenvalues-3D}}
For the asymptotic expansions of the eigenvalues claimed in Corollary~\ref{expansion-eigenvalues-3D}, we leave the proof to the reader since it is a slight adaptation of the proof of Corollary~\ref{expansion-eigenvalues-2D}.

\chapter{Spectrum of thin triangles and broken waveguides}\label{chapter-triangles}

\begin{flushright}
\begin{minipage}{0.5\textwidth}
O egregiam artem! Scis rotunda metiri, in quadratum redigis quamcumque acceperis formam, interualla siderum dicis, nihil est quod in mensuram tuam non cadat: si artifex es, metire hominis animum, dic quam magnus sit, dic quam pusillus sit.
\begin{flushright}
\textit{Epistulae morales ad Lucilium}, LXXXVIII, Seneca
\end{flushright}
\end{minipage}
\vspace*{0.5cm}
\end{flushright}

This chapter is devoted to the proof of Theorems \ref{spectrumtriangle} and \ref{spectrum-guide}.
\section{Quasimodes and boundary layer}
\subsection{From the triangle to the rectangle}
We first perform a change of variables to transform the triangle into a rectangle:
\begin{equation}
\label{Eut}
   u=x\in (-\pi\sqrt{2},0) ,\quad t=\frac{y}{x+\pi\sqrt{2}} \in (-1,1)\,.
\end{equation}
so that $\Tri$ is transformed into 
\begin{equation}
\label{E:Rec}
   \Rec=(-\pi\sqrt{2},0)\times(-1,1)\,.
\end{equation}
The operator $\mathcal{L}_{\Tri}(h)$ becomes:
\begin{equation}
\label{E:LRec}
   \mathcal{L}_{\Rec}(h)(u,t;\partial_u,\partial_t) = -h^2\Big(\dr_{u}
   -\frac{t}{u+\pi\sqrt{2}}\,\dr_{t}\Big)^2-\frac{1}{(u+\pi\sqrt{2})^2}\,\dr_{t}^2\,,
\end{equation}
with Dirichlet boundary conditions on $\partial\Rec$. The equation $\mathcal{L}_{\Tri}(h)\psi_h=\beta_h\psi_h$  is transformed into the equation 
\[
   \mathcal{L}_{\Rec}(h)\hat\psi_h=\beta_h\hat\psi_h \quad\mbox{with}\quad 
   \hat\psi_h(u,t) = \psi_h(x,y).
\]

\subsection{Quasimodes}
We want to construct quasimodes $(\beta_h,\psi_h)$ for the operator $\mathcal{L}_{\Tri}(h)(\partial_x,\partial_y)$. It will be more convenient to work on the rectangle $\Rec$ with the operator $\mathcal{L}_{\Rec}(h)(u,t;\partial_u,\partial_t)$.
We introduce the new scales
\begin{equation}
\label{E:ssigma2}
   s=h^{-2/3}u\quad\mbox{and} \quad\sigma=h^{-1}u\,,
\end{equation}
and we look quasimodes $(\beta_h,\hat\psi_h)$ in the form of series 
\begin{equation}
\label{5EAn}
   \beta_h\sim\sum_{j\geq 0}\beta_{j}h^{j/3} 
   \quad\mbox{and}\quad
   \hat\psi_h(u,t) \sim 
   \sum_{j\geq 0}\big(\Psi_{j}(s,t)+\Phi_{j}(\sigma,t)\big) h^{j/3}
\end{equation}
in order to solve
$\mathcal{L}_{\Rec}(h)\hat\psi_h=\beta_h\hat\psi_h$ in the sense of formal series.
As will be seen hereafter, an Ansatz containing the scale $h^{-2/3}u$ alone (like for the Born-Oppenheimer operator $\mathcal{H}_{\BO, \Tri}(h)$) is not sufficient to construct quasimodes for $\mathcal{L}_{\Rec}(h)$.
Expanding  the operator in powers of $h^{2/3}$, we obtain the formal series:
\begin{equation}
\label{E:Lj}
   \mathcal{L}_{\Rec}(h)(h^{2/3}s,t;h^{-2/3}\partial_s,\partial_t) 
   \sim \sum_{j\geq 0} \mathcal{L}_{2j}h^{2j/3}
   \quad\mbox{with leading term}\quad \mathcal{L}_{0} = -\frac{1}{2\pi^2}\dr^2_{t}
\end{equation}
and in powers of $h$:
\begin{equation}
\label{E:Nj}
   \mathcal{L}_{\Rec}(h)(h\sigma,t;h^{-1}\partial_\sigma,\partial_t) \sim
   \sum_{j\geq 0} \mathcal{N}_{3j}h^{j}
   \quad\mbox{with leading term}\quad \mathcal{N}_{0} = -\dr_{\sigma}^2-\frac{1}{2\pi^2}\dr^2_{t}\,.
\end{equation}
In what follows, in order to finally ensure the Dirichlet conditions on the triangle $\Tri$, we will require for our Ansatz the boundary conditions, for any $j\in\N$:
\begin{gather}
\label{5Dir1}
   \Psi_{j}(0,t)+\Phi_{j}(0,t)=0,\quad -1\leq t\leq1 \\
\label{5Dir2}
   \Psi_{j}(s,\pm1)=0,\ \  s<0 \quad\mbox{and}\quad 
\Phi_{j}(\sigma,\pm1)=0,\ \   \sigma\leq 0\,.
\end{gather}

More specifically, we are interested in the ground energy $\lambda=\frac18$ of the Dirichlet problem for $\mathcal{L}_0$ on the interval $(-1,1)$. Thus we have to solve Dirichlet problems for the operators $\mathcal{N}_0-\frac18$ and $\mathcal{L}_0-\frac18$ on the half-strip 
\begin{equation}
\label{E:Hst}
   \Hst = \R_-\times(-1,1)\,,
\end{equation} 
and look for exponentially decreasing solutions. The situation is similar to that encountered in thin structure asymptotics with Neumann boundary conditions. The following lemma shares common features with the Saint-Venant principle, see for example \cite[\S2]{DauGru98}.

\begin{lem}\label{lem-N0}
We denote the first normalized eigenvector of $\mathcal{L}_0$ on $\sH^1_0((-1,1))$ by $c_0$:
\[
   c_{0}(t)=\cos\left(\frac{\pi t}{2}\right)\,.
\]
Let $F=F(\sigma,t)$ be a function in $\sL^2(\Hst)$ with exponential decay with respect to $\sigma$ and let $G\in \sH^{3/2}((-1,1))$ be a function of $t$ with $G(\pm1)=0$. 
Then there exists a unique $\gamma\in\R$ such that the problem
$$\left(\mathcal{N}_{0}-\frac{1}{8}\right)\Phi=F\ \ \mbox{in}\ \ \Hst,\quad \Phi(\sigma,\pm1)=0,\quad  \Phi(0,t)=G(t)+\gamma c_{0}(t)\,,$$
admits a (unique) solution in $\sH^2(\Hst)$ with exponential decay. There holds
\[
   \gamma =-\int_{-\infty}^{0} \int_{-1}^1 F(\sigma,t)\,\sigma c_0(t)\,d\sigma dt -
   \int_{-1}^1 G(t)\,c_0(t)\,dt\,.
\]
\end{lem}

The following two lemmas are consequences of the Fredholm alternative.
\begin{lem}\label{lem-L00}
Let $F=F(s,t)$ be a function in $\sL^2(\Hst)$ with exponential decay with respect to $s$. Then, there exist solution(s) $\Psi$ such that:
$$\left(\mathcal{L}_{0}-\frac{1}{8}\right)\Psi=F\ \ \mbox{in}\ \ \Hst, \quad \Psi(s,\pm1)=0 $$
if and only if $\big\langle F(s,\cdot),c_{0}\big\rangle_{t}=0$ for all $s<0$. In this case, $\Psi(s,t)=\Psi^\perp(s,t) + g(s)c_0(t)$ where $\Psi^\perp$ satisfies $\big\langle\Psi(s,\cdot) ,c_{0}\big\rangle_{t}\equiv0$ and has also an exponential decay.
\end{lem}

\begin{lem}\label{lem-Ai}
Let $n\geq 1$. We recall that $z_{\Ai}(n)$ is the n-th zero of the reverse Airy function, and we denote by 
$$g_{(n)}=\Ai\big((4\pi\sqrt{2})^{-1/3}s+z_{\Ai}(n)\big)$$ the eigenvector of the operator $-\dr_{s}^2-(4\pi\sqrt{2})^{-1}s$ with Dirichlet condition on $\R_-$ associated with the eigenvalue $(4\pi\sqrt{2})^{-2/3}z_{\Ai}(n)$. Let $f=f(s)$ be a function in $\sL^2(\R_-)$ with exponential decay and let $c\in\R$. Then there exists a unique $\beta\in\R$ such that the problem:
$$\left(-\dr_{s}^2-\frac{s}{4\pi\sqrt{2}}-(4\pi\sqrt{2})^{-2/3}z_{\Ai}(n)\right)g=f+\beta g_{(n)}\ \ \mbox{in}\ \ \R_-, \mbox{ with } g(0)=c\,,$$
has a solution in $\sH^2(\R_-)$ with exponential decay.
\end{lem}

Now we can start the construction of the terms of our Ansatz \eqref{5EAn}.

The equations provided by the constant terms are:
$$
  \mathcal{L}_{0}\Psi_{0}=\beta_{0}\Psi_{0}(s,t),\quad   \mathcal{N}_{0}\Phi_{0}=\beta_{0}\Phi_{0}(s,t)
$$
with boundary conditions \eqref{5Dir1}-\eqref{5Dir2} for $j=0$, so that we choose $\beta_{0}=\frac{1}{8}$ and $\Psi_{0}(s,t)=g_{0}(s)c_{0}(t)$. The boundary condition \eqref{5Dir1}  provides: $\Phi_{0}(0,t)=-g_{0}(0)c_{0}(t)$ so that, with Lemma \ref{lem-N0}, we get $g_{0}(0)=0$ and $\Phi_{0}=0$. The function $g_0(s)$ will be determined later. Collecting the terms of order $h^{1/3}$, we are led to:
$$(\mathcal{L}_{0}-\beta_{0})\Psi_{1}=\beta_{1}\Psi_{0}-\mathcal{L}_{1}\Psi_{1}=\beta_{1}\Psi_{0},\quad (\mathcal{N}_{0}-\beta_{0})\Phi_{1}=\beta_{1}\Phi_{0}-\mathcal{N}_{1}\Phi_{1}=0$$
with boundary conditions \eqref{5Dir1}-\eqref{5Dir2} for $j=1$.
Using Lemma \ref{lem-L00}, we find $\beta_{1}=0$, $\Psi_{1}(s,t)=g_{1}(s)c_{0}(t)$,  $g_{1}(0)=0$ and $\Phi_{1}=0$.
Then, we get:
$$(\mathcal{L}_{0}-\beta_{0})\Psi_{2}=\beta_{2}\Psi_{0}-\mathcal{L}_{2}\Psi_{0},\quad(\mathcal{N}_{0}-\beta_{0})\Phi_{2}=0\,,$$
where $\mathcal{L}_{2}=-\dr_{s}^2+\frac{s}{\pi^3\sqrt{2}}\,\dr_{t}^2$ and with boundary conditions \eqref{5Dir1}-\eqref{5Dir2} for $j=2$. Lemma \ref{lem-L00} provides the equation in $s$ variable
\[
   \big\langle (\beta_{2}\Psi_{0}-\mathcal{L}_{2}\Psi_{0}(s,\cdot)),c_{0}\big\rangle_{\sL^2(\dx t)}=0, \quad s<0\,.
\]
Taking the formula $\Psi_0=g_{0}(s)c_{0}(t)$ into account this becomes
\[
   \beta_2 g_0(s) = \left(-\dr_{s}^2-\frac{s}{4\pi\sqrt{2}}\right)g_0(s)\,.
\]
This equation leads to take $\beta_{2}=(4\pi\sqrt{2})^{-2/3}z_{\A}(n)$ and for $g_{0}$ the corresponding eigenfunction $g_{(n)}$. We deduce $(\mathcal{L}_{0}-\beta_{0})\Psi_{2}=0$, then
get $\Psi_{2}(s,t)=g_{2}(s)c_{0}(t)$ with $g_{2}(0)=0$ and $\Phi_{2}=0$.

We find:
$$(\mathcal{L}_{0}-\beta_{0})\Psi_{3}=\beta_{3}\Psi_{0}+\beta_{2}\Psi_{1}-\mathcal{L}_{2}\Psi_{1},\quad(\mathcal{N}_{0}-\beta_{0})\Phi_{3}=0\,,$$
with boundary conditions \eqref{5Dir1}-\eqref{5Dir2} for $j=3$.
The scalar product with $c_{0}$ (Lemma \ref{lem-L00}) and then the scalar product with $g_{0}$ (Lemma \ref{lem-Ai}) provide $\beta_{3}=0$ and $g_{1}=0$. We deduce: $\Psi_{3}(s,t)=g_{3}(s)c_{0}(t)$, and $g_{3}(0)=0$, $\Phi_{3}=0$.
Finally we get the equation:
$$(\mathcal{L}_{0}-\beta_{0})\Psi_{4}=\beta_{4}\Psi_{0}+\beta_{2}\Psi_{2}-\mathcal{L}_{4}\Psi_{0}-\mathcal{L}_{2}\Psi_{2},\quad(\mathcal{N}_{0}-\beta_{0})\Phi_{4}=0\,,$$
where 
$$\mathcal{L}_{4}=\frac{\sqrt{2}}{\pi} \,t\dr_{t}\dr_{s}-\frac{3}{4\pi^4}s^2\dr_{t}^2\,,$$
and with boundary conditions \eqref{5Dir1}-\eqref{5Dir2} for $j=4$.
The scalar product with $c_{0}$ provides an equation for $g_{2}$ and the scalar product with $g_{0}$ determines $\beta_{4}$. By Lemma \ref{lem-L00} this step determines $\Psi_{4}=\Psi_{4}^{\perp}+c_{0}(t)g_{4}(s)$ with a non-zero $\Psi_{4}^{\perp}$ and $g_{4}(0)=0$. Since by construction $\big\langle\Psi_{4}^{\perp}(0,\cdot),\,c_0\big\rangle_{\sL^2(\dx t)}=0$, Lemma \ref{lem-N0} yields a solution $\Phi_{4}$ with exponential decay. Note that it also satisfies $\big\langle\Phi_{4}(\sigma,\cdot),\,c_0\big\rangle_{\sL^2(\dx t)}=0$ for all $\sigma<0$.

We leave the obtention of the other terms as an exercise.

\section{Agmon estimates and projection method}
Let us provide the estimates of Agmon which can be proved.

\begin{prop}\label{Agmon1'}
Let $\Gamma_0>0$. There exist $h_{0}>0$, $C_{0}>0$ and $\eta_{0}>0$ such that for $h\in(0,h_{0})$ and all eigenpair $(\lambda,\psi)$ of $\mathcal{L}_{\Tri}(h)$ satisfying $|\lambda-\frac18|\le\Gamma_0h^{2/3}$, we have:
$$\int_{\Tri}  e^{\eta_{0}h^{-1}|x|^{3/2}}\Big(|\psi|^2+|h^{2/3}\dr_{x}\psi|^2\Big)\,\dx x\dx y
\leq C_{0}\|\psi\|^2\,.$$
\end{prop}

\begin{prop}\label{Agmon2'}
Let $\Gamma_0>0$. There exist $h_{0}>0$, $C_{0}>0$ and $\rho_{0}>0$ such that for $h\in(0,h_{0})$ and all eigenpair $(\lambda,\psi)$ of $\mathcal{L}_{\Tri}(h)$ satisfying $|\lambda-\frac18|\le\Gamma_0h^{2/3}$, we have:
$$\int_{\Tri} (x+\pi\sqrt{2})^{-\rho_{0}/h}\Big(|\psi|^2+|h\,\dr_{x}\psi|^2\Big)\, \dx x \dx y
\leq C_{0}\|\psi\|^2\,.$$
\end{prop}

Let us consider the first $N_{0}$ eigenvalues of $\mathcal{L}_{\Rec}(h)$ (shortly denoted by $\lambda_{n}$). In each corresponding eigenspace, we choose a normalized eigenfunction $\hat\psi_{n}$ so that $\langle\hat\psi_{n},\hat\psi_{m}\rangle=0$ if $n\neq m$.
We introduce:
$$\mathfrak{E}_{N_0}(h)=\spann(\hat\psi_{1},\ldots,\hat\psi_{N_0})\,.$$
Let us define $Q_{\Rec}^0$ the following quadratic form:
$$Q_{\Rec}^0(\hat{\psi})=\int_{\Rec} \left(\frac{1}{2\pi^2}|\dr_{t}\hat{\psi}|^2-\frac{1}{8}|\hat{\psi}|^2\right) (u+\pi\sqrt{2})\, dudt\,,$$
associated with the operator $\mathcal{L}_{\Rec}^0=\Id_{u}\otimes\left(-\frac{1}{2\pi^2}\dr_{t}^2-\frac{1}{8}\right)$ on $\sL^2(\Rec,(u+\pi\sqrt{2})dudt)$.
We consider the projection on the eigenspace associated with the eigenvalue $0$ of $-\frac{1}{2\pi^2}\dr_{t}^2-\frac{1}{8}$:
\begin{equation}\label{Pi0}
   \Pi_{0}\hat{\psi}(u,t) = 
   \big\langle\hat{\psi}(u,\cdot),c_{0}\big\rangle_{t} \,c_{0}(t)\,,
\end{equation}
where we recall that $c_{0}(t)=\cos\left(\frac{\pi}{2}t\right)$. We can now state a first approximation result:
\begin{prop}\label{approx1-Dauge}
There exist $h_{0}>0$ and $C>0$ such that for $h\in(0,h_{0})$ and all ${\hat\psi\in\mathfrak{E}_{N_0}(h)}$:
$$0\leq Q_{\Rec}^0(\hat\psi)\leq Ch^{2/3}\|\hat\psi\|^2$$
and
$$\|(\Id-\Pi_0)\hat\psi\| + \|\dr_{t}(\Id-\Pi_0)\hat\psi\| \leq  Ch^{1/3}\|\hat\psi\|\,.$$
Moreover, $\Pi_{0}$ : $\mathfrak{E}_{N_0}(h)\to\Pi_{0}(\mathfrak{E}_{N_0}(h))$ is an isomorphism.
\end{prop}
%We let $\hat\psi(u,t)=\psi(x,y)$.  
We have already noticed that the quadratic form of the Dirichlet Laplacian on $\Tri$ is bounded from below by the Born-Oppenheimer approximation:
$$\mathcal{Q}_{\Tri,h}(\psi)\geq \int_{\Tri} h^2|\dr_{x}\psi|^2+\frac{\pi^2}{4(u+\pi\sqrt{2})^2}|\psi|^2\dx x\,,$$
so that, by convexity
$$\mathcal{Q}_{\Tri,h}(\psi)\geq \int_{\Tri} h^2|\dr_{x}\psi|^2+\frac{1}{8}\left(1-\frac{2x}{\pi\sqrt{2}}\right)|\psi|^2\dx x\,.$$
It remains to change the variables and replace $\psi$ by $\Pi_{0}\psi$ when $\psi$ is in the span generated by the first eigenfunctions and this is then enough to deduce Theorem \ref{spectrumtriangle}.

\section{Reduction of the broken waveguide to the triangle}
In this section, we prove Theorem \ref{spectrum-guide} (in fact, we restrain our attention to the first two terms). For that purpose, we first state Agmon estimates to show that the first eigenfunctions are essentially living in the triangle $\Tri$ so that we can compare the problem in the whole guide with the triangle.
\begin{prop}
\label{6Agmon}
Let $(\lambda,\psi)$ be an eigenpair of $\mathcal{L}_{\Gui}(h)$ such that $|\lambda-\frac{1}{8}|\leq Ch^{2/3}$. There exist $\alpha>0$, $h_{0}>0$ and $C>0$ such that for all $h\in(0,h_{0})$, we have:
$$\int_{x\geq 0} e^{\alpha h^{-1}x}
\Big(|\psi|^2 + |h\dr_{x}\psi|^2\Big)\, \dx x\dx y\leq C\|\psi\|^2 \,.$$
\end{prop}
\begin{proof}
The proof is left to the reader, the main ingredients being the IMS formula and the fact that $\mathcal{H}_{\BO, \Gui}$ is a lower bound of $\mathcal{L}_{\Gui}(h)$ in the sense of quadratic forms. See also \cite[Proposition 6.1]{DaLafRa11} for a more direct method.
\end{proof}
We can now achieve the proof of Theorem {\rm\ref{spectrum-guide}}. Let $\psi_{n}^h$ be an eigenfunction associated with $\lambda_{\Gui,n}(h)$ and assume that the $\psi_{n}^h$ are orthogonal in $\sL^2(\Omega)$, and thus for the bilinear form $\mathcal{B}_{\Gui,h}$ associated with the operator $\mathcal{L}_{\Gui}(h)$.

We choose $\varepsilon\in(0,\frac13)$ and introduce a smooth cutoff $\chi^{h}$at the scale $h^{1-\eps}$ for positive $x$
\[
   \chi^h(x) =\chi(xh^{\varepsilon-1}) \quad\mbox{with}\quad 
   \chi\equiv1 \ \mbox{ if }\ x\le\tfrac12,\quad
   \chi\equiv0  \ \mbox{ if }\ x\ge1
\] 
and we consider the functions $\chi^{h}\psi_{n}^h$.
We denote:
$$\mathfrak{E}_{N_0}(h)=\spann(\chi^h\psi_{1}^h,\ldots,\chi^h\psi_{N_0}^h)\,.$$
We have:
$$\mathcal{Q}_{\Gui,h}(\psi_{n}^h)=\lambda_{\Gui,n}(h)\|\psi_{n}^h\|^2$$
and deduce by the Agmon estimates of Proposition \ref{6Agmon}:
$$\mathcal{Q}_{\Gui,h}(\chi^{h}\psi_{n}^h)=
\big(\lambda_{\Gui,n}(h)+\mathcal O(h^{\infty})\big)\|\chi^{h}\psi_{n}^h\|^2\,.$$
In the same way, we get the "almost"-orthogonality, for $n\neq m$:
$$\mathcal{B}_{\Gui,h}(\chi^{h}\psi_{n}^h,\chi^{h}\psi_{m}^h)=\mathcal O(h^{\infty})\,.$$
We deduce, for all $v\in\mathfrak{E}_{N_0}(h)$:
$$\mathcal{Q}_{\Gui,h}(v)\leq
\big(\lambda_{\Gui,N_{0}}(h)+\mathcal O(h^{\infty})\big)\|v\|^2\,.$$
We can extend the elements of $\mathfrak{E}_{N_0}(h)$ by zero so that $\mathcal{Q}_{\Gui,h}(v)=\mathcal{Q}_{\Tri_{\eps,h}}(v)$ for ${v\in\mathfrak{E}_{N_0}(h)}$ where $\Tri_{\eps,h}$ is the triangle with vertices $(-\pi\sqrt{2},0)$, $(h^{1-\eps},0)$ and $(h^{1-\eps},h^{1-\eps}+\pi\sqrt{2})$. A dilation reduces us to:
$$\left(1+\frac{h^{1-\eps}}{\pi\sqrt{2}}\right)^{-2}(-h^2\dr_{\tilde{x}}^2-\dr_{\tilde{y}}^2)$$
on the triangle $\Tri$.
The lowest eigenvalues of this new operator admits the lower bounds $\frac{1}{8}+z_{\A}(n)h^{2/3}-Ch^{1-\eps}$ ; in particular, we deduce: 
\[\lambda_{\Gui,N_{0}}(h)\geq \frac{1}{8}+z_{\A}(N_{0})h^{2/3}-Ch^{1-\eps}\,.\]
For the converse inequality, it is sufficient to notice that, by monotonicity of the Dirichlet boundary condition and the min-max principle, we have, for all $n\geq 1$,
\[\lambda_{\Gui,n}(h)\leq\lambda_{\Tri,n}(h)\,,\]
and we apply Theorem \ref{spectrumtriangle}.

\chapter{Non linear dynamics in bidimensional waveguides}\label{NLWG}
\begin{flushright}
\begin{minipage}{0.65\textwidth}
Pour que le caract\`ere d'un \^etre humain d\'evoile des qualit\'es vraiment exceptionnelles, il faut avoir la bonne fortune de pouvoir observer son action pendant de longues ann\'ees.
\begin{flushright} 
\textit{L'homme qui plantait des arbres}, Giono
\end{flushright}
\vspace*{0.5cm}
\end{minipage}
\end{flushright}
 
This chapter is devoted to the proof of Theorem \ref{mainthmH2}.
\section{A priori estimates of the non linearity}

\subsection{Norm equivalences}
Let us first remark that 
$$\mathcal P_{\eps,1}=(1-\eps x_2 \kappa(x_1))^{-1}D_{x_1}-\frac{\eps x_2 \kappa'(x_1)}{2(1-\eps x_2 \kappa(x_1))}\,.$$ 
Hence, by Assumption \ref{assumption1}, there exists three positive constants $C_1$, $C_2$, $C_3$ such that, for all $\eps\in (0,\eps_0)$ and for all $u\in \sH^1_0(\mathcal{S})$,
\begin{equation}
\label{equivnorm2}
\left(1-C_1\eps\right)\|\partial_{x_1}u\|_{\sL^2}\leq \|\mathcal P_{\eps,1}u\|_{\sL^2}+C_2\eps \|u\|_{\sL^2}\leq (1+C_3\eps)\|\partial_{x_1}u\|_{\sL^2}+C_3\eps\|u\|_{\sL^2}\,.
\end{equation}
Furthermore,  the graph norm of $\mathcal{H}_\eps$ is equivalent to the $\sH^2$ norm for all $\eps\in (0,\eps_0)$, with constants depending on $\eps$. More precisely, we have the following result.
\begin{lem}
There exist positive constants $C_4$ and $C_5$ such that, for all $\eps\in (0,\eps_0)$ and for all $u\in \sH^2\cap\sH^1_0(\mathcal{S})$,
\begin{align}
&C_4\left(\left\|D_{x_1}^2u\right\|_{\sL^2}+\frac{1}{\eps^2}\left\|\left(D_{x_2}^2-\mu_1\right)u\right\|_{\sL^2}+\|u\|_{\sL^2}\right)\leq  \label{equivnorm}\\
&\qquad \leq \left\|\left(\mathcal{H}_\eps-\frac{\mu_1}{\eps^2}\right) u\right\|_{\sL^2}+\|u\|_{\sL^2}\leq C_5\left(\left\|D_{x_1}^2u\right\|_{\sL^2}+\frac{1}{\eps^2}\left\|\left(D_{x_2}^2-\mu_1\right)u\right\|_{\sL^2}+\|u\|_{\sL^2}\right)\,. \nonumber
\end{align}
\end{lem}
\begin{proof}
To prove the left inequality in \eqref{equivnorm}, we use standard elliptic estimates. For $u\in \sH^2\cap\sH^1_0(\mathcal{S})$, we let
\begin{equation}
\label{f}
f=\left(\mathcal{H}_{\eps}-\frac{\mu_1}{\eps^2}\right)u=\mathcal{P}_{\eps,1}^2u+\eps^{-2}(D_{x_{2}}^2-\mu_{1})u
\end{equation}
and taking the $\sL^2$ scalar product of $f$ with $D_{x_1}^2 u$, we get
$$\langle D_{x_{1}}\mathcal{P}_{\eps,1}^2u, D_{x_{1}}u\rangle_{\sL^2}+\eps^{-2} \|D_{x_1}\left(D^2_{x_2}-\mu_1\right)^{1/2} u\|_{\sL^2}^2\leq \|f\|_{\sL^2}\|D_{x_{1}}^2u\|_{\sL^2}\,.$$
Then we write
\begin{align*}
\langle D_{x_{1}}\mathcal{P}_{\eps,1}^2u, D_{x_{1}}u\rangle_{\sL^2}
&=\left\|\mathcal{P}_{\eps,1}D_{x_{1}}u\right\|^2_{\sL^2}+\langle [D_{x_{1}},\mathcal{P}_{\eps,1}]u, \mathcal{P}_{\eps,1}D_{x_{1}}u\rangle_{\sL^2}\\
&\quad -\langle\mathcal{P}_{\eps,1}u,  [D_{x_{1}},\mathcal{P}_{\eps,1}] D_{x_{1}}u\rangle_{\sL^2}
\end{align*}
and use
\begin{equation}
\label{commutateur}
\left\|[D_{x_{1}},\mathcal{P}_{\eps,1}]u\right\|_{\sL^2}\leq C\eps \left(\|D_{x_1}u\|_{\sL^2}+ \|u\|_{\sL^2}\right)\,,
\end{equation}
together with \eqref{equivnorm2} and the interpolation estimate $\|D_{x_1}u\|_{\sL^2}\leq C\|D_{x_1}^2u\|_{\sL^2}^{1/2}\|u\|_{\sL^2}^{1/2}$, to get
$$\langle D_{x_{1}}\mathcal{P}_{\eps,1}^2u, D_{x_{1}}u\rangle_{\sL^2}\geq (1-C\eps)\|D_{x_{1}}^2u\|_{\sL^2}^2-C\eps\|u\|_{\sL^2}^2\,.$$
It follows that
$$\|D_{x_{1}}^2u\|_{\sL^2}\leq C\|f\|_{\sL^2}+C\|u\|_{\sL^2}$$
and then, using \eqref{f} and again \eqref{equivnorm2},
\begin{align*}
\eps^{-2}\left\|(D_{x_{2}}^2-\mu_{1})u\right\|_{\sL^2}\leq \|f\|_{\sL^2}+\|\mathcal{P}_{\eps,1}^2u\|_{\sL^2}&\leq \|f\|_{\sL^2}+C\|D_{x_1}^2u\|_{\sL^2}+C\|u\|_{\sL^2}\\
&\leq C\|f\|_{\sL^2}+C\|u\|_{\sL^2}\,.
\end{align*}
This proves the left inequality in \eqref{equivnorm}. The right inequality can be easily obtained by using Minkowski inequality, \eqref{equivnorm2} and \eqref{commutateur}.
\end{proof}

\subsection{A priori estimates}
In this section, we give some results concerning the nonlinear function $W_{\eps}$ defined in \eqref{Weps}.

Let us first  recall a Sobolev inequality due to Br\'ezis and Gallou\"et (see the original paper \cite[Lemma 2]{BG80} and the recent paper \cite{OV15}).
\begin{lem}
For all $v\in \sH^2(\R^2)$, we have,
\begin{equation}
\label{soboBG}
\|v\|_{\sL^\infty(\R^2)}\leq \sqrt{2\pi}\left( \|v\|_{\sH^1(\R^2)}\left(\ln(1+\|v\|_{\sH^2(\R^2)})\right)^{\frac{1}{2}}+1\right)\,.
\end{equation}
\end{lem}
\begin{proof}
We write the classical inequality:
\[\|v\|_{\sL^\infty(\R^2)}\leq \|\hat v\|_{\sL^1(\R^2)}\,,\]
and we notice that, for all $R\geq 0$,
\[\|\hat v\|_{\sL^1(\R^2)}=\int_{|\xi|<R}|\hat v(\xi)|\dx\xi+\int_{|\xi|\geq R}|\hat v(\xi)|\dx\xi\,.\]
We let $<\xi>=\left(1+|\xi|^2\right)^{\frac{1}{2}}$ and we have
\[\int_{|\xi|<R}|\hat v(\xi)|\dx\xi\leq \int_{|\xi|<R}<\xi>^{-1}<\xi>|\hat v(\xi)|\dx\xi\leq  \left(\int_{|\xi|<R}\frac{1}{1+|\xi|^2}\dx\xi\right)^{\frac{1}{2}}\|v\|_{\sH^1(\R^2)}\,,\]
and 
\[\int_{|\xi|<R}\frac{1}{1+|\xi|^2}\dx\xi=\pi\ln(1+R^2)\,.\]
Moreover, we can write
\[\int_{|\xi|\geq R}|\hat v(\xi)|\dx\xi\leq \int_{|\xi|\geq R}<\xi>^{-2}<\xi>^2|\hat v(\xi)|\dx\xi\leq\sqrt{2}\|v\|_{\sH^2(\R^2)}\left(\int_{|\xi|\geq R}\frac{1}{(1+|\xi|^2)^2}\right)^{\frac{1}{2}}\,,\]
and 
\[\int_{|\xi|\geq R}\frac{1}{(1+|\xi|^2)^2}\dx\xi=\pi(1+R^2)^{-1}\,.\]
We deduce that
\[\|v\|_{\sL^\infty(\R^2)}\leq \|v\|_{\sH^1(\R^2)}\sqrt{\pi}\left(\ln(1+R^2)\right)^{\frac{1}{2}}+\|v\|_{\sH^2(\R^2)}\sqrt{2\pi}(1+R^2)^{-\frac{1}{2}}\,,\]
and then we take $R=\|v\|_{\sH^2(\R^2)}$ and use $\ln(1+R^2)\leq 2\ln(1+R)$.
\end{proof}
Then, we can provide some estimates on $W_{\eps}$.
\begin{lem}
\label{lemF}
For all $\eps\in (0, \eps_0)$, the function $W_{\eps}$ is locally Lipschitz continuous on $\sH^2\cap \sH^1_{0}(\mathcal S)$: there exists $C_\eps>0$ such that 
\begin{equation}
\label{lip-Weps}
\forall u_{1},u_{2}\in \sH^2\cap \sH^1_0(\mathcal S),\quad \sup_{t\in \M}\|W_{\eps}(t;u_{1})-W_{\eps}(t;u_{2})\|_{\sH^2}\leq C_\eps (\|u_{1}\|_{\sH^2}^2+\|u_{2}\|_{\sH^2}^2)\|u_{1}-u_{2}\|_{\sH^2}\,.
\end{equation}
Then, for all $M>0$ and for all $\eps\in (0,\eps_0)$, there exists a constant $C_\eps(M)>0$ such that, for all $u\in \sH^2\cap \sH^1_0(\mathcal S)$ with $\|u\|_{\sH^1}\leq M$, one has
\begin{equation}
\label{bg}
\sup_{t\in \R}\|W_{\eps}(t;u)\|_{\sH^2}\leq C_\eps(M)\big(1+\log \left(1+\|u\|_{\sH^2}\right)\big)\|u\|_{\sH^2}\,.
\end{equation}
\end{lem}
\begin{proof}
The group $e^{-i\tau \mathcal{H}_{\eps}}$, defined thanks to the Stone theorem (see Theorem \ref{theo.Stone}),  is unitary on $\sL^2(\mathcal S)$,  $\sH^1_{0}(\mathcal S)$ and $\sH^2(\mathcal S)\cap \sH^1_{0}(\mathcal S)$, if these two last spaces are respectively equipped with the norms $\|(\mathcal H_\eps u)^{1/2}\|_{\sL^2}$ and $\|\mathcal H_\eps u\|_{\sL^2}$, which are equivalent to the $\sH^1$ and $\sH^2$ norms with $\eps$-dependent constants, by \eqref{equivnorm}.

We let $v_{j}=e^{-it\mathcal{H}_{\eps}}u_{j}$ and we estimate
$$\|W_{\eps}(t; u_{1})-W_{\eps}(t; u_{2})\|_{\sH^2}\leq C_\eps \|m_{\eps}^{-1}(|v_{1}|^2v_{1}-|v_{2}|^2v_{2})\|_{\sH^2}\leq C_\eps' \||v_{1}|^2v_{1}-|v_{2}|^2v_{2}\|_{\sH^2}$$
where we have used the unitarity of $e^{-it\mathcal{H}_{\eps}}$ for the graph norm of $\mathcal{H}_{\eps}$. Then, the conclusion follows by using the embeddings $\sH^2(\mathcal S)\hookrightarrow \sL^{\infty}(\mathcal S)$ and $\sH^2(\mathcal{S})\hookrightarrow \sW^{1,4}(\mathcal{S})$. 

Let us now deal with \eqref{bg}. We first recall the Gagliardo-Nirenberg inequality\footnote{It may be proved by an integration by parts.} in dimension two (see \cite[p. 129]{Nir59}):
\begin{equation}\label{GN}
\|v\|_{\sW^{1,4}}^2\lesssim\|v\|_{\sL^\infty}\|v\|_{\sH^2}\,.
\end{equation}
By using continuous extensions from $\sH^2(\mathcal S)$ to $\sH^2(\R^2)$, one obtains the same inequality as in \eqref{soboBG} for $u\in \sH^2\cap \sH^1_0(\mathcal S)$. Hence, for all $v\in \sH^2(\mathcal S)$ with $\|v\|_{\sH^1}\leq M$,
\begin{align*}
\||v|^2v\|_{\sH^2}\lesssim \|v^3\|_{\sL^2}+\|\Delta (v^3)\|_{\sL^2}&\lesssim \|v\|_{\sL^6}^3+\|v^2\Delta v\|_{\sL^2}+\|v |\nabla v|^2\|_{\sL^2}\\
&\lesssim \|v\|_{\sH^1}^3+ \|v\|_{\sL^\infty}^2\|\Delta v\|_{\sL^2}+\|v\|_{\sL^\infty}\|v\|_{\sW^{1,4}}^2\\
&\lesssim C(M)\left(1+\log(1+\|v\|_{\sH^2})\right)\|v\|_{\sH^2}\,,
\end{align*}
where we used the Sobolev embedding $\sH^1(\mathcal S)\hookrightarrow \sL^6(\mathcal S)$, \eqref{GN} and \eqref{soboBG}. Finally, for all $u\in \sH^2\cap \sH^1_0 (\mathcal S)$ with $\|u\|_{\sH^1}\leq M$, setting $v=e^{-it\mathcal{H}_{\eps}}u$ we get $\|v\|_{\sH^1}\leq C_\eps M$ and 
\begin{align*}
\|W_{\eps}(t;u)\|_{\sH^2}\leq C_\eps \||v|^2v\|_{\sH^2}&\leq C_\eps(M)\left(1+\log(1+\|v\|_{\sH^2})\right)\|v\|_{\sH^2}\\
                                                                                             &\leq C_\eps' (M)\left(1+\log(1+\|u\|_{\sH^2})\right)\|u\|_{\sH^2}\,.
\end{align*}
This proves \eqref{bg} and the proof of the lemma is complete.
\end{proof}

\section{Lower bound of the energy and consequences}\label{sec.lbnl}

\subsection{Lower bound}
We will need the following easy lemma.
\begin{lem}\label{Sob-anis}
For all $u\in\sH^1(\M)$, we have
\begin{equation}
\label{sobo1D}\|u\|_{\sL^4}^4\leq 2\|u\|_{\sL^2}^3\|u'\|_{\sL^2}\,.
\end{equation}
For all $u\in\sH^1(\mathcal{S})$, we have
\begin{equation}
\label{sobo2D}
\|u\|_{\sL^4}^4\leq 4\|u\|^2_{\sL^2(\mathcal{S})} \|\partial_{x_{1}}u\|_{\sL^2(\mathcal{S})} \|\partial_{x_{2}}u\|_{\sL^2(\mathcal{S})}\,.
\end{equation}
\end{lem}
\begin{proof}
The proof of \eqref{sobo1D} is a consequence of the standard inequality, for any $f\in\sH^1(\M)$, $\|f\|^2_{\sL^\infty}\leq 2\|f\|_{\sL^2}\|f'\|_{\sL^2}$.
To prove \eqref{sobo2D}, let us recall the following inequality
$$\int_{\mathcal{S}} |f|^2 \dx x_{1} \dx x_{2}\leq \|\dr_{x_{1}}f\|_{\sL^1(\mathcal{S})} \|\dr_{x_{2}}f\|_{\sL^1(\mathcal{S})},\quad\forall f\in\sW^{1,1}(\mathcal{S})\,.$$
Indeed, by density and extension, we may assume that $f\in\mathcal{C}^\infty_{0}(\R^2)$ and we can write
$$f(x_{1}, x_{2})=\int_{-\infty}^{x_{1}} \partial_{x_{1}}f(u,x_{2}) \dx u,\qquad f(x_{1}, x_{2})=\int_{-\infty}^{x_{2}} \partial_{x_{2}}f(x_{1},v) \dx v\,.$$
We get
$$|f(x_{1}, x_{2})|^2\leq \left(\int_{\R} |\partial_{x_{1}}f(u,x_{2})|\dx u\right)\left(\int_{-1}^1 |\partial_{x_{2}}f(x_{1},v)|\dx v\right)$$
and it remains to integrate with respect to $x_{1}$ and $x_{2}$.
We apply this inequality to $f=u^2$, use the Cauchy-Schwarz inequality and \eqref{sobo2D} follows.
\end{proof}
Now, we prove a technical lemma on the energy functional.
\begin{lem}
\label{lemma-energy}
There exists $\eps_2\in (0,\eps_0)$ such that, for all $\eps\in (0,\eps_2)$, the energy functional defined by \eqref{Energy} satisfies the following estimate. For all $M>0$, there exists $C_0>0$ such that, for all $\phy\in \sH^1_0(\mathcal S)$ with $\|\phy\|_{\sL^2}\leq M$, one has
\begin{equation}
\label{minorE}
\mathcal{E}_{\eps}(\phy)\geq \frac{1}{4}\|\partial_{x_1}\phy\|_{\sL^2(\mathcal S)}^2+\left(\frac{3}{8\eps^2}-C_0M^4\right)\|\partial_{x_2}(\mathsf{Id}-\Pi)\phy\|^2_{\sL^2(\mathcal S)}-C_0M^2-C_0M^6\,.
\end{equation}
\end{lem}
\begin{proof}
Remark that
\begin{align*}
\mathcal{E}_{\eps}(\phy)&=\frac{1}{2}\int_{\mathcal{S}}|\mathcal{P}_{\eps,1}\phy|^2 \dx x_{1} \dx x_{2}+\frac{1}{2\eps^2}\left \langle \left(D_{x_2}^2-\mu_1\right)\phy, \phy\right\rangle_{\sL^2}+\frac{1}{2}\int_{\mathcal{S}}V_{\eps}|\phy|^2 \dx x_{1} \dx x_{2}\\
&+\frac{\lambda}{4}\int_{\mathcal{S}}m_{\eps}^{-1}|\phy|^4 \dx x_{1} \dx x_{2}\,.
\end{align*}
Next, recalling that $\Pi_1$ denotes the projection on the first eigenfunction $e_1$ of $D_{x_2}^2$, we easily get
$$\|\phy\|_{\sL^4(\mathcal{S})}^4\leq 8\|\Pi_1 \phy\|^4_{\sL^4(\mathcal{S})}+8\|(\mathsf{Id}-\Pi_1) \phy\|^4_{\sL^4(\mathcal{S})}\,.$$
We may write $\Pi_1\phy(x_{1}, x_{2})=\theta(x_{1}) e_1(x_{2})$ so that, with \eqref{sobo1D},
\begin{align}
\|\Pi_1\phy\|_{\sL^4(\mathcal{S})}^4&=\gamma \int_{\M} \theta(x_{1})^4 \dx x_{1}\leq 2\gamma \|\theta\|_{\sL^2(\M)}^3\|\theta'\|_{\sL^2(\M)}=2\gamma \|\Pi_1\phy\|_{\sL^2(\mathcal{S})}^3\|\partial_{x_{1}}(\Pi_1\phy)\|_{\sL^2(\mathcal{S})}\nonumber\\
&\leq 2\gamma\|\phy\|_{\sL^2(\mathcal{S})}^3\|\partial_{x_{1}}(\Pi_1\phy)\|_{\sL^2(\mathcal{S})}\label{normL4}
\end{align}
where $\gamma=\int_{-1}^1 e_1(x_{2})^4 \dx x_{2}$, and thus, for all $\eta\in(0,1)$,
$$\|\Pi_1\phy\|_{\sL^4(\mathcal{S})}^4\leq \eta \|\Pi_1 \partial_{x_{1}}\phy\|_{\sL^2(\mathcal{S})}^2+\eta^{-1}\gamma^2\|\phy\|_{\sL^2(\mathcal{S})}^6\,.$$
Moreover, thanks to \eqref{sobo2D}, we have, for all $\eta\in(0,1)$,
\begin{align}
\|(\mathsf{Id}-\Pi_1) \phy\|^4_{\sL^4(\mathcal{S})}&\leq 4\|\phy\|^2_{\sL^2(\mathcal{S})} \|\partial_{x_{1}}(\mathsf{Id}-\Pi_1)\phy\|_{\sL^2(\mathcal{S})} \|\partial_{x_{2}}(\mathsf{Id}-\Pi_1)\phy\|_{\sL^2(\mathcal{S})}\nonumber\\
&\leq \eta\|\partial_{x_{1}}(\mathsf{Id}-\Pi_1)\phy\|^2_{\sL^2(\mathcal{S})}+4 \eta^{-1}\|\phy\|^4_{\sL^2(\mathcal{S})}\|\partial_{x_{2}}(\mathsf{Id}-\Pi_1)\phy\|^2_{\sL^2(\mathcal{S})}\,.\label{lun1}
\end{align}
Now we remark that, if $\mu_2=\pi^2$ denotes the second eigenvalue of $D_{x_2}^2$ on $(-1,1)$ with Dirichlet boundary conditions, we have
\begin{equation}\label{spectral-gap-NL}
\left \langle \left(D_{x_2}^2-\mu_1\right)\phy, \phy\right\rangle_{\sL^2(\mathcal S)}\geq \left(1-\frac{\mu_1}{\mu_2}\right)\left\|\partial_{x_2}(\mathsf{Id}-\Pi_1)\phy\right\|_{\sL^2(\mathcal S)}^2=\frac{3}{4}\left\|\partial_{x_2}(\mathsf{Id}-\Pi_1)\phy\right\|_{\sL^2(\mathcal S)}^2\,.
\end{equation}
Therefore, using \eqref{equivnorm2}, \eqref{lun1}, \eqref{spectral-gap-NL}, using that 
\[\|V_\eps\|_{\sL^\infty}\leq C\,,\qquad 0\leq m_\eps^{-1}\leq 1+C\eps\,,\]
we obtain
\begin{align*}
\mathcal{E}_{\eps}(\phy)&\geq \frac{1}{2}(1-C\eps)\|\partial_{x_1}\phy\|_{\sL^2(\mathcal S)}^2-C\|\phy\|_{\sL^2(\mathcal S)}^2+\frac{3}{8\eps^2}\left\|\partial_{x_2}(\mathsf{Id}-\Pi_1)\phy\right\|_{\sL^2(\mathcal S)}^2\\
&\quad -2|\lambda|(1+C\eps)\left(\eta \|\partial_{x_1}\phy\|_{\sL^2(\mathcal S)}^2+4 \eta^{-1}\|\phy\|^4_{\sL^2(\mathcal{S})}\left\|\partial_{x_2}(\mathsf{Id}-\Pi_1)\phy\right\|_{\sL^2(\mathcal S)}^2\right)-C\|\phy\|_{\sL^2(\mathcal S)}^6\\
&\geq \frac{1}{4}\|\partial_{x_1}\phy\|_{\sL^2(\mathcal S)}^2+\left(\frac{3}{8\eps^2}-C\|\phy\|^4_{\sL^2(\mathcal{S})}\right)\left\|\partial_{x_2}(\mathsf{Id}-\Pi_1)\phy\right\|_{\sL^2(\mathcal S)}^2-C\|\phy\|_{\sL^2(\mathcal S)}^2-C\|\phy\|_{\sL^2(\mathcal S)}^6
\end{align*}
where we has chosen $\eta=\frac{1-2C\eps}{8|\lambda|(1+C\eps)}$, which is positive for $\eps$ small enough.
\end{proof}

\begin{proof}
It is easy now to deduce Lemma \ref{cauchy-tensorial} from Lemma \ref{lemma-energy}. Indeed, consider a sequence $\phi_0^\eps$ satisfying Assumption \ref{assumption2} and introduce the constants
\begin{equation}
\label{eps1}
\eps_1(M_0)=\min\left(\eps_2,\left(\frac{3}{16C_0M_0^4}\right)^{1/2}\right)\,.
\end{equation}
We deduce from \eqref{minorE} that, if $\eps\in (0,\eps_1(M_0))$, we have
\begin{align}
\frac{3}{16}\left(\|\partial_{x_1}\phi_0^\eps\|_{\sL^2}^2+\frac{1}{\eps^2}\|\partial_{x_2}(\mathsf{Id}-\Pi_{1})\phi_0^\eps\|^2_{\sL^2}\right)&\leq  \frac{1}{4}\|\partial_{x_1}\phi_0^\eps\|_{\sL^2}^2+\left(\frac{3}{8\eps^2}-C_0M^4\right)\|\partial_{x_2}(\mathsf{Id}-\Pi_{1})\phi_0^\eps\|^2_{\sL^2}\nonumber\\
&\leq \mathcal{E}_{\eps}(\phi_0^\eps)+C_0M_0^2+C_0M_0^6\nonumber\\
&\leq M_1+C_0M_0^2+C_0M_0^6\,.\label{i1}
\end{align}
The conclusion \eqref{conf} stems from \eqref{i1} by remarking also that
\begin{equation*}
\|\partial_{x_2}\Pi_{1} \phi_0^\eps\|_{\sL^2}=\|\langle \phi_0^\eps, e_1\rangle_{\sL^2((-1,1))} \partial_{x_2}e_1\|_{\sL^2}\leq \frac{\pi}{2}\|\phi_0^\eps\|_{\sL^2}\leq \frac{\pi}{2}M_0
\end{equation*}
and by using the Poincar\'e inequality
\begin{equation*}
\|(\mathsf{Id}-\Pi_{1})\phi_0^\eps\|_{\sL^2(\M,\sH^1(-1,1))}\leq \frac{\sqrt{1+\pi^2}}{\pi}\|\partial_{x_2}(\mathsf{Id}-\Pi_{1})\phi_0^\eps\|_{\sL^2}\,.
\end{equation*}
\end{proof}

\subsection{Global existence}\label{sec.ge}
\begin{prop}\label{wp}
Let $\phi_0^\eps\in \sH^2\cap \sH^1_0(\mathcal S)$ and let $\eps\in (0,\eps_0)$. Then, the following properties hold:
\begin{enumerate}[(i)]
\item The problem \eqref{CNLS''} admits a unique maximal solution $\phy^\eps\in \mathcal{C}([0,T_{\rm max}^\eps);\sH^2\cap \sH^1_0(\mathcal S))\cap \mathcal{C}^1([0,T_{\rm max}^\eps);\sL^2(\mathcal S))$, with $T_{\rm max}^\eps\in (0,+\infty]$ that satisfies the following conservation laws
\begin{align}
\|\phy^\eps(t;\cdot)\|_{\sL^2}&=\|\phi^\eps_{0}\|_{\sL^2}\quad \mbox{(mass)},\label{consmass2}\\
\mathcal E_\eps(\phy^\eps(t;\cdot))&=\mathcal E_\eps(\phi^\eps_{0})\quad \mbox{(nonlinear energy)}\,,\label{consenergy2}
\end{align}
where $\mathcal E_\eps$ is defined in \eqref{Energy}.\\
\item There exists a constant $C_1>0$ such that, if $\eps<\eps_2$ (given in Lemma \ref{lemma-energy}) and if $\eps \|\phi_0^\eps\|_{\sL^2}^2\leq C_1$, then $T_{\rm max}^\eps=+\infty$.
\end{enumerate}
\end{prop}
\begin{proof}
\begin{enumerate}[(i)]
\item Let us fix $\eps\in(0,\eps_{0})$ and analyze in a first step the well-posedness in $\sH^2\cap\sH^1_0(\mathcal S)$. For $\phi^\eps_0\in  \sH^2\cap \sH^1_0(\mathcal S)$, we consider the conjugate problem of \eqref{CNLS''} (given in \eqref{CNLS''-conj}) in its Duhamel form
$$\widetilde \phy^\eps(t)=\phi^\eps_{0}-i\int_{0}^t \left(e^{is\mathcal{H}_{\eps}}(V_{\eps}-\eps^{-2}\mu_{1})e^{-is\mathcal{H}_{\eps}}\widetilde \phy^\eps(s)+\lambda W_{\eps}(s ; \widetilde \phy^\eps(s))\right)\dx s=\mathcal{W}_{\eps}(\widetilde \phy^\eps)(t)\,.$$
For $M, T>0$, we consider the complete space
$$\widetilde G_{T, M}=\{\mathcal{C}([0,T] ; \sH^2\cap \sH^1_0(\mathcal S)) :  \forall t\in[0,T],\quad \theta(t)\in \overline{\mathcal{B}}_{\sH^2}(\theta_{0},M)\}\,.$$ 
The application $\mathcal{W}_{\eps}$ is a contraction from $\widetilde G_{T,M}$ to $\widetilde G_{T,M}$ for $T$ small enough. Indeed, thanks to \eqref{lip-Weps}, there exists $C_\eps>0$ such that for all $T>0$, $M>0$, $t\in[0,T]$ and $\phy_{1}, \phy_{2}\in \widetilde G_{T,M}$,
$$\|\mathcal{W}_{\eps}(\phy_{1})(t)-\phy_{0}\|_{\sH^2}\leq C_\eps T+C_\eps TM^3\,,$$
$$\|\mathcal{W}_{\eps}(\phy_{1})(t)-\mathcal{W}_{\eps}(\phy_{2})(t)\|_{\sH^2}\leq (C_\eps T+C_\eps TM^2)\sup_{t\in[0,T]}\|\phy_{1}(t)-\phy_{2}(t)\|_{\sH^2}\,,$$
where we have again used the unitarity of $e^{it\mathcal{H}_{\eps}}$ with respect to the graph norm of $\mathcal{H}_{\eps}$ and the equivalence between the graph norm of $\mathcal{H}_{\eps}$ and the $\sH^2$-norm, for each fixed $\eps$. Therefore the Banach fixed point theorem insures the existence and uniqueness of a local in time solution and thus of \eqref{CNLS''} for each given $\eps\in(0,\eps_{0})$. In fact, it is not difficult to deduce the existence of a maximal existence time $T_{\rm max, \sH^2}^\eps\in (0,+\infty]$ such that $\phy^\eps\in C([0,T_{\rm max, \sH^2}^\eps);\sH^2\cap \sH^1_0(\mathcal S))\cap C^1([0, T_{\rm max, \sH^2}^\eps);\sL^2(\mathcal S))$ and such that we have the alternative
\begin{equation}
\label{alternative2}
T_{\rm max, \sH^2}^\eps=+\infty \quad \mbox{or}\quad \lim_{t\to T_{\rm max, \sH^2}^\eps}\|\phy^\eps(t)\|_{\sH^2}=+\infty\,.
\end{equation}
The conservation of the $\sL^2$-norm  is obtained by considering the scalar product of the equation with $\phy^\eps$ and then taking the imaginary part. For the conservation of the energy, we consider the scalar product of the equation with $\partial_{t}\phy^\eps$ and take the real part.
\item Thanks to the energy conservation and Assumption \eqref{ass3} and by using a Sobolev embedding, we can bound uniformly w.r.t. $\eps$ the initial energy. Then, we deduce from Lemma \ref{minorE} that $\phy^\eps(t;\cdot)$ is uniformly bounded in $\sH^1$. 

From \eqref{CNLS''} and \eqref{bg} we get
$$\|\dr_{t}\phy^\eps\|_{\sH^2}\leq C_\eps\big(1+\log\left(1+\|\phy^\eps(t;\cdot)\|_{\sH^2}\right)\big)\|\phy^\eps(t;\cdot)\|_{\sH^2}\,.$$
It remains to use an argument \textit{\`a la} Gronwall. Given a Banach space $G$, let us consider a function $\phy\in\mathcal{C}^1([0,T^*),G)$ such that for, $t\in[0,T^*)$,
$$\|\phy'(t)\|\leq C(1+\log(1+\|\phy(t)\|))\|\phy(t)\|\,.$$
We easily get
$$\|\phy(t)\|\leq F(t),\qquad \mbox{ with }\quad F(t)=\|\phy_{0}\|+C\int_{0}^t (1+\log(1+\|\phy(\tau)\|))\|\phy(\tau)\|\dx\tau$$
and
$$\frac{\dx}{\dx t}F(t)=C(1+\log(1+\|\phy(t)\|))\|\phy(t)\|\leq C(1+\log(1+F(t)))F(t)\,,$$
so that
$$\frac{\dx}{\dx t}\log\left(1+\log(1+F(t))\right)\leq C\,.$$
Consequently, we find an estimate of the form
$$\|\phy(t)\|\leq F(t)\leq e^{ae^{b t}}\,.$$
Applying this inequality to $\phy^\eps$ with $G=\sH^2(\mathcal S)$, one gets a bound for the $\sH^2$ norm of $\phy^\eps$ on the interval $[0,T_{\max,\sH^2}^\eps)$, which is a contradiction.
\end{enumerate}
\end{proof}

The conservation of the energy and Lemma \ref{lemma-energy} imply Theorem \ref{mainthmH2}.

\bibliographystyle{mnachrn}

\bibliography{BIB}

\end{document}